**The Light Dark Matter eXperiment**

Torsten Akesson[1] Layan Alsaraya[*,2] Stephen Appert[3] Charles Bell[4] Elizabeth Berzin[2]
Nikita Blinov[5] Léo Borrel[3] Cameron Bravo[6] Liam Brennan[7] Lene Kristian Bryngemark[1]
Pierfrancesco Butti[6] Riccardo Catena[†,1] Anthony Chavez[7] Owen Colegrove[7] Giulia Collura[7]
Patill Daghlian[3] Filippo Delzanno[7] E. Craig Dukes[8] Valentina Dutta[9] Bertrand Echenard[3]
Ralf Ehrlich[8] Thomas Eichlersmith[10] Jonathan Eisch[5] Einar Elén[1] Eric Fernandez[8]
Erich Frahm[10] Cooper Froemming[10] Andrew P. Furmanski[10] Gréta Gajdán[1] Majd Ghrear[2]
Niramay Gogate[11] Victor Gomez[3] Matt Graham[6] Taylor Gray[†,1] Josh Greaves[8]
Chiara Grieco[7] Craig Group[8] Peter György[1] Carsten Hast[6] Axel Helgstrand[1] Hannah Herde[1]
Christian Herwig[4] David G. Hitlin[3] Tyler Horoho[8] Cameron Huse[7] Joseph R. Incandela[7]
David Jang[7] Nathan Jay[3] Asahi Jige[7] Joseph Kaminski[5] Wesley Ketchum[5]
Matthew Kilpatrick[7] Kai Kristiansen[7] Gordan Krnjaic[‡,5] Susanne Kyre[7] Travis Lange[6]
Juan Lazaro[7] Amina Li[7] Shirley Li[5] Shujin Li[9] Yuze Li[7] Ziqi Lin[7] Hongyin Liu[7] Megan Loh[2]
Lisa Andersson Loman[1] Zihan Ma[7] Jeremiah Mans[10] Sanjit Masanam[7] Phillip Masterson[7]
Steven Metallo[10] Sophie Middleton[3] Noah Moran[3] Omar Moreno[6] Geoffrey Mullier[1]
Joseph Muse[10] Akshay Nagar[7] Timothy Nelson[6] Gavin Niendorf[7] Rory O'Dwyer[2]
William Ortez[7] Lennart Österman[1] Leo Östman[1] James Oyang[3] Pritam Palit[9]
Dhruvanshu Parmar[11] Jessica Pascadlo[8] Emrys Peets[6] Arnaud Pele[3] Ruth Pöttgen[1]
Huilin Qu[7] Melissa Quinnan[7] Jaida Raffelsberger[1] Benjamin Reese[6] Chelsea Rodriguez[10]
Eduardo Sanchez[3] Luis G. Sarmiento[1] Philip Schuster[6] Chris Sellgren[7] Harrison Siegel[7]
Pradyun Solai[9] Matthew Solt[8] Francisca Stedman[9] Cristina Mantilla Suarez[8]
Lauren Tompkins[2] Natalia Toro[6] Nhan Tran[5] Jason Trevor[3] Tamas Almos Vami[7]
Kieran Wall[8] Erik Wallin[1] Zhengyu Wan[7] Yuxuan Wang[7] Andrew Whitbeck[5]
Duncan Wilmot[7] Xinyi Xu[7] Jihoon Yoo[7] Danyi Zhang[7] Guanglei Zhao[§,3] Weilong Zhou[3]

[1]*Lund University, Department of Physics, Box 118, 221 00 Lund, Sweden*
[2]*Stanford University, Menlo Park, CA 94025, USA*
[3]*California Institute of Technology, Pasadena, CA 91125, USA*
[4]*University of Michigan, Ann Arbor, MI 48109, USA*
[5]*Fermi National Accelerator Laboratory, Batavia, IL 60510, USA*
[6]*SLAC National Accelerator Laboratory, Menlo Park, CA 94025, USA*
[7]*University of California Santa Barbara, Santa Barbara, CA 93106, USA*
[8]*University of Virginia, Charlottesville, VA 22904, USA*
[9]*Carnegie Mellon University, Pittsburgh, PA 15213, USA*
[10]*University of Minnesota, Minneapolis, MN 55455, USA*
[11]*Texas Tech University ¶, Lubbock, TX 79409, USA*

[*]Affiliated with Stanford through San Francisco State University
[†]Affiliated with Lund through Chalmers Technical University
[‡]Also affiliated with University of Chicago and the Kavli Institute for Cosmological Physics
[§]Affiliated with Caltech through Reed College
¶Texas Tech University is no longer a member of the LDMX Collaboration

**Abstract**

The Light Dark Matter eXperiment (LDMX) is an electron fixed-target experiment optimized to search for sub-GeV dark matter production through the missing momentum signature. LDMX is designed to operate in End Station A at SLAC, using an 8 GeV electron beam accelerated alongside the LCLS-II drive beam. The design of the apparatus is strongly motivated by the performance requirements of a high-rate missing momentum search and leverages detector technologies and designs from other experiments along with existing facilities at SLAC. LDMX will improve on previous results by up to three orders of magnitude, enabling broad sensitivity to dark sector scenarios including the dark matter interaction strengths motivated by freeze-out of MeV–GeV mass dark matter to the observed relic abundance. With hermetic forward coverage, LDMX also has sensitivity to visible signatures of dark sectors and provides a unique probe of electron-nuclear interactions important to interpreting data from accelerator-based neutrino experiments. This report encompasses the technical design of the LDMX Detector, its simulated performance, and the physics capabilities of the experiment.

# Contents

# Chapter 1

# Introduction

The search for the fundamental nature of Dark Matter is organized around a few paradigms that are able to explain its origin in the early universe and naturally accommodate its production in the abundance implied by astronomical and cosmological observations. In the range of masses where dark matter would behave like a particle – having localizable interactions – the paradigm is thermal relic dark matter, where the observed dark matter density arose from thermal interactions with ordinary matter. The most familiar example of thermal dark matter is the WIMP, where an additional particle with weak-scale mass and only weak nuclear interactions is generated via thermal equilibrium in the hot early universe, and has its relic density set by the mechanism of thermal freeze-out as the universe cools. There are many attractive features of thermal freeze-out dark matter. It is simple and compact, requiring little additional structure to the Standard Model. It is natural, requiring only the addition of features similar to known particles and interactions. It is predictive, making clear projections regarding how and where to search for these particles.

Thermal freeze-out dark matter is also more *generic* than the specific example of the WIMP. Although the conjecture of a WIMP, with perhaps only a single new particle interacting through a known force, is the most compact realization of thermal relic dark matter, it can only explain the relic abundance if the mass of the dark matter particle is heavier than roughly the proton mass. Otherwise, freeze-out would produce more dark matter than is observed. However, if dark matter interacts with regular matter through a new force, dark matter particles can be much lighter, down to $\mathcal{O}(10)$ keV. Such particles would evade existing searches for WIMPs in direct detection experiments and high-energy colliders.

In recognition of the importance of searching for sub-GeV thermal relic dark matter, the DOE-commissioned report *Basic Research Needs for Dark-Matter Small Projects New Initiatives* identified this as a focus of two different Priority Research Directions (PRDs), where PRD 1 recognizes the unique capabilities of fixed-target accelerator experiments. In particular, it highlights the ability of a missing-momentum search using multi-GeV electron beams on a fixed target to detect the production of dark matter through the kinematics of the production reaction, where such experiments have generic sensitivity to the interaction strengths allowed by the observed relic abundance for freeze-out dark matter with MeV-GeV masses. In this experiment, an incoming electron radiates dark matter, losing most of its energy in the process, resulting in an apparent momentum imbalance between the initial and final states due to the invisibility of dark matter to the apparatus. The Light Dark Matter eXperiment (LDMX) is a realization of this concept designed to search for dark matter with up to $10^{16}$ incoming electrons with little or no background.

LDMX uses detector technologies and designs from other experiments (CMS, Mu2e, HPS) and beam from an existing accelerator facility (LCLS-II) to minimize the effort required to execute a missing momentum search for sub-GeV dark matter on a short timescale and modest budget. The apparatus has four principal detector systems; charged particle tracking, electromagnetic calorimetry, hadronic calorimetry, and a trigger hodoscope. These detector systems are tied together by a trigger and data acquisition system, computing and software, and beamline hardware that interfaces the detector to the accelerator infrastructure.

Electrons for the experiment are provided by the Linac to End Station A (LESA) facility at SLAC, which sends bunches from the LCLS-II drive beam that cannot be used by the photon science program to End Station A. This facility, which is ideal for LDMX, leverages major elements of laboratory infrastructure to enable the experiment. Similarly, the detector systems leverage large DOE investments in technology and instrumentation made over the last few decades. The tracking systems on either side of the target measure momentum of the incoming and recoiling electron, and help identify background events with more particles in the final state. These trackers use designs from the Heavy Photon Search Silicon Vertex Tracker (HPS SVT) based on technologies developed for the LHC experiments. The Electromagnetic Calorimeter (ECal) measures the energy of the final state, including the recoiling electron as well as radiated photons, and identifies rare events where much of that energy is carried away by particles that escape the detector. The ECal is a Si-W calorimeter using technology and designs developed for the CMS HGC, the Phase II upgrade of the CMS endcap calorimeter. The Hadronic Calorimeter (HCal) helps identify background events with penetrating hadrons and muons in the final state, where sensitivity to neutral hadrons is especially important. The HCal uses designs and technology from the Cosmic Ray Veto of the Mu2e experiment coupled with the same readout as the ECal. Finally, the Trigger Scintillator (TS) hodoscope counts the particles in each incoming bunch to set an appropriate threshold for the missing energy trigger. This information is used by the Trigger and DAQ (TDAQ) to select and record events with anomalously low energy in ECal given the number of incoming electrons for readout and storage. The TS uses standard plastic scintillator and readout similar to that for the HCal, the TDAQ uses hardware developed at FNAL for CMS and at SLAC for range of smaller experiments, and computing infrastructure for storage and data processing is provided by the S3DF facility at SLAC.

# Chapter 2

# LDMX Overview

## 2.1 Background

On April 17, 2019, the DOE Office of Science issued a Funding Opportunity Announcement (FOA) calling for proposals to develop the design and execution plans for small projects to carry out dark matter particle searches, making use of DOE laboratory infrastructure and/or technology capabilities. These small projects were to support the dark matter science Priority Research Directions (PRDs) defined in the BRN workshop on Dark Matter New Initiatives (DMNI) [1]. The Light Dark Matter eXperiment (LDMX) was approved as one of these development projects.

The scientific justification for LDMX is based on the priorities laid out in the BRN report [1], which identified three Priority Research Directions (PRD). The first is:

> **PRD 1: Create and detect dark matter particles below the proton mass and associated forces, leveraging DOE accelerators that produce beams of energetic particles**.... Interactions of energetic particles recreate the conditions of dark matter production in the early universe. Small experiments using established technology can detect dark matter production with sufficient sensitivity to test compelling explanations for the origin of dark matter and explore the nature of its interactions with ordinary matter.

In discussing this PRD, the report highlights a strong motivation for "10- to 1000-fold improvements in sensitivity over current searches" for dark matter (DM) production (Thrust 1) and discusses the unique capability of the missing-momentum technique to meet or exceed this goal over most of the MeV-GeV mass range. The report also notes a second emphasis on "explor[ing] the structure of the dark sector by producing and detecting unstable dark particles" (Thrust 2).

LDMX is a small experiment that realizes this missing-momentum concept. As shown in Fig. 2.1, dark matter is produced in electron fixed-target collisions and detected through the use of tracking and calorimetry to identify events where an incoming electron loses most of its energy to DM production. Operating at a high rate in a continuous-wave (CW) electron beam for only a few years, this approach can fully address Thrust 1 by achieving a 1000-fold improvement in sensitivity to dark matter production, while also broadly searching for unstable dark particles that are the objective of Thrust 2. In addition to the primary dark matter motivation, LDMX data can also provide measurements of electron-nucleon interactions at large momentum transfers that are of critical importance to interpreting the data from the flagship neutrino program at Fermilab.

The LDMX experiment can achieve new sensitivity for sub-GeV dark matter with fewer than $10^{12}$ electrons on target (EoT), requiring only weeks of operation. The full potential of the experiment to cover the thermal targets of all possible DM candidates not excluded by the CMB bounds on DM annihilation [2] is reached with effective event yields as high as $10^{16}$ EoT on a thin target[1]. Ensuring the needed high purity for the missing momentum signature requires the ability to correctly associate all of the particles belonging to each individual event, so that each incoming electron is correctly tagged to its interaction products. For large event yields, this requires a near-CW beam and granular detectors with high-rate capability and ns time resolution. Furthermore, because the experiment is an active beam dump, some elements of the detector must be relatively radiation tolerant.

[1] This can also be achieved with more modest yields with a thicker target

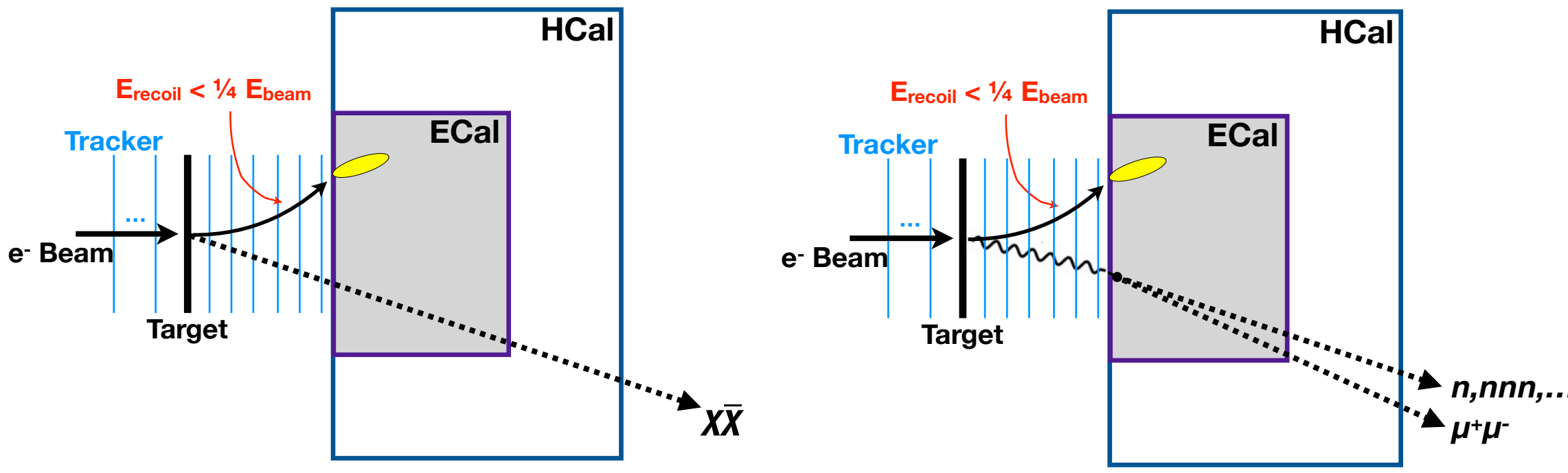

Figure 2.1: The conceptual layout of the LDMX apparatus demonstrating the missing-momentum technique of searching for dark matter in fixed target production (left) and key backgrounds that the detector is designed to reject (right). In signal events, an incoming beam electron loses most of its energy and experiences a hard kick in transverse momentum in the production process, producing a soft, high-angle recoil and no other detectable interaction products in strong contrast to the Standard Model scattering processes. However, in cases that are rare but relevant to a high-statistics experiment, a hard bremsstrahlung photon is produced that undergoes muon conversion or photo-nuclear reactions in the detector material that are difficult to detect. These backgrounds define the required veto performance of the detector.

Most of the technologies required to meet these challenges are well-established, and none of them are beyond the current state of the art. High repetition rate electron beams are available within the DOE complex at both SLAC and JLab, as well as at other labs worldwide. Charged particle tracking systems with the granularity and rate capability required for the experiment have existed for at least 20 years. The technologies needed for electromagnetic and hadronic calorimetry, triggering, and data acquisition are also mature or have recently been realized for other experiments such as Mu2e and upgrades for the LHC experiments.

As a result, little fundamental development is required. The task is to adapt technologies, designs, and hardware that already exists in a way that optimizes the performance while minimizing the technical risk, cost, and effort involved in mounting the experiment. In particular, LDMX re-purposes designs from the HPS Silicon Vertex Tracker for tracking, the CMS upgrade HGC for the ECal, and the Mu2e Cosmic Ray veto for the HCal, and plans to use the LCLS-II drive beam at SLAC in a parasitic mode that is invisible to the primary photon science program.

The rest of this section describes the physics objectives of the experiment in more detail, and places the experiment in the context of other experiments searching for sub-GeV dark matter. It briefly introduces the LDMX physics potential beyond sub-GeV dark matter and lays out the scope necessary to meet the objectives of the experiment. It ends with a summary of the project organization, management, and the cost and schedule for construction.

## 2.2 Physics Objectives

A compelling explanation for the origin and abundance of dark matter is the idea that it arose as a thermal relic from the hot early Universe. This paradigm is viable over the MeV to TeV mass range and requires a small non-gravitational interaction between dark and familiar matter. Any such interaction implies a DM production mechanism in accelerator-based experiments, and in most sub-GeV realizations, couplings of the DM to electrons are key to the thermal DM origin. Consequently, measurements sensitive to these couplings are a priority in the global DM program. LDMX's missing momentum measurement is designed to directly explore this coupling, and at the same time greatly expands the sensitivity to hadron-DM couplings via missing-energy signals of invisible decays of Standard Model mesons into dark matter [3].

Scalar, Majorana, or Pseudo-Dirac particle DM can be thermally produced through contact interactions with Standard Model leptons $f$ (for example, $\frac{1}{\Lambda^2}\bar{\chi}\sigma^\mu\chi\bar{f}\sigma_\mu f$ for the Majorana fermion $\chi$). All three scenarios are consistent with CMB bounds on DM annihilation [2]; the fermion models in particular are compatible with a small DM mass (i.e., technically natural) and are poorly constrained by existing terrestrial experiments. Thermal freeze-out predicts the interaction scale $\Lambda$ for a given DM mass $m_\chi$, or equivalently $y \approx 0.9 m_\chi^4/\Lambda^4$ shown in Fig. 2.2. These predicted couplings define an important sensitivity milestone [4, 5, 1]. Most of their parameter space falls within a factor of approximately 10 to 1000 of the interaction strengths that have been explored previously by both beam-dump experiments [6, 7] and fixed-target missing-energy searches [8] Accelerator experiments are often capable of resolving the mediator particle responsible for the interaction, in addition to the dark matter, making the contact operator description incomplete. For this reason, we quote results (in Figures 2.2 - 2.5) using the canonical model where the interactions arise from a dark photon of mass $m_{A'}$, coupling $\epsilon e$ to SM charged matter, and coupling $g_D = \sqrt{4\pi\alpha_D}$ to dark matter. In this case, the parameter predicted by thermal freeze-out abundance is typically $y = \epsilon^2\alpha_D(m_{DM}/m_{A'})^4$, and this is commonly used to quote experimental sensitivity. Fig. 2.3 illustrates the modest dependence of LDMX's sensitivity on the dark photon mass for fixed DM mass — a theme more thoroughly explored in [9].

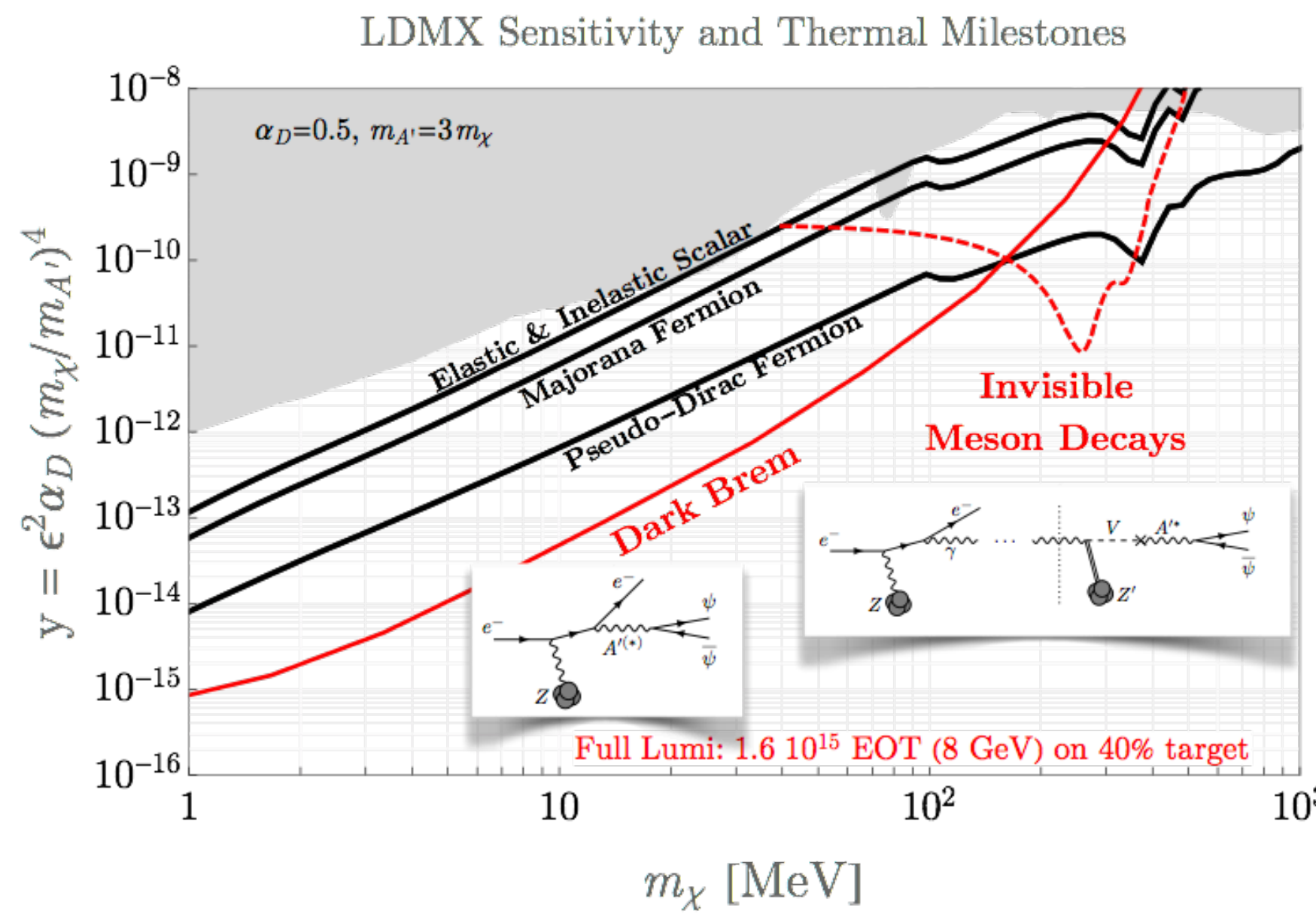

Figure 2.2: Thermal dark matter milestones (black curves), existing constraints (gray), and LDMX projected reach, in the conventions of [5]. LDMX improves over previous DM production searches by three orders of magnitude or more, which is required to robustly explore the thermal freeze-out scenarios highlighted by PRD 1. The insets show two DM production reaction observables at LDMX: dark bremsstrahlung (left inset, with sensitivity indicated by the solid red curve) and exclusive standard model meson photo-production followed by rare decay to DM (right inset, with sensitivity indicated by the dashed red curve).

Figures 2.2 - 2.4 illustrate the power of LDMX to explore the commonly discussed thermal freeze-out scenarios, including challenging milestones such as Pseudo-Dirac dark matter [4] over a wide range of mediator masses, and thermal freeze-out with a mediator that is near-resonance [10] or below the DM pair threshold [4, 12]. Furthermore, by exploring deep in the coupling parameter space, LDMX probes models such as secluded annihilation [13] of light DM into scalar mediators, motivated parameter space for DM-electron couplings through other types of mediators [12], and SIMP and ELDER models [14, 15, 16] where DM interactions with ordinary matter maintain kinetic equilibrium while DM self-interactions deplete its abundance. The exclusive production of SM vector mesons, whose subsequent decay to DM leads to a missing energy/momentum signal, was highlighted in Ref. [3] as a highly sensitive probe of *hadronically* coupled DM achievable at LDMX as illustrated by the red dashed line in Fig. 2.2.

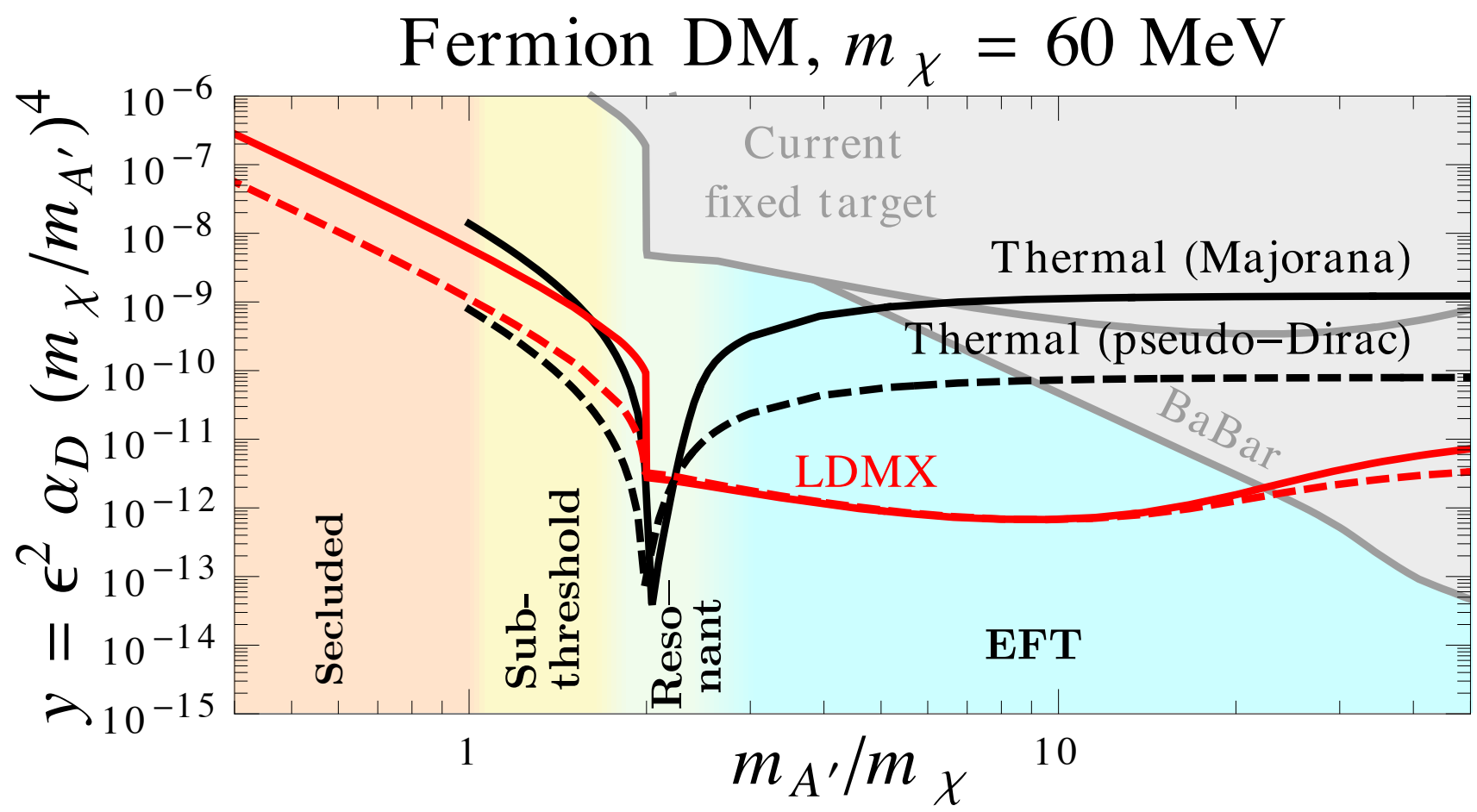

Figure 2.3: Milestones, LDMX reach (dark bremsstrahlung only), and constraints (as of 2019) for a fixed DM mass of 60 MeV, as a function of the mediator-to-DM mass ratio, for Majorana (solid) and pseudo-Dirac (dashed) DM. Colored bands indicate different qualitative domains of DM annihilation (see also [9] for a discussion).

## 2.3 Relation to Other Experiments

In this section, we give an overview of searches for sub-GeV DM, and describe LDMX's place in the field.

### 2.3.1 Accelerator-Based Experiments

The expected sensitivity of LDMX, compared with other accelerator-based experiments (completed, ongoing, and proposed), is illustrated in Fig. 2.5. At low masses, LDMX is uniquely capable of 1000-fold improvements in sensitivity, with sensitivity unrivaled by other experiments. The most relevant accelerator searches to compare with LDMX include:

- *Collider missing-mass searches (Belle-II)* are most relevant to thermal DM above $\sim 100$ MeV produced through an on-shell mediator, complementing LDMX's sensitivity to lower-mass DM and production through off-shell mediators. A Belle II 20 fb$^{-1}$ study [17] projects a factor of 10 sensitivity improvement over BABAR's [18] missing mass search, constraining $y \gtrsim 10^{-9}$. Neglecting background and systematic uncertainties, which dominate in the sub-GeV mass range and are known to be the limiting factor, studies suggest up to 10-100 further improvement might be possible in the upper end of the GeV mass range with the full Belle II dataset by the end of the 2020's.
- *Beam dump based searches (e.g. CCM [19], MiniBooNE-DM[7], and the COHERENT detectors at ORNL [20])* are limited by signal rate – their sensitivity scales with 4 powers of the small interaction coupling, compared to only 2 powers for missing mass/energy/momentum searches. Therefore, even with substantially increased current or geometric/kinematic acceptance, state-of-the-art proposals [21, 22, 23, 24, 25, 26, 27, 5, 20] achieve only 10-fold sensitivity increases (for clarity, only CCM and the COHERENT detectors are shown in Fig. 2.5; others are comparable). These generally test baryonic couplings, while thermal freeze-out of DM below $\sim 100$ MeV relies on electron couplings; these couplings are of similar strength in the hidden-photon model assumed in the comparison.
- *Missing energy searches (NA64)* also compete with DMNI projects such as LDMX and CCM. The only such experiment is CERN's NA64 [28, 8, 29], situated in the H4 beamline [30], which delivers a secondary 100 GeV electron beam to the front of NA64's calorimeters. NA64's most recent published results with an electron beam use a sample of $9.37 \cdot 10^{11}$ EoT [29], with a $0.51 \pm 0.13$ event background estimate. Long-term NA64 projections reflect a goal of integrating $3 \cdot 10^{12}$ EoT before LS3 (mid-2026) and around $10^{13}$ EoT by the end of LHC run 4 [31], close to their setup's irreducible neutrino-background floor [32]; the projection for $3 \cdot 10^{12}$, shown as a thin dotted line in Fig. 2.5(left), relies on a 10-fold improvement in background rejection over [8]. In addition to electron data, the NA64 experiment is also collecting data with a 160 GeV muon beam [31, 33, 34]. A muon-beam configuration offers similar sensitivity for dark-photon models, and is interesting for testing explanations of the $(g-2)_\mu$ anomaly, but does not directly probe the electron-DM coupling relevant to thermal dark matter below O(100) MeV. Nonetheless, we show a zero background sensitivity projection (in the dark photon mediator model) for NA64 with the assumption of $2 \times 10^{13}$ muons on target, corresponding to their goals until the end of run 4. If these goals are realized before completion of LDMX, NA64 could probe much of the scalar and Majorana benchmark models outside of the resonance region. LDMX would still explore new, well-motivated parameter space, including the resonance regions of these models and the pseudo-Dirac benchmark. *Moreover, a few-week LDMX pilot run could compete with NA64's most aggressive sensitivity projections, motivating accelerated preparations.*

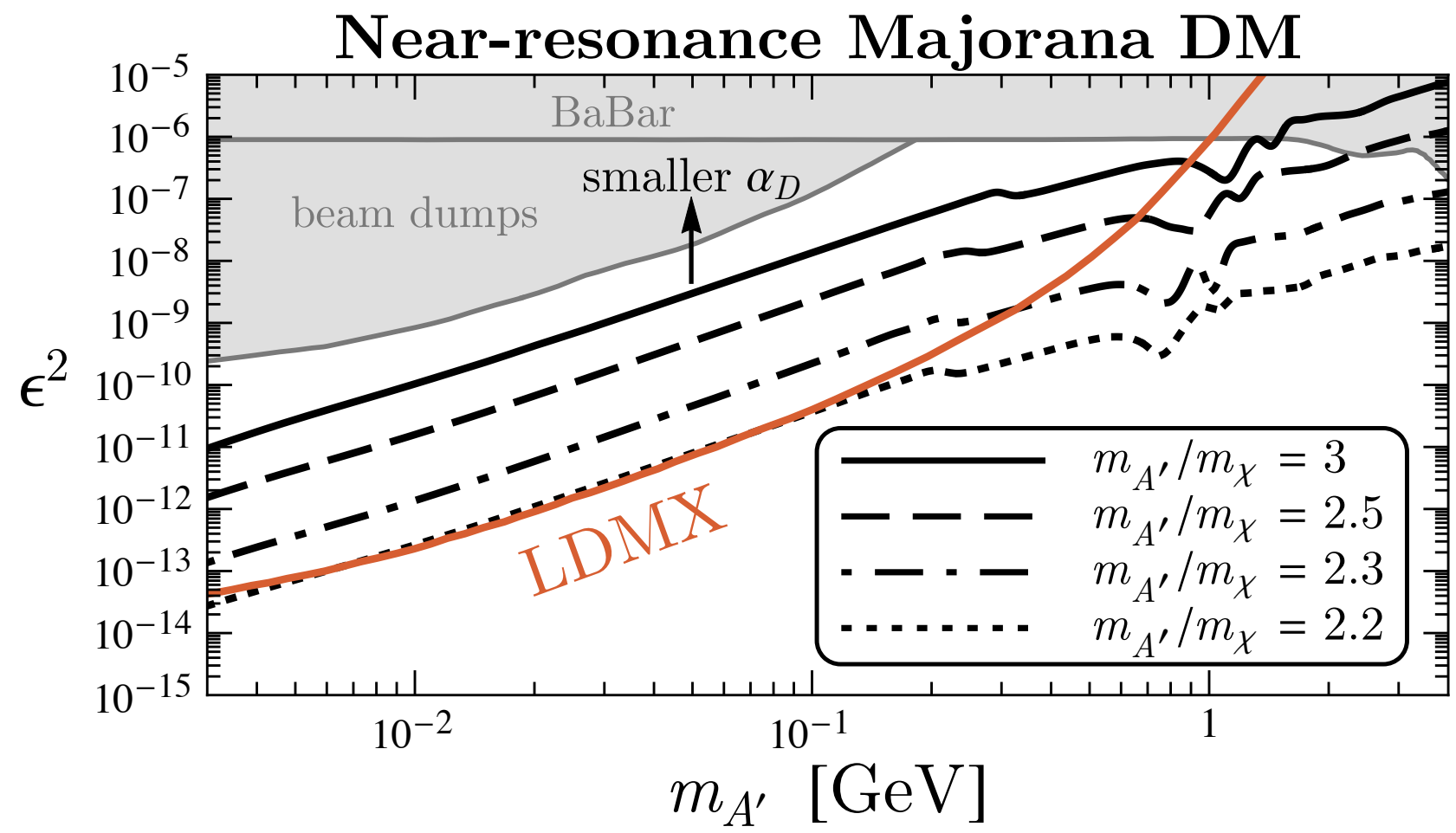

Figure 2.4: Milestones in the resonant region, plotted as in [10] for up to 10% mass tunings, LDMX dark-bremsstrahlung sensitivity and constraints (as of 2019). Figure adapted from [11].

### 2.3.2 Non-Accelerator-Based Experiments

Advances during the past years in low-threshold direct detection in both semiconductor and noble-liquid detectors [44, 45, 46, 47] offer a complementary window on sub-GeV DM. Broad comparisons are difficult to make because accelerators probe interactions of semi-relativistic dark matter while direct detection involves scattering with much lower momentum transfer. In general, thermal DM predictions for accelerators have mild dependence on DM spin because both early-universe thermal DM production and accelerator production probe similar momentum scales. By contrast, direct detection probes very different kinematics and so thermal DM predictions for scattering cross-sections span 20 orders of magnitude (with only the best-case scenario appearing on the scale of most projections). The full range of predictions is shown in Fig. 2.5(right), along with model-dependent mappings of LDMX's sensitivity to the direct detection parameter space. For specific benchmark models discussed in the BRN Report [1]:

- (elastic scalar dark matter benchmark) is comparably accessible to DMNI sub-GeV direct detection and LDMX, because the dark matter scattering is velocity-independent in such models.
- (Majorana fermion dark matter benchmark) is accessible to LDMX, but direct detection cross-section is suppressed by CM-frame $v^2$ (a $10^{-6} - 10^{-10}$ cross-section suppression).
- (inelastic scalar or fermion, aka pseudo-Dirac, dark matter benchmarks) are accessible to LDMX, but the leading direct-detection reaction is a one-loop diagram rather than tree-level (suppressing the cross-section by $\sim 10^{-15} - 10^{-20}$).

Thoroughly exploring thermal freeze-out for all DM spins is important, and requires a powerful accelerator-based search such as LDMX. Indeed, fermionic models may even be theoretically favored over the scalar benchmark since they are (at the effective operator level) technically natural.

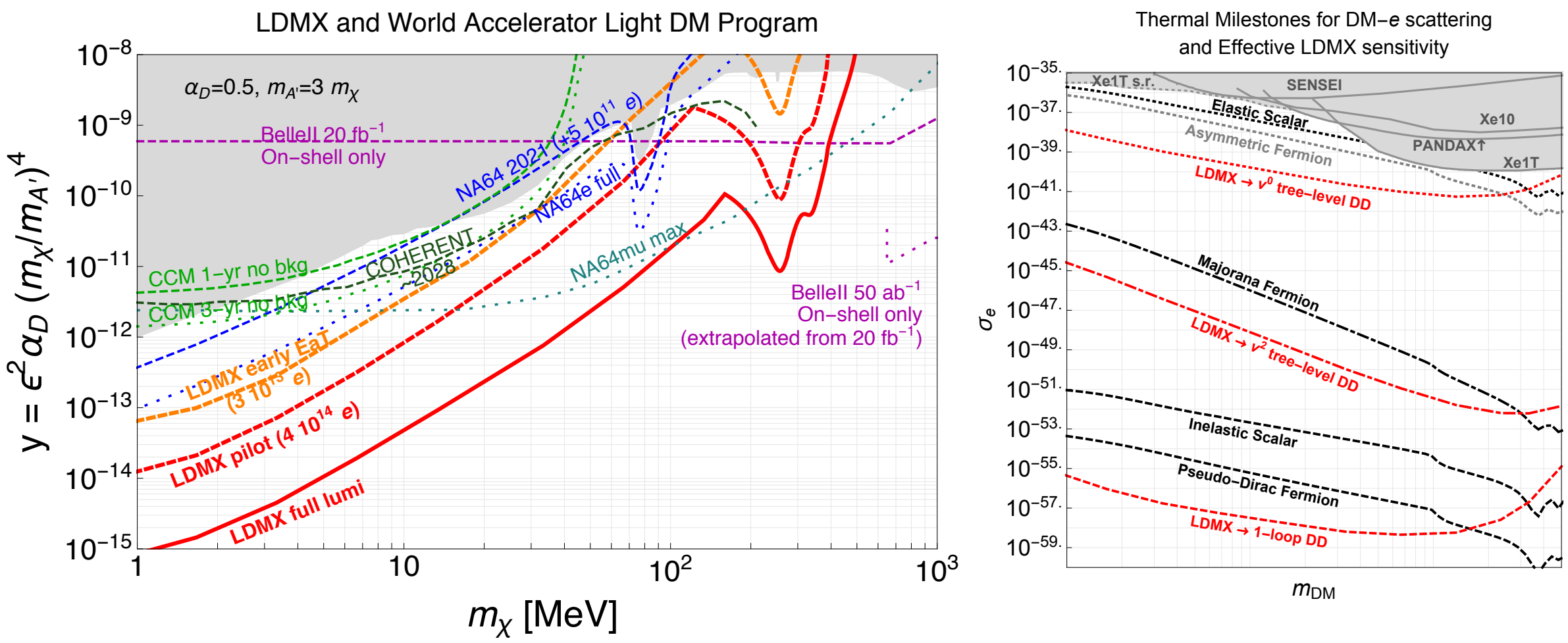

Figure 2.5: **Left:** The expected sensitivity of LDMX (thick red), a $4 \cdot 10^{14}$ EoT pilot run (thick dashed red), and an early running EaT analysis in $3 \cdot 10^{13}$ EoT (thick dashed orange) compared with constraints (shaded) from [35, 18, 36, 8, 6, 37, 20, 38], with expected near-term analyses (dashed) and long-term projections (dotted) from ongoing experiments NA64 (blue) and Belle II (magenta), DMNI project CCM (green), and COHERENT detectors at Neutrino Alley (dark green). NA64 projections are scaled from [37] assuming zero background and unchanged acceptance for the maximum $3 \times 10^{12}$ EoT that can be obtained before LS3. We also include a background free projection for a possible muon beam run currently under discussion at CERN for $2 \times 10^{13}$ MOT (possible by early 2030's), though this does not probe the electron interactions that LDMX measures; CCM projections are scaled from [19] to 2.8 events in 1 and 3 years respectively; COHERENT projections are from [20] and Belle II curves from the Snowmass whitepaper [39] (the dotted line terminates at masses below which unknown systematics dominate and increasingly degrade sensitivity). We note that while there are more recent results e.g. from NA64 [29] and COHERENT [40] they do not surpass the projections shown here. **Right:** Milestones for thermal relic freeze-out and asymmetric fermion DM in electron-recoil direct detection, which vary over 20 orders of magnitude depending primarily on whether the structure of DM-SM interactions induces velocity-independent (dotted), $v^2$-suppressed (solid), or 1-loop (dashed) scattering. LDMX sensitivity can also be mapped onto this parameter space for each scenario (for LDMX, only dark brem sensitivity is shown here). Direct detection constraints are taken from [41] and the solar reflection constraint from [42]. We note that while more recent direct detection results exist, e.g. Ref. [43], they do not change the overall message of this figure.

The broad complementarity between low-threshold direct detection and LDMX extends beyond the thermal freeze-out paradigm. For example, IR freeze-in through an ultra-light mediator [48, 49, 50] has couplings well below LDMX's sensitivity, but can be observable in direct detection because low-momentum scattering is enhanced. By contrast, UV-dominated freeze-in [51, 52, 11] produces observable signals at LDMX without a direct detection signal. **The overarching conclusion from these model-specific comparisons is that the physical parameter spaces for dark matter detection in accelerators and direct detection are fundamentally different and highly complementary.**

## 2.4 Additional Physics Opportunities with LDMX

### 2.4.1 Broader Potential of the Missing Momentum Signature

#### 2.4.1.1 Spin-1 Dark Matter

Although considerably less studied than scalar, pseudoscalar, or fermionic DM, models for light spin-1 DM with a mass in the MeV to GeV range have attracted considerable attention over the past few years [53, 54, 55, 56, 57, 58, 59, 60]. Here, we consider both a set of simplified models for light spin-1 DM that extends the SM by a spin-1 DM candidate and a spin-1 particle mediator only, and an ultraviolet complete model based on a non-abelian gauge group where DM is a strongly interacting massive particle (SIMP). For details

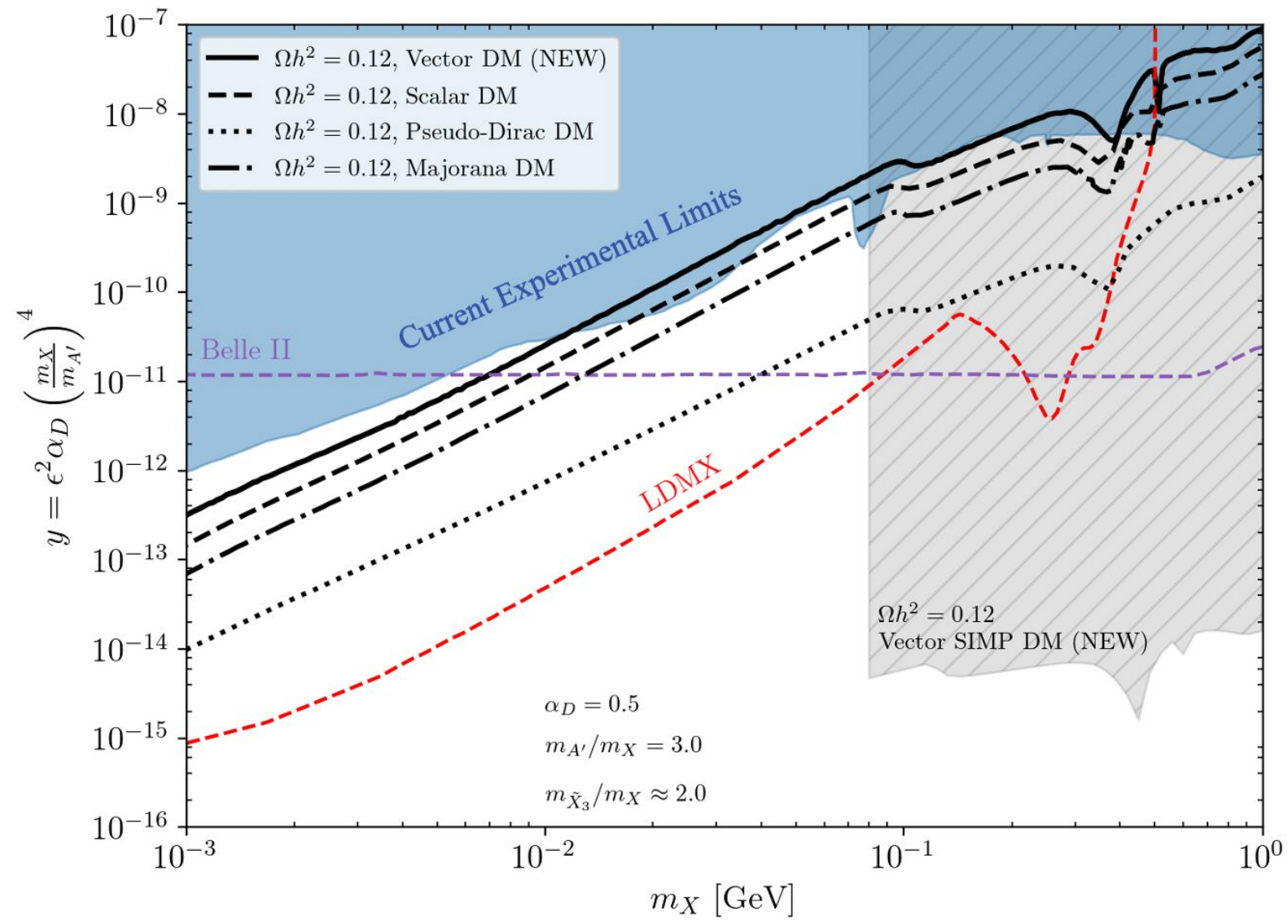

Figure 2.6: Summary plot with relevant current and future projected experimental exclusion limits at 90% C.L. for three models studied previously: scalar, pseudo-dirac, and majorana, in addition to the two new spin-1 DM models. Relic targets are denoted as black curves of various line styles, and as a shaded region for the vector SIMP DM due to its unique early universe production. The simplified spin-1 DM model plotted here is what is referred to as the $\Re[b_6]$ model in Ref. [61]. The current experimental limits are only an approximate shading, which in reality varies between models, thus should only be used as rough guides. Figure taken from Ref. [61].

on the models and the methodology we refer to Ref. [61].

Fig. 2.6 illustrates the parameter space corresponding to a thermal target in these new models in relation to other thermal targets, current experimental limits and the ultimate LDMX reach. The relic target for SIMP DM is completely different in shape compared to those of the simplified models due to its unique production in the early universe. These production processes are only dependent on $\alpha_D$, and are independent of $\epsilon$, resulting in a vertical line, in fact an entire region, consistent with the relic density constraint. We refer to [61] for more details on the relic density. The blue shaded region is based on results from LSND [25, 62], MiniBooNE [63], and NA64 [8]. The Belle-II line was derived from the expected 90 % confidence level limits for phase 3 of Belle-II reported in [64]. LDMX will have sensitivity to these models already at very early stages.

#### 2.4.1.2 Additional Dark Sector Signatures

In addition to the above-mentioned dark matter searches, LDMX makes several notable contributions to searches for unstable dark sector particles beyond DM. Some of these scenarios — such as sub-MeV axions, millicharged particles, and $B - L$ gauge bosons decaying to neutrinos — are tested directly by LDMX's standard missing-momentum analysis. For example, even early missing-momentum data will test much of the remaining sub-MeV "cosmological triangle" parameter space for QCD axions coupled to either photons or electrons, as considered in e.g. [65, 66, 67]. LDMX will also improve constraints on invisible $\phi$ and $\omega$ meson decay by 4–5 orders of magnitude [3]. These examples are both shown in Fig. 2.7. Likewise, LDMX's excellent sensitivity to millicharged particle production has substantial overlap with the parameter space motivated by the EDGES anomaly [12], exceeds existing limits by up to an order of magnitude, and surpasses the sensitivity of dedicated millicharge detector proposals in the $< 100$ MeV mass range [68].

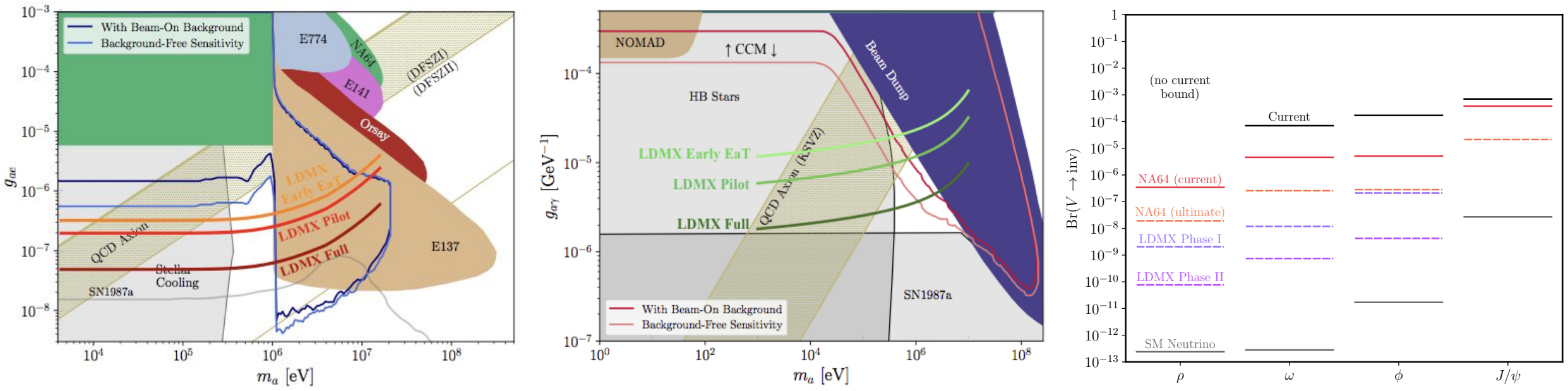

Figure 2.7: Three examples of the broader physics potential of LDMX (see [12] for more): **Left:** Projected sensitivity of LDMX missing-momentum analysis to QCD axions and axion-like particles via electron coupling, overlaid on figure from [67]. The three phases of LDMX running from Fig. 2.5(left) are shown in shades of red. Even early running EaT data could close the remaining sub-MeV QCD axion parameter space. **Center:** Projected sensitivity of LDMX missing-momentum analysis to QCD axions and axion-like particles via photon coupling, overlaid on figure from [67]. The three phases of LDMX running from Fig. 2.5(left) are shown in shades of green. **Right:** Figure from [3] illustrating the sensitivity of the LDMX missing-momentum analysis to fully-invisible decays of various vector mesons, relative to current bounds (black) curves, NA64 capabilities, and the expected neutrino-decay signal in the Standard Model.

### 2.4.2 Search for Visible Signatures of Dark Sectors

In addition to the invisible missing-momentum signature, distinct analyses can leverage the unique design of LDMX to search for long-lived dark sector particles decaying deep in the LDMX calorimeters [12]. Visible decays can occur for scenarios in which $m_{A'} > 2m_e$ that are strongly motivated by models of sub-GeV dark matter. Since LDMX is a hermetic detector with capabilities to fully reconstruct final states, it can effectively operate as an active beam dump experiment to search for long-lived particles that decay into SM particles (usually $e^+e^-$ pairs). Specifically, this signal would be the sudden appearance of SM particles in either the ECal or Back HCal as illustrated in Fig. 2.8.

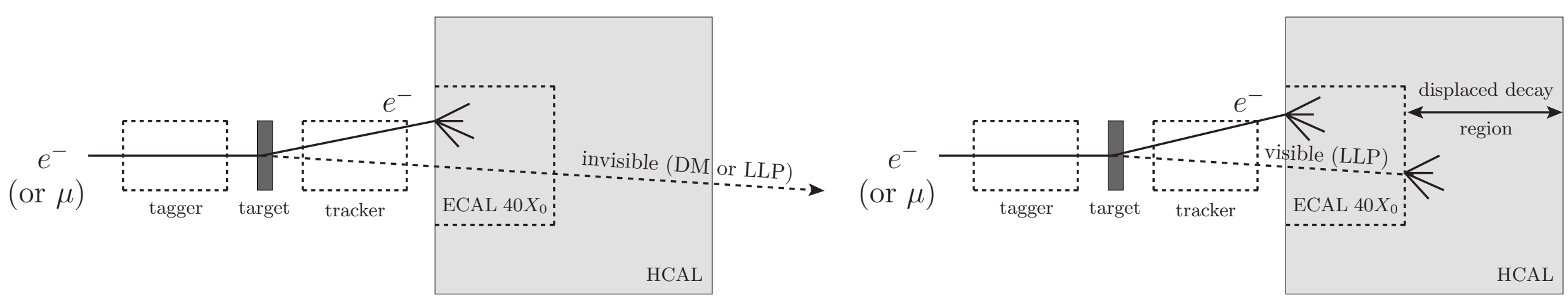

Figure 2.8: A schematic comparing a signature from a dark photon from Left: an invisible decay and Right: a visible decay. A long-lived particle with a visible decay would have a soft recoil electron accompanied by a sudden appearance of SM particles in either the ECal or the Back HCal. Figure reproduced from [11]

The specific models that involve long-lived particles that can be produced at LDMX and decay visibly are the minimal dark photon model, axion-like particles (ALPs), strongly interacting massive particles (SIMPs), milli-charged particles, and B-L gauge bosons. For a variety of these models, LDMX has projected sensitivity in currently unexplored parameter space. This is discussed in more detail in Sec. 4.6.

### 2.4.3 Measurements of Electron-Nucleon Cross Sections

Beyond dark sector physics, LDMX can make powerful measurements [69] of electron-nuclear scattering, which address key systematics for DUNE and other neutrino oscillation experiments (see [70, 71]). LDMX complements other experimental efforts in this direction (mainly at JLab [72, 73, 74, 75, 76, 77]) that can be used to improve neutrino scattering models in generators such as GENIE and GiBUU. These generators have been found to differ from *inclusive* electron-scattering cross-sections by up to $\mathcal{O}(50\%)$ [71], reflecting a large level of uncertainty in the modeling of lepton scattering off of heavy nuclei, including with respect to the initial nuclear state, interaction processes themselves, and final state interactions of hadrons exiting the nucleus.

Compared to the experiments and proposals above, LDMX is unique for its broad forward coverage, its ability to tag incoming electrons and reduce uncertainties on initial state radiation, its low reconstruction energy threshold in various hadronic final states, and its ability to detect neutrons with high efficiency. These will allow LDMX to perform a suite of inclusive and semi-exclusive measurements of electron scattering, and an ability to probe hadron multiplicity and kinematics within its phase space in addition to electron kinematics [69]. With the expected incident electrons on target, millions of interactions are expected to be recorded, allowing for multi-dimensional cross section measurements with little statistical uncertainty, even for many semi-exclusive or exclusive channels. Moreover, due to LDMX's use of an 8 GeV beam, LDMX data used to search for dark matter will be taken in a range of momentum and energy transfer that closely overlaps the resonant and inelastic scattering regions of phase space that are most relevant to neutrino interactions in DUNE [69].

For these reasons, the LDMX collaboration has a dedicated effort to include an electronuclear trigger for data taking and is engaged in ongoing studies with neutrino physicists to refine our understanding of LDMX capabilities in this area. The goals of the electro-nuclear scattering program in LDMX include:

- inclusive electron scattering multi-dimensional cross-section measurements, both as a function of outgoing lepton kinematics (angle and momentum), but also as a function of interaction kinematics (momentum and energy transfer);
- semi-exclusive electron scattering multi-dimensional cross-section measurements, including final states with identified protons, neutrons, charged and neutral pions, and kaons;
- reconstruction of more complex, composite interaction kinematics variables, like energy and momentum imbalance quantities in both transverse and 3D; and,
- exploration of neutrino energy reconstruction techniques as applied to electron-scattering measurements, including tests of initial energy reconstruction resolution.

These measurements will include comparisons against commonly used interaction models and event generators used in neutrino experiments like DUNE, and can include detailed tuning of and constraints on various model parameters, particularly those related to initial nuclear state, final state interactions, and the vector component of lepton scattering cross sections. The resulting measurements will thus offer not only a large and detailed set of unique data to improve future lepton interaction models and event generators, but can also provide constraints on systematic uncertainties in current and future neutrino cross-section measurements.

Studies to refine LDMX's approach to these measurements are ongoing. These include:

- comparisons of interactions on different targets, particularly consideration of a titanium target, in addition to the tungsten target optimized for dark matter searches;
- development of inclusive electron scattering triggers, and consideration of additional triggers or high-level event filtering that can target additional final states of interest;
- development of track and calorimetric clustering tools for electron-scattering interactions, like those that can handle high-particle-multiplicity final states that are of particular interest for studies of deep inelastic scattering interactions and effects of final state interactions, or detailed energy resolution studies for proton and neutron hadronic clusters; and,
- studies of particle identification techniques using a combination of tracker and calorimetric reconstruction.

These studies, while essential for the electro-nuclear scattering measurements, also provide input into detector calibration and background estimation studies relevant for the dark matter search (e.g., through reconstruction of photons from neutral pion decay, or identification of charged kaons). For a more detailed discussion of the electro-nuclear measurements, see Sec. 4.9.

# Chapter 3

# Design, Construction, and Operations

## 3.1 Detector System Overview

The missing momentum signature exploited by LDMX to search for Dark Matter has three components:

1. substantial energy loss by the incoming beam electron, leaving the recoiling electron with a small fraction (e.g. less than 30%) of its initial energy.
2. a large transverse momentum kick of the electron, which, together with the degraded energy, means the recoiling electron is ejected at a large angle with respect to the incoming beam.
3. the absence of any other visible final-state particles that could carry away the significant energy lost by the electron.

These three observables, and the ability to utilize them at high rates for up to equivalent $10^{16}$ incoming electrons on a 10% $X_0$ target to search for only a few signal events, define the composition and layout of the apparatus.

Taken together, the first two elements of this signature require estimation of the change in vector momentum of individual electrons across a thin ($10 - 40\%$ $X_0$) target, where multiple scattering in the target determines the useful precision. Although the beam energy is known, the beam can be contaminated with off-energy electrons or other particles, so the momentum of each incoming electron must be robustly measured. This can be accomplished with a narrow, low-mass tracker upstream of the target in a magnetic field optimized for measuring beam-energy electrons. The same technology may be used for measuring downstream recoils, but the low energy and wide angles of signal recoils demand wider acceptance in a lower magnetic field. The third element of this signature requires a highly sensitive veto for additional outgoing particles, suggesting hermetic, large-acceptance calorimetry placed directly in the beamline behind the target. Because the vast majority of outgoing particles are scattered electrons or bremsstrahlung photons, the central part of this calorimeter must be optimized for electromagnetic showers (an "ECal"). Furthermore, because the ECal signal rate is of the same order as the repetition rate of the beam, the ECal must be fast and have good spatial and temporal resolution to distinguish energy deposits from different events. Indeed, the ability to resolve the ECal responses to individual electrons sets the overall ceiling on the beam repetition rate ($\lesssim$ 40 MHz) as well as limiting each bunch to a few electrons spatially separated within the beam spot. The LESA beamline (see Sec. 3.3.3) can operate at these limits, allowing LDMX to accumulate an equivalent $10^{16}$ incoming electrons on a 10% $X_0$ target in a reasonable few-year running period. Meanwhile, the most pernicious potential backgrounds involve a hard bremsstrahlung photon that carries away most of the electron energy, followed by a highly atypical muon conversion or photo-nuclear reaction that happens to leave little energy in the ECal (see Fig. 3.3). Identification of these events calls for a large and highly sensitive hadronic calorimeter (HCal) surrounding the ECal to veto events with any significant in-time energy deposit.

A first concept of the LDMX apparatus, a compact realization of this approach, has been presented in detail in [78] and is shown in Fig. 3.1. Following the beam, the detector subsystems in the magnet region are a silicon tagging tracker (STT) inside a dipole magnet and a silicon recoil tracker (SRT) in the fringe field of the magnet, with a thin tungsten target interposed between them. Behind the SRT is a compact and highly segmented Si-W electromagnetic calorimeter (ECal) with excellent MIP sensitivity that is surrounded by a large scintillator-based hadronic veto system (HCal) with low energy thresholds. Two important details follow from this picture. First, because the beam passes directly through the trackers and into the ECal, these detectors must contend with high radiation doses corresponding to $10^{16}$ EoT. To mitigate this issue, and also reduce the peak occupancies in these devices, a large, rectangular beamspot with an area of 16 $\text{cm}^2$ is used. As a result, only the ECal has challenging requirements for radiation tolerance. Second, the rates in the detector prohibit streaming readout: a fast trigger is required. Because signal events have unusually large missing energy in the ECal, and such events are very rare, the simplest strategy is to trigger on low energy in the ECal. To set an appropriate energy threshold for this trigger, the number of incoming electrons in each beam bunch must be known. This can be accomplished with an array of small scintillator bars – a Trigger Scintillator (TS) system – placed in the path of the beam to count the number of incoming electrons in each bunch.

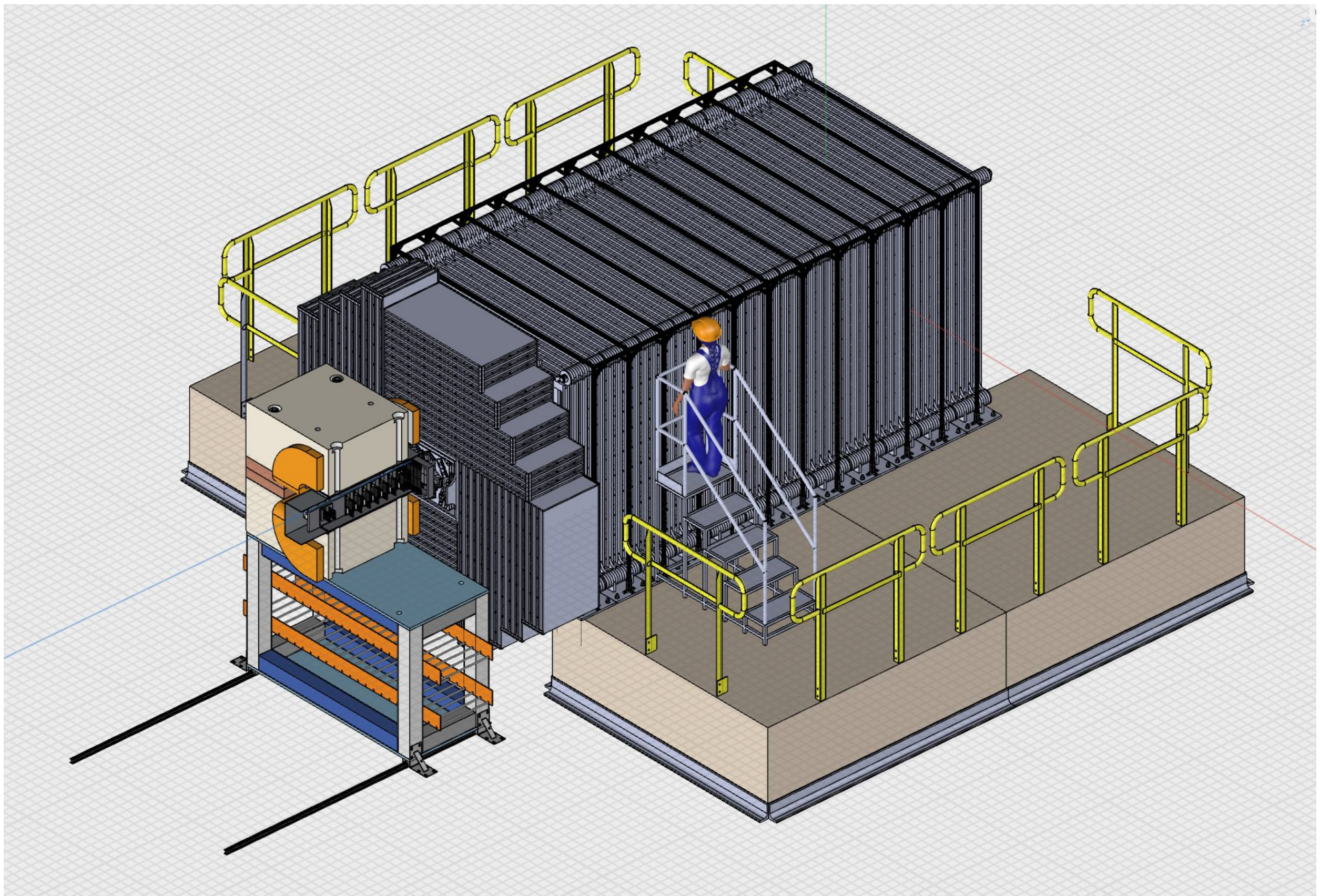

Figure 3.1: Overview of the LDMX detector solid model.

These detector subsystems – Beamline and Magnet, Trigger Scintillator, Tracking, ECal and HCal – along with the trigger and data acquisition electronics (TDAQ) and the software and computing environment required for simulation and analysis of the data define the scope of the technical systems for the experiment that have been developed under the DMNI project in preparation for construction. The following provides an overview of the technical details of these systems.

## 3.2 Requirements

To achieve the objective of recording high-missing energy events, the ability to resolve the ECal responses to individual electrons sets the overall ceiling on the beam repetition rate ($\lesssim$ 40 MHz) as well as limiting each bunch to a few electrons spatially separated within the beam spot. The LESA beamline (see Sec. 3.3.3) can operate at these limits, allowing LDMX to accumulate equivalent $10^{16}$ electrons on a 10% $X_0$ target in a reasonable running period. LDMX can accumulate roughly $4{\times}10^{14}$ EoT with a 10% $X_0$ target and one electron per sample on average, with twelve months' worth of operations. Such a dataset will be sufficient to generate a 100-fold increase in sensitivity to 1 MeV dark matter over current experimental constraints, assuming a background-free experiment and at least 40% signal efficiency. LDMX can reach ten times beyond that with 3.5 years of operation and a 40% $X_0$ target, i.e. provide as much as a 1000-fold improvement over current experimental constraints. The ability to realize this higher sensitivity will depend on detector performance, availability of beam time, and whether backgrounds observed in data are not drastically different from expectations based on current simulations and calculations. Therefore, we have designed the detector to facilitate a search for thermal relic dark matter using a $4{\times}10^{14}$ EoT dataset, even though we expect to achieve higher levels of sensitivity with more data and the knowledge gained from analyzing the $4{\times}10^{14}$ EoT dataset. The radiation exposure to reach the 1000-fold improvement will produce non-negligible doses in the active elements of the ECal. As such, sufficiently radiation-hard technology has been chosen.

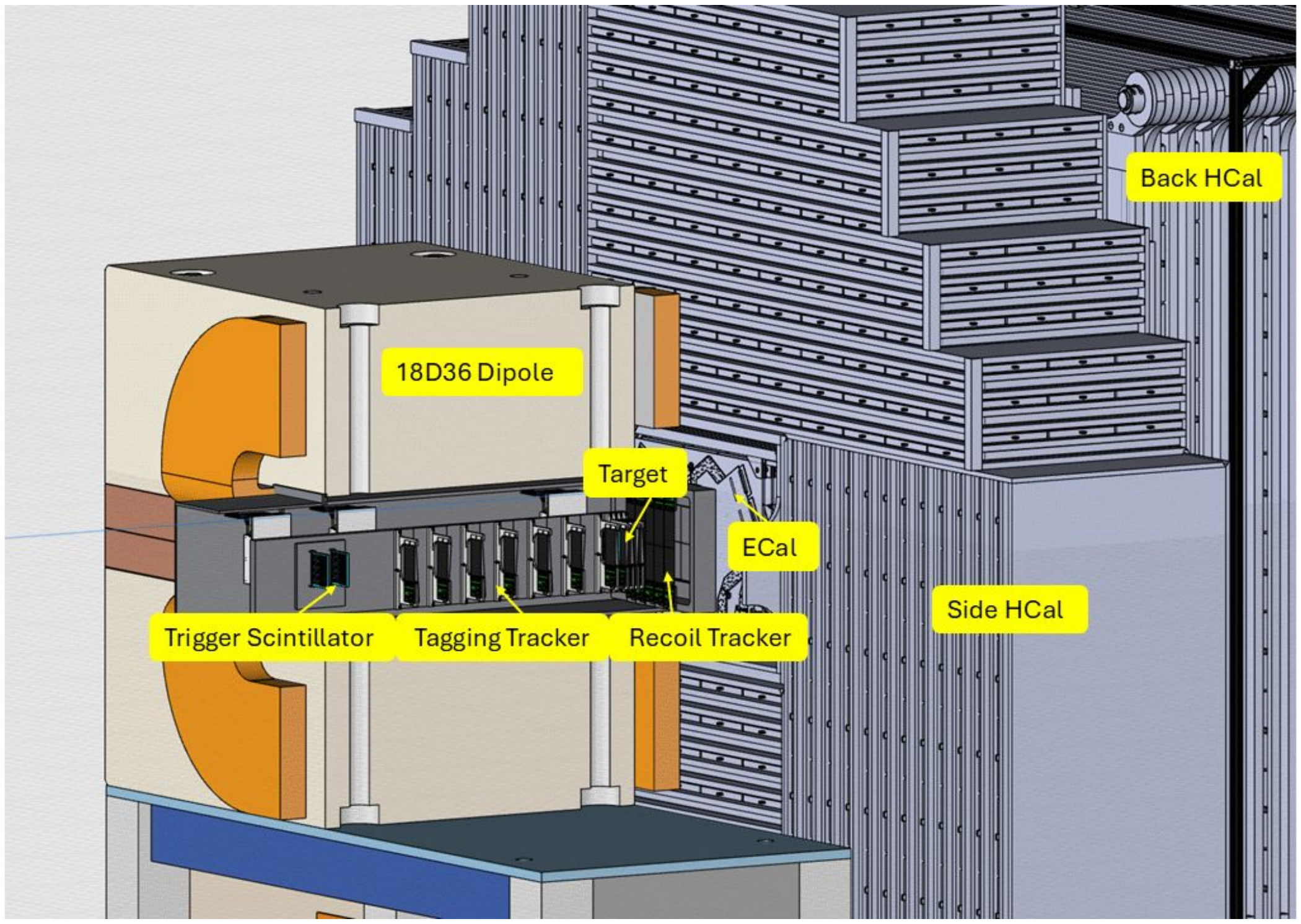

Figure 3.2: Cutaway view of the LDMX detector showing, from left to right, the trigger scintillators, the tagging tracker, the target inside the spectrometer dipole, the recoil tracker, the ECal, and the side and back HCal.

The rate at which electrons traverse the target demands a trigger-based data acquisition system that ensures trigger decisions are made quickly enough that all subsystems can operate with minimal dead time. The tracking system's ability to buffer data places a maximum latency requirement on the trigger system and, in turn, all systems providing inputs to trigger algorithms. The data acquisition system must also be capable of supporting sufficiently high readout rates to ensure signal events are saved with high efficiency and ancillary datasets are sufficiently large to support the physics program.

For performing measurements and searches, sufficient resources are required to maintain data stores, analyze data, and generate simulated datasets. LDMX has built various tools and leveraged technology developments from other projects to ensure we can operate and calibrate the detector and produce timely measurements of our datasets. This work is summarized in Sec. 3.10.

The key backgrounds in a missing momentum search drive the requirements for each detector subsystem design. The combined capabilities of all subsystems must be such that backgrounds can be reduced to negligible rates in a $4\times10^{14}$ EoT dataset while retaining at least 40% of signal events over the entire mass range. Based on simulations and calculations of key processes, we have enumerated the key backgrounds for a missing momentum search. Fig. 3.3 shows the relative rate of potential missing-momentum backgrounds that must be efficiently rejected to reach the 100-fold increase in sensitivity. In addition, Fig. 3.3 illustrates the relevant detector signatures that may be leveraged to do so.

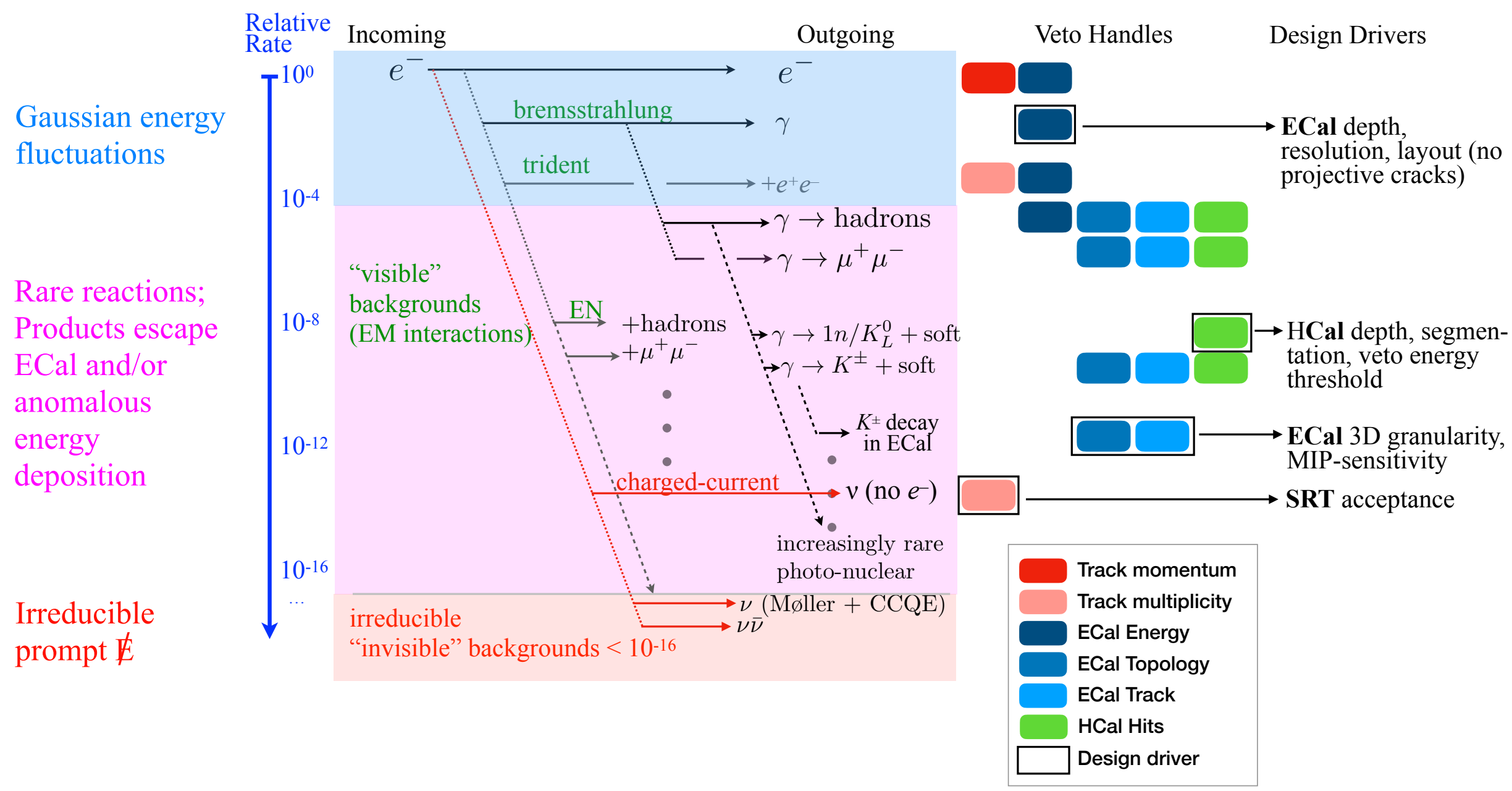

Figure 3.3: An illustration of the potential background sources from Standard Model interactions for the missing-momentum search. Relative rates of each process are shown, in addition to detector subsystems capable of distinguishing each process from a dark matter signal.

Pure electromagnetic processes, such as bremsstrahlung and tridents are the most copious backgrounds. Such events can be rejected through sufficiently precise measurements of electron momenta and the energies of electromagnetic showers. More rare processes involve bremsstrahlung photons interacting with nuclei within the detector, which can produce muon pairs or hadrons, resulting in a reduced detector response compared to pure electromagnetic showers. Such interactions can be reduced by vetoing signals in the HCal and exploiting activity patterns in the ECal. Track measurements provide an additional identification handle when such interactions occur in the target.

The backgrounds are categorized to help delineate subsystem requirements. Broadly, they are categorized by electro-nuclear and photo-nuclear interactions. These categories are each further subdivided based on whether they produced hadrons or muons. In some cases, additional subclasses of backgrounds are singled out; for example, neutral hadrons are particularly pernicious and drive key requirements for the HCal, while charged Kaon production drives key ECal requirements. In both of these cases, dedicated simulated datasets have been studied to understand how effective our detectors will be at rejecting such backgrounds in datasets as large as equivalent $10^{16}$ electrons on a 10% $X_0$ target.

Weak interactions involving neutral currents can mimic the DM signature but are rarer than the $10^{-16}$ target; tens of charged current events with energetic neutrinos should ultimately be collected. However, the lack of an outgoing, low-momentum electron in the tracker would clearly distinguish such events from the DM signal.

In cases where more events are observed than expected, missing momentum provides a critical diagnostic. Standard Model backgrounds generally have a much softer recoil electron $p_T$ spectrum, while the spectrum for the signal will depend on the mediator mass. The recoiling electron $p_T$ spectrum, thus, provides additional background discrimination and characterizes any signal observed.

Given the above, the high-level requirements are:

- **LDMX Requirements**

LDMX1 Have at least 40% efficiency to thermal relic signals
LDMX2 Reject beam-related backgrounds
LDMX3 Reject fake missing energy from EM showers
LDMX4 Reject muon production from PN
LDMX5 Reject hadron production from PN
LDMX6 Reject charged Kaon production from PN
LDMX7 Reject muon production from EN
LDMX8 Reject hadron production from EN
LDMX9 Reject trident backgrounds
LDMX10 Reject charged-current neutrino production
LDMX11 Provide systems necessary for measuring the transverse momentum spectrum of electrons recoiling off of the target
LDMX12 Provide operational capabilities for 6 years, including sufficient performance after doses expected from $4 \times 10^{15}$ EoT
LDMX13 Provide a trigger system capable of identifying events with large missing energy
LDMX14 Provide sufficient readout bandwidth for candidate signal events and ancillary datasets for the LDMX physics program
LDMX15 Provide sufficient storage to facilitate long-term preservation of all beam data and sufficiently high-statistics simulated datasets for supporting operations and the LDMX physics program
LDMX16 Provide sufficient software and hardware infrastructure for generating simulated datasets, processing all data within a few months, and facilitating measurements.

Each subsystem section below will provide more specific requirements based on these requirements. In Sec. 4, we demonstrate that when used in simulation studies, our detector designs provide sufficient background rejection power to have a nearly background-free search region and sufficient signal efficiency to provide roughly a 100-fold improvement in sensitivity to thermal targets.

## 3.3 Beamline and Magnet

### 3.3.1 Introduction and Overview

LDMX will operate at SLAC in End Station A using the Linac to End Station A (LESA) beam facility. [79] LESA takes bunches of electrons accelerated by the LCLS-II drive beam that are unused for generating x-ray pulses to operate experiments in End Station A (ESA) in parallel with the ongoing photon science program. While LESA, and not LDMX, is responsible for delivering the beam to ESA with the required attributes, LDMX provides a small number of components that are necessary for and tightly coupled to the needs of the experiment. These include a large-diameter section of beampipe terminating in a thin vacuum window, additional vacuum pumping and monitoring for this final section of beampipe, installation of a Point Beam Loss Monitor (PBLM) for the LESA Beam Containment System (BCS), and a set of simple beam background monitors to be installed in front of the magnet as diagnostics for beam quality issues that may arise in steering and defocusing the large beamspot by LESA. Also included in this subproject, and comprising the vast majority of the scope, is the spectrometer dipole that enables the momentum measurement by the tracking systems and spreads out the final state particles in the downstream calorimeters. This work includes the refurbishment of an existing magnet, along with the power, control, cooling, and support systems required to operate it. The following sections describe the requirements, design, performance validation, and project plan for these components and systems.

### 3.3.2 Requirements

There are only a few requirements to be met by the beamline and magnet components to ensure that the apparatus delivers the projected sensitivity. These requirements flow down from the low-level performance requirements of the tracking systems, ensuring that the resolutions and backgrounds in the tracking system have been correctly simulated in validating the tracking design.

- **Beamline & Magnet Detector-Physics Requirements**

  BM1 A field map with the precision of 1% in value and 1 mm in position throughout the tracking volume shall be sufficient to ensure that tracking performance requirements are met.

- **Beamline & Magnet Technical requirements**

  BM2 The magnet shall be capable of achieving 1.5T field at the center of the magnet.

  BM3 The vacuum window shall be less than 0.5% $X_0$ thick.

### 3.3.3 Design

LCLS-II drive beam is transported into End Station A (ESA) via the Sector 30 Transfer Line (S30XL), which extracts beam from the LCLS-II dump line and carries it to the A line terminating in ESA. This Accelerator Improvement Project (AIP), called Linac to End Station A (LESA), is well underway and expects to deliver beam to End Station A in early FY26. For LDMX, LESA will deliver bunches at 37.14 MHz at an average of 1-2 $e^-$ per pulse, in periodic trains of bunches between the high charge 929 kHz bunches being extracted for the LCLS-II undulators. The beam current will be controlled by a combination of a dedicated laser system at the injector to fill the bunches, and a set of spoilers and optics downstream to allow precise and stable control of the current in ESA. Optics upstream of ESA will focus the beam and collimate it to a rectangular spot, and a pair of quadrupoles installed at the upstream end of ESA will enlarge the beam to create the large luminous region on the LDMX target needed to reduce occupancies and spread out radiation doses. An overview of the attributes of the LESA beam is shown in Table 3.1 along with the parameters to be used for operation of LDMX.

| **LESA and LDMX Beam Parameters** | | |
|---|---|---|
| Parameter | LESA | LDMX Requirements |
| Energy | 8 GeV | 8 GeV |
| Train Repetition Rate | 1 MHz | 1 MHz |
| # Bunches/Train | 1-100 | 20 |
| Bunch Spacing | 5.39 ns – 1 $\mu$s | 26.9 ns |
| Charge/Bunch | $< 1 - 3 \times 10^5$ | $1 - 2$ |
| Beam Spot | < 1 mm – > 1 m, variable shape | 20 mm (x) × 80 mm (y), rectangular |

Table 3.1: Beam parameters available at the LESA facility and those to be used by the LDMX experiment.

While LESA is responsible for delivering the beam as described above, the LDMX project includes some components that are classified as beamline because they are tightly coupled with, and are the interface to, the LESA facility and draw on resources and expertise at SLAC engaged on the LESA project and within the Accelerator Directorate (AD). These elements are described in the following sections.

#### 3.3.3.1 Beamline Elements

**3.3.3.1.0.1 Beampipe and Vacuum Window** Some distance downstream of the defocusing quadrupoles that produce the large beamspot for LDMX, the beam will become too large to pass through the four-inch diameter beampipe that currently terminates the beamline in ESA. LDMX will install a larger six-inch diameter aluminum beampipe, approximately twenty meters in length, to transport the enlarged beam to the front of the detector. Roughly two meters in front of LDMX, this beampipe will end with a thin vacuum window, launching the beam through air into the apparatus. Standard titanium and aluminum vacuum windows are suitable and meet the material budget, where the design of a commercially available titanium vacuum window assembly available with a sufficient aperture is shown in Fig. 3.4. The thickness of the titanium foil in this design is 0.005" or roughly 0.35% $X_0$, roughly equivalent to the thickness of a single silicon sensor of the tagging tracker.

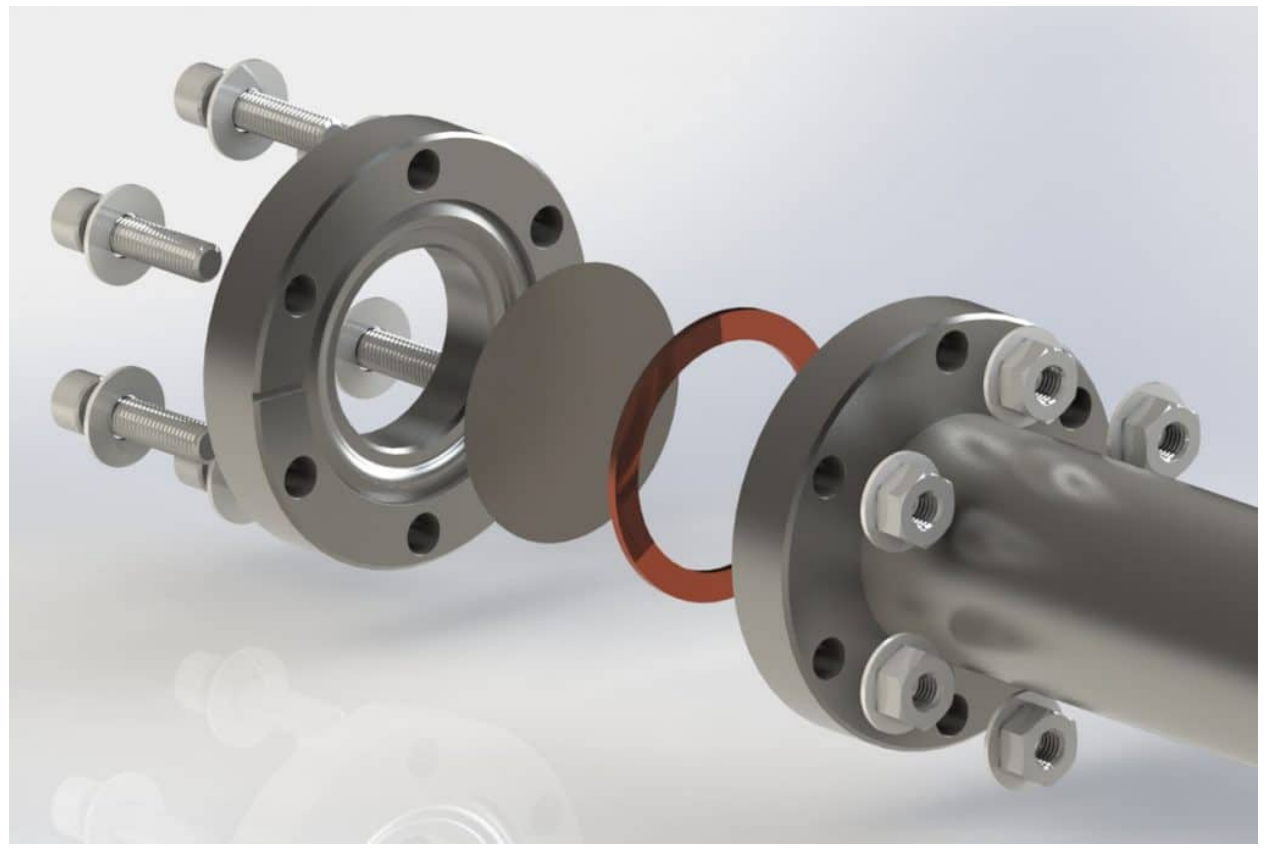

Figure 3.4: A commercially available titanium vacuum window assembly in a design suitable for terminating the beampipe for LDMX in End Station A.

**3.3.3.1.0.2 Vacuum Pumping and Monitoring** Although LESA provides vacuum pumping of the beamline in ESA, LDMX will install additional pumping capacity in this final section of the beamline to minimize beam-gas interactions at larger distances upstream of the detector, where scatters have a larger lever arm to spread out the beam. LDMX will purchase two of the high-capacity pumping stations similar to those used elsewhere in the LCLS-II project for installation in ESA. Vacuum gauges will be installed at these locations that send data to the EPICs-based slow-control and monitoring for the experiment.

**3.3.3.1.0.3 Beam Halo Monitoring** LESA includes diagnostics for the beam current, location, and shape, but only upstream of the spoiler and defocusing quadrupoles where the currents are larger and the beamspot is smaller. LDMX will use detector data from the Trigger Scintillator, Trackers, and ECal to monitor the spatial and temporal distribution of the beam within the detector acceptance. In addition, to provide a diagnostic tool for potential focusing and steering problems upstream, LDMX will monitor a widely scattered beam with a set of four scintillator counters installed on the front face of the magnet and arranged around its bore. These counters will be simple scintillator paddles with SiPM readout for tolerance to the magnetic field, and will be read out by a dedicated system that provides the rate of counts to the EPICS-based slow control and monitoring system for the experiment. This information will be used in real time by shift workers to identify and communicate issues to LESA operators, and recorded so that any periods with suspected bad beam conditions can be correlated with this data.

**3.3.3.1.0.4 Radiation Protection** The radiation environment in End Station A has been studied by SLAC Radiation Protection as part of the LESA project for 5 W of beam impinging on a thick target in the location where the LDMX apparatus will be installed. The radiation field for this beam current, which is two orders of magnitude larger than the LDMX beam current at 1 electron per bunch, is shown in Fig. 3.5. Operation in this configuration has been preliminarily approved, so that no issues are expected in approving the radiation protection plan for LDMX operation. During normal operations, there exists the possibility that a misfire of the LESA kicker could result in bringing LCLS-II high-charge (order 1E8 $e^-$) bunches to ESA at 1 MHz. The risk of such an event to the experiment and the radiation protection plan is mitigated by the Beam Containment System (BCS) which can perform a fast shutoff of the kicker system within two LCLS-II pulses.

The placement of the LDMX detector changes the environment in End Station A, requiring a modification to this beam containment system. In particular, LDMX will shield some radiation monitoring installed on the downstream wall of ESA. To address this, a diamond Point Beam Loss Monitor (PLBM) of the type widely used in LCLS-II, will be installed on the front face of the LDMX magnet and tied into the BCS.

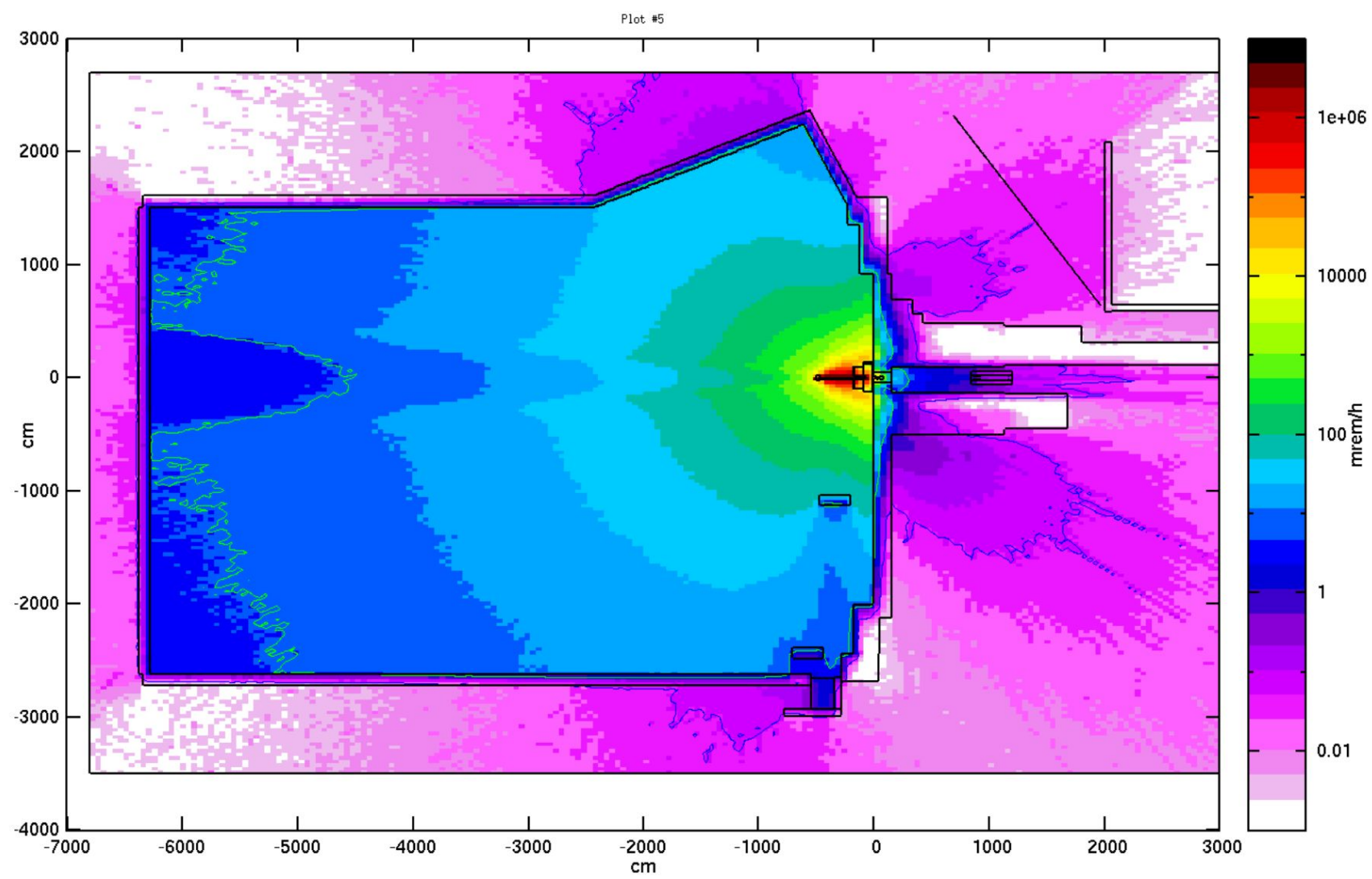

Figure 3.5: The radiation field for a beam power of 5 W on a thick target in ESA, where LDMX will be located, which has been preliminarily approved by the SLAC Radiation Protection department.

#### 3.3.3.2 Spectrometer Magnet

High-purity tracking of incoming beam energy electrons suggests a large uniform field for the tagging tracker in the region upstream of the target. Meanwhile, high acceptance measurement of recoiling tracks down to two orders of magnitude lower than the beam energy requires a much lower field in the region beginning at the target. The simplest solution to these simultaneous requirements is the operation of both trackers in a single magnet, with the tagging tracker in the high, uniform central field and the recoil tracker in the rapidly falling fringe field, with the target positioned at the border between these two regions.

This idea is realized in LDMX with the use of a standard normal-conducting dipole magnet. A significant number of suitable magnets with designation 18D36 – having an 18-inch wide and 36-inch long bore – were built during the 1970s and 1980s and are widely available. A partial map of the field produced by this magnet at a central field of 1 T is shown in Fig. 3.6, along with the key regions of the LDMX detector. A few magnets of this type are in storage at SLAC, along with a stock of magnet iron suitable for shimming to achieve the pole separation required to accommodate the LDMX tracking systems. Further, one such magnet in storage at SLAC already has a 14-inch vertical gap in the central bore, which combines sufficient space for the tracking systems with the ability to achieve the required central field for the tagging tracker. This magnet is shown in Fig. 3.6, and is an 18D36 MKII with SLAC Property Control #19221.

This magnet was most recently measured at SLAC during October and November of 1978, where data were taken for pole tip fields of 0.4 T, 0.5 T, and 1.0 T at the center of the magnet to evaluate fringe fields, as well as longitudinal and transverse field uniformity. The magnet resistance was obtained from the data at 1.0 T for an excitation current of 1530 A as 0.085 Ω, with ohmic losses of approximately 200 kW. For a 30 $^{\circ}C$ temperature rise of the cooling water, a flow rate of roughly 25 gpm would have been required, and there is evidence of a small amount of saturation at the maximum tested current of 1550 A. An expert at SLAC familiar with these magnets and their operation at higher currents has examined this data and is confident that with sufficient cooling flow, a field of 1.5 T is achievable, which requires approximately 450 kW and 55 gpm of cooling flow. In addition to the mechanical refurbishment of the magnet, the cooling system will be upgraded to handle the increased flow. Cooling water for the magnet is available in ESA with house-supplied cooling water, using auxiliary pumps to provide the increased pressure required to achieve the necessary flow.

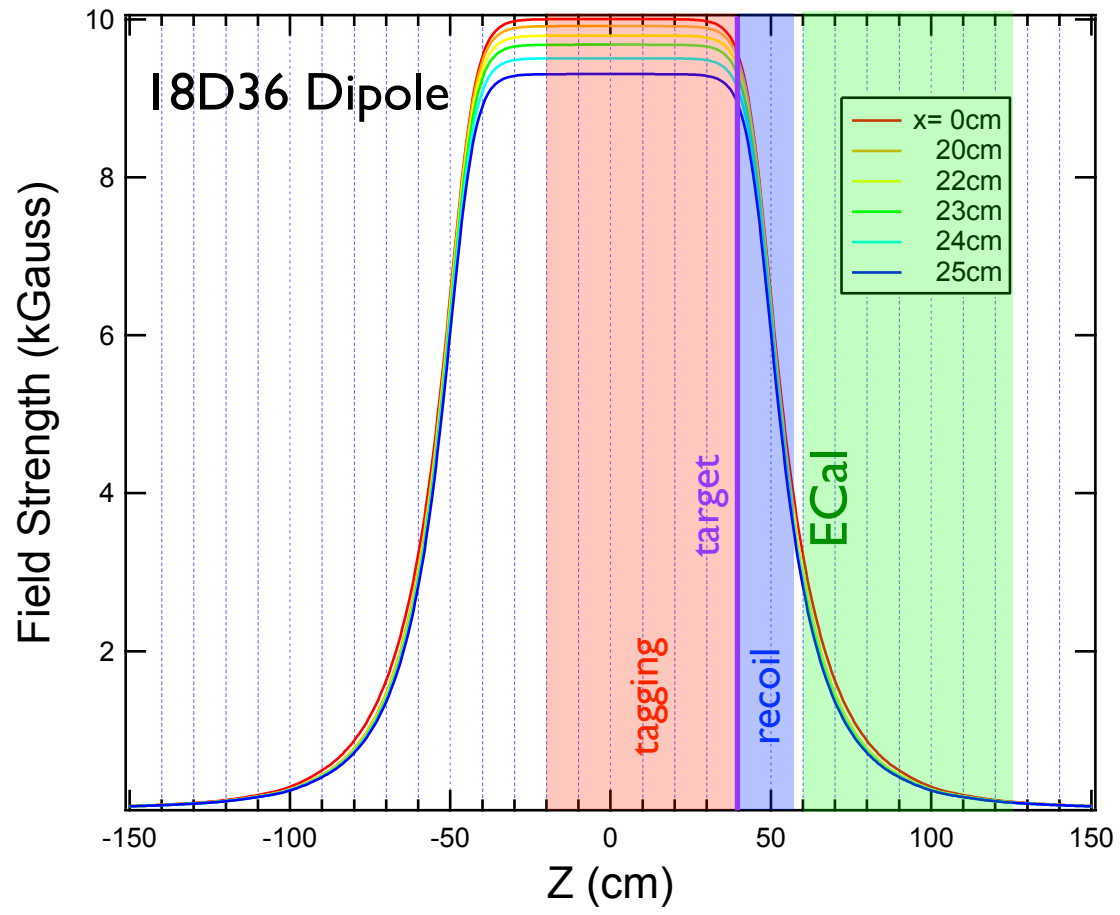

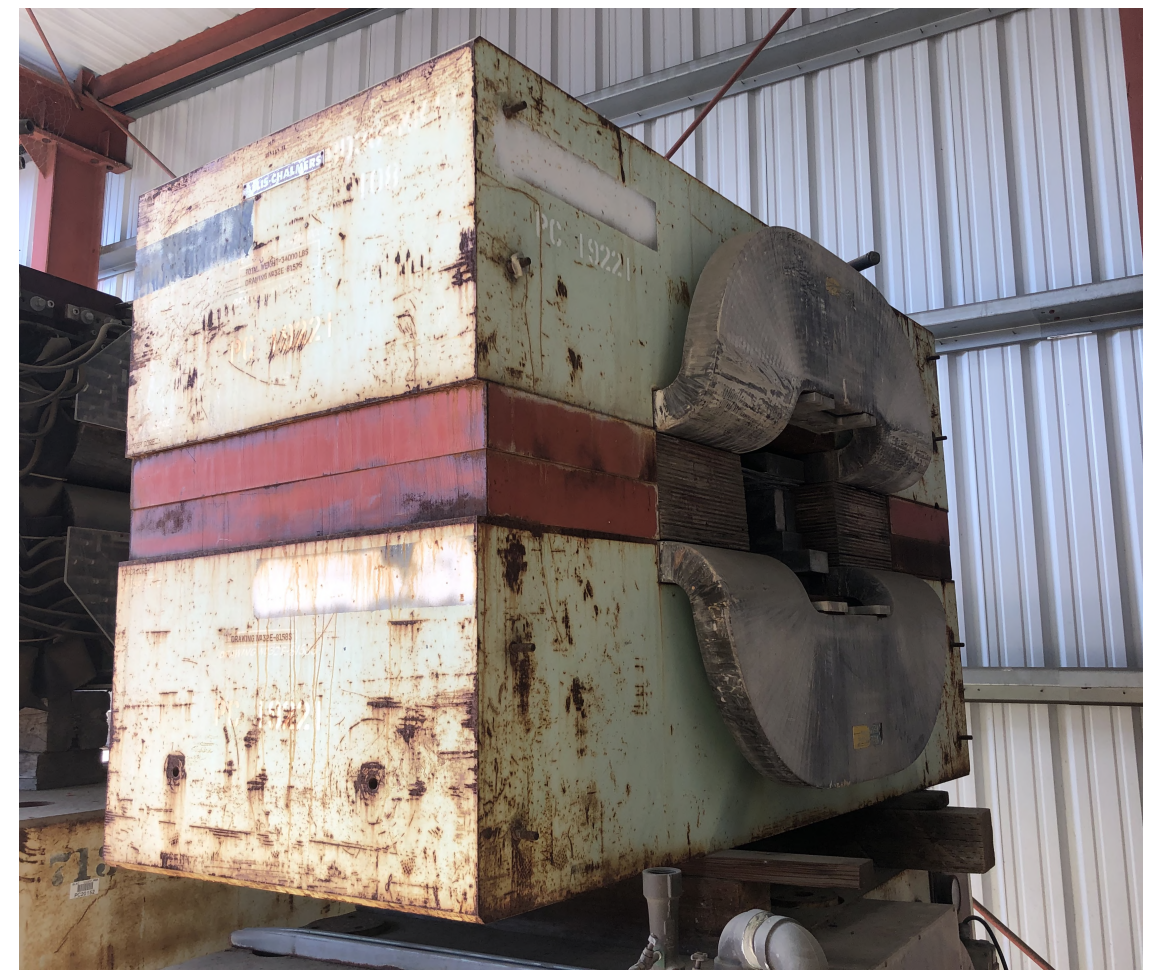

Figure 3.6: Left: The vertical component of the magnetic field for a central field of 1 T in an 18D36 dipole along the center bore of the magnet ($z$=0) and at some distances lateral to the central axis. The location of the target and the different detector volumes are shown. Right: The 18D36 magnet in storage at SLAC, which will be refurbished for the LDMX experiment.

The magnet will be supported on a stand fabricated from rectangular steel tubing. To install and service the ECal, which will be supported from the magnet iron, the magnet and HCal must be moved apart. This is achieved by mounting the magnet stand on a pair of rails and retracting the magnet/ECal from the HCal. Commonly available rail systems are available that can support this load with a large safety factor. In addition to being bolted to the support stand, earthquake bracing secured to the concrete floor of ESA will lock the magnet in place when installed and fully retracted. A small hydraulic ram will be used to move the magnet along the rails between the open and closed positions.

#### 3.3.3.3 Magnet Power and Control System

There are several power supplies servicing End Station A that can power the magnet. Some refurbishment and modernization of these systems along with thorough testing and qualification, is required. A 1.2 MW supply will be tested and equipped with new control systems. Some internal high-power capacitors might need replacement, as might some of the cooling hoses. Finally, the upstream switchgear, consisting of 12 kV substation breakers, 12 kV feeder cables, power supply transformers, and disconnect switches, require maintenance and testing. The substation breakers have been maintained recently and are in good working order. All other equipment has not been used for many years.

#### 3.3.3.4 Monitoring and Interlocks

As with other subsystems, the operation of the beamline and magnet components will be monitored via EPICS. This includes monitoring of the magnet current, coolant flow, and coil temperature, and monitoring the count rates of the beam halo counters. While the magnet coils are protected by Klixons, magnet power will also be interlocked to the coil temperature and operation of the cooling plant using a hardware (PLC) interlock.

### 3.3.4 Project Plan

There are two level three task groups in the beamline and magnet project plan, which have very different scope. Accordingly, these are described separately in the following subsections.

#### 3.3.4.1 Beamline Components

The beamline components project plan covers the design, procurement, fabrication, assembly, and integration required for the beamline components of the LDMX experiment in End Station A. The items covered by this part of the project plan include the large diameter beampipe and vacuum window, the vacuum pumping and monitoring, the beam halo monitoring system, along with reconfiguration of the beam containment system in End Station A, including all necessary design, engineering work and technical work, plans, and approvals. The scope of this work is small, so it is far from being on the critical path for the LDMX Project.

After design work in coordination with LESA engineers to determine the length of the required beampipe, along with the fittings and flanges required, the components for the beampipe and vacuum window will be ordered and delivered to SLAC. Following assembly and bakeout, the beampipe will be moved to End Station A for installation. The vacuum pumping stations will be ordered and integrated with the EPICS monitoring system. Components for the beam-halo monitoring system will be ordered, assembled, and tied into the EPICS monitoring system for the experiment in the lab before installation in ESA. A PBLM currently deployed in Sector 30, where LESA curently terminates, will be moved to End Station A, installed on the upstream face of the magnet, and tied into the BCS.

#### 3.3.4.2 Magnet System

Refurbishment of the magnet system leverages core competencies at SLAC, and personnel experienced with these specific magnets have defined the scope of and produced the project plan for this work. The scope of this work is relatively small so that it is far from being on the critical path for the LDMX Project, and consists of the following steps.

##### 3.3.4.2.1 Design

The project plan for the magnet system includes several design tasks that precede work in some areas. These include the design of the upgraded magnet cooling, the design of the updated power supply controls, the design of the magnet support stand and rail system, and the engineering and approvals of the support components and seismic bracing. Also included is planning and approvals for work on the upstream switchgear that feeds the magnet power supply.

##### 3.3.4.2.2 Procurement and Fabrication

Procurement and fabrication for the magnet system includes materials and components for magnet support and seismic bracing, mounting hardware to be installed in the magnet for support of the tracker support box and ECal, materials and components required for refurbishment of the magnet, the water-cooled cables required to connect the magnet to the power supplies, and fixturing and components for fiducializing the magnet and performing precision field mapping. Also included is subcontracting for maintenance work required on upstream switchgear that feeds the magnet power supply.

##### 3.3.4.2.3 Assembly and Integration

The magnet will be moved from storage to the SLAC Heavy Fabrication Building where it will be photographically documented with a focus on the existing electrical terminals and cooling supply and return lines. In addition, all pre-existing documentation of operation and measurement of the magnet will be recorded electronically.

There are two coils, upper and lower of the drop-eared variety. Each coil has 16 independent circuits with supply and return cooling water hoses attached via double hose clamps. Groups of 8 conductor ends are tied together by mounting bars to guarantee proper spacing and electrical isolation. All hose ends will be removed, as well as the 64 Klixon temperature switches.

After all attachments are disconnected, the magnet will be split into two halves, and the coils removed. All surfaces of each core half, with the exception of the two parting planes, will be cleaned and repainted. The Radiation Protection (RP) department at SLAC has examined the magnet and taken measurements of residual activity, and no restrictions are anticipated on cleaning the steel core, removing loose paint, or removing rust from the outside or parting planes of the magnet. The coils are in very good condition for their age, and will be lightly cleaned before receiving a new coat of epoxy paint.

Prior to re-assembly, modifications for the installation of the tracker support box and support of the ECal – described in Sections 3.3.6 and 3.7 respectively – will be made. After basic assembly, all new cooling hoses – using fittings instead of the existing hose clamps – will be attached, and new Klixons will be installed along with RTD's to allow monitoring of the magnet coil temperature via EPICS. The magnet terminal flags will be securely reattached to the magnet core. At that point, a cursory electrical checkout will be done: resistance to ground, polarity, and testing for shorts.

Next, the magnet core will be fiducialized by the SLAC metrology department, both to aid in field positioning as well as for magnetic measurements. Since the magnet was completely disassembled, stretched wire measurements will be performed of integrated strength vs. current along the axis of the bore ($x = 0$, $y = 0$) in steps of 100 A. A set of survey monuments will be installed on the magnet to allow the positioning and survey of the Tracker and ECal with respect to the magnet, and the magnet position with respect to the beamline in ESA.

#### 3.3.4.3 QA

QA testing for the Beamline and Magnet takes place throughout the construction process to ensure the performance of the completed system. The QA steps for each WBS element are described in the following subsections.

##### 3.3.4.3.1 Beamline Component Testing

Following assembly and bakeout, the beampipe and vacuum window will undergo pump down and helium leak checks. The vacuum pumps will be blanked off and tested before installation in End Station A. The Beam Halo Monitors will be assembled as a cosmic telescope in the lab for testing. The PBLM will be tested after being moved from the beam switchyard and installed on the magnet.

##### 3.3.4.3.2 Magnet System Testing

The magnet components will undergo a more thorough radiation survey after disassembly and before cleaning and machining. Measurements will be taken of the magnet iron and coils. The coils and cooling circuits will be tested both prior to refurbishment and afterwards for cooling performance, current capacity, and magnetic field capability.

The power supply will undergo testing to assess its condition, including tests with a dummy load at low power for AC and DC operation, interlock function, and cooling performance. Testing with the magnet at full power will follow, using local controls. The new remote control system and interlocks will be bench tested after assembly and before testing the controls with the refurbished power supply.

#### 3.3.4.4 ES&H

ES&H for all work at SLAC will be managed by the SLAC ES&H Department through the Fundamental Physics Directorate ES&H coordinator. A critical element of ES&H for the beamline system is the Radiation Protection plan for the experiment, which establishes the safe operating parameters and procedures of the beam for the experiment. Starting with existing studies performed for LESA, documents governing the operation of LDMX will be produced, reviewed, and approved.

Work on the Beamline Components involves lifting, heavy fabrication, electrical work, and industrial solvents, and will be permitted and managed in accordance with established processes for similar work often undertaken at SLAC by the same experienced personnel.

There are a number of significant hazards in working with the Magnet, its power and control systems, the upstream switchgear, and the cooling plant for the magnet. These include the potential for discovering radioactive contamination to be removed, major lifting activities, media blasting and solvent use, heavy machining and fabrication, welding, high-power electrical work, and work around high magnetic fields. While they are typical hazards at SLAC, many of them are among the most dangerous classes of work undertaken at the lab, and will undergo extensive planning and scrutiny by SLAC ES&H personnel before work can begin. The earthquake hazard presented by the magnet is significant, and work planning and control will be needed to mitigate this hazard during periods when the magnet and its larger components are not locked down due to ongoing work.

### 3.3.5 Performance Validation

#### 3.3.5.1 Physics Performance Validation

Due to the recoil tracker's placement in the rapidly changing part of the magnetic field, it will be important to know its precise placement with respect to the field. We have done studies to quantify the precision needed and the systematic effects that would arise. This was done by generating Monte Carlo data with the tracker and field in their nominal positions and orientations while reconstructing the tracks with the field either displaced or rotated (about the center of the magnet).

In this study, we independently scanned the field displacement $(x, y, z)$ and rotation (roll, pitch, yaw) and reconstructed the tracks in the recoil tracker for the same generated events. We looked for changes in the track parameters, $x, y$ position, and $dx/dz, dy/dz$ at the target, and the magnitude of the momentum, versus the displacement or rotation. After scanning through wide ranges of each variable, the only significant dependence was in the momentum for the $z$ displacement and roll, and pitch rotations. To quantify the dependence, we fit the scattered beam electron peak and plot the mean versus the displacement or rotation angle. These are shown in Fig. 3.7.

The reconstructed momentum peak in the recoil has a roughly linear dependence on the $z$-displacement with a slope of $\sim 50$ MeV/mm. This helps us to set the tolerance needed for aligning the tracker along the beamline and possibly make corrections based on the observed momentum. While there is some dependence on rotations of the tracker relative to the magnetic field, as shown in Fig. 3.7, the rotation would need to be $>\sim 100$ mrad to have a noticeable effect.

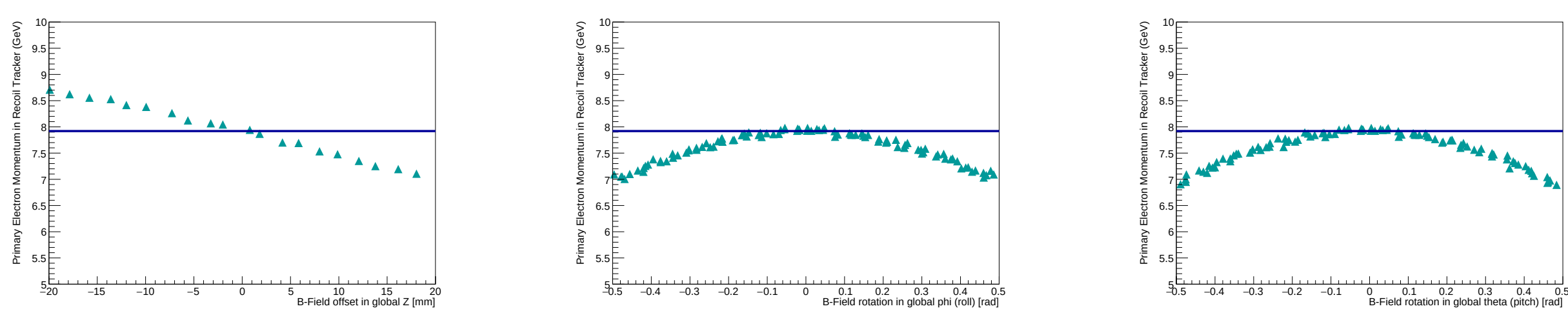

Figure 3.7: The reconstructed scattered beam electron peak momentum vs magnetic field $z$-displacement (left), roll angle (middle), and pitch angle (right). The horizontal line corresponds to the nominal peak momentum.

We anticipate mapping the magnetic field with a spatial precision of 1 mm or better and 1% or better in field value. Based on the above studies, this would result in distortions in momentum of roughly 1% throughout the acceptance, where an overall offset would be easily removed via calibration. The resulting shifts in momentum scale and degradation of resolution have a negligible impact on the physics performance of the apparatus, so we consider the physics performance requirement to be satisfied by the design.

#### 3.3.5.2 Technical Performance Validation

The performance studies for the experiment, and in particular the momentum resolutions of the tagging and recoil trackers, depend on achieving the design field intensity with the spectrometer dipole. This is characterized by a central field, at the center of the magnet bore, of 1.5 T. There are a number of identical 18D36 Mk II magnets that have been operated at this field, including the magnet used by the HPS experiment at JLab, so we consider this technical requirement of the design to be well demonstrated.

Similarly, studies of the backgrounds assume that the vacuum window has negligible thickness compared to the planes of the Trigger Scintillator system and Silicon Tagging Tracker upstream of the target. Typical vacuum window designs, in both aluminum and titanium, are well known to meet this requirement within standard safety factors, so we consider this technical requirement to be well demonstrated.

### 3.3.6 Risks and Opportunities

The key opportunity in the Beamline and Magnet sub-project is the opportunistic use of LCLS-II beam and End Station A – LESA itself – a unique capability at SLAC without which the experiment cannot be done. Another major opportunity is the presence at SLAC of magnets that can be refurbished for LDMX, and power supplies and cooling in End Station A that are capable of operating the magnet.

There are also several risks in the Beamline and Magnet sub-project that are included in the risk register with mitigations. First, there is a risk that the magnet we plan to use for the experiment will be found to be inoperable and unfeasible to repair. In this case, one of the other two magnets of the same type could be used. Second, until there is a completed and approved radiation protection plan for the experiment, there remains the possibility that changes or additions to the design will be needed to meet the RP requirements. Finally, there is the possibility that the upstream switchgear maintenance is more involved than expected, increasing the scope and cost of that work.

## 3.4 Trackers

### 3.4.1 Introduction and Overview

The missing momentum concept for detecting the production of dark matter relies upon precisely estimating the contents and kinematics of the initial and final state for each event. Charged-particle tracking plays two important roles in the LDMX realization of this concept. First, tracking upstream of the target defines the initial state of each beam pulse – the number of incoming beam electrons and their positions and four-momenta at the target – while minimally interfering with those incoming particles. Second, tracking downstream of the target defines the trajectories and four-momenta of the outgoing particles, necessary to estimate the momentum change across the target and determine the charged particle content of the final state associated with each incoming electron.

These two tracking spectrometers, called the Silicon Tagging Tracker (STT) and Silicon Recoil Tracker (SRT), are shown together with the Trigger Scintillation and Target hardware in Fig. 3.8. The STT and SRT have layouts that are optimized for very different purposes, but employ the same sensors and readout components, and share mechanical, cooling, power, and data acquisition infrastructure. In particular, the STT is long and narrow, with measurement stations arranged along the incoming beam trajectory through the center of the bore of the spectrometer magnet. This layout, which places the layers in a large uniform field, selects against low-energy beam impurities and is ideal for providing a robust and precise estimate of incoming electron momentum. In contrast, the SRT sits in the downstream fringe field of the magnet. It is as short and wide as possible to enable a high acceptance measurement of low-momentum particles at wide angles in a compact space. This layout, in turn, allows the ECal to be placed close to the target, maximizing its acceptance for a given size.

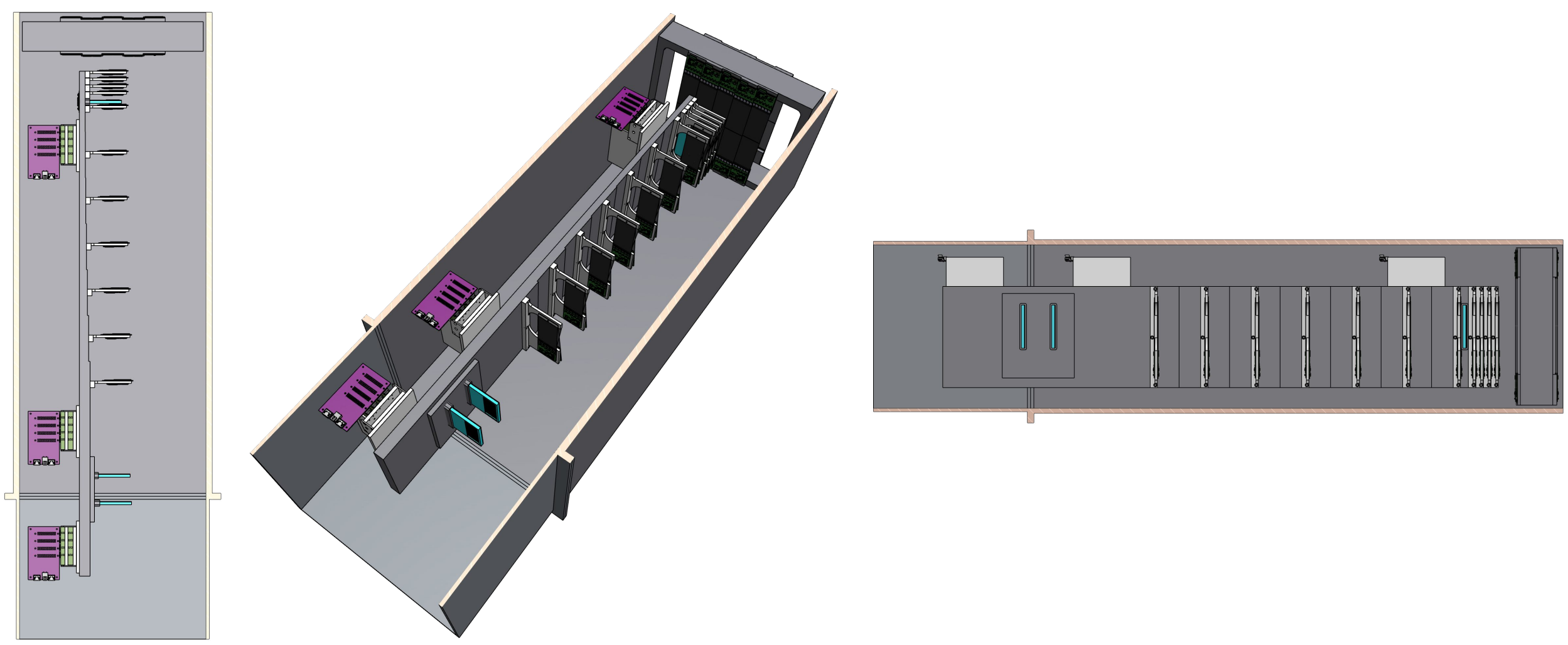

Figure 3.8: An overview of the trackers (STT and SRT), the Trigger Scintillator (TS), and target systems inside the Tracker Support Box from above (left), a quarter view from upstream (middle), and the side (right). The beam enters from the bottom/left, passing through the two upstream TS stations (light blue), the seven STT stations, the TS target station and target (light blue), and the layers of the SRT before passing into the ECal. The readout electronics for the TS are also shown on the back side of the Stereo Support Plate of the tracker.

The STT and SRT utilize technologies developed for the LHC experiments, and recast designs from the Silicon Vertex Tracker of the Heavy Photon Search experiment, the HPS SVT, which was designed and built at SLAC. These designs include the sensors, the front end readout electronics, the support and cooling structures, the control and data acquisition electronics, and the power, cooling, monitoring and interlock systems. The requirements of these components for the LDMX trackers are in all cases easily met or exceeded by the HPS designs, where the SVT has been in operation at JLab since 2015.

### 3.4.2 Requirements

There are a relatively small number of well defined requirements for the STT and SRT, which ensure that the LDMX apparatus is capable of meeting the physics goals of the experiment.

First, because the purity of the incoming beam cannot be absolute, it is critical for the STT to correctly identify incoming particles that would be in the low-energy signal region if correctly reconstructed by the SRT, ECal, and HCal. In particular, an event where an incoming electron with energy less than 30% of the beam energy is mistakenly reconstructed by the STT with an energy and trajectory consistent with a full energy beam electron is a background with no other veto.

Second, the SRT should assist the ECal in correctly identifying non-interacting beam electrons as background when they have greater than 30% of the incoming beam energy. While the ECal alone must perform this missing energy veto in events where the majority of the outgoing energy is in a hard bremsstrahlung photon (order 1% of incoming electrons), the tracker can help provide the last 1-2 orders of magnitude of rejection in the case where the electron retains most of its energy.

Third, in order to reject any backgrounds remaining after ECal and HCal vetoes, and to allow for the reconstruction of the mediator mass in case of a signal observation, the combination of the STT and SRT should reconstruct the momentum kick imparted to the electron by the hard interaction in the target as precisely as possible. The useful precision is limited by the irreducible smearing from multiple scattering in the target.

Finally, the STT and SRT must have sufficient acceptance and efficiency for signal events to allow the experiment to reach its goals for sensitivity in a reasonable period of operations.

The physics requirements above flow down into a set of technical requirements. These are established and validated by simulation studies so that when met, the proposed designs for the STT and SRT meet the aforementioned physics requirements. In addition, there are some technical requirements for the tracker that must be met with respect to the entire LDMX apparatus, in particular requirements on readout latency and speed. Formal statements of these requirements for the tracking detectors are as follows:

- **Tracker Detector-Physics requirements**

TRK1 The STT shall have full acceptance for non-interacting beam electrons with a reconstruction efficiency of at least 95%.

TRK2 The STT shall have a fake rate for identifying incoming electrons with less than 30% of the beam energy as full energy beam electrons of less than $10^{-13}$, with the understanding that off-energy beam impurities will be less than one per mil.

TRK3 The SRT shall have at least 45% acceptance$\times$efficiency for signal recoils for the full range of mediator masses from 1 MeV to 1 GeV.

TRK4 The SRT shall have a fake rate for reconstructing non-interacting electrons as having less than 30% of the beam energy of less than 1%.

TRK5 The STT and SRT shall have a combined resolution in the transverse momentum at the target ($\delta_{p_T}$) smaller than the smearing in this quantity due to multiple scattering in the target, which will be at least 10% $X_0$ for all configurations.

- **Tracker Technical Requirements**

TRK6 The planes of the STT and SRT shall have a material budget averaging less than 0.5% $X_0$ in the tracking volume.

TRK7 The single-hit efficiency for all planes in the STT and SRT to reconstruct hits from individual minimum-ionizing particles and assign them to the correct beam bunch with time information shall be at least 98%.

TRK8 The single-hit position resolution in all planes of the STT and SRT shall be 15$\mu$m RMS or smaller.

TRK9 The STT and SRT shall be capable of sustaining an average readout rate of 25 kHz for random triggers with less than 5% dead time.

TRK10 The STT and SRT shall allow a minimum of 3.8 $\mu$s latency for readout of triggered events.

### 3.4.3 Design

The next two subsections provide an overview of the geometry and main components of the SRT and STT. These are followed by more detailed descriptions of the key sub-components of STT and SRT, along with the support, cooling, readout electronics, power, monitoring, and controls for the tracking systems.

#### 3.4.3.1 Silicon Tagging Tracker (STT) Geometry and Overview

Ideally, every incoming beam electron would have precisely the energy of the LCLS-II drive beam and be along a trajectory precisely defined by the beam optics. In practice, beam halo produced by passage through upstream beamline components will produce a small population of incoming particles that do not have the expected beam energy (from bremsstrahlung), are somewhat off-trajectory (from scattering), or perhaps are not even electrons (from hard interactions). Therefore, LDMX employs the Silicon Tagging Tracker (STT) to measure the energy and trajectory of each incoming charged particle, where that information can be used to eliminate suspect events and ensure the precision of the missing momentum measurement.

Because the STT alone must confirm the incoming energy of a high statistics sample, the design emphasizes robustness of this measurement above all else. Towards this end, the STT layout, shown in Fig. 3.9, and summarized in Table 3.2 has the following key attributes:

| **STT Layout** | | | | | | | |
|---|---|---|---|---|---|---|---|
| Layer | L1 | L2 | L3 | L4 | L5 | L6 | L7 |
| $z$ w.r.t. target (cm) | -61.5 | -51.5 | -41.5 | -31.5 | -21.5 | -11.5 | -1.5 |
| stereo angle (mrad) | +100 | -100 | +100 | -100 | +100 | -100 | +100 |
| $x$-resolution (horiz.) | $\approx 7\mu$m | $\approx 7\mu$m | $\approx 7\mu$m | $\approx 7\mu$m | $\approx 7\mu$m | $\approx 7\mu$m | $\approx 7\mu$m |
| $y$-resolution (vert.) | $\approx 100\mu$m | $\approx 100\mu$m | $\approx 100\mu$m | $\approx 100\mu$m | $\approx 100\mu$m | $\approx 100\mu$m | $\approx 100\mu$m |

Table 3.2

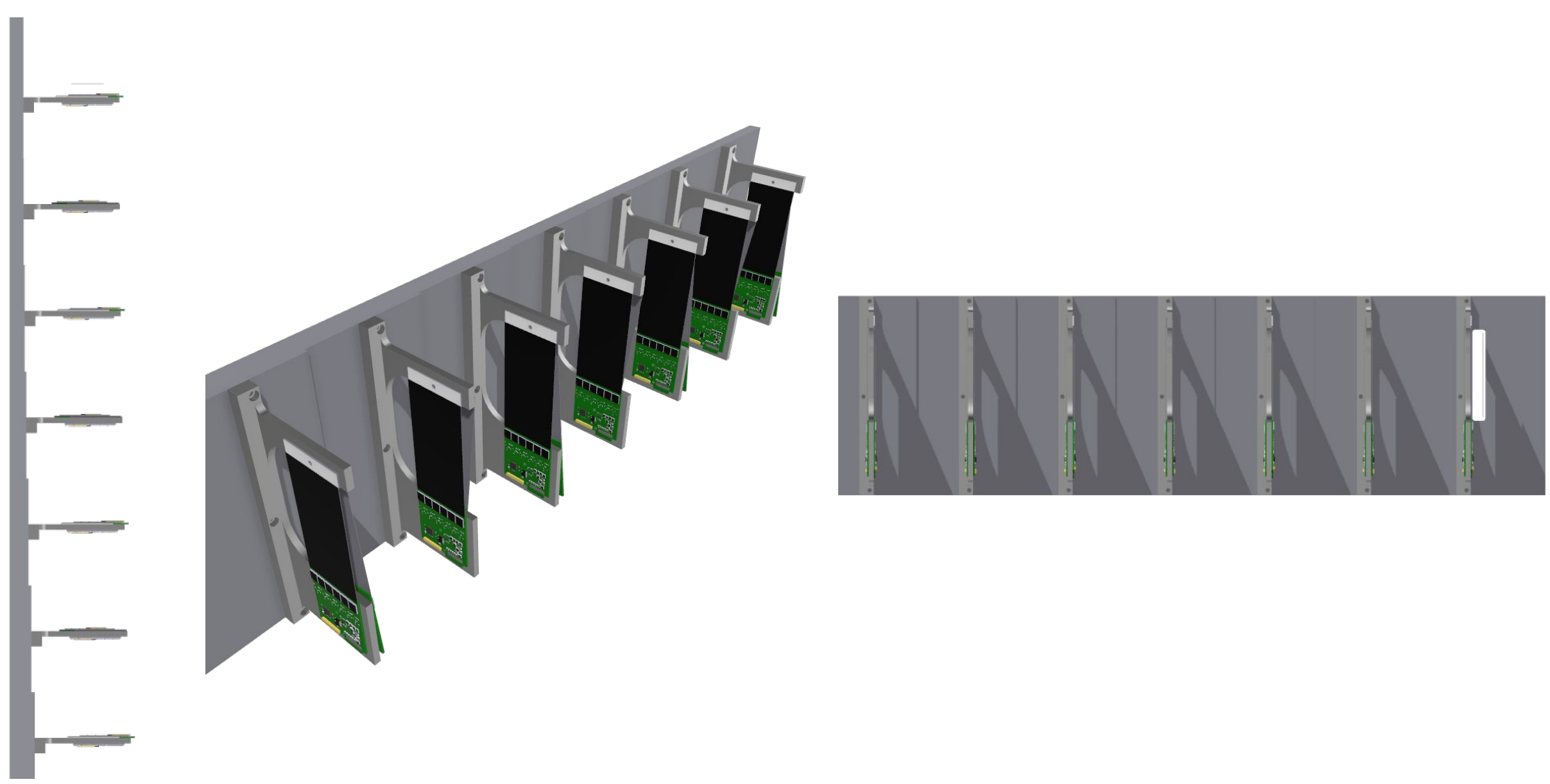

Figure 3.9: The seven Stereo Stations of the STT mounted on the stepped Stereo Support Plate that centers them along the trajectory of the 8 GeV beam in the vertical 1.5 T central field of the dipole magnet, (left) from above, (middle) the quarter view from upstream, and (right) from the side. The slot in the Stereo Support Plate accommodates insertion of the target TS assembly from the back side of the plate.

1. The STT is a long and narrow telescope, so that any particle with energy near the signal region for recoils – $< 30\%$ of the beam energy – are bent out of the acceptance.
2. With no material on the "electron side" – the side towards which degraded low-energy beam electrons are bent in the magnetic field – the STT cannot scatter or shower particles back into the tracking volume or downstream detectors.
3. The first layer of the STT is downstream of the leading edge of the magnetic field, so that lower energy particles are bent away before reaching it.
4. The STT consists of seven 3-d measurement stations, spaced far apart in the full central field of the dipole magnet, to provide a precise and robust measurement of momentum.
5. The last layer of the STT is positioned very close to the target, to minimize the impact parameter and transverse momentum uncertainty at the target.

The spacepoint measurements of the STT are made by Stereo Stations that consist of a pair of silicon microstrip modules placed back-to-back with one "axial" sensor, having strips along the direction of the magnetic field (vertical), and a stereo partner with strips at 100 mrad angle to the axial layer. The sign of the stereo angle alternates layer by layer to break the degeneracy that would create ghost tracks. The axial measurements provide the curvature measurement for a precise estimate of the energy. The stereo measurements provide an estimate of the pitch of the helical trajectory, necessary to determine the 3-d direction of the track at the target.

The design of these Stereo Stations descends from the design of similar components built for the HPS tracker and are assembled by fastening a pair of Stereo Modules on either side of an aluminum support and cooling frame. The Stereo Modules consist of a single sensor with a hybrid circuit board at one end that hosts the front-end readout ASICs, and a ceramic mounting tab at the other end. With all readout material and heat loads beyond the end of the sensor, there is no cooling or support material in the tracking volume, only the silicon sensor itself. This configuration, enabled by the choice of strips instead of pixels, gives a significant advantage for low-momentum tracking at high repetition rates, where more massive, complicated, and expensive hybrid pixels would often be used. With low occupancies per bunch, ghosting and poor track purity – the major drawbacks of strips relative to pixels in high-occupancy environments – are not significant issues.

The Stereo Stations are mounted on a stepped Stereo Support Plate that places them so that they are centered along the ideal trajectory of the incoming beam. The dimensions of the sensors are such that they extend $\approx$1 cm beyond the extent of the nominal beamspot on all sides, to ensure that they fully contain the beamspot and keep dead material away from the beam halo. The Stereo Support Plate has a groove on the back side. A copper cooling tube is pressed into the groove. Water flows through the cooling tube to cool the modules. This support and cooling plate also supports the stations of the Trigger Scintillator system and the tungsten target assembly, described in Sections 3.5 and 3.6.

The Stereo Support Plate is in turn mounted inside of a Tracker Support Box, which holds the entirety of the STT and SRT, along with the target and the planes of the Trigger Scintillator system. The Tracker Support Box is itself mounted inside the bore of the Spectrometer Magnet.

#### 3.4.3.2 Silicon Recoil Tracker (SRT) Geometry and Overview

Providing good acceptance for signal events across the full mass range requires tracking down to the lowest possible momentum and at the widest possible angles relative to the incoming electron. The Silicon Recoil Tracker (SRT) provides this capability in a small space by arranging the same modules as the STT, along with others that are similar, in the fringe field downstream of the STT and target and directly in front of the ECal. Layers in the rapidly falling fringe field are placed so that the field integral ($\sim \int zB(z)\mathrm{d}z$) between successive layers is similar throughout the SRT, with closely spaced layers near the target, and wider spacing towards the back. The design of the SRT is shown in Fig. 3.10 and summarized in Table 3.3.

The first four layers of the SRT use the same Stereo Stations as the STT, mounted to the same Stereo Support Plate as the STT stations. They are centered along the direction of the nominal beam at the target to provide charge-symmetric coverage for outgoing particles. These stations are spaced as close together as feasible to provide 3-d tracking information at the target with the largest angular acceptance in the shortest possible system. Placed in a region where the field is nearly the full central field of the dipole, these closely spaced layers also provide curvature information for very low momentum tracks that could curl out of the acceptance before reaching the axial layers.

| SRT Layout | | | | | | |
|---|---|---|---|---|---|---|
| Layer | L1 | L2 | L3 | L4 | L5 | L6 |
| $z$ w.r.t. target (cm) | +1.5 | +3.0 | +4.5 | +6.0 | +10.0 | +18.0 |
| stereo angle (mrad) | +100 | -100 | +100 | -100 | — | — |
| $x$-resolution (horiz.) | $\approx 7\mu$m | $\approx 7\mu$m | $\approx 7\mu$m | $\approx 7\mu$m | $\approx 7\mu$m | $\approx 7\mu$m |
| $y$-resolution (vert.) | $\approx 100\mu$m | $\approx 100\mu$m | $\approx 100\mu$m | $\approx 100\mu$ | — | — |

Table 3.3

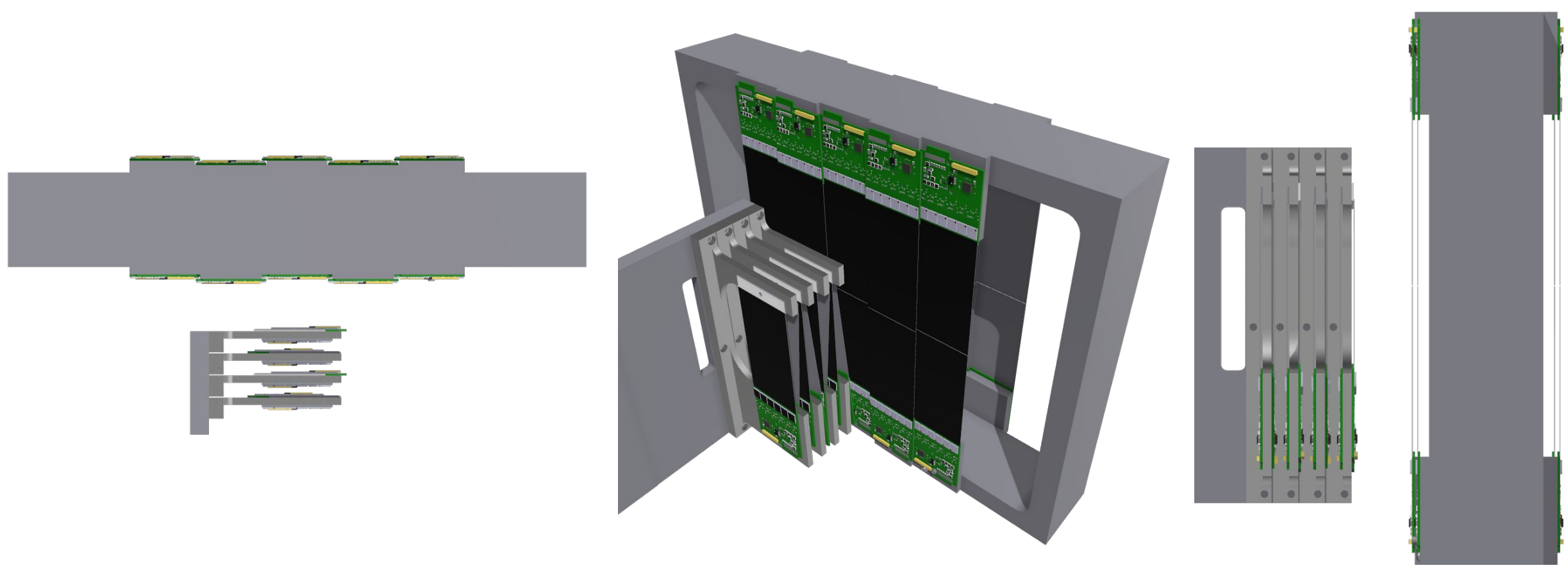

Figure 3.10: The six measurement stations of the SRT, shown from above (left), the quarter view from upstream (middle), and the side (right). The first four Stereo Stations are mounted on the Stereo Support plate immediately behind the TS target assembly and the last two axial-only layers are mounted on either side of a separate castellated support ring. The rear most layer of the SRT is 2 cm away from the front face of the ECal.

The last two layers of the SRT use so-called Axial Modules of a design that is similar to longer modules used in the back layers of the HPS SVT. These modules use the same sensors and readout hybrids as the stereo modules, but place a pair of sensors end-to-end, mechanically joined but electrically isolated, with a hybrid at each end to read out the two sensors separately, as shown in Fig. 3.14. These modules are placed with their strips parallel to the axial strips of the Stereo Stations, and do not have stereo partners. They are responsible only for improving the curvature measurement for higher-momentum tracks while minimizing multiple scattering of lower-momentum tracks – down to 50 MeV/$c$ – that point into the ECal.

The Axial Modules of the SRT are mounted on either side of the rectangular Axial Support Ring on castellated surfaces to provide overlaps between adjacent modules. Like the Stereo Support Plate, the Axial Support Ring has an embedded copper cooling pipe to remove heat produced by the hybrid electronics. It is mounted at the downstream end of the Tracker Support Box that sits inside the magnet bore.

#### 3.4.3.3 Silicon Sensors

As in HPS, LDMX will use standard, radiation-tolerant $p^+$-in-n bulk silicon microstrip sensors manufactured by the Hamamatsu Photonics Corporation (HPK). The attributes of this design, to be implemented on 6-inch wafers, are shown in Table 3.4. The sensors can also be efficiently fabricated on 8-inch wafers, depending on the standard process technology at the time an order is placed, as shown in Fig. 3.11. These parameters – both electrical and mechanical – are patterned after and nearly identical to the sensors that are used in HPS. The HPS sensors were designed and fabricated on 4-inch wafers for the RunII upgrade of the DØ experiment over twenty years ago. The main difference is that the LDMX tracker sensors have 767 readout strips rather than 639, allowed by fabrication on larger wafers. While the radiation tolerance of this technology is modest, with a maximum exposure of $10^{16}$ electrons in a 200 mm×400 mm beam spot and minimal secondaries, the peak flux is $6 \times 10^{14}$ electrons/cm$^2$. The resulting bulk damage corresponds to a flux of roughly $2 \times 10^{13}$ 1 MeV $n_{eq}/cm^2$, roughly 2 orders of magnitude less exposure than similar sensors experience in the HPS experiment, and low enough that the silicon itself does not require cooling to deal with $I^2R$ heating from leakage current or very high voltages to fully deplete the silicon after irradiation.

| Tracker Sensor Design | |
|---|---|
| attribute | Value |
| type | $p^+$-in-n bulk |
| full depletion voltage | ≈65 V |
| breakdown voltage | >250 V |
| bad channel rate | < 1% |
| process | HPK 6 or 8 inch |
| strip pitch | 60 μm |
| # strips | 767 |
| active area | 46.02 mm × 94 mm |
| diced dimensions | 48.02 mm × 96 mm |

Table 3.4: Design specifications for the silicon microstrip sensors used throughout the STT and SRT.

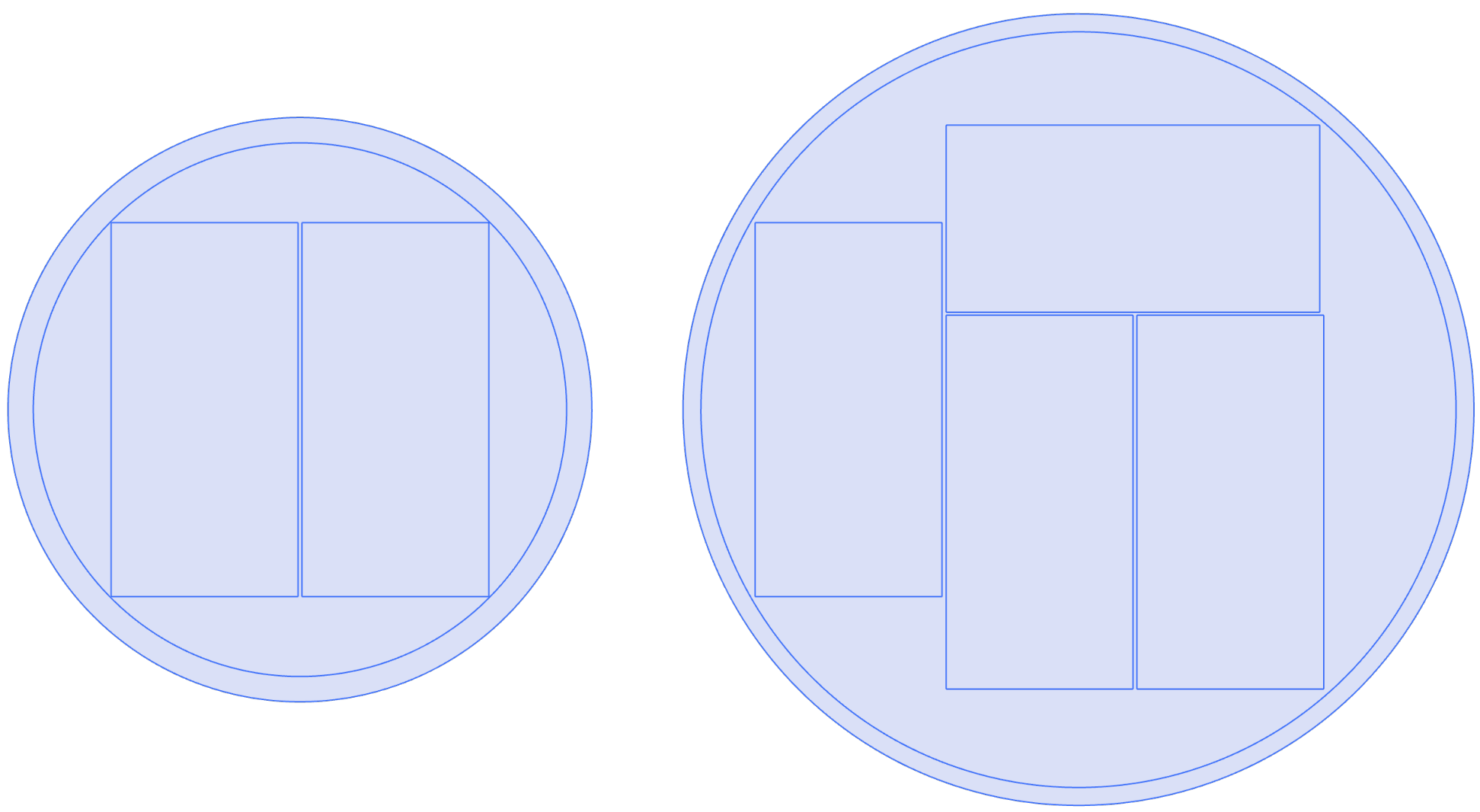

Figure 3.11: The layout of sensors for the trackers within the usable region of 6"(left) and 8"(right) wafers.

#### 3.4.3.4 Front End Readout Hybrids

The front-end readout hybrids for the tracker modules are patterned after the most recent HPS designs, using the exact same schematic blocks, components, connection scheme, and geometry. The only obvious difference is that the LDMX hybrids host six APV25 [80] readout chips rather than five to accommodate connection to the wider LDMX sensors, resulting in a slightly more efficient design. Fig. 3.12 compares the layout of the LDMX hybrid to a similar design from the HPS SVT. The hybrid connects to the rest of the readout chain with a pair of Hirose connectors for power, slow control, and data, both of which connect through cables to the Front End Board (FEB) described in Sec. 3.4.3.9. Along the front edge of the hybrid, a set of six APV25 readout ASICs are mounted that connect to the sensor via standard 1 mil, wedge-bonded, Al-1% Si wirebonds. The sensor mounts to a recessed ledge in front of the chips that places the top of the sensor at the same height as the mounting surface for the APV25 chips, where the pads for biasing the chip are. This ledge has a large-area pad to supply bias voltage to the sensor backside via conductive adhesive.

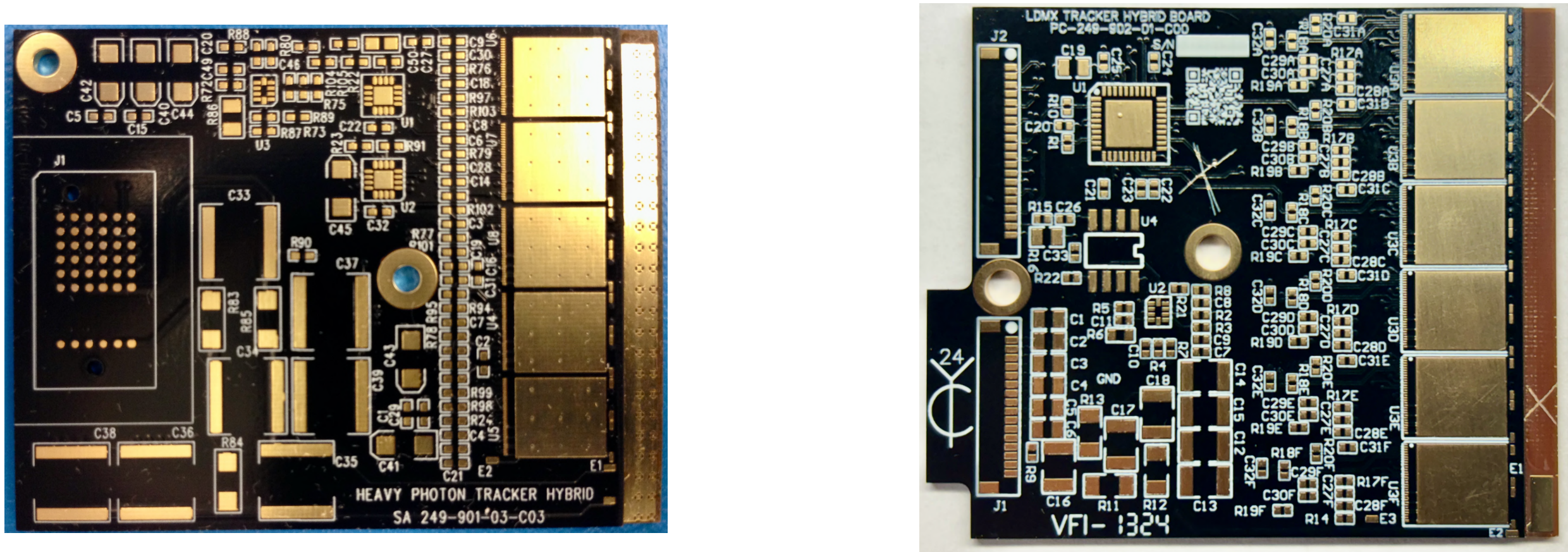

Figure 3.12: Left: The most recent design of APV25 readout hybrids used in the HPS SVT. Right: The prototype of of the readout hybrids for the STT and SRT, shown after QA testing of the solder sample.

Selection of the APV25 readout ASIC leverages extensive design work and experience at SLAC gained through its use in the HPS SVT, where it was chosen due to its availability, low cost, and flexible readout capabilities. With a pipeline depth of 192 cells and operating a 37.14 MHz, trigger latencies of up to 3.8 $\mu$s can be supported. The APV25 has a so-called “multi-peak" readout mode, where successive samples of the front-end amplifier output can be read out from the pipeline, in groups of three, by a single trigger signal. Fitting these samples with the known pulse shape, obtained via channel-by-channel calibration, allows for reconstruction of both the amplitude of the signal and the hit time with much greater precision than the sample (clock) period. In LDMX, this capability offers a useful handle in eliminating events resulting from out-of-time dark current electrons. The APV25 chips – left over from construction of the CMS trackers – are extremely inexpensive ($\sim$\$10 each), and SLAC has extensive experience and well-developed designs for the supporting electronics, data acquisition, calibration, and reconstruction infrastructure, including detailed Monte Carlo simulation for their operational characteristics.

#### 3.4.3.5 Stereo Stations and Modules

The Stereo Station design, shown in Fig. 3.13, is used throughout the STT and for the front layers of the SRT and descends directly from designs used in the HPS SVT. A Station is composed of a pair of identical Stereo Modules sandwiched around an aluminum support and cooling plate. The modules are mounted to these cooling plates with screws, allowing for easy replacement and rework of the tracking system after initial assembly.

The Stereo Modules are the smallest indivisible functional units of the system, and are comprised of three main components: the sensors and front-end readout hybrids described in the previous sections and a small, alumina ceramic mounting plate with a mounting hole. The sensor is glued to the ledge along the front end of the hybrid using a combination of electrically conductive silver-filled epoxy (Epoxies Inc. 40-3905) and structural epoxy (Hysol 9396): A small dot of silver epoxy connects the sensor backside to the HV bias pad and the structural epoxy fills the rest of the glue gap. At the end of the sensor opposite the hybrid, the ceramic plate – having the same thickness as the hybrid ledge (1 mm) – provides a flat, non-conductive mounting surface for the half-module to the aluminum support and cooling plate.

The Stereo Module Support, shown in Fig. 3.13 is machined from a single piece of aluminum, and has a set of tapped holes for mounting the modules. These module supports also have features for precision mounting to the Stereo Support Plate that positions the stereo layers along the beamline. There are a set of three counter-bored clearance holes through the base, and a precision-bored hole and elongated slot in the mounting surface that mate with positioning dowel pins pressed into precisely-positioned holes in the Stereo Support Plate. The Stereo Modules are precisely positioned relative to these mounting features during assembly using a fixture described in section 3.4.4. These features fully constrain all six degrees of freedom in mounting the modules to the stepped surfaces of the support plate that define the precise positions of the sensors along the beamline.

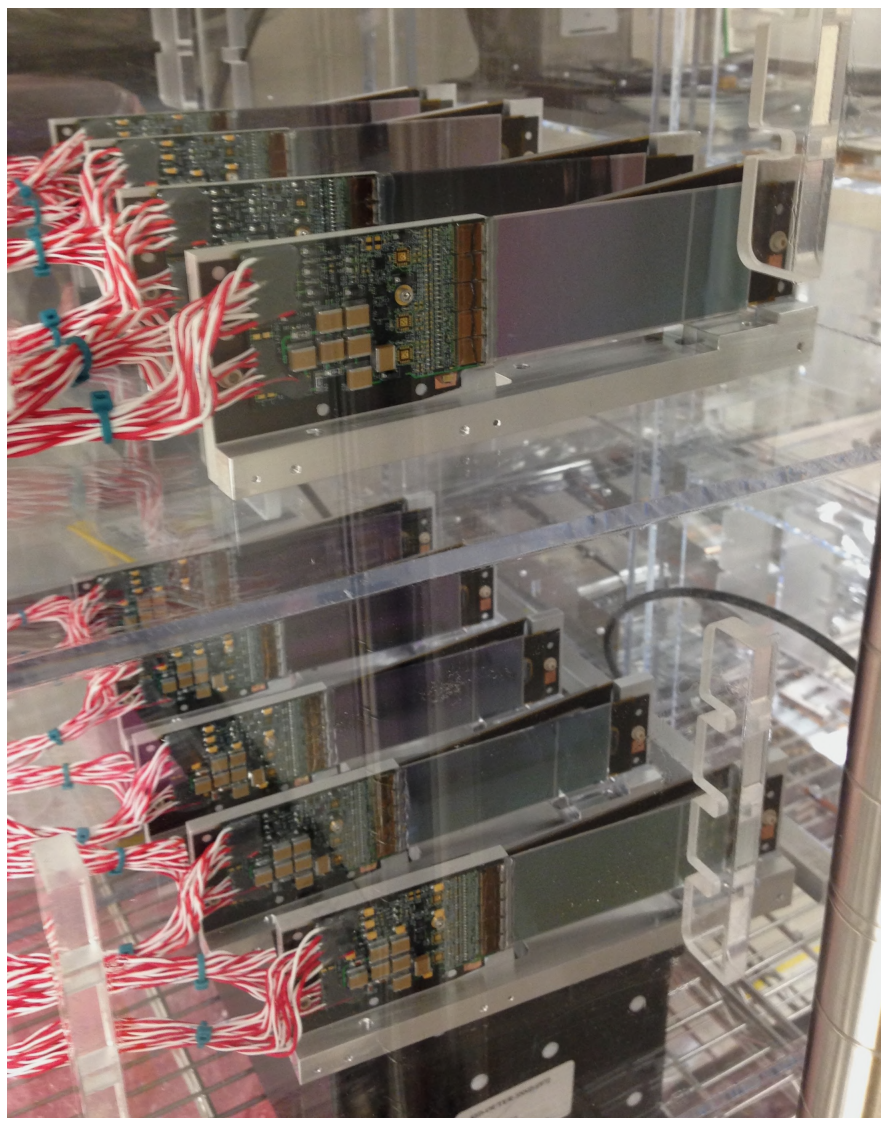
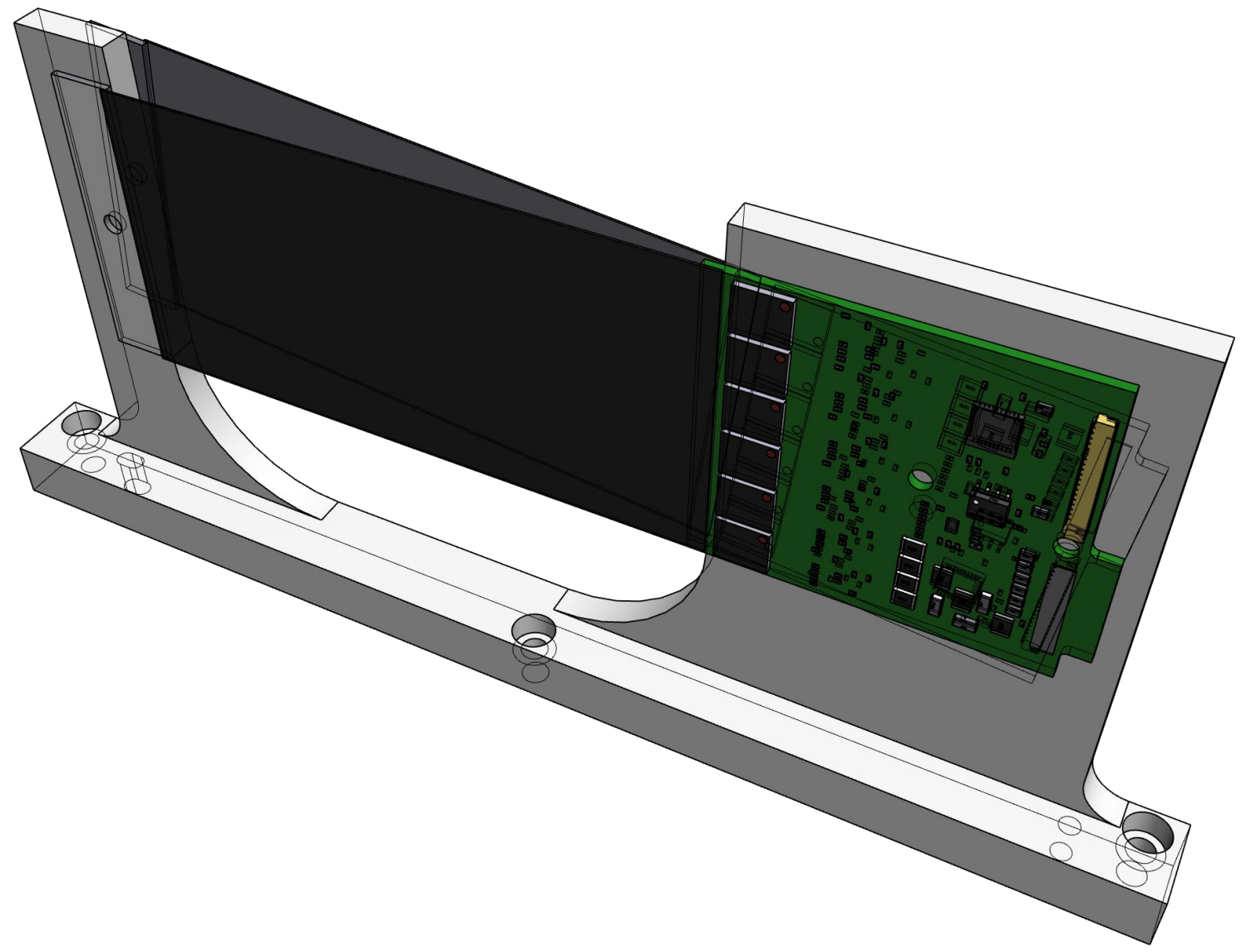

Figure 3.13: Left: The shorter, 100 mrad Stereo Stations of the HPS SVT. Right: The similar design of Stereo Stations for LDMX.

#### 3.4.3.6 Axial Modules

The design of the axial modules borrows from the longer modules used in the rear layers of the HPS SVT, as shown in Fig. 3.14. Using the same sensors and readout hybrids as the stereo half-modules, two of the half-modules are placed end-to-end and coupled mechanically with structural epoxy (Hysol 9396) and a thin (two mil) underlayment of Kapton under the glue joint. A gap of 0.5 mm between the sensors keeps them electrically isolated so that they may be operated individually at different bias voltages.

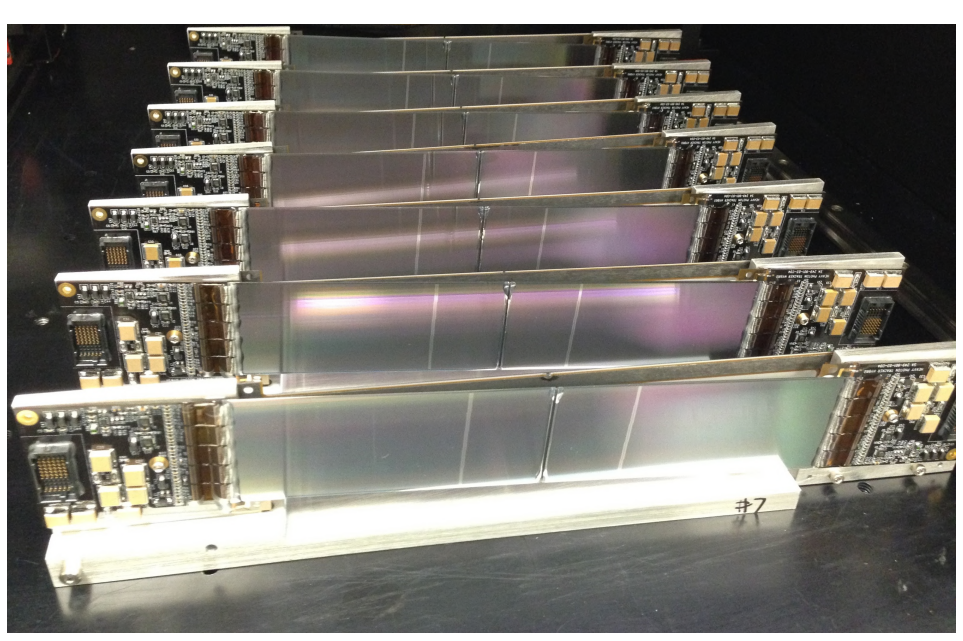
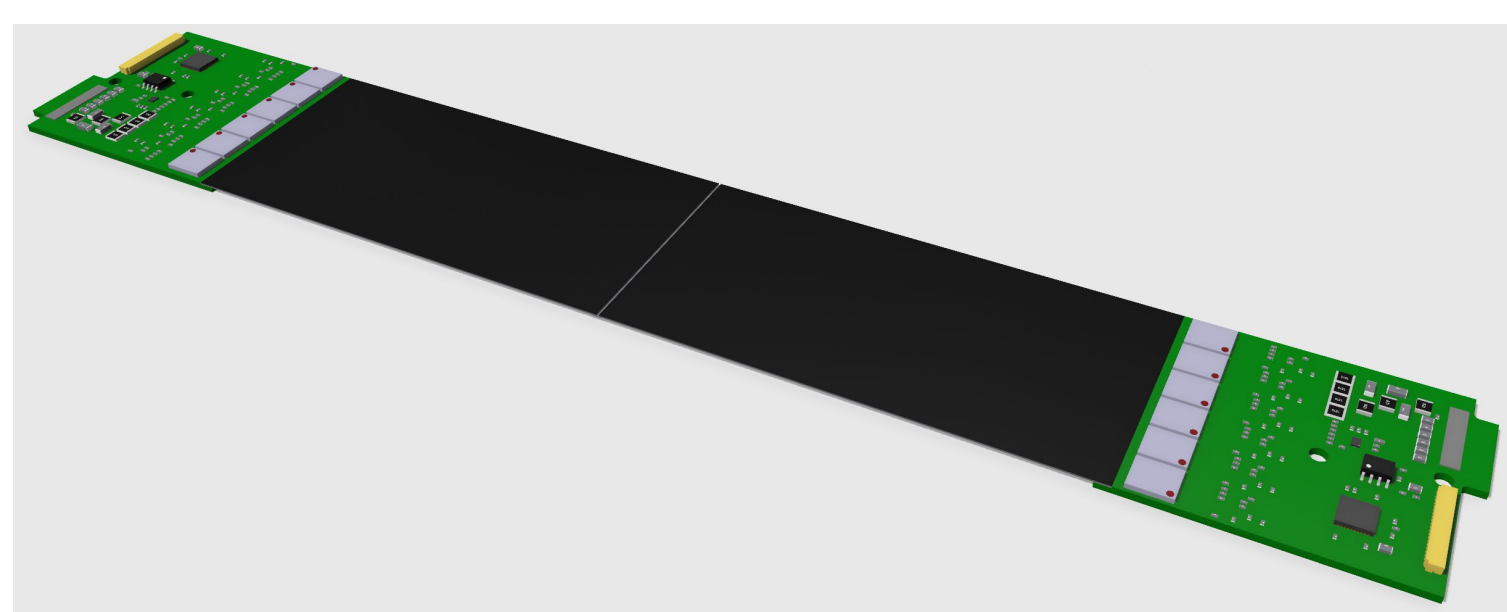

Figure 3.14: Left: the longer, 50 mrad Stereo Stations of the HPS SVT. Right: the design of similar modules for the LDMX SRT axial layers.

#### 3.4.3.7 Module Support and Cooling Structures

The modules of the tracking systems are supported by two separate structures, a Stereo Support Plate for all of the Stereo Stations of both the STT and SRT, and an Axial Support Ring for the Axial Modules of the SRT.

#### 3.4.3.7.1 Stereo Support Plate

The Stereo Support Plate, shown in Fig. 3.15 is a single aluminum plate, stepped on one side to position the stereo modules of the STT along the trajectory of incoming full energy electrons so that the nominal beamspot illuminates the same central regions of the sensors in each layer. At the mounting position for each sensor module is the same pattern of tapped holes and precisely located positioning pins that engage the hole and slot on the mating surface of the module support. Around the perimeter of the flat back side of the plate is a machined groove into which a copper cooling tube is pressed, deploying a technique used in the support structures of the HPS SVT as well as other detector systems built at SLAC for both HEP and photon science. This copper cooling tube is fabricated in standard 1/4" flexible copper tubing terminated with compression fittings. Along the bottom edge of the plate are PTFE bearings for sliding the support plate into the detector along a channel, and a pair of precision ball-ended screws that support the plate in kinematic mounts and allow for adjustment of the pitch of the plate relative to the beam. Along the top edge is a similar PTFE bearing and adjustment screw that allows for roll adjustment with respect to the beam.

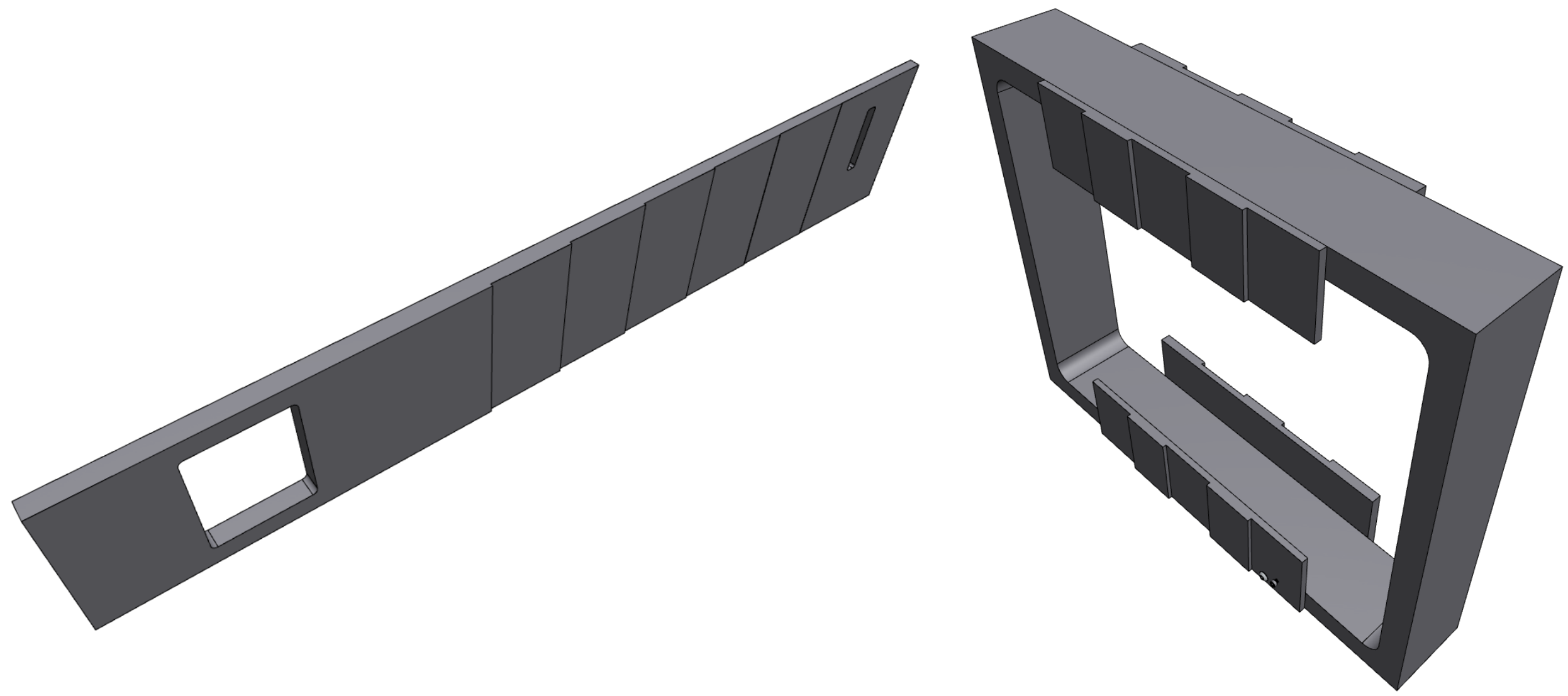

Figure 3.15: Left: The stepped Stereo Support Plate, which supports and cools the Stereo Stations of the STT and SRT, along with the stations and readout electronics of the TS and target system. Right: the Axial Support Ring of the SRT, which includes the ring itself, and four castellated plates mounted with screws and positioning pins to which the Axial Modules attach. Details of the hardware that aid in installing and precisely positioning the Stereo Support Plate and Axial Support Ring in the Tracker Support Box are not shown.

In addition to supporting and cooling the stereo modules, the Stereo Support Plate also supports the Trigger Scintillator system and Target Assembly and precisely positions them relative to the tracker modules, as described in Sec. 3.5 and shown in Fig. 3.8. This includes support of the Trigger Scintillator modules themselves, and the readout electronics, which are mounted on the back side of the Stereo Support Plate, which also provides cooling to those components.

#### 3.4.3.7.2 Axial Support Ring

The Axial Support Ring, shown in Fig. 3.15, is a rectangular ring with castellated mounting surfaces on both sides that support, cool, and precisely position the Axial Modules in the rearmost two layers of the SRT. The modules are mounted with screws and positioned by precisely machined mounting surfaces and positioning pins that contact the hybrids in the same positions as the assembly fixtures for the Axial Modules so that relative sensor positions in the full assembly are well controlled. Like the Stereo Support Plate, the Axial Support Ring features a machined groove around the periphery into which a 1/4" copper cooling loop terminated with compression fittings is pressed. The bottom of the ring has press-fit tungsten carbide kinematic mounts at each corner – one cone and one v-groove – to position the axial layers and constrain all but one degree of freedom. The top of the ring has a press-fit tungsten carbide flat, which positions the final degree of freedom, the pitch relative to the beam.

#### 3.4.3.8 Tracker Support Box

The large-scale support for the STT and SRT, along with the Trigger Scintillator system and the Target Assembly, is provided by the Tracker Support Box, which also serves as an environmental enclosure, keeping both systems in a dark, clean, dry, and RF-shielded environment. This enclosure, shown in Fig. 3.16, is patterned after the HPS SVT support box, with four sides of machined aluminum and aluminized polyimide sheets on the upstream and downstream ends to allow for passage of beam particles.

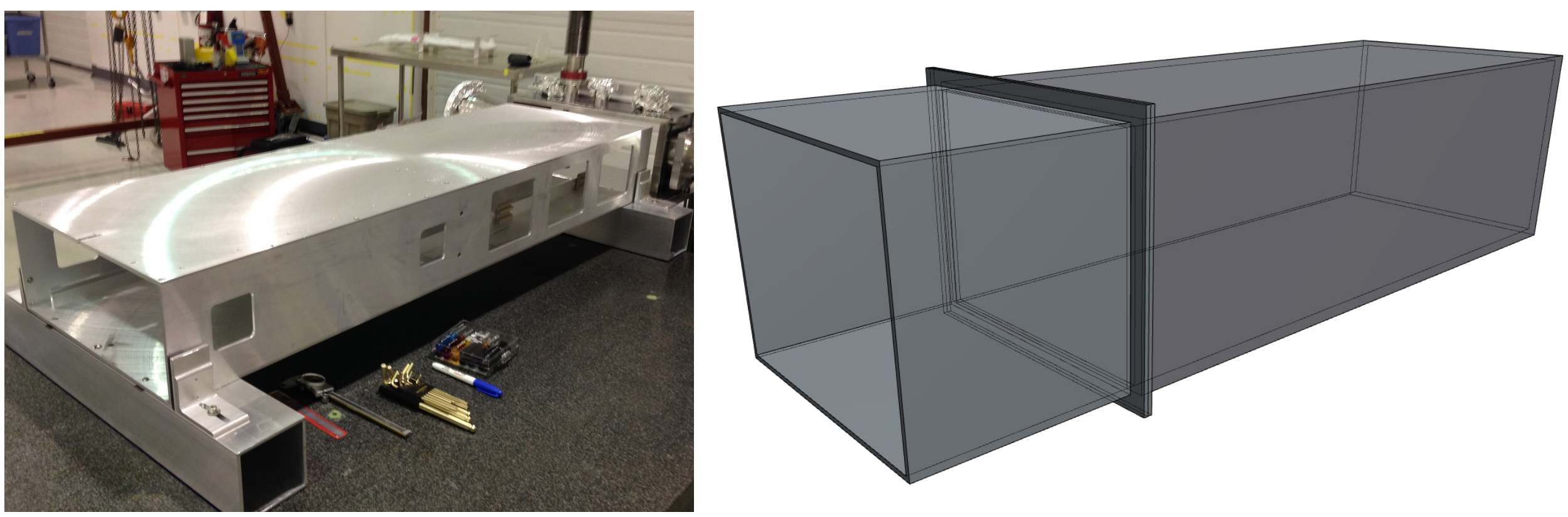

Figure 3.16: Left: The support box that holds the HPS SVT inside the bore of an 18D36 dipole in Hall B at JLab. Right: The similar Tracker Support Box for LDMX that installs inside of the LDMX magnet and holds the Stereo Support Plate and Axial Support Ring. This box has an extension at the upstream end to fully enclose the Stereo Support Plate while allowing for cabling and connection work inside the box from the upstream end with the extension removed. Details, including support and mounting hardware and aluminized polyimide sheets at the upstream and downstream ends, are not shown.

The inside of the box has features that facilitate the installation of and provide support for the Stereo Support Plate and Axial Support Ring. For the Stereo Support Plate, there are a pair of grooves machined in the top and bottom of the box in which the PTFE bearings of the plate slide for installation from the upstream end of the box. Inside of the bottom groove are press-fit kinematic mounts that define the position and orientation of the support plate in all but one degree of freedom. At the top is a press-fit flat that defines the final degree of freedom, roll with respect to the beam. The elevation, pitch, and roll of the plate can be adjusted using the precision ball-ended screws in the support plate that engage these kinematic mounts. A set screw in the support box engages a clearance hole in the plate to lock it safely in position, ensuring that the plate cannot be removed during transportation and installation procedures.

The Axial Support Ring installs from the downstream end of the box and mounts on a pair of precision ball-end screws through the bottom of the box that engage the kinematic mounts in the bottom of the ring, defining the position and orientation of the ring in all but one degree of freedom. At the top of the box, a third such screw contacts the carbide flat, defining the final degree of freedom, the pitch with respect to the beam. As with the stereo support plate, the elevation, pitch, and roll of the ring can be adjusted using these screws and a set screw locks the ring safely in position.

On the “positron side" of the box – the side away from which degraded low-energy electrons curve – are a set of feedthroughs for power, data, and cooling to the tracking and Trigger Scintillator systems inside. At the upstream end are feedthroughs for cooling water and connections for the inlet and outlet of dry air.

Further back along the side of the box are Mini-DSub bulkhead connectors for the tracker modules. Short pigtail connectors connect the modules to these connectors on the inside of the box and from the nearby Front End Boards (FEBs), described in the next section, to these connectors on the outside.

A set of three precision ball-ended screws support the box itself inside of the magnet bore; two at the rear on either side, and one in the front in the center of the box. These locate the box precisely within the magnet bore and allow for fine tuning the position of the box within the magnet in elevation, pitch, and roll to align the tracking systems to the magnet and the ECal.

#### 3.4.3.9 Data Acquisition and Power

There are three main components of the on-detector readout electronics and data acquisition for the trackers: the 42 front end readout hybrids described in Sec. 3.4.3.4, the six Front End Boards (FEBs), and a single Optical Transition Board (OTBs). A high-level block diagram of these components is shown in Fig. 3.17. Since the readout hybrids use the APV25 chip, a very simple front end ASIC which outputs an analog stream, the FEBs and Optical Transition Boards (OTBs) that connect them to the back end DAQ must be placed in close proximity to the modules. As with the modules themselves, the design of these DAQ components descends from HPS SVT designs, as described below.

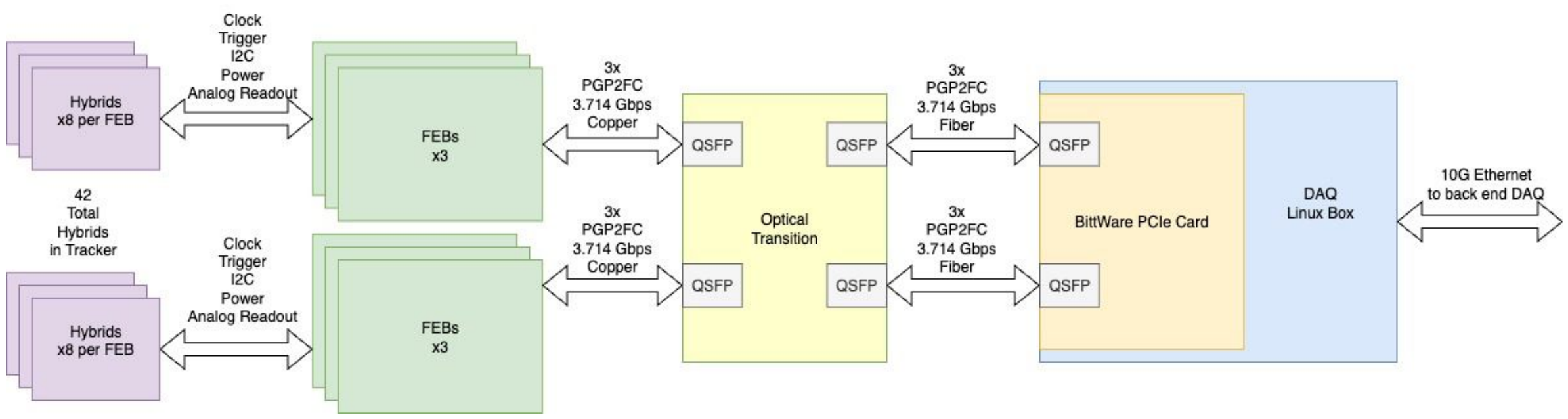

Figure 3.17: A high level block diagram of the readout and on-detector DAQ components of the trackers.

##### 3.4.3.9.1 Front End Boards (FEBs)

The Front End Board (FEB) distributes power and provides clocking and control to the APV25 chips, as well as amplifying, digitizing, and processing the outgoing data streams. In HPS, the FEBs must be located inside of the vacuum chamber along with the rest of the SVT, which places very challenging constraints on the geometry and construction techniques, requiring very dense (20 layer) PCBs, a layout lending itself to active liquid cooling, and highly customized connectors and cabling, while prohibiting the design from offering any spare capacity. The LDMX FEBs are not subject to these constraints, so they can be much larger, less dense, use simpler connectors and cabling, be air cooled, and allow re-mappable allocation of channels with ample spare capacity. Each FEB services up to eight hybrids and sends out data to a single OTB channel. The FEBs connect to the hybrids via custom cables using twisted pair polyamid-imide (PAI) coated magnet wire, and to the OTBs via SFP passive direct-attach copper cables.

Fig. 3.18 shows a block diagram of the FEB. The FEBs are built around a single Artix Ultrascale+ FPGA and take low-voltage power input, with configurable power regulation and distribution of the digital and analog voltages for the individual hybrids as well as for the FEB itself. The FEB configures and controls the hybrids through $I^2C$ interface and includes amplification and digitization for the analog streams of the connected APV25 chips. Simple signal processing and zero-suppression is performed by the FPGA before sending out data via SFP connections to the back end via the OTBs. The FEB also receives high-voltage bias for the sensors and passes them through to the connected modules.

Fig. 3.19 shows the design of the HPS and LDMX FEBs, where the schematic of an individual channel is the same, yet the LDMX FEBs have eight channels instead of the four channels of an HPS FEB. The lack of a severe space constraint and operation in vacuum as in HPS, allow for a much more conservative design for the FEBs. The LDMX FEBs are supported between the magnet coils on the “positron side" of the SVT box, and connect to the trackers via short cables connected to the bulkhead connectors on the Tracker Support Box.

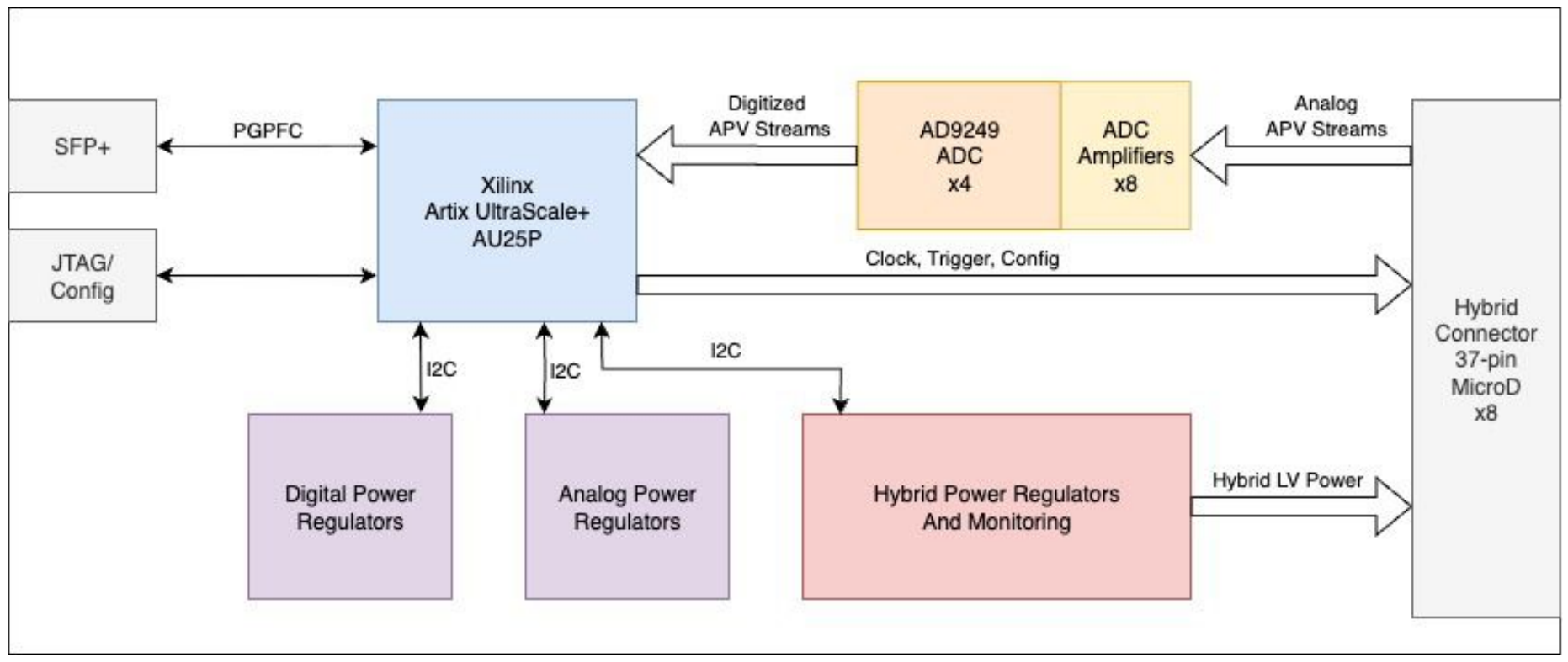

Figure 3.18: Block diagram of Tracker FEB components

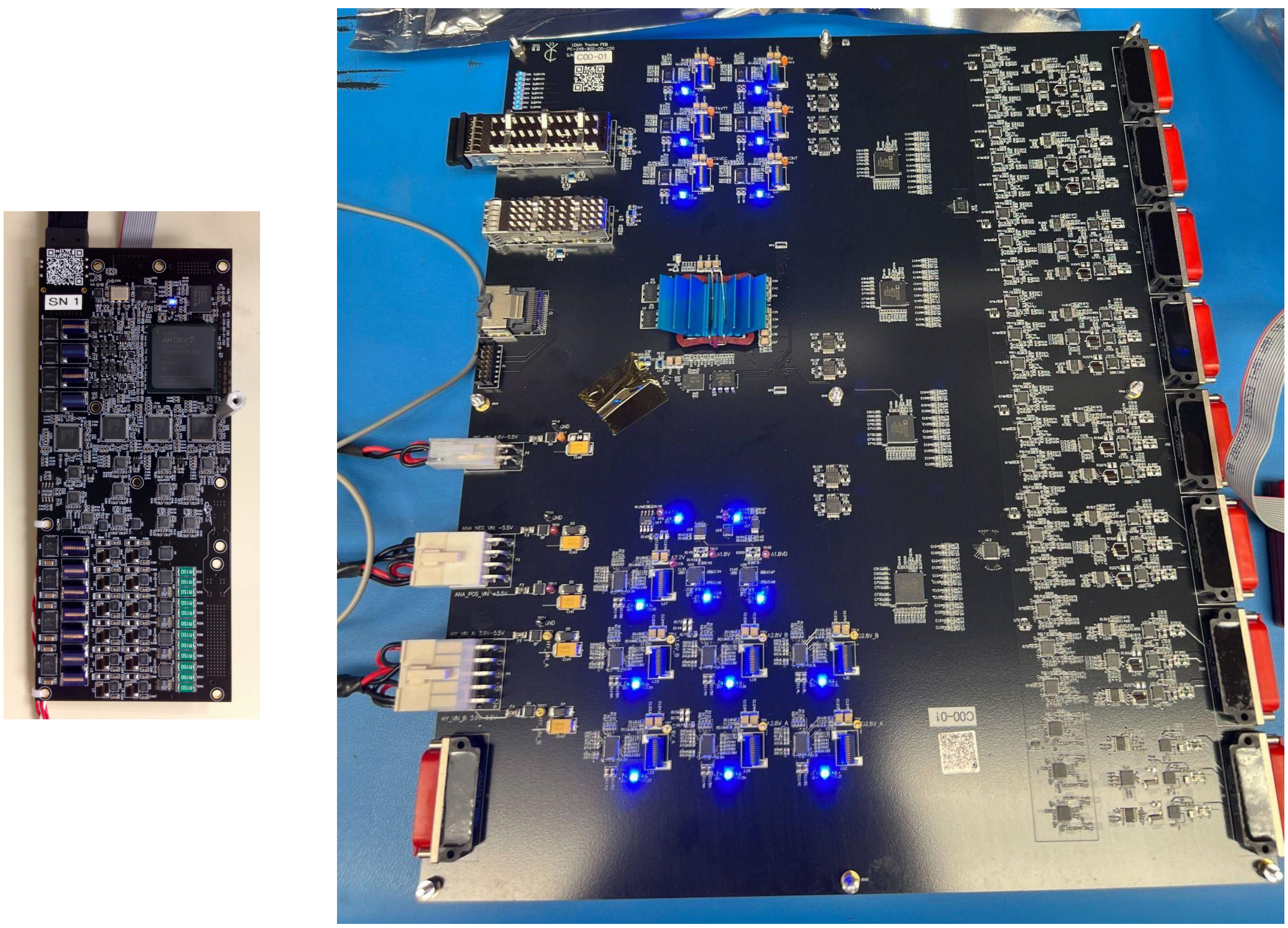

Figure 3.19: Left: The four-channel Front End Board (FEB) of the HPS SVT, which must be installed along with the SVT inside the beam vacuum. Right: Shown at roughly the same scale, the ten-channel FEB for LDMX, which installs outside the Tracker Support Box on the “positron side" between the magnet coils. The lack of severe space constraint, along with operation in air, allows for a simpler and more robust design with greater redundancy and flexibility.

#### 3.4.3.9.2 Optical Transition Boards (OTBs)

In order to communicate with the back end DAQ over long distances, optical transmission is necessary. However, the field within the magnet is too large for reliable operation of optical transceivers. So, a pair of Optical Transition Boards (OTBs), supported outside of the magnet just upstream of the FEBs, hosts optical transceivers that connect to the FEBs via copper cables and to the back end DAQ via optical fibers, as in HPS. The OTB is essentially a dual QSFP pass-through, with connections to FEBs via copper direct attach QSFP cables on one side, and an optical QSFP transceivers to the back end on the other. QSFP to 4x SFP+ fanout cables will be utilized for the connections from the OTB to the FEBs, so that the 6 FEB SFP+ connections are aggregated through 2 QSFP connectors. The design of the OTB is shown in Fig. 3.20 alongside the design from the HPS SVT.

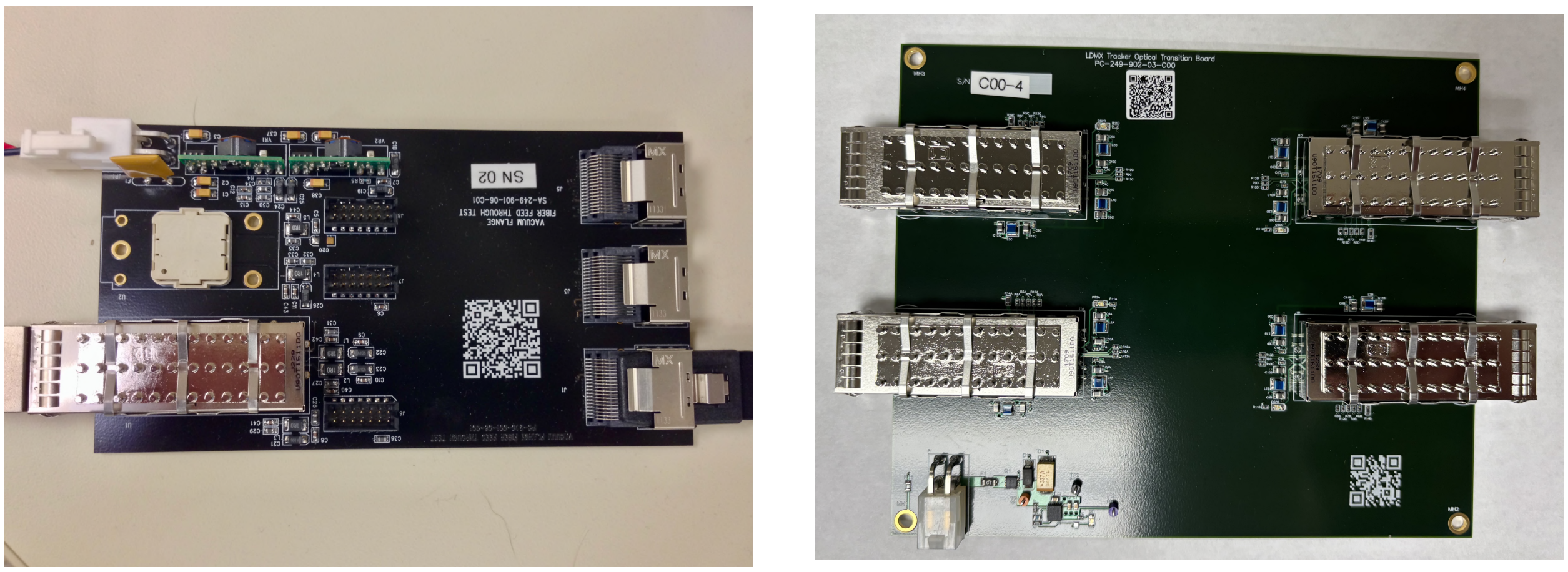

Figure 3.20: Left: The optical transition board for the HPS SVT. Right: An updated OTB optimized for the LDMX trackers.

#### 3.4.3.9.3 Power Supplies

The tracking system requires both low voltage power for the hybrids, FEBs, and OTBs and high voltage power to bias the silicon sensors. Low voltage and high voltage supplies on racks alongside the detector provide power to the FEBs and OTBs, where the FEBs regulate and distribute low voltages and directly pass through high voltage to the hybrids through the analog cables, as in the HPS readout system. As with the HPS SVT, Wiener MPOD power supplies are used for both low voltage and high voltage sensor bias.

#### 3.4.3.10 Cooling and Environmental Control

The tracking and Trigger Scintillator systems inside the tracker support box both require active cooling for heat removal. Fortunately, neither system's performance or radiation tolerance needs require operation at low temperatures, so distilled water may be used. Together, the two systems inside the Tracker Support Box generate a heat load of approximately 200 Watts. Limiting the temperature rise in the cooling loop to 1°C requires roughly 3 L/min cooling flow, and in turn roughly 1 atmosphere of pressure drop for water at 18°C in the cooling circuit.

A Julabo F1000 chiller meets these specifications and supplies cooling water to the detector through flexible copper cooling lines, passing through a UV light system to prevent algae growth and a small manifold to split water between lines for the Stereo Support Plate and the Axial Support Ring. Feedthroughs on the support box have ceramic isolators to electrically isolate the cooling lines from the support box. The support box is purged with dry nitrogen, which is important for operation of the silicon sensors.

The Tracker Support Box is an RF shield and the central ground for both the tracking and Trigger Scintillator systems. The modules of the STT and SRT, as well as the readout electronics for the Trigger Scintillator, are grounded to the Stereo Support Plate and Axial Support Ring, which in turn have grounding straps that connect to the Tracker Support Box. The Tracker Support Box is in turn grounded to the magnet yoke. Meanwhile, as with the cooling, all of the feedthroughs for power and data are isolated from the Tracker Support Box, and the grounds for power supplies are floating at the rack end, making the magnet the single grounding point for the entire tracking and Trigger Scintillator systems. This grounding and shielding scheme mirrors the arrangement used successfully in the HPS experiment.

Unlike HPS where the FEBs are in a vacuum, the FEBs for LDMX can be air-cooled. However, the heat dissipated by the FPGA does require active air cooling to prevent overheating. To provide air cooling for the FEBs, they are mounted in a box with an intake air filter and an extraction fan that ensures sufficient airflow over the board.

#### 3.4.3.11 Control and Interlocks

The trackers are monitored and controlled through standard EPICS interfaces, used widely for detector systems at SLAC and elsewhere. This allows the remote monitoring and control and monitoring of the SVT power and environment, as well as the state of the readout and data acquisition components.

The system includes a number of interlocks that ensure the operational safety of the detector. These include software interlocks implemented via EPICS, as well as hardware interlocks provided through Programmable Logic Controllers (PLC) for the most critical elements. Table 3.5 summarizes the quantities monitored and interlocked for operation of the tracker.

| **Tracker Monitoring and Interlocks** | | |
|---|---|---|
| Monitored Quantity | Interlock Type | Disabled by Interlock |
| chiller state | hard, soft | LV and HV power |
| chiller temperature | hard, soft | LV and HV power |
| coolant flow switch | hard, soft | LV and HV power |
| FEB fan state | hard, soft | LV power |
| FEB board temperatures | soft | LV power |
| FEB FPGA temperatures | soft | LV power |
| beam state | hard | LV and HV power |
| Sensor HV currents | soft | HV power |
| Hybrid temperatures | soft | LV and HV power |
| Hybrid voltages and currents | none | — |
| FEB voltages and currents | none | — |
| OTB voltages and currents | none | — |

Table 3.5: Quantities that are monitored and in many cases interlock the operation of various components of the tracking systems. Monitoring and software interlocks are performed using EPICS while hardware interlocks are implemented via PLCs.

### 3.4.4 Project Plan

The project plan for construction of the tracking systems is based upon a very similar plan carried out for construction of the HPS SVT, which developed the architectures, designs, and construction techniques to be used, and has the same overall scale: the HPS has 40 readout hybrids whereas the LDMX STT and SRT have a total of 42. Construction of the HPS SVT, including much of the design work already completed for LDMX, took roughly two years, and a similar timeline is planned for construction of the LDMX trackers. While the tracking systems are not on the critical path for construction of the LDMX apparatus, it is within a few months of being a significant schedule constraint, so some attention must be paid to the critical path. For the tracking systems, the critical path flows from long-lead procurement of the silicon sensors, through production of the sensor modules, to the final assembly and integration of the STT and SRT prior to installation.

The work breakdown structure (WBS) for the tracker largely factorizes into four sections that can proceed simultaneously and in parallel before final assembly and integration. These four legs of the project are the sensor modules; the support and cooling systems; the readout electronics and power systems; and the control, monitoring, and interlocks. The work plan for the project is summarized in the following sections, where there are generally three types of activity: Design, Procurement and Fabrication of components, and Assembly. QA testing will be discussed separately in Sec. 3.4.4.9, and subsequent installation is discussed in Sec. 3.11.

#### 3.4.4.1 Sensors

The silicon sensors have a relatively long lead time - typically on the order of one year from the beginning of design discussions with HPK to delivery – and are the first element on the critical path for construction of the trackers. As a result, the design and procurement process for the sensors is the highest priority task at construction start. A draft design and specifications for the sensors already in hand will be finalized and sent to HPK for a quote, where sole source procurement on the basis of prior work is anticipated. After basic acceptance testing, the final steps of module production can begin.

#### 3.4.4.2 Hybrids

The design of the hybrid readout board is well developed, and will need only formal acceptance testing and review before the start of production. Other components of the hybrid include the APV25 ASIC, already in hand, and other components available in stock or on short lead times.

Hybrid assembly will take place at SLAC, with some steps subcontracted to assembly labs at UCSC/SCIPP, where much of the wirebonding for the ATLAS ITk upgrade takes place under the supervision of long-time HPS collaborators. After loading, cleaning, and initial testing of the PCB, the hybrids will be sent to UCSC for mounting and back-end wirebonding of the APV25 chips. The hybrids will then be tested and characterized prior to building them into completed modules with the attachment of sensors.

#### 3.4.4.3 Stereo Stations

The first step in producing Stereo Stations is finalizing the design, together with the designs of the sensors, hybrids, and support structures, the set of assembly, storage, and wirebonding fixtures and shipping containers needed to facilitate production. Similar fixtures for HPS are shown in Fig. 3.21. Procurement of the ceramic mounting tabs, hardware, and adhesives required for the modules will follow, along with procurement of the stereo module supports and fixtures, which will likely be contracted to a commercial vendor for CNC machining.

Assembly of modules will take place in the cleanroom at SLAC in an area established for similar work on the HPS SVT, and wirebonding of the sensors to the APV25 will take place at UCSC. After testing and selection, stations will be assembled and surveyed on the OGP Flash 500 multisensor CMM in the cleanroom at SLAC, prior to final assembly and integration on the Stereo Support Plate.

#### 3.4.4.4 Axial Modules

The production of Axial Modules is the same as for the Stereo Modules, beginning with finalizing the design of the Modules and their components, and proceeding through assembly in parallel with the Stereo Modules with the same sequence of steps. While the Stereo Modules are completed by assembly on the aluminum module supports to make Stereo Stations, the Axial Modules remain individual units until they are assembled on the Axial Support Ring during Final Assembly and Integration.

#### 3.4.4.5 Support and Cooling

Support and Cooling includes the Stereo Support Plate, the Axial Support Ring, and the Tracker Support Box, along with all mechanical and cooling components, including the cooling flanges, cooling lines, and chiller. Design work on these components will be completed in coordination with final design of the Trigger Scintillator and Magnet mechanics which have interfaces with these components. Fabrication of larger components such as CNC machining will take place either in shops at SLAC or a commercial vendor depending upon costs and lead times. These large scale support elements will be dry-assembled without sensor modules prior to final assembly and integration.

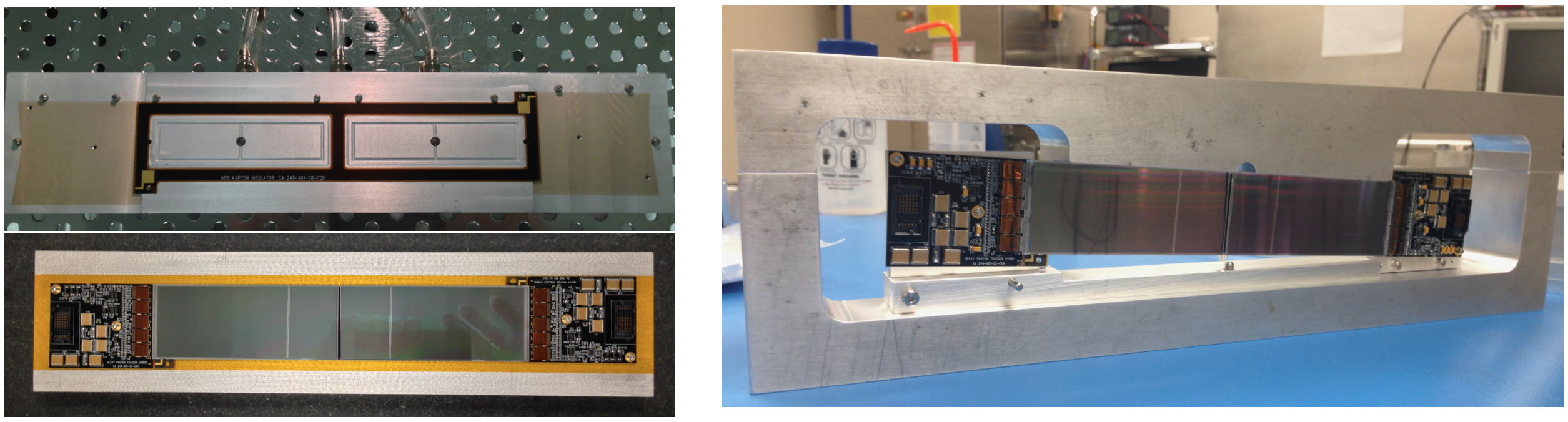

Figure 3.21: Fixtures used in the assembly of the HPS double-sensor modules which will have direct analogues in the assembly of both the Stereo and Axial Modules for the LDMX STT and SRT. Top left: Fixture employing vacuum chucks used for precision positioning when gluing sensors and hybrids together to produce a module. Bottom left: A storage, shipping, and wirebonding plate (shown without cover). Bottom right: Precision assembly fixture used to position the modules on either side of a stereo support during assembly of Stereo Stations and for survey on the CMM.

#### 3.4.4.6 Readout and Power

Readout and Power includes the FEBs, the OTBs, the Power Supplies, the mechanical support for the FEBs and OTBs alongside the Tracker Support Box, and all cables and connectivity for distribution of power and data. In addition to the hardware, this WBS includes the development of all firmware and software necessary to operate the trackers in coordination with the TDAQ.

At project start, the designs of the FEB and OTB will undergo final verification and review, and the boards will be fabricated, along with long-lead procurement of the power supplies. Firmware and software development will be completed to allow thorough testing of the readout chain using prototype sensor modules. Components for the FEB and OTB supports will be ordered and cables will be fabricated to connect the full system prior to Final Assembly and Integration.

#### 3.4.4.7 Control, Monitoring, and Interlocks

The design of the control, monitoring, and interlock system and hardware will be completed and components ordered. The system will be mocked up and bench tested to allow development of the slow control software and hardware interlock logic prior to installation in End Station A.

#### 3.4.4.8 Final Assembly and Integration

Final assembly and integration brings all of the major subcomponents of the STT, SRT, and the Trigger Scintillator system into a completed assembly in preparation for installation of these subsystems and their services in End Station A.

STT and SRT assembly begins with installation of the Stereo Modules and components of the Trigger Scintillator on the Stereo Support Plate in the cleanroom at SLAC. Similarly, the Axial Modules are mounted on the Axial Support Ring. A high-level survey of both is completed before installing the Stereo Support Plate and Axial Support Ring into the Tracker Support Box and survey of their relative positions is performed. All cables and cooling lines will be connected and tested, along with final DAQ testing of all elements. The front and back faces of the Tracker Support Box will then be closed in preparation for moving the completed

Tracker and Trigger Scintillator subsystems to End Station A for installation into the Magnet. Finally, the Tracker Support Box and other components will be moved to End Station A for installation.

#### 3.4.4.9 QA

QA testing for the STT and SRT involves a thorough set of tests throughout the production process that ensure the performance of the completed system. These tests include electrical, mechanical, and cooling tests, and set a baseline for the key operational attributes of the detector prior to installation. The QA steps for each L3 WBS element are described in the following subsections.

##### 3.4.4.9.1 Sensor Testing

Thorough testing of the sensors will be performed by HPK, including IV and CV testing to determine full depletion and breakdown voltages and a list of non-working channels, with the test data delivered alongside the sensors. Upon receipt, IV testing will be repeated to verify the breakdown voltages prior to sensor selection.

##### 3.4.4.9.2 Hybrid Testing

After receiving boards, all components except the APV25 chips will be loaded and cleaned and HV testing will be performed. After mounting and wirebonding APV25 chips at UCSC, full baseline testing of pedestals, gains, and noise will be performed to verify the functionality of all channels and provide a performance baseline prior to attachment and wirebonding of sensors.

##### 3.4.4.9.3 Stereo Module Testing

The Stereo Module Supports and all assembly fixtures will undergo survey and mechanical acceptance testing after fabrication to ensure that they meet the specifications necessary to produce modules within the required tolerances.

After wirebonding of sensors to the APV25 front ends, pedestals, gains, and noise will be tested with sensor bias to verify the functionality of all channels and provide a performance baseline. Wirebonds for any bad channels will be removed and those modules will be retested.

The completed Stereo Modules will be mechanically surveyed to establish the precise location of the active region of each sensor relative to its mounting points.

##### 3.4.4.9.4 Axial Module Testing

All assembly fixtures will undergo survey and mechanical acceptance testing after fabrication to ensure that they meet the specifications necessary to produce modules within the required tolerances.

After wirebonding of sensors to the APV25 front ends, pedestals, gains, and noise will be tested with sensor bias to verify the functionality of all channels and provide a performance baseline. Wirebonds for any bad channels will be removed and those modules will be retested.

##### 3.4.4.9.5 Support and Cooling Testing

The Stereo Support Plate, Axial Support Ring, and Tracker Support Box will undergo survey and mechanical acceptance testing after fabrication to ensure that they meet the specifications necessary to precisely position the detector modules.

Survey of the Stereo Support Plate and Axial Support Ring will establish the precise locations of the module mounting points relative to the survey monuments on these structures. Similarly, survey of the Tracker Support Box will establish the mounting points for the Stereo Support Plate and Axial Support Ring relative to the survey monuments at the ends of the Tracker Support Box.

The Stereo Support Plate and Axial Support ring will be installed and tested inside the Tracker Support Box, and the Tracker Support Box inside of the Magnet to ensure proper fit and positioning prior to Final Assembly and Integration.

Cooling components will be tested for flow and leaks, individually and as a completed system with the chiller.

#### 3.4.4.9.6 Readout and Power Testing

Each FEB will undergo full acceptance testing of all channels with a prototype module, a tested OTB, and the back end DAQ. Similarly, each OTB will undergo full acceptance testing of all channels with a prototype module, a tested FEB, and the back end DAQ. All cables will be tested to ensure proper pinouts and mapping and again tested after connection to ensure full functionality.

The FEBs and OTBs will be assembled and tested on their supports prior to installation to ensure full function of the integrated readout subsystems.

#### 3.4.4.9.7 Control, Monitoring, and Interlocks Testing

The individual monitoring sensors and components will be tested prior to integration with the cooling and support systems. The control and monitoring software, and the hardware and software interlock logic will be bench tested prior to installation.

#### 3.4.4.9.8 Final Assembly and Integration Testing

After mounting the Stereo Modules on the Stereo Support Plate, a final survey of the completed support plate will provide a cross-check of the final sensor positions within. In the case of the Axial Support Ring, the formal survey of sensor positions is done at this stage, fully defining the active regions of all sensors in the axial layers relative to the survey monuments of the Axial Support Ring. These surveys will be performed on the OGP Flash 500 multisensor CMM in the SLAC clean room, as shown in Fig. 3.22 for a similar structure of the HPS SVT.

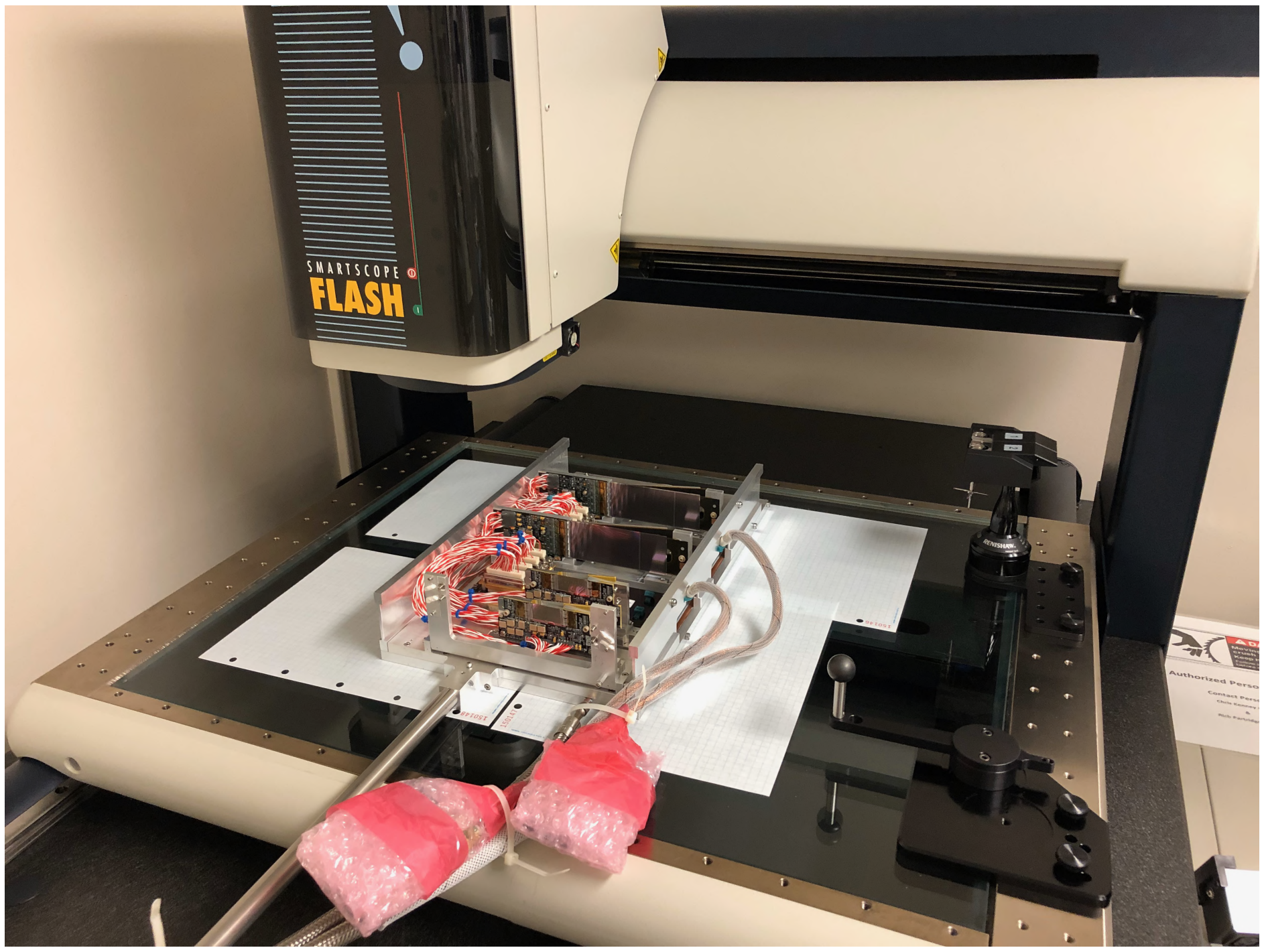

Figure 3.22: One of the support structures of the HPS SVT undergoing final survey on the OGP Flash 500 at SLAC.

After installation of the Stereo Support Plate and Axial Support Ring in the Tracker Support Box, all three elements will be surveyed with a Faro arm on an optical table to adjust and establish their precise relative positions and orientations, which is the final step in defining the positions of all sensors relative to the Support Box.

Prior to installation in End Station A, all cables and cooling lines will be connected and tested, and all modules of the completed system will be tested one-by-one for full functionality with a FEB, OTB, and the back-end DAQ.

#### 3.4.4.10 ES&H

ES&H for all work at SLAC will be managed by the SLAC ES&H Department through the Fundamental Physics Directorate ES&H coordinator. In addition to safety issues typical for laboratory work – housekeeping, slips/trips/falls, etc – there are only a few notable hazards in the construction of the tracker. First is electrical safety, where some low voltage supplies have current capabilities which make them potentially hazardous. These supplies will always be operated at lower voltages and current limits as a mitigation for this hazard. By contrast, the high voltage supplies to be used have such low current limits that they do not pose a significant hazard. Second, in the construction of the sensor modules, a number of chemicals are used that can be hazardous, including epoxy adhesives, ethanol, and toluene. Mitigation for these hazards involves the use of PPE, ventilation, and proper storage and processing. Finally, the OGP machine moves under computer control and poses a mechanical hazard. A number of engineering and administrative controls mitigate the hazard of OGP operation, which is only performed by trained operators.

### 3.4.5 Performance Validation

The Monte Carlo simulation and reconstruction tools of `ldmx-sw`, described in Sec. 3.10 have been used to validate that the design described above meets the physics performance requirements for the STT and SRT outlined in Sec. 3.4.2. The results of these studies are presented in the following sections. The technical performance requirements have all been shown to be met by this design by the performance of the HPS SVT, and will also be discussed.

All LDMX studies below, unless explicitly mentioned, were performed using an 8 GeV beam energy simulated such that the beam electrons hit uniformly and perpendicularly across the face of the target.

#### 3.4.5.1 STT Physics Performance

From Sec. 3.4.2 there are two physics requirements on the STT. The first (TRK1) is that the reconstruction efficiency of non-interacting beam electrons be $> 95\%$. This was tested by generating a sample of beam electrons, propagating and simulating them through the STT and reconstructing the tracks.

The beam-electron track efficiency is calculated as the fraction of generated beam electrons whose truth energy at the target is $> 7.8$ GeV that are reconstructed with a momentum between $7.6 < p_{e^-} < 8.4$ GeV. With this definition, we measure the STT reconstruction efficiency to be greater than 97%, satisfying this requirement.

The second requirement (TRK2) on the STT is that the tracker can reject off-energy beam electrons by a factor of, at least, $10^{-13}$. To study this, we conservatively assume that all of the off-energy beam impurities ($10^{-3}$ of the total beam) are at 2.4 GeV, 30% of nominal beam energy. We quantify the rate for misidentifying off-energy electrons from the beam as full-energy, 8 GeV electrons by generating a large sample of 2.4 GeV electrons well upstream of the analyzing magnet. We then run the electrons through the simulation and reconstruction chains and measure the fraction of those electrons that are reconstructed near the beam energy. In order to speed up processing we reject simulation-level electrons that leave a simulated energy deposit in fewer than 8 layers. The track seeding and reconstruction requirements are intentionally left quite loose.

Fig. 3.23 shows the fraction of generated electrons as a function of the angle of the electron at the target in the horizontal (bend) direction, $\phi$, (left) and the reconstructed momentum (right) at the target for $1 \times 10^7$ generated 8 GeV electrons (red, as a reference) and $1 \times 10^{12}$ generated 2.4 GeV electrons (blue). We do not observe any 2.4 GeV-generated electrons reconstructed near 8 GeV. In fact, by making reasonable cuts on the reconstructed track $\phi$ at the target, we observe zero 2.4 GeV-generated electrons at all momenta.

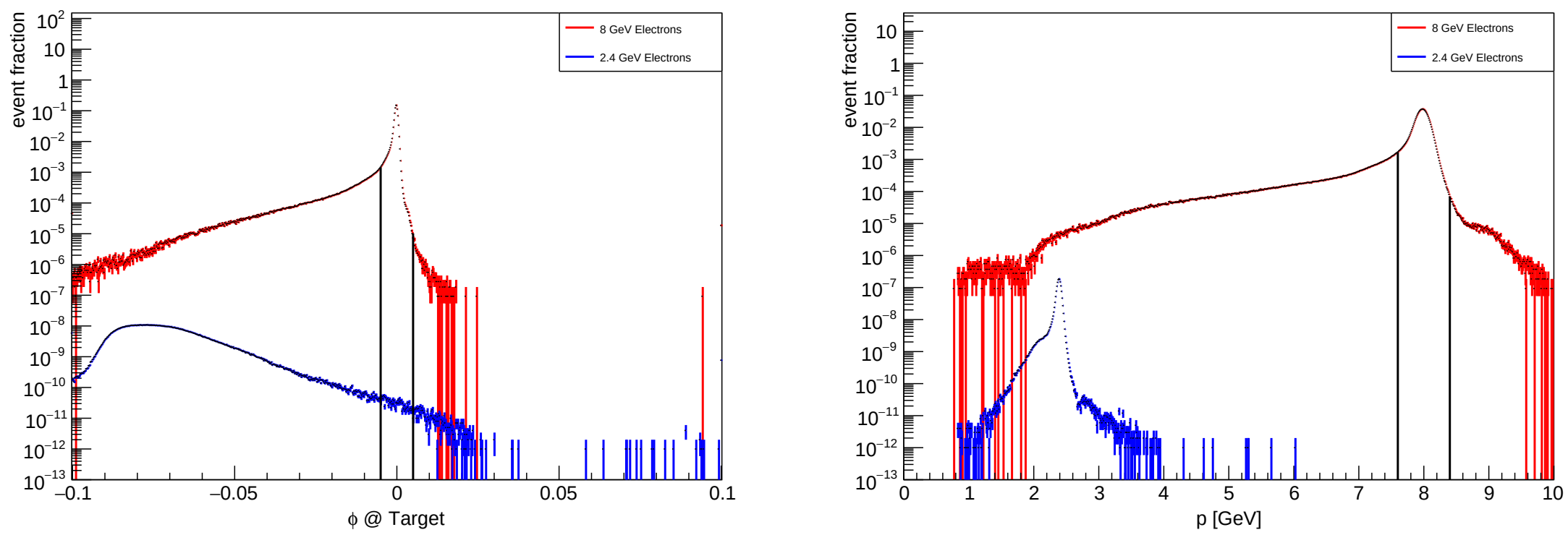

Figure 3.23: The reconstructed track $\phi$ (left) and momentum (right) distributions for 2.4 (blue) and 8 (red) GeV electrons in the tagger. The black vertical lines show the regions where we would define a good beam electron.

With the current number of EoT, this study only shows we can reject off-energy electrons at the $< 2.3 \times 10^{-12}$ level. However, we do have many orders of magnitude of additional discrimination between true beam-energy electrons and false positives in such variables as: the angle $\phi$ at the target (as shown), the reconstructed angle $\theta$ between the the track and z-axis at the target, the reconstructed x-position at the target, and the layers of the hits included in the track.

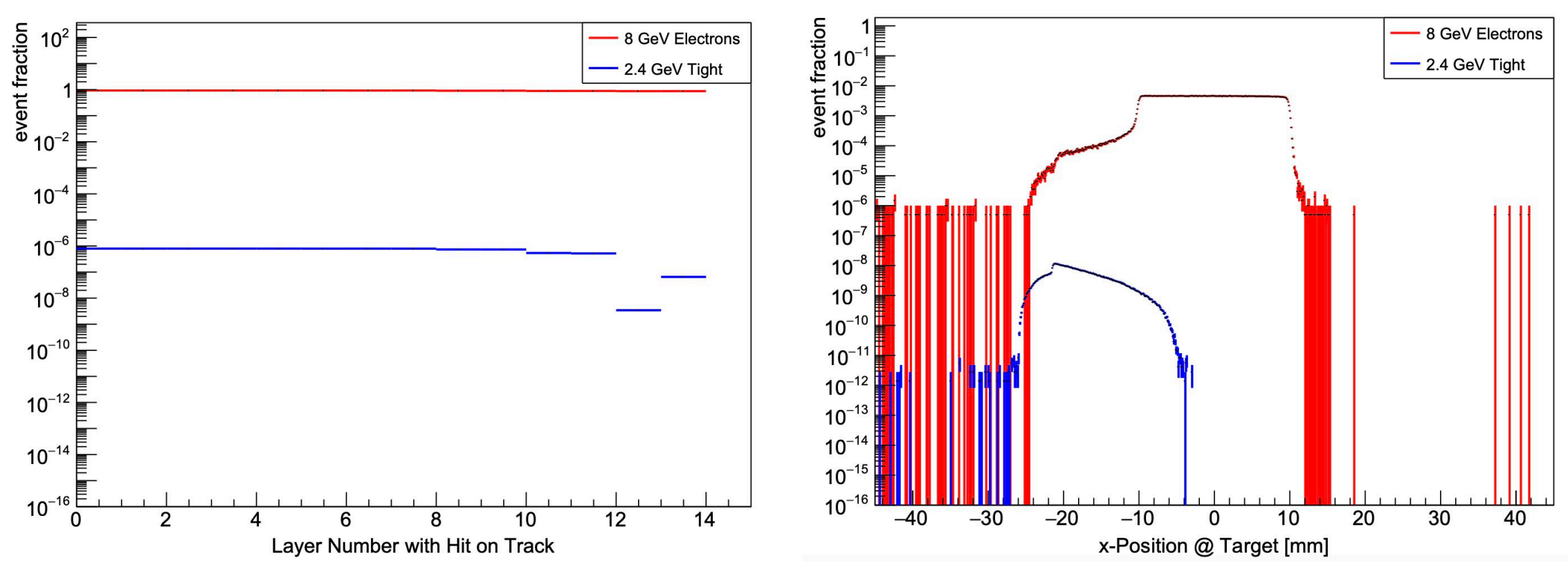

Figure 3.24: The reconstructed track $\phi$ (left) and momentum (right) distributions for 2.4 (blue) and 8 (red) GeV electrons in the tagger. The black vertical lines show the regions where we would define a good beam electron.

#### 3.4.5.2 SRT Physics Performance

Moving to the recoil tracking detector, the requirements are that the acceptance×efficiency is greater than 45% for mediator masses between 1 MeV and 1 GeV (TRK3) and that the probability to reconstruct a non-interacting beam electron (i.e. full beam energy) as having less than 30% of the beam energy is less than 1% (TRK4). The left-hand plot of Fig. 3.25 shows the acceptance×efficiency for a range of mediator masses obtained from reconstructed signal MC as a function of the recoil track momentum. This shows that, except at very high mass and low recoil momentum, we observe reconstruction efficiencies of about ∼95%. We find that the acceptance×efficiency, integrated over recoil momentum, ranges from 94% at 1 MeV to 54% at 1 GeV.

The SRT is used to help eliminate non-interacting electrons from our sample so we need a low rate for reconstructing these electrons with low momentum, in this case defined at <30% of the beam energy.

To estimate the rate of this mis-reconstruction, we reconstruct the recoil momentum for simulated beam electrons that left at least 7.6 GeV in the ECal. The distribution for recoil momentum is shown in the right hand side of Fig. 3.25 demonstrating that we reconstruct non-interacting electrons with energy <2.4 GeV at a rate of $\sim 1 \times 10^{-5}$.

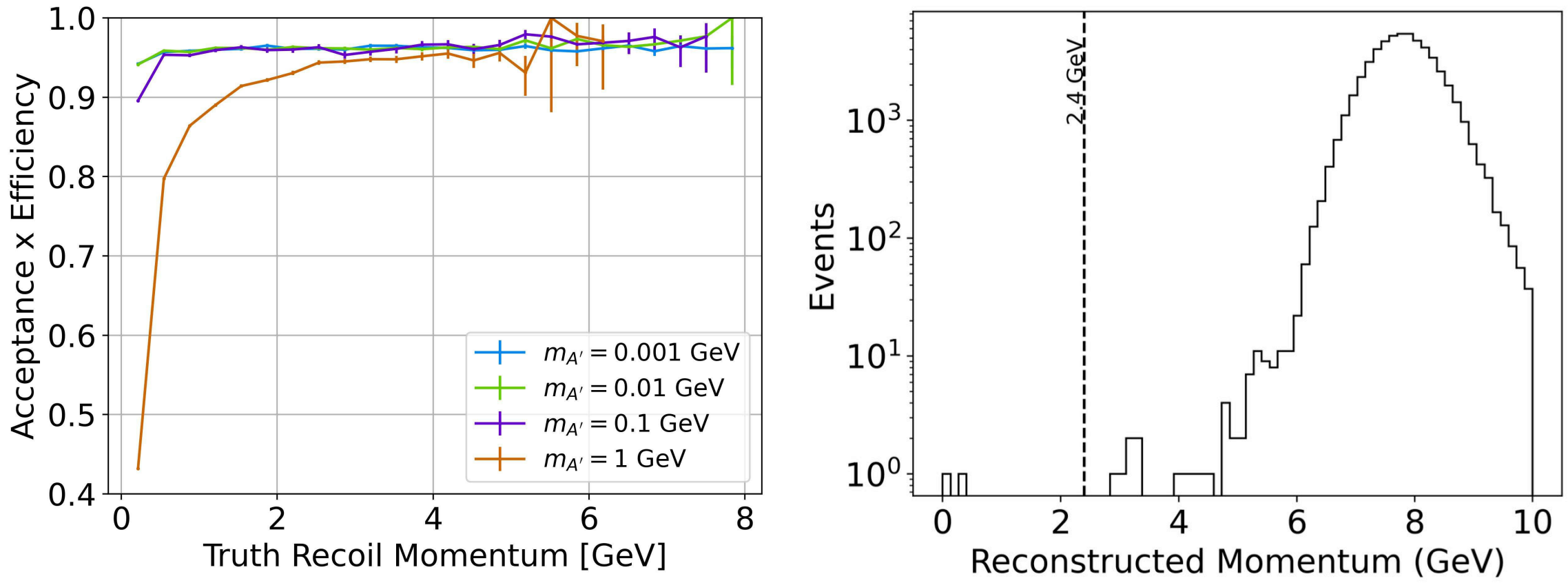

Figure 3.25: The track acceptance×reconstruction efficiency versus truth recoil momentum for mediator masses ranging from 0.01–1.0 GeV. Right: The reconstructed momentum distribution in the SRT for non-interacting beam electrons.

#### 3.4.5.3 STT/SRT Combined Performance

The final requirement for the silicon tracking system is the precision of the change in electron transverse momentum at the target ($\delta_{p_T}$). The $\delta_{p_T}$ across the target can be used as another discriminant between signal and background and to probe the mediator physics in the case of a signal observation, and thus we do not want the measurement uncertainty to be the primary contribution to the overall uncertainty on this quantity.

We used an inclusive sample simulated through the tagger detector (beam electron), target, and recoil detector (recoil electron) giving a range of recoil momentum in the SRT. The reconstructed electron $p_T$ is measured at the target for both the tagger and recoil detectors and subtracted to give $\delta_{p_T}$. This is then compared to the true value of the $\delta_{p_T}$ obtained from the MC generator.

The distribution of the resolution vs the true recoil electron momentum is shown in Fig. 3.26. The requirement (TRK5) is that this resolution is less than the RMS on $\delta_{p_T}$ due to multiple scattering in the target, about 4 MeV. We see that for recoil tracks we are below this mark for recoil electrons below ∼4 GeV and below 5 MeV up to the beam energy. For signal events, the $\delta_{p_T}$ values we expect to observe are all <2.4 GeV.

#### 3.4.5.4 Technical Performance Validation

The technical performance requirements for the trackers are met and can be validated from first principles or prior results as follows.

R6 The planes of the STT and SRT shall have a material budget averaging less than 0.5% $X_0$ in the tracking volume. **The design places only silicon in the tracking volume, which will have 0.35%$\mathbf{X_0}$ per plane.**

R7 The single-hit efficiency for all planes in the STT and SRT to reconstruct hits from individual minimum-ionizing particles and assign them to the correct beam bunch with time information shall be at least 98%. **In regions of the HPS detector where occupancies are similar to those in LDMX, the single-hit efficiency is in excess of 98% and the time resolution is 2 ns, so the efficiency in LDMX operating at 37.14MHz will be at least 98%.**

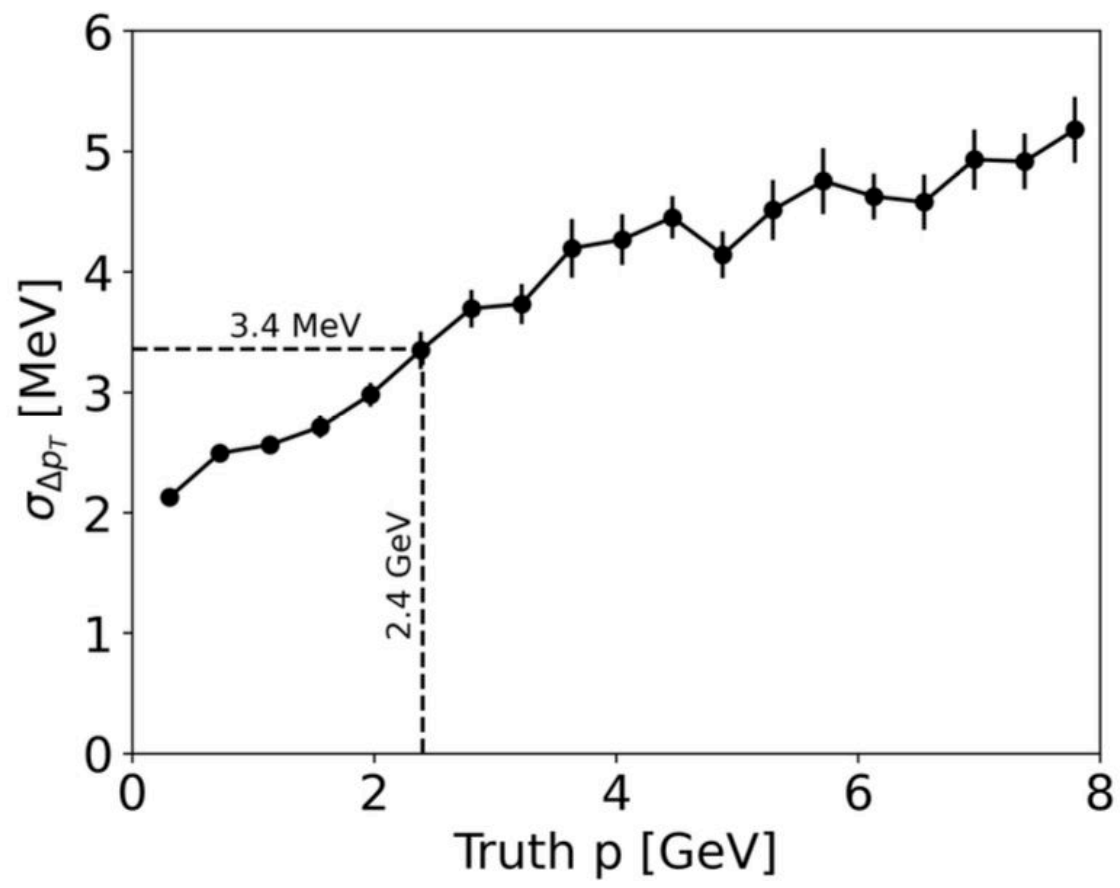

Figure 3.26: The resolution on the change of $p_T$ as measured from the STT and SRT.

R8 The single-hit position resolution in all planes of the STT and SRT shall be 15$\mu$m RMS or smaller. **This requirement is exceeded by a large factor with APV25 readout, and is only stated to ensure that the strip geometry and noise performance of alternate readout chips do not degrade the tracking performance to unacceptable levels. In the case of the VFAT ASIC, which has similar noise characteristics to the APV25, but binary readout, the strip geometry of sensors being used in the trackers has been shown to meet this requirement by previously documented studies. [81]**

R9 The STT and SRT shall be capable of sustaining an average readout rate of 25 kHz for random triggers with less than 5% dead time. **In HPS, the same readout meets this dead-time requirement at 37 kHZ for six sample APV25 readout. With three-sample readout being planned for LDMX, this rate limit roughly doubles.**

R10 The STT and SRT shall allow a minimum of 3 $\mu$s latency for readout of triggered events. **The APV25 ASIC meets this by design, and other readout options considered as risk mitigations exceed it by large factors.**

### 3.4.6 Risks and Opportunities

The most important item is the opportunity presented by the re-use of designs and equipment from the HPS Silicon Vertex Tracker, where the LDMX trackers are designed to gain as much leverage as possible from HPS. In addition to design, assembly, and testing techniques, a great deal of firmware and software is reused directly, and spare HPS components can in many cases be used to test parts of the LDMX readout chain. A related opportunity is leveraging the existing collaboration with UCSC, which greatly reduces the cost and risk of some assembly steps for the trackers.

The most important risk is the limited supply of APV25 ASIC and the fact that no more will be produced. In the case of a catastrophic loss of the stock of APV25 chips, the hybrids and readout would need to be redesigned to accommodate another ASIC. While there are a few possible options, the VFAT chip, designed at CERN and currently available, has been considered a likely replacement. As a mitigation for this risk, the project plan includes WBS 1.3.9, the development of readout designs using the VFAT chip, which will begin at project start and proceed in parallel with construction until the risk is retired.

## 3.5 Trigger Scintillator

### 3.5.1 Introduction and Overview

The TS consists of three arrays of scintillator bars that are used to estimate the number of beam electrons within each beam pulse. One array of 48 bars will be placed directly upstream of the target while two other identical arrays will be upstream of the tagger tracker. Requiring a coincidence of hits consistent with an electron track provides an accurate determination of the number of electrons in each beam pulse.

Each scintillator bar will be connected to a light pipe of the same dimension and read out with an independent Silicon Photomultiplier (SiPM). The SiPM signals will be digitized by deadtimeless readout electronics developed for the CMS HCal high luminosity LHC upgrades. These boards produce low noise charge integrating amplitude measurements and pulse arrival time measurements with 0.5 ns precision. The front-end electronics will be controlled via a Kria[1]-based interface that will deliver both fast and slow control signals to the front end electronics and SiPM boards. Data will be continuously streamed to ATCA trigger boards over 5 GHz fiber optics, where trigger primitives for the TS are computed and data pipelines are managed.

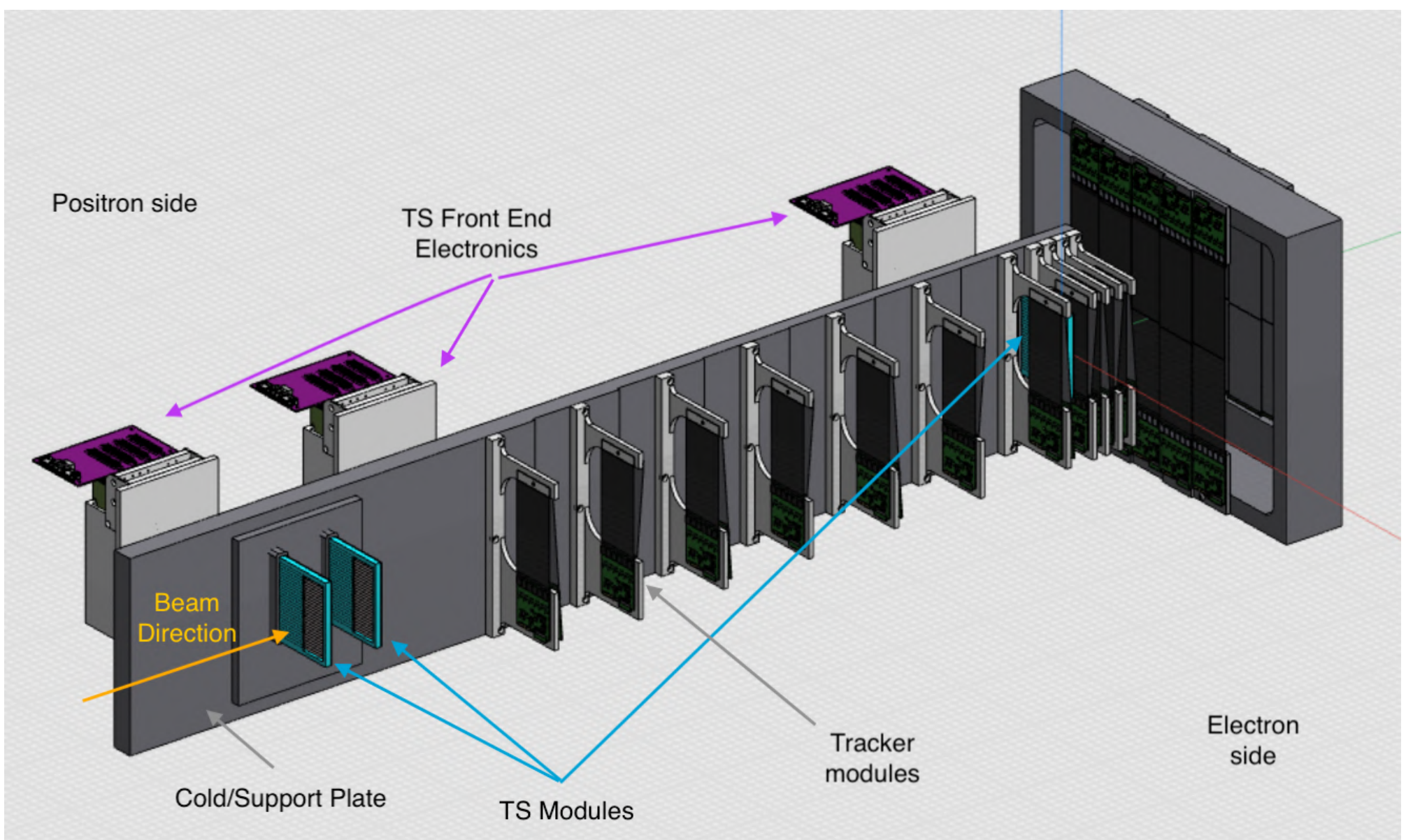

Figure 3.27: Model of the tracker, trigger scintillator and target system. The TS modules are shown in cyan, their front end electronics are in light gray and magenta.

The TS trigger primitives determine the number of electrons incident on the target with high fidelity and low latency. This is a primary input to online missing energy triggers. The missing energy will be computed by subtracting the measured energy in the ECal from the $nE_{\text{beam}}$, where $n$ is the number of incoming electrons the TS has inferred and $E_{\text{beam}}$ is the typical energy of a beam electron.

$$E_{\text{miss}} = nE_{\text{beam}} - E_{\text{ECal}} \tag{3.1}$$

In cases where the TS measures more than the true number of electrons, the online missing energy will be large, having a significantly higher chance of being triggered. These events appear to have "fake" missing energy and will cause background events to pass the missing energy trigger at an increased rate. In cases where the TS measures less than the true number of electrons, the online missing energy will be smaller, making such events fail the trigger thresholds. These events contribute to a signal inefficiency.

[1] https://www.amd.com/en/products/system-on-modules/kria.html

### 3.5.2 Requirements

The TS functionality translates into a set of requirements on the overall performance of the TS system, with the key goal of ensuring fake missing energy events do not dominate the trigger bandwidth. To achieve this, the rate at which fake missing energy events are triggered should be less than 1 kHz. For a bunch clock of 37.1 MHz, at most 1 in 3700 events should over-predict the true number of electrons. This top level requirement, along with practical design constraints and overall LDMX physics goals translates into a set of requirements summarized below.

- Detector-Physics Requirements
  - TS1 Maintain 90% single track efficiency.
  - TS2 Fake additional track rate of $< 1$ kHz for average number of electrons of 1
  - TS3 Must have $> 95\%$ single hit efficiency
- Technical Requirements
  - TS4 The analog pulse should have a fall-time of less than 54 ns, or the period of two RF buckets.
  - TS5 The peak of the MIP distribution must be at least 3 $\sigma$ away from the pedestal.
  - TS6 The detector should be tolerant for LCLS-II calibration pulses ( .5nA at S30XL kickers)
  - TS7 Rate of additional hits over threshold due to cross-talk is $< 1\%$ per true hit
  - TS8 All hits must be available to the trigger system within 1 $\mu$s.
  - TS9 Detector should withstand at least 1e15 EoT before replacement is necessary
  - TS10 Be able to replace a TS module in a short (a few days) shutdown period
  - TS11 The TS material should be $< 1\%$ of a radiation length in each module such that it is a small fraction of the overall radiation length.

### 3.5.3 Design

As shown in Fig. 3.27, the TS consists of three planes of double-layer scintillating bar arrays, or electron counting modules. Two are located directly upstream of the tagging tracker array, at approximately 800mm from the target, with a separation of 60mm. A third array is placed directly in front of the target. The target is held in the same mechanical structure as the last TS plane. The modules, electronics, control system, and mechanics are described below.

#### 3.5.3.1 Electron counting modules

The electron counting modules, shown in Fig. 3.28, consist of 2mm×3mm thick polyvinyltoluene (PVT) bars covering the active region of the detector, glued to light pipes of the same cross-sectional dimensions which transmit the scintillation light to SiPMs. The light pipes are directly coupled to $2 \times 2$ mm$^2$ SiPMs, producing roughly 80 photoelectrons per MIP. We have designed the system such that each module is comprised of $2 \times 24$ rectilinear bars arranged in two layers. The bars in the two layers are staggered with respect to each other by a half a bar width. The bar active area measures $2 \times 3 \times 30$ mm with the long dimension in the $x$-direction and $z$ and $y$ directions aligned along the 3 mm and 2 mm dimensions respectively. The light pipes are $\simeq$50 mm long. The bars in each layer are stacked along the $y$-direction with roughly a 0.3 mm gap in between adjacent bars. The two layers are separated by 2 mm. The total thickness of the module that the beam will traverse is 0.01 $X_0$. The active area of each module will be roughly 20x80 mm, covering the same area as the target. To increase the light collection efficiency and reduce optical cross-talk, a thin strip of reflective film, ESR, will be placed on each side of the bars. The SiPMs are placed on the positron side of the beam to minimize the number of low-energy, off-trajectory electrons that will traverse the SiPMs. The light pipes are utilized to keep the active area of the detector the same size as the target. The digitizing electronics will be hosted near the modules, also on the positron side of the beam. An aluminum cooling plate will separate the SiPMs and electronics from the beam.

#### 3.5.3.2 Common SiPM Electronics

The S13360-2050VE SiPM will be used. These SiPMs have a spectral response range from 330 nm to 900 nm and a peak response at 450 nm, which is well matched to PVT scintillators whose peak emmission is at 425 nm. SiPM gains are typically $2 \times 10^6$. Given this gain, and assuming a most probable value of 80 PEs will be produced per MIP passing through the active target, these SiPM signals can be digitized by the frontend electronics without saturating the ADC; more information on the dynamic range of the readout is provided in section 3.5.3.3.

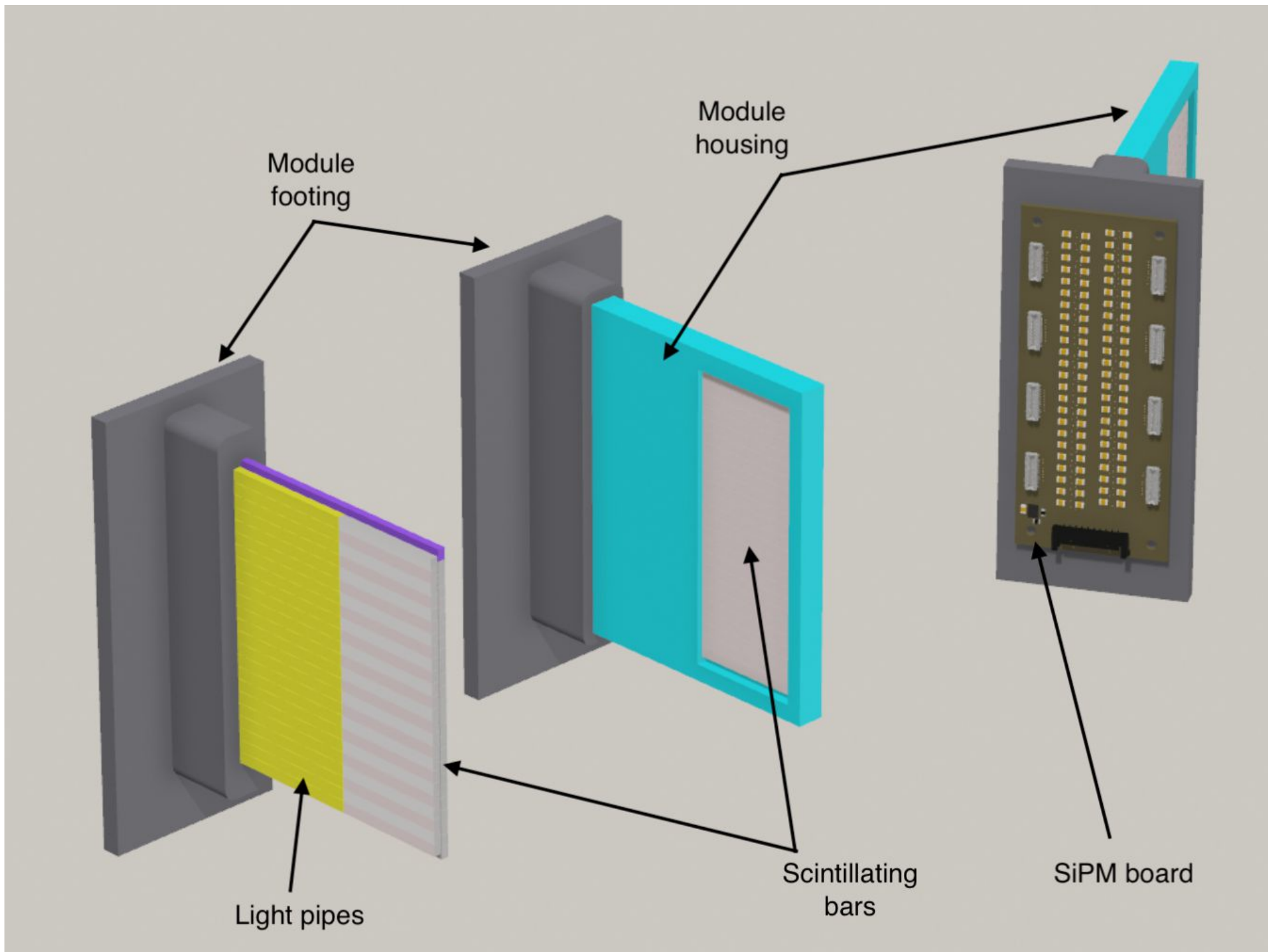

Figure 3.28: Three views of the TS modules. Left most view shows the scintillator bars in alternating shades of gray, the yellow shows the light pipes, the dark grey shows the footing which attaches to the TS-Tracker cold plate. The purple bar is an inactive bar. Middle view shows the scintillator housing in blue. The right most view shows the back of the SiPM board which is directly coupled to the light bars via physical contact with the light pipes.

For the electron counting modules, arrays of 2×24 SiPMs will be mounted to a single printed circuit board (SiPM board) and arranged in a one-to-one mapping to the scintillator bars. Each board will have a temperature sensor and bias voltages for groups of 12 SiPMs. The analog signals from groups of 6 SiPMs will be connected to a 32 pin connector and a short flat custom cable will carry the signal to the frontend electronics. Each SiPM mounting board will thus contain 8 connectors for analog signals and one connector for power and temperature sensing. The rear side of the SiPM board, which includes the 8 cable connectors, is shown in the right most view of Fig. 3.28.

#### 3.5.3.3 Readout electronics

The TS readout will be built on the CMS HCal Readout modules (RMs)[82]. SiPMs will be placed close to RMs to reduce the effect of cable capacitance on their pulse shape. Data from RMs will be transmitted to the ATCA-based APx board, where trigger primitives will be computed and data pipelining will be managed. A separate Zynq-based system will manage slow and fast control signals and distribute these signals to RMs over Cat6 cables. The control signals will be received by a set of fanout boards, referred to as backplanes. A summary of the architecture of the TS readout is shown in Fig. 3.29.

The RMs utilize the QIE11 ASIC, which performs deadtime-less digitization at 37.2 MHz. Each clock cycle, QIEs output an 8-bit ADC and a 6-bit TDC. A RM is shown in Fig. 3.30. Each RM hosts 12 individual channels each, necessitating a set of four for each TS plastic scintillator module (12 RMs total).

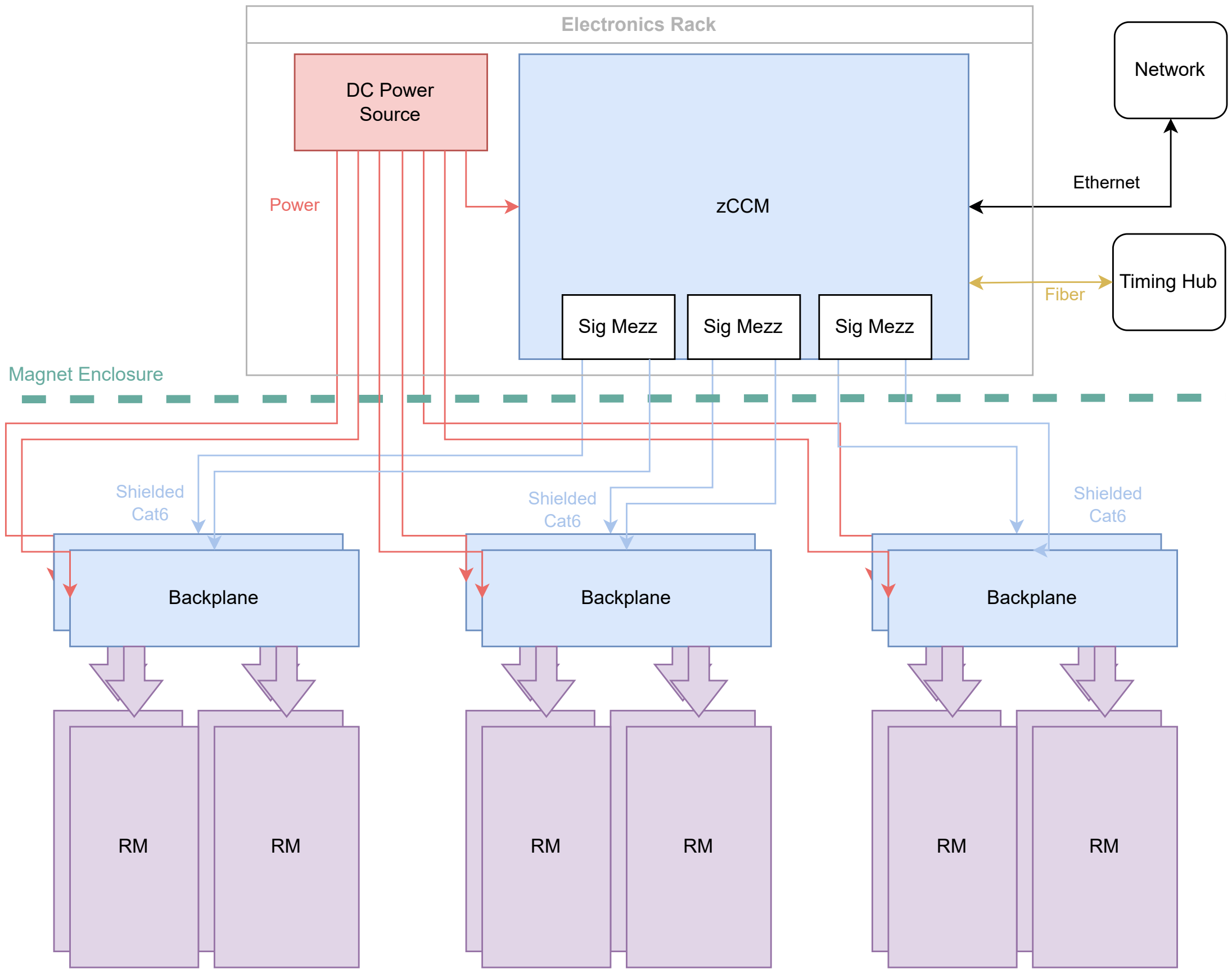

Figure 3.29: TS readout and control system. See text for a description of the components.

The QIE11 is a charge integrating ADC with dynamic gain that implements an approximate logarithmic response, resulting in a dynamic range of $10^5$. Input currents can also be shunted, allowing the absolute range to be increased by a factor of 12. The response of the ADC is shown in Fig. 3.31 (right) for the nominal shunt value. The minimum precision of the ADC is 3 fC and the typical noise is half that. The quantization error depends on the input charge, but over most of the dynamic range, the relative quantization error is less than 1.5%, see Fig. 3.31 (left). For the typical SiPM gains used in this system, a single photoelectron deposits roughly 300 fC. At this gain, a minimum ionizing particle in 2 mm of plastic scintillators produces a charge distribution whose most probable charge is roughly 24 pC.

The TDC measures the arrival time of the pulse by using a current discriminator with a programmable threshold. The least significant bit of the TDC is 0.5 ns.

An FPGA on the RM serializes data from 12 QIEs and sends the serialized data to a pair of optical transmitters. Data is continuously streamed at 5 GHz with no zero-suppression in the frontend electronics. Data is received by an APx where trigger primitives and a data pipeline are held until the global trigger requests data to be read out. Details about the trigger implementation are described in Sec. 3.9.5.2.

A group of 4 RMs is connected to a backplane that distributes power and control signals to each RM. A schematic of the backplane's main functions is shown in Fig. 3.32 (left). A picture of a prototype backplane is shown in Fig. 3.32 (middle and right).

Each backplane receives control signals over a shielded Cat6 cable. These control signals include an I2C bus for communicating with components on the backplane and with the RMs. The backplane also receives several fast signals for timing the RMs, including the beam-synchronized clock. These signals are replicated on the backplane before being forwarded to the RMs. Finally, an ADC, a UART bridge and an EEPROM

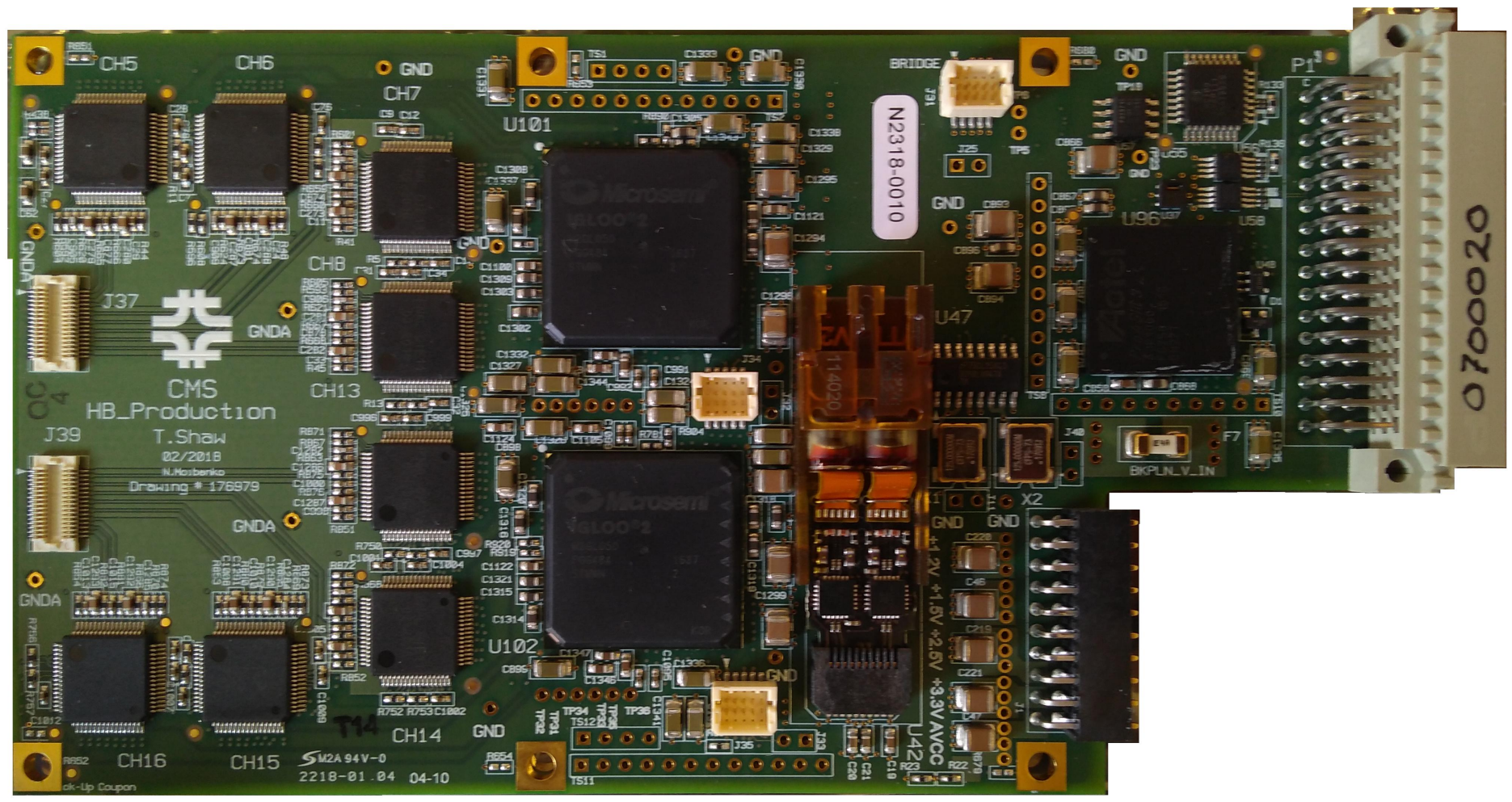

Figure 3.30: Picture of a CMS readout module developed for the barrel hadron calorimeter. The eight chips on the left are half of the QIE11 ASIC, which digitize data from SiPMs. The two larger chips in the middle are FPGAs which, align and serialize data from QIE11s. The orange housing to the right of the FPGAs is a dual optical transmitter. Finally, the chip on the right right of the board is an FPGA which manages control signals.

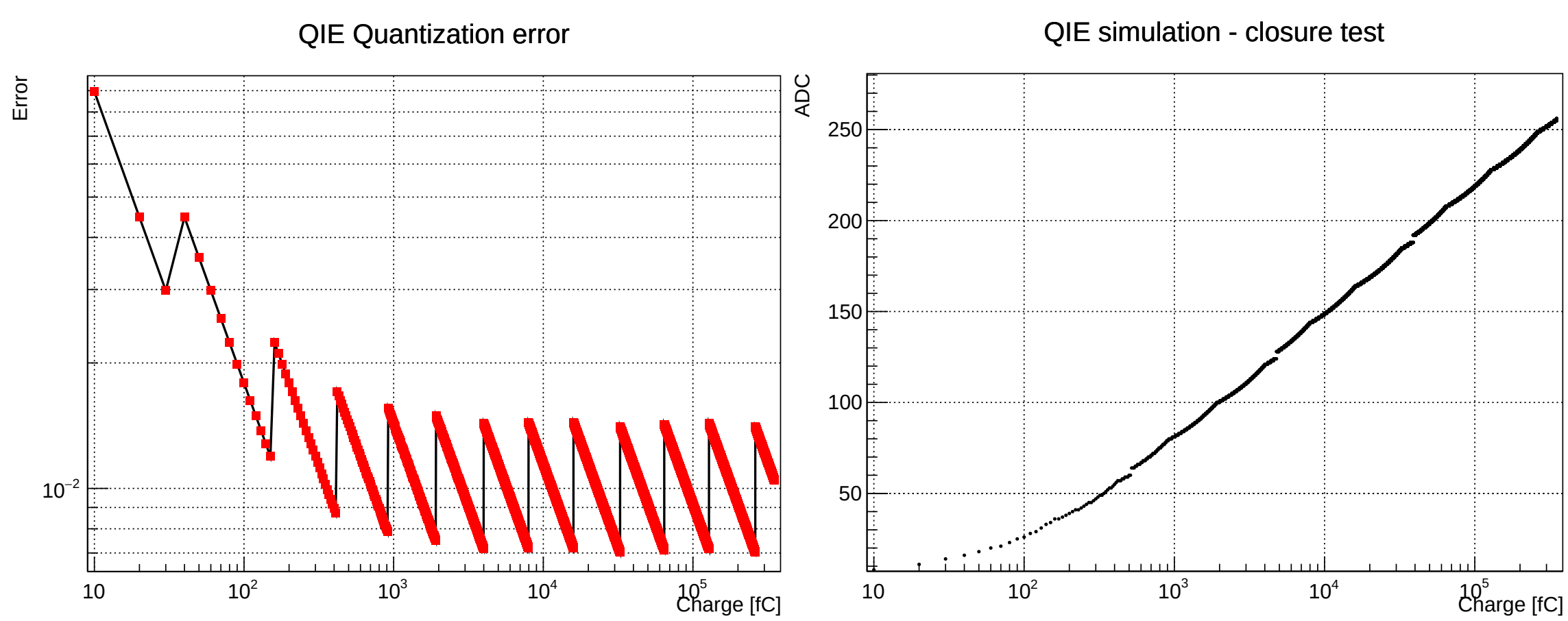

Figure 3.31: Left: The relative quantization error for the QIE11 as a function of the input charge. The zig-zag nature is a result of various ranges in which the gain of the digitizer is changed. Within a given range the response of the ADC is linear. Right: The transformation implemented in the QIE11 between input charge and output ADC codes. The ranges observed in the quantization error are reflected in the piece-wise nature of the chip's response.

enable the backplane to support external peripherals, such as temperature sensors.

Power is received on each backplane, which is regulated for powering its active components. Power is also passed to each RM, where a series of DCDC converters are used to regulate power for each of the five power rails on the RMs. The DCDC converters can operate in a magnetic field, and the chip regulating the output power is tolerant to high radiation doses as necessary for the LHC. Switching and linear regulators provide 3.3 V and 5 V power for components on the backplane.

#### 3.5.3.4 Control system

The frontend electronics will require I2C communication, a beam-synchronized clock for controlling ADC measurement boundaries, and a few other synchronization signals. The commercial zynq-based Kria

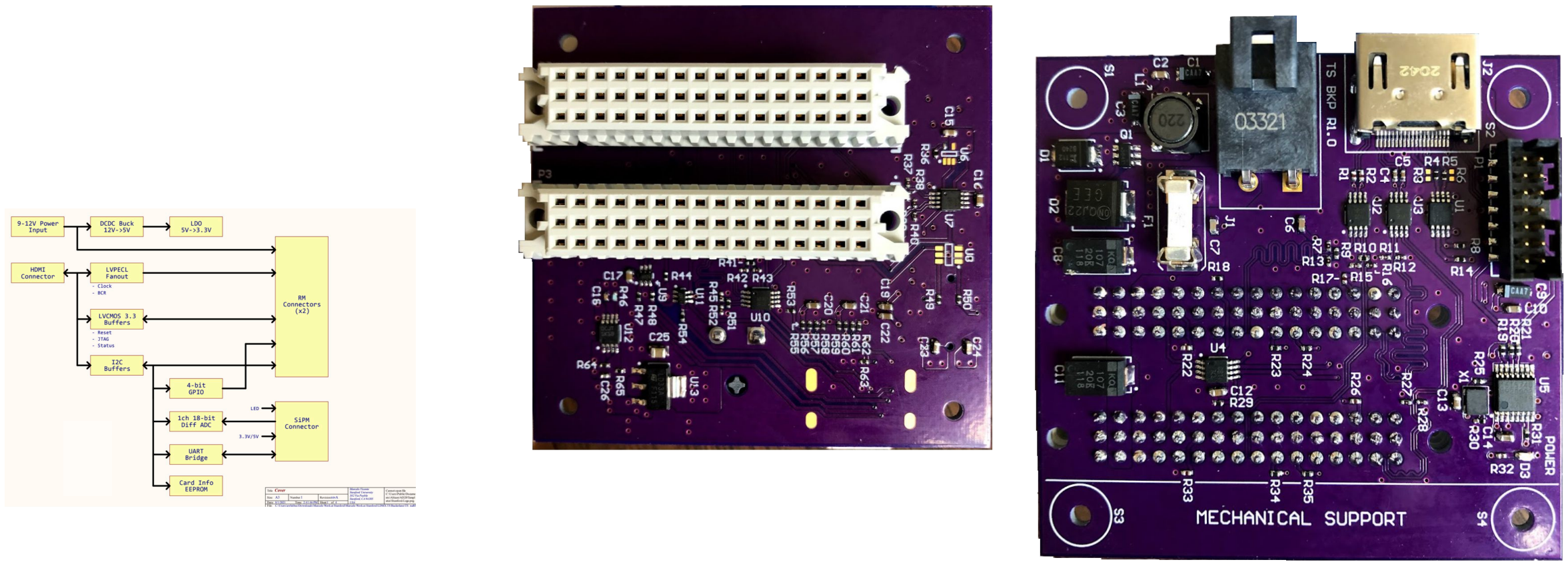

Figure 3.32: TS Backplane. Left: Block diagram for the TS backplanes. Middle: Top view of TS backplane prototype. The two 48-pin connectors mate with a pair of RMs. Right: Backside of TS backplane prototype including power and HDMI connectors.

k26 system on module (SoM) will be hosted on a custom baseboard shown in Fig. 3.33. The baseboard will provide custom interfaces between the TS frontend electronics and the SoM. Each SoM and baseboard pair will be referred to as a Zynq Clock and Control Module (zCCM). All together, two zCCMs will enable up to 12 total backplanes, allowing for expansion of the system if necessary.

The clock for the frontend electronics must be synchronized with the LCLS-II RF frequency. The centralized system in LDMX for distributing such a clock is the Fast Control (FC) Hub (see Sec. 3.9.4 for details on the fast control system). In addition to a clock, the TS frontend electronics will need several fast-timing signals; an example is a reference signal for phase-aligning all of the frontend electronics and the data received on different serial links in the backend electronics. These signals will all be derived from timing messages encoded into frames of the FC Hub links. To enable this, a high-speed fiber interface is required to receive the FC link. A standardized IP block will be used to control the Zynq transceivers, extract the FC commands, and synchronize clocks with the LCLS-II RF.

The Zynq will also manage a variety of slow-control signals, including signals for toggling power and I2C. The I2C interfaces will enable monitoring and configuring of all peripherals in the frontend, including the digitizing ASIC. The I2C controllers will be hosted on the Zynq.

The physical interfaces between the zCCM and the backplanes are LVDS over twisted pairs. The Readout Interface Mezzanines (RIMs) on the zCCM implements that interface. LVDS signals ensure that signal integrity over long cables; lengths of 10 meters or more are expected if zCCMs are mounted in an LDMX electronics rack.

The Zynq will run a distribution of Linux. A network interface will provide the means to communicate with the board remotely. Standardized Linux drivers will be used to enable remote applications to configure and monitor the zCCM, the backplanes, and the frontend electronics.

#### 3.5.3.5 Mechanics, power, and cooling

The TS modules, SiPM boards, RM and backplanes will be mounted on a joint aluminum cold plate with the tracker modules. This design prioritizes protecting the electronics from beam radiation and minimizing the capacitive coupling of the SiPM readout cables by minimizing the cable length, which will be less than 10 cm.

The scintillator housing, shown in Fig. 3.28 will enable plastic scintillators in the two different layers to be aligned, and will allow for the SiPM to be aligned with the scintillators. The housing is formed from two halves that allow for scintillators to be easily loaded. The two halves are then bolted together. For one of the scintillator modules, it will be possible to mount the passive target to one side.

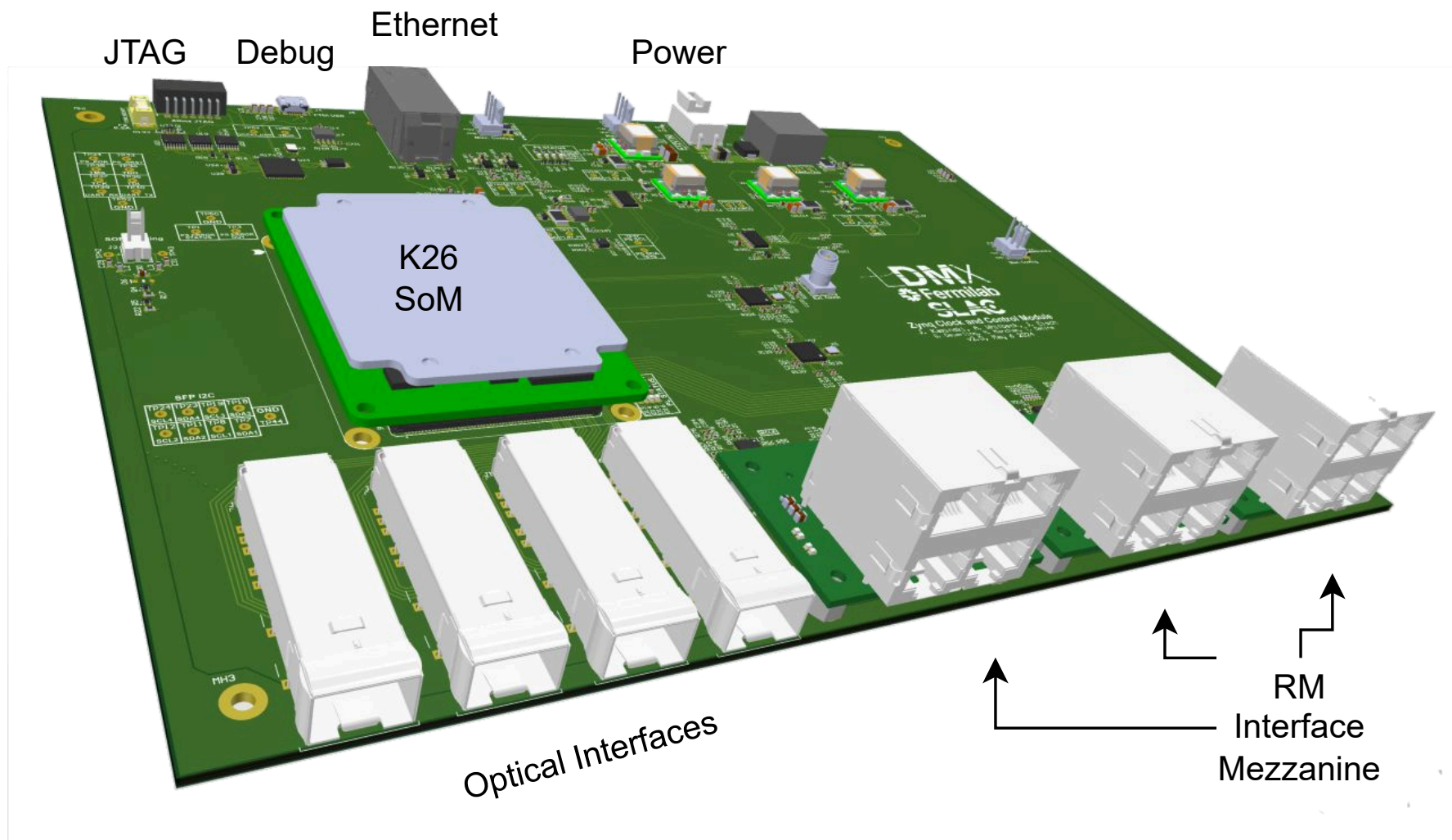

Figure 3.33: The prototype zynq-based clock and control module (zCCM).

The frontend electronics will be housed in an aluminum enclosure. Each enclosure will support the mating of a backplane, a DCDC converter board, and 4 RMs. Thermal coupling of the backplanes and the RMs will transfer heat to the cold plate. This housing forms an independent unit for the electronics, which will support reading out 24 channels, half of one module. The prototype electronics housing can be seen in Fig. 3.34.

Power usage estimates are based on built prototypes: the backplane consumes roughly 4 W of power and each RM will consume about 12 W of power. All 12 RMs will thus draw roughly 168 W, supplied by 6V power supplies. The heat generated from this power consumption will be conducted to the aluminum cold plate to which they are mounted.

Power to the SiPM boards will be provided to each group of 12 SiPM independently. Three 16-wire cables will be used to carry SiPM bias voltage and enable reading temperatures of the SiPM. The SiPMs will be biased at roughly 53 V and will draw minimal power.

### 3.5.4 TS Performance in simulation

#### 3.5.4.1 Description of TS simulation and benchmark samples

As part of the simulation in `ldmx-sw`, the TS is implemented as three arrays of $2\times24$ plastic scintillators. There is no modeling of support structures, SiPM, or electronics. The output of the `Geant4` simulation step is a list of `SimHits`, with a position, timestamp, and energy deposited for each hit, as well as information about which simulated particle made the energy deposition resulting in the hit.

Given the TS location upstream of the target, and its electron counting role, the most important samples for TS performance studies are inclusive samples of multiplicity 1-4 $e^-$. The generic samples described in 4.1 are used. Custom samples with variations in TS geometry used for design studies were produced and validated against the official samples.

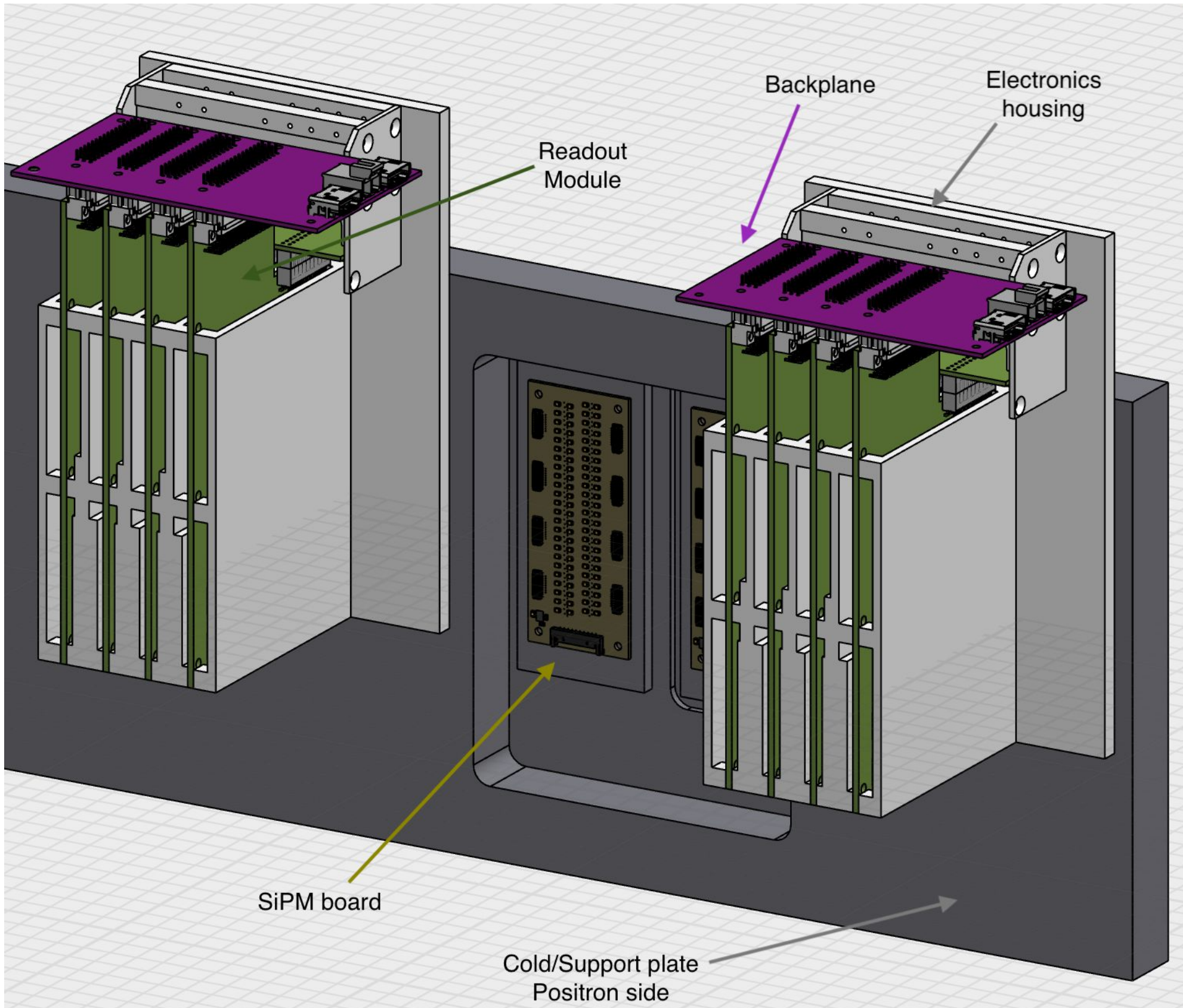

Figure 3.34: P

#### 3.5.4.2 Digitization of Simulated Samples

The TS digitization emulation, or *digi* production, takes `SimHits` as input, and produces a list of ADC and TDC values that model the performance of the SiPM and the digitizing electrons for the primary time sample of the beam, and several time samples before and after the primary time sample. This emulation is intended to allow MC simulated data and real collected detector data to be treated on equal footing. All later reconstruction steps are based on this so called *digi* collection. Hit reconstruction starting from digis was validated by comparison to a simpler reconstruction process, which converted the collected energy deposited in all `SimHits` in a bar to a photoelectron count. The count was taken as the most probable value of a Poisson distribution smeared by the expected resolution, including electronics noise. The implemented digitization was confirmed to reproduce the same response and resolution.

The digitization emulation is shown schematically in Fig. 3.35. The energy deposited in a bar is translated into an amplitude of a simulated pulse with the functional form of a piece-wise exponential, representing the rising and falling edge, respectively. This pulse shape is motivated by considering the SiPMs to be perfect capacitors. The `SimHit` time is preserved by translation to the pulse peaking time, which allows for studies of out-of-time pileup. The integrated charge in the bar is sampled at 37.1 MHz to produce a digitized (ADC) value for each channel and time sample. The 8-bit ADC follows a quasi-logarithmic scale over four partially overlapping subranges from $0-255$. This ensures optimal precision over a relatively large dynamic range.

Furthermore, the 6-bit TDC represents the subsample (range 0-49) in which the pulse crossed the TDC threshold (with special codes 62 for already being above threshold at the start of the time sample, and 63 for never crossing it).

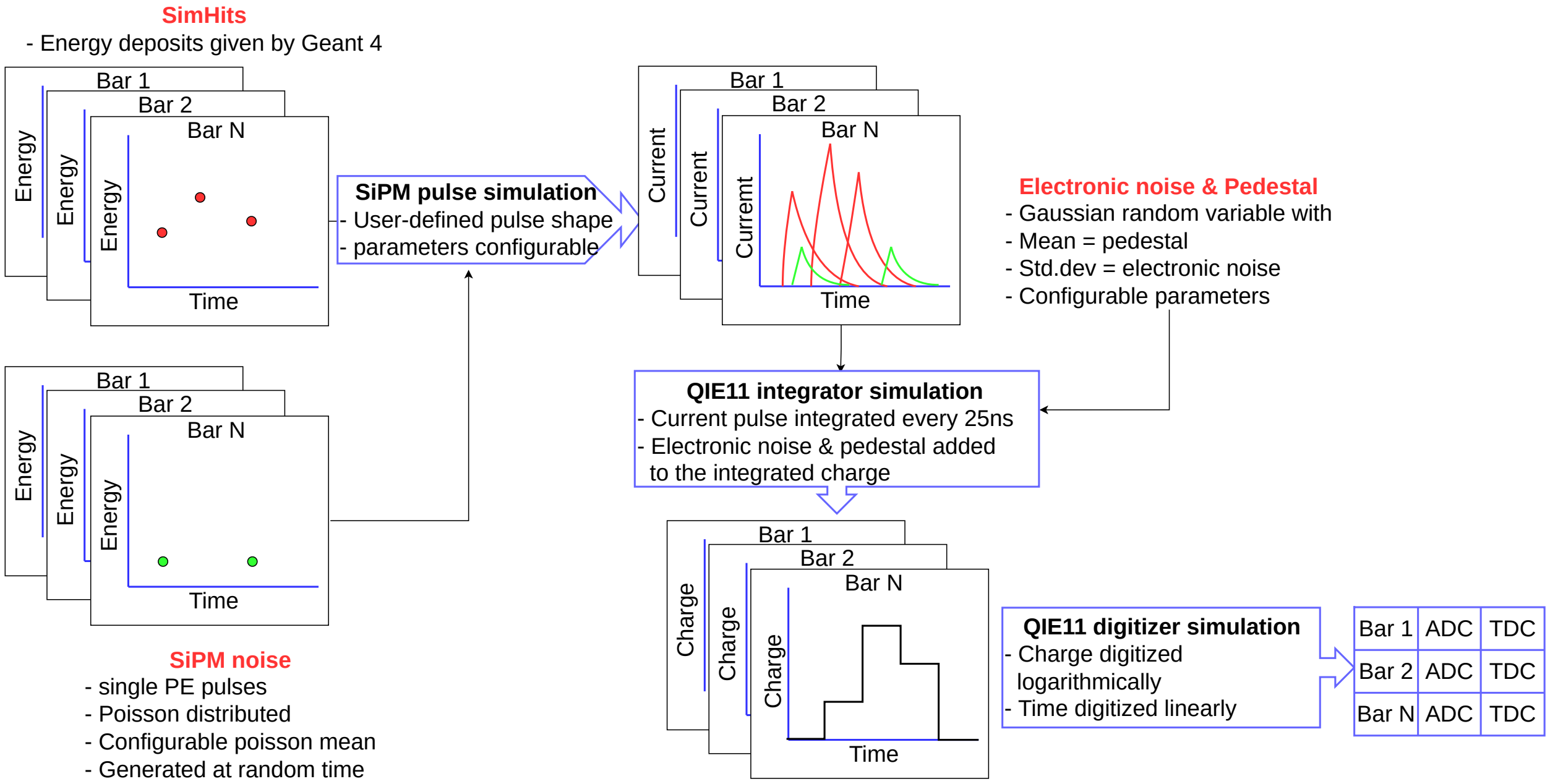

Figure 3.35: SimHits to digis workflow

#### 3.5.4.3 Firmware Based Electron Counting

The TS electron counting algorithm consists of four sequential components: the Hit Maker, Cluster Finder, Tracking Engine and Duplicate Remover. The algorithmic flow of each step is described below. Details of the hardware implementation are in Sec. 3.9.5.

**Hit Maker.** The first step in the TS processing is the Hit Maker. It determines if the pulse has a TDC coincident with beam, converts the received ADCs in the arrived pulse and the 4 subsequent pulses to charge. The ADC to charge relationship is shown in Fig. 3.36. The total charge of the hit is the 7-bit sum of these amplitudes over 5 time samples. The time of the hit is the first sample TDC. In firmware all channels are processed in parallel by separate Hit Maker instances.

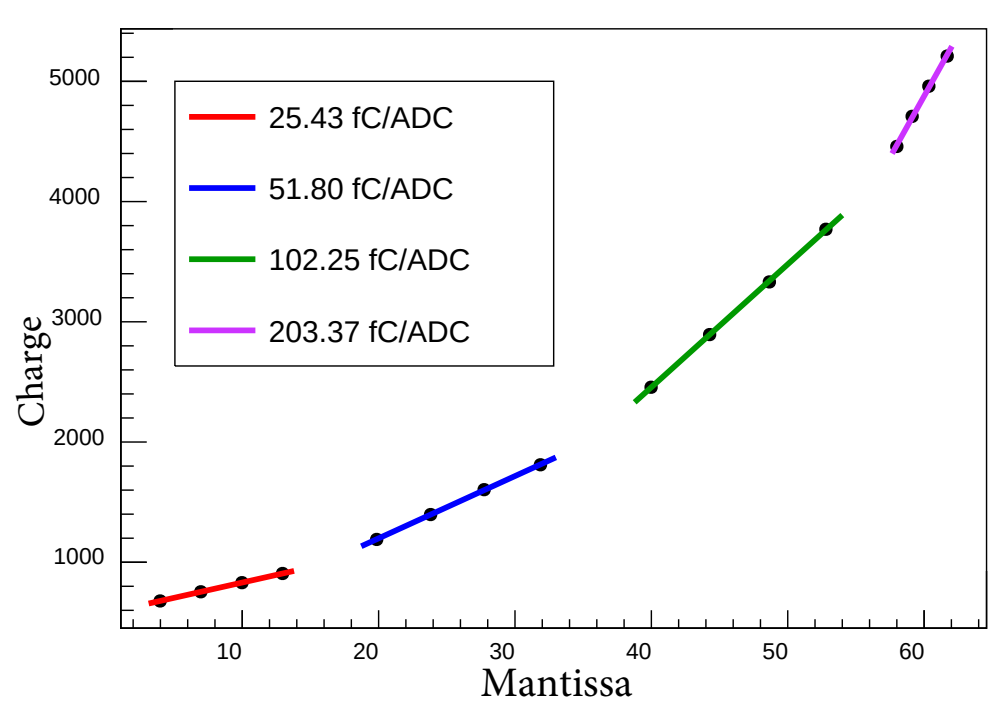

Figure 3.36: The Charge vs. ADC Curve in the QIE Card. Fig. from [82].

Pile-up is identified by a TDC which indicates that the pulse was over threshold in the preceding beam pulse. Currently the hit is flagged as a pile-up hit and removed. This type of pile-up is termed out-of-time pileup and its effects on the performance of the TS electron counting is described below.

**Cluster Finder.** Hits are passed to a cluster making engine, which functions on all channels in a module. TS clusters consist of one or two channels in subsequent rows of a module, as indicated in Fig. 3.37. Clusters are formed by finding adjacent channels with hits with a charge compatible with at least 50 PE. Clusters may have at most 2 channels.

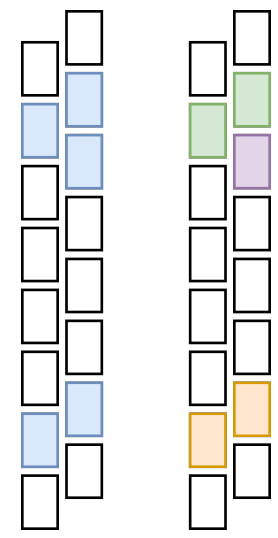

Figure 3.37: Left: hits above threshold in portion of TS module. Right: colors represent clusters derived from hits on left.

**Tracking Engine.** The tracking engine uses a look up table (LUT) of clusters from possible electron trajectories to form track candidates. A cluster centroid in Pad 1 may propagate to at most 3 centroid values in pad 2 given the beam constraints. These centroid positions are captured in a LUT which associates triplets of cluster centroids, one in each pad, with an allowed track. A diagram of track to cluster association for pad 1 and 2 is shown in Fig. 3.38.

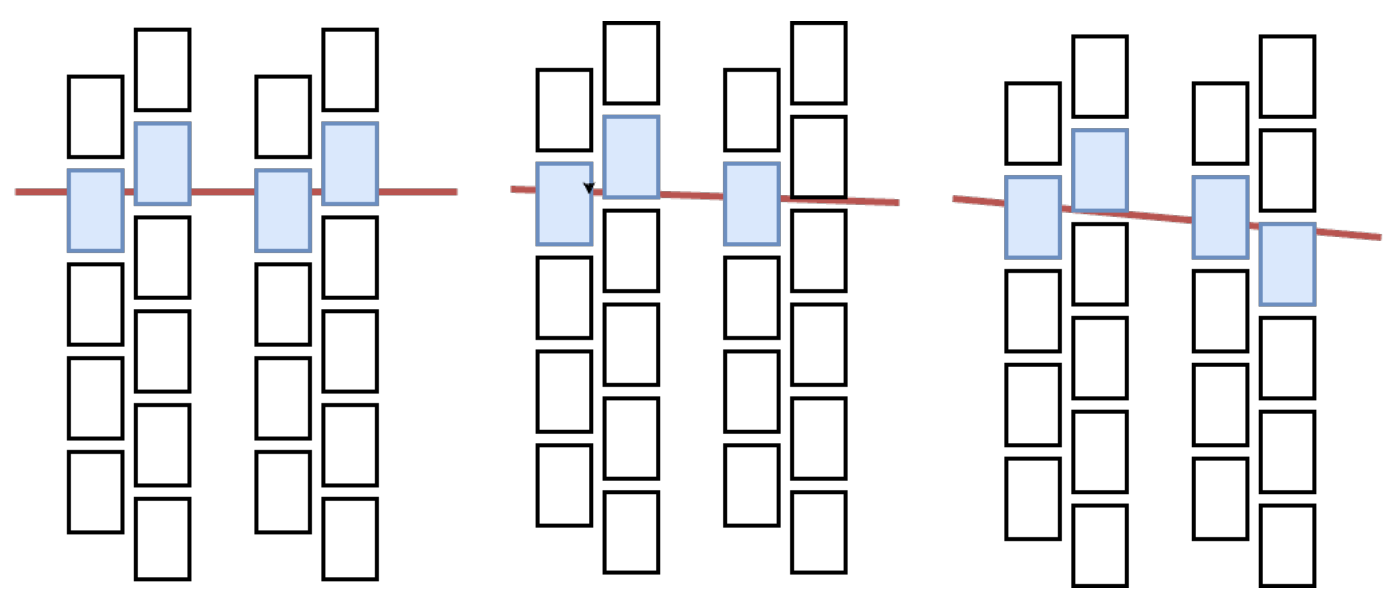

Figure 3.38: Possible cluster centroid pairings for tracks between pad1 and pad2.

**Duplicate Remover.** Duplicate tracks are defined as tracks sharing hits. All reconstructed tracks hit positions are compared, layer by layer. If tracks share a hit, only the track with the lowest residual is kept. The residual is defined as the sum of the squared difference of each cluster position to the track position, which is the unweighted average cluster position. By construction tracks cannot share hits in the first layer.

#### 3.5.4.4 Electron Track Finding Performance

Because the primary purpose of the TS is to accurately count the number of electrons, the performance is characterized by the over- and under-counting rates as a function of the number of incident beam electrons. In particular, over-counting, where the TS firmware identifies more tracks than true incident beam electrons, can lead to falsely triggered events. Minimizing over-counting is a strong design driver for the system.

The left plot in Fig. 3.39 shows the number of true beam electrons on the $x-$axis and the number of reconstructed TS tracks on the $y-$axis for samples with up to 4 incident beam electrons. The nominal system geometry is shown on the left. Efficiency to reconstruct 1-electron is 96%, note this includes a 2% inefficiency due to electrons which scatter and do not reach the target. The overcounting rate at 1 electron is $1.2\times10^{-5}$. The efficiency to reconstruct a 4 electron event as having 4 electrons is 65% while the over counting rate is $4.2\times10^{-5}$.

Signals from prior RF buckets, called out-of-time pile-up, can degrade the performance of the TS system. The right plot of Fig. 3.39 shows the performance of samples where there was an average of 1 electron in the preceding readout window. In this figure the 0 electron case is included, because tracks reconstructed from previous beam crossings would lead to a false trigger rate. The efficiency is decreased by 3% for a 1 electron sample due to the existence of extra hits. We believe we can mitigate this effect with a simple pile-up hit identification algorithm. The over-counting rate is not affected significantly as we reject hits which have a TDC which indicates the bar is already above threshold in this sample. There is no observed over-counting in the 0 electron case. We additionally studied the cases where there was on average 0.001 electrons 5 ns before the beam arrived. This led to no change in the over-counting rate.

We varied the size of the gap between scintillator bars in simulation to understand the effect of the geometry on the TS track finding performance. Reducing the gap size from the nominal 0.3mm spacing slightly reduced the efficiency and slightly increased the over counting rate. Increasing the gap size slightly reduced the over counting rate, and slightly increased the efficiency up to gap sizes of 0.7mm, after which the efficiency reduced. Fig. 3.40 summarizes the contents of the confusion matrices. We note that each electron has a 2% chance of not hitting the target, contributing to the under counting rate.

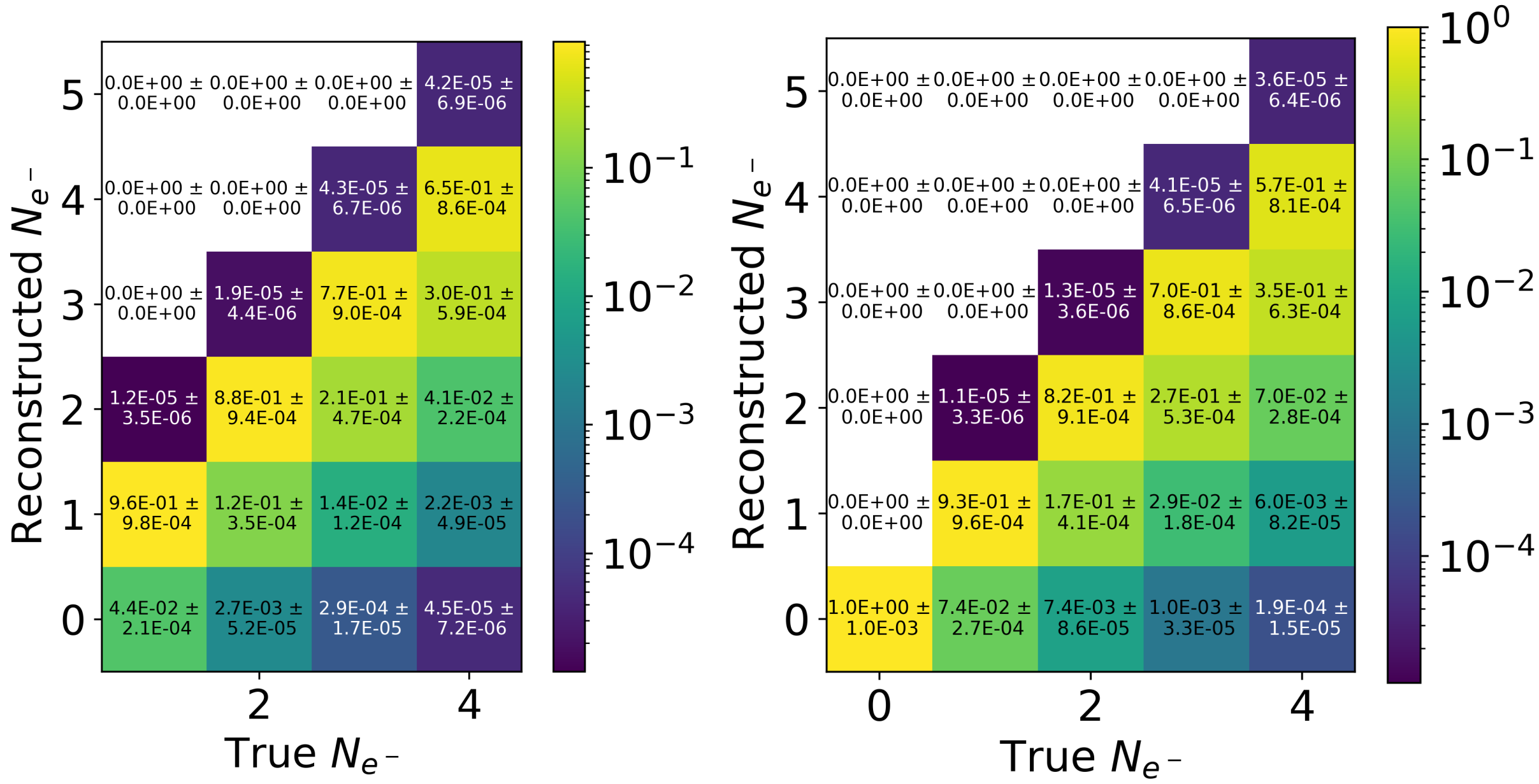

Figure 3.39: (Left) Reconstructed versus true electron count for nominal geometry with no out of time pile-up. (Right) Reconstructed versus true electron count with out of time pile-up.

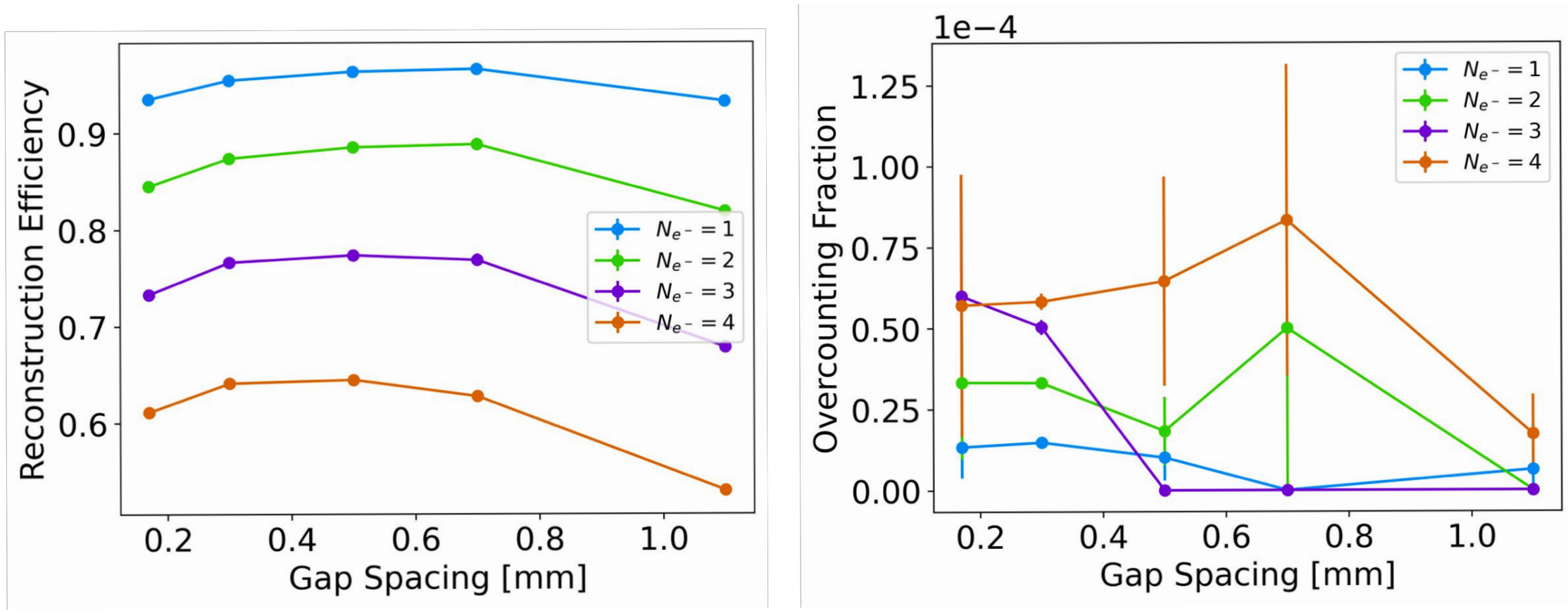

Figure 3.40: Left: Fraction of events where the number of reconstructed electrons matches the true number of simulated electrons as a function of the bar spacing in simulation. Right: Over-counting fraction as a function of the bar spacing in simulation.

We additionally studied the resiliency of the TS performance to cross-talk and single bar inefficiencies.

The impact of cross-talk is simulated by adding an additional hit with at least 50 PE adjacent to existing hits at a fixed probability (1 to 5%). Note, this is a conservative study as the likelihood of light leakage or electrical coupling generating a 50 PE response is low. The most likely cause of cross-talk at this rate would be misalignment between the bars and SiPM. Studies from a test beam at CERN, which are detailed in the next section, suggest an upper limit on additional hits due to cross-talk of 1.5%.

Fig. 3.41 shows the impacts of cross-talk on the efficiency and overcounting rate. It can be seen that the efficiency is not significantly impacted but cross-talk does lead to a significant increase in the over-counting rate. This motivates keeping the cross-talk rate low through good optical isolation and alignment of the bars and SiPMs.

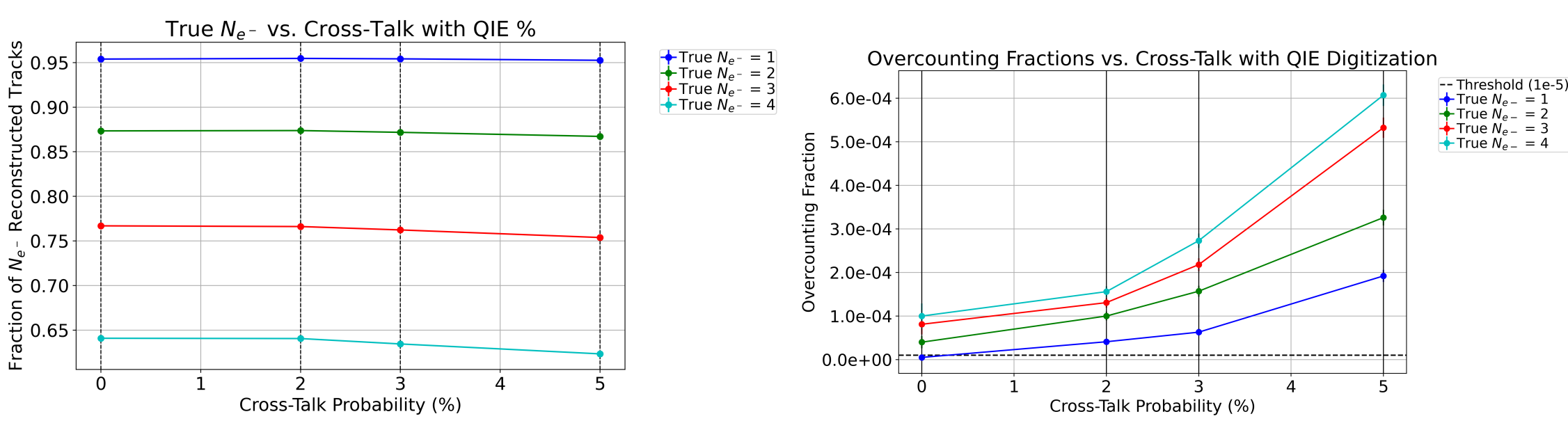

Figure 3.41: Left: Fraction of events where the number of reconstructed electrons matches the true number of simulated electrons as a function of the percentage of additional hits due to cross-talk. Right: Over-counting fraction as a function of the percentage of additional hits due to cross-talk

### 3.5.5 Prototyping and performance in beam tests

#### 3.5.5.1 CERN Testbeam

A TS prototype was tested in the CERN T9 testbeam in April of 2022. It consisted of one TS module with a SiPM board prototype instrumenting 12 adjacent channels of the 48 total. The scintillators were mounted to the electronics assembly using a 3D-printed housing. The SiPM board was screwed directly into the scintillator housing, which aligned the scintillators with the SiPMs. Signals from the SiPMs were digitized by one QIE RM. The QIE RM, its backplane and power regulators were assembled in an aluminum enclosure. In order to dissipate the power drawn from the electronics a fan blew air over the chips on the RM.

A zCCM prototype and a single backplane enabled power and control for the RM. Data from the RM was sent to a custom DAQ, which was based on the CAPTAN+X development board. Data, encoded with 8b10b, was transmitted over a pair of optical fibers at 5 GHz. Data from all channels was transmitted from the RM to the CAPTAN+X. Once received, data was decoded and a fixed number of time samples around the arrival of a trigger signal was buffered.

The prototype was built by Texas Tech University for the beam test as shown in Fig. 3.42. As illustrated in the drawing at the right of Fig. B.2 , only the top six scintillator bars of each layer were coupled to the SiPMs. Fig. 3.42 shows the TS prototype and other LDMX components which were tested at the same time (LYSO target, see section 3.6.2.4 and the HCAL, see Sec. 3.8.6).

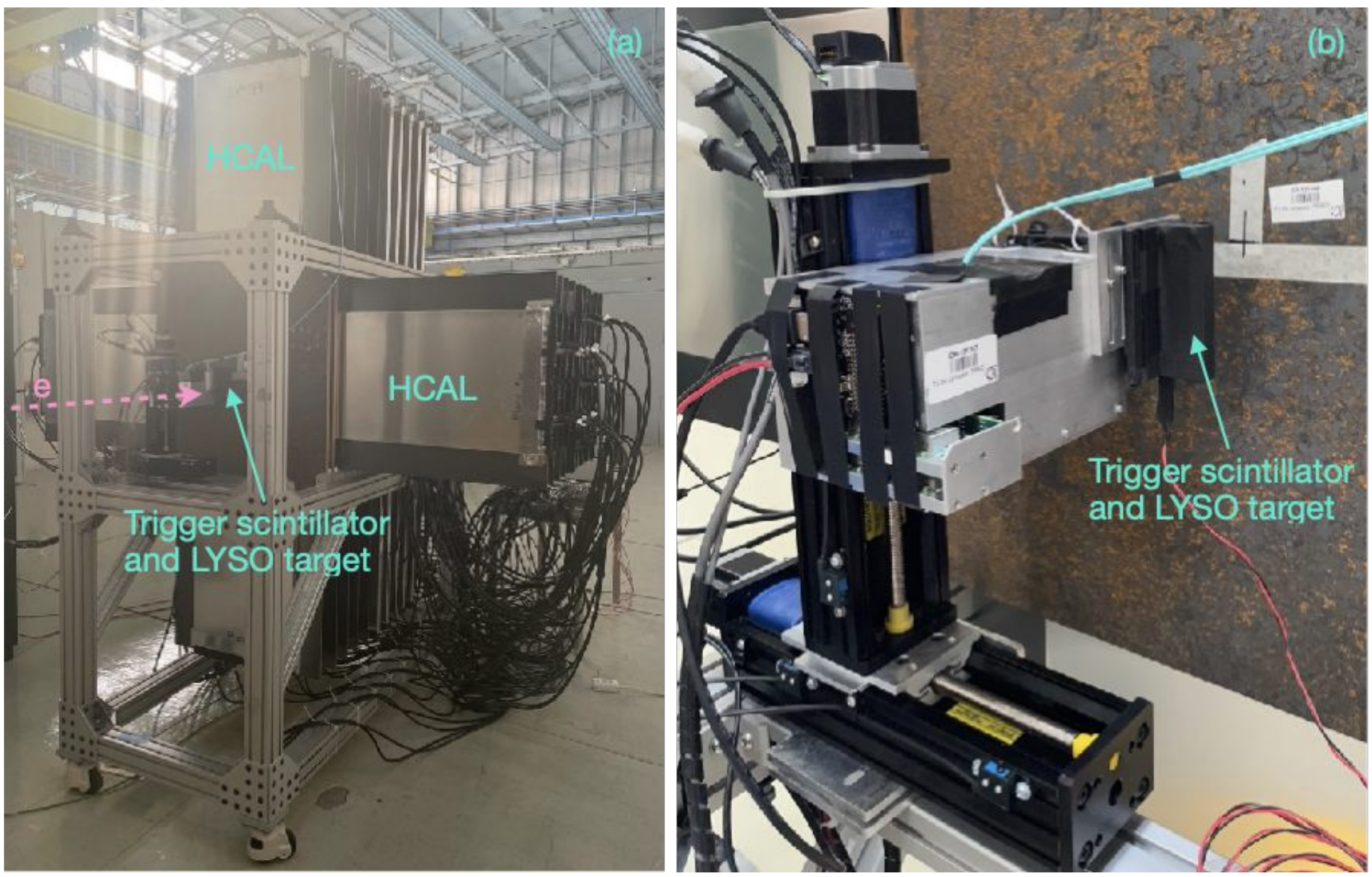

Figure 3.42: (a) The prototype trigger scintillator, LYSO target(see Sec. 3.6.2.4), and HCal (see Sec. 3.8.6) in the CERN beam-test area T9. The pink dashed line shows the beam direction. (b) The prototype trigger scintillator.

**Testbeam datasets and reconstruction:** There were two different periods of data taking. In total, approximately 8 million events from good runs with electrons were recorded. Testbeam data was triggered by a combination of beam scintillators, indicating the arrival of a beam particle, and information from two Cherenkov detectors, providing particle identification. The trigger was used to isolate either electrons, muons, or pions for each individual run. For the purposes of the TS studies, most particles studied are minimum ionizing and are not expected to interact differently with the TS.

The data buffered in the CAPTAN+X was extracted over Ethernet via the run control software and saved to disk. A fraction of the data was used to plot distributions of ADC, TDC, and linearized charge values to ensure that the system was functioning as expected.

A dedicated testbeam prototype simulation was used to produce Monte Carlo simulations to compare to testbeam data. The simulations use a right-handed coordinate system with $z$ along the incoming beam axis and positive $y$ upward, and origin in the center of the HCal. The TS was centered on $x = 0$, $y = 0$ and positioned 7 cm upstream of the front of the HCal. A particle gun of 4 GeV electrons was positioned at $z = -1100$ mm (about 50 cm upstream of the TS) with an illumination which covered the entire TS prototype.

Testbeam data reconstruction used the same reconstruction steps as simulation, up to and including clustering. This required converting testbeam data to `ldmx-sw` format for processing and intermediate inspection. To this end, a bitstream decoder, producing `ldmx-sw` "QIE Digis" was implemented, and the hit reconstruction was adapted to take a longer integration window, individual channel pedestals and SiPM gains into account. A mapping translates from electronics ID (the order of the ADCs from the individual channels are written in the bitstream) to physical SiPM location (bar ID).

**Results:** Test beam data is used to characterize the TS's sensitivity to noise/transient systematic errors, its response to MIPs, as well as more composite performance metrics (hit efficiency, secondary rate, etc). This section will address each of these results in order of dependence.

First we determined the thermal noise and pedestals, shown in the left plot of Fig. 3.43. This distribution is a combination of PE counts due to thermal PEs as well as cross-talk from the SiPM pixels and after pulsing. The distribution was fit against a Poisson distribution with parameter $\lambda$ for the DCR (dark current rate) as well as a Borel distribution with parameter $\epsilon$ corresponding to sources of cross-talk. The $\lambda$ and $\epsilon$ parameters drawn from this fit depend on the allowed time window $\tau$ of charge integration; for a value of $\tau = 10$ ns they converge to $\lambda \sim .04(1)\text{ns}^{-1}$ or a DCR of $\sim .20(5)$ MHz and $\epsilon \sim .05(1)$. These values are consistent with manufacturer estimates. The plot on the right of Fig. 3.43 is the number of peaks vs. fitted average charge. The slope of this plot determines the gain of an individual channel, while the intercept corresponds to pedestal.

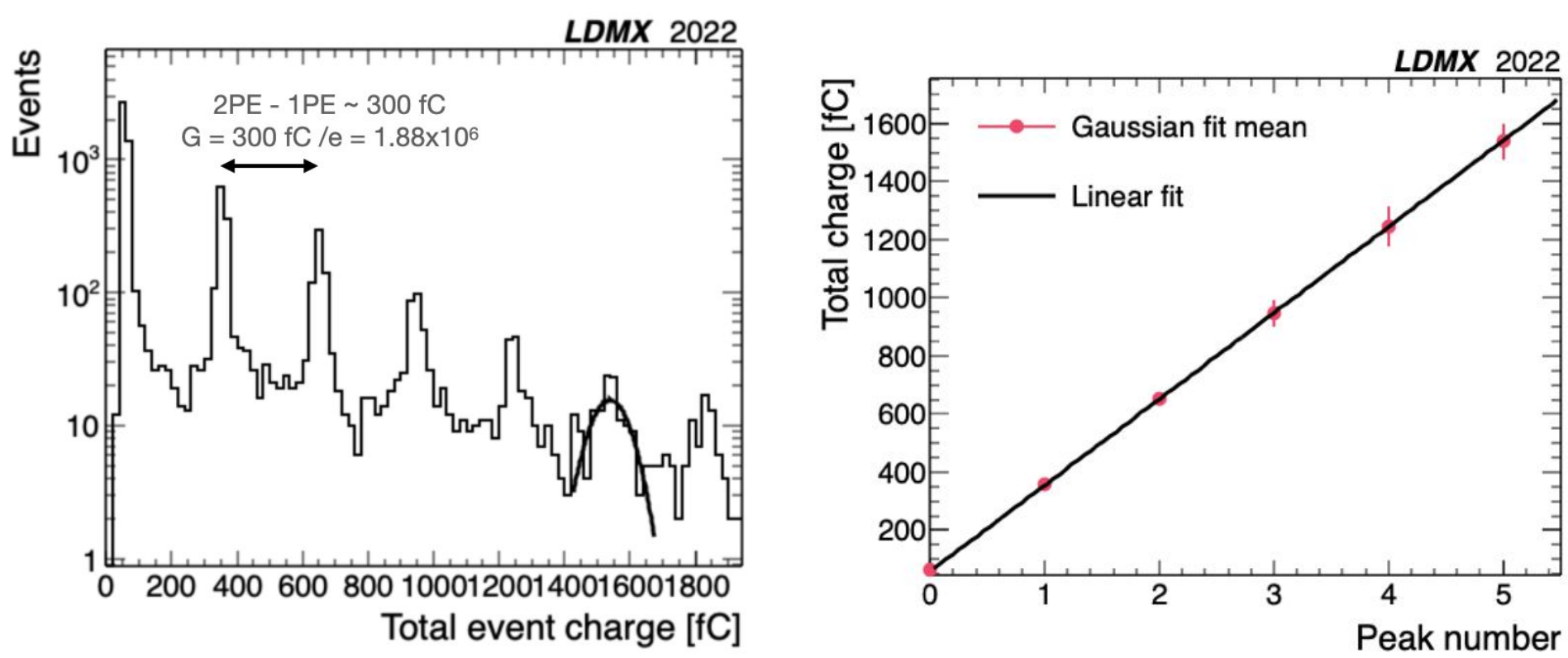

Figure 3.43: **Left**: Distribution of the integrated charge of a single run and a single channel. The black curve represents an example fitted Gaussian, which is used to extract the mean value of each peak. **Right**: Fitted mean of single PE peaks versus their index and a fitted line.

The fitted pedestals and gains were observed to be stable throughout the April 2022 Testbeam runs. Fig. 3.44 includes a run indexed graph of the fitted pedestals and gains for channel 4 in the prototype detector; the measured stability is typical of all non-dead bars. Channel gains were typically on the order of $\sim 1.9e^6$ with a channel-based variation of no more than $\sim 1e^5$. Events were removed which had pulse shapes consistent with noise or other electronics artifacts.

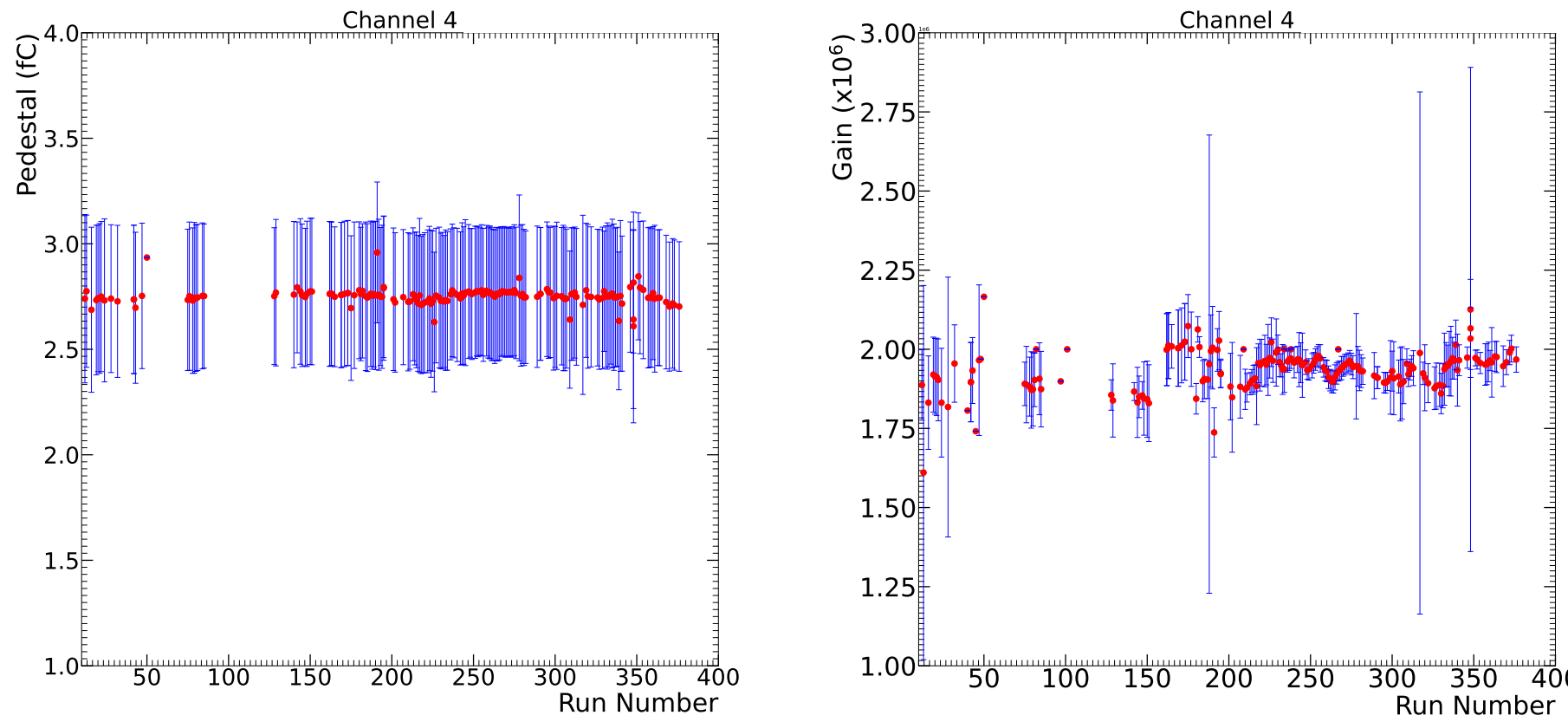

Figure 3.44: **Left**: Distribution of measured pedestals versus run for channel four. **Right**: Distribution of measured gain versus run for channel four.

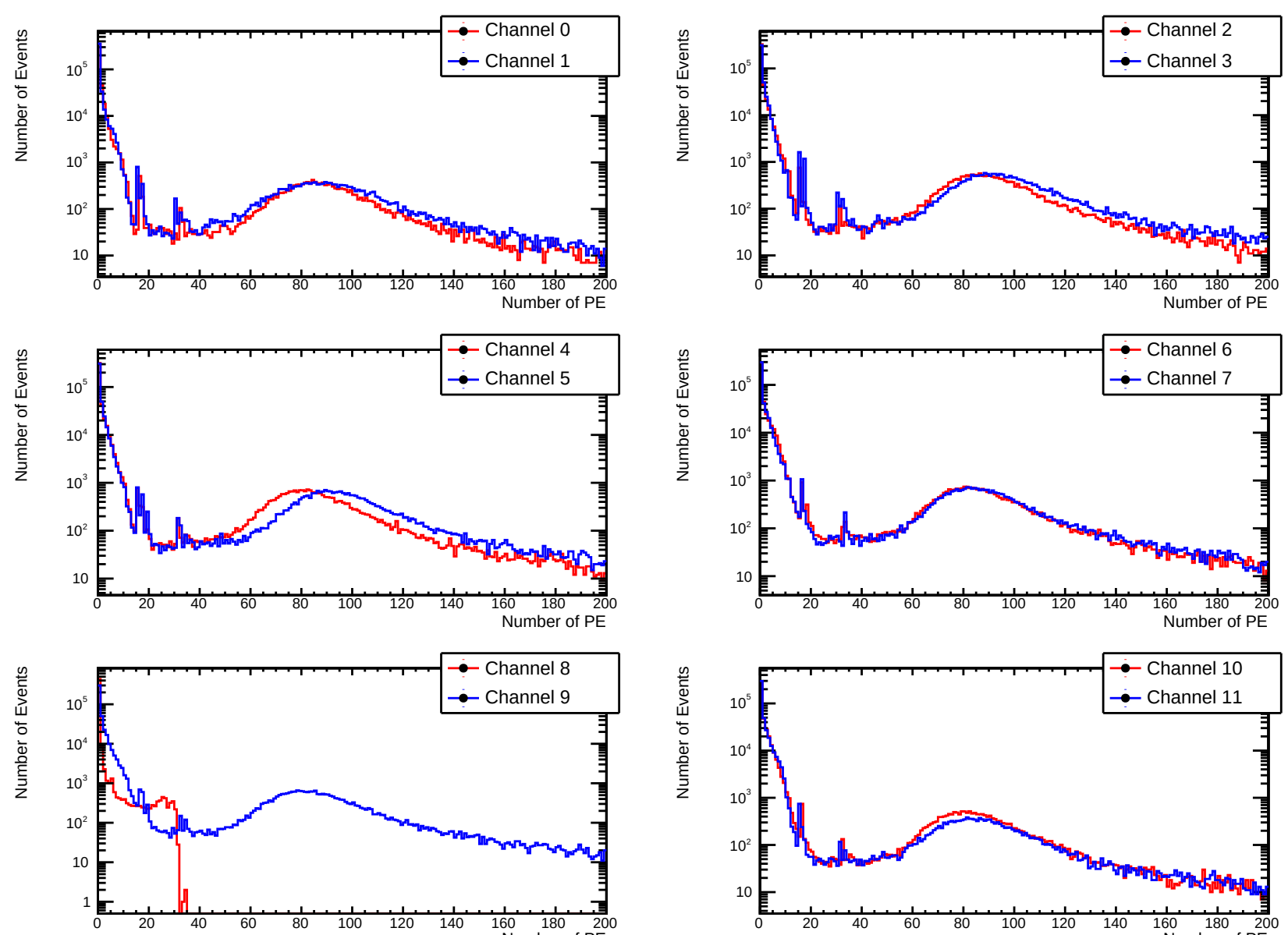

Figure 3.45: The PE distributions for the 12 channels in the testbeam. Channel 8 had a dead SiPM.

To determine the position and stability of the MIP peak, charge distributions were fit with a Langau distribution, the convolution of a Landau and Gaussian distribution. Fig. 3.46 demonstrates the Langau distribution overlaid with a quadratic polynomial background. The fit constrains the Gaussian mean to be aligned with the most probable value (MPV) of the Landau distribution. This MPV is plotted with respect to time on the right in Fig. 3.46. The MPV in any run lies within $\sim 5\%$ of its average value; run to run variations are small and do not require inter-run calibration. However, the MPV varies significantly between channels; therefore individual channel $PE$ values are scaled to ensure uniform response.

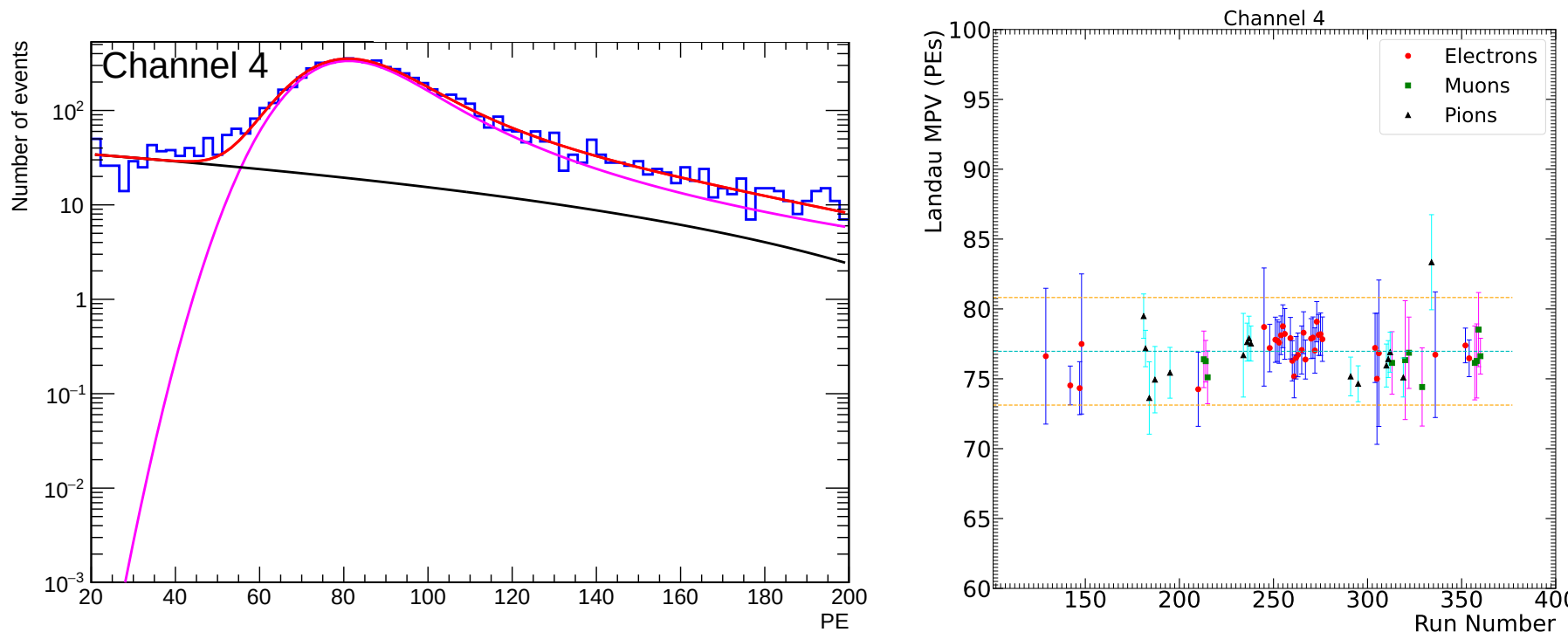

Figure 3.46: **Right**: Fit functions used to fit the MIP region properly. The functions shown here involve the Langau (pink), polynomial background (black) and the sum of those two (red). The plot is shown in log scale.**Left**: Spectra of Landau MPVs over time for channel 4. The cyan dotted line corresponds to a weighted average of the MPVs measured, weighted by the number of MIPs for the run. The dotted orange lines represent 5% uncertainty.

The MPV with respect to which the MIP profiles were scaled was the channel and run averaged MPV, $\sim 80$ PE. Each channel was multiplied by this over its own run averaged MPV; the resultant MIP distributions for a single run are shown in Fig. 3.47.

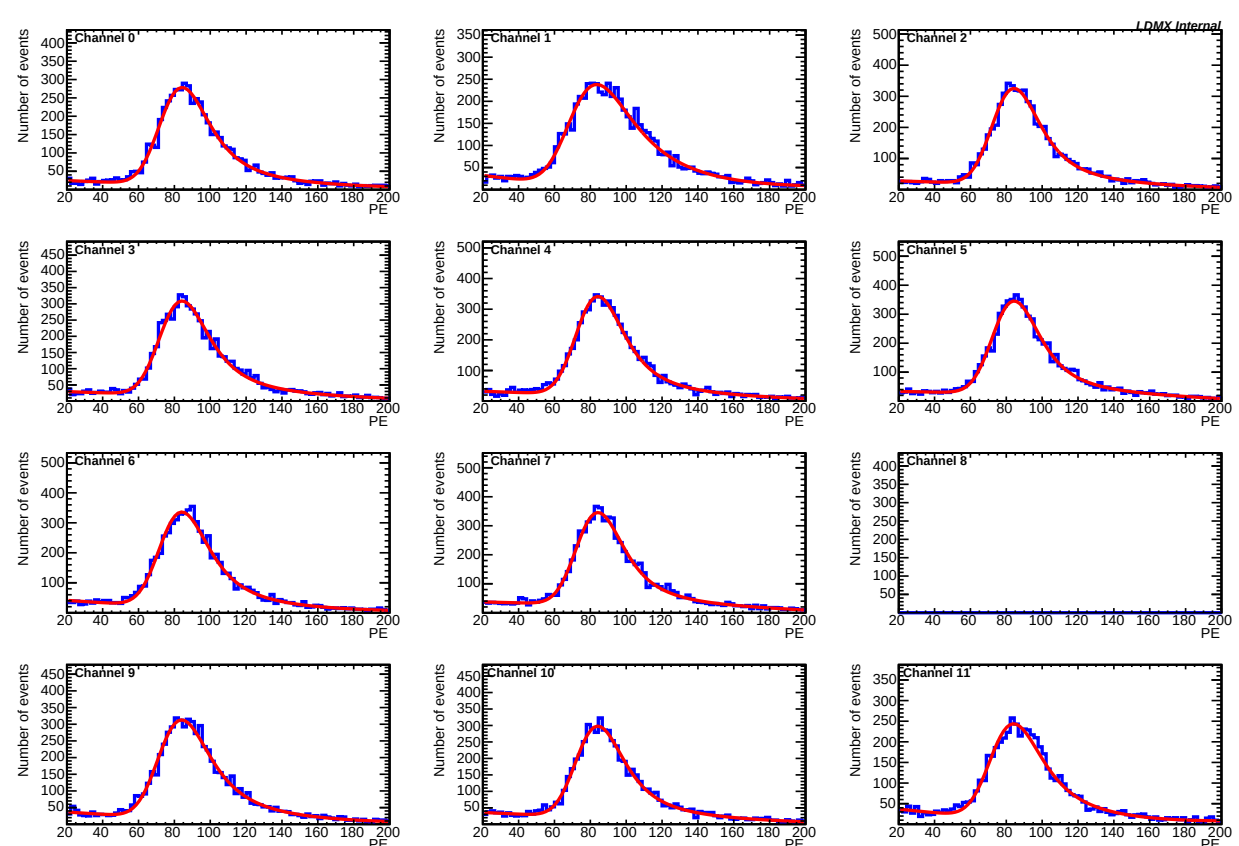

Figure 3.47: MIP response summary for all the instrumented channels of Run 236 after inter-calibration.

With the single photon and MIP distribution of each TS bar evaluated and calibrated, we determine composite performance estimates of the TS prototype. Calibrated bars are used to determine a signal to noise ratio (SNR). If $\sim 50$ PE is chosen as a threshold for the signal PE distribution, then the ratio between the integral of charge above threshold over the total integrated charge for the profiles in Fig. 3.47 is $> .99$, or a SNR greater than 100.

In order to estimate the prototype detectors' hit efficiencies we use a metric based on the tag and probe method:

$$\epsilon_i = max(\frac{N_{i \text{ and } i\pm1 \text{ and not } i\pm2}}{\mathcal{N}_{i\pm1 \text{ and not } i\pm2}}) \tag{3.2}$$

In Equation 3.2, $N_{i+1 \text{ and not } i+2}$ are events where a bar exceeds 50 PE in channel $i+1$, does not in $i+2$. This is consistent with an electron passing through $i+1$ and then either through channel $i$ or the gap between $i$ and $i+2$; this ratio ought to be $\sim 1$ times the fractional gap between bars. This efficiency was simulated in MC against a prototype geometry whose dimensions were measured from the prototype, but several months after the run. The resultant hit efficiency metrics are given on the left in Fig. 3.48. The plot on the right in Fig. 3.48 is the time averaged data over MC ratio.

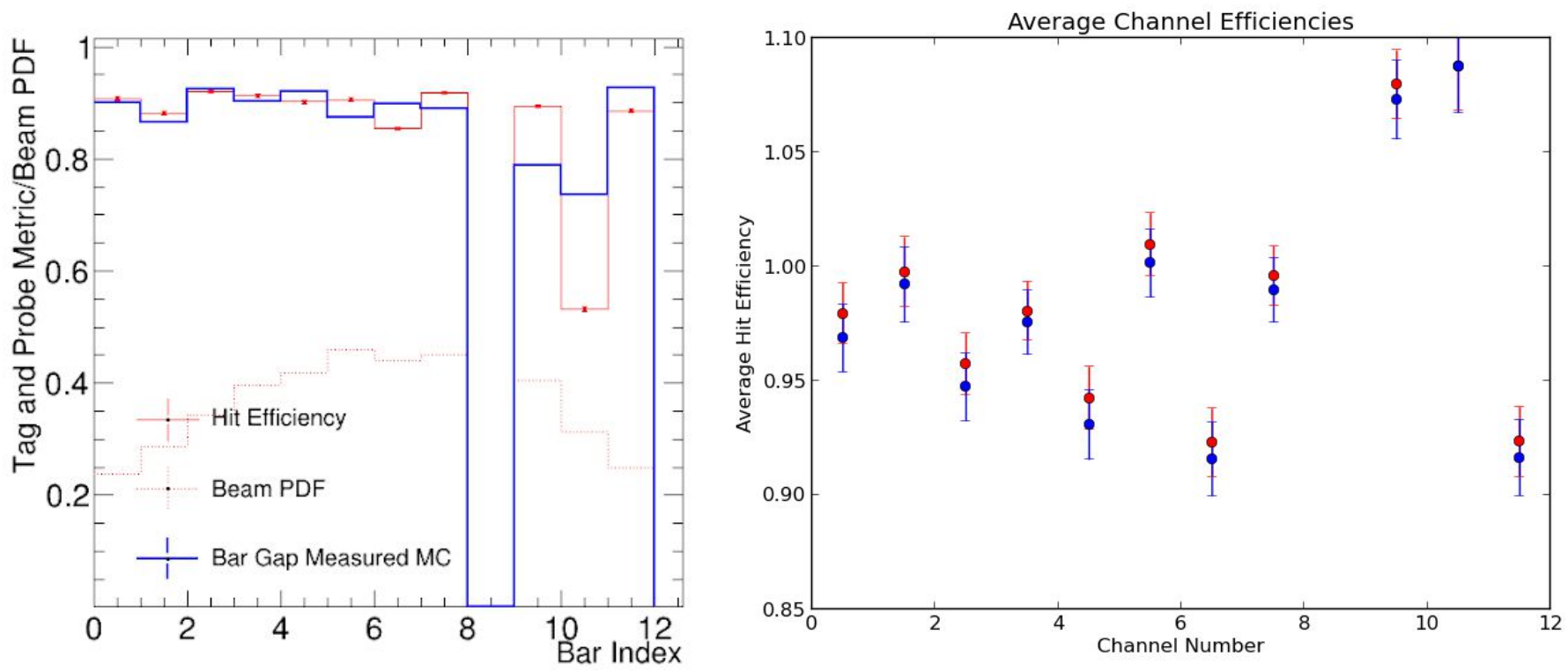

Figure 3.48: **Right**: Hit efficiency metric for run 354, **Left**: Run averaged hit efficiency metric per channel.

For evenly indexed channels, there were runs with noted electrical connectivity issues; these runs may account for the apparent drop in efficiency. The points in the right plot of Fig. 3.48 exclude these runs in the 'dip' region. For channels 9 and 10, the hit efficiency metric exceeds 100%. indicating that the bars in the prototype likely shifted after the run and before measurements of the assembly were taken. Bar 8 had a dead SiPM.

#### 3.5.5.2 Tests at the S30XL dump

An evolved version of the TS prototype described above was installed in front of the S30XL beam dump in the beginning of 2025. The TS bars and housing are the same as in the CERN test beam, but new prototypes of the SiPM board, SiPM cable, backplane, and zCCM were produced. Additionally, the readout module board was the same as will be in the final design, allowing for higher precision in the TDC measurement. The APx was used to read out the device, store data on triggers, and synchronize with the LCLS-II timing system. Significant core TDAQ firmware was deployed and tested, described further in 3.9.4. Fig. 3.49 shows the device installed in the S30XL beamline.

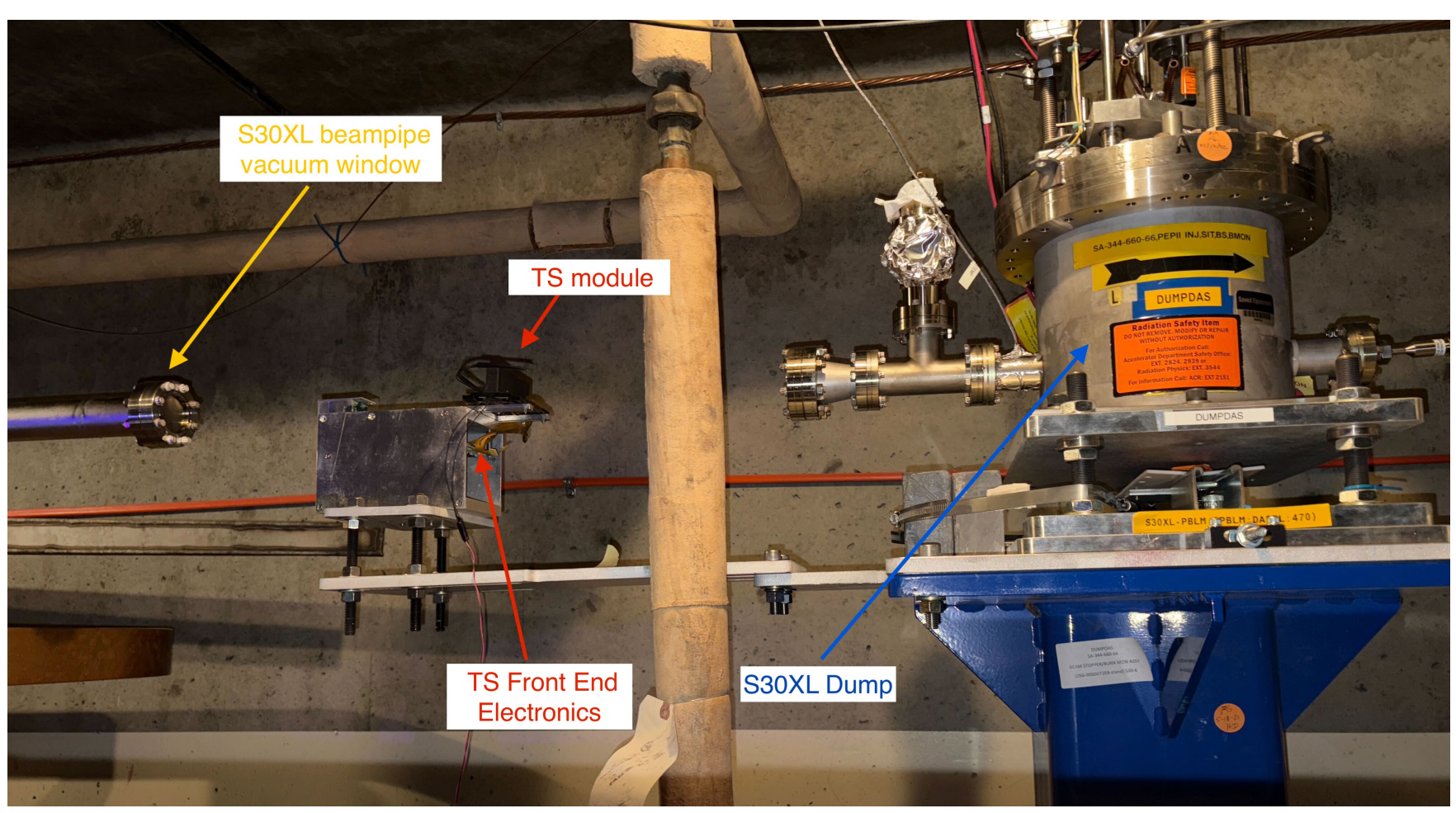

Figure 3.49: The TS prototype installed in the S30XL beamline. The beamline vacuum window, TS prototype and electronics, and dump are indicated in the figure.

The goal of this test beam was to measure the dark current from the LCLS-II cathode when the S30XL

kickers are sending beam. The kicker flat top is approximately 600 ns, and dark current is expected to populate the 5 ns LCLS-II RF buckets during this period. The value of the dark current was unknown prior to this test with an expectation that it was 1-10 pA. Additional goals were to test the new prototype TS hardware and develop and test the TDAQ hardware, firmware and software.

The dark current was successfully measured in several runs. The APx readout was configured to read out 100 readout samples, which encompassed the entire 10Hz kicker pulse window, on every LCLS-II timing signal. We additionally tested self-triggering by the TS prototype, however, the beam background rate was ≃1kHz which made the self triggering infeasible for a 10Hz dark current measurement. However, the ≃1kHz rate was low enough that the beam background rate in the 10Hz kicker window was negligible. This was verified by taking data when the kickers were off but we were still receiving the timing signal.

Fig. 3.50 shows the recorded PE as a function of 26.7 ns readout samples for one kicker pulse window, which is indicated by a large gray band in each plot. The colored dots indicate individual electron peaks, where those with the same color came in the same timesample. This particular event has 9 individual electrons in the 600 ns window.

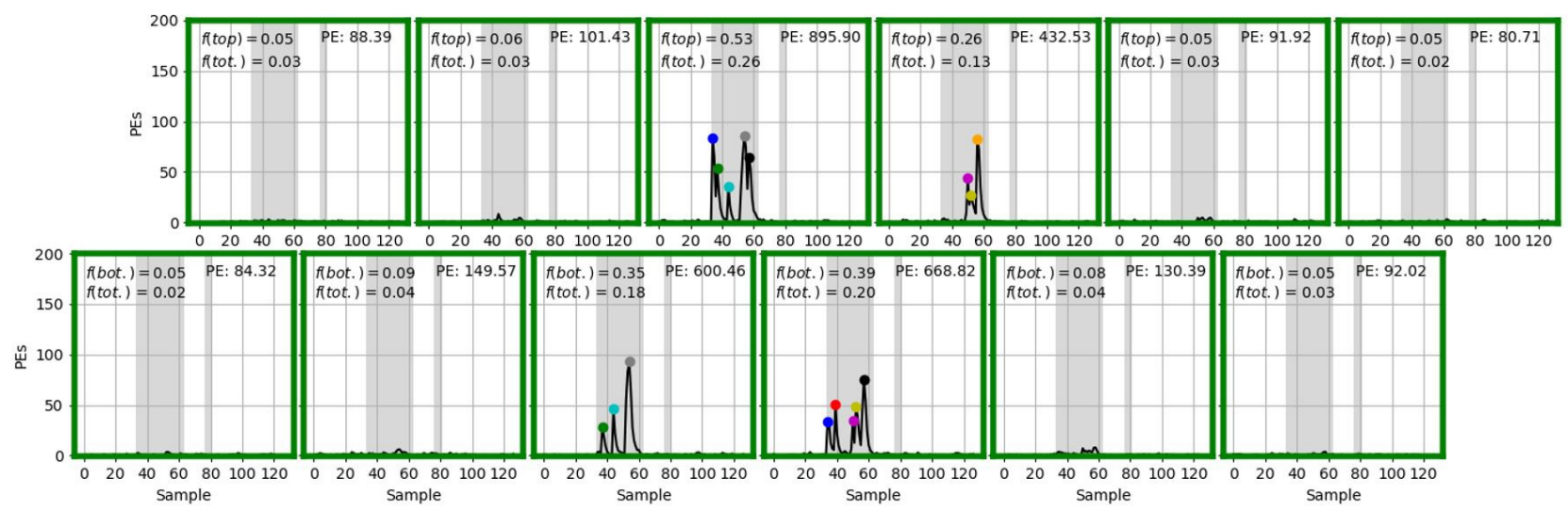

Figure 3.50: Example readout for one kicker pulse window for all instrumented bars in the TS prototype. The x-axis is the 26.7 ns time sample number. The y-axis is the number of read-out PEs. The large gray region indicates the kicker pulse window. The small gray region indicates an area of high beam background. The colored dots indicate peaks in the same time sample. Plots are arranged in the same order as the bars are physically installed.

Fig. 3.51 left shows the PE for all bars fit to a Poisson distributed sum of Landau peaks. The right plot shows the number of electron peaks found in each kicker window and the distribution is fit to a Poisson distribution. In this particular run, a Poisson average of 6.98 electrons was observed, leading to dark current estimate of 1.6 pA.

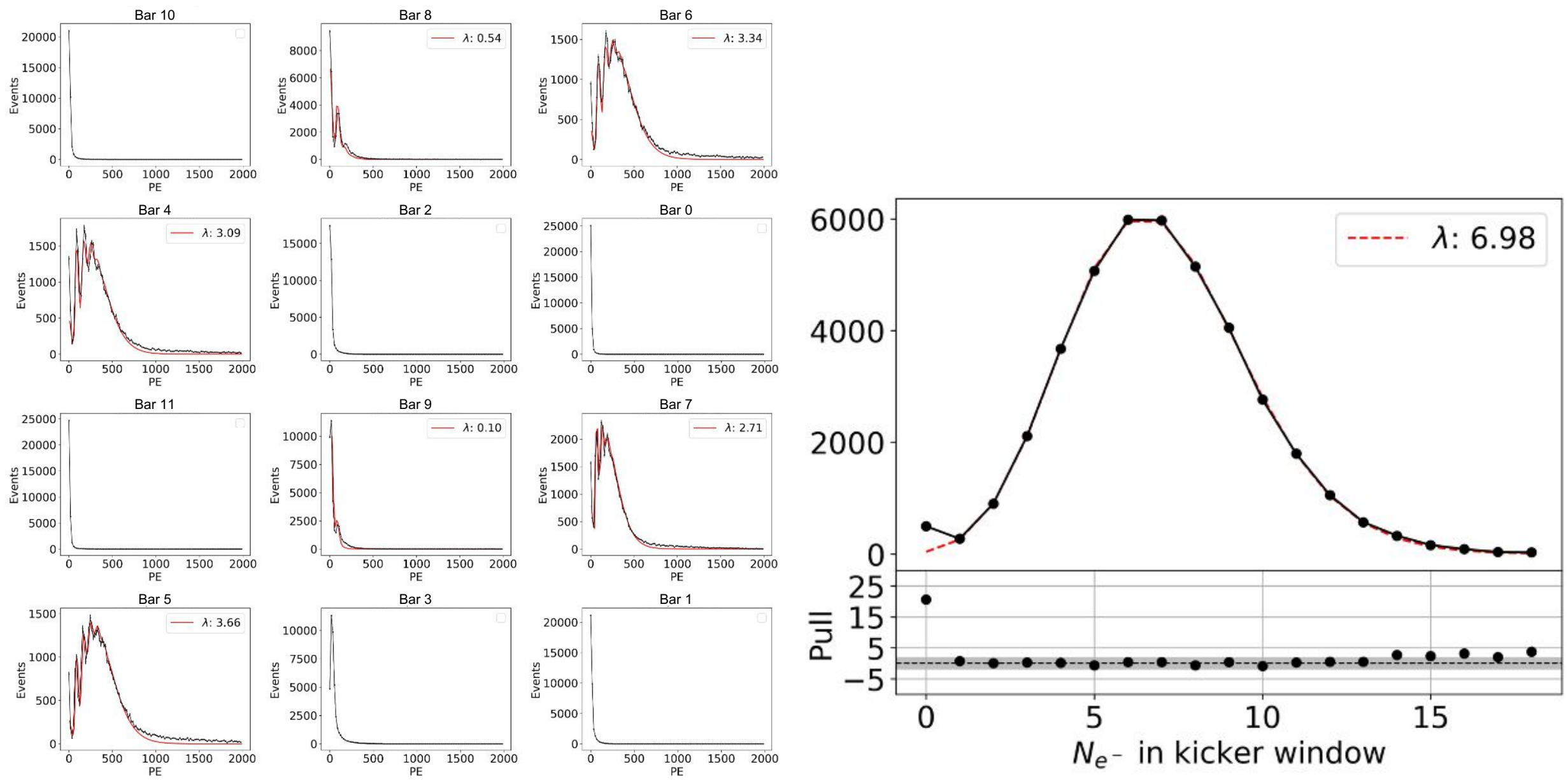

Figure 3.51: (Left) Histogram of PE for each bar in the kicker flattop window fit with the sum of Poisson distributed Landau peaks. (Right) Number of electrons measured in the kicker window, fit with a Poisson distribution (red). The bottom ratio plots shows the pulls of the fit, with the gray band indicating 2 $\sigma$ uncertainty on the fit.

TDC information was used to measure precise electron arrival times. Fig. 3.52 shows the arrival times of the electron pulses in ns with respect to the start of the readout window, clearly indicating the 5 ns RF structure of the LCLS-II beamlines. The right plot of Fig. 3.52 shows the arrival time in ns versus the bar number on the $y$-axis, showing the acceptance of the kicker magnets changing as they undergo their 150 ns ramp.

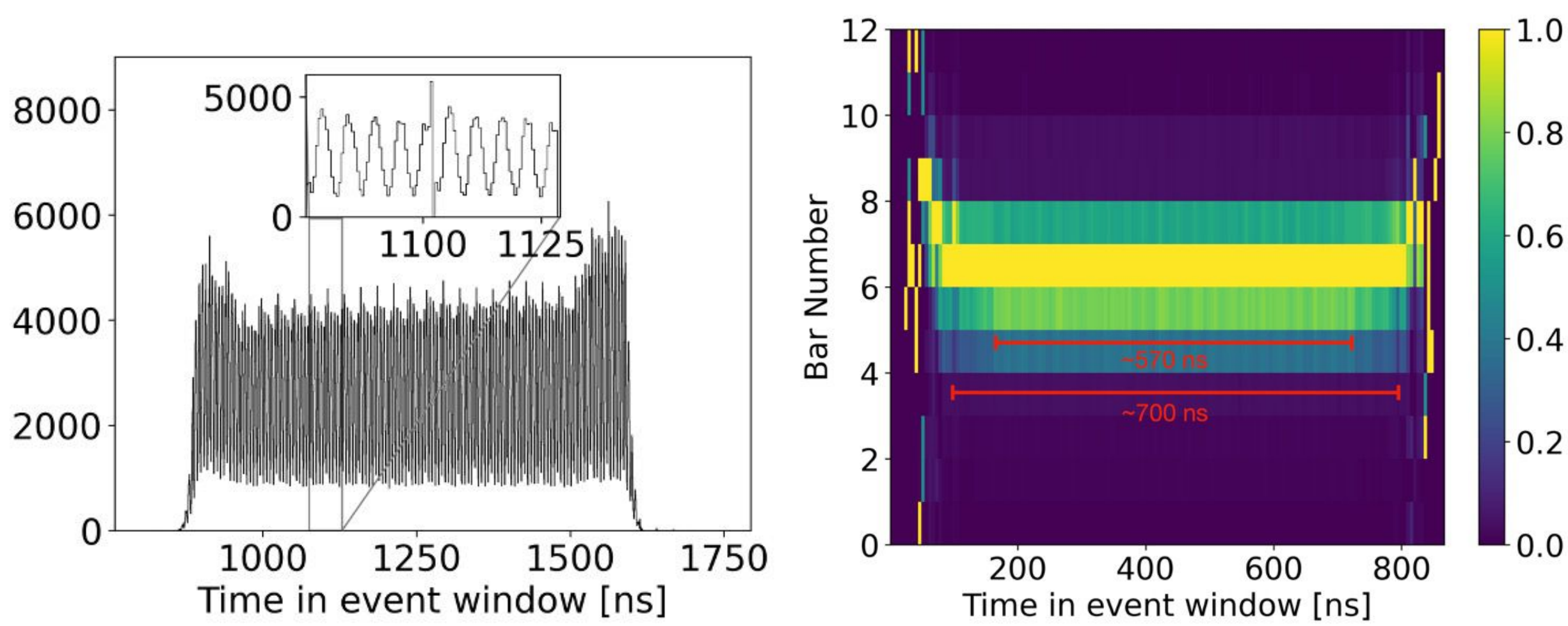

Figure 3.52: (Left) Electron pulse arrival time in ns in the readout window summed over all bars. (Right) Arrival time versus bar number for electron pulses.

This test beam has successfully demonstrated the basic capabilities of the TS hardware and the TDAQ firmware and software, as well as providing precise information on the characteristics of the LCLS-II dark current.

### 3.5.6 Construction and Quality Control

**TS Modules:** The PVT bars and light pipes will be manufactured by an external company. The prototype used bars manufactured by Elgin with adequate performance. The ESR wrapping and module housing will be made by Caltech with a custom jig.

**Electronics:** The manufacturing of all electronics components is to be coordinated by FNAL. FNAL will be responsible for electrical tests upon receipt of the boards and demonstrate basic functionality at FNAL before shipping to Stanford for assembly.

**Assembly:** Assembly of the modules with the SiPM boards will take place at Stanford. The cold plate and mounting of the TS front end electronics will be done at SLAC. Precise measurements of the as built system will be done at SLAC prior to installation.

**Quality Assurance:** Each module will be calibrated and tested prior to installation using our methods developed to test the TS modules for the S30XL beam test. We will measure and archive the gain and pedestals in our test stand.

### 3.5.7 Risks and Opportunities

Given the relative simplicity of the TS system, there are few risks to enumerate. The primary risk for the TS system is that the over-counting rate is significantly higher than in simulation. If this is the case we have the option to add a 4th layer to the TS system to require an additional level of coincidence and reduce the fake rate. There is a risk that the PVT to light pipe connection fails. We will test with prototypes during the construction phase to assess the severity of this risk. We can mitigate this risk with spare bars. There is a risk that the primary beam dumps in the front end electronics. They are fairly radiation hard but could be replaced with spares if damaged.

There is an opportunity to increase the spatial resolution of incoming electrons at trigger level by adding a layer of vertically oriented bars to each TS module. Preliminary studies indicate that the resolution in the horizontal direction, which currently equals the bar length at O(cm), could be reduced to a few mm while conserving the total number of readout channels. 2D information at trigger level would in particular resolve in-time pileup electrons which are close in $y$, improving electron counting accuracy at higher multiplicity. Furthermore this information could be used in the high-$p_{\mathrm{T}}$ trigger (currently limited to $p_y$, see Sec. 3.9.5.5) dedicated to study electronuclear processes in the target, which is further discussed in Sec. 4.9.

## 3.6 Targets

### 3.6.1 Overview

The target is situated between the Silicon Tagging Tracker and the Silicon Recoil Tracker in a common mechanical structure with the last TS module. The baseline is a passive target made up of a 4 cm by 10 cm sheet of tungsten with a thickness corresponding to 0.1 radiation lengths ($X_o$). This thickness has not been optimized, but is expected to provide a good balance between signal rate and transverse momentum resolution from multiple scattering. We also discuss alternative passive targets, and Sec. 3.6.2 investigates LYSO as an alternative active target [83].

Following Refs. [84, 85], for a given kinetic mixing parameter ($\epsilon$) and $A'$ mass ($m_{A'}$), the cross-section for an electron with energy $E_e$ to produce an $A'$ with energy $E_{A'} \equiv xE_e$ is

$$\frac{d\sigma}{dx} = 4\alpha^3\epsilon^2\chi(E_e, m_{A'}, Z, A)\sqrt{1 - \frac{m_{A'}^2}{E_e^2}}\frac{1 - x + x^2/3}{m_{A'}^2\frac{1-x}{x} + m_e^2 x}, \tag{3.3}$$

where $m_e$ is the mass of an electron, $Z$ is the atomic number and $A$ [g/mol] is the atomic mass of the target atoms. Above, $\chi$ is the effective flux of photons given by

$$\chi(E_e, m_{A'}, Z, A) = \int_{(m_{A'}^2/(2E_e))^2}^{m_{A'}^2} \frac{t - (m_{A'}^2/(2E_e))^2}{t^2} G_2(t)dt, \tag{3.4}$$

where the atomic form factor, $G_2 = G_2^{\text{el}} + G_2^{\text{in}}$, has an elastic and inelastic component. The expressions for $G_2^{\text{el}}$ and $G_2^{\text{in}}$ are provided in Refs. [86, 87, 88]; note that the expression for $G_2^{\text{in}}$ in Ref. [84] has a typographical error where the second bracketed term is squared. The $A'$ yield for $N_e$ electrons with initial energy $E_o$ incident on a target of $T$ radiation lengths is then given by

$$N_{A'} = N_e \frac{N_o X_o}{A} \int_{m_{A'}}^{E_o - m_e} \int_{E_{A'}+m_e}^{E_o} \int_0^T I(E_e; E_o, t)\frac{1}{E_e}\left(\frac{d\sigma}{dx}\right)_{x=E_{A'}/E_e} dt dE_e dE_{A'}, \tag{3.5}$$

where $N_o$ [1/mol] is Avogadro's number and $X_o$ [g/cm$^2$] is a unit radiation length of the target material. Above, $I(E_e; E_o, t)$ is the energy distribution of an electron after traversing $t$ radiation lengths. For thin targets ($T << 1$), $I(E_e; E_o, t) \approx \delta(E_e - E_o)$ and Eq. 3.5 becomes

$$N_{A'} = N_e \frac{N_o X_o}{A E_o} T 4\alpha^3\epsilon^2\chi(E_o, m_{A'}, Z, A)\sqrt{1 - \frac{m_{A'}^2}{E_o^2}} \int_{m_{A'}}^{E_o - m_e} \frac{1 - E_{A'}/E_o + (E_{A'}/E_o)^2/3}{m_{A'}^2\frac{1-E_{A'}/E_o}{E_{A'}/E_o} + m_e^2 E_{A'}/E_o} dE_{A'}. \tag{3.6}$$

To compare relative $A'$ production in thin targets at fixed $E_o$, we need only consider

$$N_{A'} \propto \frac{X_o T Z^2}{A}\frac{\chi(E_o, m_{A'}, Z, A)}{Z^2}, \tag{3.7}$$

The factor $\chi/Z^2$ is plotted versus $m_{A'}$ for aluminum and tungsten targets in Fig. 3.53. Noting that $\chi/Z^2 \approx 10$ for $m_{A'} < 0.1$ GeV, the first term in Eq. 3.7 gives simple but useful quantities to compare thin targets at fixed $E_o$ and small $m_{A'}$. First, $\frac{X_o Z^2}{A}$, can be used to compare targets at fixed $T$. Alternatively, we can substitute $X_o T = \rho l$ into Eq. 3.7, where $\rho$ [g/cm$^3$] and $l$ [cm] are the target density and thickness. This gives a second quantity for comparing targets at fixed $l$, $\frac{\rho Z^2}{A}$. Both quantities are plotted for several elements in Fig. 3.54. For larger $A'$ mass, the $\chi/Z^2$ factor becomes more important.

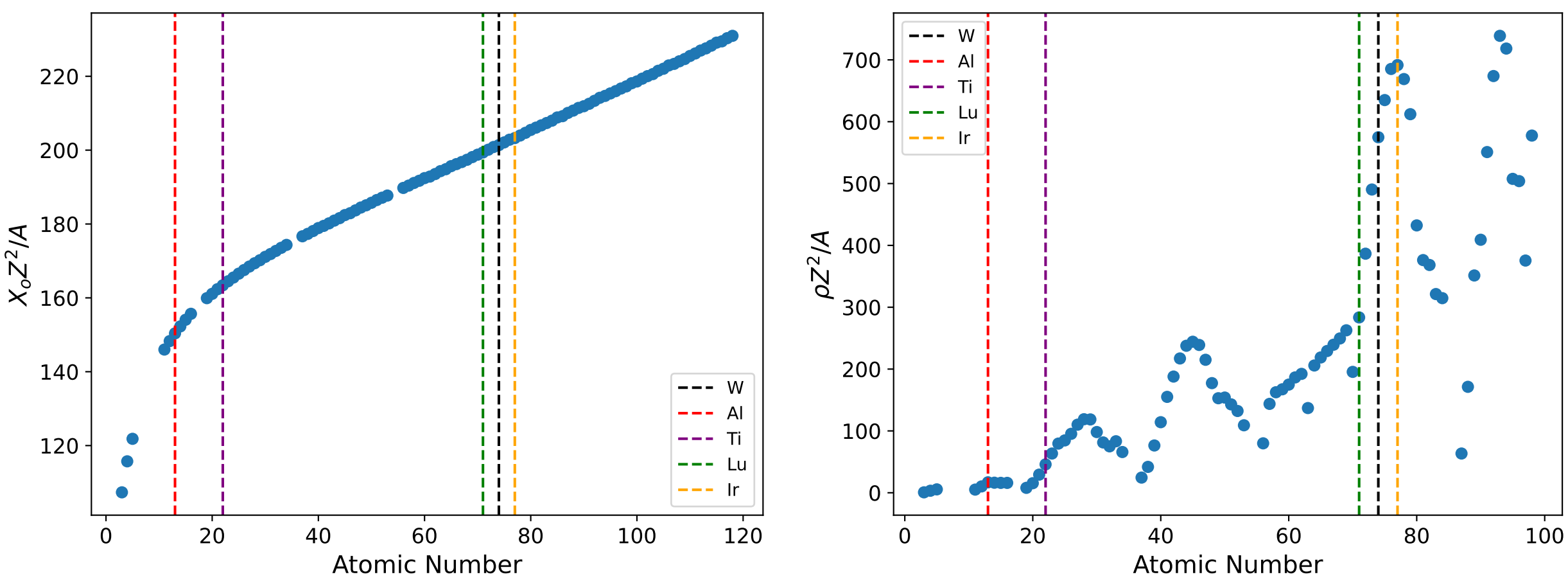

(a) (b)

Figure 3.54: (a) $X_oZ^2/A$ and (b) $\rho Z^2/A$ are plotted versus atomic number for several elements. The dashed red, purple, green, black, and orange vertical lines indicate aluminum, titanium, lutetium, tungsten, and iridium, respectively. All data is obtained from Ref. [89]. Elements without a listed density in Ref. [89] are excluded, and for gases, the density is taken at 20°C and 1 atm.

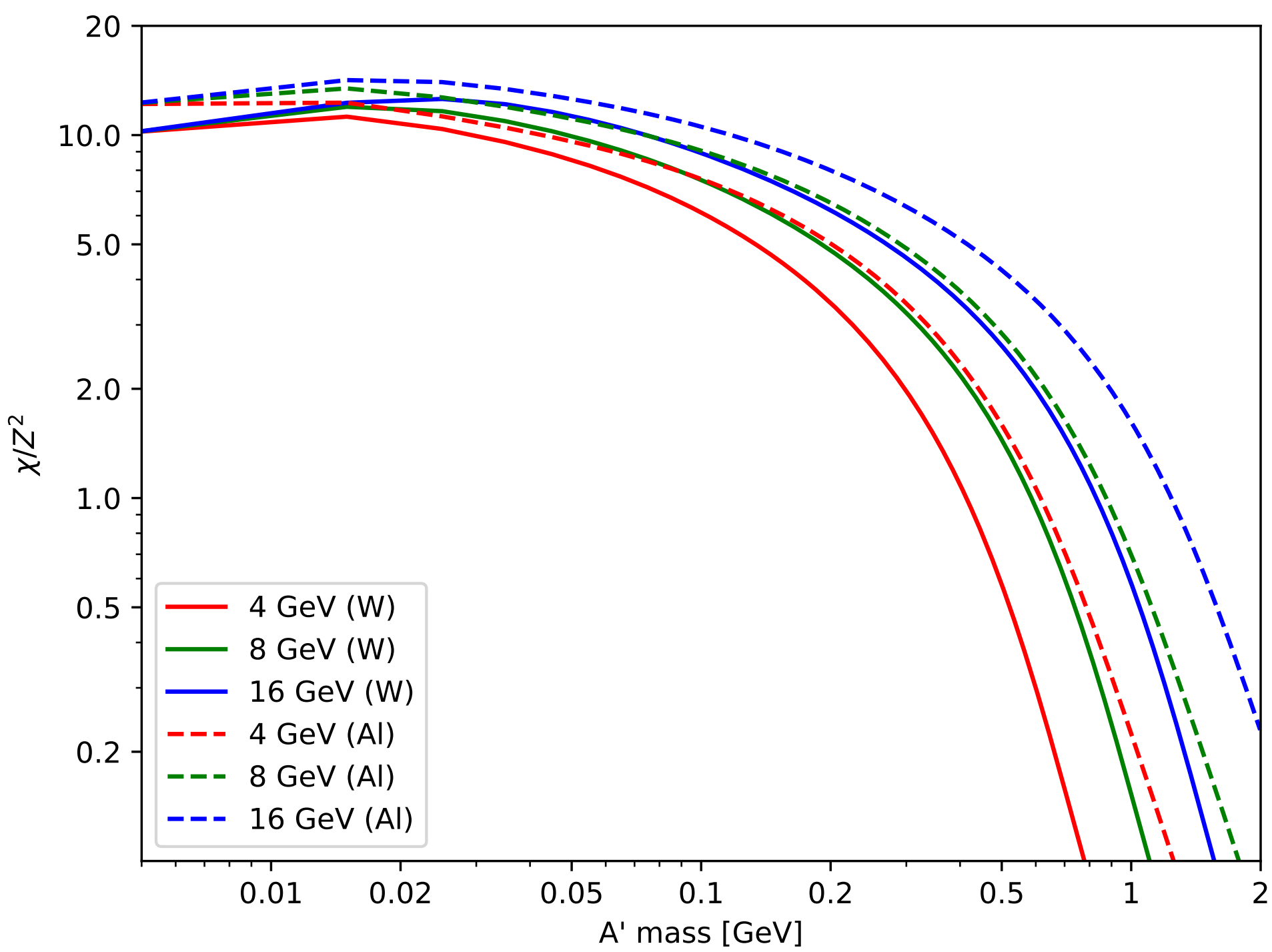

Figure 3.53: The effective flux of photons, Eq. 3.4, is divided by $Z^2$ and plotted versus $A'$ mass for aluminum and tungsten targets at an incident beam energy of 4, 8, and 16 GeV.

Some important quantities for tungsten as well as three potential alternative targets, aluminum, titanium, and LYSO ($Lu_{1.8}Y_{0.2}SiO_5$), are presented in Table 3.6. Each target is assumed to have a thickness of $T = 0.1$ radiation lengths. Using the thin target approximation, the relative yield of these targets with respect to tungsten, at a beam energy of 8 GeV is illustrated in Fig. 3.55. For thin targets, the energy loss of the electron is negligible, so that an electron traversing $T$ radiation lengths of LYSO can be approximated as traversing $T_i = T\frac{X_o \rho_i}{\rho X_{o,i}}$ radiation lengths of each component, successively, where $i = \{\mathrm{Lu}, \mathrm{Y}, \mathrm{Si}, \mathrm{O}\}$ and $\rho_i$ is the contribution of each component to the overall density of LYSO. For $T = 0.1$ in LYSO, $T_{\mathrm{Lu}} = 0.0895$, $T_{\mathrm{Y}} = 0.0034$, $T_{\mathrm{Si}} = 0.0025$, and $T_{\mathrm{O}} = 0.0046$.

The yield of low-Z targets, such as aluminum and titanium, is suppressed with respect to tungsten at low $A'$ mass. This is easily explained by the factor $\frac{X_o Z^2}{A}$ in Equation 3.7. However, the minimum momentum transfer increases with $A'$ mass. As the wavelength associated with the momentum transfer becomes comparable to the size of the nucleus, coherence in the scattering degrades and the cross section is suppressed. Coherence degrades at larger $A'$ mass for low-Z targets, giving aluminum and titanium an enhancement over tungsten when $m_{A'} > 200$ MeV. The yield of LYSO is similar to that of tungsten. This is expected because $T_{\mathrm{Lu}}$ is close to 0.1, and Lu has an atomic number close to that of W. In addition, LYSO contains low-Z components giving it a slight enhancement over tungsten at larger $A'$ mass.

| Target | Z | A [g/mol] | $X_o$ [g/cm$^2$] | $\rho$ [g/cm$^3$] | l [cm] |
|---|---|---|---|---|---|
| W | 74 | 183.84 | 6.76 | 19.30 | 0.035 |
| Al | 13 | 26.98 | 24.01 | 2.70 | 0.889 |
| Ti | 22 | 47.87 | 16.16 | 4.54 | 0.356 |
| $Lu_{1.8}Y_{0.2}SiO_5$ | 71, 39, 14, 8 | 174.97, 88.91, 28.09, 16.0 | 8.67 | 7.10 | 0.122 |

Table 3.6: Important properties of tungsten as well as aluminum, titanium, and LYSO, three potential alternative targets. A thickness of T = 0.1 radiation lengths is assumed for each target.

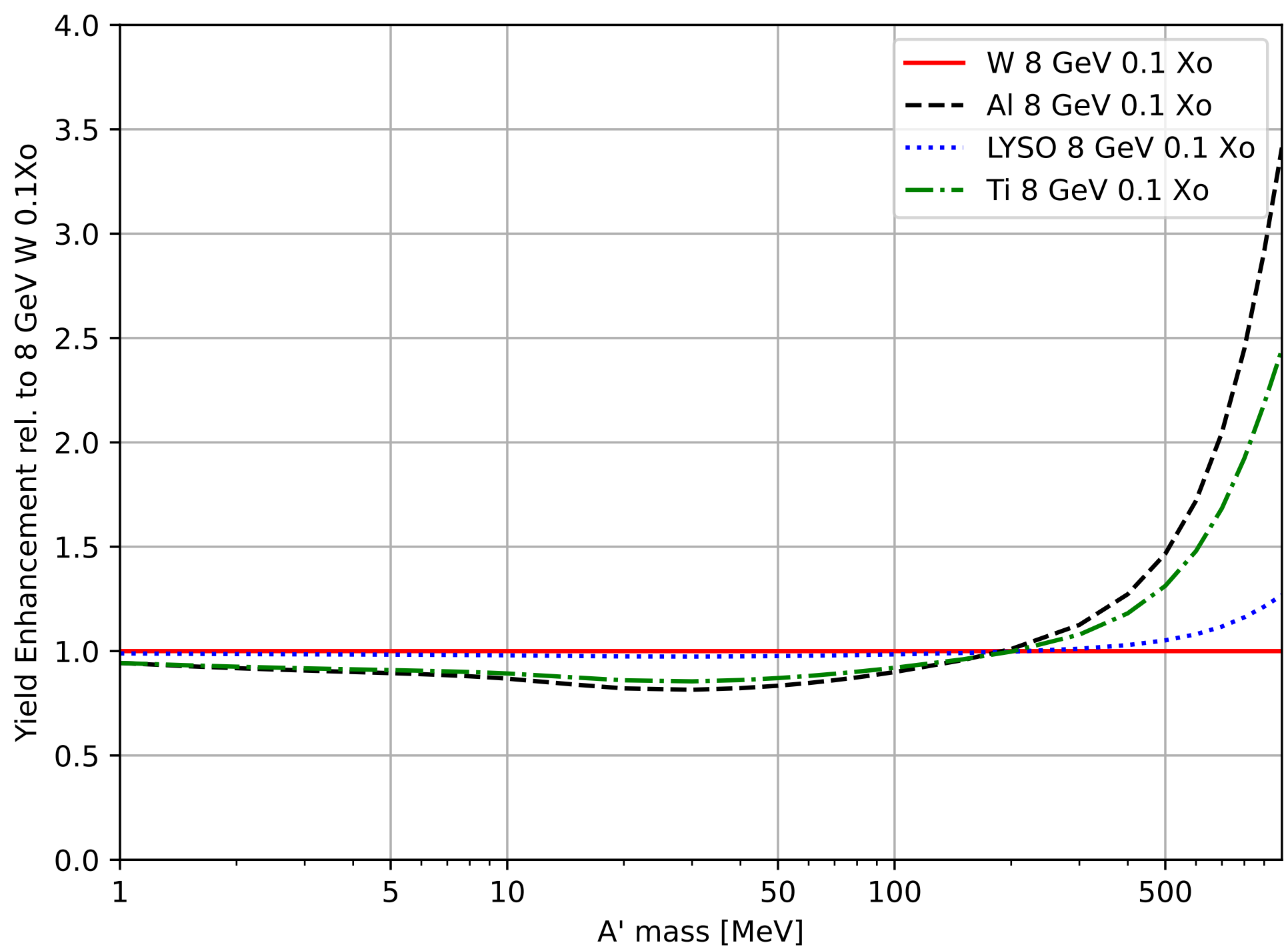

Figure 3.55: The relative yields of 8 GeV electrons incident on 0.1 radiation length of tungsten, aluminum, and LYSO, compared to tungsten, in the thin target approximation.

High-Z targets such as tungsten offer advantages such as an enhanced production cross-section at lower

$A'$ mass and reduced photonuclear and electronuclear backgrounds at a fixed target thickness, which scale roughly as $A/Z^2$ [78]. Alternative low-Z targets such as aluminum and titanium could provide an easily accessible extension to LDMX, enhancing its sensitivity in targeted regions of parameter space while diminishing sensitivity in others. For example, in Fig. 2.2, which illustrates the $\alpha_D = 0.5$ and $m_{A'} = 3m_\chi$ benchmark, switching the target from tungsten to aluminum would increase the sensitivity of the experiment in the $m_\chi > 66.7$ MeV region. In this case, the experiment would cover more of the thermal dark matter milestones illustrated in Fig. 2.2 via dark bremsstrahlung.

### 3.6.2 Active Target

#### 3.6.2.1 Introduction

In the case that photonuclear and electronuclear backgrounds prove to be more challenging to reject than our simulation studies suggest, we can pursue the usage of an active target. A high density LYSO scintillator ($Lu_{1.8}Y_{0.2}SiO_5 : Ce$) as an active target (AT) can provide an effective means of detecting photonuclear interactions in the target and thereby additional rejection of this background, especially when recoil hadrons or nuclei deposit most of their energy in the target. Active targets have been used in a variety of nuclear and particle physics experiments and have proven to be an extremely useful tool [83]. Accordingly, we have developed an active LYSO target as a potential mitigation for the case where we see events passing signal criteria which have a recoil electron momentum spectrum consistent with photo- or electronuclear production in the target. This class of backgrounds, so called *target PN* and *EN*, are discussed more fully in Chapter 4.

LYSO ($Lu_{1.8}Y_{0.2}SiO_5 : Ce$) is a cerium-doped, lutetium-based scintillation crystal that has high density (7.1 g/cm$^3$), short decay time ($\sim$36 nsec) and an exceptionally high photon-emission rate ($\sim$33 photons/keV) [90]. An active LYSO target can provide substanial rejection of photonuclear background.

The AT design is a module of the same thickness of LYSO (10% $X_0$) as the tungsten target and the same transverse dimensions. It is designed to replace the tungsten target in a "plug-and-play" fashion, utilizing the same readout electronics of the TS system. The backplane on the TS/Tracker mounting plate will accommodate the additional two additional readout module boards. The SiPMs and readout scheme used for the LYSO are identical to those used by the Trigger Scintillator, as detailed in section 3.5.3.

#### 3.6.2.2 Design

The active LYSO target consists of two layers of LYSO bars, as shown in Fig. 3.56, and provides complete geometric coverage of the effective LDMX trigger area. The first layer consists of thirty one 0.6x3.0x40 (mm) LYSO bars, each wrapped with 0.1 mm thick 3M™ Enhanced diffuse Reflector (ESR) film; the second layer has thirty two bars. Each LYSO bar unit is 0.8 mm thick, 3.2 mm wide, and 40.1 mm long. The two layers are offset by 1.6 mm so that incident electrons will pass through at least one LYSO bar. This two-layer LYSO array provides an active target area approximately the same as that of the tungsten target (100.0 x 40.0 (mm)), and is about 10.3% $X_0$ thick.

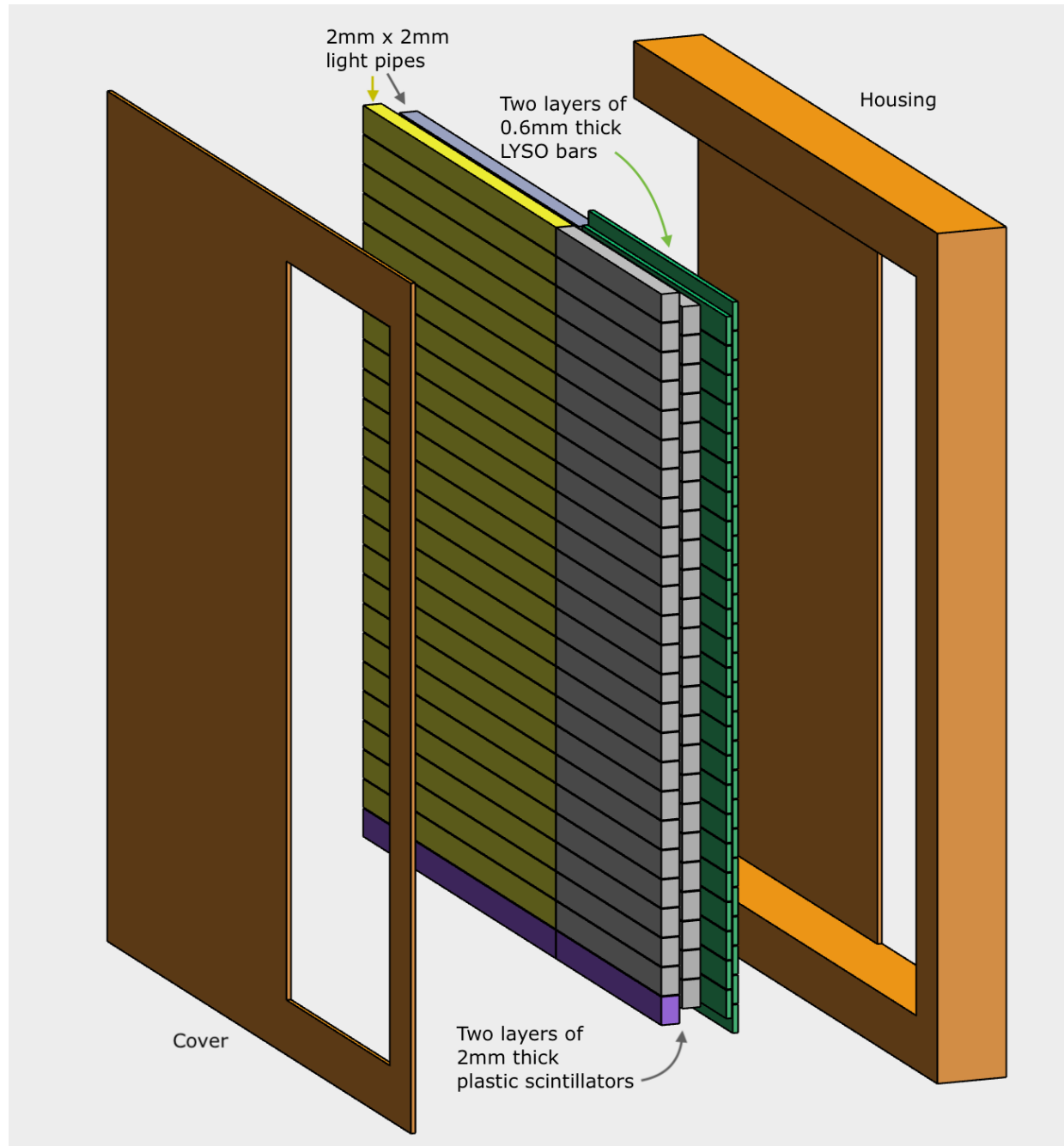

Figure 3.56: The proposed integration of the active LYSO target with the third trigger scintillator module.

A triplet readout scheme will be used to collect photons produced in the active LYSO target. Twenty-four Hamamatsu S13360-2050VE SiPMs are used to read out twenty-four central LYSO bars in the first layer and twenty-five LYSO bars in the second layer. As shown in Fig. B.1, each SiPM reads three LYSO bars. This design ensures 100% coverage and provides 1.6 mm positioning capability for incident electrons while requiring half the number of photosensors compared to the traditional one-to-one readout configuration. Details of the triplet readout scheme are in Appendix B.

.

#### 3.6.2.3 Performance in Simulation

A standalone Geant4 Monte Carlo simulation was performed to study the photon yield in the LYSO target when 8 GeV electrons pass through the active LYSO target. The modified implementation of Birks' law [91] was used in the Geant4 simulation to correct for the photon yields for heavy recoils. A total of 354,879 target photonuclear (target PN) events and 295,363 regular electromagnetic (EM) events were selected from 10 million events generated with a boost of a factor of 10,000 for the photonuclear interaction. GEANT4 predicts that this type of photonuclear event leaves, on average, 66.4 MeV in the LYSO target - about 40 times more than a regular EM event.

Fig. 3.57(a) shows the normalized photon yield spectra of the 8 GeV electrons that interact with the LYSO target. The spectra were generated with considerations of (1) the Landau distribution of energy depositions in the LYSO, (2) the efficiency of energy-to-photon conversion according to the modified Birks' law [91], (3) an overall geometric efficiency for photons reaching the SiPMs by comparing with beam test results, (4) SiPM photon detection efficiency, (5) electronic noise (0.14 PE), and (6) the Poisson distribution of detected photoelectrons. As one can see, the target PN events generate more light in the LYSO target than the EM events. The LYSO target can thus be used to reject target PN background events at the cost of losing some signal events. To determine how the LYSO target's veto affects the signal detection efficiency, the light yield spectra were converted to cumulative spectra, and the LYSO veto power was plotted as a function of detection efficiency, as shown in Fig. 3.57(b).

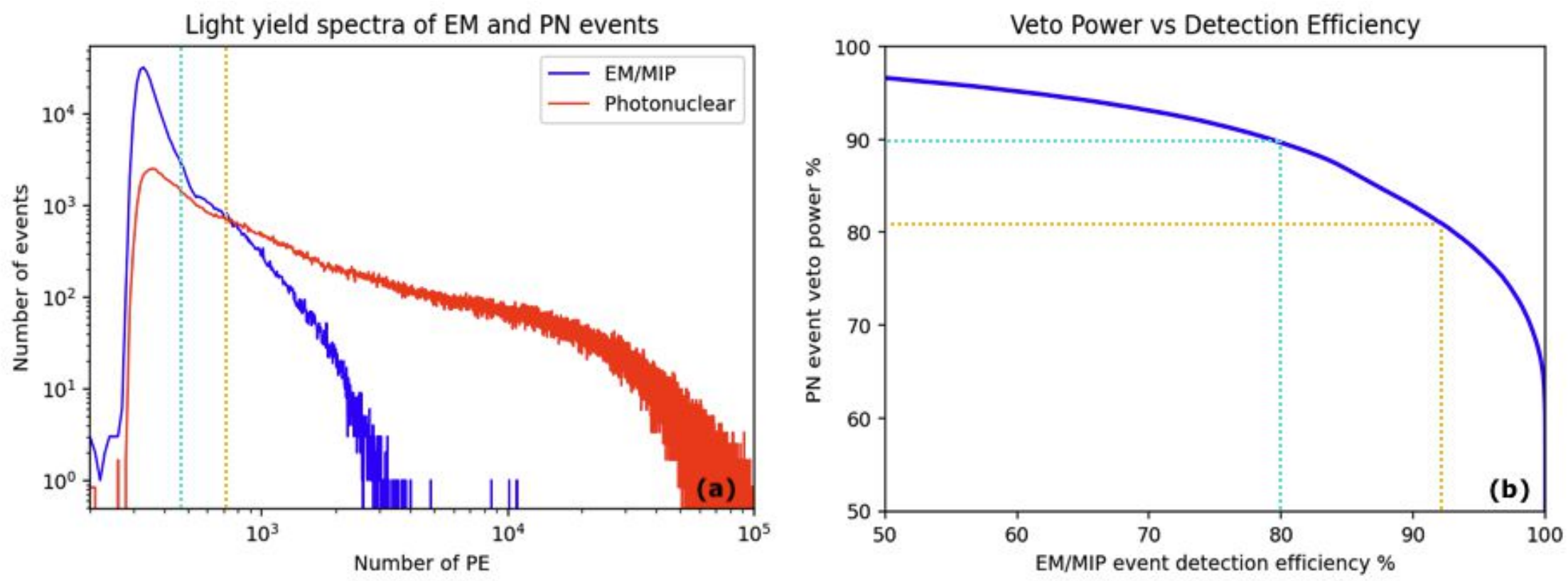

Figure 3.57: (a) Light yield spectra of EM events and target PN events. (b) Illustration of LYSO target's veto power for target PN events and corresponding detection efficiency for signal events (EM events).

This GEANT4 simulation demonstrates that the LYSO target can be used to effectively distinguish target PN events from EM events and thereby provide additional background rejection. According to the LDMX's selection criterion $E_{e,out} < 3.16\,\text{GeV}$ for an outgoing scattered electron, a cut at 450 (700) photoelectrons will reject approximately 90% (81%) of the target PN background events while retaining 80% (92%) of signal EM events, as shown in Fig. 3.57(b). As illustrated, choosing different minimum photoelectron requirements results in a varying veto power for target PN events and corresponding detection efficiencies for signal events (EM events) in the LDMX experiment.

#### 3.6.2.4 Prototype Performance

The LDMX Collaboration conducted a beam test at CERN in April 2022. The performance of LDMX's prototype trigger scintillator, active LYSO target, and Hadronic Calorimeter (HCAL) were studied using electron, muon, pion, and proton beams at various energies ranging from 100 MeV to 4 GeV. Fig. 3.42(a) shows LDMX's prototype in CERN's beam-test area T9. The trigger scintillator prototype and the active LYSO target prototype were combined to form a hybrid TS-LYSO module and positioned directly in front of the first HCAL iron plate, as shown in Fig. 3.42(b).

The hybrid TS-LYSO module consists of one layer of 2.0 x 3.0 x 30 (mm) plastic trigger scintillators followed by a double-layer LYSO array. Each array of LYSO bars consisted of six 0.6 x 3.0 x 30.0 (mm) LYSO bars, covered by 0.1 mm ESR on the top and bottom with 0.2 mm gaps between the bars. As no ESR reflector was inserted between the LYSO bars of the hybrid TS-LYSO module, crosstalk was expected due to the light leakage between adjacent LYSO bars. As illustrated in Fig. 3.58, only the top six scintillator bars of each layer were coupled to the SiPMs.

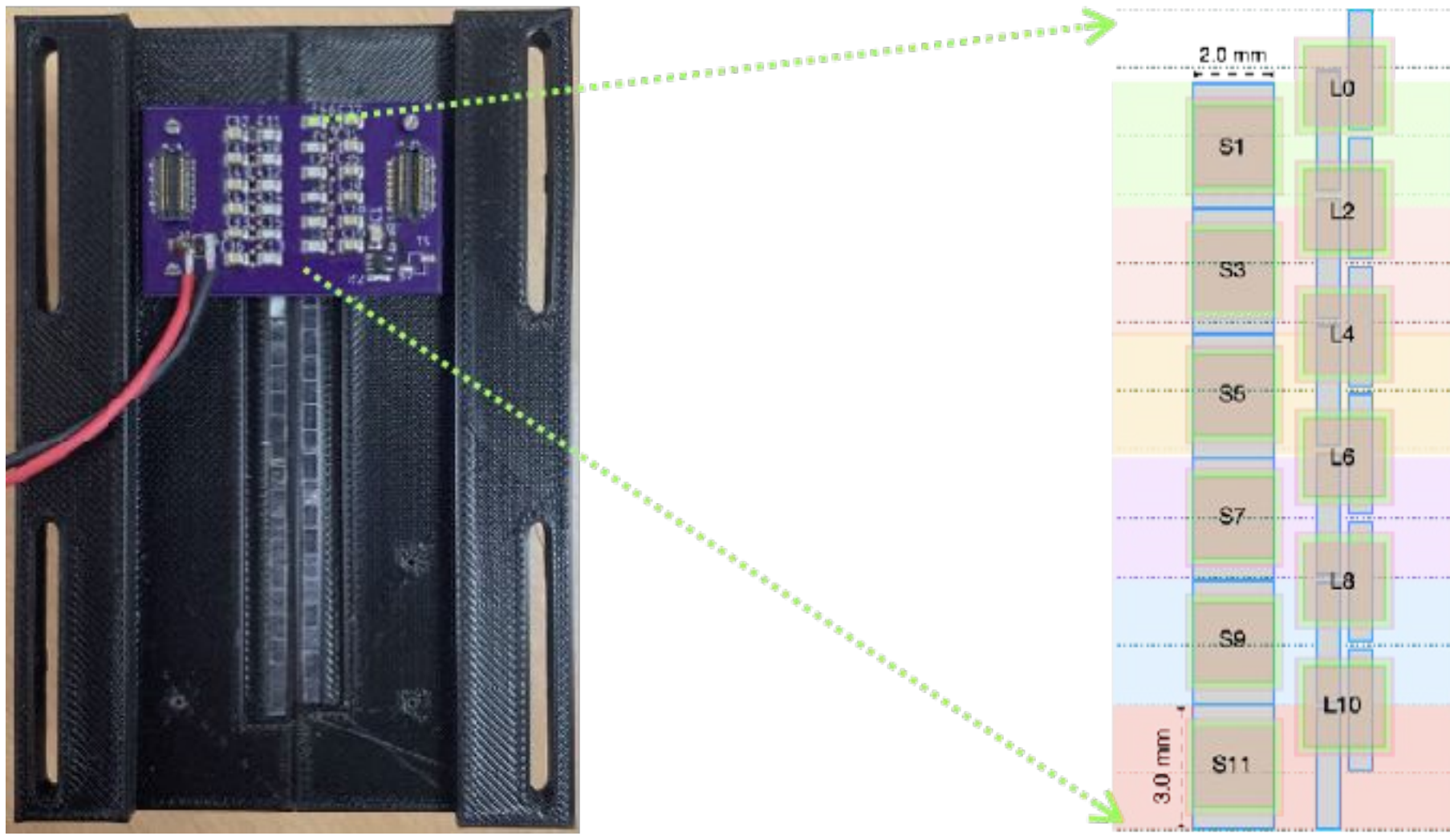

Figure 3.58: (a) The hybrid TS-LYSO module used in the Spring 2022 beam test. (b) Illustration of a two-layer LYSO array with a triplet readout design (L0, L2, ..., L10) positioned behind a trigger scintillator array (S1, S2, ..., S11).

The results of the hybrid TS-LYSO prototype obtained from the 4 GeV electron beam test are discussed below. In our study, trigger scintillator signals were used to identify the passage of the electrons and determine which LYSO bars were hit. In total, 20,000 events were collected and analyzed.

Fig. 3.59(a) shows the light-yield spectrum of one LYSO bar when 4 GeV electrons pass through after event selection. The spectrum is then fitted with a Landau distribution, and the Most Probable Value (MPV) of the fit is used to estimate the number of photoelectrons (PE). The ADC-to-PE conversion factor of $300 \pm 5$ fC per PE is calibrated using a light-yield spectrum produced by random trigger events, shown in Fig. 3.59(b).

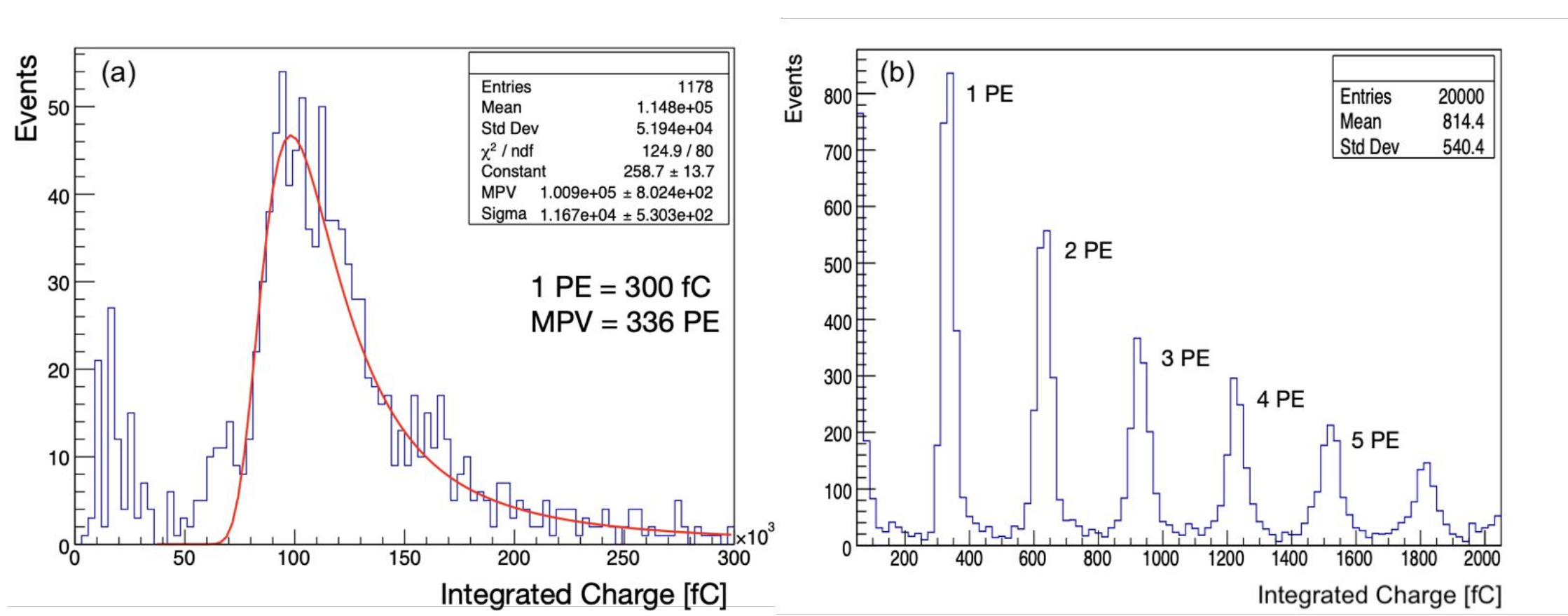

Figure 3.59: (a) Light yield spectrum generated by 4 GeV electrons in the LYSO bar L6. The trigger scintillator was used to provide triggers and identify MIP events. (b) Light yield spectrum of random trigger events, which was used for SiPM calibration.

Additional studies, including those which document the positioning ability of the system are shown in Appendix B.

## 3.7 Electromagnetic Calorimeter - ECal

### 3.7.1 Overview

The need for granularity, very high efficiency, radiation hardness, compactness, and speed led to the selection of high granularity silicon-tungsten technology for the LDMX ECal. To that end, we adopt designs from the CMS High Granularity Calorimeter (HGC) for the CMS phase 2 upgrade [92]. The ECal, shown in figures 3.1 and 3.2, is a sampling calorimeter comprised of tungsten absorber planes having a total depth of 40 radiation lengths interleaved with silicon planes that are paired into bilayers. Note that the final design will have 32 silicon layers paired in 16 bilayers while for the studies presented in this report, 34 layers paired in 17 bilayers were used in the simulation. The change will be achieved by increasing the absorber in the deepest layers of the ECal with no anticipated change in performance. Fig. 3.60 provides a schematic view of the detailed structure of the ECal, as described in more detail below.

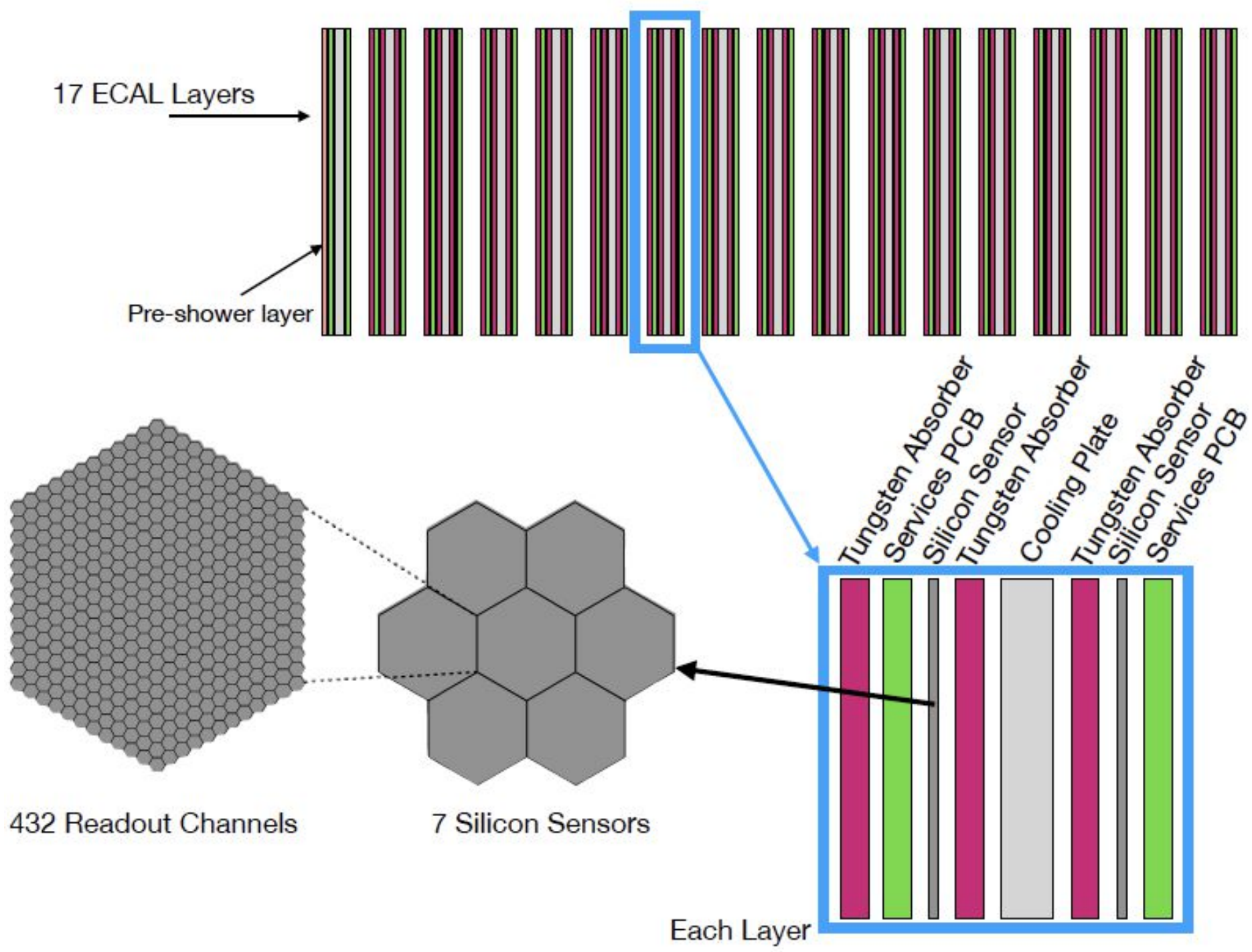

Figure 3.60: Schematic diagram of the LDMX ECal system with the components of a bilayer highlighted.

Each ECal plane contains seven hexagonal modules as shown in Fig. 3.60. The modules are constructed with a carbon fiber baseplate covered by a Kapton-insulated copper shield layer to which the bottom of the silicon sensor is glued. On top, the sensor is bonded to a hexagonal readout PCB ('Hexaboard') that carries a variety of components - most notably the CMS HGCROC front-end ASIC developed by the Laboratoire Leprince-Ringuet (LLR) Omega group. The LDMX ECal trigger and DAQ rely on the CMS HGCROC. We use the CMS high-density (HD) module design. The total number of modules in the LDMX ECal amounts to less than 1% of the total number to be built for CMS and less than 5% of the total number of CMS HD modules. In CMS the HD modules use sensors with active depths of 200 $\mu$m and 120 $\mu$m. Given the lower radiation environment of LDMX as discussed below in Appendix A, we will use 400$\mu$m silicon to increase the signal-to-noise ratio. For LDMX, we plan to order sensors from Hamamatsu who is currently producing the HD sensors for CMS. This will allow the use of existing HD masks on standard 800 $\mu$m wafers that are then ground to 400 $\mu$m. CMS chose the hexagonal sensor geometry to enable uninterrupted tiling while maximizing the use of an 8" diameter silicon wafer. The sensors have a span of $\sim$17 cm between any pair of parallel edges. The HD sensor is divided into 432 individual hexagonal readout cells, 420 of which have an area of 0.52 cm$^2$. The remaining 12 readout cells are much smaller and are used to calibrate the response to a minimum ionizing particle (MIP).

There are planes of seven silicon modules with readout and a tungsten absorber sheet on either side of a cooling plane, thus defining a bilayer. A separately-supported tungsten absorber plane is located between each pair of consecutive bilayers. The seven modules in each silicon plane are arrayed in a "flower configuration" with a central *core* module surrounded by a ring of six others as seen in Fig. 3.60. Motherboards carrying trigger and data signals overlay and connect the 7 modules in each silicon plane. The motherboards host radiation-tolerant data-processing concentrator ASICs and control mezzanines based on the CERN lpGBT ASIC. Control and readout data are carried on optical fibers to electronics hosted in the experiment's common ATCA trigger crate and Bittware-based DAQ processors.

| **ECal Tungsten Absorber Layers by bilayer Type** | | | | | | |
|---|---|---|---|---|---|---|
| Type | Single or Pair | W $\Delta z$ | W area | W mass | Total for FEA | Count |
| | | [mm] | [cm$^2$] | [kg] | [kg] | |
| PS | Pair | - | - | - | 20 | 1 |
| | Single | 1.0 | 1855 | 3.6 | | |
| A | Pair | 1.0 + 1.0 | 1850 | 7.2 | 30.8 | 1 |
| | Single | 2.0 | 1855 | 7.2 | | |
| B | Pair | 1.5 + 1.5 | 1850 | 10.8 | 38.0 | 1 |
| | Single | 3.5 | 1855 | 12.5 | | |
| C | Pair | 1.75 + 1.75 | 1850 | 12.5 | 45.0 | 9 |
| | Single | 7.0 | 1855 | 25.0 | | |
| D | Pair | 3.5 + 3.5 | 1850 | 25.0 | 70.0 | 5 |

Table 3.7: Baseline ECal specifications for W-absorber layers used for Monte Carlo simulations and FEA calculations. For each bilayer type, the thickness, area, mass of W and total weight used for FEA calculations are specified. There are one each of the pre-shower (PS), A, and B bilayers; 9 C bilayers and 5 D bilayers. The PS layer has no W absorber while others have three planes of W absorber; a single layer that stands between silicon layers on neighboring bilayers and a pair that bracket the cooling plane at the core of the bilayer and thus make up the total W-absorber between the two silicon layers of the bilayer itself. As noted in the text, the plan is to build the ECal with 16 rather than 17 bilayers. All other items, such as the silicon, carbon fiber and cooling planes, motherboards and PCB layers are included in the simulation but not specified in the table. For FEA studies, these additional items were very conservatively represented by an added 20kg per layer as reflected in the "Total for FEA" column.

The sampling plane, (a flower with 2 motherboards) is ∼6 mm thick. The cooling layer has ∼4 mm outer diameter thin-wall stainless steel cooling tubes embedded in, and covered by, thick and thin C-Fiber sheets, respectively. The bilayer thickness, excluding the $W$ absorber thickness, which varies with depth in the ECal, is ∼2.3 cm. The various layers and their specific details are shown in Table 3.7. The depth of the entire device, including $W$-absorbers and gaps between bilayers is ∼55 cm. The sensitive detection volume is thus contained in a region of roughly $55 \times 55 \times 55$ cm$^3$, while the full system occupies a volume with a transverse cross-section of $95{\times}65$cm$^2$ and depth of 65 cm. It is small and compact, but dense, with a mass of ∼825 kg. The support structure retains this large mass while precisely positioning bilayers and absorber planes. The ECal is designed to facilitate disassembly and replacement of damaged components, or to reposition absorber planes, if necessary. The latter could be deemed advantageous to investigate rare backgrounds, systematic uncertainties or signal sensitivity. Small transverse offsets between flowers on either side of the bilayer improves resolution of charged particle tracks and avoids the alignment of small dead regions between modules. The ECal structure includes manifolds for the distribution of coolant and cooled dry air. Solutions for low-voltage power using radiation-tolerant DC/DC converters and for the necessary feed-throughs and cabling harnesses have been identified.

An ECal based on multi-cell silicon sensors is well-suited to the identification of photons and electrons with high efficiency and good energy resolution. It is also extremely powerful for the rejection of photonuclear backgrounds with very small energy deposits by means of shower shape variables and particle tracking. The ECal energy resolution has a constant term of $\sim \pm4\%$ and stochastic term of $\sim 21\%/\sqrt{E}$. The Moliére radius is ~2.5 cm but the radius of containment of 68% of the energy in EM showers is less than 1 cm in the first ~15 layers of the ECal, enabling discrimination of individual electrons and photons with small angular separation, ultimately limited by cell size. It also provides efficient detection of charged hadrons that range out in a single silicon layer and it can be used to track those that traverse multiple planes with excellent per-cell efficiency for MIPs. The large ECal depth — $40X_0$ of $W$ absorber — is driven by the need to ensure detection of EM showers whose energy deposition lies in the very low tail of the distribution, with probabilities below $10^{-15}$. The large depth also improves the ability to catch late-developing showers and photonuclear interactions that produce muons and charged hadrons.

The ECal provides important input to the LDMX trigger. For each module, the HGCROCs produce a per-bunch trigger sum over fixed groups of nine cells, which are transmitted to an associated CMS ECON-T ASIC. A full-module energy sum is created in the ECON-T and transmitted off-detector along with the highest-energy trigger cells in the module. The motherboard is designed to transmit four trigger cells from each of the outer six modules of each layer and eighteen trigger cells from the inner module. More details on the trigger and DAQ are presented below.

### 3.7.2 Requirements

Ideally the ECal should detect and enable a veto for every case in which the incident electron produces a hard bremstrahlung photon. In reality, the ways in which bremsstrahlung photons are manifested lead to a variety of challenges but also some opportunities. For instance, the vast majority of these photons produce easily identified electromagnetic showers. However, in the case of $10^{14}$–$10^{16}$ EoT, even the easy rejection of 99.999% of these cases would still leave $10^{8}$–$10^{10}$ events, exceeding signal expectations by many orders of magnitude. As a result, the choice of technology and design of the ECal was driven to a large extent by rare events, which are discussed in more detail in Sec. 3.2 above.

Challenging background events involve showers in which significantly less than average beam energy is produced in the ECal as a result of a rare interaction. For example, the photon could undergo a photonuclear (PN) interaction that results in a handful of energetic charged hadrons or a pair of muons. These would deposit little energy but could easily be detected if they could be isolated from the recoil electron shower. In the case of hadron production, the problem gets more difficult as few, or even just one charged hadron is produced, and even more difficult if that hadron is very soft and traverses very little of the sensing region of the detector. These cases require high granularity and extremely high efficiency to detect and precisely locate small energy depositions. Of course, it is also possible that only neutral hadrons are produced, in which case the ECal has little it can do and the full responsibility for rejecting these events is passed to the HCal, which is described in a later section of this report. Moreover, beam electrons that undergo little to no interactions in the target, represent the majority of events, and ultimately lead to a fairly high radiation dose for parts of the central core of the detector. Finally, while there will be of order one or two electrons per bunch, the rate of arrival of these bunches will be comparable to the bunch crossing rate at the LHC.

Taking all of the above into account represents a very formidable challenge. It was not *a priori* certain that any calorimeter could meet the challenge, but the technology represented by the CMS HGC appeared to give the greatest chance of success. It comes with the high rate operation we need to manage $\sim 10^{14}$–$10^{15}$ EoT/year, and is radiation tolerant. The high granularity has the potential to individuate the electron and photon showers down to small separation distances. Moreover, the silicon has 100% efficiency for minimum ionizing particles (MIP) traversing any live region of a sensor and the granularity can be used to reconstruct MIP tracks. The technology thus has the potential to be ideal for the task of managing most of the backgrounds of concern. Of course, it must do all of this while maintaining high signal efficiency. The requirements on the ECal design are driven by 5 general physics requirements summarized in Sec. 3.2 above, which can be translated into detector requirements that in turn dictate a set of technical requirements, both of which are summarized below.

- **ECal Detector-Physics Requirements**

ECAL1 Energy resolution and acceptance adequate to efficiently trigger on missing momentum and detect particles coming from the target out to ~30° from the beam axis.

ECAL2 Granularity adequate to detect photons near the trajectory of recoiling electrons and to produce trigger primitives for everything from MIP tracks to multi-GeV showers.

ECAL3 Reconstruction of MIP tracks with high efficiency even when crossing very few silicon layers.

ECAL4 Negligible false veto rate for noise and cosmic rays.

- **ECal Technical Requirements**

ECAL5 40 radiation lengths of absorber along any trajectory from the target center to an angle of ∼20°.

ECAL6 Signal-to-Noise Ratio (SNR) $\geq 5$ (10) for MIPs within (outside) central core modules of layers 5-10 where the shower energy deposition is at a maximum (a.k.a. shower-max).

ECAL7 Pad thresholds at $\geq 3$ (5) $\sigma$ of noise inside (outside) the shower-max region of the central core.

ECAL8 Maintain cell-to-cell calibration at the 5% level, including both sensor and electronics effects.

ECAL9 Maintain $\leq 20\%$ Relative Humidity (RH) and keep sensors (electronics) at temperatures as low as -20°C without exceeding 0°C for >3 weeks/year (below ∼30°C).

ECAL10 Time resolution for hits better than ∼1 ns and produce trigger primitives with a latency $< 1\ \mu s$.

ECAL11 6 years operation for a total radiation exposure and fluence of up to 1 MRad and $10^{14}$ 1 MeV neutron equivalent/cm$^2$. Be able to replace a complete bilayer in a short shutdown.

### 3.7.3 Components developed for CMS HGC

The ECal incorporates a number of components that were developed in the context of the CMS HGC upgrade. The CMS HGC makes use of silicon for radiation tolerance and for high granularity to cope with much higher rates of simultaneous proton-proton collisions in the high luminosity era of the LHC (HL-LHC). While radiation will not be as extreme in LDMX, it will be high enough to merit a radiation-tolerant system as discussed in Appendix A. Additionally, granularity is extremely important for isolating and detecting bremsstrahlung photons in the most difficult extremes, such as rare photonuclear interactions that leave a minimal trace of an interaction. Indeed, the primary role of the ECal is that of a photon veto, with energy resolution being a close second. The main CMS components that are relevant to LDMX are the silicon modules, which are the basic units from which larger sensor planes are tiled. The modules are multilayered devices with every layer being a hexagon of the same size. Starting from the bottom there is a baseplate followed by a thin Cu shield layer insulated in Kapton, a sensor, and finally a PCB referred to as a hexaboard that houses the front-end HGCROC ASIC. The LDMX module will differ only slightly from that of CMS in that the sensors will have deeper active regions and the baseplate will be carbon fiber rather than copper-tungsten or titanium. Other important CMS components that will be needed in LDMX are the concentrator ASICs (ECON-D and ECON-T).

#### 3.7.3.1 Silicon Sensors

The sensors are planar DC-coupled hexagonal silicon fabricated on 8” wafers. The hexagonal shape of the sensors makes more efficient use of the available area of the circular wafers, as compared to other tileable polygonal shapes, while minimizing the ratio of periphery to active surface. The vertices of the hexagonal sensors are truncated to provide clearance for the mounting hardware. The truncations, called mousebites, make possible a slight further increase in the use of the 8" wafer. The design of the sensors has been developed and validated through careful study by the CMS collaboration[93].

It has been determined that p-type sensors are preferred, as they produce less non-Gaussian noise due to radiation-induced surface-charge effects. Each LDMX sensor has 432 individual diodes, called *cells*. This is the High Density (HD) sensor layout of CMS, which uses p-type epitaxial wafers of 300 $\mu$m thickness that have an active depth of 120 $\mu$m achieved via back-side deep diffusion. The CMS HGC also has a Low Density (LD) sensor version of the same dimensions with 192 cells that is used for the vast majority of the system. For LDMX, the higher granularity CMS HD mask-set will be used with p-type float zone sensors that have both active and physical thickness 400 $\mu$m. Thicker sensors would not increase cost but do significantly increase the signal-to-noise ratio. Standard cells in HD sensors have a nominal area of 0.56 cm$^2$. Cells are arranged to produce a “threefold diamond” layout, shown schematically in Fig. 3.61. This facilitates a convenient definition of 3 symmetric sets of 16 groups of 9 neighboring cells that form coarse trigger cells, shown as different color groupings in the figure. The subdivision of the module into 3 symmetric domains simplifies the layout of the Hexaboard, with each domain having two dedicated HGCROCs. A photograph of a CMS HD sensor is shown at the right in Fig. 3.61.

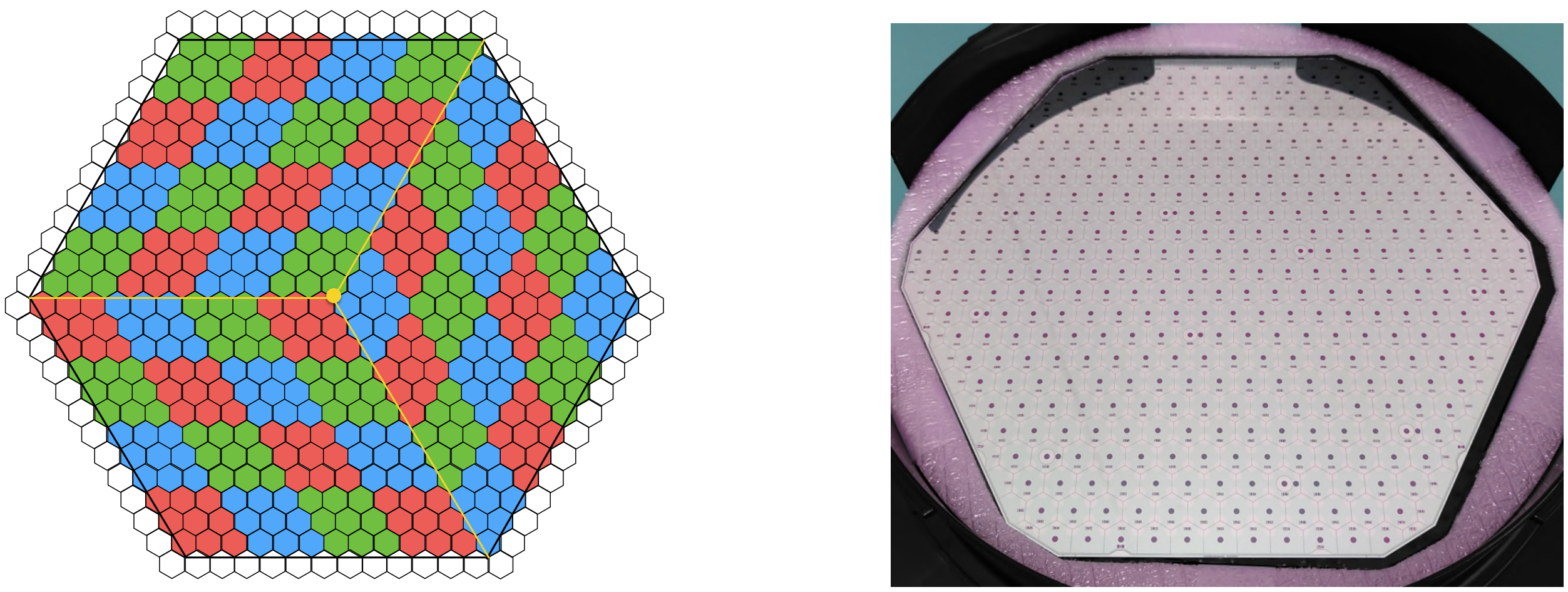

Figure 3.61: Design of the 8" HD silicon sensors: schematic illustration of the threefold diamond configuration with colored groups of 9 cells that get summed to form coarse trigger cells (left), and a photograph of a high density sensor (right). (*Images courtesy of the CMS Collaboration*)

The choice of the HD layout with smaller cell size is driven by physics performance considerations, such as separation of the showers from the recoil electron and the bremsstrahlung photon, better localization of the lateral spread of showers, and electrical stability considerations in the periphery of the mask design. For 400 $\mu$m HD sensors being considered for use in LDMX, the channel capacitance is $\approx 20$ pF. Twelve of the cells in each sensor are further segmented to include a much smaller calibration pad with correspondingly lower capacitance and noise. These cells provide a more precise calibration of the charge produced by a MIP.

CMS has irradiated 300 $\mu$m p-type sensors to 3 fluence levels. For the lowest level of $\sim 6 \times 10^{14}$ 1-MeV-neq/cm$^2$ they find that charge collection efficiency at 600 V bias is $\sim 80\%$ of the unirradiated value [92]. This remains true for annealing periods of more than 100 days at a sensor temperature of 0°C. As seen below in Appendix A, for $10^{15}$ EoT, the peak fluences in LDMX are in the region of shower-max (silicon layers 5-10) at $\sim 8 \times 10^{13}$ 1 MeV-neq/cm$^2$. On either side of shower-max, peak fluences drop roughly an order of magnitude every 4-5 layers, starting from $\sim 10^{13}$ 1 MeV-neq/cm$^2$ at layers 4 and 11. Moreover, the peak fluences are entirely concentrated in the core silicon modules centered on the beam axis, with 2 to 4 orders of magnitude lower values in the 6 surrounding modules. LDMX silicon layers should therefore remain efficient for operational scenarios up to $\sim 10^{16}$ EoT.

Should this turn out to be not the case because of any number of reasons, including the fact that LDMX will not operate nearly as cold as CMS, then one could reach $10^{16}$ EoT in a second phase of operation by periodically replacing the 6 core modules that receive the highest radiation at shower-max. For this reason, the ECal is designed to facilitate swapping in new bilayers in short shutdown periods. The alternative of cooling ECal to an operating temperature of $-35$°C as planned for CMS would require an enhancement of the cooling plant and significant modifications of the cooling system, thermal insulation and external heating layer.

#### 3.7.3.2 Baseplate and Shield

The design of the CMS HGC ECal HD module uses a 1.4 mm thick 25%-75% Cu-W baseplate as seen in Fig. 3.62. Two non-through holes along the central vertical axis of the bottom of the baseplate position the completed module via pins on the cooling plate. A shield made up of a thin Cu layer sandwiched between two 100 $\mu$m thick layers of Kapton is laminated to the top of the baseplate. The sensor is glued onto the Kapton of the shield layer which insulates it while at bias voltage from the baseplate at ground. The hexaboard is then glued onto the sensor and its ground is tied to the shield layer by wirebonds to gold pads on the hexaboard and shield. The bonds are placed in notches in the Kapton found near all 6 corners as seen in Fig. 3.63.

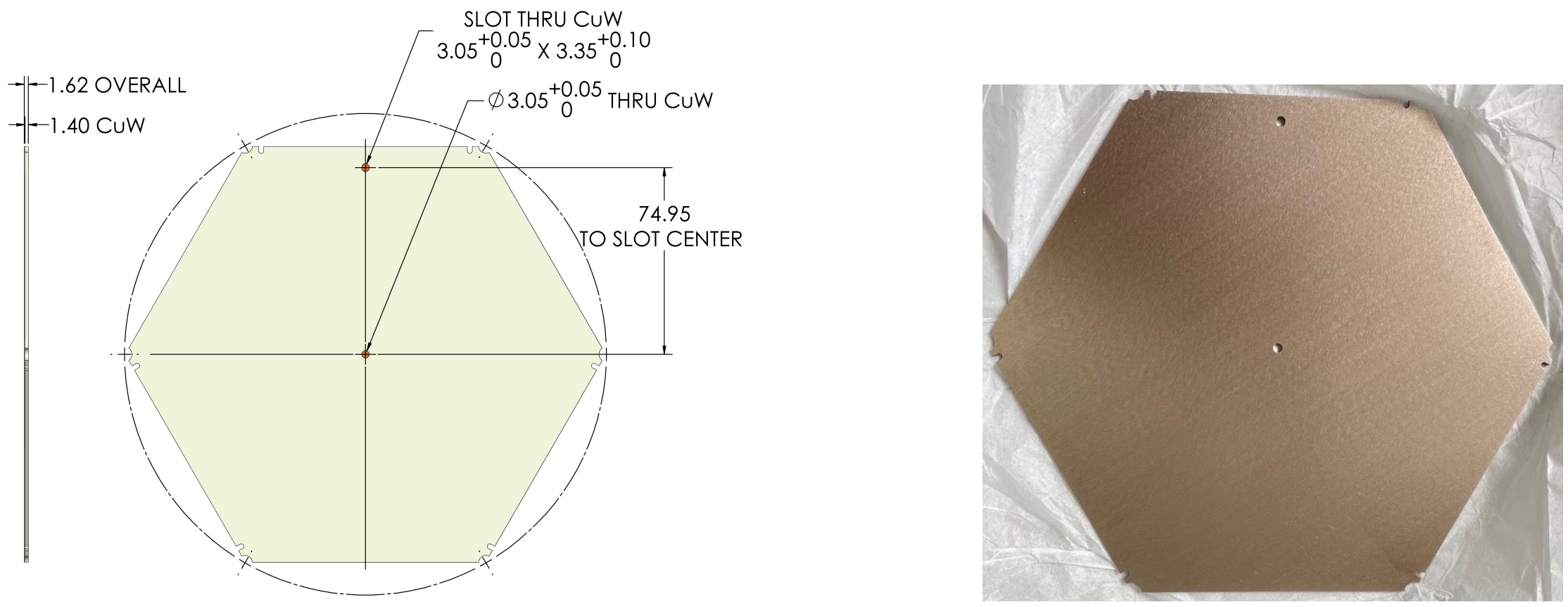

Figure 3.62: Design and an actual CMS copper-tungsten baseplate.

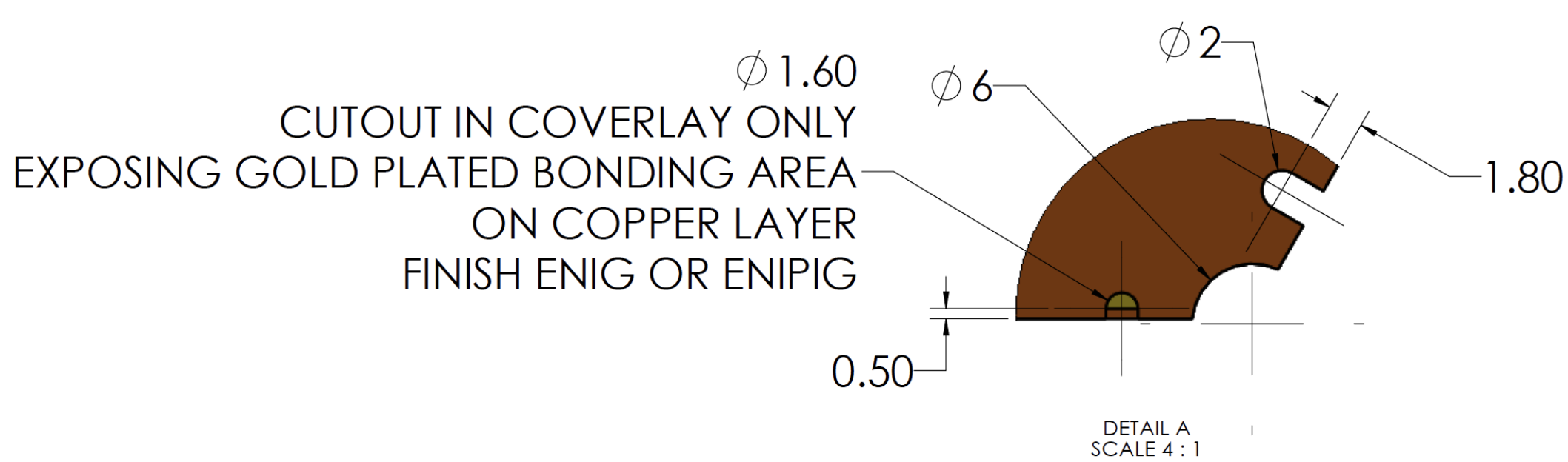

Figure 3.63: Detail of Au bond area, mousebite, and notch in shield where wirebonds connect the bias voltage lines on the hexaboard to the back of the sensor.

The biasing of the sensor is also achieved via wirebonds from the base of the hexaboard to the back of the sensor at 6 places via notches near the corners of the Cu-W and shield as seen in Figs. 3.62 and 3.63.

For LDMX, we will replace the Cu-W baseplate by a C-fiber baseplate with the same geometry and shield. The modules will then be attached to W absorber planes on either side of the cooling plane. The W absorber is a continuous sheet without gaps. The C-fiber used for the cooling planes and the module baseplate is chosen to have very high in-plane thermal conductivity to help minimize thermal gradients in the sensors that could result from the highly concentrated heat sources, such as the HGCROCs on the hexaboard.

#### 3.7.3.3 HGCROC

The HGCROC ASIC was designed by the Omega group for the HL-LHC upgrade of the CMS detector[94]. The ASIC was designed to operate at up to 40.08 MHz to match LHC conditions, providing a substantial operational margin at the 37.5 MHz operation required for LDMX. The PLL of the HGCROC has a wide enough bandwidth to easily accommodate the operational frequency of LDMX. Each HGCROC serves 72 standard cells as well as two additional calibration cells and four channels for common-mode noise subtraction. The latter are not connected to sensor pads. The HGCROC performs the following functions:

- Digitization of the signal amplitude and time of arrival
- Buffering of hits for up to 20 $\mu$s while waiting for a Level 1 Accept (L1A) signal
- Threshold comparison for time of arrival

- Summation of amplitudes of groups of 9 cells as inputs to the trigger primitive generator (TPG)
- Transmission of TPG inputs to the trigger electronics at the bunch rate
- Transmission of full granularity data upon receipt of an L1A at rates up to 750 kHz

For the sensors, after pre-amplification, the measurement for charges up to 100 MIP-equivalent is performed with a low-power 10-bit Successive Approximation Register (SAR) Analog-to-Digital Converter (ADC) in 130 nm technology. For charges above $\sim$ 80 MIP-equivalent (overlapping the SAR range) the HGCROC uses the time-over-threshold (ToT) technique with a 12-bit Time-to-Digital Converter (TDC) (50 ps bin size, 200 ns full range). The ToT dynamic range extends the measurement up to 10 pC and its bin size of 2.5 fC corresponds to less than 4% of the charge at the switch between ADC and ToT. The pre-amplified pulse also goes to a discriminator and the time of arrival (ToA) is measured with a 10-bit TDC with $<$ 26.7 ps step Least Significant Bit (LSB) and 26.7 ns full range. After alignment, the ADC, ToT and ToA data are stored at 37.5 MHz in a 512-column circular buffer to wait for a potential L1A signal. If an L1A signal arrives, the data are formatted and transmitted to the ECON-D ASIC via 1.2 Gb/s electrical links.

For trigger purposes, the HGCROC converts any TOT hits onto a common scale with the ADC using a programmable multiplier. Each channel has independent ADC and TOT pedestal subtraction constants. The HGCROC can be programmed to sum groups of either four channels or nine channels. For groups of four channels, 64 channels of the HGCROC can be included in the trigger sums, for a total of sixteen 19-bit trigger sums. For groups of nine channels, all 72 primary channels of the HGCROC are included in the trigger sums, for a total of eight 21-bit trigger sums. After summing, the trigger results are compressed into a seven-bit floating point scale with four exponent bits and three mantissa bits. Trigger sums are packed in groups of four into 1.2 Gb/s electrical links. Therefore, when the HGCROC runs in four-channel-summing mode as for the LDMX HCal, the HGCROC produces four 1.2 Gb/s electrical links and when the HGCROC runs in nine-channel-summing mode as for the HD sensors of LDMX, the HGCROC produces two 1.2 Gb/s electrical links.

The HGCROC is configured using an I2C bus which can operate at up to 1 MHz. The chip contains 1178 programmable eight-bit registers which allow substantial tuning of the behavior of the amplifiers, uniformization of pedestals, adjustment of trigger calculations, and other key parameters. The HGCROC receives a clock at eight times the bucket clock (therefore 300 MHz for LDMX) and a fast-control stream with the same rate (single-data-rate capture). The HGCROC extracts the bucket clock by aligning to the idle frame of the fast-control stream. The fast-control format is specified by CMS and is different from the standard LDMX fast-control. The adaptation between the protocols is handled by the Bittware off-detector cards for transmission over the lpGBT optical links.

The production class of the HGCROC is version 3, heretofore referred to as "HGCROCV3" for which there have been CMS-specific refinements labeled HGCROCV3a to HGCROCV3e with everything from HGCROCV3c onward being adequate for LDMX. HGCROCV3 has been tested for radiation tolerance, showing no latchup behavior and only moderate performance degradation up to CMS HL-LHC peak radiation levels of 150 MRad and $10^{16}$ 1-MeV-neq/cm$^2$. The rate of single-event effects (e.g. bit flips in data) is limited and will be much lower in LDMX than in CMS given the use of an electron beam rather than a hadron beam.

#### 3.7.3.4 Hexaboards

The HGCROCV3 ASICs are mounted on the PCB hexaboard, which is glued to the silicon sensor as discussed earlier. As the sensor has 432 primary cells, including 12 calibration cells, six 72-channel HGCROCs are required for the readout of the module. The hexaboard also contains three low-voltage regulators which provide 1.2V power for the HGCROC from the 1.5V input power provided to screw points on the module. The hexaboard has been developed as part of the CMS HGC project. A photograph of a CMS HD hexaboard is shown in Fig. 3.64.

The hexaboard contains a large number of holes where the sensor surface is exposed. These are the locations where the hexaboard is wirebonded to the sensor. To reduce noise pickup, the traces carrying the analog signal are shielded by ground planes in the hexaboard and have vias to the surface only at the HGCROCs. The wirebond pads are recessed in “stepped holes” as seen in Fig. 3.68, which helps to protect the bonds from damage. The hexaboard also contains bond pads for the sensor guard ring and for the bias voltage. The bias voltage is carried through the PCB on vias to pads on the bottom side of the PCB. These pads are wirebonded to the back side of the sensor to provide biasing.

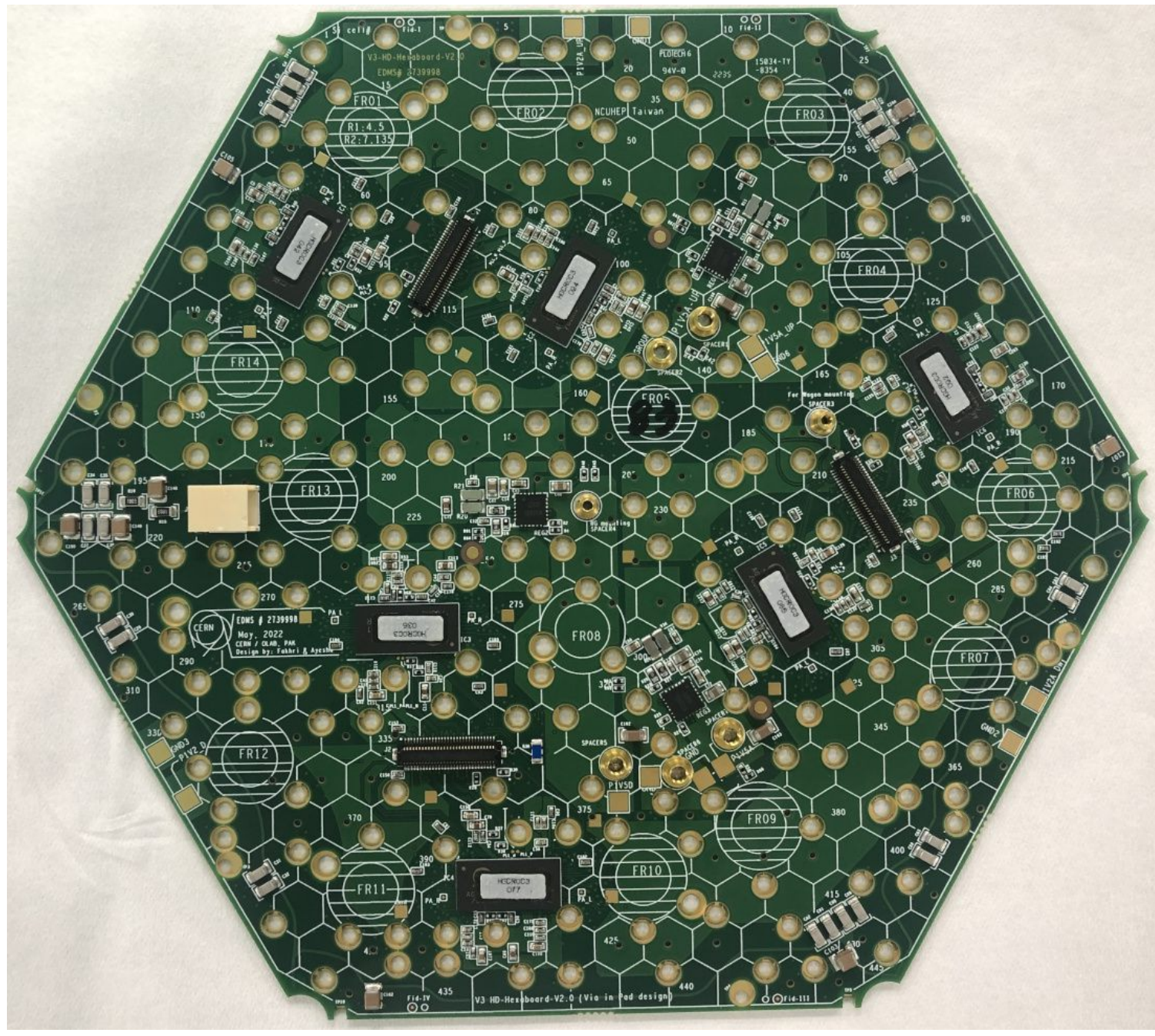

Figure 3.64: Photograph of CMS HD hexaboard, showing the six HGCROC ASICS and three data connectors.

The electrical controls and readout are carried through three Hirose DF12 connectors. Each connector carries signals for two HGCROCs, which are located on either side of the connector. The signals include the gigabit CLPS trigger and DAQ electrical links discussed above, I2C communications for HGCROC configuration, resets and error signals, and power-enable/power-good signals for the regulators. There is an RTD temperature sensor on each hexaboard as well.

#### 3.7.3.5 Module Design & Construction

The LDMX module design is based on that of the HD module for the CMS HGC upgrade. The modules are made by laminating four hexagonal layers with a hybrid combination of 3M transfer tape to quickly set the connection and a 24-hour-cure Araldite epoxy to ensure adhesion after significant irradiation. Modules built in this way at UCSB have been irradiated to 50 MRad and thermal-cycled at least 50 times without issue. The module layers are shown in Fig. 3.65. Note that Cu-W and Ti baseplates are being used in the Electromagnetic and Hadronic sections of the CMS HGC, respectively, while LDMX will use thin, high thermal conductivity C-fiber that will be mounted to the $W$-absorbers on the cooling planes of the bilayers. The high TC C-fiber helps to distribute heat from the electronics layer to avoid localized hot spots.

UCSB has substantial experience with automated module assembly. For the existing CMS tracker, UCSB built $\sim$ 4300 modules, with a surface are of roughly 75m$^2$ in 13 months. For the CMS HGC, UCSB has built over 100 modules as of March 2025, and is planning to build 4500 modules in a period of roughly 14-18 months starting in late 2025. As was the case for the earlier tracker modules, automated assembly is carried out with an Aerotech pick-and-place gantry while wirebonding is performed with an automated wedge-bond machine. The latter is a Hesse BondJet820 with significantly better performance and reliability than the previous generation machines. Many of these modules have been used in CMS test beam runs at FNAL and CERN [95]. They also include 20 HD modules, about half of which are built on C-fiber baseplates, making them very similar to what is planned for LDMX. Fig. 3.66 shows some of the tooling of the gantry during automated assembly of an HD module at UCSB.

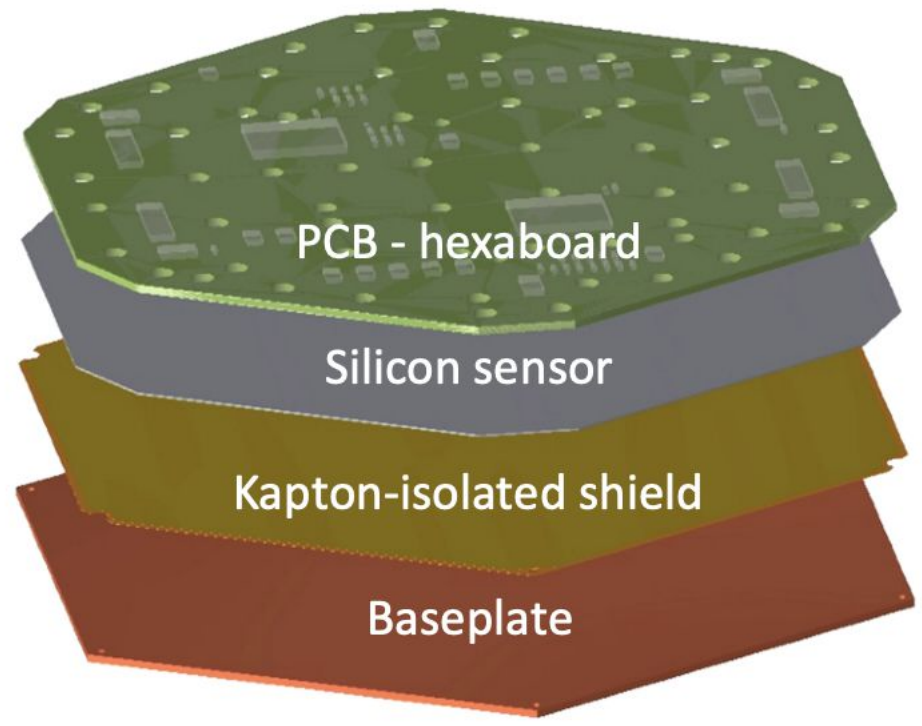

Figure 3.65: Schematic of the four layers that make up an ECal module.

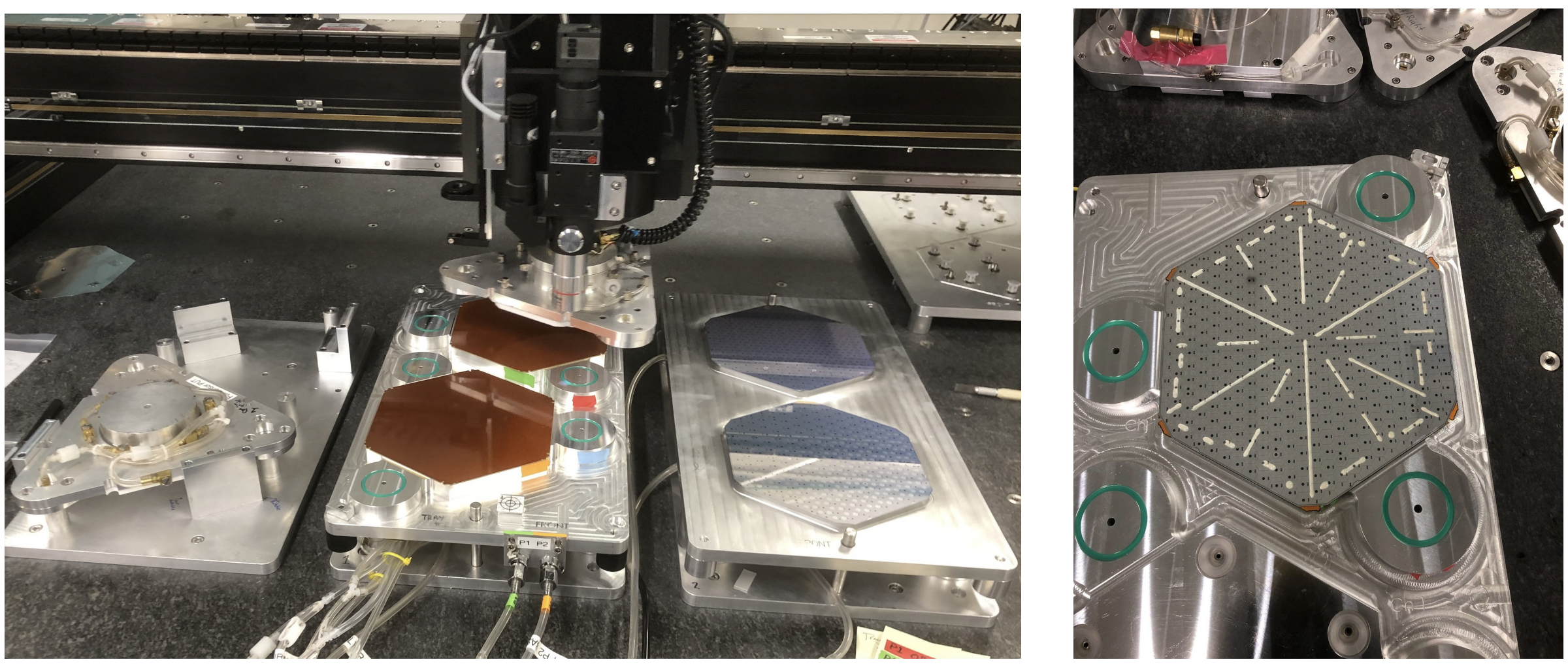

Figure 3.66: Gantry operation during module assembly. Two baseplates with Kapton-clad shields and two silicon sensors are seen on assembly trays with a sensor pickup tool attached to the gantry-head for pick-and-place after glue is applied (Left); and epoxy is applied to the topside of a sensor to attach the hexaboard (Right).

A large number of studies have been done to verify the robustness of the module design:

- Unirradiated modules were thermal-cycled 100 times between +20C and -40C without damage or impact on operation.
- Ten half-hexagonal modules (50%/50% with C-fiber/titanium baseplates) have been irradiated with $\gamma$ rays to 50 MRad and then thermal-cycled 100 times between +20C and -40C without damage or impact on operation.
- Extensive studies were done to see whether arcing could occur between the sensor backplane biased to as high as 1000V and the baseplate at ground. The Kapton isolation was found to be adequate to prevent arcing with as little as 25 $\mu$m excess beyond the end of the sensor to exposed baseplate metal.
- Whether epoxy that occasionally spreads to the sensor guard ring area might become problematic (e.g. conductive) after significant irradiation was tested by generously covering mini-sensors with Araldite epoxy in the guardring region and irradiating to 200 MRad. Tests showed that these sensors behaved very similarly to a control group that were irradiated without epoxy on the surface. However, it was later found that HGC modules can have noisy channels next to the guardring if there is an Araldite connection when there is humidity present. This was successfully addressed by moving the epoxy slightly further away from the guard ring.
- Similarly, the Sylgard 186 encapsulant used to protect wirebonds was tested to see if it could lead to

wirebond breakage after significant irradiation when thermal-cycled. Highly irradiated encapsulant can lose elasticity and upon cooling, may contract in a way that damages wirebonds. Test structures were fabricated with the same geometric configuration of encapsulated bonds and irradiated to a fluence and dose about 50% above the level expected for the most irradiated modules at the end of HL-LHC running. They were then thermal-cycled 50 times at CERN. There was only one wirebond out of 375 that broke. Note that fluence and dose levels relevant to LDMX are lower than those sustained by previous silicon detectors that had encapsulated wirebonds and continued to function without problems, such as the CDF Run-II detector, including Layer 00 installed at a radius of 1.4cm [96] and the current CMS outer tracker.

For LDMX modules we have specified an assembly accuracy of 100 $\mu$m . When laminating with epoxy alone, we found this accuracy can be difficult to meet when hexaboards are warped more than about 500 $\mu$m. This was an issue because we frequently receive hexaboards that are non-flat by as much as 1.5-2.0 mm, and indeed, manufacturers have said that they cannot guarantee better than this. This was the principal reason that we began using acrylic transfer tape sheets together with Araldite epoxy. Tests have shown that this hybrid dual-adhesive lamination method makes it possible to build high-quality modules that are flat to better than 500 $\mu$m, with uniform glue gaps set by the thickness of the transfer tape, even for extremely warped hexaboards, while Araldite then assures adhesion after significant irradiation.

#### 3.7.3.6 Module Wirebonding & Encapsulation

Automated wirebonding is performed with a Hesse BondJet820 wedge wirebond machine seen in Fig. 3.67. Programmable with pattern recognition, the wirebonder uses 25 $\mu$m aluminum wire with 1% silicon. Bond strength is measured by a destructive pull test using a Royce 610 pull test machine. Welding parameters are tuned for optimal connections to the pads at both ends of the connection. We require $> 8$ gram-force (gf).

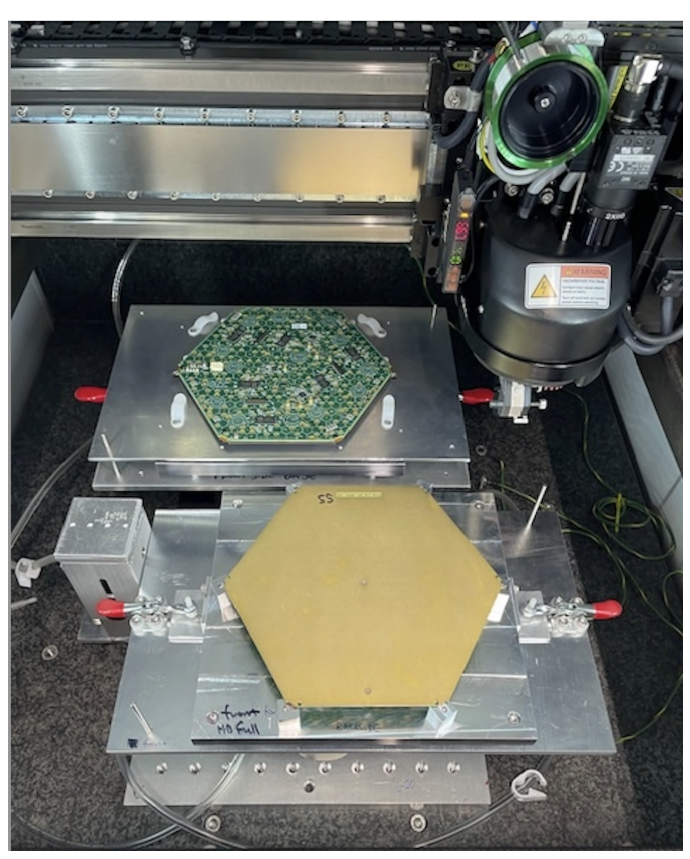

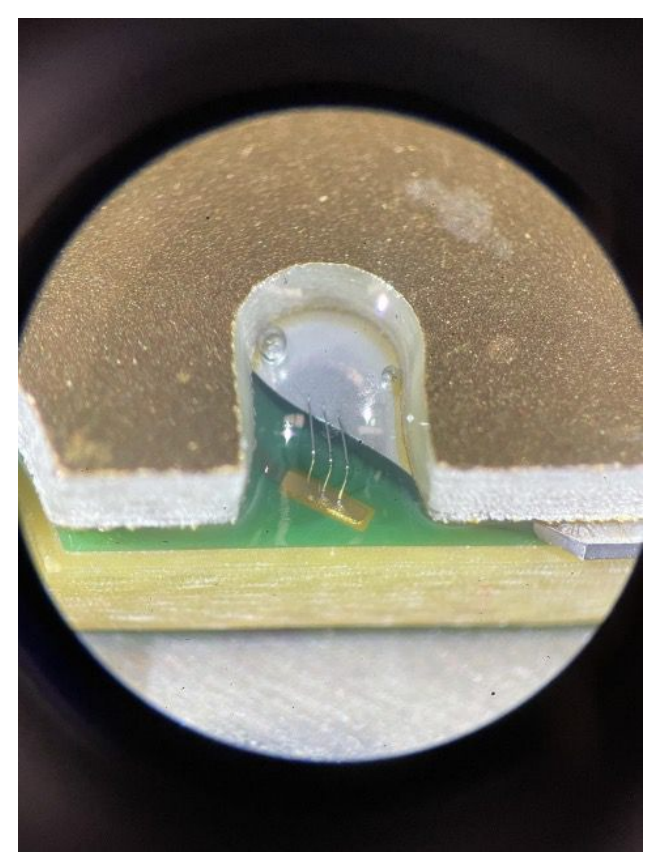

Figure 3.67: Left: Wirebonder setup for making either backside or frontside wirebonds; Right: 3 bias bonds between sensor and hexaboard after encapsulation.

Wirebonds are protected with Sylgard 186 encapsulant, which is dispensed with a pneumatic pump by a table-top programmable robot. Curing is accomplished in less than an hour at 40°C. After backside bias bonds are complete, the module is mounted on a carrying tray, facing upward, and remains in this position for the remainder of wirebonding, encapsulation, and testing.

Wirebonding of the front side of the module creates three types of microelectronic connections as seen in Fig. 3.68: (*i*) bonds are made in stepped holes to the conductive guard ring running around the sensor to ground it in order to shape the field near the edge of the sensor, which can be at some intermediate voltage up to the full bias value, (*ii*) bonds are made in notches in the hexaboard to ground the Kapton-clad shield between the sensor and the baseplate, and (*iii*) connections are made in stepped holes to all of the individual readout cells. The front side automated wirebond program for HD modules completes all 1368 bonds in ∼15 minutes. To check the integrity of the wirebonds, the module is scanned by an OGP coordinate measuring machine where a photo is taken of each stepped hole and notch under magnification. A machine learning algorithm is used to find missing or damaged wires. Broken or missing wirebonds detected by the QC survey are repaired before encapsulation and testing.

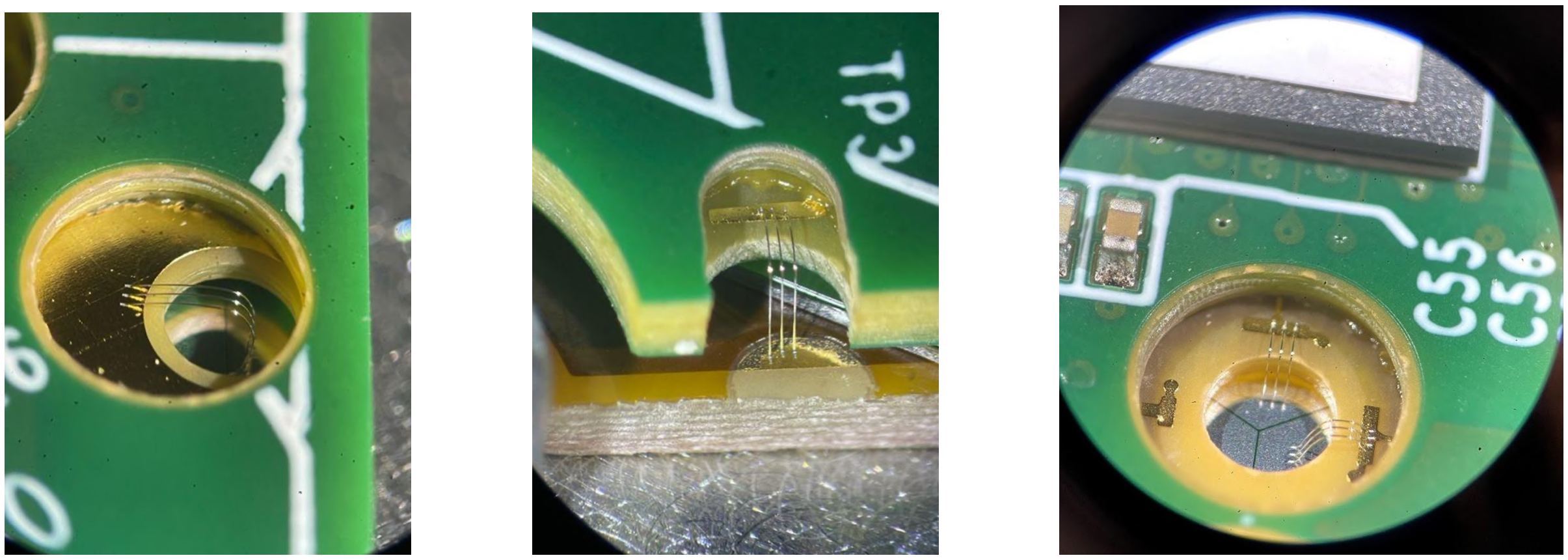

Figure 3.68: Frontside bonds: to guard ring (left), to shield layer (middle), and to signal cells (right).

#### 3.7.3.7 Trigger Concentrator ASIC (ECON-T)

The trigger concentrator ASIC (ECON-T) was developed by the US-CMS group at FNAL for the CMS HGC. The ECON-T is designed to carry out trigger calculations using inputs from the HGCROC ASIC. As discussed in Sec. 3.7.3.3, the HGCROC produces elink outputs which each carry the trigger sum information for four trigger sums. In the ECal, each trigger sum is formed from nine detector cells, while in the HCal, the sums are formed from four detector channels. The ECON-T is capable of receiving up to twelve elinks, which is sufficient for a full ECal module.

Within the ECON-T, the elink data is captured and aligned so that all channels for a given bunch crossing are processed together. The ECON-T linearizes the compressed floating-point scale to 21 bits and allows the application of an individual 12-bit multiplicative calibration constant to each channel, representing a scaling by a value between 0.0 and 1.99951, which is sufficient to provide intercalibration between the channels of a module. After calibration, the trigger cell energies can be passed to one of several algorithms.

The ECON-T applies a user selectable compression algorithm for all aligned elink data. One algorithm, the threshold algorithm, is designed to transfer all cells above a programmable energy threshold to the backend electronics. This algorithm produces a variable-length data packet which is difficult for most backend electronics to handle reliably within reasonable limits of processing time and FPGA resource utilization. The other algorithms are all fixed-format, providing a defined set of information for each bunch.

For the ECal, the baseline algorithm is the "Best Choice" (BC) algorithm. Under the BC algorithm, the ECON-T will transmit a sum of energy across the module as well as the location and energy for the highest trigger cells. The energy sum across the module is encoded into an eight-bit floating-point format. The energies for the trigger cells are encoded into a seven-bit floating-point format. The number of trigger cells sent depends on the number of active output elinks. With two output trigger elinks, four trigger cells can be transmitted, while with six trigger elinks a total of eighteen trigger cells can be transmitted. The usage of different configurations is discussed in Sec. 3.7.4.2.

The ECON-T has been tested for algorithmic correctness and operation in both simulation and test stands. It has been tested for radiation tolerance, showing no significant impact to operation up to 600 Mrad and no evidence of single-event latchups or other disruptive impacts in high-fluence proton tests [97]. Single-event effects do occur, but are expected to be low-impact for the CMS case and will be substantially lower for the LDMX ECal.

#### 3.7.3.8 DAQ Concentrator ASIC (ECON-D)

The DAQ concentrator ASIC (ECON-D) was also developed by the US-CMS group at FNAL for the CMS HGC. The ECON-D is designed to carry out zero-suppression on the DAQ data from the HGCROC and combine the output of multiple HGCROC readout elinks into a smaller number of elinks for transmission off of the detector. The ECON-D receives up to twelve HGCROC readout elinks (enough for a full ECAL module) and can drive a programmable number of elinks up to six.

In the LDMX case, where the specified readout rate (50 kHz) is more than an order of magnitude below the specification for CMS (750 kHz), the zero-suppression functionality is not necessary for front-end operation purposes. However, to meet the requirements for the recorded data volume, the ECal system must carry out zero-suppression at some point in the system. A full readout of ECal without zero-suppression requires 125 kB, while the computing budget assumes an event size of 1.25 kB for ECal events which pass the primary physics trigger within the $\mu = 1.0$ case. The ECON-D ASICs can be configured to carry out zero-suppression, which will reduce the volume to approximately 14.1 kB, of which 13.1 kB will be headers and checksums.. Further processing in software before archival storage would be able to remove the majority of the headers and accounting data to meet the data storage target.

The zero-suppression algorithm in the ECON-D is highly-configurable, allowing for techniques to carry out common-mode noise subtraction and out-of-time pileup energy subtraction. In the LDMX case, the out-of-time pileup energy subtraction may be of particular value, given the high occupancy of the core of the detector with beam-energy electrons. Without the out-of-time energy subtraction, the readout data volume could be easily dominated by the residual tails from previous bunches with full-energy electrons.

The ECON-D has been tested for algorithmic correctness and operation in both simulation and test stands. It has been tested for radiation tolerance, showing no evidence of single-event latchups or other disruptive impacts in high-fluence proton tests [97]. Recently, some data corruption has been observed, particularly when the chip is operated at voltages below 1.20 V, at low temperatures (particularly below $-10°$C), and at elevated total ionizing radiation doses. Quality-control screening appears to allow these effects to be minimized, and the final design layout of the motherboard will be adjusted to keep the ECON-D away from the highest-dose regions.

For the use in HCal (Sec. 3.8.4.3), which operates at room temperature and which will not experience significant ionizing radiation, sufficient ECON-D chips are already available. For the more-challenging case of the central modules of the ECal, the yield of ECON-D ASICs that meet this requirement is lower. Sufficient "high-radiation-tolerance" ECON-D ASICs for these regions are expected to be available by Q1 2026, which is well in advance of the construction need.

### 3.7.4 Bilayers and Full Assembly

In this section, we describe the structural design of the ECal. The ECal is made up of bilayers separated by $W$-absorber planes, all of which are suspended like hanging file folders in a stainless steel support box as seen in Fig. 3.73. This layout minimizes the distortion of layers, some of which are very heavy, and enables straightforward assembly as well as the removal of individual layers for repair or modification and re-installation. The latter could be necessary if any modules need to be replaced or if it is decided to take data with an altered distribution of absorber and sensing layers for a specific type of measurement, such as might be needed to better understand systematic uncertainties. The ECal is designed to be as longitudinally compact as possible in order to maximize angular acceptance. The choice of tungsten absorber facilitates a compact design while also providing good thermal and mechanical properties for a system that is expected to undergo many thermal cycles over its lifetime.

#### 3.7.4.1 ECal Bilayer

Compactness of the ECal is an important driver for the bilayer concept by enabling one cooling plane to cool two sensor planes and their electronics. This is inspired by the CMS HGC where the cooling planes are Cu plates with embedded cooling lines for a $CO_2$ evaporative cooling system. For LDMX we have replaced the Cu with C-fiber. The cooling plane is made up of a thick C-fiber plate with milled grooves to route stainless steel cooling tubes. The tubes are glued in place and then enclosed by a thin high-TC C-fiber sheet. This assembly is attached to a stainless steel strongback by means of screws at the corners and the strongback then spans the width of the ECal support structure where its ends are pinned to support ledges. This is the core of the bilayer that supports all other parts of the assembly.

The layering of one side of the bilayer is shown in Fig. 3.69. The cooling plane is attached to a stainless steel strongback to eventually install it in the support structure. In the bottom corners of the cooling plane one sees the DC-DC converters and other ancillary electronics (blue). Next, a thin W-absorber layer (grey) is screwed to the cooling plane. This W-absorber is cut to be just larger than that of the 7 module flower array

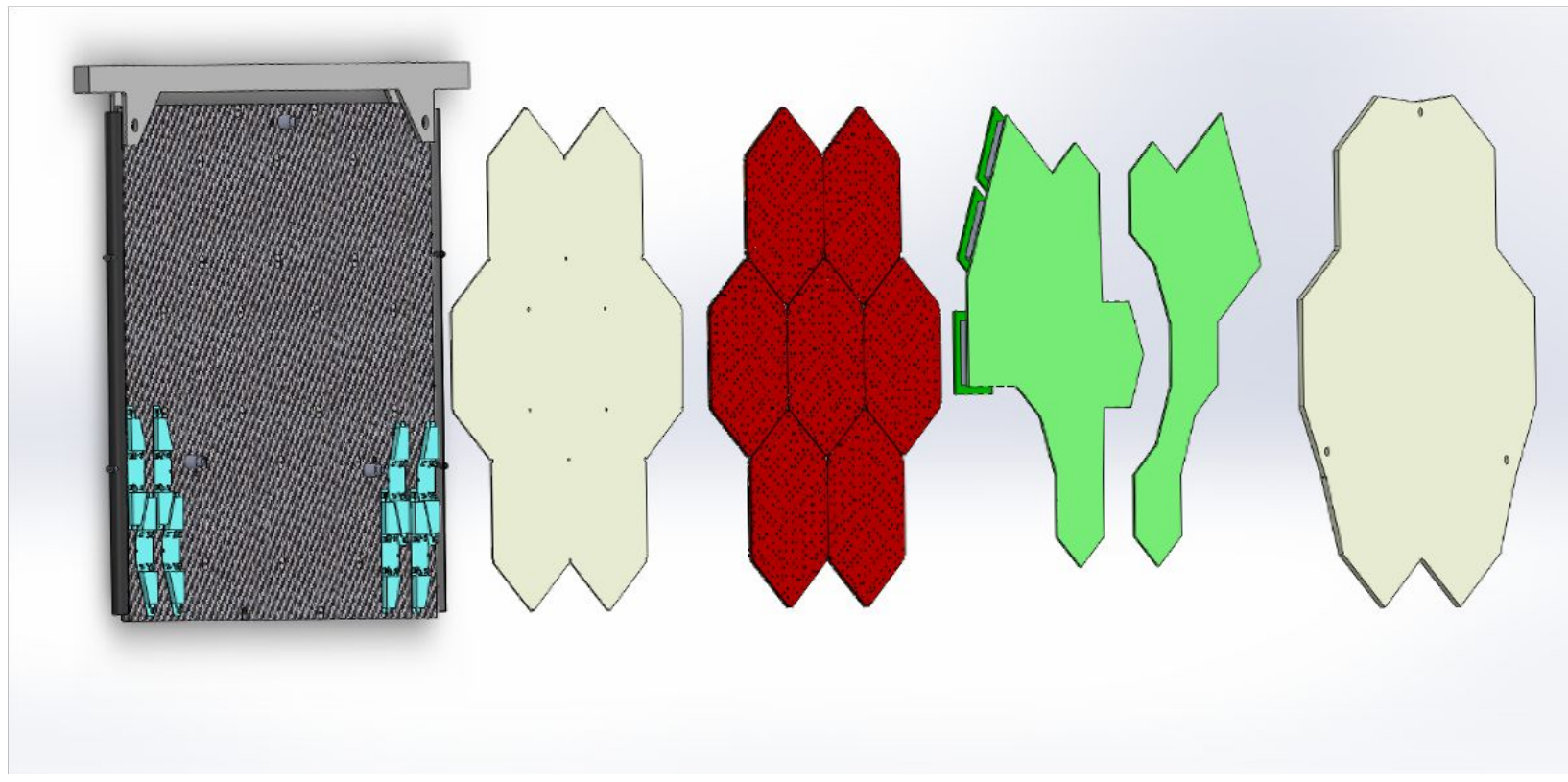

Figure 3.69: Bilayer strata (left to right): Strongback suspending the cooling plane (black) with thin W absorber (grey) and DC-DC converters (blue); 7-module flower (red); motherboards (green); thick W-absorber (grey).

(red) which is attached to the W-absorber layer by screws with washers that overlap the C-fiber baseplates. Like all other module layers the baseplates have corner mousebites, but ones that are less deep, leaving some material exposed. In this design the W-absorber is a continuous sheet without gaps and the C-fiber of the baseplate is chosen to have very high in-plane thermal conductivity to help minimize thermal gradients on the sensors that would otherwise occur as a result of the highly concentrated heat sources of the HGCROCs on the HB. Two motherboards with mezzanine cards at their periphery (green) are then connected to the modules. Finally, on the downstream side of each bilayer a W-absorber, whose thickness is typically twice the thickness of the thin W-absorber connected to the cooling plane, is positioned between the sensing planes of two neighboring bilayers. Other than this additional W-absorber, the layers on the two sides of the bilayer are identical.

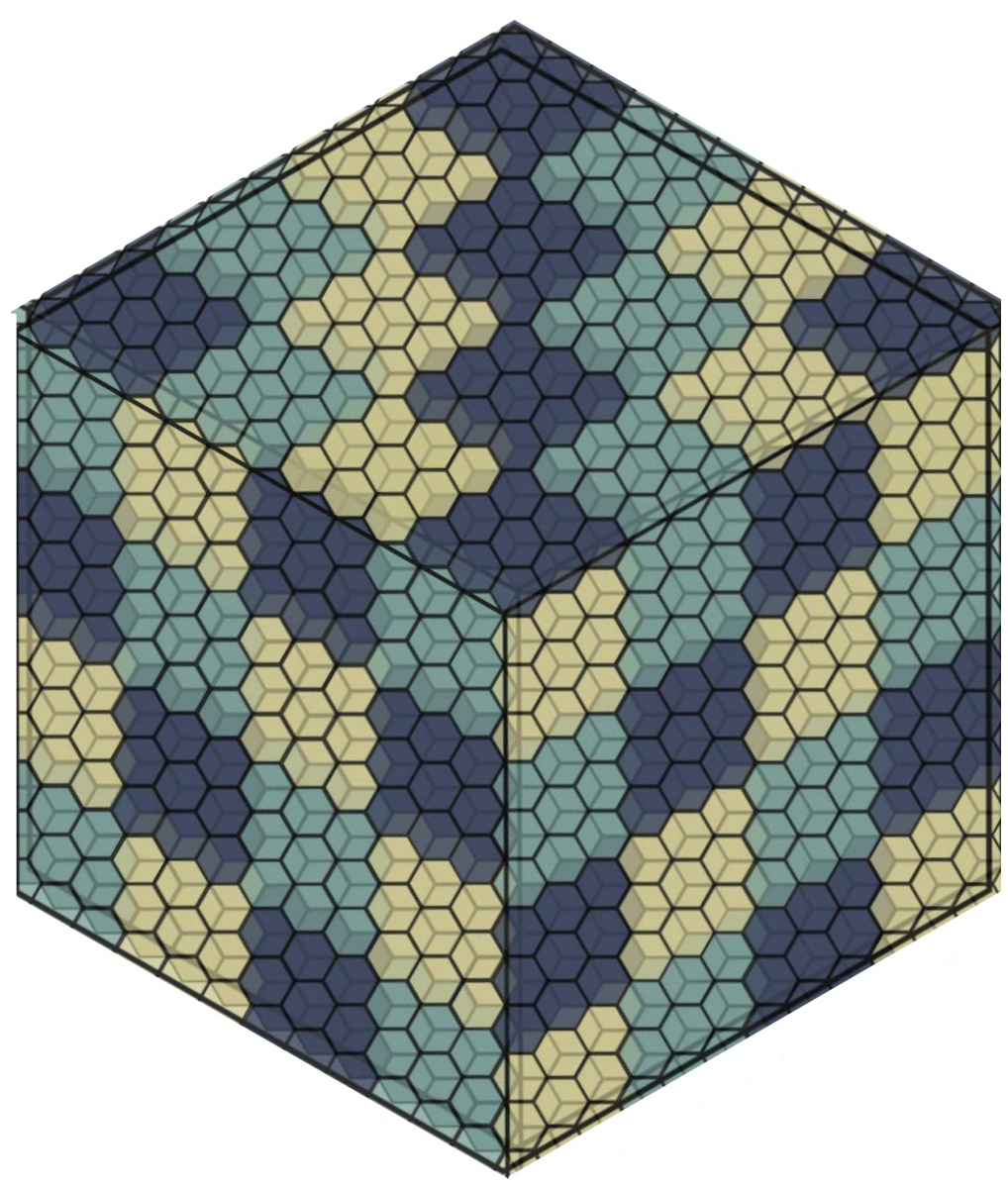

Figure 3.70: Schematic of cell positions of sensors on either side of a bilayer: Transparent top layer cells show overlap with 3 bottom layer cells. Also, non-aligned sensor edges to avoid dead regions are also seen.

The module planes on either side of the bilayer are offset relative to one another by one-half the size of a readout pad along an axis perpendicular to the bottom left edge, as seen in Fig. 3.70. This positions the center point of pads in the sensor plane on one side of the cooling plane to be aligned with the intersection point of three pads in the sensor plane on the other side. This serves two purposes: (*i*) isolated charged particles traveling through the ECal can be located with smaller uncertainty and (*ii*) the dead regions from the physical gaps between modules and the 500 $\mu$m of guard ring region at the edge of each sensor do not align with any dead regions on the other side of the bilayer. The total width of the dead region will depend on how tightly the modules can be placed relative to one another. We estimate the physical gap to be 500 $\mu$m at -20°C so that the full dead region between modules will be 1.5 mm.

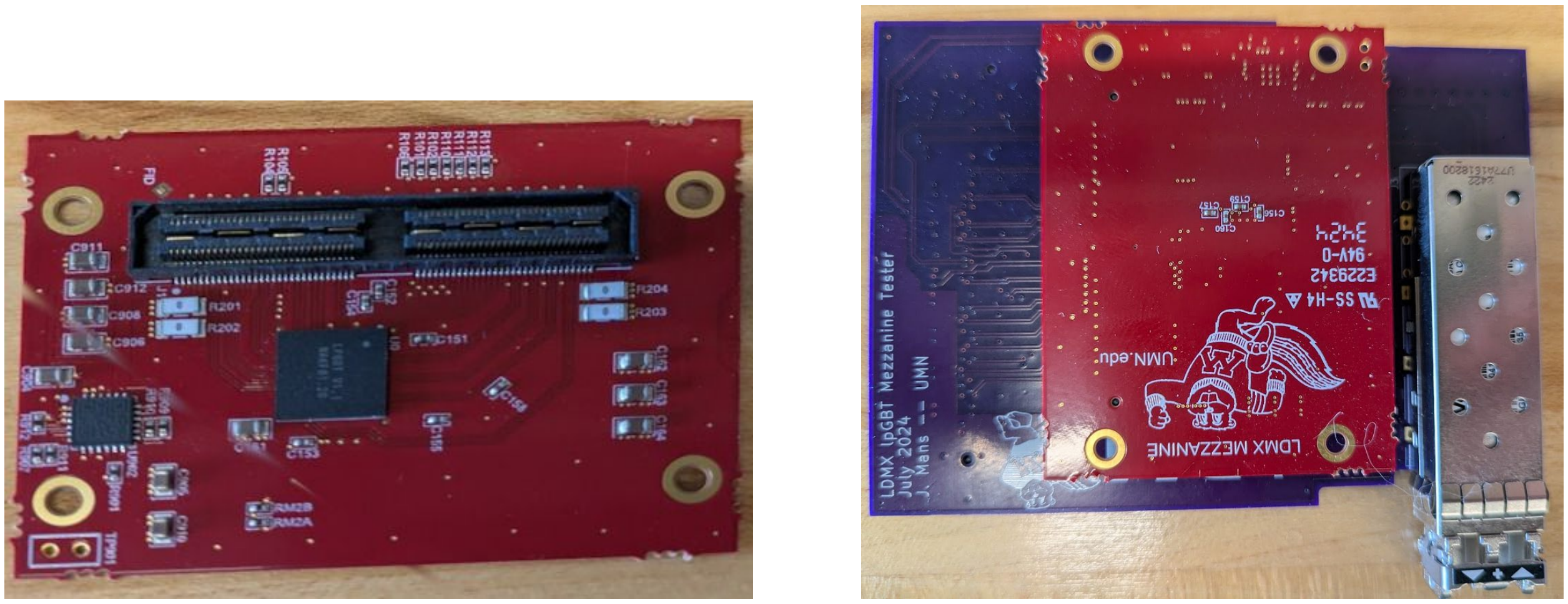

Figure 3.71: (Left) Image of a prototype LDMX lpGBT mezzanine card (Right) The lpGBT mezzanine mounted on a quality control readout FMC adapter.

#### 3.7.4.2 Motherboards and Readout

Each module of the ECal has three connectors for control and data signals, as discussed in Sec. 3.7.3.4. The modules are connected to motherboard PCBs which carry signals between the modules and the lpGBT data link ASICs which are located at the edge of the layer. There are two motherboards per plane to keep the circuit board size within a reasonable manufacturing size. One motherboard connects to the central module and three outer modules, while the second motherboard connects to the other three outer modules. These designs are informed by experience from the CMS HGCAL engine and HD wagon PCB boards, but they are specific designs appropriate for the LDMX geometry, which will be finalized in the second half of 2025.

The lpGBT ASIC is a fine-pitch BGA, which makes its integration on a large-format PCB very difficult. Instead, the lpGBT ASICs are each mounted on a separate small mezzanine which is plugged into the motherboard. In addition, the design effort can be shared between ECal and HCal, both of which plan to use this mezzanine. A photo of the mezzanine and its tester are shown in Fig. 3.71. This mezzanine has been prototyped and successfully tested with dedicated QC hardware as part of the LDMX prototyping effort.

Two or three of these mezzanines are used on each motherboard, with the third mezzanine required for the trigger data of the central module of the flower. The arrangement of signals from the mezzanines on the motherboard in the three-module case is shown in Fig. 3.72. The DAQ lpGBT serves as the source of the clock and fast-control streams for the full module, including for the Trigger lpGBTs, via its 2.5 Gbps optical downlink. The optical uplink and downlink formats include extensive forward-error-correction which protect against data corruption.

Both Trigger and DAQ lpGBTs participate in the control and monitoring of the front-end via I2C buses, GPIO digital signals, and ADC channels. The lines for these signals are distributed via the motherboard PCBs. The motherboards host connectors for the modules, the lpGBT mezzanines, and optical components. The motherboards also host Rafael clock-fanout ASICs [98] (two per module) and the ECON-T and ECON-D ASICs which concentrate the trigger and data streams that are transmitted via 1.2 Gbps data links to the lpGBTs for encoding and combination to produce 10 Gbps DAQ and trigger readouts. The power for the Rafael and ECON ASICs is provided through the module connectors.

The readout data from the HGCROCs flows to the ECON-D ASIC associated with a module through twelve 1.2 Gbps data links. The ECON-D ASIC carries out zero-suppression to the extent necessary and combines the twelve inputs into a programmable number of outputs. At the LDMX specified readout rate of 25 kHz, no zero suppression is required in the front-end even for the combination into a single DAQ elink per module. However, zero suppression will be required to achieve the target readout volumes for final storage. The ECON-D allows for the collection of non-zero-suppressed data at any time using a special fast command, and pedestal will be regularly acquired during operations to track any change in detector behavior.

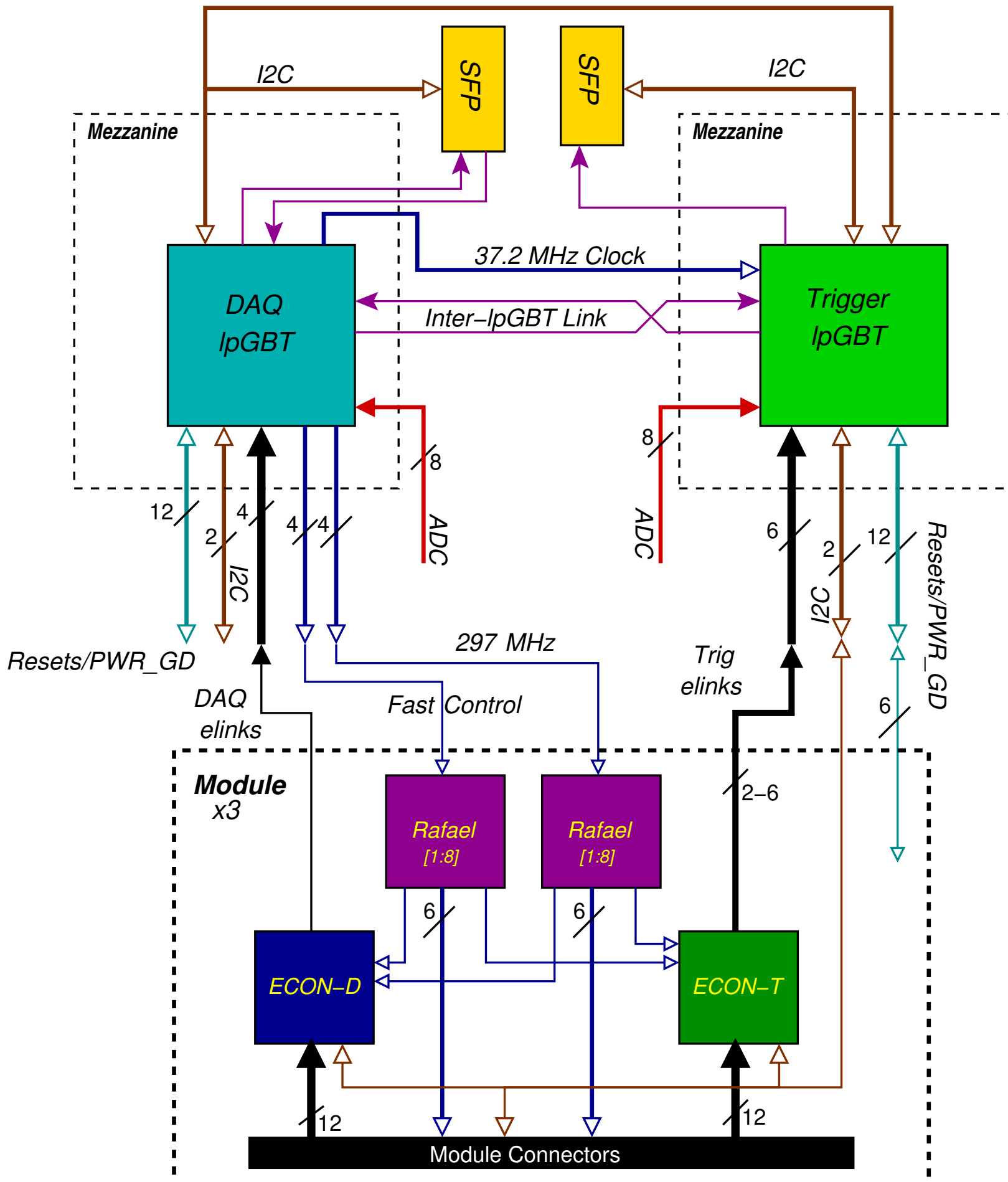

Figure 3.72: Signal arrangement for the triple-module configuration.

The trigger data from the HGCROCs flows to the ECON-T ASIC associated with the module through twelve 1.2 Gbps data links. The ECON-T applies a basic calibration to the trigger cells from the HGCROC and then applies one of several different algorithms to the trigger cells. The baseline for the LDMX ECal is the "Best Choice" algorithm, which provides the energy sum for the full module along with the energy and location of the highest cells in the module. For the outer modules of each layer, two elinks will be used, which will allow the highest four trigger cells per module to be sent. For the central module, up to six elinks will be available, allowing up to eighteen cells to be sent, which exceeds the requirement. The data formats from these ECON-T ASICs are given in Table 3.8. The total energy in the module (SumTC) is transmitted in a floating point format with three mantissa bits and five exponent bits covering an effective range of 22 bits, while the individual trigger cells are encoded in a floating point format with three mantissa bits and four exponent bits covering an effective range of 18 bits.

#### 3.7.4.3 ECal Full Assembly

The ECal, with a total weight $>$ 800 kg, will be mounted directly to the face of the magnet by means of forward protrusions ('ears') that are part of the sidewalls of the support structure, shown in Fig. 3.73. As

**Two-elink format (peripheral modules)**

<table>
<tr><th>Elink</th><th>31</th><th>30</th><th>29</th><th>28</th><th>27</th><th>26</th><th>25</th><th>24</th><th>23</th><th>22</th><th>21</th><th>20</th><th>19</th><th>18</th><th>17</th><th>16</th><th>15</th><th>14</th><th>13</th><th>12</th><th>11</th><th>10</th><th>9</th><th>8</th><th>7</th><th>6</th><th>5</th><th>4</th><th>3</th><th>2</th><th>1</th><th>0</th></tr>
<tr><td>0</td><td>BC1</td><td colspan="3">BC[2:0]</td><td colspan="8">SumTC[7:0]</td><td colspan="6">AddrTC1[5:0]</td><td colspan="6">AddrTC2[5:0]</td><td colspan="6">AddrTC3[5:0]</td><td colspan="2">ATC4[5:4]</td></tr>
<tr><td>1</td><td colspan="4">AddrTC4[3:0]</td><td colspan="7">TC1[6:0]</td><td colspan="7">TC2[6:0]</td><td colspan="7">TC3[6:0]</td><td colspan="7">TC4[6:0]</td></tr>
</table>

**Six-elink format (core module)**

<table>
<tr><th>Elink</th><th>31</th><th>30</th><th>29</th><th>28</th><th>27</th><th>26</th><th>25</th><th>24</th><th>23</th><th>22</th><th>21</th><th>20</th><th>19</th><th>18</th><th>17</th><th>16</th><th>15</th><th>14</th><th>13</th><th>12</th><th>11</th><th>10</th><th>9</th><th>8</th><th>7</th><th>6</th><th>5</th><th>4</th><th>3</th><th>2</th><th>1</th><th>0</th></tr>
<tr><td>0</td><td>BC1</td><td colspan="3">BC[2:0]</td><td colspan="8">SumTC[7:0]</td><td colspan="20">ChMap[47:28]</td></tr>
<tr><td>1</td><td colspan="28">ChMap[27:0]</td><td colspan="4">TC1[6:2]</td></tr>
<tr><td>2</td><td colspan="3">TC1[2:0]</td><td colspan="7">TC2[6:0]</td><td colspan="7">TC3[6:0]</td><td colspan="7">TC4[6:0]</td><td colspan="7">TC5[6:0]</td><td>TC6[6]</td></tr>
<tr><td>3</td><td colspan="6">TC6[5:0]</td><td colspan="7">TC7[6:0]</td><td colspan="7">TC8[6:0]</td><td colspan="7">TC9[6:0]</td><td colspan="5">TC10[6:2]</td></tr>
<tr><td>4</td><td colspan="2">TC10[1:0]</td><td colspan="7">TC11[6:0]</td><td colspan="7">TC12[6:0]</td><td colspan="7">TC13[6:0]</td><td colspan="7">TC14[6:0]</td><td colspan="2">TC15[6:5]</td></tr>
<tr><td>5</td><td colspan="5">TC15[4:0]</td><td colspan="7">TC16[6:0]</td><td colspan="7">TC17[6:0]</td><td colspan="7">TC18[6:0]</td><td colspan="6">0</td></tr>
</table>

Table 3.8: Data link contents for each bunch arrival for the signals sent from the ECal ECON-T ASICs to the trigger electronics. The BC1 bit is set for the first bunch of a pulse train, based on the settings of the HGCROC and the fast control system, which distributes the BCR signal used by the HGCROC and ECON-T.

such, the ECal will be cantilevered, requiring a strong and stiff support box. The baseline design makes use of 0.75″ (0.5″) thick stainless steel plates for the side (back and bottom) walls. The sides, bottom and back plates have openings cut in them to provide access to the detector as seen at right in Fig. 3.73. The 2″ wide ledges from which the layers are suspended are also 0.75″ thick stainless steel. Stainless steel strongbacks bridge the ledges to support the suspended bilayers, which are attached to the strongbacks at their two top corners. The strongbacks are 33″ long by 1″ tall and 0.5″ wide for the PS, A, and B type layers, while for the heavier, thicker, and more numerous C and D layers they have a profile of 1.5″ tall by 1″ wide. The Finite Element Analysis (FEA) of deflections and stresses in the baseline design, with all strongbacks fully loaded, is shown at right in Fig. 3.73. The ears (dark blue) are fixed to the magnet and do not deflect. The heaviest bilayers produce a maximum deflection of 161 $\mu$m in the strongbacks (red).

Fig. 3.74 shows the ECal layers with labeling of groups, together with the distances from the front of the detector to layers PS, A, and B as well as to the first layer of 9 in group C and the first layer of 5 in group D. The spacing of the layers is also called out. In its final configuration, thin aluminum plates will seal the box. All sides of the box will be thermally insulated.

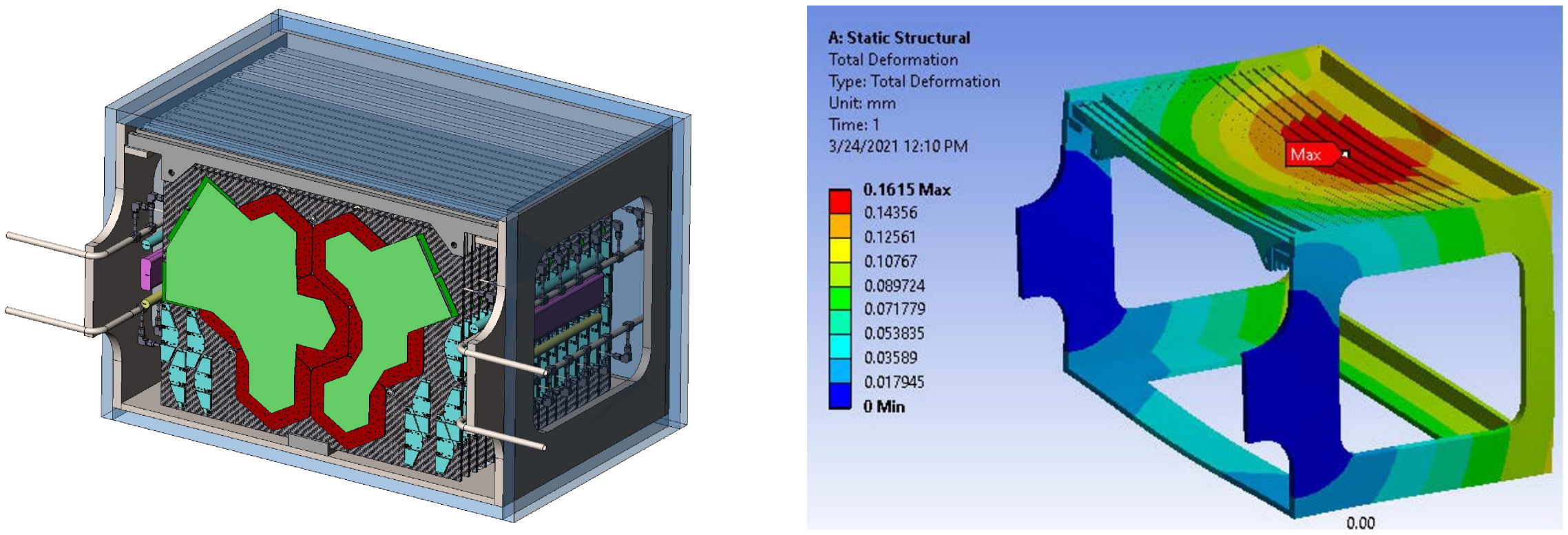

Figure 3.73: Left: hanging file-folder design. Right: deflections with strongbacks fully loaded.

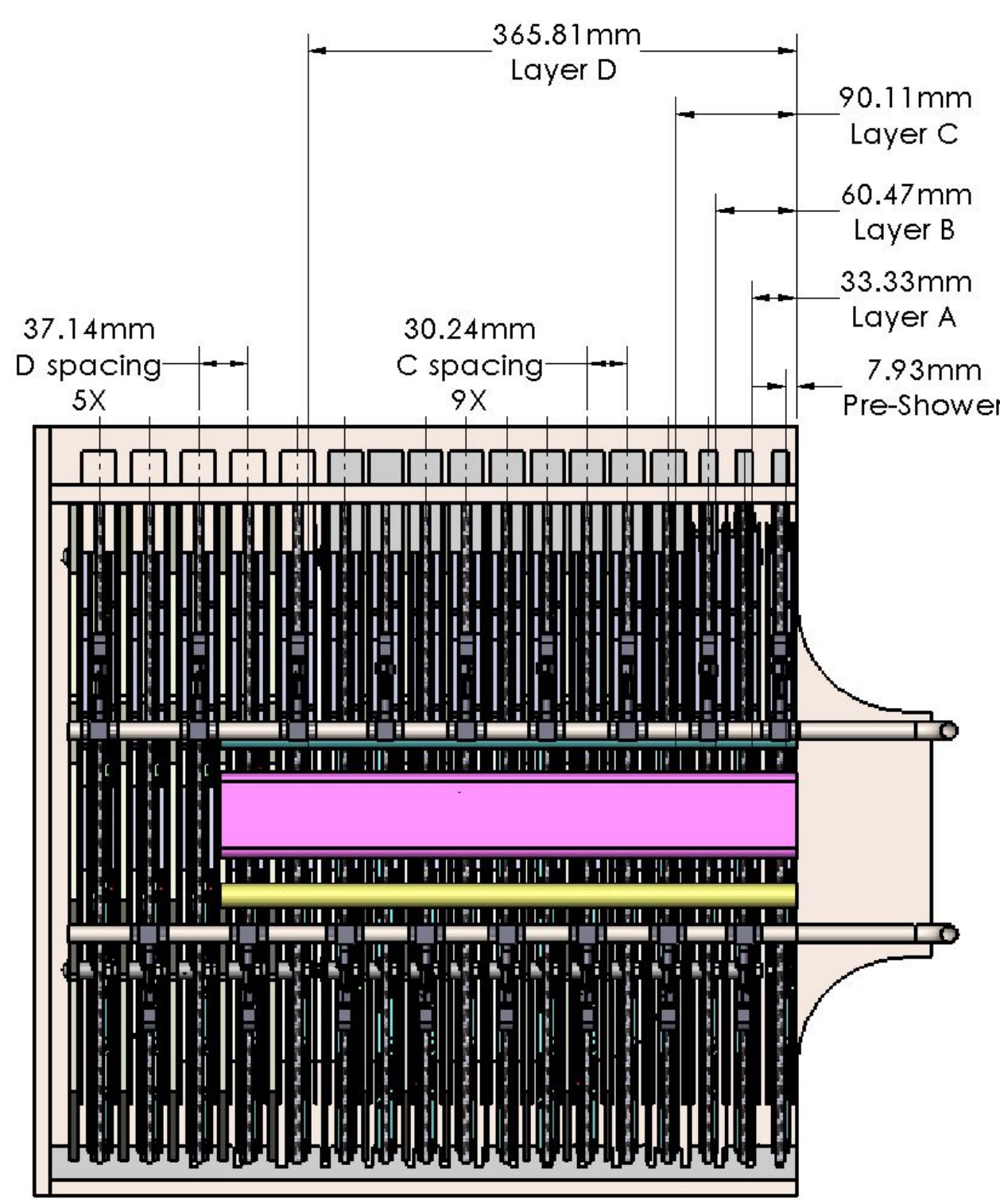

Figure 3.74: Side view of ECal indicating layer types as well as their positions and spacings.

Fig. 3.75 provides dimensions of the support box and insulation in the transverse view and also shows some aspects of the bilayer including the cooling plane (black), placement of main inlet/outlet cooling lines, their fittings, and the estimated minimum cross sectional space required for – but not the final locations of – the optical fibers, HV and LV cables in yellow, magenta, and blue, respectively. Also shown are the connections to the strongbacks, and the flower shaped W absorber covering the motherboards. It appears on only one side of the bilayer, and is fastened at the two points in red at the top.

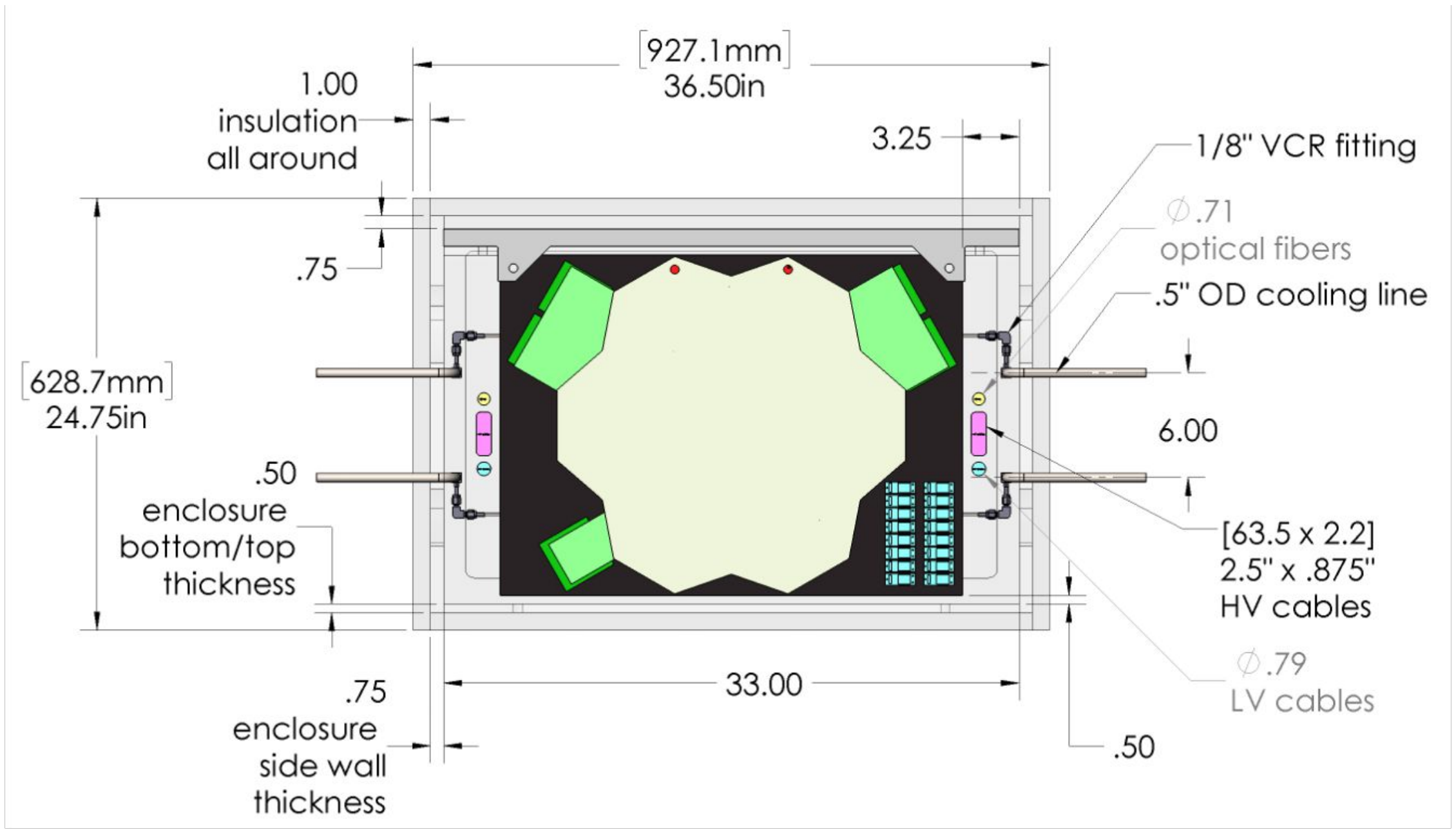

Figure 3.75: Front view of ECal with locations and dimensions of various elements labeled.

For the baseline design shown in Figures 3.74 and 3.75 as well as a half dozen variations of this layout, FEA studies were performed to determine the locations and magnitudes of the maximum deflections and stresses with all strongbacks fully loaded. Table 3.7 summarizes the size and weight specifications for layers by group along with their multiplicities as used in the baseline FEA calculation. Tungsten makes up the majority of the mass but for each layer there is also the strongback, C-fiber cooling plane with embedded stainless steel cooling tubes, connectors and coolant, module baseplates, sensors, hexaboards, DC-DC converters, mezzanine cards, and motherboards. The mass of these items is a few kg per layer but we have conservatively set this to 20 kg in the FEA. Design variations had different thicknesses of walls and ledges, some with additional bracing under the latter, as well as different amounts of material removed from the sides, bottom and back. One considered closed sidewalls. Another trial included 1″ tall strongbacks for type C & D layers. The baseline design performed well, with maximum deflection of 161 $\mu m$ as seen at right in Fig. 3.73. The maximum stress was 3030 psi (21 MPa). With solid sidewalls the deflection was reduced to 90 $\mu m$. Alternatively, reducing the strongback height to 1″ for type C and D layers increased the maximum deflection to 226 $\mu m$. The baseline, with good access to the interior, is close to optimal.

## 3.7.5 Services

### 3.7.5.1 Cooling system

While the ECal modules contain efficient electronics, many channels of the detector are read out for every arriving accelerator pulse and the bunch spacing is slightly smaller than that at the LHC. This means that all modules will be continuously powered and active. Meanwhile, the target for the maximum temperature of any point on the silicon sensors is -20°C to ensure that leakage currents and noise are kept to low enough values to maintain high signal-to-noise ratio throughout the system, even after considerable fluences have been integrated (see Appendix A for more details).

Based on CMS experience, we know that individual HD modules will each dissipate ∼8 W of power.

Conservatively estimated, the additional electronics boards on each LDMX ECal bilayer will produce up to an additional ∼60 W. This brings the total power dissipated by a bilayer to ∼200 W. The total heat produced by the ECal as a whole is thus around 3.5 kW. The design of the bilayer cooling plane, including the dimensions of the piping and results for tests of prototype systems, are presented in section 3.7.6. In addition, the cooling system must address environmental heat that circumvents the enclosure and insulation layers. This will add about 250W.

To meet the requirement of sensor operation at a maximum temperature of -20°C, we plan to use a Julabo W56 chiller with P60 refrigerant. This system is capable of removing ∼7 kW at an operating temperature of −30°C. The W56 chiller is water-cooled and would couple to the large-capacity process water flow which is available in ESA. The chiller includes both an electronic control system which can be linked to the SLAC-standard EPICS slow control system and a set of interlock outputs and inputs, which will be connected to the detector PLC safety system discussed below.

#### 3.7.5.2 Dry air supply and heat shield

Due to the low-temperature operation of the ECal, it will be necessary to have a dry air supply to flush humidity from the detector enclosure in order to keep the service interfaces clear and to install a heat shield around the ECal. If we conservatively assume an internal temperature of -25°C and a 25 mm thick solid foam insulating layer with thermal conductivity of 2W/K-m$^2$, then for a total surface area of the ECal of $\sim 2.5\text{m}^2$, not including the front surface, a heat shield would need to compensate ∼250 W to maintain an external surface temperature of 20°C. Thin, flexible heaters with capacities of up to 350 W/m$^2$ at 20°C, are relatively inexpensive and could be deployed on the sides, back, top and bottom, but probably not on the ECal face with full exposure to the beam. Dry air flow would be needed in this region.

#### 3.7.5.3 Low-voltage system

Each plane of the ECal hosts seven modules, five lpGBT mezzanines, and optical components for five optical transmitters and receivers. The power for these components will be provided from low-profile radiation-tolerant DC/DC converters using the CERN-developed BPOL12 ASIC capable of accepting input voltage up to 11V and providing up to 4A of current each with a total power limit of 10W. For safety purposes, we will assume that power capability is 3A for each converter. The converters are integrated in triple units, such that a single physical converter board can provide three different voltages. A photograph of such a unit is shown in Fig. 3.76. A total of eight converter boards (24 DC/DC converter channels) would be used for each plane as shown in Fig. 3.77.

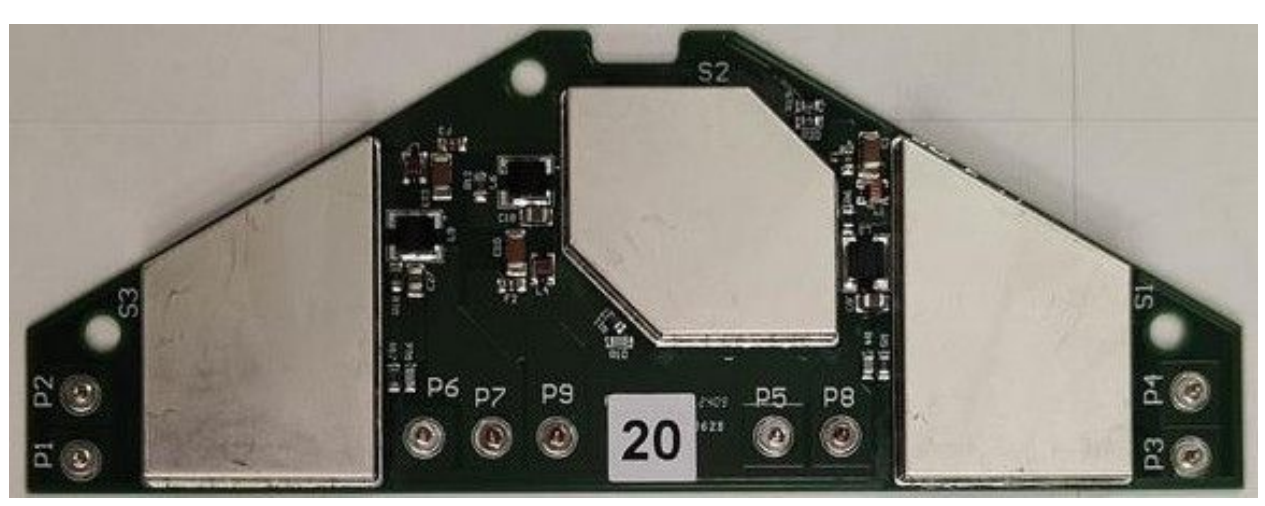

Figure 3.76: Three-channel DC/DC converter module making use of the BPOL12 ASIC and a custom low-profile toroidal coil. The coils and other key components are located underneath the shields which contain induced noise in sensitive electronics such as the silicon modules.

The CMS collaboration has measured the power utilization for the HGCROC and projects a per-module requirement for two channels of 2.0A each for the analog portion of the HGCROCs and one connection of 0.92A for the digital portion of the HGCROCs. The projections here are conservative/worst-case for HGCROC utilization, which is dependent on the operating configuration. In addition, the digital power supply of the module provides power to the ECON-T, ECON-D, and Rafael ASICs associated with the module, requiring an additional 1.0A of current supply from the digital regulator. These voltages would be supplied at 1.5V from the DC/DC converter unit, such that a total of seven three-channel DC/DC converters will be required for the modules.

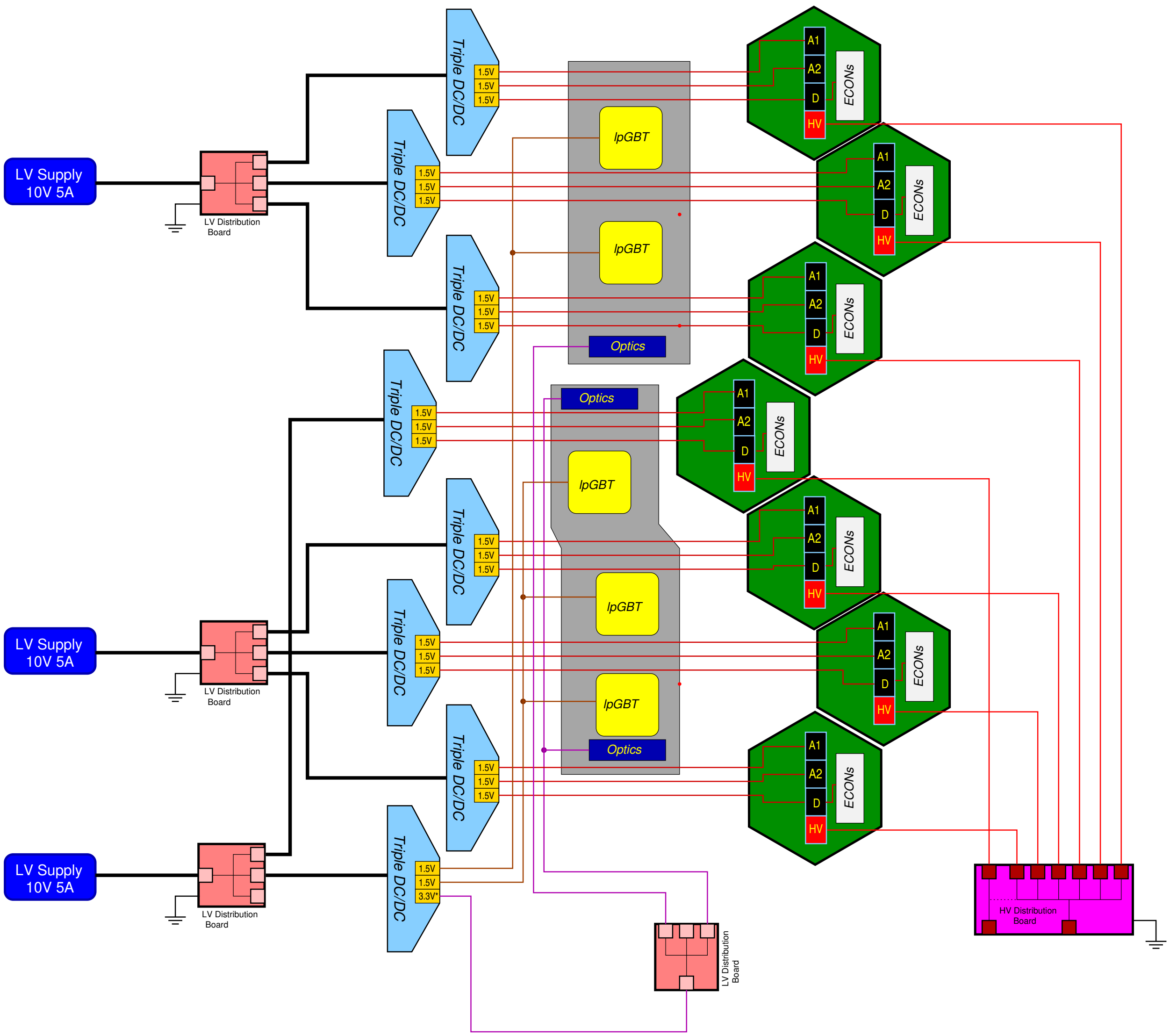

Figure 3.77: Power and bias voltage distribution for a layer of the ECal. The seven modules are shown in a line rather than in the physical flower arrangement to allow for clearer line drawing.

Based on measurements from the CMS collaboration, each lpGBT mezzanine will require 0.5 A of 1.2V power. We expect to supply the two motherboards from separate 1.5V channels of the eighth DC/DC converter unit, with linear regulators to provide the 1.2V current needed by the lpGBT. The eighth DC/DC converter will be modified so that the third channel will provide the necessary voltage for the optical transceiver, either 3.3V or 2.5V, depending on the final choice of optical transceivers.

Given the 70% typical efficiency of the DC/DC converters, the total input current for a triple converter module servicing HGCROCs is 1.27A at 10V, and for the lpGBTs and optics would be 1.1A at 10V. A full plane therefore consumes 100W (10A at 10V). The planned off-detector LV supply modules for the ECal, in common with the rest of LDMX, are MPV8016I modules which contain eight channels of 50 W each, allowing up to 5A at 10V. The planned configuration is to use three channels per plane. Given the use of eight converters and three channels, it is not possible to exactly balance the loads. One channel will need to supply 3.8A, the second channel 3.6A, and the third just 2.5A. In all cases, the channels of the MPV8016I will have a substantial margin of at least 30% of rated power.

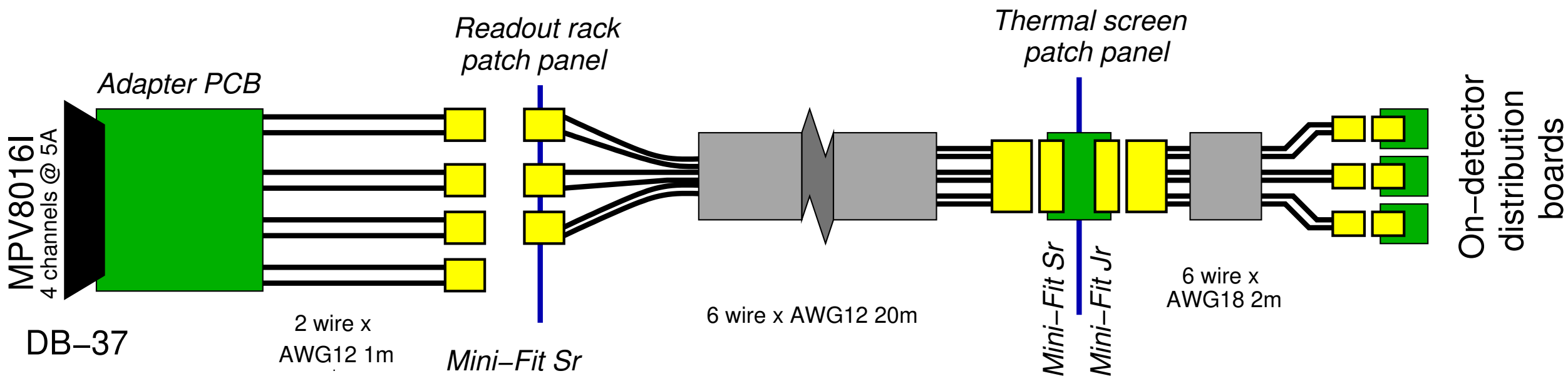

Figure 3.78: Connection structure between ECal low-voltage power supplies and DC/DC converters through patch panels and warm/cold transition.

The details of the cabling for the low-voltage from the power supplies to the per-layer low-voltage distribution boards is shown in Fig. 3.78. A breakout PCB will connect the four-channel connectors of the MPV8016I to 12 AWG individual channel cables. The breakout PCB includes a crowbar circuit to limit the output voltage of the power supplies (which can be up to 16V) to the maximum 12 V at which the DC/DC converters can operate. These cables connect to a patch panel which maps the MPV8016I channels to six-wire 12 AWG trunk cables containing the three channels required for a layer. The 12 AWG cables are chosen to limit the resistance and voltage drop over the $\approx 20$ meter cables between the power supplies and the detector (0.4V for the highest-current channel). The trunk cables connect to feed-through circuit boards, which are part of the warm/cold interface between the external volume and the internal ECal volume.

On the opposite side of the feed-through circuit board, the power wiring shifts to 18 AWG, which is rated to 5 A for tightly-packed configurations. These wires connect to per-layer breakout boards which split the power three ways for distribution to the DC/DC converters as shown in Fig. 3.77. Given the configuration, no fusing is necessary within the detector volume, as all wires are sufficiently rated for the full maximum current of the supply.

The control of the power supplies will be implemented through standard EPICS interfaces, used widely for detector systems at SLAC and elsewhere. The modules in question already have standard EPICS drivers.

#### 3.7.5.4 Sensor Bias Voltage

The silicon sensors of the ECal each have an individual bias supply wire for positive high voltage and a floating return line which returns the current to the bias supply. The return path is through the HGCROC ASIC, so the bias ground and the low-voltage ground are common at the module. The maximum bias voltage expected during LDMX operation is 800V, with all wiring specified for 1kV to provide operating margin. The bias current in the central modules, which have the highest irradiation, is expected to be as large as 2mA, while the outer modules are expected to have currents below 500uA.

The bias voltage will be supplied by ISEG EHS F6 10x modular power supply units, which are 16-channel floating-ground units capable of up to 8mA per channel. Each plane of the ECal will have two bias supplies per layer – one for the central module and one for the outer group of six modules. As a result, the full ECal will be handled by a total of four supply units.

Distribution of the bias voltage within a plane will be provided by a bias/filter PCB such that each plane will require two pairs of bias/return wires which will be carried to the ECal bulkhead feed-throughs. The bias and return lines will be shielded outside the ECal volume.

The control of the bias power supplies will also be implemented through standard EPICS interfaces.

#### 3.7.5.5 Fiber Optics

As discussed in Sec. 3.7.4.2, each plane of the detector will require two bidirectional control/DAQ optical links. There will also be a requirement for three trigger uplinks for a total of seven fibers per plane. These fibers will be organized onto a standard 12-fiber cable with an MTP optical connector on the external ECal bulkhead and LC fiber terminations within the plane.

From the bulkhead, MTP/LC trunk fibers will carry the signals from the detector to a patch panel, where the signals will be reorganized and directed to the appropriate off-detector electronics – either the control/DAQ cards or the trigger APx modules.

#### 3.7.5.6 Detector Interlocks and Safety

The ECal will make use of the common LDMX interlock systems based on Siemens PLC technology to provide interlocks and automatic actions to protect the detector against specific operational situations which could result in detector damage.

1. Cooling – The ECal operation temperature ($-20°$C – $0°$C) does not pose cryogenic risks, but could result in condensation which could damage either the ECal or another subsystem. Within the detector volume, the dry air system will provide a safe operating environment. The cooling system will be interlocked through the detector safety system to require that the dew point in the detector volume be below $-30°$C, as measured by several sensors in the ECal volume.
2. Low-voltage – The ECal requires substantial low-voltage power for operation (3.2 kW), which could pose a risk to the detector. However, the large thermal mass of the absorber tungsten substantially mitigates this risk. Without cooling, the thinnest layers of the absorber will heat at 5 s/°C, while the thickest will heat four times slower. The safety system will interlock the low-voltage power supplies to the operational signals of the cooling system and, as a backup, to RTDs on the detector structure.
3. High voltage – Depletion of the silicon sensors will require the use of moderate high voltage – between 300V and 600V, depending on the level of irradiation that a sensor has experienced. The currents involved may be as high as 2 mA and will be supplied by supplies which are incapable of more than 8 mA of current. The detector safety system will be configured to cut the high voltage to the full detector in the case of a short detected on any single channel.

Each layer of the detector will include two RTD sensors which are configured in a two-wire readout mode. In addition, fourteen sensors will be read out with a precision four-wire configuration: four sensors on the support structure, six on the faces of the insulation, and four on the cooling pipes. Four humidity sensors will also be installed in the volume to track the dew point of the system. These operate in a three-wire configuration.

Table 3.9: Summary of cabling and PLC module channels required for ECal interlock system.

| Item | Count | Wires | Channels |
|---|---|---|---|
| On-plane PT1000 RTDs (two-wire configuration) | 64 | 128 | RTD |
| Support-structure, insulation RTDs (four-wire configuration) | 10 | 40 | RTD |
| Cooling pipe RTDs (four-wire configuration) | 4 | 16 | RTD |
| Humidity sensor (three-wire configuration) | 4 | 12 | Analog |

### 3.7.6 Prototype Support Structure, Cooling Layer, and Cooling Studies

Prototypes of the ECal support structure and cooling planes have been built and studied to provide feedback to the final design. The space requirements of the unit, sensor positions, and sensor operation dictate the placement of the electrical and cooling service lines. Moreover, the ability to extract and reinstall a bilayer containing 3 layers of W and 14 modules adds constraints on the placement of the cables, cooling lines and connectors.

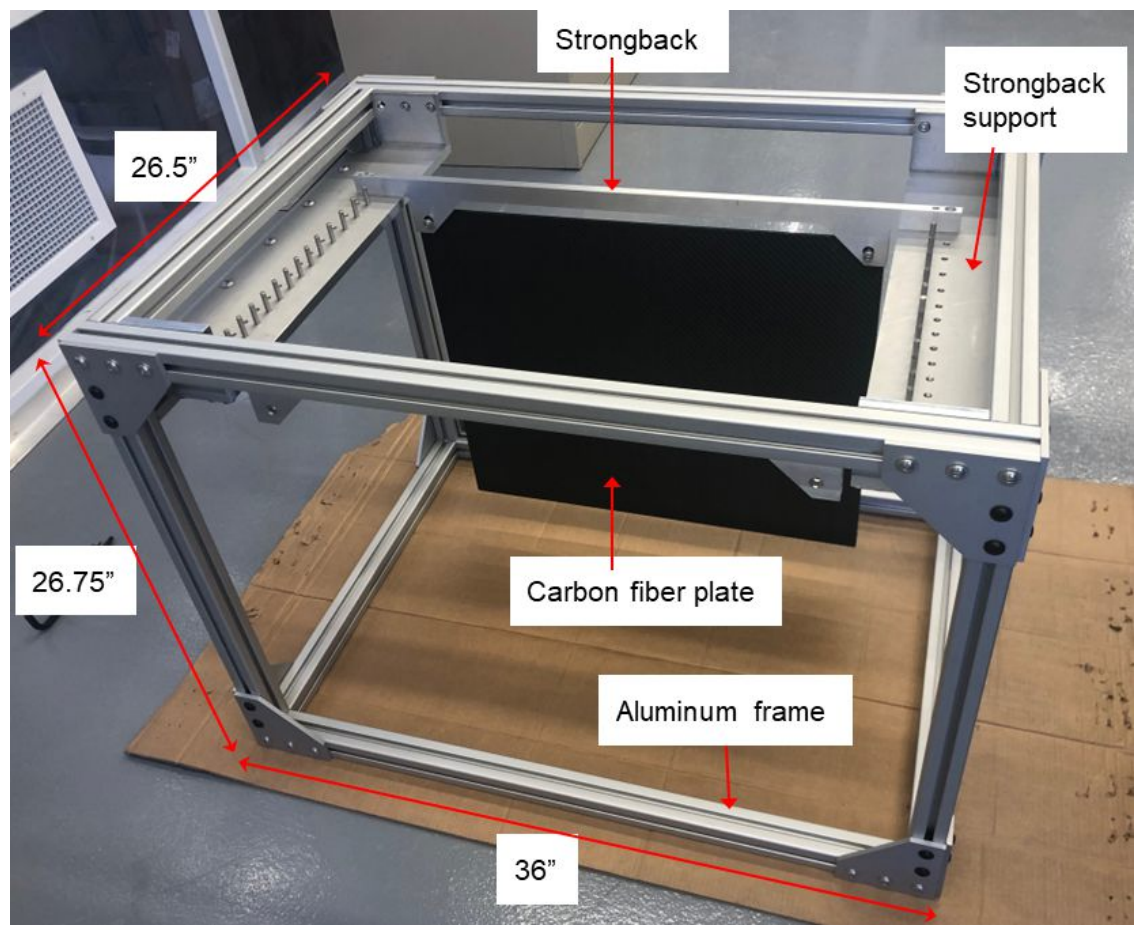

Figure 3.79: Aluminum prototype support box with 24" wide x 20" tall C-fiber plate suspended by a strongback fastened to support ledges. The strongback support ledge has dowel pins and screw holes for 17 suspended bi-layer assemblies.

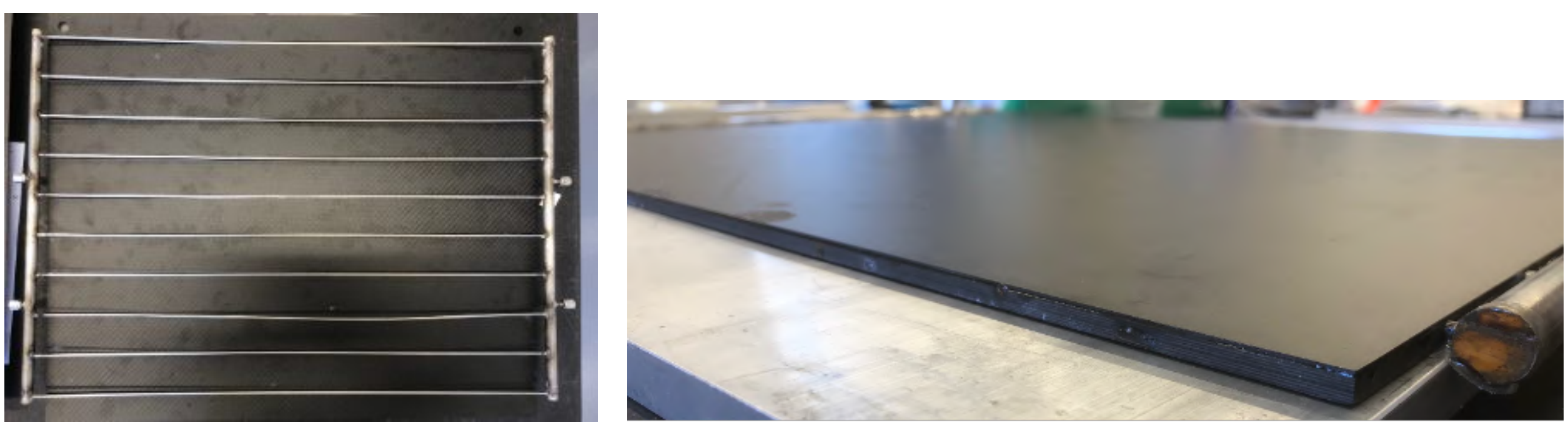

Figure 3.80: Left: Ten cooling lines brazed to manifolds with 2 fittings each to connect to the chiller flow loop. Right: An edge view of the cooling plane after the cooling tubes have been press fit and glued into grooves milled in a thick C-fiber plate and then covered by a thin C-fiber top plate.

The prototype support structure shown in Fig. 3.79 is built from sturdy 1.5" aluminum T-frame, which is less expensive than the machined stainless steel plates that will be used to support the full weight of the system without distortion, as discussed in Sec. 3.7.4. The aluminum frame is adequate for studies of the cooling capabilities of different cooling plane designs and to determine how to distribute services. Support ledges, each with 17 dowel pins spaced according to bilayer locations, which vary with depth according to tungsten layer thicknesses, are mounted near the top of the sides of the frame. Strongbacks span from ledge to ledge, holding the C-fiber cooling plane vertically by means of mounting screws at the top corners of the plate. The C-fiber plate is the core of a multilayer with each side having (i) a W plane and ancillary electronics, followed by (ii) 7 HD modules that nearly cover the W, then (iii) motherboards connecting the modules for readout, and finally (iv) a third W layer on one side to maintain the alternating pattern of absorber and sensing layers when combined with a neighboring bilayer.

The electronics, particularly the HGCROC front-end chips on the modules and components on the mezzanines and motherboards, will generate heat. For the module alone, the electronics components combine to dissipate roughly 8W. It is important to maintain the sensors at low temperature to reduce noise and potentially mitigate radiation impacts for those that see significant fluences near the shower maximum. The heat will be removed using a C-fiber plane in which cooling lines are embedded to carry coolant in a loop with a refrigerated bath. A prototype of the cooling lines and their end manifolds, as well as the completed cooling, are shown in Fig. 3.80. There are 10 horizontal stainless steel cooling lines spaced by 1.9″ with a common flow direction. The tubes are brazed to 0.5″ diameter stainless steel manifolds positioned vertically. The size of the manifolds and cooling tube fittings have been reduced from an earlier prototype design, creating more space for service lines.

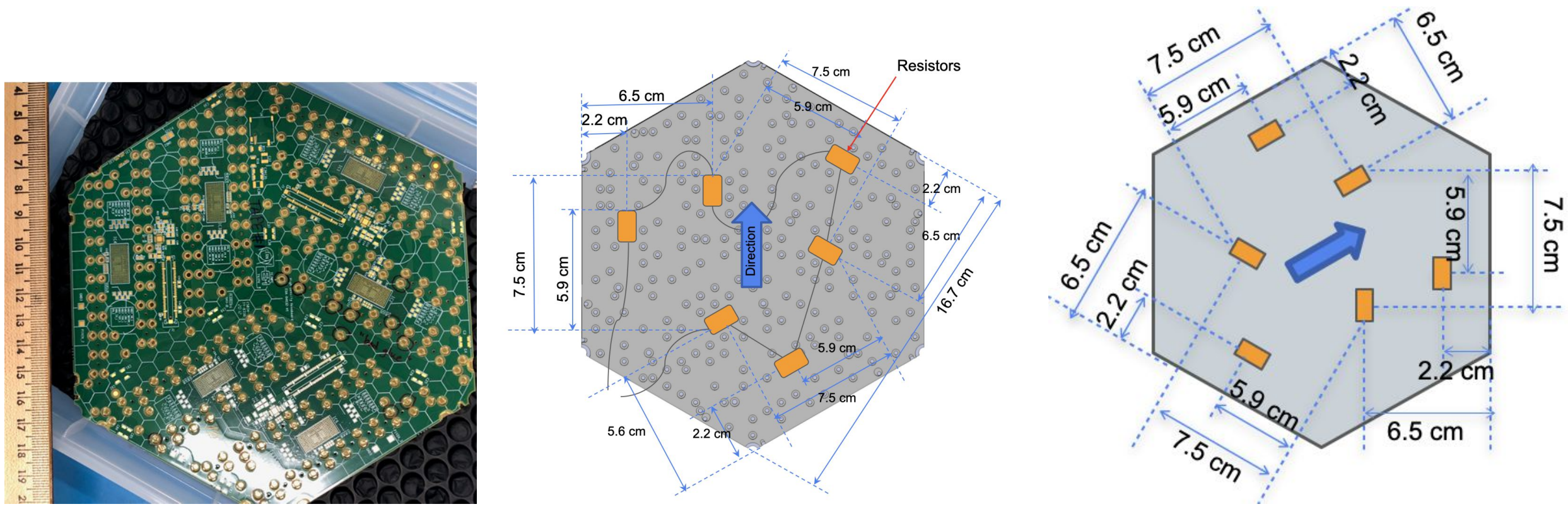

Figure 3.81: Left: photo of a High Density hexaboard. Middle and Right: Locations of 6 HGCROCs in vertical and $60^o$ rotated orientations.

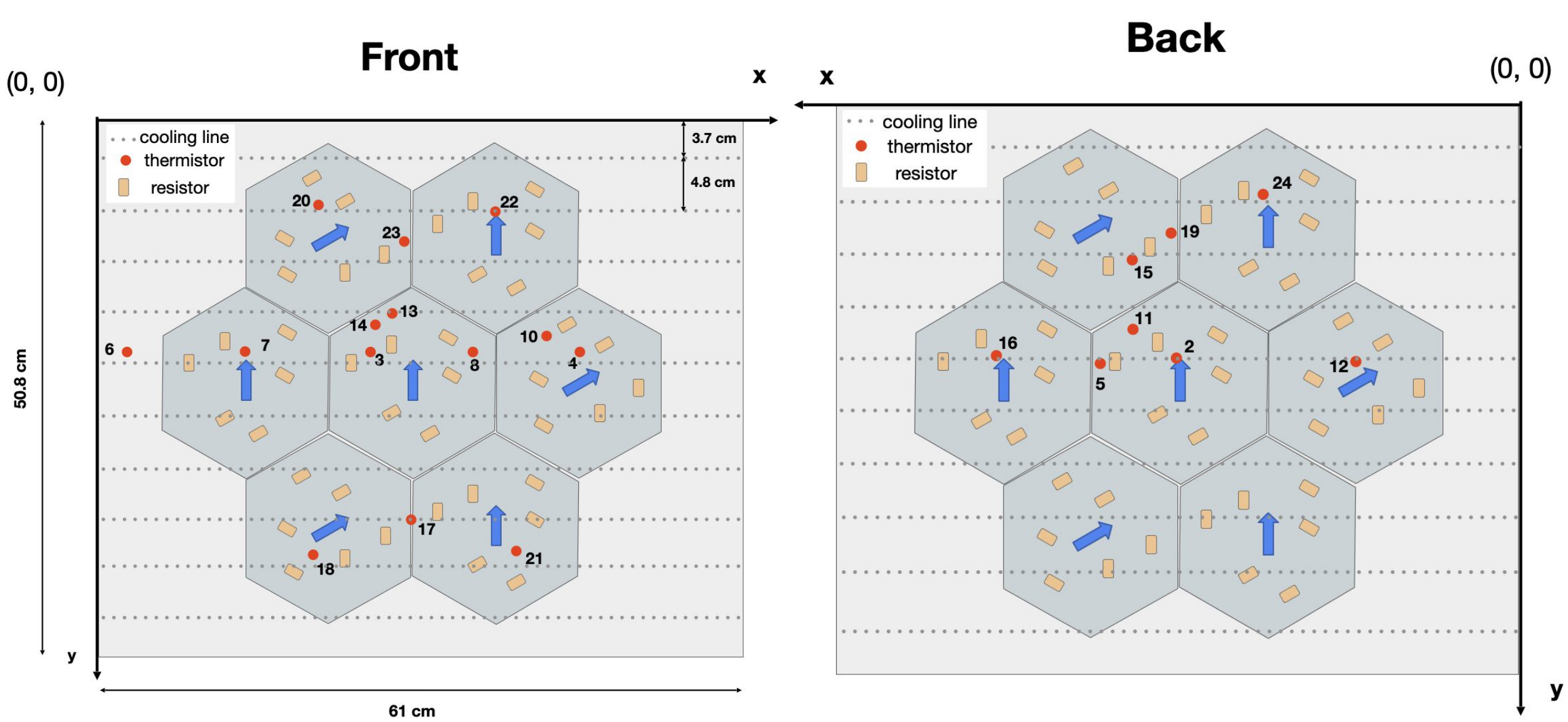

Figure 3.82: Schematic of the positions and orientations of modules used to determine the placement of thermistors on front and back of the cooling plane.

To simulate the heat produced by the electronics, 3 Ω resistors (Vishay: 71-RE60G3R00, mil. spec.) are placed at the location of the 6 HGCROCs of the 7 modules on each side. The HGCROCs each produce about 1W of heat. The resistors in our setup are all connected in series, and we reach 1W on each by applying 10.4 V and 8.1 A to the string of 84. There are additional components on the modules and motherboards that will produce additional heat that could reach the silicon for a total of around 8-10W per module. Two different module orientations are implemented in our test setup as seen in Fig. 3.81. Resistors and thermistors are taped down on the C-fiber surfaces using double-sided thermally conductive tape. On the front side, 11 thermistors are placed directly above the positions of cooling lines and 3 between cooling lines, while on the backside, 6 are located above and 2 between cooling lines. The module orientations and locations of the resistors and thermistors on either side of the bilayer are shown in Fig. 3.82 and Fig. 3.83 shows a photo of the top of the prototype bilayer instrumented for cooling studies. Additional thermistors are placed on the interior and outside of the insulated box as well as on the entering and exiting cooling lines.

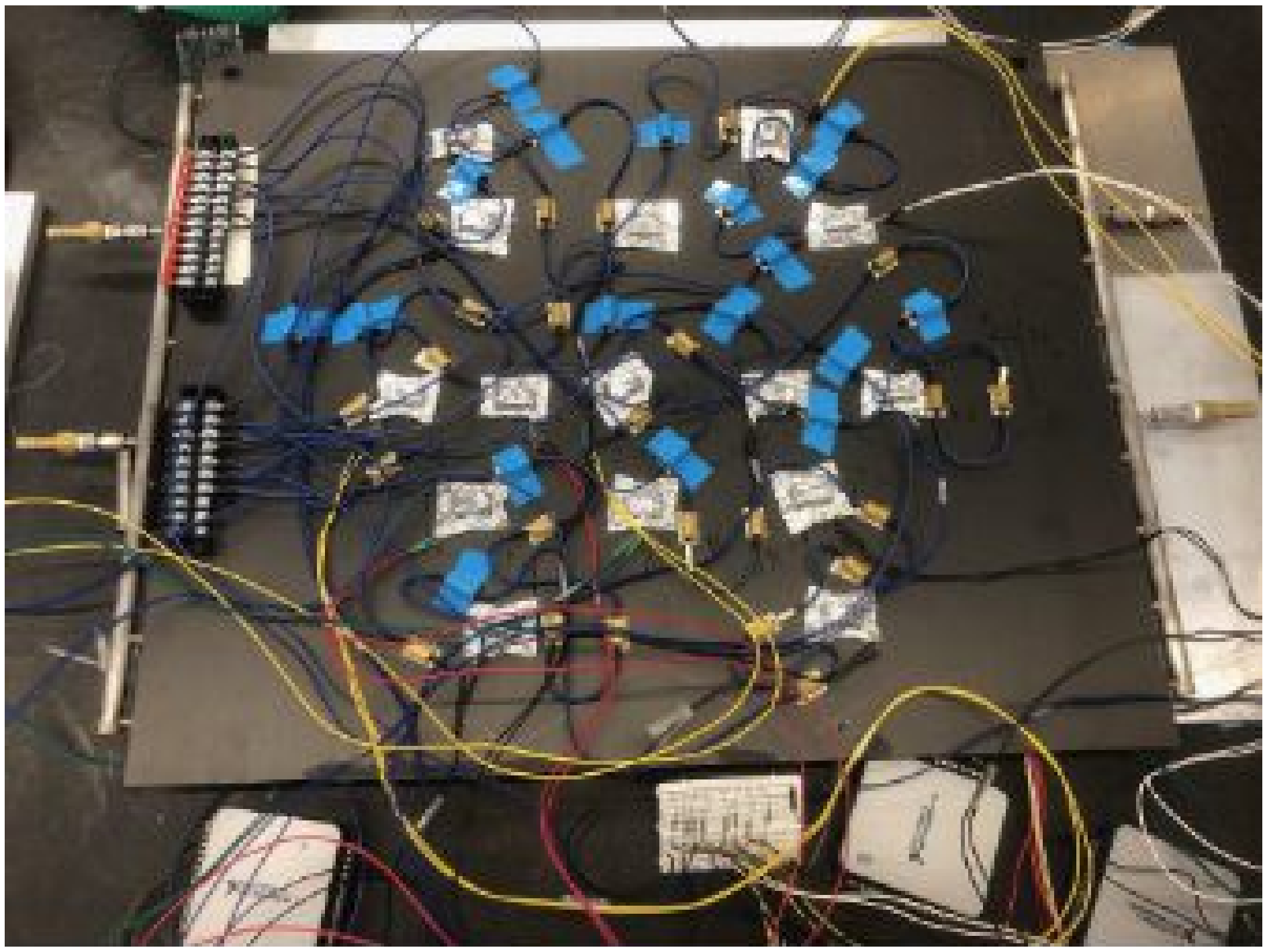

Figure 3.83: Resistors (brass) and thermistors (under Al tape) on front-side of prototype cooling plane.

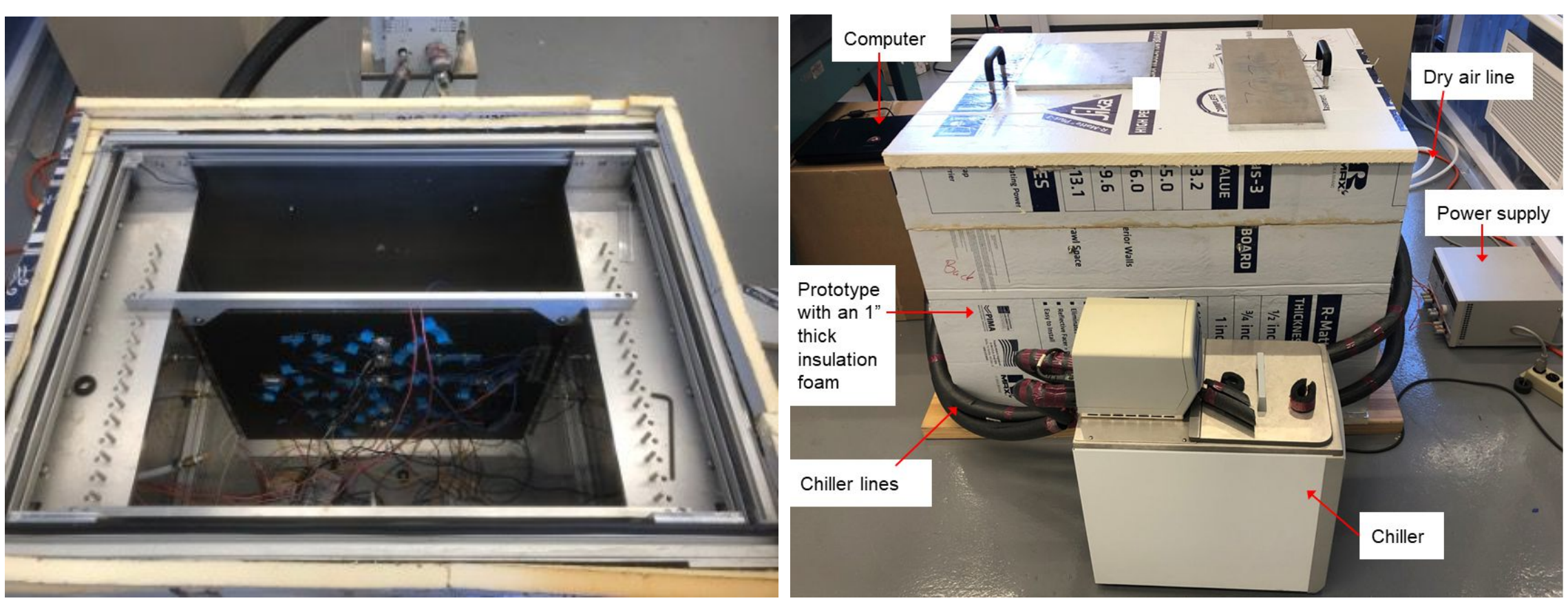

Figure 3.84: Left: Cooling plane for cooling studies inside the prototype support box. Right: Prototype support box fully enclosed by thin aluminum sheets and covered with 1″ thick insulation.

Low-noise, reliable read-out of the modules is achieved with the sensor held relatively uniformly at a temperature as near to the coolant temperature as possible. For CMS, the coolant temperature will be -35°C. For LDMX, which sees far less radiation and for which we expect to use thicker silicon, the coolant temperature can be higher. We expect to operate the chiller at -15 to -20°C, but the system will have the capacity to run at -30°C. The carbon fiber used in this test has a uniform high TC coefficient of >100 $W/m \cdot K$ in the plane but $\approx$ $1W/m \cdot K$ through the plane. As planned for the final system, we sealed the support structure with thin aluminum plates to be airtight. All service lines passing through the aluminum sheets are sealed off with rubber grommets. A dry environment inside the prototype support structure is ensured by a constant flow of dry air. A Neslab RTE10 chiller was used to circulate water-glycol coolant through the cooling lines embedded in the C-fiber cooling plane. The chiller has a heat removal capacity of 400 W at 0°C. For the final detector, the heat load will be several kW, requiring a larger capacity chiller. We plan to use a Julabo PRESTO W56 capable of removing 7kW at -30°C. Once this chiller is available, we will repeat our studies to verify that we can operate with the silicon at -20°C and even -30°C to have some contingency. The prototype support box discussed here is completely encased in 1″ thick foam insulation with R-value of 6.0 as seen at right in Fig. 3.84. The thermistors attached to the cooling plane provide temperature readings at various representative positions that are probed regularly via computer. The power supply seen in the figure provides current to the resistors.

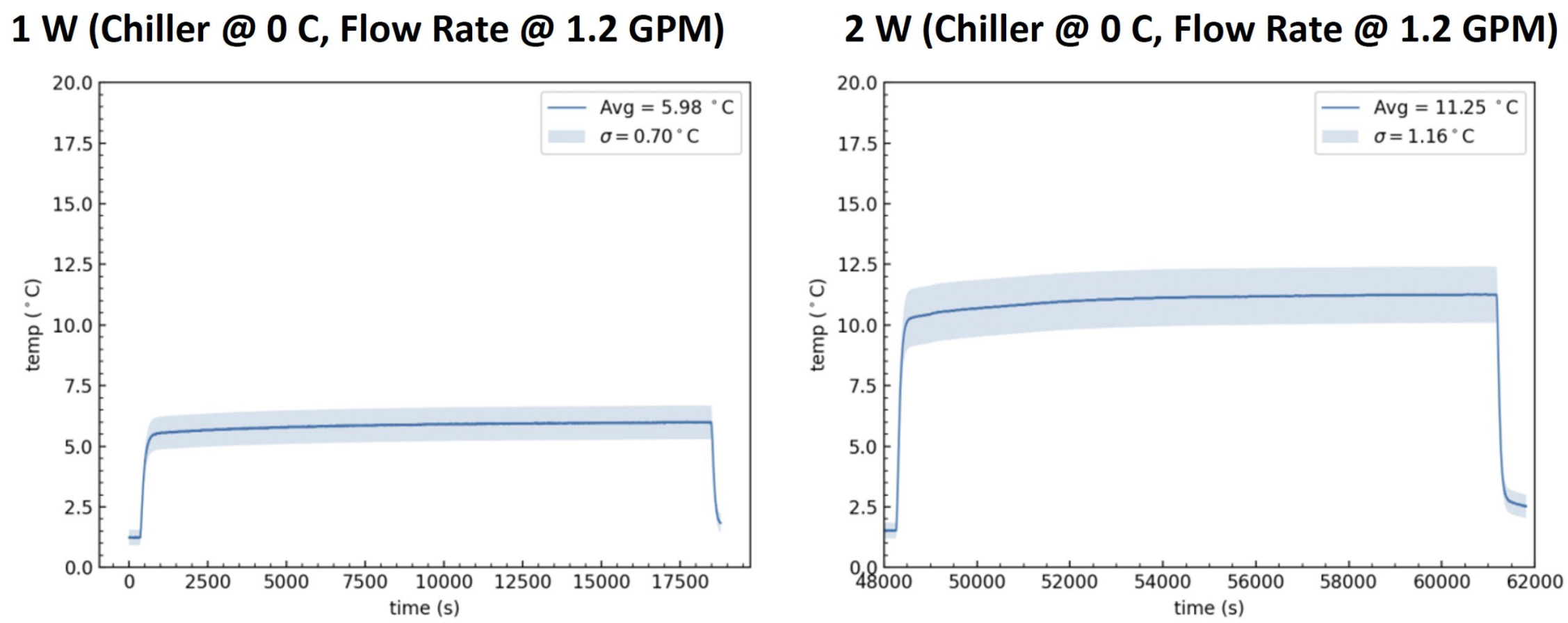

Figure 3.85: Cooling performance of the high TC C-fiber cooling plane for 1W and 2W heating per resistor, (6W and 12W per module) with chiller set to 0°C and a flow rate of 1.2 GPM.

A first prototype of the cooling plane performed less well than desired. Indications were that the heat was not being efficiently transferred to the circulating coolant. This could either be the result of the thermal conductivity of the C-fiber that was used, or the coolant itself was not adequately turbulent to absorb the heat. Non-uniformly coiled thin wires were inserted into the tubes to increase turbulent flow and found to improve the cooling by less than 5%. We then built a second prototype cooling plane as described and shown in figures above. In addition to switching from generic to high TC C-fiber, we also improved the contact of the embedded cooling pipes to the C-fiber and we obtained a much more uniform and tight lamination of the top C-fiber plane. The results for operation at 1W and 2W per resistor (6W and 12W per module) are shown in Fig. 3.85 for chiller operation nominally set to 0°C and a flow rate of 1.2 gallons per minute (GPM). The solid lines represent the average values of all thermistors on module locations, while the shaded band represents the 1 $\sigma$ spread in their values. The temperature levels off asymptotically for the two cases at values of $6.0 \pm 0.7$°C and $11.3 \pm 1.2$°C, respectively, after a few hours of operation in both cases. From this, it appears reasonable to expect the sensors in LDMX to be maintained at around 8°C above the chiller temperature indicating that the a chiller operating temperature of -30°C will be more than adequate to hold the sensors below -20°C.

### 3.7.7 Detector-level Performance

#### 3.7.7.1 Energy resolution

An MC particle gun was used to produce samples to study energy reconstruction for electrons and photons in ECal. True energies ranged from 0.25 GeV to 8 GeV in variable increments with 50,000 particles generated for each simulated energy.

For each electron incident along the +z-axis we obtained the reconstructed energy and divided it by the true energy. Distributions of these ratios are plotted in Fig. 3.86. Most are centered at unity (vertical line), with deviations seen for 0.25 GeV and 0.5 GeV.

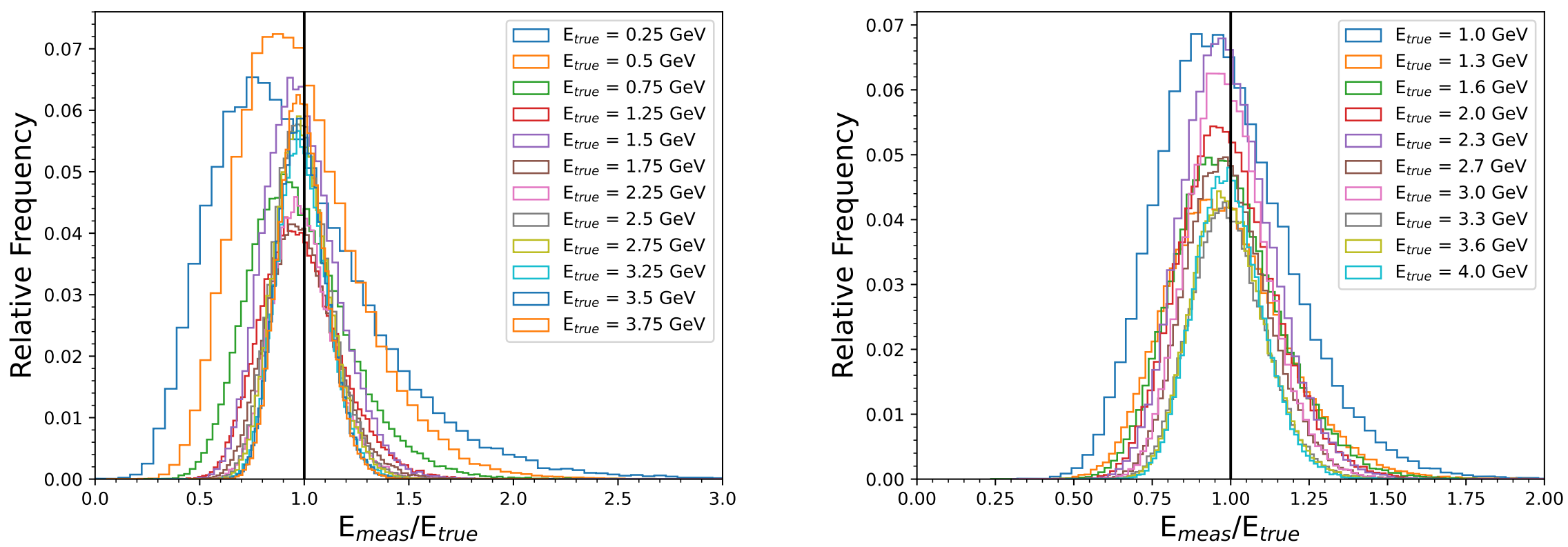

Figure 3.86: Distributions of $E_{\text{reco}}/E_{\text{true}}$ for electrons with $\theta = 0°$.

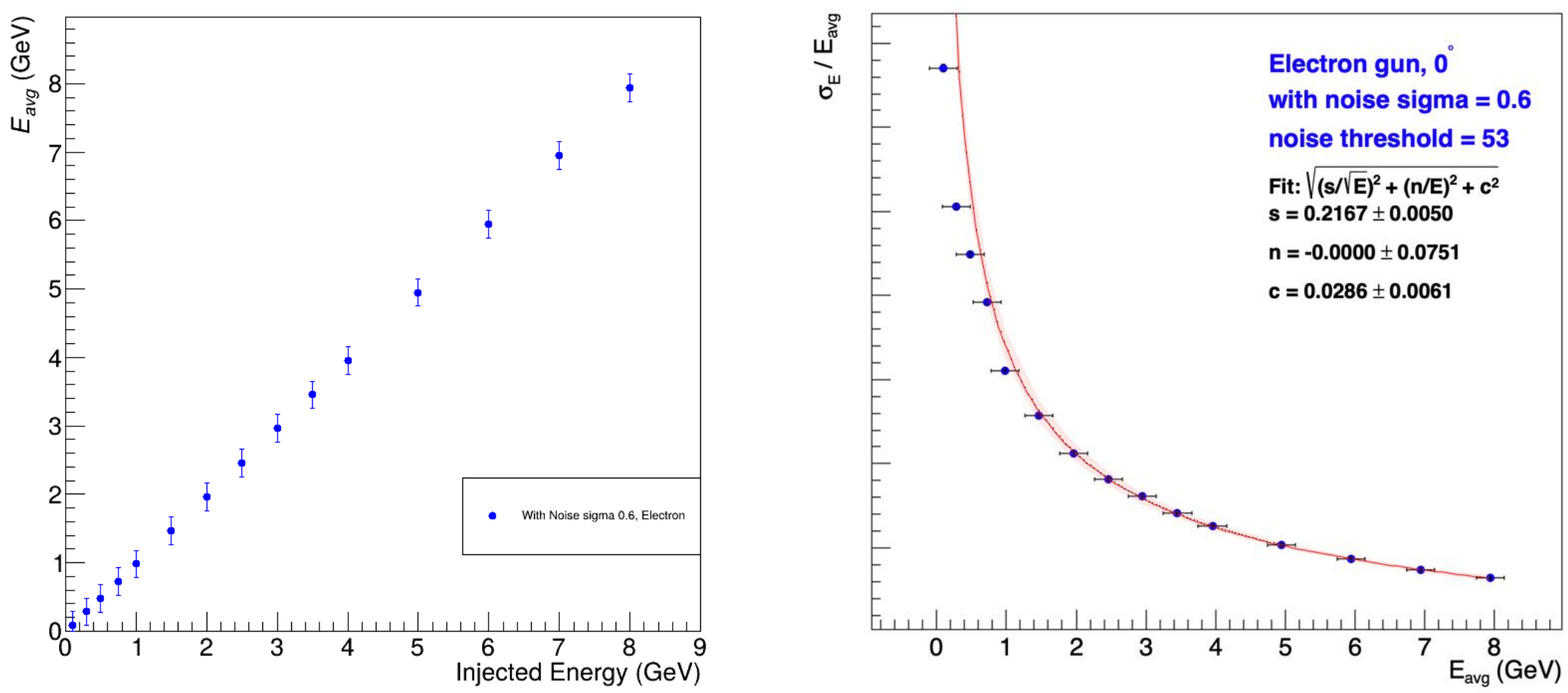

Figure 3.87: Left: Measured vs. true energy for electrons traveling along the +z-axis. Right: Standard deviation of the measured energy over the average of the measured energy with fit as described in the text.

For each distribution, we calculate an average and standard deviation. The uncertainty in the mean is used for the statistical uncertainty in the average reconstructed energy. We plot average measured energy versus $E$ and $\sigma/E$ vs. $E$, where $E$ is the true (injected) energy in Fig. 3.87 (left) and the average measured energy (right). For the latter we include the fit to the energy resolution function involving 3 terms added in quadrature: $\frac{\sigma}{E} = \frac{s}{\sqrt{E}} \oplus c \oplus \frac{n}{E}$, where $s$ parameterizes the stochastic term, $c$ the constant term, and $n$ the noise term. We perform a $\chi^2$ minimization to the data yielding the parameters shown in Table 3.10. The fit agrees with an alternative fit obtained with the CERN Minuit Function Minimization tool. Analogous studies were performed for electrons incident on the ECal at an angle of 15° as well as photons incident at 0° and 15° with fits to the parameter values also reported in Table 3.10. Finally, we also studied the impact of reasonable variations of per-channel noise and a slightly larger gap between successive W-absorber layers and found negligible difference with the results reported in the table.

| Term | Electrons 0° | Electrons 15° | Photons 0° | Photons 15° |
|---|---|---|---|---|
| $s$ | $0.217 \pm 0.005$ | $0.221 \pm 0.005$ | $0.220 \pm 0.005$ | $0.225 \pm 0.005$ |
| $c$ | $0.028 \pm 0.007$ | $0.026 \pm 0.007$ | $0.024 \pm 0.007$ | $0.024 \pm 0.007$ |
| $n$ | $0.00 \pm 0.07$ | $0.00 \pm 0.06$ | $0.00 \pm 0.07$ | $0.00 \pm 0.06$ |

Table 3.10: Fit values for the parameters in the energy resolution function for simulated electrons and photons.

#### 3.7.7.2 Trigger object performance

The ECal is responsible for calculating inputs to the LDMX trigger calculations for a range of trigger algorithms, including the missing energy trigger and cluster counting triggers. The trigger primitives are calculated by the HGCROC and ECON-T ASICs as discussed in Sections 3.7.3.3 and 3.7.3.7, respectively. This process involves summing and compression steps in the HGCROC, followed by calibration and further summing in the ECON-T ASIC. The primary step that could affect trigger performance is the summing and floating-point compression of the ECal energy in the production of the trigger cells by the HGCROC.

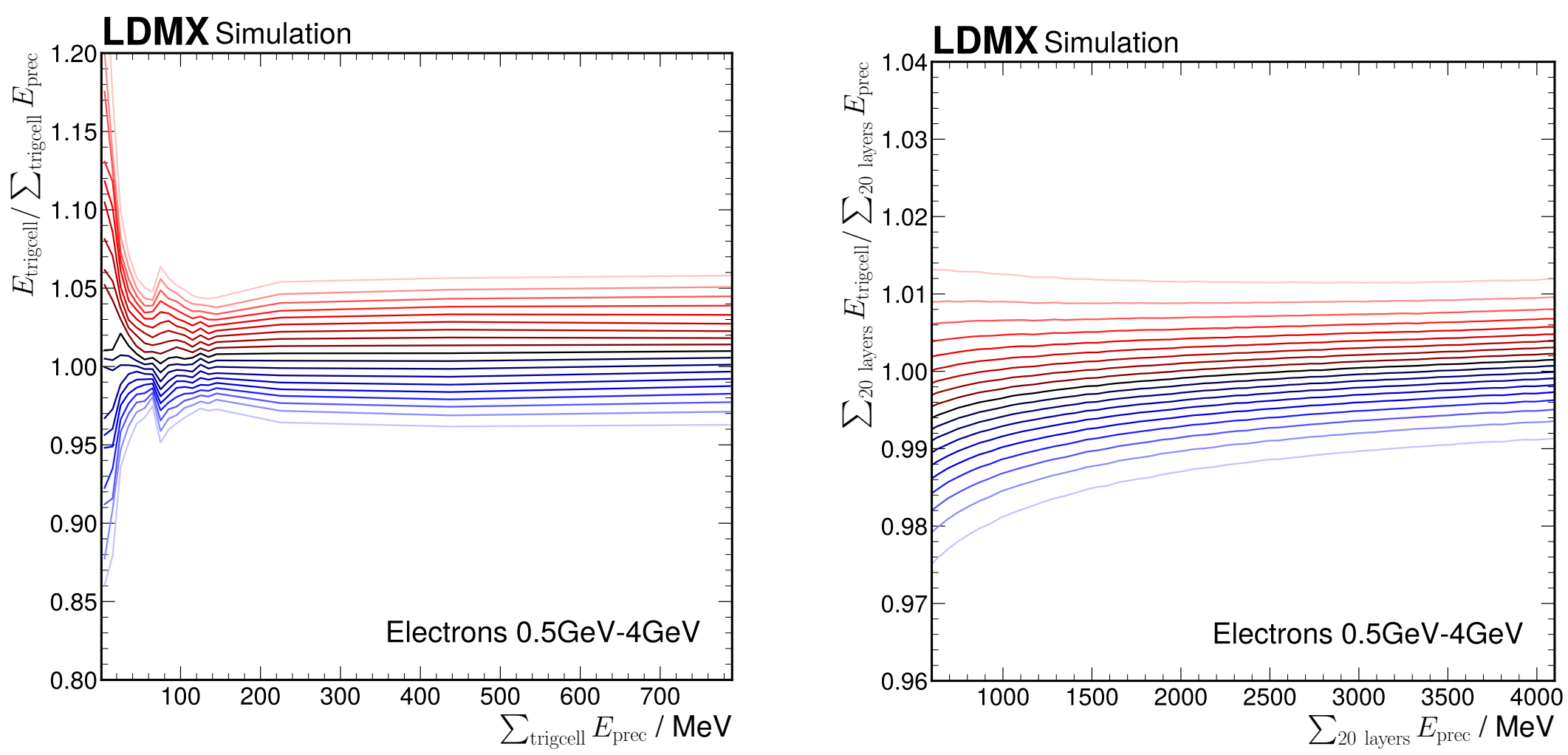

(a) (b)

Figure 3.88: Comparisons between ECal full-resolution (precision) and compressed (trigcell) energy measurements. Figure (a) compares the response of individual trigger cells, while figure (b) shows the comparison for the sum of all trigger cells to the sum of all full-resolution deposits over the first 20 layers, which is a key component of the missing energy trigger. Each curve in each figure represents 5% of the total distribution, with the top red curve representing the 5% most over-estimated cells and the bottom blue curve the 5% most under-estimated cells.

Fig. 3.88 summarizes the performance of the summing and compression algorithm for individual trigger cells and for the overall event energy sum, which is a key component of the primary missing energy trigger. The comparison between the full-resolution (precision) sum over the nine channels that make up a trigger cell shows that the typical resolution of an individual trigger cell sum after compression is 3% for energies above 50 MeV. Below 50 MeV, digitization effects become significant (up to 10%). While there are a substantial number of such cells in a typical shower, the digitization effects tend to cancel out as can be seen in Fig. 3.88(b). The trigger digitization contributes 1% to the trigger shower resolution near the typical missing energy trigger threshold, which is small compared with the overall shower resolution due to calorimeter sampling effects.

### 3.7.8 Quality Control

The quality control of the integrated ECal before delivery to SLAC is described in Sec. 3.7.10. In this section we describe the basic quality control for Modules, Mezzanines and Motherboards, Mechanics, and Tungsten Absorber.

#### 3.7.8.1 Quality Control of Modules

To verify high-quality assembly, each module is surveyed by an Optical Gauging Products (OGP) machine after lamination of the sensor to the baseplate and again after the lamination of the hexaboard to the sensor. The OGP is capable of measuring the relative placements of the baseplate, sensor, and hexaboard to $\pm 10\mu$m. Examples of survey data for 13 modules recently built at UCSB are shown at left in Fig. 3.89.

The rare occurrence of damage to the sensor or hexaboard during assembly, wirebonding or electronic testing can lead to high leakage currents, high noise, sensor heating, and potentially also the inability to operate up to a high bias level that will be required for those modules exposed to the highest levels of radiation at shower maximum. Currents are measured for bias voltages from 0V to 500V in steps of 10V per second for all modules shortly after they are assembled and wirebonded, and again after the wirebonds are encapsulated. Once modules have been conditioned via multiple thermal cycles and electronics tests, the leakage current is measured for bias voltages up to 800V. An example of an I-V plot is shown at right in Fig. 3.89 for a CMS module with $120\mu$m-active-thickness silicon at both $20^o$C and -$40^o$C. Reliable results require a stable measurement environment, including good control of the humidity and dust levels in the clean room and substantial electrostatic discharge precautions for all tasks that involved handling of modules or module components. All of these requirements are now actively maintained and monitored at UCSB.

Module testing also involves operation of the front-end electronics on the hexaboard to validate the full channel-by-channel functionality of the readout chips and to assess whether there are channels that are exceptionally noisy, dead or potentially not wirebonded properly. The noise varies with increasing bias voltage for properly connected channels as a result of the falling load capacitance associated with the increasing depletion depth of the sensor. A photo of a module in a single-module test stand is shown at left in Fig. 3.90. An example of a channel-by-channel noise plot is shown at right. Failed or disconnected channels appear as having abnormally small noise levels and/or unusual pedestal values. A low and stable noise level is important for sensitive measurements and calibration, as well as for stable data volume and detector operation.

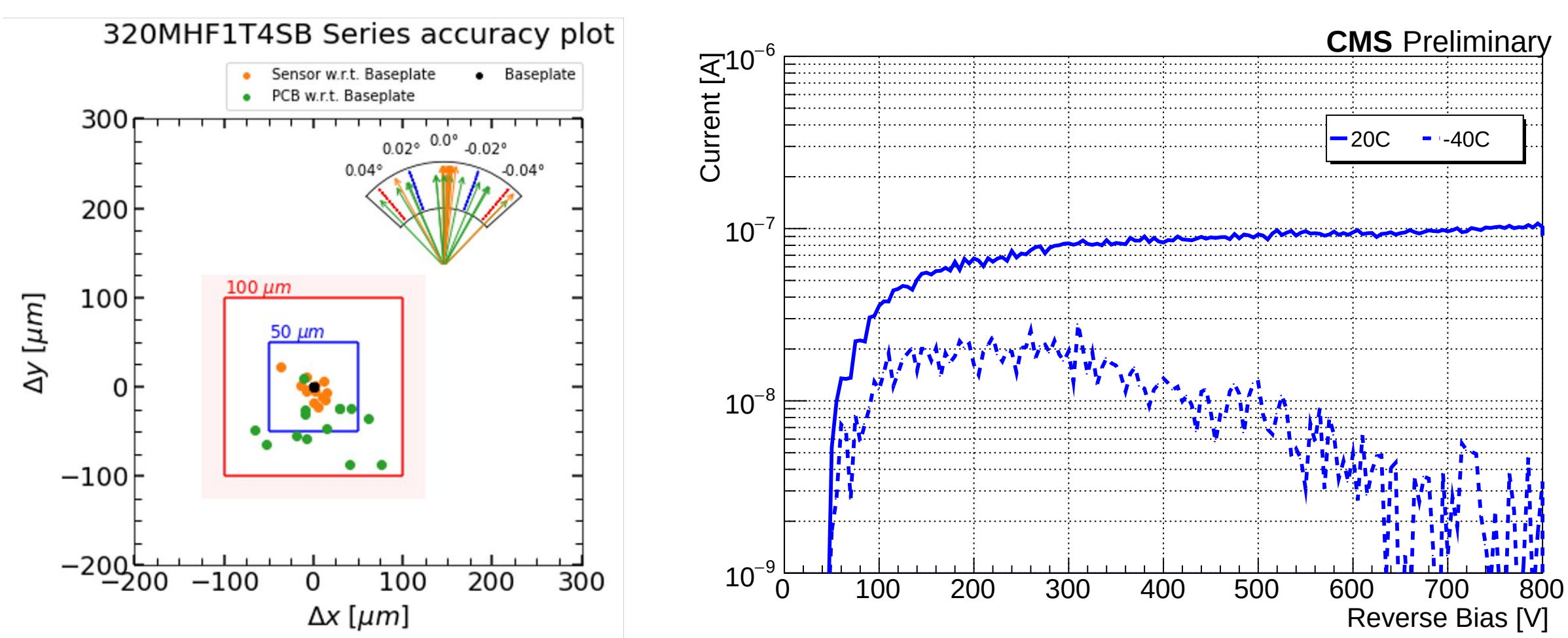

Figure 3.89: Examples of quality control results for silicon modules. (Left) The alignment quality control results from OGP surveys of 13 HD modules recently built at UCSB. (Right) A representative current-versus-voltage plot covering the operating voltage range from 0V to 800V.

Module assembly and testing for LDMX will benefit from the procedures established and exercised for the more than 4200 CMS Modules that will be built at UCSB. Acceptance criteria will likely evolve somewhat with CMS experience. For now, mechanical placement criteria are that linear offsets be less than 50 (100) $\mu$m for the sensor (hexaboard) and angular offsets be less than 0.04 degrees for both the sensor and hexaboard. Modules will undergo 10 thermal cycles, after which they should not have any signs of delamination or broken wirebonds. Finally, 10% of modules will be operated at -40$^o$C in a box with transparent cover while stationed on the OGP to survey for significant distortions in the module's shape and run an automated program that takes photographs to be later reviewed to look for separation of encapsulant from the sensor in the wirebond holes.

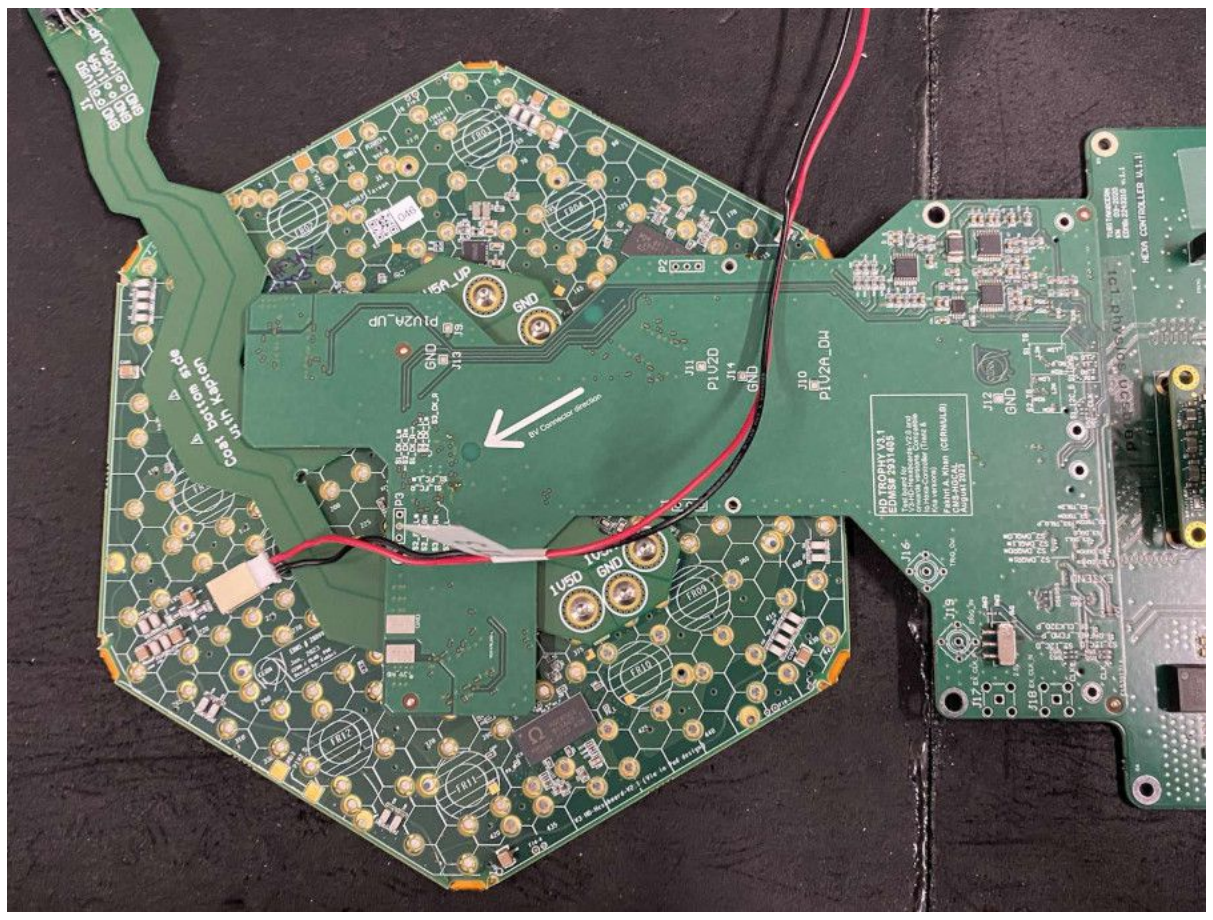

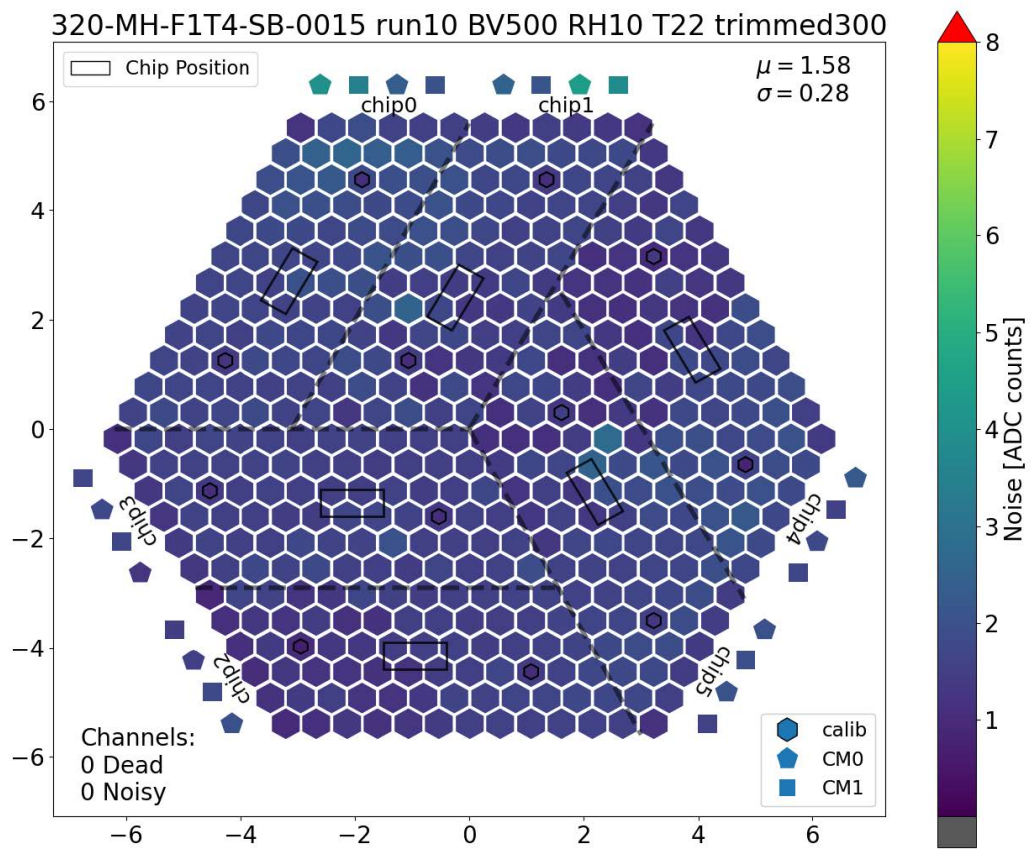

Figure 3.90: (Left) Module in a single-module readout test stand and, (Right) Channel-by-channel noise plot showing uniform, low noise across a CMS HD-full module.

For electrical operation, the general specifications will also likely evolve somewhat, depending on experience with CMS modules. For LDMX at present, we plan to require that a module have no more than 5 bad channels defined as noisy, dead, or unable to be wirebonded to the readout electronics, and current below 100$\mu$A while operated at 20$^o$C and biased to 500V. Once the modules are conditioned by thermal cycles, they will be biased to 800 volts and the current will be required to be below 1 mA while operated at 20$^o$C and biased to 800 volts. The modules at the center of the 7-module flower in each plane will see most of the activity in operation and play a bigger role in identifying bremsstrahlung photons to veto background events. For these core modules, we will select modules with zero bad channels and current below 10 (100)$\mu$A when operated at 20$^o$C and biased to 500V (800V).

#### 3.7.8.2 Quality Control of Mezzanines and Motherboards

The lpGBT mezzanines must be tested for successful assembly of the BGA, mounting of the connectors, and associated aspects. Each mezzanine will be tested using a fixture that allows the 1.2 Gbps signals to be provided from a pseudo-random bitstream source. Similar fixtures have been constructed for CMS HGCAL engine/wagon qualification using a ZCU102 AMD/Xilinx evaluation board. Lower-speed signals will be qualified using loopback tests. The optical signals will be sourced from the ZCU102, allowing for a coherent test.

Once the mezzanines have been quality-controlled, they will be mounted to motherboards. Each module connection site and its associated lpGBT ASICs will be checked using a Zynq-based checkout unit, which has been developed for the CMS HGCAL HD wagon production. After confirming good connectivity for all signals and bit-error-rate for the high-speed signals, the motherboards will be operated with their mated mezzanines for a burn-in period of two weeks, and then checked again before integration into the ECal array.

#### 3.7.8.3 Quality Control of Mechanics and Absorber

The stainless steel pieces used to fabricate the support box for the ECal will be surveyed for flatness and uniformity before and after machining. Similarly, all of the machining of the components used to support the ECal bilayers such as the ledges and strongbacks will be carefully checked against the drawings and specified tolerances.

The tungsten sheets, which are the primary absorber layers, will be weighed and their thickness and flatness will be measured on the OGP system at UCSB to ensure adequate control of the material budget of the calorimeter and to ensure mechanical assembly will be possible as designed. The OGP is capable of thickness measurements with a resolution of 50 $\mu$m, which is perfectly adequate given the overall constant term requirement for the calorimeter.

After each cooling plane is assembled with its cooling pipes, a flow-check, pressure-check, and leak-check will be performed to confirm that the flow impedance of each plane matches expectations and that there are no leaks or weak points in the pipework.

### 3.7.9 Module Noise and Response of CMS Prototype ECal to Positrons

In the plot on the right in Fig. 3.90, the average noise in the 432 channels of this CMS HD full module is noted as 1.58 ADC counts with an rms of 0.28 ADC counts. This is typical and representative of what we can expect for modules before significant radiation damage when operated at 20$^o$C. It corresponds to an equivalent input of approximately 2000 electrons. A minimum ionizing particle (MIP) deposits 3.9 MeV/cm on average as it passes through silicon. At normal incidence, a MIP will thus deposit 46.8 keV in a 120$\mu$m sensor while electron-hole pair formation in silicon has a threshold of 3.6 eV, so that up to 13,000 e-h pairs may be formed. Typically, the number formed and collected is lower, and CMS measures a MIP signal-to-noise ratio of 4.8 for this case as seen in Fig. 3.91, which is not far from the value of 5.0 obtained from a more sophisticated calculation [99].

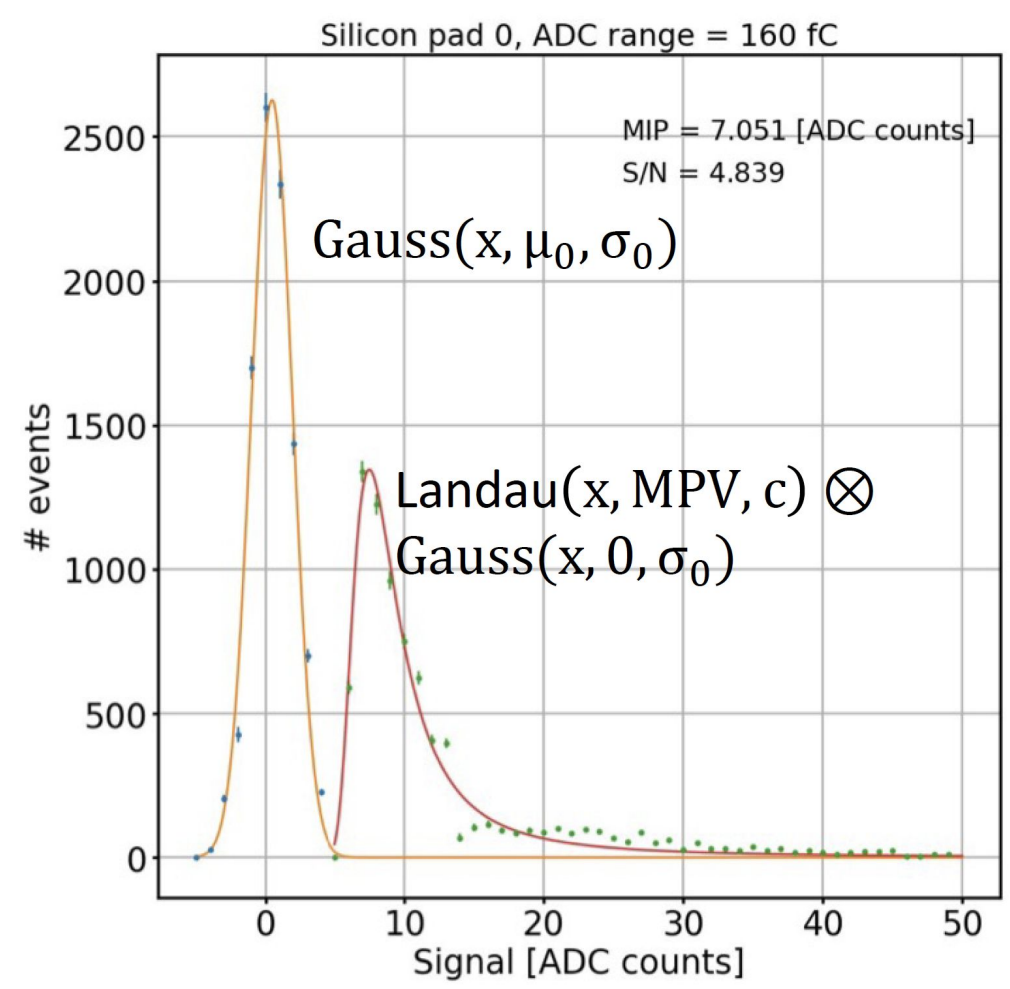

Figure 3.91: Response of a high density, 120$\mu$m-active-thickness sensor module to a muon test beam in 2023. Figure courtesy of CMS [99].

For the much lower radiation environment of LDMX, we plan to take advantage of thicker sensors for which the signal increases proportionately. The noise would be expected to have a dependence on temperature, sensor thickness and radiation dose. The former is anticipated as a result of the temperature dependence of dark currents characterized by a factor of 2 increase for every 7$^o$C rise in temperature near room temperature to a doubling every 5$^o$C in the vicinity of -30$^o$C. For the second, pad capacitance is inversely proportional to the active-sensor thickness, which determines a part of the noise associated with the front-end electronics. However, given the design and operation of the HGCROC on the hexaboard, neither the lower capacitance nor the reduction in temperature has much of an impact on per-channel noise. For example, CMS measures noise of 1900 electrons for LD full modules with either 200$\mu$m or 300$\mu$m sensors [100] while measurements we have made at UCSB indicate no change in per-channel noise for operation at -40$^o$C versus +20$^o$C. A third source of increased noise is radiation damage but here studies indicate that there is no significant impact until well above the maximum radiation doses anticipated for LDMX. Indeed,

HGCROC single-channel noise is observed to increase by only 20% for radiation doses of 345 MRad [101], which is more than 2 orders of magnitude beyond the worst that we expect for LDMX, as discussed in Appendix A.

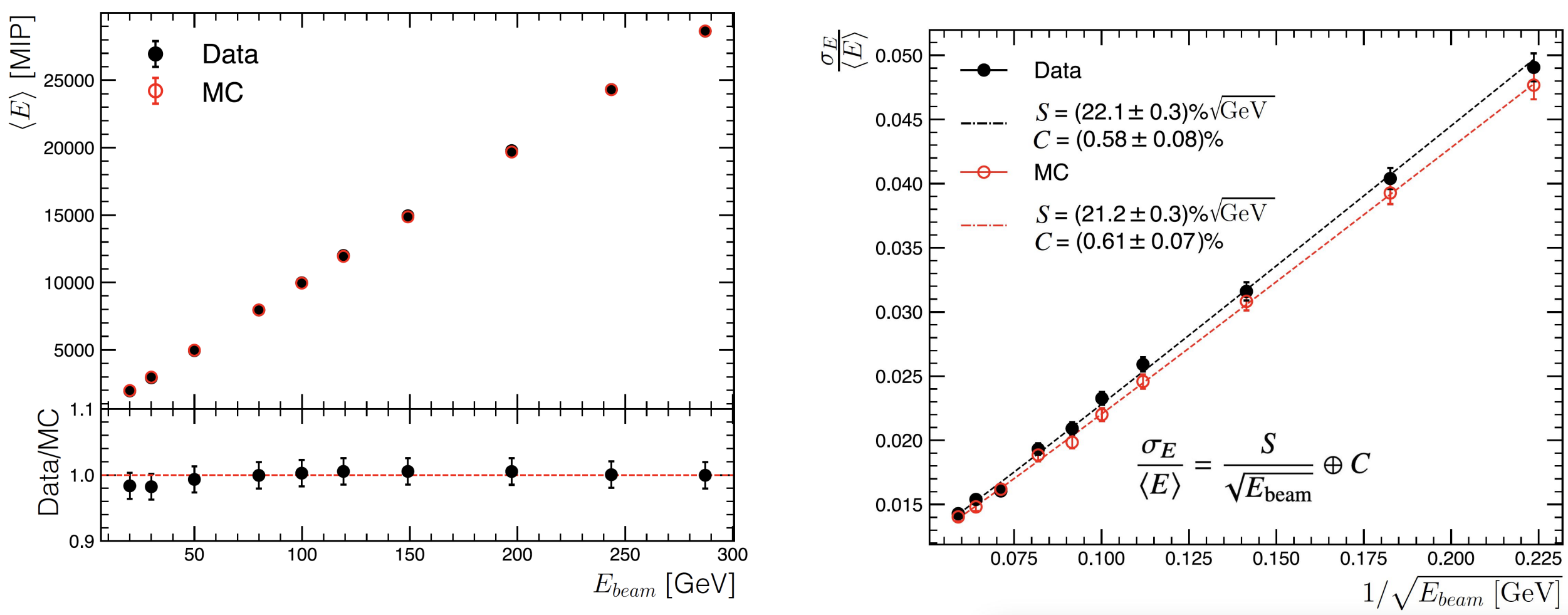

Figure 3.92: Comparisons of measured and MC simulated energy reconstruction for a prototype of the CMS HGC electromagnetic section in response to positron beams ranging in energy from 20 to 300 GeV: The reconstructed versus beam energy (left) and the linearity of the relative resolution $\sigma_E / < E >$ versus $1/\sqrt{E_{beam}}$ (right). Figures courtesy of CMS [102].

Use of thicker silicon will yield a proportionate increase in signal while noise would remain fairly constant in the course of LDMX operation. Increasing the sensor thickness to 400$\mu$m will yield a signal-to-noise ratio of 20, providing a good separation of MIP signals from noise. As an example, taking for the noise $\sigma = 2$ (i.e. 1.7 + 0.3) ADC counts, one could set a threshold as high as $\sim$ 12 ADC counts, or 6$\sigma$, with no loss in MIP sensitivity.

Having now covered per-channel noise and response to MIPs, and noting the good match with expectation, one naturally wonders how well the ECal can be expected to match the energy resolution for electrons and photons as predicted with simulated data as presented in Sec. 3.7.7.1. To that end, we note that the CMS HGC collaboration carried out a beam test with 20-300 GeV positrons in 2018 in which they arrayed sensor and absorber layers as planned for the CMS HGC ECal configuration of 28 sensing layers interleaved with 27 radiation lengths of absorber [102]. The results are shown in Fig. 3.92. The left plot in the figure demonstrates good linearity. The response of the prototype versus energy is consistent between MC and data and, in particular, the ratio of data to MC is consistent, within uncertainties, with unity across the full range of energies. The central value of the ratio is however low for the 3 lowest beam energies. The resolution $\sigma_E / < E >$ versus $1/\sqrt{E_{beam}}$ displayed in the plot at right in the figure indicates a systematic, albeit small, discrepancy between data and MC that grows with $1/\sqrt{E_{beam}}$ - i.e. as the energy is decreased, with the resolution in data being slightly worse than the MC estimate. This could be due to any number of things, including the typically limited level of detail in the MC regarding the items contained in the actual device. The discrepancy appears to be negligible for the 300 GeV beam (leftmost point) and increases linearly to $0.01/\sqrt{E_{beam}}$ for the rightmost point for a 20 GeV beam. Extrapolating to 8 GeV, as is relevant to LDMX, we would thus expect a discrepancy of roughly $0.015/\sqrt{E_{beam}}$. The linearity of the ECal's response to electrons and the electron energy resolution will be measured in-situ in the experiment. Based upon the CMS study presented here, we can expect the results to match MC simulation well.

### 3.7.10 Integration of the ECal

The integration of the ECal will be carried out at UCSB in advance of delivery to SLAC. The sequence of steps expected for the integration is described below.

As a first stage, the ECal support frame will be assembled and surveyed. The cooling manifolds and wire support harnesses will be connected to the support frame to prepare the frame for the insertion of layers. Before installation of any components, the sensors for the detector control system which connect to the cooling manifolds and the humidity-monitoring sensors will be installed with wiring to the detector bulkhead and confirmed functional.

The cooling planes, which support a layer of modules on each side, will be produced and quality-controlled as discussed elsewhere in this chapter. The active and service components of each detector will be added to the planes, one side at a time. After one side is completed, the plane will be flipped over using tooling that avoids putting any significant load on the modules or service components.

The integration will begin with the mounting of the high-voltage and low-voltage distribution boards and the DC/DC converter sets. The silicon modules will then be mounted and the bias voltage wiring will be carried out one module at a time, with an IV curve taken after each module is mounted to confirm that there is no observed damage to the detector.

After the mounting of the modules is confirmed, the motherboard, lpGBT mezzanines, optical components, and low-voltage supply cabling will be completed. A thorough checkout of the readout of the modules through the motherboard will be carried out using a testing DAQ, including measurements of noise as well as confirmation of control and trigger paths and signal mapping. The testing will be carried out using air cooling of the modules and base plate. The full process will be repeated for the second side of the bilayer.

After the full bilayer is completed and tested, it will be rotated to the vertical configuration and lowered into the detector support frame. The plane will be connected to the cooling manifolds which will have all other unused ports stoppered allowing for leak checking to be carried out. Once verified as leak-free, the low-voltage, bias-voltage, monitoring cables, and optical fibers will be dressed into the layer up to the bulkhead connectors at the warm/cold interface of the ECal. The detector readout for the bilayer will then be checked through the final cables and fibers. This process will be repeated for all the bilayers.

Once the full set of layers is installed, a full-scale flow check of the calorimeter cooling will be carried out by enabling individual planes and tracking the temperature behavior. The thermal insulation of the assembled detector will then be applied and sealed. The full detector will be subjected to a cooling check to confirm that the thermal leakage is as-expected. At this point, the full detector will be placed in a crate and transported to SLAC for installation at ESA.

#### 3.7.10.1 ECal Configuration as Delivered to ESA

The ECal will be delivered to ESA as a fully-integrated detector from the warm/cold bulkheads through the full set of detector layers. The trunk cables, fibers, and cooling pipes to the magnet cart will be brought to ESA separately.

### 3.7.11 Calibration

The ECal is a sampling calorimeter with silicon active material and tungsten as its primary absorber material. Other materials are present in the calorimeter volume, including readout circuit boards, the cooling pipework, and supply wires for low-voltage and bias voltage.

The calibration strategy for the detector is based on cell-to-cell inter-calibration using single minimum-ionizing particles and absolute energy scale calibration using beam electrons that have experienced minimal energy loss in the tracking system. The overall calibration requirements on the ECal (5% intercalibration and 1% scale) are dominated by the impact of miscalibration on the trigger, where calibration effects could smear the electromagnetic mismeasurement tail into the trigger region, requiring a tightened trigger threshold.

#### 3.7.11.1 Pre-calibration

An initial calibration of the detector modules will be established during construction using cosmic-ray muons. The standard production testing process for silicon modules includes a test sequence which is aimed to provide an initial calibration of each cell at the 5% level, which will require cosmic ray operation for a period of approximately three days. The uniformity of charge collection across the silicon sensor is quite good, so that the primary channel-to-channel variation can arise from small differences in the shaping and other response of the front-end ASIC. The pre-calibration will provide an initial calibration basis from which the first trigger configuration can be prepared and which will provide a foundation for subsequent in-situ channel calibration.

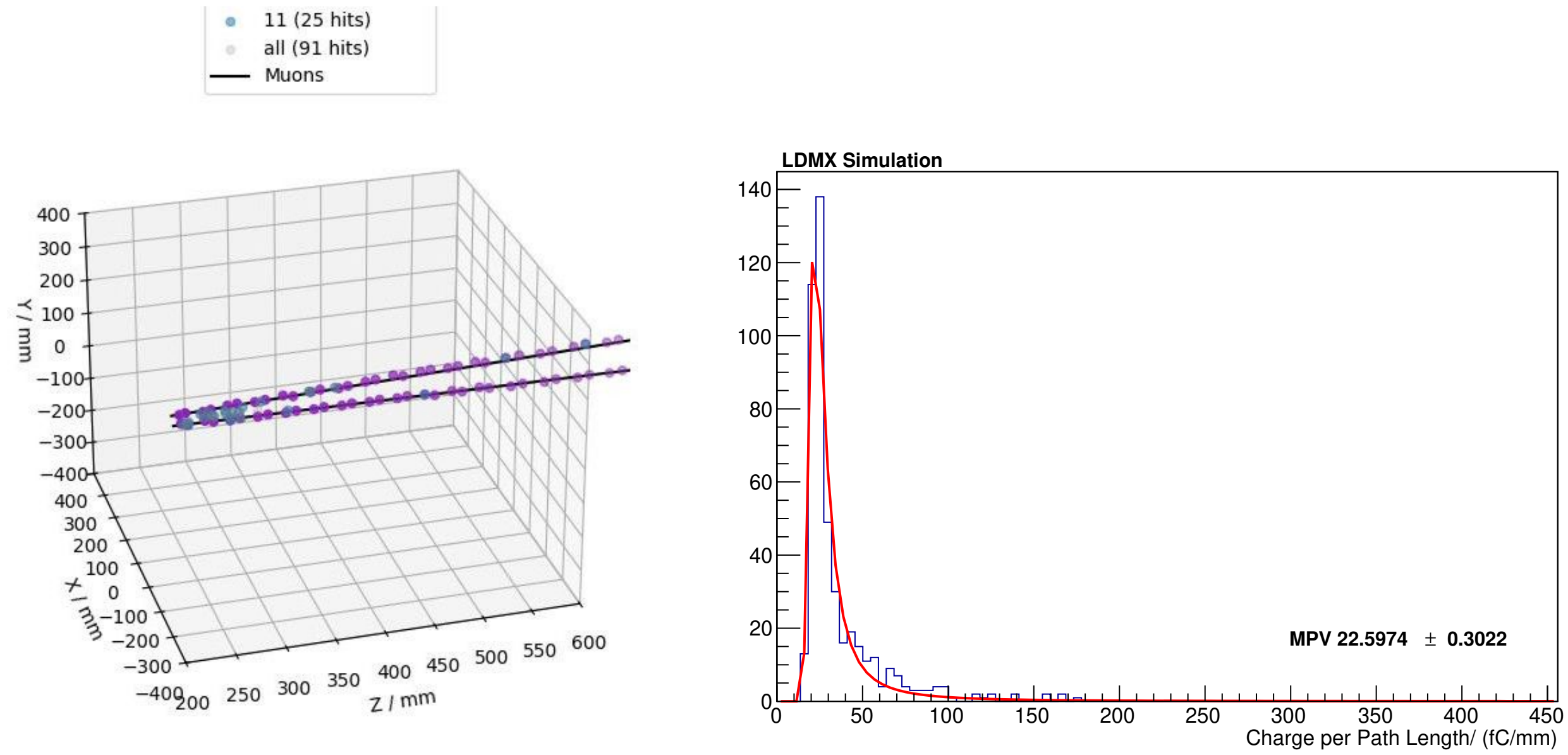

Figure 3.93: (Left) Event display of a candidate calibration dimuon event in LDMX simulation, showing the muon tracks in the calorimeter as well as electromagnetic shower debris which must be ignored by the algorithm. (Right) Calibration fit result for a representative channel.

#### 3.7.11.2 Internal Calibration

The detector will collect regular internal calibration data using functionalities built into the HGCROC and data collection chain. Dedicated noise/pedestal runs will be taken daily, and a low rate of events without zero suppression will be taken during regular operations. The HGCROC includes capabilities for charge-injection runs as well, which allow for calibration of the charge scale and for intercalibration of the ADC and TOT modes of chip operation.

#### 3.7.11.3 In-situ Inter-Calibration

The in-situ cell-to-cell calibration of ECal will be carried out using muon-conversion events. These are primarily events where the incoming beam electron has experienced a QED bremsstrahlung interaction, with the resulting photon subsequently converting into a dimuon pair through an interaction with a nucleus [103]. The muons in such events are excellent calibration tools, as the muons in LDMX have a typical $\frac{dE}{dx}$ within $\pm 2.5\%$ given the narrow range of momenta produced in LDMX.

The calibration selection identifies isolated tracks in the calorimeter, such as in the example event shown on the left of Fig. 3.93. The algorithm then removes from consideration any layers where there is substantial shower energy nearby the track, avoiding any impact of recoil electron shower remnants. The resulting energy measurements are then corrected for track angle and fit on a channel-by-channel basis to extract inter-calibration values. An example of such a fit is shown on the right of Fig. 3.93.

During regular beam operation, dimuon events that are appropriate for calibration are selected efficiently by the standard LDMX missing energy trigger. Calibration-candidate dimuon events typically have more than 70% of the beam energy transferred to the muons, resulting in high trigger efficiency. The rate of such events for the nominal $\mu = 1$ LDMX beam is approximately 7500 per day. The positional distribution of dimuon events produced in nominal beam running is shown in Fig. 3.94. As can be seen, the dimuon production during normal operation is sufficient to provide a daily calibration sample for the most-irradiated portions of the ECal where charge-collection effects may change most rapidly. However, the illumination of the cells away from the beam axis is substantially smaller, with over 200 beam-days required to achieve O(50) events/cell in the detector periphery.

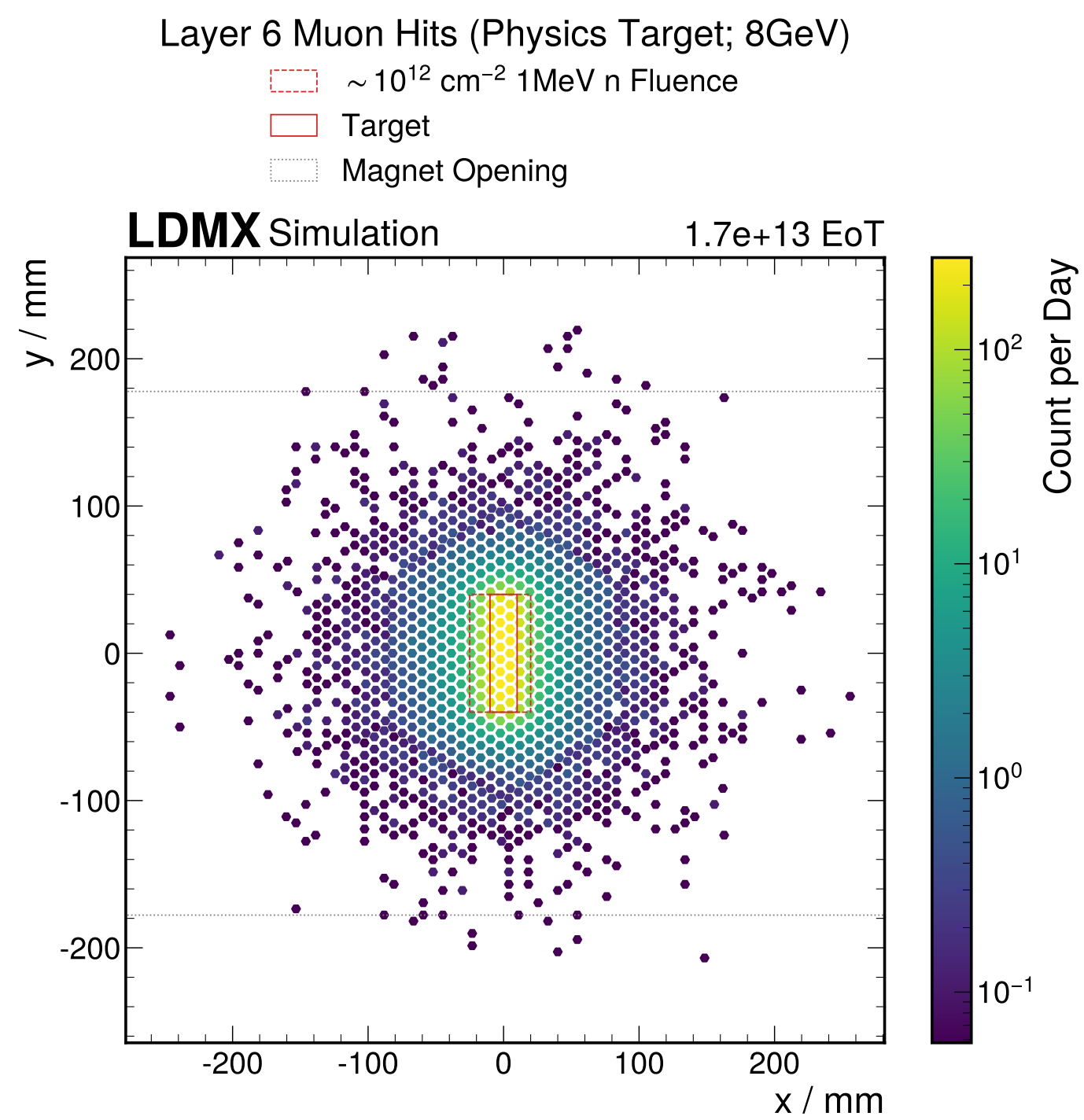

Figure 3.94: Distribution of calibration-candidate muons in the normal beam-operation configuration, normalized to events-per-day as determined by the two-dimensional distribution of the muon-induced hits in layer 6, near the shower maximum for full-energy beam showers. The target location and the location of the highest-irradiation cells are indicated. Due to the smaller size of the cells near the boundary of the sensor, the number of events per cell near the module boundary is reduced.

To provide sufficient illumination for full-detector inter-calibration, LDMX plans to take data with a beam-stopping calibration target at the upstream end of the magnet. The calibration target is expected to be a tungsten block sufficiently thick (15 cm or $\approx 40X_{\circ}$) to fully absorb electromagnetic showers, allowing the delivery of substantially higher-charge bunches (e.g. $\mu = 1000$) without filling the calorimeter with electron-induced shower energy. In addition, the thick target increases the dimuon yield per incoming electron by a factor of $\approx 200$. The calibration target location upstream of the magnet would allow a much larger distance for the muons to spread out to illuminate the calorimeter, compared with the physics target. The distribution of muon-induced hits in the calorimeter with the calibration target configuration is shown in Fig. 3.95 for a total of $1.4 \times 10^{12}$ electrons-on-target, with an effective observed rate of $10^{-6}$ calibration muons per incoming electron. Given the 25 kHz design specification of the LDMX DAQ system (3.9.2), the data sample visualized in the figure could be acquired in a day of operation, even taking into account substantial setup and tuning time. Such a calibration campaign could be carried out twice per year or even more often if necessary. The resulting data samples would be of use for HCal calibration and possibly for tracker alignment as well.

#### 3.7.11.4 Energy scale calibration

The design of the LDMX ECal maximizes the uniformity of the calorimeter through the use of sheet absorbers rather than segmented per-module absorbers as used in the CMS HGC. As a result, the primary energy scale calibration is expected to be highly uniform across the calorimeter as the absorber mass variation is expected to be small. The primary energy scale calibration will be performed using isolated electrons. The initial energy scale will be seeded using the measured absorber masses as well as the simulated impact of the modules, support mechanics, cooling infrastructure, and readout electronics present in the beam path. The simulated materials will be compared with the measured masses for the aspects in question.

Full-beam-energy electrons will provide the primary calibration, providing corrections to the initial mass-based sampling factors. For layers deeper in the calorimeter, where typical electromagnetic showers do not effectively illuminate, the corrections to the sampling factor will be carried from the well-illuminated layers. A sample of electron recoil events with well-reconstructed hard bremsstrahlung photons will provide an additional calibration source to help disentangle ambiguities arising from a monoenergetic calibration

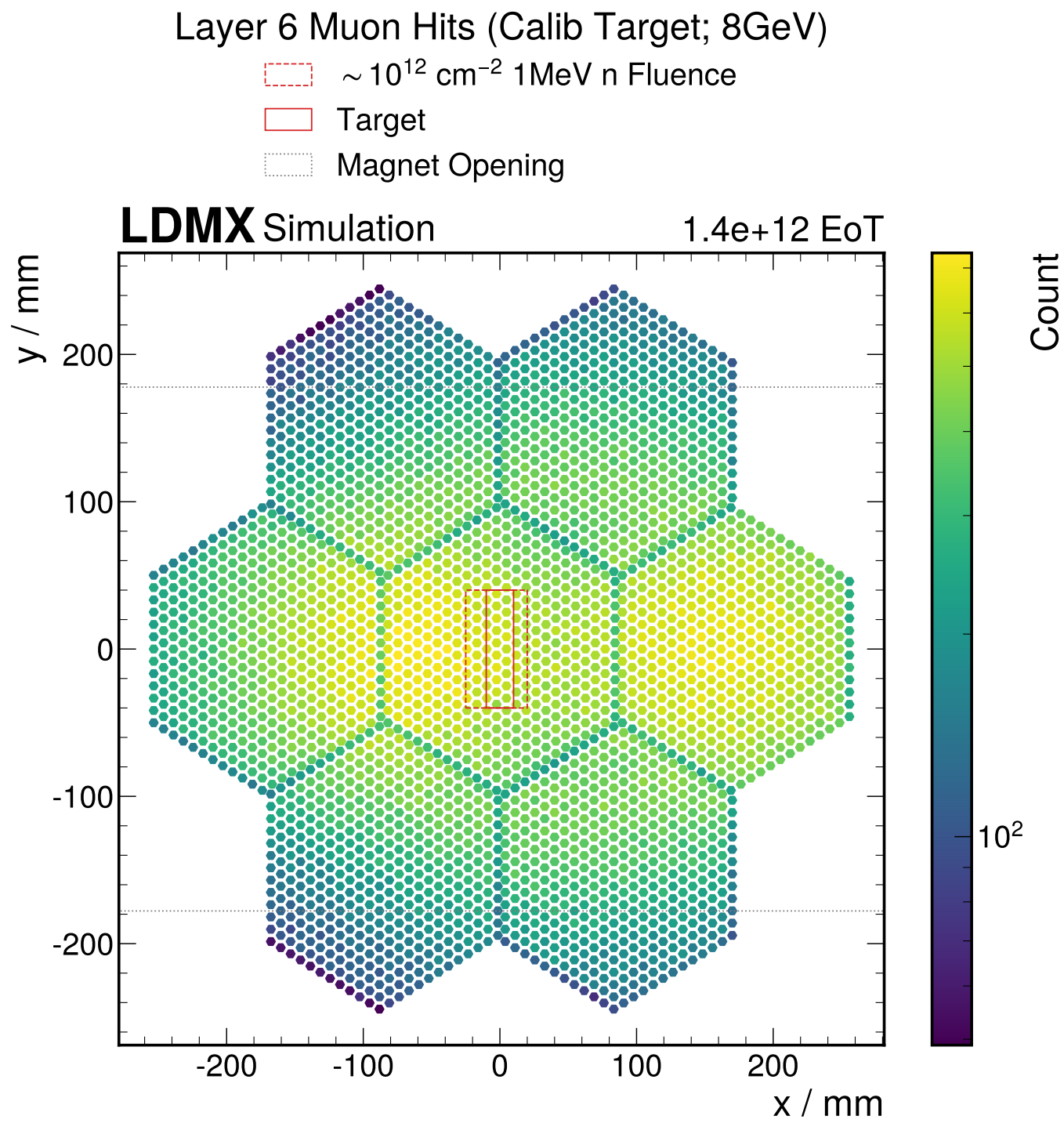

Figure 3.95: Distribution of calibration-candidate muons in the calibration-target beam-operation configuration, normalized to events-per-day assuming a calibration run delivering a total charge of $1.4 \times 10^{12}$ EOT, which is equivalent to a full-day run at $\mu = 1.5$, or a shorter run at a higher $\mu$.

source.

### 3.7.12 ES&H

The ECal subsystem will have several aspects which require consideration from an environment, safety, and health point of view both during installation and during operation. Several aspects related to detector safety are discussed in Sec. 3.7.5.6 on interlocks and the detector safety system.

1. Mechanical – Despite its relatively small physical size, the ECal is quite a heavy object due to the use of substantial volumes of tungsten absorber in the construction of the detector. As a result, handling of the detector during installation will require appropriately trained riggers and careful verification of the mounting of the detector. The individual calorimeter components will also require careful handling, though the individual absorber layers are much lighter. For work at UCSB, approval of necessary procedures will be overseen by the UCSB safety team, while at SLAC, the calorimeter installation will be handled by the trained SLAC team.
2. High voltage – Depletion of the silicon sensors will require the use of moderate high voltage – between 300V and 600V, depending on the level of irradiation that a sensor has experienced. The currents involved may be as high as 2 mA and will be supplied by supplies which are incapable of more than 8 mA of current. All wiring and connectors will be rated to a minimum of 1 kV. The wiring system will be inspected for safety before operation, including the arrangement of detectors and safety grounds to ensure that no one can come in contact with high voltage.
3. Cooling – The ECal operation temperature ($-20$°C – 0°C) does not pose cryogenic risks, but could result in condensation which could damage either the ECal or another subsystem. After irradiation of the ECal sensors, it becomes necessary to control the amount of time when the silicon sensors are at temperatures above 0°C.
4. Dry air – To avoid condensation on the components of the detector, which can result in corrosion or other detector damage, the detector volume must be kept dry through the continual flow of dry air. The dry air system is the most-critical part of the ECal detector services in the case of a power cut and should be maintained through the use of a battery-backup and (for long power outages) through the use of a generator or other service.

5. Low-voltage – The ECal requires substantial low-voltage power for operation (3.2 kW), which could pose a risk to the detector in the case of a cooling failure. However, the large thermal mass of the absorber tungsten substantially mitigates this risk. Without cooling, the thinnest layers of the absorber will heat at 5 s/°C, while the thickest will heat four times slower. This operational risk will be managed by the safety interlock system (Sec. 3.7.5.6) which can disable the low-voltage supplies when the cooling is inoperable or the temperature rises. The individual channels of the power supply are configured to trip at the 5 A level, which controls the risk for short-circuits.

## 3.8 Hadronic Calorimeter - HCal

### 3.8.1 Overview

The primary role of the hadronic calorimeter (HCal) is to provide a high-efficiency veto of neutral hadrons, mostly neutrons and $K_L$, in the energy range from hundreds of MeV to several GeV. Given the expected low occupancy, the transverse segmentation is not very demanding; however, a design with minimal cracks or dead spaces is important to achieve excellent single-particle detection efficiency. A sampling calorimeter with appropriate absorber thickness and plastic scintillator readout can satisfy these requirements. The use of readily available steel plates facilitates a straightforward and robust design, wherein the absorber itself functions as the structural framework, directly supporting the sensitive detector elements. Furthermore, technologies developed for the Mu2e Cosmic Ray Veto (CRV) [104] and the HL-LHC upgrade of the CMS endcap calorimeter [105] can be easily recast to the LDMX case, significantly reducing the need for R&D.

The HCal is a scintillator sampling calorimeter comprising two sections: a Back HCal located behind the ECal, and the Side HCal, a smaller device surrounding the ECal. A model of the device is shown in Fig. 3.96, and its characteristics are summarized in Table 3.11. The Back HCal consists of 96 layers of steel absorber plates, divided into twelve modules of eight layers each. The plates are 2 m x 2 m wide with 25 mm thickness. The sensitive elements consist of 20 mm thick $\times$ 50 mm wide polystyrene bars co-extruded with a $TiO_2$ reflector. The extrusion has a through hole into which a wavelength-shifting fiber is inserted. The scintillator bars are assembled into units of four bars called quad-counters. Ten quad-counters are glued to the absorber plates in a X,Y configuration on each alternating layer. A thin Al cover is added on top of the quad-counter to ensure light tightness. The front-end readout electronics are directly mounted at each end of the quad-counter, improving the light efficiency collection compared to a design in which fibers have to be routed to photo-sensors located further away. Four ears with a through hole are bolted behind each corner, serving both as a structural element for module assembly and a spacer to ensure that the gap between plates is maintained. A set of eight fully equipped layers are combined into a module, fastened by steel pipes inserted through the ears and secured with a threaded flange. The module is finally attached to floor-mounted concrete blocks by feet captured by the threaded pipes at the bottom of the module. The feet are secured to the blocks by Hilti anchors. A light aluminum frame surrounds the device to mount the readout electronics and provide support for the cables.

The Side HCal consists of four stepped modules arranged in a pinwheel-like fashion around the ECal to approximate a uniform amount of absorber in azimuth. Each module is a self-supporting welded structure containing 24 layers of 20 mm absorber with scintillator bars arranged in a X,Z or Y,Z configuration. Given the mechanical constraints, the bars are only read out at one end. The modules are fastened to the first absorber layer of the Back HCal and further supported by a series of steel posts secured to the floor by Hilti anchors.

The scintillation light produced by particles crossing a bar is collected by the wavelength-shifting fiber and read out at each bar end by a silicon photo-multiplier (SiPM). Four SiPMs are mounted on a Counter Mother Board (CMB) providing the bias voltage, a temperature monitor, and flasher LEDs to calibrate each bar independently. The SiPM signals are transmitted to a High Granularity Calorimeter Read Out Chip (HGCROC) board via an HDMI cable. A single HGCROC Board is designed to operate and read out the signals from 16 CMBs, i.e. 64 SiPMs. Four HGCROC boards are housed on a large backplane board, together with mezzanine cards hosting the CMS ECON-T and ECON-D ASICs for data concentration in the DAQ and trigger path, as well as the lpGBT ASIC for serialization and transmission.

A 384 channel prototype of an earlier version of this design, described below, has been built and operated in a CERN test beam to validate the full hardware chain and benchmark the Monte Carlo simulation against the calorimeter single particle response.

| Scintillator | |
|---|---|
| bar width | 50 mm |
| bar height | 20 mm |
| bar length | 2000 mm |
| number of fiber / bar | 1 |
| number of bars / quad-counter | 4 |
| **Mechanics - Back HCal** | |
| number of quad-counter / layer | 10 |
| number of layer / module | 8 |
| number of modules | 12 |
| number of quad-counters | 960 |
| **Mechanics - Side HCal** | |
| number of modules | 4 |
| number of layers / module | 24 |
| number of layers @ 9 quad-counter | 3 |
| number of layers @ 8 quad-counter | 3 |
| number of layers @ 7 quad-counter | 3 |
| number of layers @ 6 quad-counter | 3 |
| number of layers @ 3 quad-counter | 12 |
| number of quad-counters | 504 |

| Electronics - Back HCal | |
|---|---|
| number of channels / CMB | 4 |
| number of CMB / quad-counter | 2 |
| number of CMB / HGCROC board | 16 |
| number of channels | 7680 |
| number of CMBs | 1920 |
| number of HGCROC boards | 120 |
| number of backplane boards | 30 |
| number of ECON-T/D mezzanine boards | 30 |
| number of lpGBT mezzanine boards | 60 |
| **Electronics - Side HCal** | |
| number of channels / CMB | 4 |
| number of CMB / quad-counter | 1 |
| number of CMB / HGCROC board | 16 |
| number of channels | 2016 |
| number of CMBs | 504 |
| number of HGCROC boards | 32 |
| number of backplane boards | 8 |
| number of ECON-T/D mezzanine boards | 8 |
| number of lpGBT mezzanine boards | 16 |
| **Electronics - Full HCal** | |
| number of channels | 9696 |
| number of CMBs | 2424 |
| number of HGCROC boards | 152 |
| number of backplane boards | 38 |
| number of ECON-T/D mezzanine boards | 38 |
| number of lpGBT mezzanine boards | 76 |

Table 3.11: Summary of HCal characteristics.

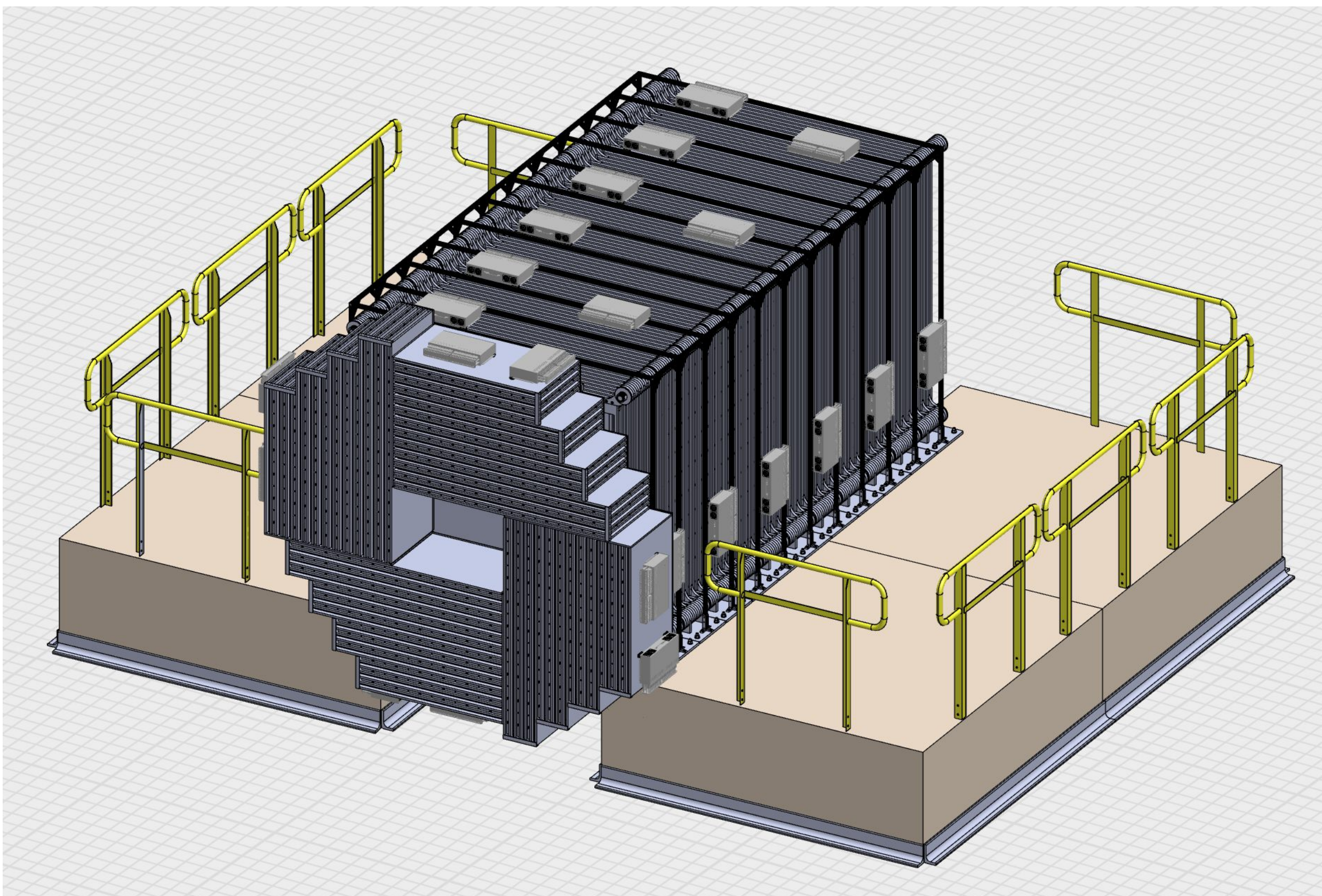

Figure 3.96: SolidWorks model of the HCal.

### 3.8.2 Requirements

The requirements on the HCal design are driven by the general physics considerations outlined in Sec. 3.2. A major background in the experiment arises from the emission of hard bremsstrahlung photons carrying away most of the incoming electron energy, followed by a photo-nuclear reaction in the target or the ECal. This process results in a wide variety of final states, most of which leave a significant amount of energy in the ECal. However, a fraction of these events are characterized by the emission of a single energetic forward neutral hadron or a few O( GeV) neutral hadrons emitted at moderate angles. These neutrons escape the ECal undetected and must be identified by the HCal with excellent efficiency. The same reactions can occur in the target, albeit at a reduced rate (the tracker also provides additional information to reject these topologies). Alternatively, bremsstrahlung photons could convert into muon pairs, typically crossing a large fraction of the detector volume. These muons can be almost entirely rejected by the tracker and the ECal, and only a few remaining events need to be vetoed by the HCal. Muon pairs can also be produced in the absence of preceding hard bremsstrahlung by trident reactions, and similar techniques are used to suppress them.

Wide-angle bremsstrahlung (WAB) photons present another source of background. The emission of a bremsstrahlung photon in the target can result in an outgoing electron with significantly lower energy than the incoming electron, a signature similar to the signal if the photon is emitted outside the ECal acceptance. Similarly, the incoming electron could undergo a wide-angle scatter in the target, followed by the emission of a softer bremsstrahlung photon at a large angle. These events must therefore be vetoed by a hadronic calorimeter surrounding the ECal.

In addition to dark matter searches, LDMX could also contribute to searches for light new physics or electro-nuclear measurements. Several scenarios leave large signals in the HCal without any significant activity in other sub-systems, and the hadronic calorimeter should provide a trigger(s) to identify these signatures. Triggers to collect calibration samples are also needed, including filters to identify cosmic ray muons or single particles produced in photo-nuclear and electro-nuclear reactions.

These physics requirements are translated into detector requirements, from which technical requirements are derived. Both are summarized below:

- **HCal Detector-Physics Requirements**

HCAL1 Detect neutrons and $K_L$ emitted at a polar angle less than 45 degrees in photo-nuclear reactions producing a single or a few neutral particles with an inefficiency less than $10^{-7}$.

HCAL2 Detect neutrons emitted in photo-nuclear reactions producing only low-energy ($< 1\,\text{GeV}$) neutrons with a combined inefficiency less than $10^{-3}$.

HCAL3 Detect wide-angle bremsstrahlung photons with an inefficiency less than $10^{-5}$.

HCAL4 Produce trigger primitives to identify MIP tracks (e.g., cosmic rays and muon pairs) and energy deposits from MIP to multi-GeV showers for calibration purposes.

HCAL5 Limit the false veto rate below 1% for noise and cosmic ray muons.

- **HCal Technical Requirements**

HCAL6 Contain 16 nuclear interaction lengths of absorber along the beamline downstream of the ECal, and 3.5 nuclear interaction lengths along the direction perpendicular to the beamline surrounding the ECal.

HCAL7 Provide a signal of at least 50 photo-electrons for an incident MIP crossing a scintillating bar.

HCAL8 Maintain the electronics noise RMS below 5% of the signal produced by a MIP.

HCAL9 Monitor the SiPM temperature with 1° Celsius accuracy.

HCAL10 Produce trigger primitives with a latency less than 1 $\mu$s.

HCAL11 Operate the detector 6 years and maintain performance after a total radiation exposure of 1 krad and a total neutron fluence of $10^9$ 1 MeV neutron equivalent/cm$^2$.

### 3.8.3 Structural components

In Fall 2021, we devised and executed a procedure for the successful fabrication of a prototype sampling hadronic calorimeter. We have adapted this concept, incorporating lessons learned, for the construction of the full calorimeter. The quad-counter design draws heavily on the di-counter manufactured for the Mu2e Cosmic Ray Veto, with fabrication techniques optimized to the HCal requirements. The module design has drawn from the experience acquired during the assembly of detector elements for the MINOS experiment. The HCal structural elements are reviewed below, while the fabrication and integration procedures are described in Sec. 3.8.8.

#### 3.8.3.1 Scintillator and Fiber

The scintillator bars are made of STYRON 665 W polystyrene, with 2,5-diphenyloxazole (PPO, 1% by weight) as primary dopant, and 1,4-bis(5-phenyloxazol-2-yl) benzene (POPOP, 0.05% by weight) as secondary dopant. A co-extruded $TiO_2$ reflective coating of 0.25 mm nominal thickness surrounds the core. This outer reflective coating is added through material injected from a second extrusion machine (co-extruder) that mixes the polystyrene and $TiO_2$ pellets. Each counter also has one co-extruded hole at the bar center of a nominal 2.6 mm diameter into which wavelength-shifting (WLS) fibers are to be placed. A cross-sectional view of a bar is shown in Fig. 3.97, where the shape of the counter, the holes, and the $TiO_2$ coating are visible.

The WLS fiber is a 1.8 mm diameter Kuraray double-clad Y11 fiber doped with 175 ppm K27 dopant [106]. Approximately 10 km of fiber is required by the HCal, and about 11 km of fiber delivered on spools will be purchased to account for wastage during the fabrication process. Quality tests are performed on the first 25 m of each spool to compare the absolute light yield and attenuation of light in the fiber. Similar measurements for the fiber in the Mu2e CRV are documented in Ref. [107].

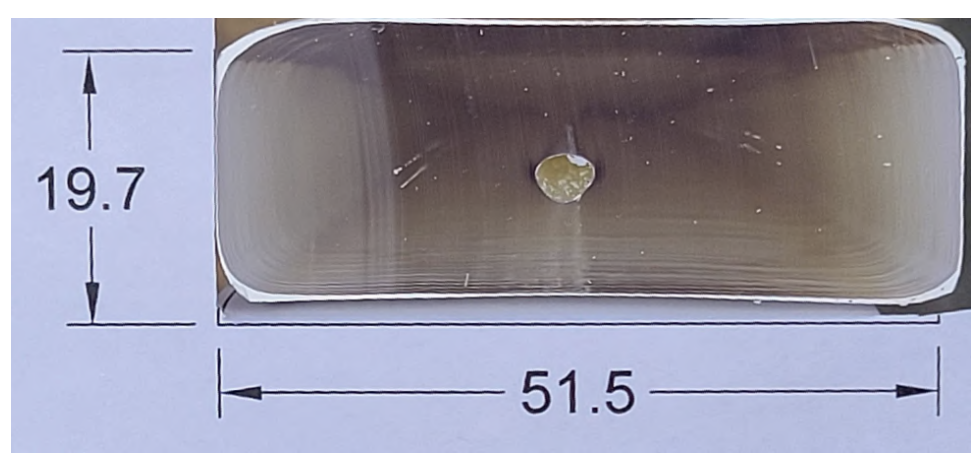

Figure 3.97: Photo from the end of a typical counter. The shape of the counter, the hole, and the $TiO_2$ coating are visible.

#### 3.8.3.2 Quad-counter

Four scintillator bars are assembled into units called quad-counters. The extrusions are first cut to length simultaneously using a chop saw fixture. An adjustable stop for tuning the length and squareness of the cut is placed at one end of the apparatus. Each cut counter is prepared for bonding by scoring the bonding surfaces using Scotch-Brite. Excess dust is removed, and the surface is cleaned using isopropanol. A vertical fixture is employed to hold the counters, and the ends of each counter are made flush with one another, while 3M DP420 epoxy cures to bond them together. Springs apply pressure to the stack to ensure adequate contact between the bonding surfaces and to minimize the overall curvature of the assembly. Fig. 3.98 shows four counters in the vertical fixture.

While the quad-bar epoxy cures, wavelength-shifting fibers are inserted into the hole of every extrusion. Approximately 50 mm to 100 mm of fiber is left protruding from both ends of the extrusions, which aids in mounting the Fiber Guide Bars (FGBs). A fast-curing epoxy, 3M DP100, is used to fix the fibers within the holes of the FGB as well as to bond the FGB to the scintillator. The FGB is designed with a funnel-shaped fiber guide hole to allow for variation in the extrusion fiber hole while registering the fiber position precisely relative to the SiPM. Each fiber guide bar is checked with go/no-go jigs to ensure that fiber holes are within size and position tolerances before installation. The installation of an FGB is shown in Fig. 3.98. Additionally, #4-20 3/8" Plastite screws are used to affix the FGB to the scintillator. The presence of these screws does not impact the light yield near the ends of the scintillator.

The excess fiber protruding from the fiber guide bars in the previous step is cut off using a nichrome-wire hot knife. The use of light pressure prevents cracks from developing that would negatively impact light transmission through the fiber. The surface of the fibers is subsequently polished using an automated flycutting machine. To ensure quality cuts, a total of three passes are made: two 0.5mm deep roughing cuts and a final 0.1mm deep polishing cut. A photo of the prototype flycutter is shown in Fig. 3.99. The control mechanism consists of commonly available CNC machine mechanics/software with rack and pinion drives. Several quality checks are finally conducted before crating and shipping the quad-counters to the module fabrication facility. The QC and QA performed at the quad-bar factory are described in Sec. 3.8.9.

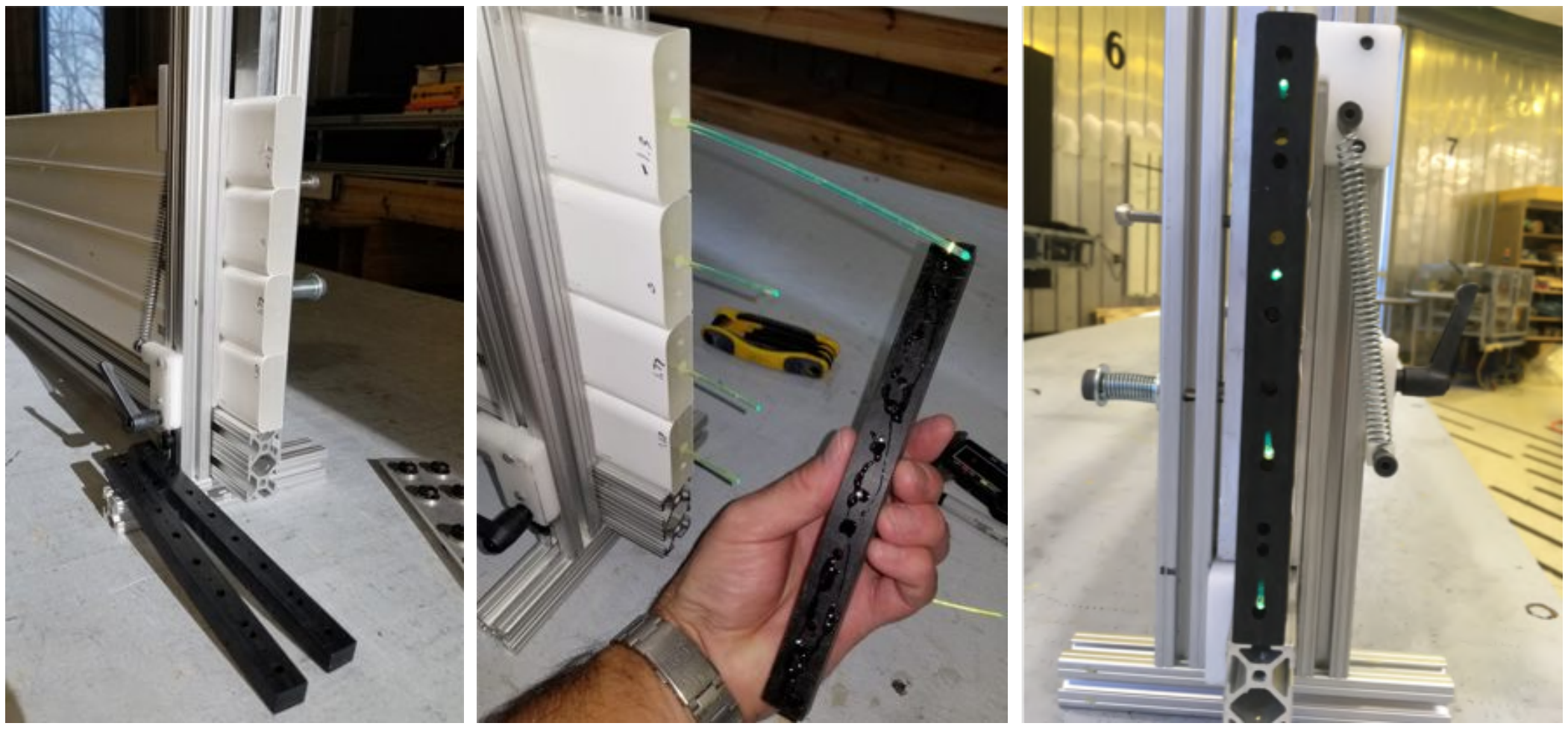

Figure 3.98: Left: Four counters clamped in the vertical fixture while epoxy cures. Center: Counters with fiber inserted and FGB being installed. Right: Quad-bar with FGB installed.

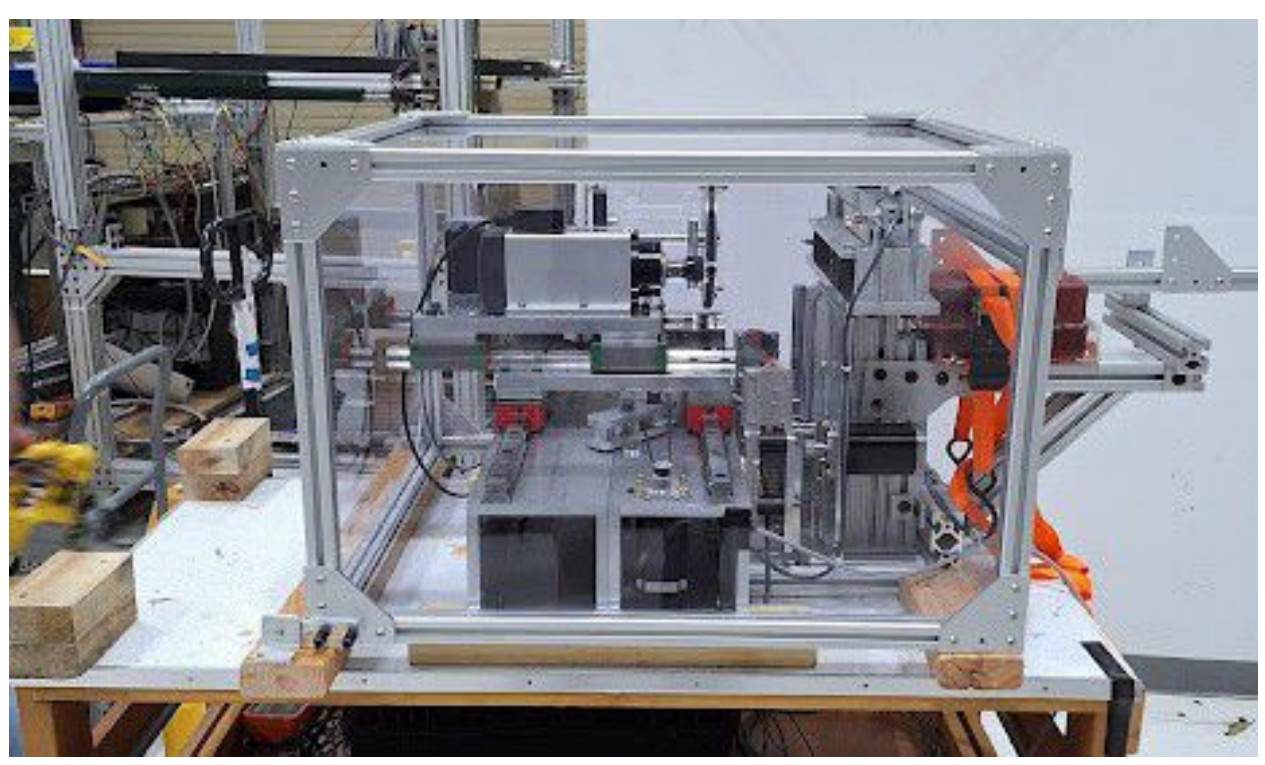

Figure 3.99: A photo of the prototype flycutter for polishing the quad-bar ends. A series of rollers is used to guide the counter into approximate alignment. A pneumatic cylinder compresses the quad-counter in place for fly-cutting. Once a run is ordered, this alignment system will retract away and allow the flycutter to approach.

#### 3.8.3.3 Back HCal modules

The Back HCal comprises 96 layers, divided into twelve modules of eight planes each to facilitate construction and installation. A plane contains ten quad-counters attached to the absorber plate. The prototype was assembled using two-component structural epoxy to glue the quad-counters onto steel plates. We also made a few planes using 3M Extreme mounting tape, which did not require extensive setting time. We will do comparison tests of the two approaches with a full quad-counter plane before making a final decision between epoxy and tape. A plane cover, made of 0.5 mm aluminum, is glued to the assembled quad-counters and made light-tight with a black silicone bead. Four ears with a through hole are bolted behind each corner, serving both as a structural element for module assembly and a spacer to ensure that the gap between plates is maintained. With the exception of the first plane of the first module, which is adapted to support the Side HCal modules, all planes are identical.

A set of eight planes is assembled into a module, orienting quad-counters horizontally and vertically in an alternating pattern, as displayed in Fig. 3.100. This requires the insertion of additional spacers to level the planes, the alignment of the four holes in the plane ears, and the insertion of the two-inch diameter threaded pipes to connect the eight steel planes. At each end of the module, the pipes are threaded into a flange that is half the inter-plane thickness. The mounting feet are inserted in place of some of the spacers and are also threaded through the pipe. The flanges are tensioned to a predetermined value and then screwed into the plane. The front-end electronics are then mounted at the end of each quad-counter. The module is finally tested using cosmic rays before being moved to the storage area and readied for transport by truck to SLAC.

#### 3.8.3.4 Side HCal modules

The Side HCal consists of four identical welded structures mounted in pinwheel fashion on the first layer of the Back HCal (see Fig. 3.96). The arrangement of Side HCal modules forms a cavity in which the ECal is mounted. The box-like structures are stepped in units of one quad-counter to better approximate a uniform amount of absorber in azimuth, as shown in Fig. 3.100. Each module contains 24 planes divided into four groups of six planes each, with a length varying between six and nine quad-counters in the long direction and three quad-counters in the short direction. To improve the sampling granularity, an absorber thickness of 20 mm has been selected. The quad-counters are oriented in an alternating fashion along the long and short directions. Due to the mechanical constraints, they are read out only at one end; the opposite side is mirrored to increase the light collection efficiency.

The module assembly proceeds slightly differently from the method described previously. The external steel frame (bottom, rear, and side plates) is first assembled before welding 17 mm thick steel plates into the structure at regular intervals, leaving an opening for the insertion of the quad-counters. In parallel, quad-counters are epoxied to 3 mm thick steel plates, followed by the installation of a 0.5 mm aluminum cover to ensure light tightness. These plates are finally glued onto the steel absorber plates and secured by corner bars that allow access to the front-end electronics. The four modules will be fastened together with angle brackets to improve mechanical stability once they are attached to the front plate of the Back HCal. Adjustable steel posts bolted to the floor will provide additional support.

#### 3.8.3.5 Seismic Evaluation

The seismic stability of the HCal structure[2] has been evaluated using SolidWorks. The study has been performed using an earlier HCal design, in which the ears are directly cut into the absorber plates, and the Side HCal is bolted onto the first absorber plate without being supported by additional steel posts. This analysis will be updated with the final design once available, but the qualitative conclusions are expected to be similar to those described hereafter.

The structure deformation and associated von Mises stresses are estimated by building a finite element model and then subjecting it to horizontal and vertical accelerations of 0.8g and 0.6g, respectively. The results are displayed in Fig. 3.101. A deformation below 3 mm for the Back HCal and up to 5 mm at the top of the Side HCal is expected, indicating that the structure is indeed stable and that the stresses on the fasteners that connect the steel structure to the EX4 blocks and the blocks to the floor meet the seismic requirements. The Hilti online calculator will be employed as a confirmation of these results, as it is explicitly called out in the seismic document.

### 3.8.4 Electronics

The readout electronics chain, schematically shown in Fig. 3.102, is adapted from the Mu2e Cosmic Ray Veto (CRV) system and the HL-LHC upgrade of the CMS endcap calorimeter. The fast scintillation light collected by the wavelength-shifting fiber is read out by silicon photo-sensors (SiPMs) located directly at the end of the bar, improving the light efficiency collection compared to a design in which fibers have to be routed further away from the scintillator. The SiPMS are mounted on front-end boards providing the bias voltage, a temperature monitor, and flasher LEDs to calibrate each bar independently. The SiPM signals are transmitted to a High Granularity Calorimeter Read Out Chip (HGCROC) board via an HDMI cable. A single HGCROC Board is designed to operate and read out the signals from 16 quad-counters. Four HGCROC boards are housed on a large backplane board, together with ECON and lpGBT mezzanine cards providing the logic and communication with the DAQ, trigger, clock, and controls. The arrangement shares similar components with the ECal readout, facilitating the testing, commissioning, and operations of the electronics. Each component is described in detail hereafter.

[2] The SLAC seismic requirements can be found at http://www-group.slac.stanford.edu/esh/documents/techbas/seismic.pdf.

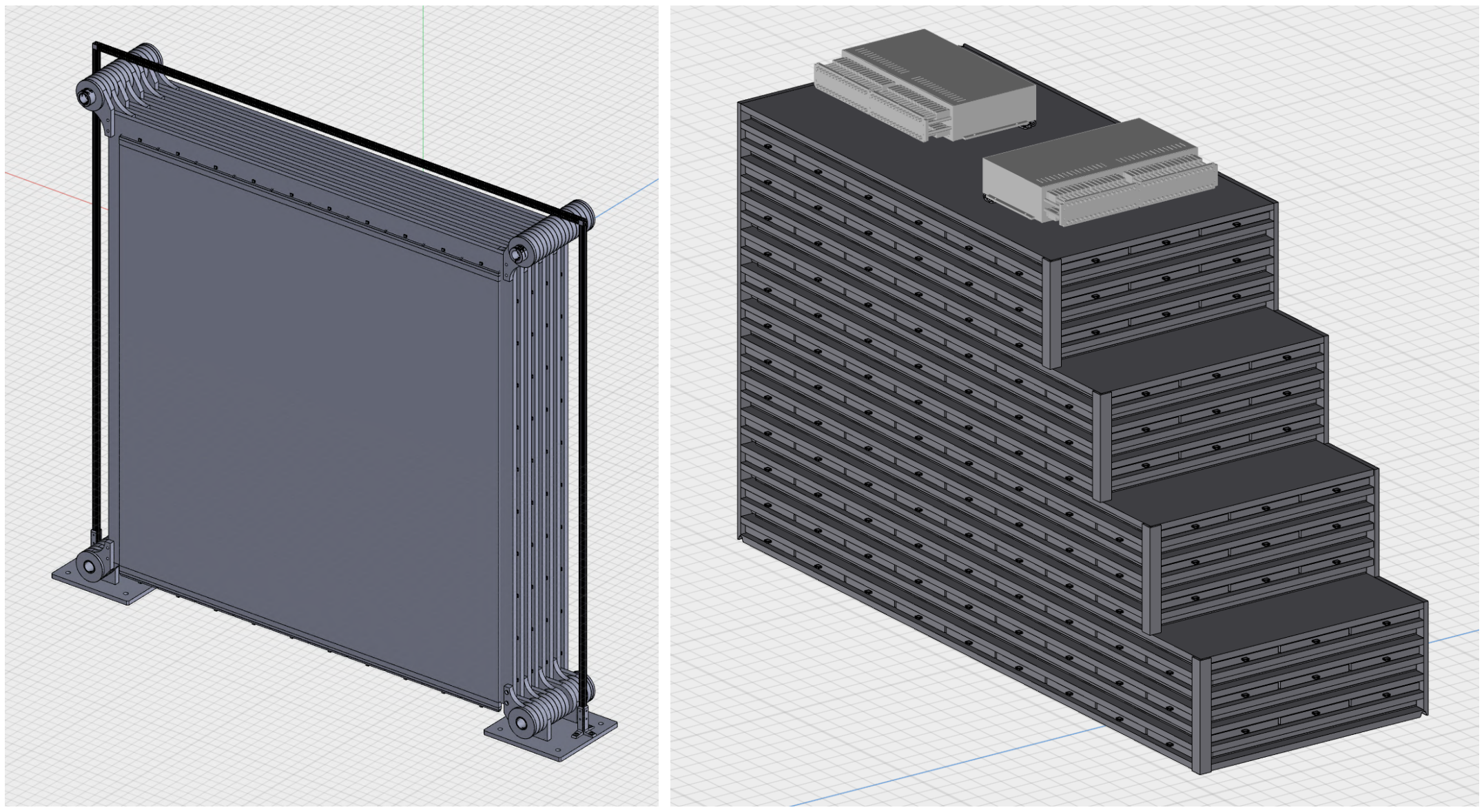

Figure 3.100: Left: SolidWorks model of the eight-absorber module of the Back HCal. The active elements (scintillating bars) are covered by a thin Al cover to ensure light tightness. Right: One of the four Side HCal modules with crates holding HGCROC boards. The individual CMBs are visible on the front and side of the module.

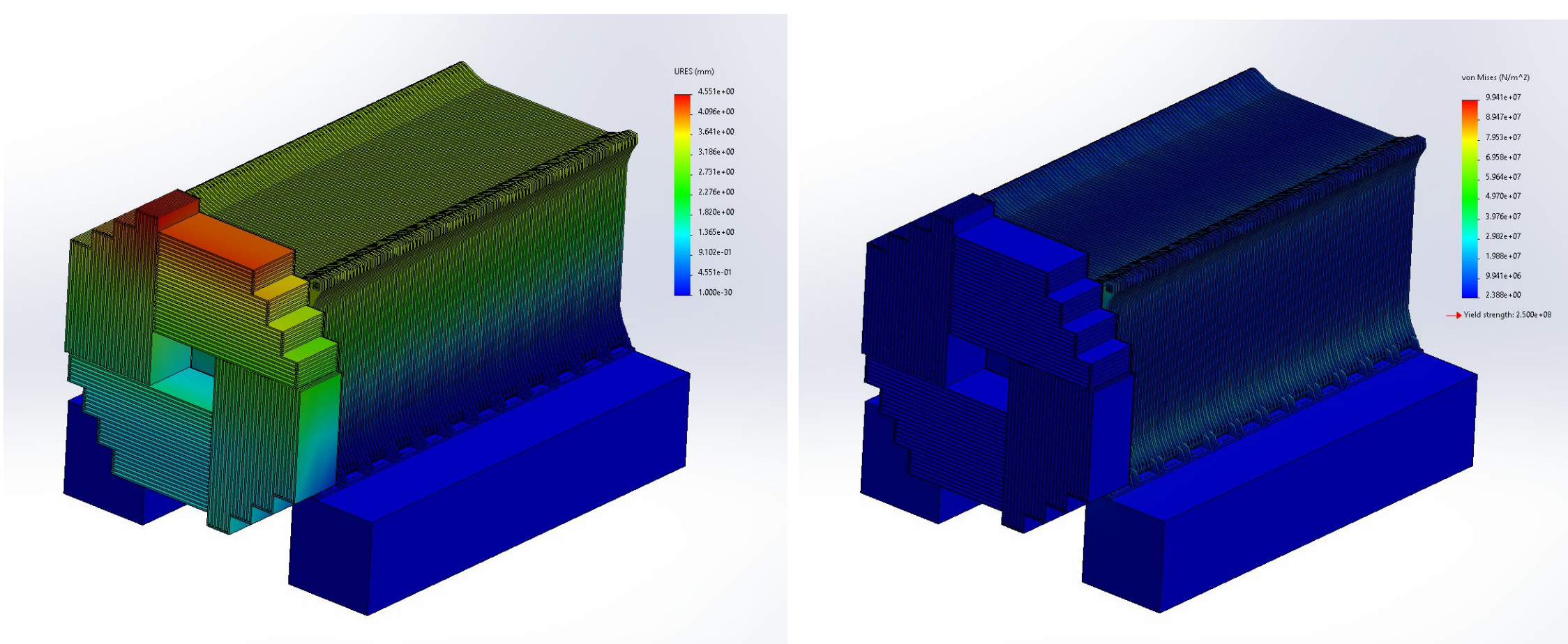

Figure 3.101: Left: Displacement of the HCal under 0.8g vertical acceleration. Right: Von Mises stress on the HCal structure under 0.8g vertical and 0.6g horizontal accelerations.

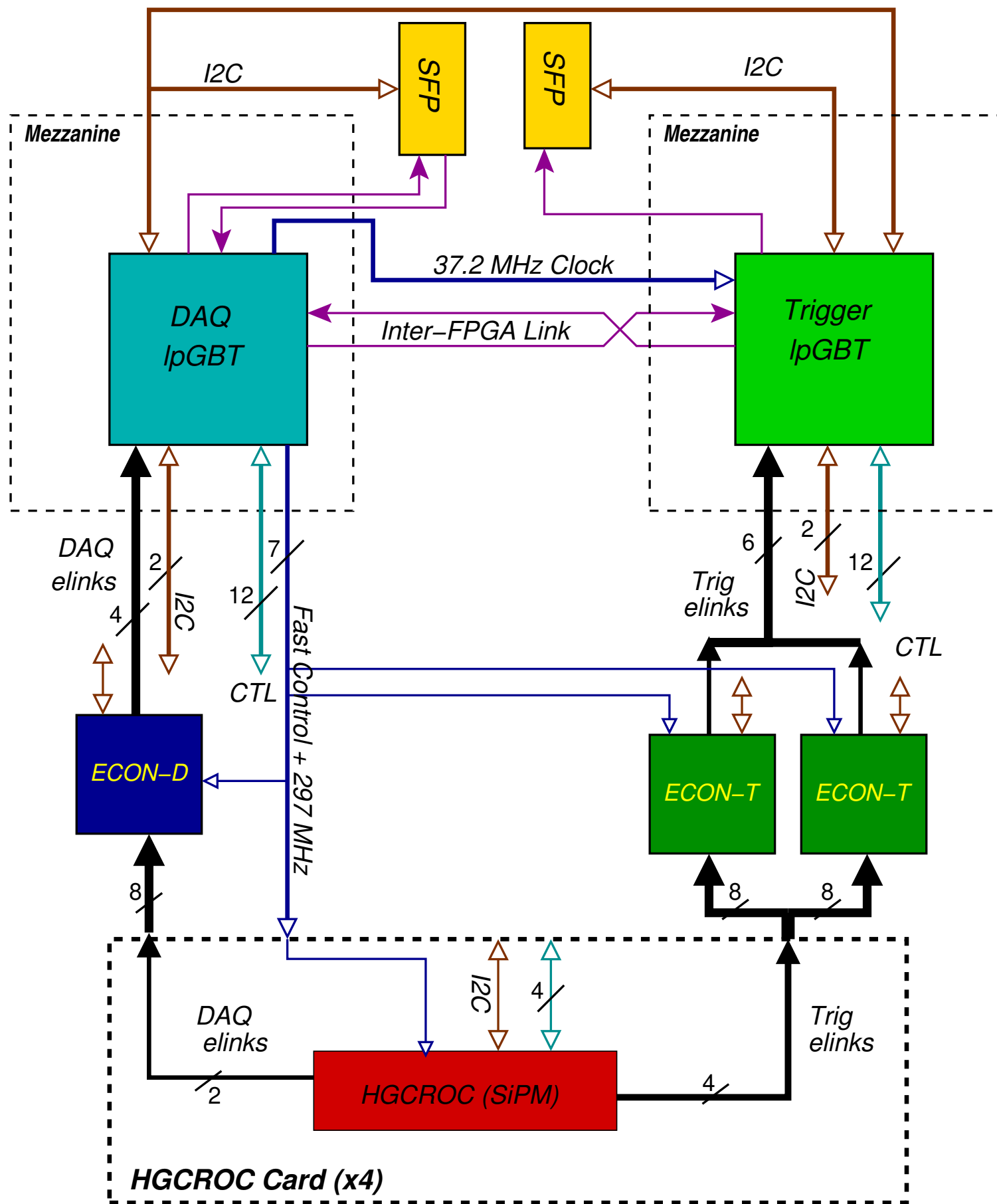

Figure 3.102: HCal signal arrangement for 64 quad-counters

#### 3.8.4.1 Counter Mother Board

The Counter Mother Board (CMB) is mounted directly on each end of a quad-counter. A CMB is an assembly of a support board for the four SiPMs[3], to which a HDMI cable[4] is connected, and a mechanical support. The CMB connects to the SiPMs with spring pins. A CMB also has four flasher LEDs and a temperature sensor. A circuit diagram of the CMB is shown in Fig. 3.103 together with a photograph of the support board with its four SiPMs, and a CMB. The SiPM signals, the temperature sensor readings, the bias voltages for the SiPMs, the bias for the LEDs, the Low Voltage Differential Signalling (LVDS) signals for flashing the LEDs, and the power for the temperature sensor, are all transmitted over the HDMI cable.

#### 3.8.4.2 HGCROC board

The HGCROC Board is shown in Fig. 3.104 together with a block diagram of its components. Its central component is the HGCROC ASIC [108] which measures and digitizes the charge deposited in the SiPMs and is shown as the rectangular gray component in the middle of the board on the photograph. The HGCROC digitizes 72 channels, of which 64 are physically connected to SiPMs, making regular synchronized readouts with the $\approx$ 40 MHz bunch frequency of the beamline. The bottom of Fig. 3.105 schematically shows the analog pulse amplification and shaping and digital parts of the digitization, as well as additional timing and calibration circuits. ADC and TDC measurements of the sampled pulses are continuously saved at $\approx$ 40 MHz and aligned in a circular buffer DRAM, where 8 sequential time samples around the event of interest

[3]Hamamatsu S14160-3015PS - $3 \times 3mm^2$ photo sensitive area, $3.6 \times 10^5$ gain, 700 kcps dark count rate with a 0.5 p.e. threshold.
[4]Ethernet HDMI cable.

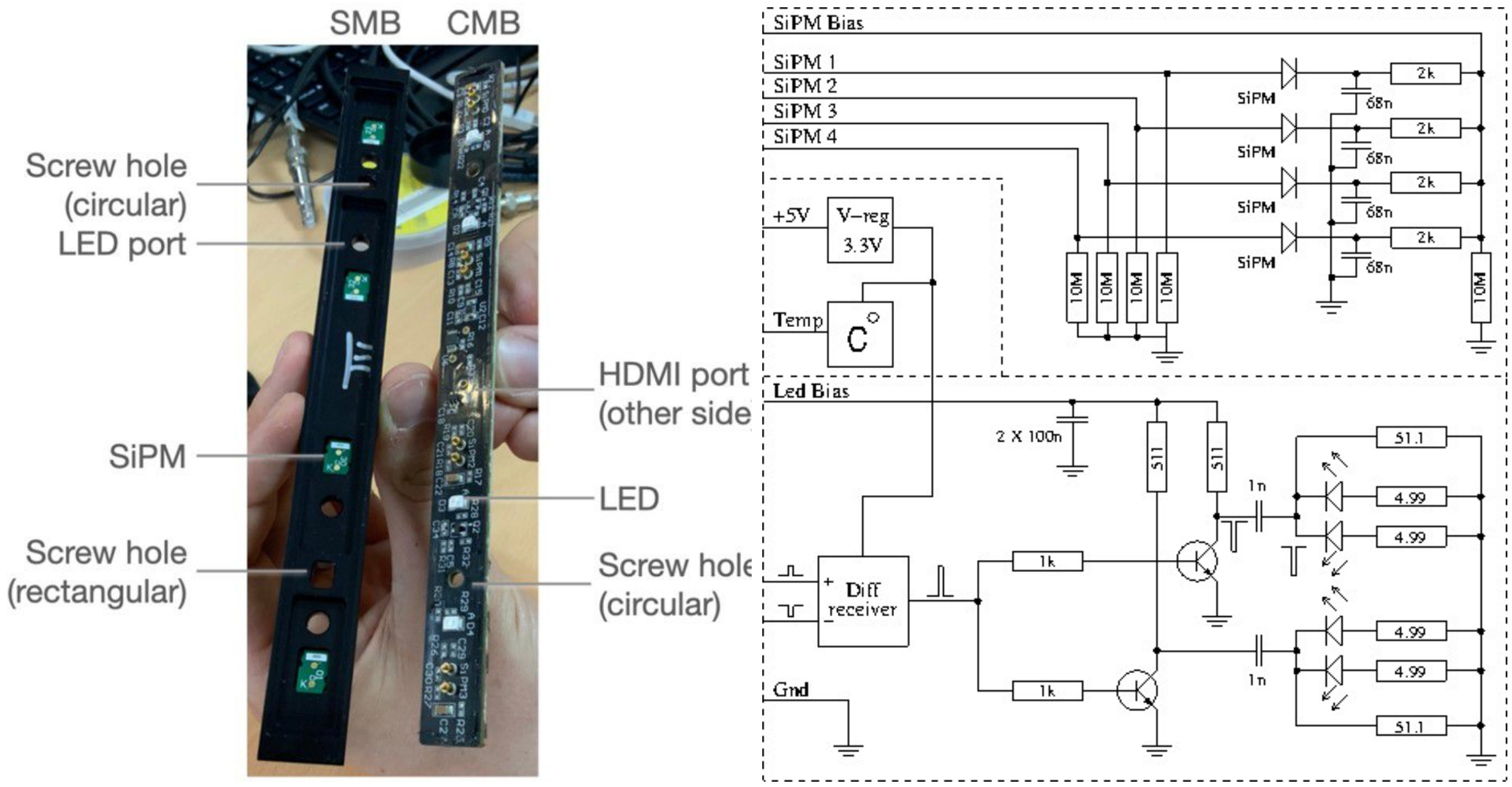

Figure 3.103: Left: Photograph of the SiPM mounting block (SMB) and the Counter Mother Board (CMB). Right: Schematics of the CMB.

may read out when followed by an L1 fast command.

In the SiPM version of the HGCROC, an additional circuit denoted the *conveyor* is placed before all analog electronics to attenuate the large signal produced in the SiPM, which is not in place in the silicon version. The strength of the attenuation is variable, from keeping 5% of the signal amplitude in beam conditions, up to keeping 37.5% when the single photon spectrum of the SiPM is to be resolved. The pre-amplifier (denoted PA in Fig. 3.105) provides the charge-to-voltage conversion of the incoming pulse, and the time-constant of the amplifier is responsible for the initial shaping of the pulse. A dedicated shaping circuit before the ADC is used to further reduce out-of-time pile-up by adjusting the shaping time by ±20%. Furthermore, the pulse that exits the shaper should ideally be sampled at its maximum to help with pulse reconstruction. A phase shift may therefore be applied to the timing of the recurring ADC measurements, to sample the ADC once at the pulse maximum.

The successive approximation analog-to-digital converter provides a 10-bit ADC measurement of the pulse height every time sample, allowing reconstruction of the signal pulse shape from 8 samples read out by an L1 trigger signal. For large pulses where the pre-amplifier has a non-linear behavior or if the ADC has saturated, a TDC measuring the time-over-threshold (TOT) of the pulse is available, counting between 2 and 200 ns, with 10-bit precision. The TOT is linearly correlated to the incoming charge for large pulses, and may therefore be used to extend the dynamic range of the measurement far beyond the ADC's saturation, necessary for reconstructing shower energies in the HCal. An additional TDC measuring the time-of-arrival (TOA) is available to assist in measuring the pulse timing, which the sparse sampling by the ADC is not suited for, by timestamping the rising edge of the pulse with 25 ps precision. The TOA is especially useful for determining the hit position along a scintillator bar in the Back HCal.

After the attenuating conveyor, a charge injection circuit can inject charges of known amplitude synchronized in time with the readout into individual channels. This feature allows the calibration of the ADC response as a function of charge. The charge is controlled by an internal 11-bit DAC, and can choose between a small capacitor producing charges up to 0.5 pC for aiding fine measurements in the low ADC range, while the more coarsely tuned 8 pC circuit will bring the readout into the TOT range, which in conjunction can be used to calibrate a linear response of the chip over both the ADC and TOT region. In the trigger path, the HGCROC is configured to transmit sums of four channels; these four channels correspond to one quad-bar or one CMB. After summing, the trigger results are compressed into a seven-bit floating point scale with four exponent bits and three mantissa bits. The trigger sums are packed in groups of four into 1.2Gb/s electrical links so that one HGCROC produces four 1.2Gb/s trigger electrical links.

The voltage on the SiPMs is set equal for each set of four SiPMs served by an HDMI cable. However, the voltage level on the anode of each SiPM can be adjusted on the HGCROC, allowing individual voltage settings over each SiPM. The HGCROC Board sets the bias on the LED on each CMB, and sends flash pulses to it triggered from the HGCROC ASIC. Finally, the temperature sensor on each CMB is read through the HGCROC Board. These data also include a unique numerical identifier for each CMB.

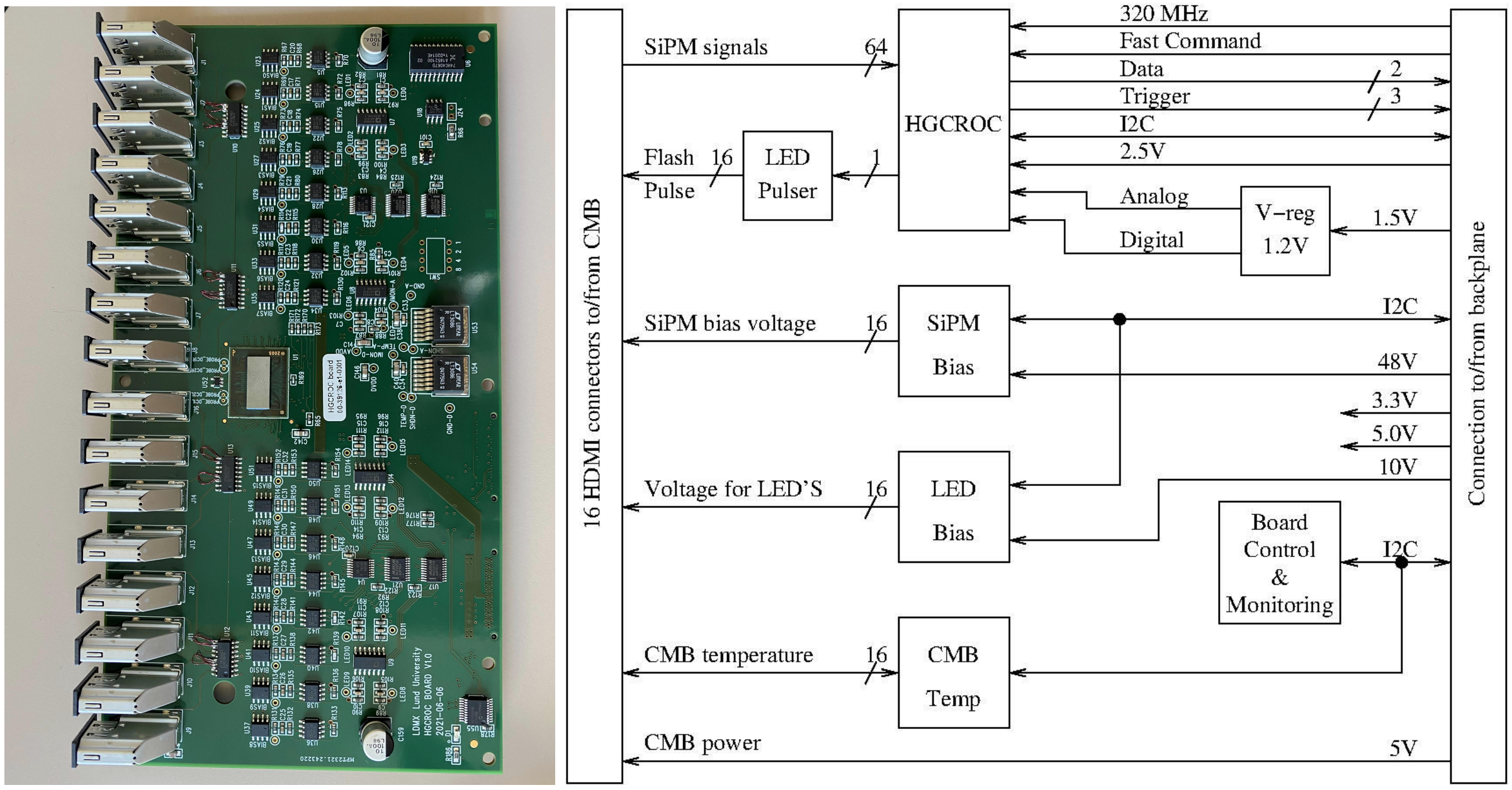

Figure 3.104: The HGCROC board hosting the HGCROC and connecting to the CMB to receive the SiPM signals, powering the SiPMs and the LEDs and reading the temperatures.

#### 3.8.4.3 Backplane Board and components

The backplane board is shown in Fig. 3.106 together with a block diagram of its components. The trigger and readout data from four HGRCOCs flow to the ECON-T and ECON-D ASICs, respectively. The trigger electrical links (16 in total, 4 per HGCROC) are read out by ECON-T ASICs, up to 12 eLinks per ASIC. Therefore, two ECON-T ASICs are used to read out four HGCROC boards. The data eLinks (8 in total, 2 per HGCROC) are readout by a single ECON-D ASIC.

The ECON ASICs functionality is described in Sec. 3.7.3.7. The ECON-D ASIC receives and transmits DAQ data and applies zero suppression if requested. The ECON-T ASIC compresses trigger data with an internal user-selectable algorithm. For the HCal, the baseline algorithm is the "Super Trigger Cell" (STC). The STC algorithm performs fixed sums of groups of four or sixteen cells and transmits the results in one of several floating-point encodings, optionally along with the identity of the highest-energy cell within the sum. The charge sums are sent on up to thirteen active output eLinks. The configuration of the HCal backplane is shown in Fig. 3.102 with two HGCROC ASICs feeding each ECON-T ASIC, each of which has three eLinks available for output. In this case, the optimal configuration of the ECON-T is the "STC-4 5E+4M" algorithm, where the ECON-T will sum four trigger cells (CMBs) and transmit the result as a nine-bit floating value, along with two bits indicating which trigger cell had the highest energy. The values sent to the HCal trigger processor therefore represent the sum of sixteen channels or four CMBs (all four channels). Given the mechanical constraints of the detector, these signals are necessarily always taken from a single end of the double-ended bars in the Back HCal and therefore represent the partial energy sum over sixteen bars.

A single ECON mezzanine board hosts the 2 ECON-T and 1 ECON-D ASICs. The mezzanine has been custom-designed for the HCal, following closely previous designs by CMS. Each ECON mezzanine board will be tested before the assembly of the backplane using a test system to check the functionality of the slow control and basic communication with the ASIC with signals sourced from an FPGA.

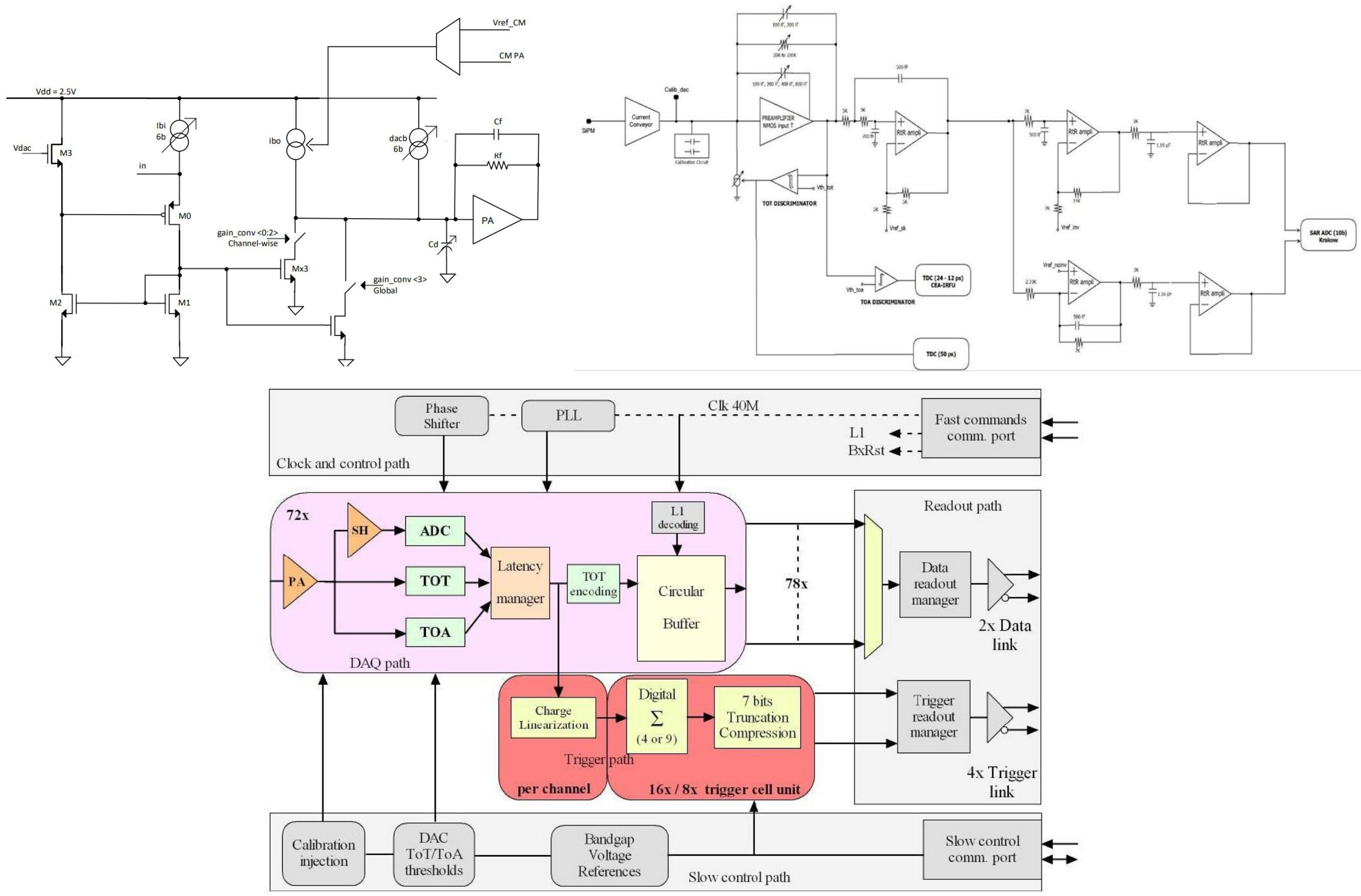

Figure 3.105: The integrated readout circuit HGCROC. Top left: The signal-attenuating input stage (the conveyor). Top right: The analog channel architecture. Bottom: The block diagram for the HGCROC3 ASIC. The pictures are from the circuit datasheet H2GCROC_datasheet_1_4.pdf.

The output signals from the ECON ASICs are directed to the Trigger and DAQ lpGBT ASICs, with each lpGBT hosted on a mezzanine. The lpGBT mezzanine is a design shared with ECal. The DAQ lpGBT serves as the source of the clock and fast control streams for the backplane, including for the Trigger lpGBT. The control and monitoring of the ECON, HGCROCs, and the HGCROC board signals are handled via I2C buses, GPIO, and ADC signals. The design of the backplane is custom to the HCal. In addition to the 4 HGCROC Boards, 1 ECON mezzanine, and 2 lpGBT mezzanines, the backplane hosts two SFP connectors for data transmission.

#### 3.8.4.4 Crates

Each readout electronics unit will be housed in a crate, as shown in Fig. 3.107. These crates will be distributed around the HCal consistent with the length of the HDMI cables from the CMBs. A possible scheme is displayed in Fig. 3.108 and 3.109. The crates will be attached to a mechanical structure around the HCal, which will also be used to relieve the cable strain.

#### 3.8.4.5 Board mapping and trigger primitives

In order to reduce the load on the DAQ, a hardware trigger system will perform fast event selection based on detector information with reduced granularity. Decisions can include information from the HCal, where individual channels are grouped together to reduce granularity. This algorithm works by combining spatially-adjacent channels and summing their energy deposition into a single trigger primitive object. If multiple primitive objects in consecutive layers are above a certain energy threshold, then it is likely that a particle of interest has passed through the HCal. A schematic of the channel grouping is shown in Fig. 3.108 and 3.109. In this layout, each trigger primitive corresponds to 2x8 channels (8 channels in each of 2 consecutive layers), and each block must record at least 100 PEs. The particular threshold has been set for the reconstruction of single MIP tracks, but is a tunable parameter that may be adjusted to accommodate generic trigger algorithms.

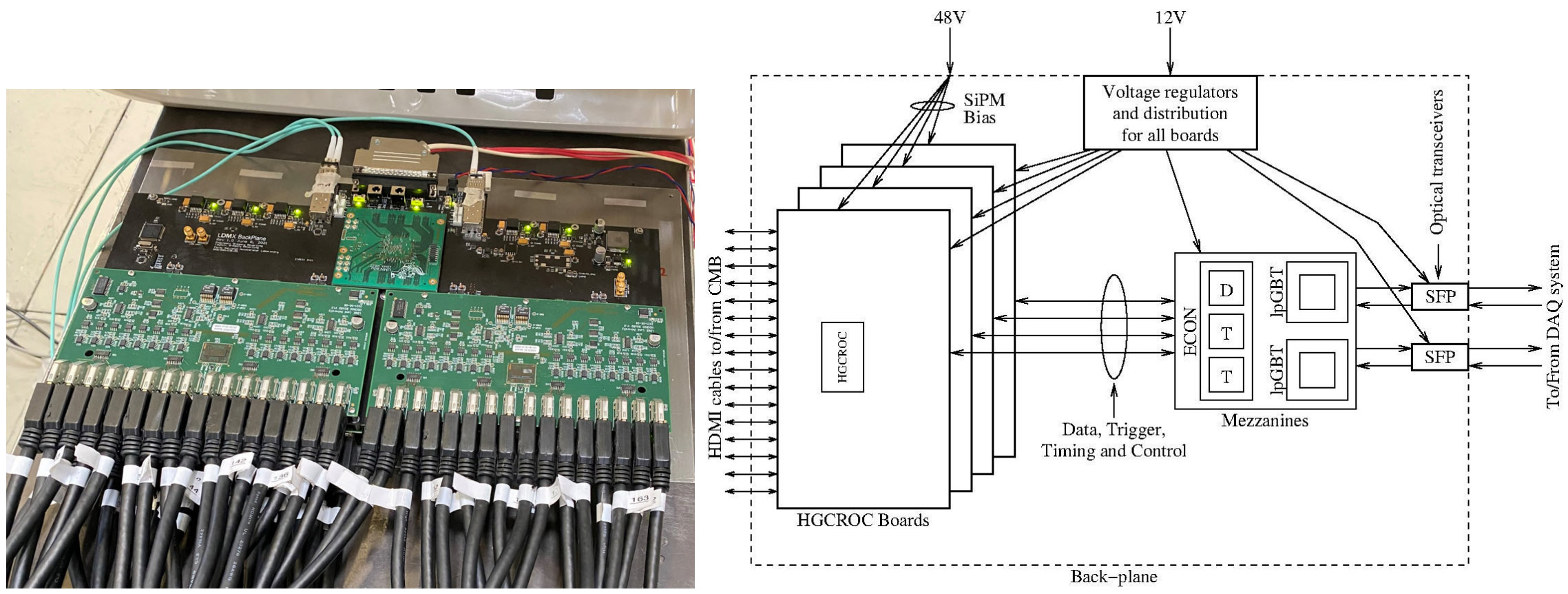

Figure 3.106: Left: Backplane board with two HGCROC boards on the top and two on the bottom, all connected with HDMI cables to CMBs. A Mezzanine card with an FPGA, communicating with optical fibers to the DAQ and the trigger, is also shown. Right: Block diagram of the backplane board.

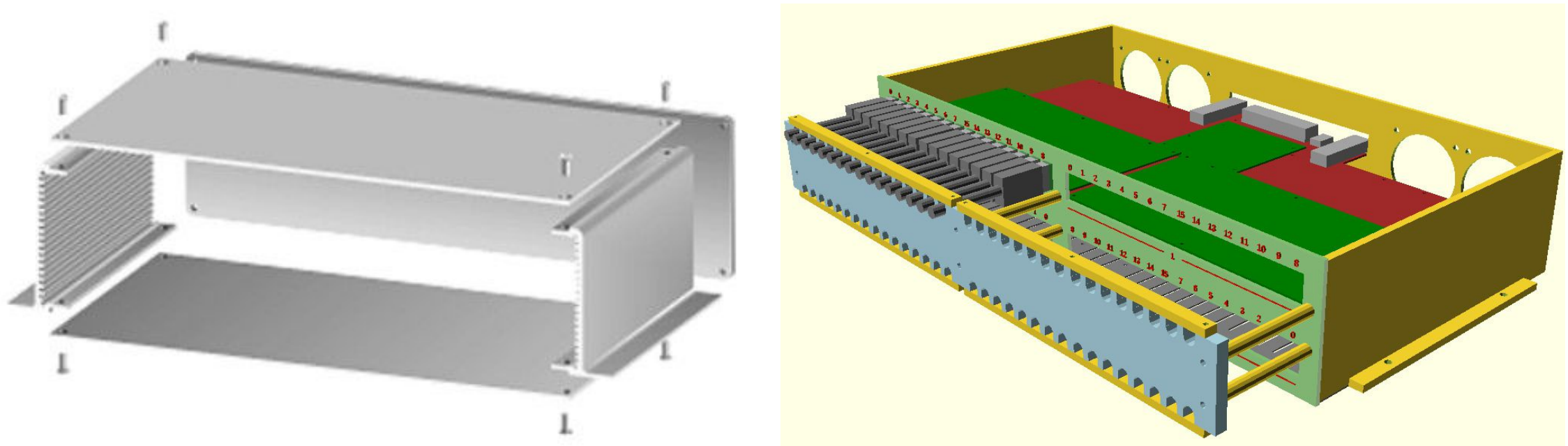

Figure 3.107: Left: Crate for one readout electronics unit of 256 channels. Produced by Enclosures & Cases (https://enclosuresandcasesinc.com/). Right: Orientation of the electronics unit in the crate, and the connections with HDMI cables with strain relief supports.

### 3.8.5 Services

#### 3.8.5.1 Low-voltage system

Each 256-channel readout electronics unit is supplied by a single 12V line and one 48V line. The 12V line takes $\approx$ 16W, but instantaneous consumption is 50% higher during switch-on, so a sequential initialization will be applied when powering up all electronics units. Furthermore, the 12V line needs a small amount of additional power to drive fans for the crate. The 48V line supplying the SiPM bias takes $\approx$ 5W (most of which is dissipated on the HGCROC Board).

The total power of the HCal electronics will be about 1 kW. As an example, this can be supplied by 5 Wiener MPV8016I modules for the 12V and 5 Wiener MPV8060I modules for the 48V. All modules can be hosted by 2 Wiener Mpod EC crates with controllers.

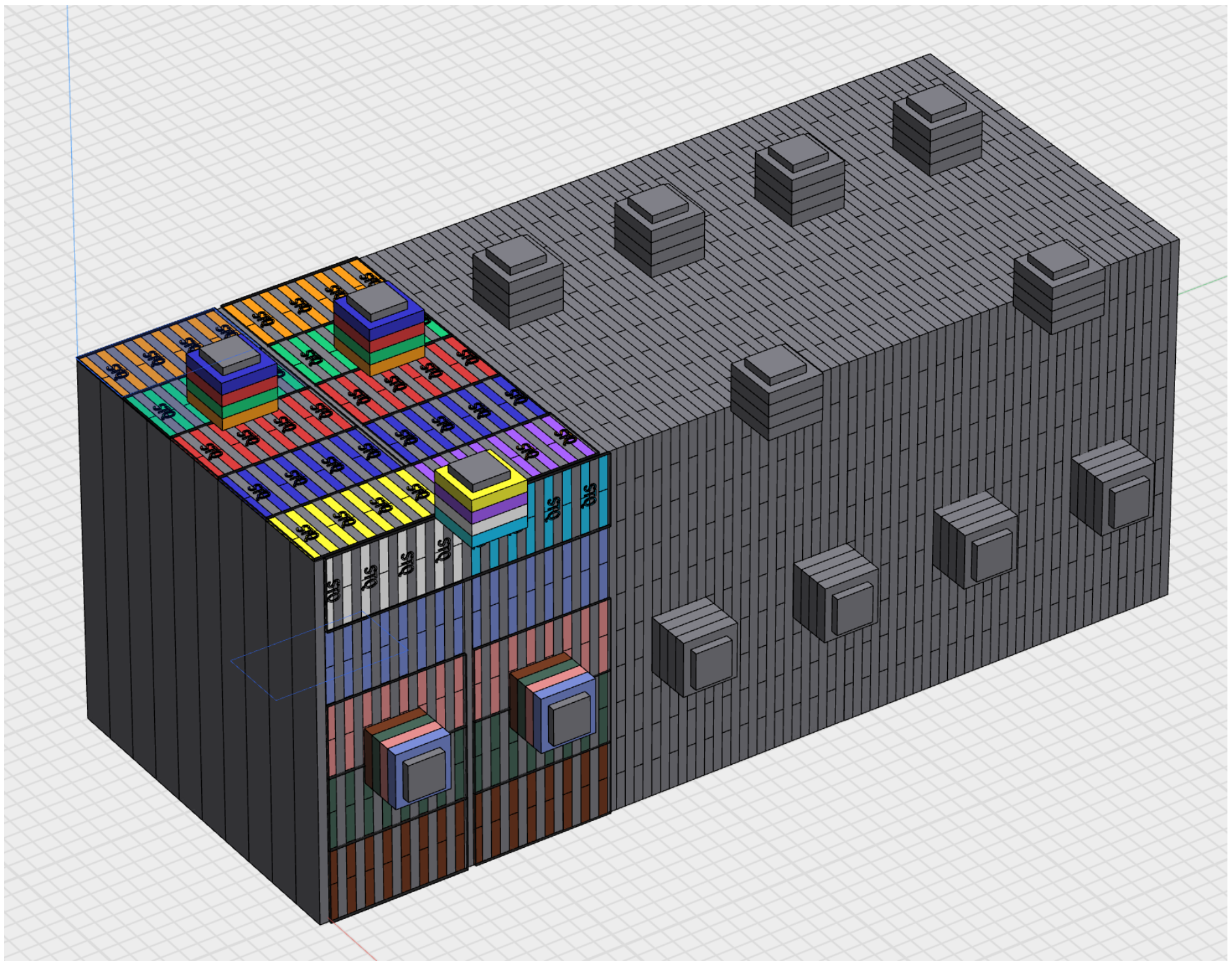

Figure 3.108: Schematic of the grouping of bars for the Back HCal trigger. Four groups of 16 CMBs (orange, green, red, and blue for vertical bars, and brown, olive green, pink and slate blue for horizontal bars) are associated with one backplane board. A grouping of horizontal and vertical bars (yellow, magenta, gray and cyan) is connected to another board, serving two contiguous sides of the Back HCal. This grouping covers 32 planes. Three identical groups serve the 96 planes. The arrangement is repeated on the two sides that are not visible in the figure, reading out the other ends of the bars.

### 3.8.6 Prototype and test beam

A focal point of R&D activities was the operation of Trigger Scintillator and HCal prototypes at CERN in April 2022. In addition to developing and operating fully integrated subsystems, test beam data provides valuable information to tune the Monte Carlo simulation as well as optimize the HCal design. Several million events were collected using hadron, muon, and electron beams with energies varying between 300 MeV and 4 GeV. The calibration procedure, veto capabilities, and detector performance are summarized below. With the successful completion of the CERN test beam program, the HCal prototype will be re-assembled at SLAC to enable further testing and integration work.

#### 3.8.6.1 CERN East Hall and T9 beam line

The test beam was conducted at the newly renovated T9 beam line, located in the East area experimental Hall at CERN. An overview of the Experimental Hall and the T9 beam line is shown in Fig. 3.110. The experimental area is roughly 5m $\times$ 12m, located next to the control room. Secondary beams of hadrons and electrons with energies up to 15 GeV at a production angle of zero degrees are derived from the interaction of the 24 GeV primary beam from the Proton Synchrotron (PS) with a target. The beam is directed to the HCal prototype and delivered uniformly in time over a burst of 0.4 s, provided once or twice every PS cycle. The particle intensity is controlled by a series of collimators, reaching up to $10^6$ particles per burst. The particle content can be modified by selecting different secondary target heads: low-Z material for purer hadron beams and high-Z material for a larger electron component.

The beam line is instrumented with two Cherenkov detectors, two sets of fiber tracker stations to measure the trajectory of the incoming particles, and scintillator paddles at the exit of the beam line to produce a fast trigger signal. The pressure and type of gas in the Cherenkov detectors can be adjusted to discriminate between electrons, pions, and heavier particles. The information recorded by the fiber tracker is recorded on a dedicated server, timestamped by a clock system synchronized over the full experimental area. While the Cherenkov detectors ensured a clean selection of high-energy electrons and pions, substantial electron contamination was found in the low-energy pion sample after data taking, and this sample was discarded from the final data analysis.

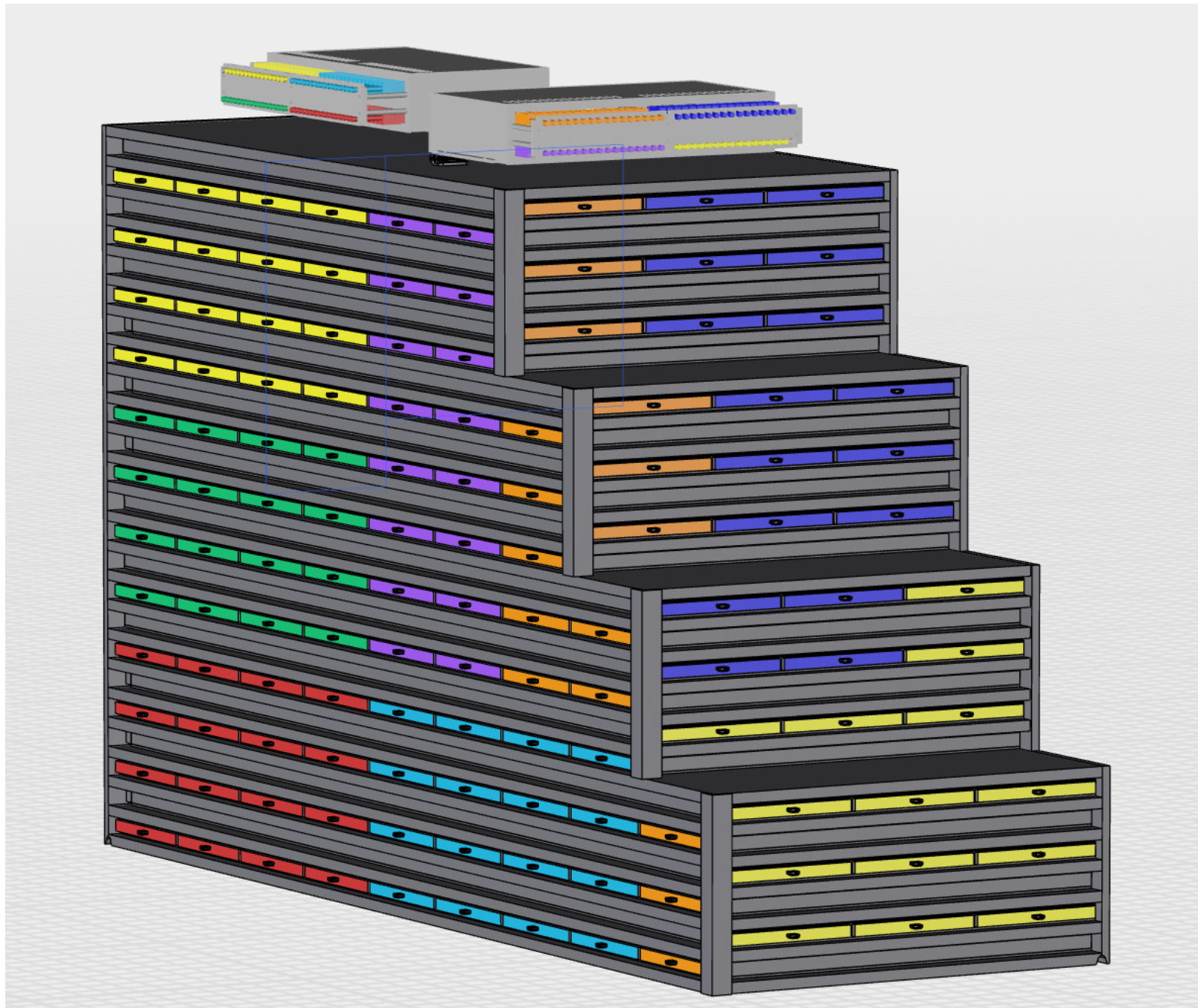

Figure 3.109: The grouping of HDMI cables for the Side HCal bars, which have single-ended readout. One grouping of 16 HDMI cables colored as yellow, red, green, and magenta is connected to a single HGCROC crate. The second grouping (blue, orange, yellow, and cyan) is connected to a second crate.

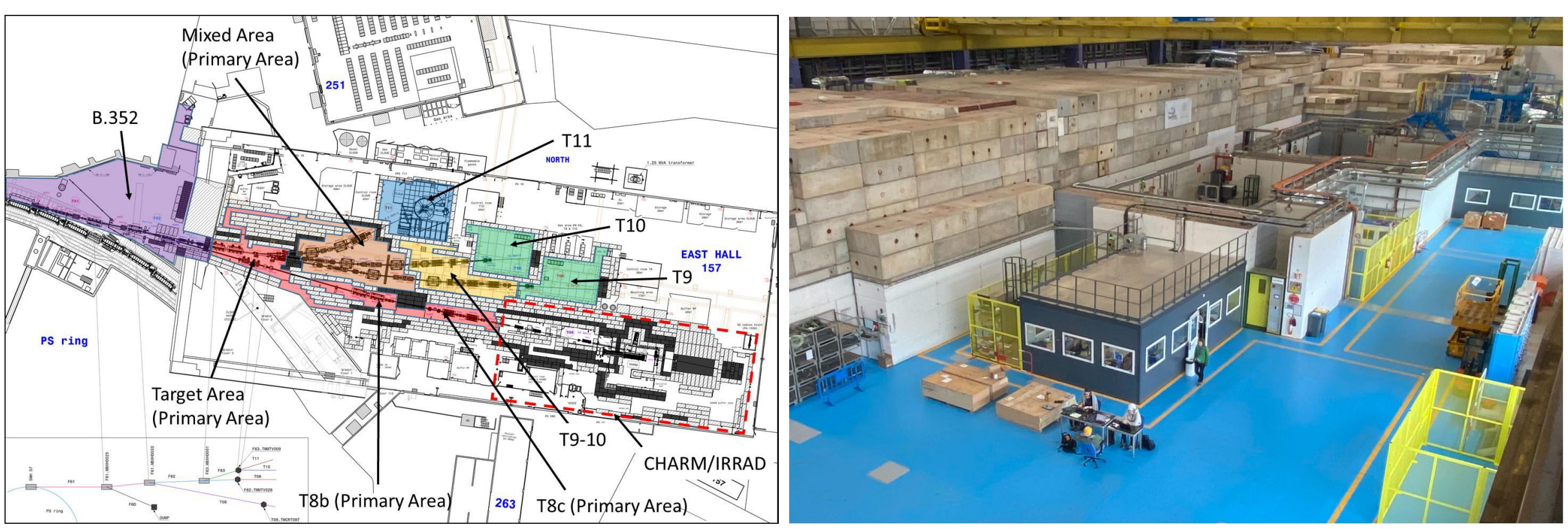

Figure 3.110: Left: Overview of the East Hall experimental area and T9 beam line at CERN. Right: bird's-eye view of the T9 experimental area.

#### 3.8.6.2 HCal prototype

The HCal prototype consists of 19 layers of 25-mm steel absorber (about 3 nuclear interaction lengths), sufficient to fully contain electromagnetic showers and a reasonable fraction of hadronic showers. The first nine layers comprise two quad-counters, while the remaining 10 layers contain three quad-counters. The layers are alternating between vertical and horizontal orientations.

The production of extruded scintillator and the quad-counter assembly are described in Sec. 3.8.3. The scintillator was extruded at the FNAL-NICADD Extrusion Line Facility with a co-extruded $TiO_2$ reflector. Wavelength-shifting fibers of 1.8 mm diameter were inserted into each bar hole without any additional epoxy. Four scintillating bars were glued into a quad-counter assembly, as shown in Fig. 3.111. A four-channel on-detector electronics unit comprised of four SiPMs[5] was mounted directly on the end of each quad-counter. A calibration with a $^{90}$Sr source was performed to determine the light yield of each scintillating bar. An average value of $\sim 14$ photo-electrons/ MeV was determined with an average fluctuation of 0.7 photo-electrons/ MeV.

The SiPM signals from each quad-counter were transmitted to a High Granularity Calorimeter Read Out Chip (HGCROC [108] version 2) board via an HDMI cable. One large backplane board hosted four HGCROC boards plus a mezzanine card providing communication with the DAQ, trigger, clock, and control. Two backplane boards (see Fig. 3.111), each with three HGCROC boards, were used for the prototype readout, for a total of 384 channels. A slightly different system based on the ECON-D and lpGBT cards will be used for the final system, as described in Sec. 3.8.4. The trigger scintillator prototype, discussed in Sec. 3.5.5, is mounted at the front of the prototype. The full apparatus installed in the T9 experimental area is shown in Fig. 3.112.

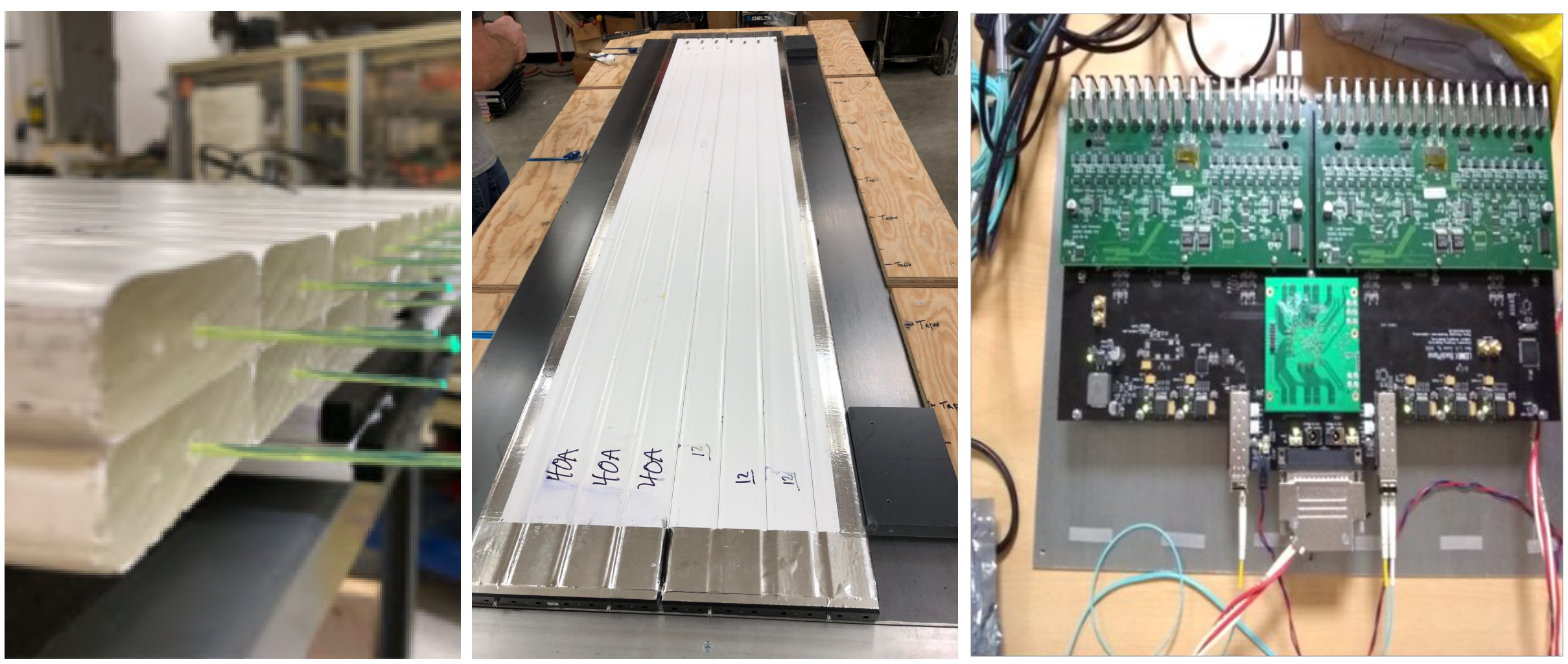

Figure 3.111: Left: Wavelength shifting fiber inserted into each scintillating bar. Middle: Assembly of two quad-counters on a 3 mm steel plate. Right: Readout electronics with one backplane board (dark green), two HGCROC boards (green), and one mezzanine board (light green).

#### 3.8.6.3 ADC-TOT calibration

The HGCROC preamplifier converts the input charge to output voltage with an amplitude proportional to the input charge until a saturation point is reached. Above that point, the width of the preamplifier pulse is used to estimate the input charge. The HGCROC maximizes the dynamic range by providing both an ADC response in the pre-saturation region and a time-over-threshold (TOT) response above it.

A charge injection circuit can inject charges of known amplitude synchronized in time with the readout into individual channels to cross-calibrate the ADC and TOT responses (see Sec. 3.8.4). Fig. 3.113 shows the HGCROC output as a function of the injected charge for a representative channel, controlled by selectable capacitors internal to the HGCROC. The sum of the ADC values, the maximum ADC value, and the TOT value from the event are displayed. The step-like structure in the sum of ADC measurements is due to progressive saturation across all ADC values, increasing with the injected charge. The preamplifier starts saturating around 0.6 pC, where the TOT “turn-on” occurs. Both the ADC and TOT exhibit a nonlinear behavior near this region. Correction factors are obtained by performing a linear fit of the sum of the ADC as a function of the injected charge in the range $0 - 0.5\,\text{pC}$, as shown in Fig. 3.113. The TOT range is then split into two regions to account for the non-linear turn-on regime and the linear behavior at larger injected charges. In both regions, the TOT is fit as a function of the ADC sum, obtained by converting the injected charge into ADC units. The calibrated TOT, expressed in ADC units, is shown in Fig. 3.113. A continuous, monotonically increasing response between the sum of ADC and calibrated TOT is observed.

[5] Hamamatsu S14160-3015PS

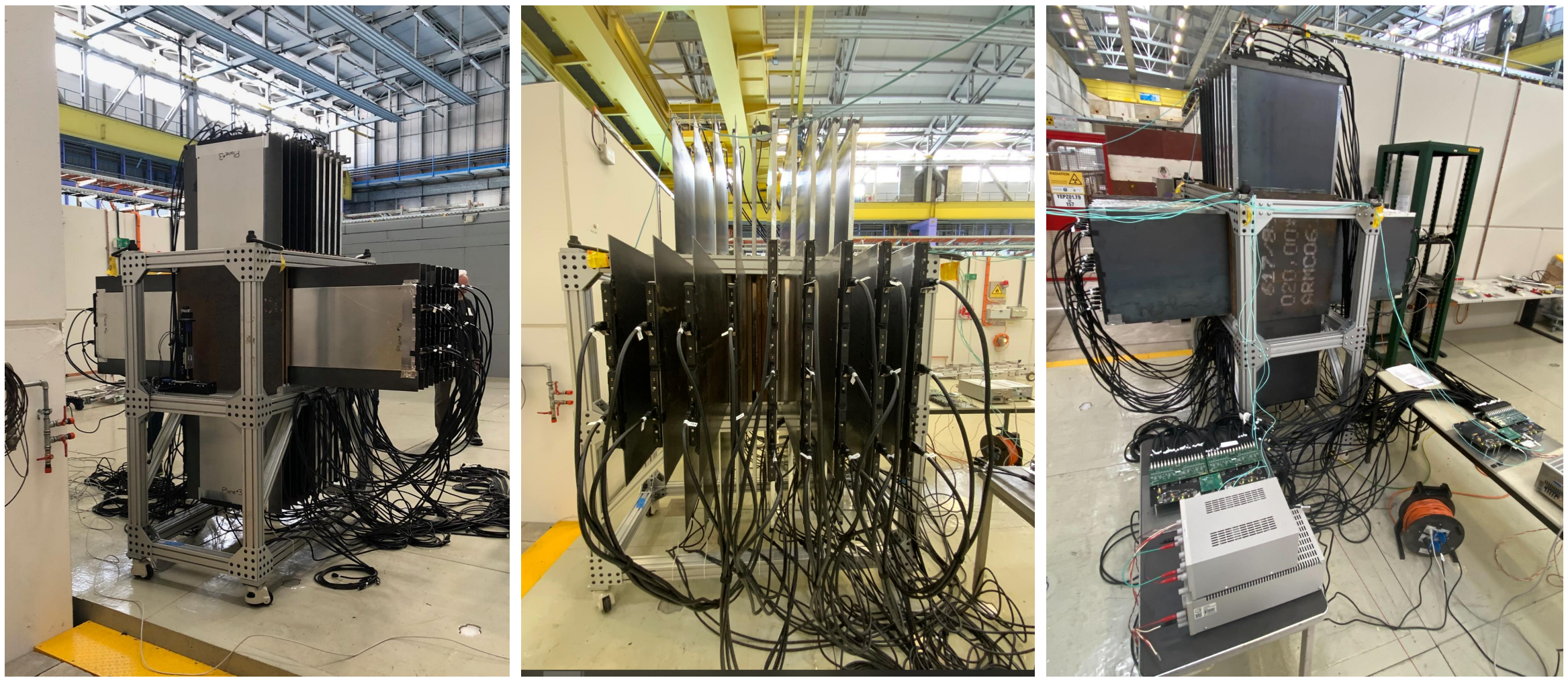

Figure 3.112: Front, lateral and rear view of the HCal prototype installed in the T9 experimental area, together with the readout electronics in the back and the trigger scintillator in the front of the HCal.

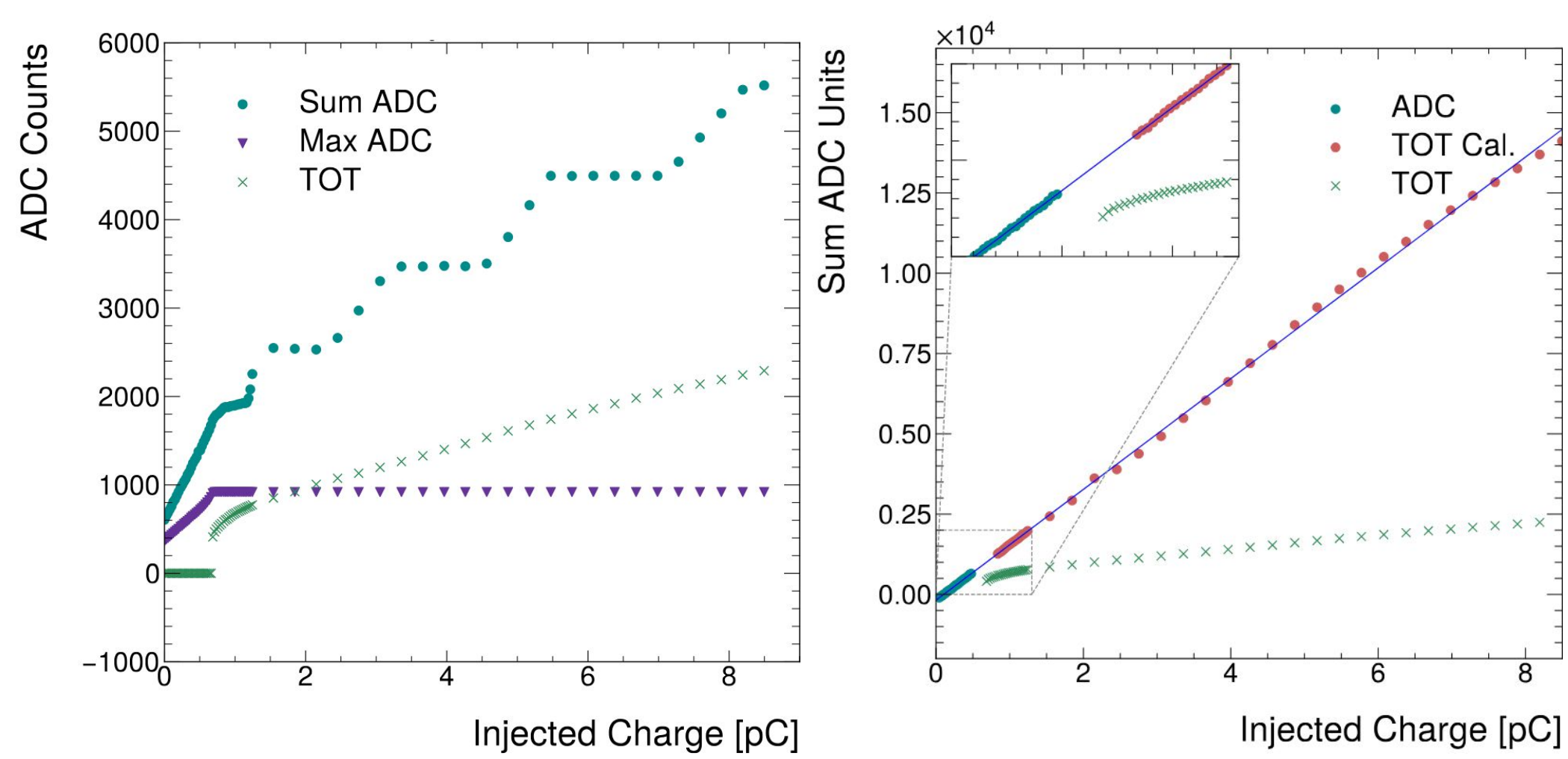

Figure 3.113: Left: The HGCROC output as a function of the injected charge. The sum of the ADC values (blue dots), the maximum ADC value (purple triangles), and the TOT value (green cross) are shown. Both the ADC and TOT exhibit non-linear behavior in the vicinity of the saturation region ($\sim 0.6$ pC). Right: The sum of ADC, pre-calibrated TOT (TOT), and post-calibrated TOT (TOT Cal) as a function of the injected charge.

#### 3.8.6.4 Pedestals and noise level

The noise level of the HCal electronics is studied with a dedicated data set collected without any beam delivered to the detector, as well as a data set with a $4\,\text{GeV}$ muon beam. The pedestals and the noise are determined for each channel by fitting a Gaussian function to the distribution of ADC in the beam-free data, or the mean ADC of the readout waveform for the muon data. Two sample fits are shown in Fig. 3.114. The distribution of the noise level for all channels, expressed in MIP equivalents, is in figure 3.115. The average of the distribution is found to be approximately 2.2% MIP equivalent, with a standard deviation of approximately 0.5% MIP equivalent. Furthermore, the peak-to-peak separation of the noise pedestal and the MIP peak expressed noise pedestal standard deviations $\sigma$ is shown in the same figure. A large separation between the pedestal and MIP peak and a low noise level is seen. Systematic uncertainties were assessed by analyzing pedestal runs taken over several days and found to be at the level of 0.3% MIP equivalent, demonstrating the excellent stability of the system. A high-statistics measurement was also performed with a laboratory test setup to test the stability of the pedestal over a long period and measure the Gaussian behavior of the tails. Non-Gaussian tails are found to be suppressed by at least five orders of magnitude, sufficient to meet the requirements.

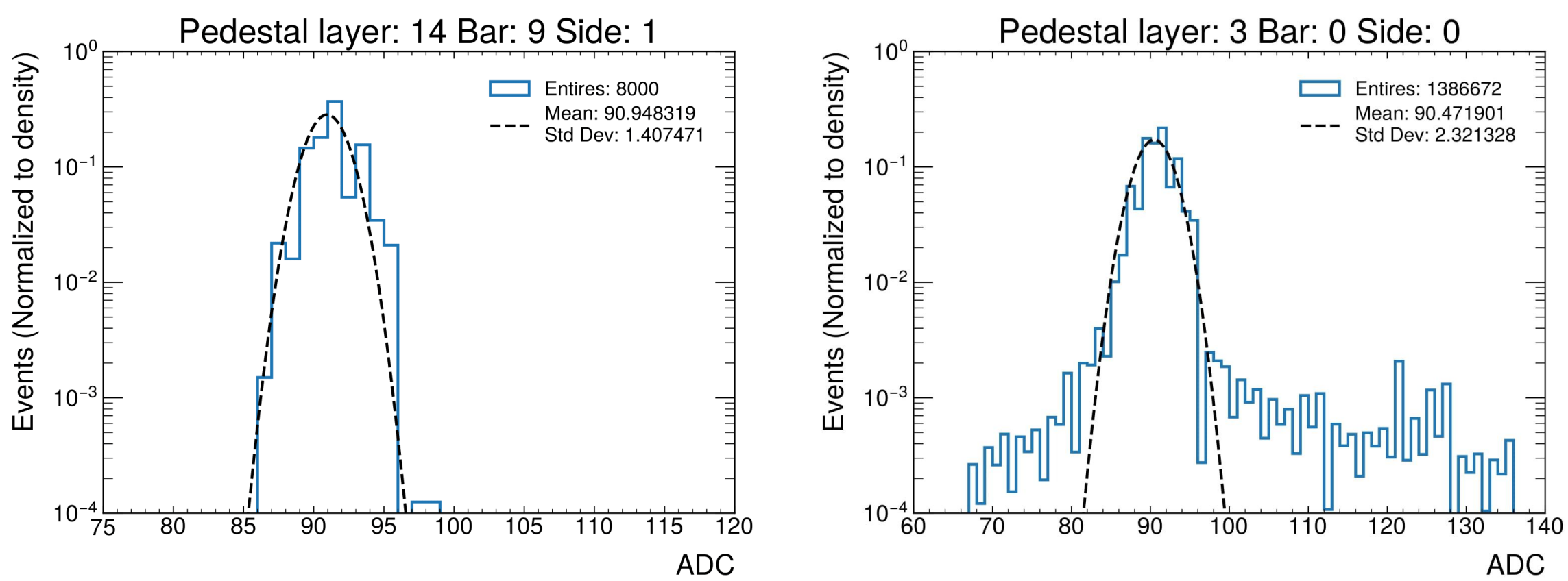

Figure 3.114: Left: Distribution of ADC values for a sample channel for data taken without a beam delivered to the detector showing a Gaussian noise distribution up to $\sim 4\sigma$ tails. Right: Distribution of the mean of ADC values over a readout series in the noise pedestal region for a sample channel for data taken with a 4 GeV muon beam.

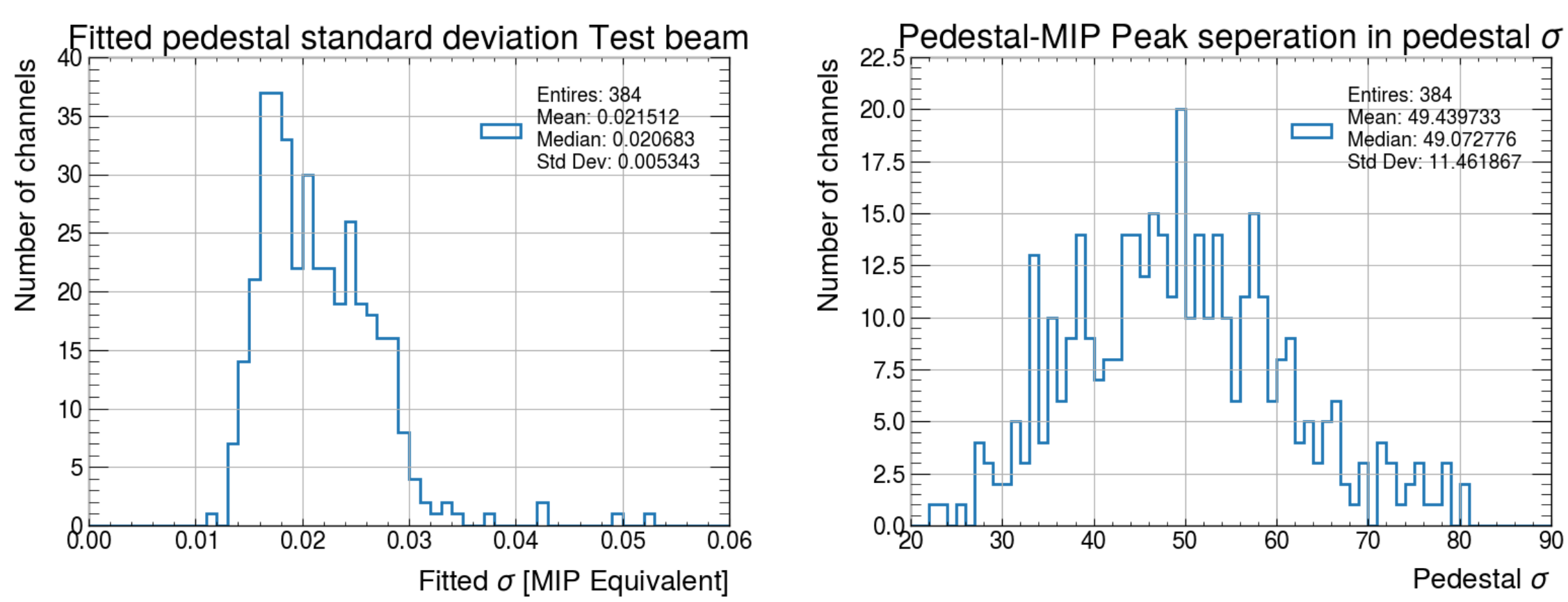

Figure 3.115: Left: The distribution of the noise level for all channels of the prototype in MIP equivalents. The mean and uncertainty of the distribution are $2.2\% \pm 0.5\%$ MIP equivalent. Right: Distribution of the peak-to-peak separations of the pedestal and the MIP-peak for data taken with a 4 GeV muon beam.

#### 3.8.6.5 Energy scale calibration

The response of each scintillator bar is calibrated using a 4 GeV muon beam, tuned so that the probability of having several muons crossing the same scintillator bar is negligible. The response of each channel is reconstructed by summing the pedestal-subtracted ADC values and fitting the resulting distribution with a Landau distribution convoluted with a Gaussian function. The left plot in Fig. 3.116 displays the observed shape of a MIP distribution. For each channel, the most probable value (MPV) represents the shortest possible path for a MIP transiting the bar, defining the "MIP equivalent" of the ADC sums (4.83 MeV). The calibration is applied to a separate data set collected with a 4 GeV muon beam. The distribution of ADC counts expressed in MIP equivalents for a single channel is shown in the right plot in Fig. 3.116. The calibration is found to be in excellent agreement with expectations.

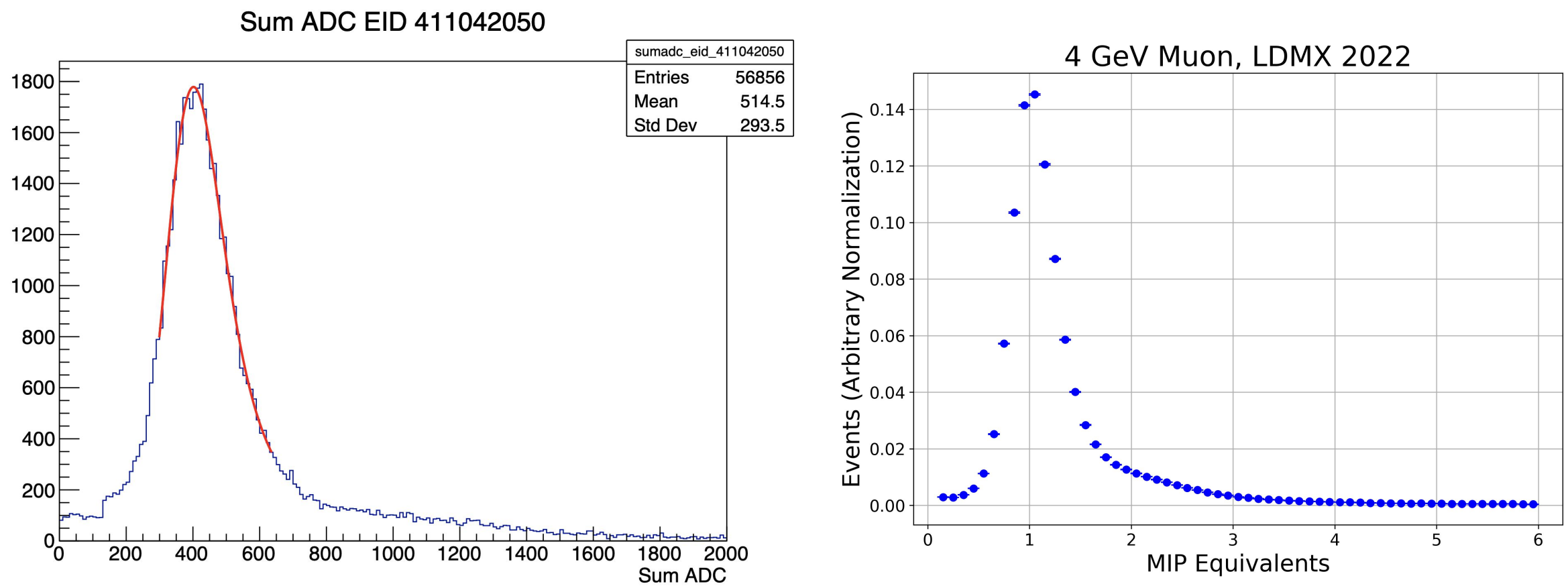

Figure 3.116: Left: Distribution of the sum of ADC values for a sample channel, together with a fit to a Landau distribution convoluted with a Gaussian function. The MIP calibration constant (the "MIP equivalent") is defined as the most probable value. Right: The distribution of ADC counts expressed in MIP equivalents using an independent data set collected with a muon beam.

#### 3.8.6.6 MIP detection efficiency

The MIP detection efficiency per bar is measured using data collected with the 4 GeV muon beam. By selecting MIP-like tracks emitted in the forward region, the efficiency of the central bars can be measured. To avoid bias, the selection criteria are applied only to bars in front of and behind the one measured. The MIP efficiency for each bar, as a function of the threshold on pulse amplitude (in mean ADC counts), is then defined as the fraction of MIP events that fall above that threshold. The amplitudes from both read-out ends are considered together, such that the efficiency is representative of the bar as a unit. The MIP efficiency for a sample bar is shown in Fig. 3.117. A Monte Carlo simulation was performed by setting the noise to the values measured from the test beam (see Sec. 3.8.6.4). The simulated MIP efficiency, also shown in Fig. 3.117, reproduces the data reasonably well.

The distribution of MIP efficiencies of the central bars is displayed in Fig. 3.117 for both data and Monte Carlo simulations. An efficiency greater than 98% per bar can be obtained by setting the threshold to five standard deviations of the noise. This level of efficiency is sufficient to reconstruct MIP tracks with the full HCal with an inefficiency well below the requirements.

#### 3.8.6.7 Energy resolution and linearity

The calorimeter response is determined by adding the energy deposited in each scintillator bar, using the calibration constants previously defined. The resulting distributions are fit with Gaussian functions to extract the mean reconstructed energy and the resolution ($\sigma_E/E$). The results are shown in Fig. 3.118. The calorimeter response is found to be almost linear up to 1 GeV, with the response at 4 GeV dropping slightly below the expectation obtained by fitting the low-energy measurement with a linear function. The resolution is well described with a function of the form $\frac{\sigma}{E} = \frac{a}{\sqrt{E}} \oplus \frac{b}{E} \oplus c$ where $\oplus$ denotes the sum in quadrature. A resolution of $\sim 15\%$ at $4\,\mathrm{GeV}$ ($\sim 30\%$ at $3\,\mathrm{GeV}$) is observed for electrons (pions), in line with the expected performance of a steel-scintillator calorimeter with this sampling structure . Dedicated Monte Carlo samples will be generated to study and tune the simulation to these results.

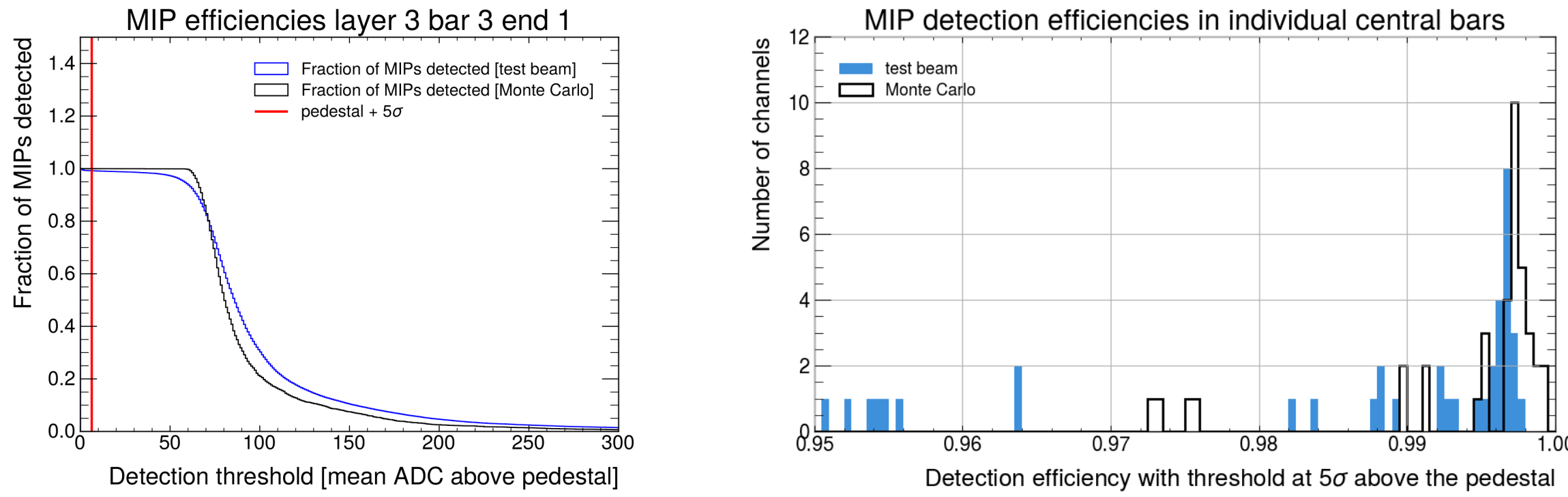

Figure 3.117: Left: The MIP efficiency as a function of the threshold on the pulse amplitude for a sample bar from test beam data and Monte Carlo simulations. Right: Distribution of MIP efficiencies for the central bars of the HCal prototype.

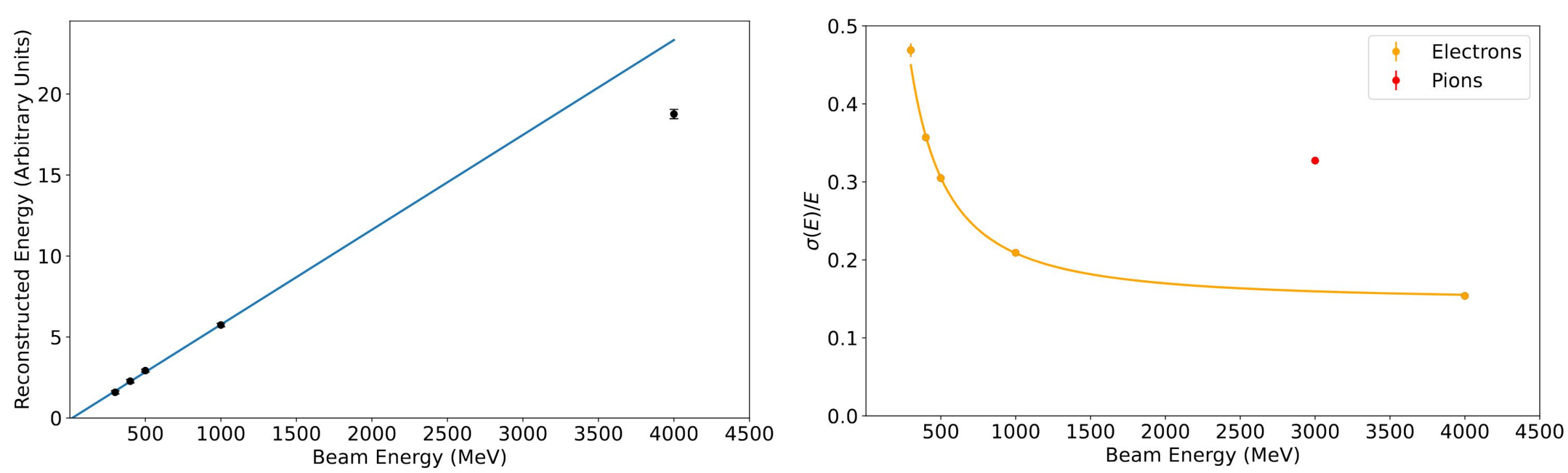

Figure 3.118: Left: the energy response as a function of the beam energy for electrons, together with a linear fit performed in the region $300\,\mathrm{MeV} - 1\,\mathrm{GeV}$. Right: the energy resolution as a function of the beam energy for electrons and pions, together with a fit of the form $\frac{\sigma}{E} = \frac{a}{\sqrt{E}} \oplus \frac{b}{E} \oplus c$ for electrons. Only statistical uncertainties are shown.

#### 3.8.6.8 Lateral and longitudinal shower development

Shower development in a calorimeter depends on the species of particle incident on the calorimeter and its energy. Particle identification with the HCal is valuable if a dark mediator decays back into visible SM particles. As the HCal dominates the LDMX volume, that decay stands a good chance of appearing in the HCal. Shower development characteristics extracted from the testbeam data set will validate and improve the HCal simulation.

Electron shower profiles were determined from the average deposited energy in each scintillator bar (each layer) for transverse profiles (longitudinal profiles). Only bars registering hits with ADC counts $5\sigma$ above the pedestal value were included in the calculation. To remove pion contamination, events with hits in the last five layers of the prototype were removed. Events with multiple bars registering hits in Layer 1, indicating that the electron showered before hitting the prototype, were also discarded. Critically, the HCal prototype includes a steel absorber layer before the first scintillator layer (marked as Layer 1), meaning the HCal prototype does not see the incident electron before its first showering. For 300 MeV electrons, the shower maximum is expected between Layers 1 and 2 of the HCal prototype.

As shown in Fig. 3.119, the longitudinal shower profile for 300 MeV electrons resembles that of an electromagnetic shower. The strong peak in deposited energy in Layers 1 and 2 is consistent with the expectation. By Layer 3, each particle's energy is below the critical energy and the shower is slowly absorbed. The two ends of the scintillator bar are read out separately, and both results are consistent. The transverse shower profile in four different layers of the HCal prototype is also shown in the same figure. The results agree with the expectation that the EM shower widens in the transverse plane the deeper it goes in the detector (indicated by increasing layer number). The measurements from the ends of each bar are internally consistent. It is expected that 99% of shower energy is contained within 3.5 Molière radii, which in the HCal prototype corresponds to roughly the width of two scintillator bars. This is supported by transverse profile measurements, where it can be seen that the two central bars consistently have the most deposited energy. Overall, the electron shower shapes are consistent with expectations.

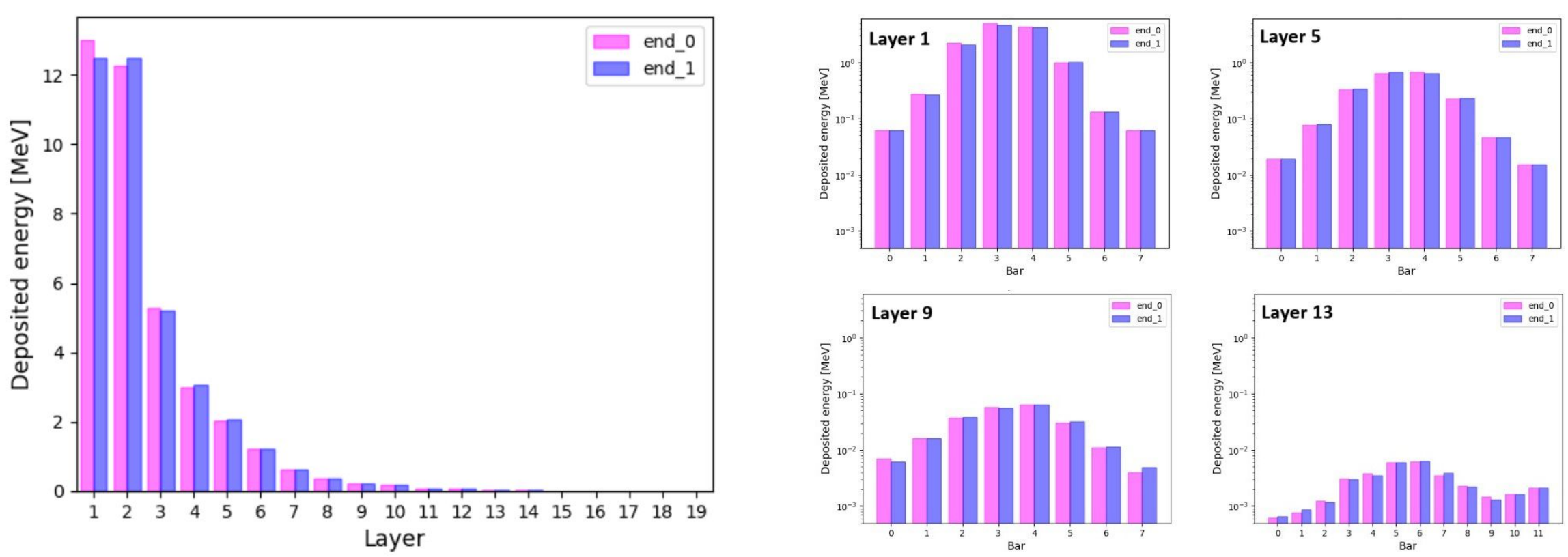

Figure 3.119: Left: Longitudinal shower profile of a 300 MeV electron run. Right: Selected transverse shower profiles from a 300 MeV electron run. The two readout channels are treated separately and can be seen in pink and purple.

### 3.8.7 Performance

#### 3.8.7.1 Photo-nuclear veto performance

A major source of background arises from the emission of hard bremsstrahlung photons followed by a photo-nuclear (PN) reaction in the target or ECal. Most events are vetoed using ECal information, but a subset of topologies in which only a few neutral particles are produced in the final state need to be identified by the HCal. As described in Sec. 4.2, about 70,000 ($7 \times 10^6$) PN events are expected to pass the trigger, tracker, and ECal BDT requirements out of a sample corresponding to a total of $10^{14}$ ($10^{16}$) EoT.

The veto performance as a function of the HCal dimensions is studied using information on the secondary particle(s) produced in the PN interaction. For a given HCal size, an event is considered vetoed if a secondary particle initiates a hadronic interaction within the corresponding fiducial volume. The interaction position is determined from the MC information and does not account for the development of the hadronic shower, which typically spans about 60-80 cm. The number of non-vetoed events as a function of the width (length) of the Back HCal, assuming a nominal length (width), is shown in Fig. 3.120. The tails of both distributions are fit with exponential functions and extrapolated up to $10^{16}$ EoT. Including a buffer space to contain the hadronic shower, a transverse size of 2m and length of 4.7m should prove adequate to veto all PN events and provide an energy measurement with reasonable resolution. Only a few PN events are vetoed solely by the Side HCal, with vetoable hits produced in the first layers closest to the beam by low-energy neutrons. Extrapolating to $10^{16}$ EoT, the combined veto inefficiency must be $\mathcal{O}(10^{-3})$. Assuming that only two neutrons are detected by the Side HCal modules, the inefficiency per neutron must be below 3%, requiring a module depth of 3.5 nuclear interaction lengths.

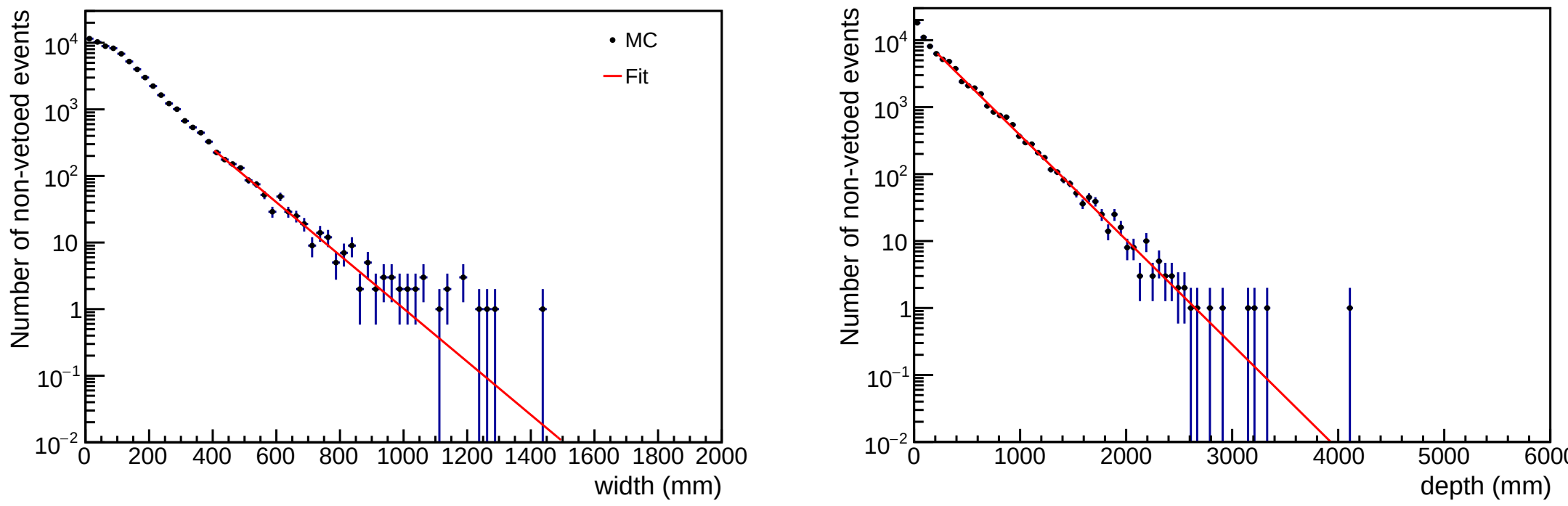

Figure 3.120: The number of non-vetoed events as a function of the Back HCal (left) width and (right) depth for $10^{14}$ EoT, together with the exponential fit described in the text. The position of the interaction point in the Back HCal is determined using MC information and does not account for the development of the hadronic shower, which typically spans about 60-80 cm.

#### 3.8.7.2 Wide-angle Bremsstrahlung veto performance

A comprehensive discussion of the wide-angle bremsstrahlung (WAB) background is given in Sec. 4.3.2.2. About $10^6$ WAB events are expected for a total of $10^{16}$ EoT equivalent. To better study the wide-angle background characteristics, events are studied for two distinct kinematic regions:

- the primary electron is contained within the ECal and the photon deposits energy in the Side HCal ($\sim 2 \times 10^4$ WAB events for $10^{16}$ EoT equivalent);
- the primary electron enters the Side HCal, and deposits most of its energy there.

The end position of the primary electrons and bremsstrahlung photons are shown in Fig. 3.121 for the two kinematic regions, together with the location of all hits in the scintillator bars. In both cases, the Side HCal is necessary to identify the WAB photons or recoiling electrons and reconstruct the corresponding electromagnetic showers. In particular, the Side HCal is critical for vetoing electrons emitted outside the ECal acceptance - a region containing up to $\sim 10\%$ of the signal for large A' masses.

Fig. 3.122 shows the number of hits generated from wide-angle backgrounds as a function of the layer in the Side HCal in which the hit is located. Although most hits are close to the central region, significant information is present within the outer layers of the Side HCal. A dedicated BDT classifier has been developed to identify wide-angle backgrounds, which requires reconstruction of events in the Side HCal. The hits in the outer layers are instrumental in maximizing the veto performance while maintaining good signal efficiency.

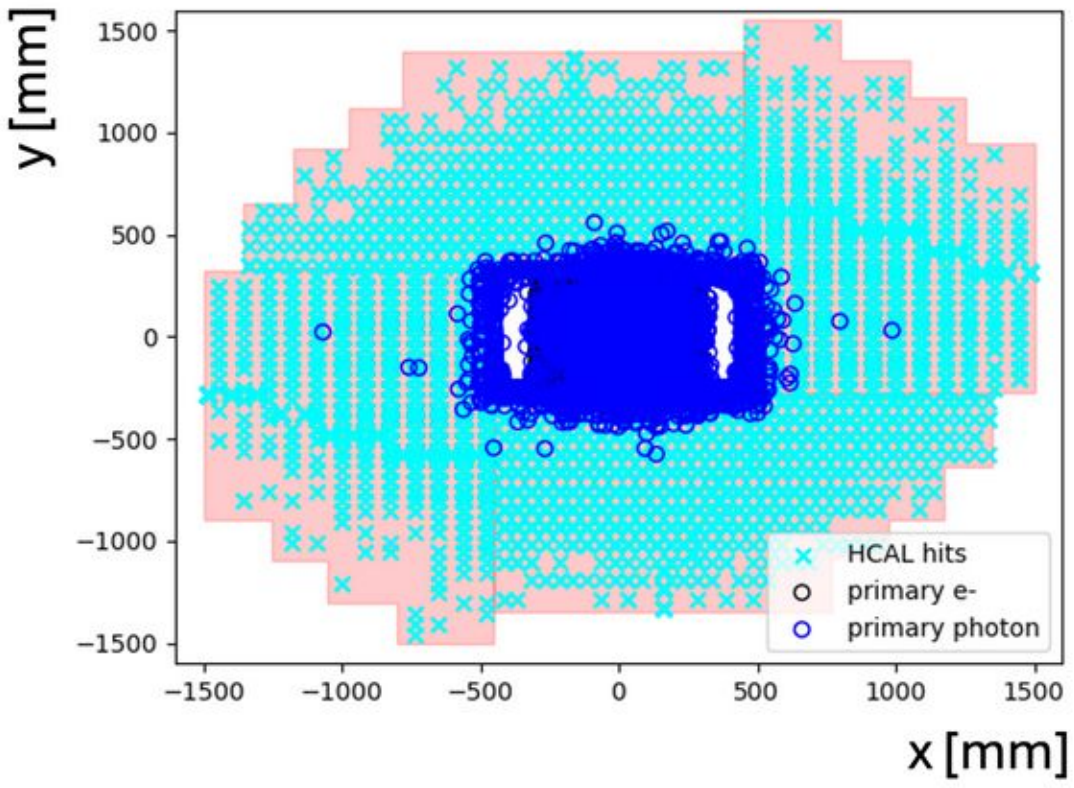

Figure 3.121: The end position of the primary electrons and the bremsstrahlung photon for (left) events in which the primary electron is contained within the ECal and the photon deposits energy in the Side HCal, and (right) events where the electron enters the Side HCal and deposits most of its energy there. Hits reconstructed in the scintillator bars are also shown. A total of 100K events are simulated for both processes.

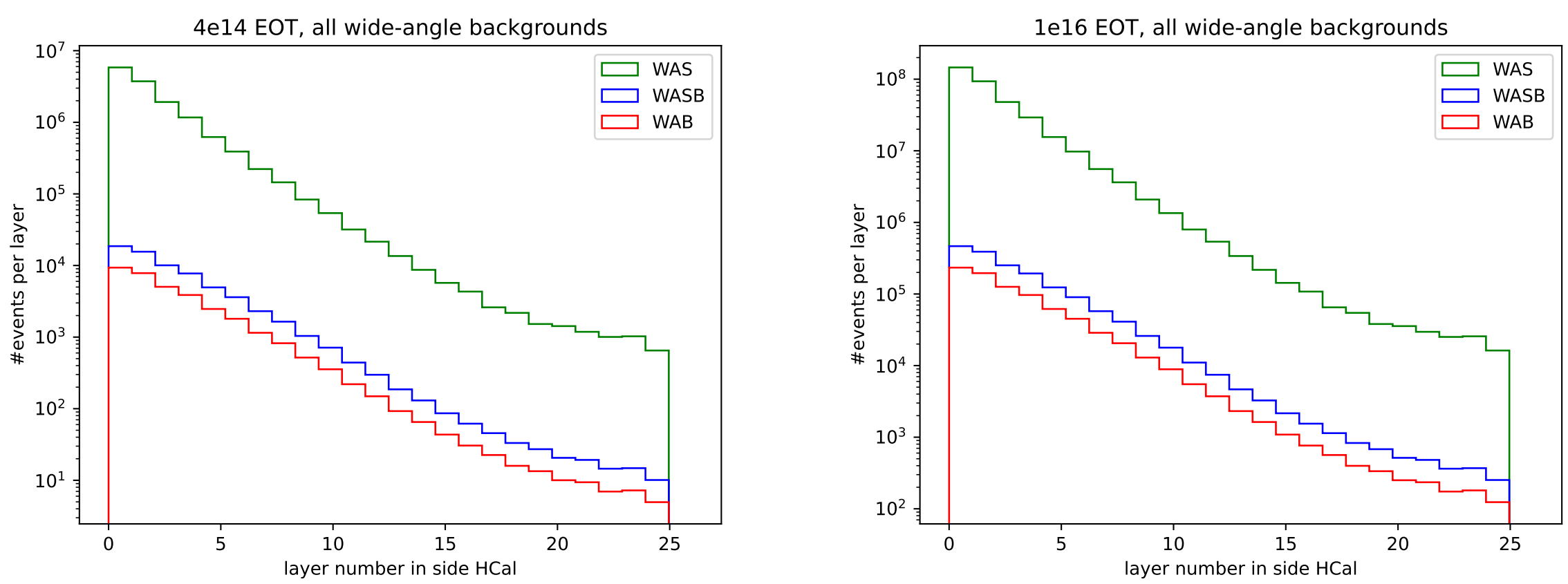

Figure 3.122: Distribution of hits in the Side HCal in each layer for all wide-angle background events from a 10% $X_0$ W target for (left) 4e14 EoT and (right) 1e16 EoT.

#### 3.8.7.3 Light yield performance

The HCal prototype was tested using cosmic rays and a Sr(90) source to determine the light yield of the scintillator bars. The Sr(90) tests were conducted with bare quad-counters to minimize the energy loss of beta electrons in the Al cover, while cosmic rays were used to test the full assembly. The clearly separated photo-electron (PE) peaks for the Sr(90) spectrum allowed for an in-situ calculation of the conversion factor between ADC and PE for each channel. The corresponding light yield spectra are shown for a sample channel in Fig. 3.123. The light yields are determined by fitting a Gaussian function to both the cosmic ray light yield spectra and the Sr(90) beta ray light yield spectra. Taking into account the average energy deposited in the scintillator by the cosmic rays (Sr(90) beta rays), a light yield of $15 \pm 1$ PE per MeV ($14 \pm 1$ PE per MeV) is observed per photosensor, corresponding to a value of $68 \pm 5$ PE per MIP.

Recently, a series of studies were initiated to increase the light yield of the scintillator bars. Various polystyrene materials, reflective paints, and WLS fibers with different diameters are studied through simulations and laboratory tests. Simulations indicate that the Taita 808 polystyrene could produce 50% more light than the AmSty 665 polystyrene currently used by FNAL-NICAAD. A $BaSO_4$ reflector could also increase the light yield by 30% compared to a $TiO_2$ coating, and an R&D program is underway to co-extrude scintillator with the new reflector. Depending on the results, these improvements will be adopted for future production runs.

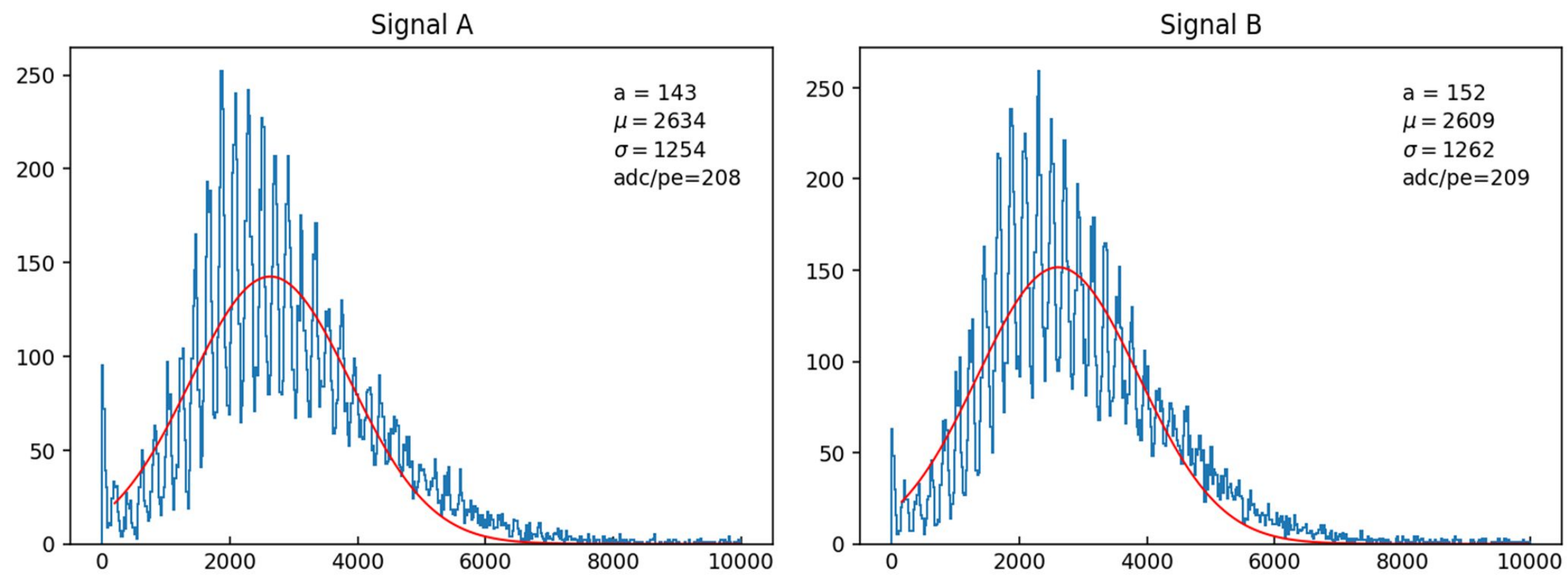

Figure 3.123: Light yield spectra measured using a Sr(90) source for a sample scintillator bar for the left (signal A) and right (signal B) photosensor. The photo-electron peaks are clearly visible. A light yield of 14 PE per MeV is observed.

#### 3.8.7.4 Noise performance

The noise level has been determined in situ using data acquired during the test beam (see Sec. 3.8.6.4). A median level of 2.2%±0.5% MIP equivalent has been observed. These results are used as input to simulate the noise level of the hadronic calorimeter and determine the signal inefficiency induced by false vetoes. The simulation assumes a uniform Gaussian noise level across the 7980 channels of the Back HCal using 1 MIP equivalent = 4.66 MeV = 68 PEs (as seen in Sec 3.8.7.1, the background arising from photo-nuclear reactions is overwhelmingly vetoed by the Back HCal). The noise per scintillating bar is obtained by averaging the value of the readouts at each end. The distribution of the maximum number of PE in the HCal is shown in Fig. 3.124 for noise levels varying between 1% − 3% MIP equivalent. The false veto rate as a function of the noise level is also shown. For example, a noise-induced dead time of 1% requires a veto threshold at the level of 8 PE for a nominal noise level of 2.2%.

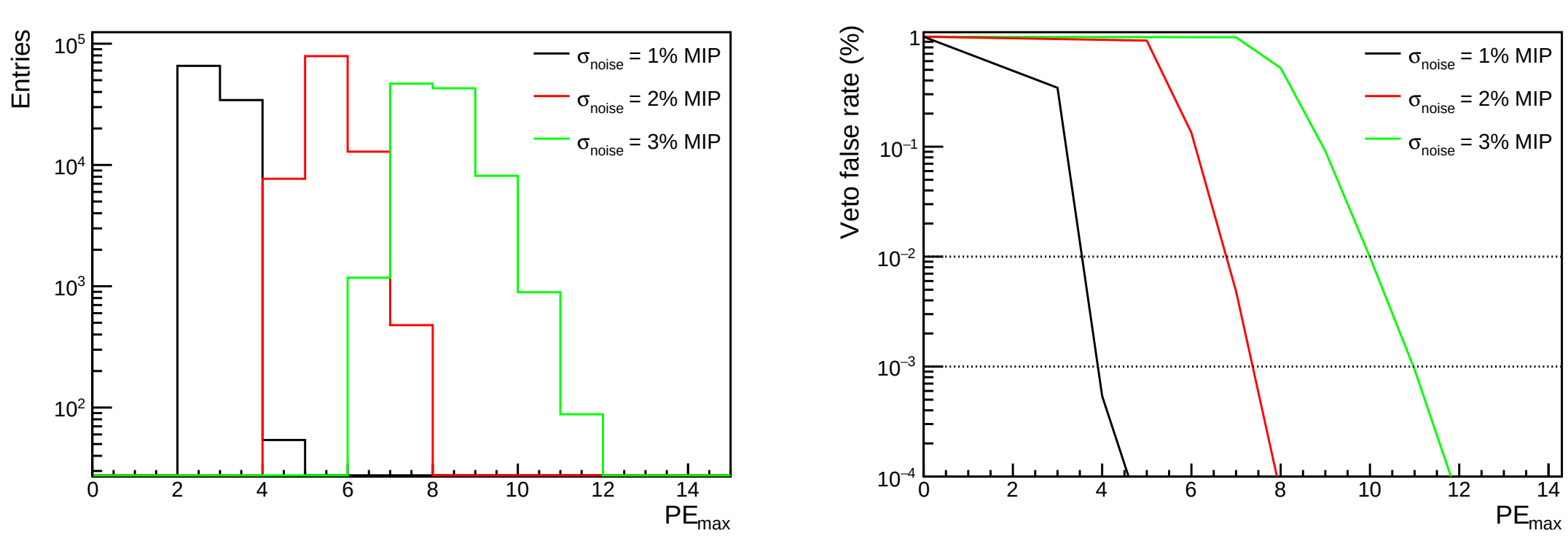

Figure 3.124: Left: The distribution of the maximum PE for noise levels varying between 1% − 3% MIP equivalent. Right: The false veto rate as a function of maximum number of PEs.

#### 3.8.7.5 Energy resolution

The energy resolution is studied using Monte Carlo simulation and the test beam data discussed previously. The MC simulations are performed separately for electrons, neutrons, and charged pions for energies ranging from 0.1 – 10 GeV. Particles are generated at the center of the stopping target location. For the Back HCal, they are emitted at incident angles towards the center of the front face of the absorber plates. For the Side HCal, particles are generated from the stopping target location towards the center of the lateral side of the device, i.e. the side facing the target (see Fig. 3.96), to reflect the topology of electro-nuclear reactions and wide-angle bremsstrahlung emission.

The following quantities are studied for the different particle species: the linearity, the e/h ratio, and the energy resolution of the Back HCal and Side HCal separately. The deposited energy as a function of the beam energy is shown in Fig. 3.125, and a reasonably linear response is observed for all particles. The ratio of the response for electrons and hadrons, the e/h ratio, is displayed in Fig. 3.126 as a function of the beam energy. While the large absorber thickness impacts the e/h ratio below $\sim 2\,\text{GeV}$, its value remains roughly constant around 1.1 (1.3) above this energy for the Back (Side) HCal. This behavior is typical for a hadronic sampling calorimeter [109]. The difference between the Back and Side HCal arises from the angle at which the incident particle hits the calorimeter, and the relative orientation of the device.

Finally, the energy resolution is shown in Fig. 3.127. The different curves are fitted with a function of the form $\frac{\sigma}{E} = \frac{a}{\sqrt{E}} \oplus \frac{b}{E}$ for electrons and $\frac{\sigma}{E} = \frac{a}{\sqrt{E}} \oplus c$ for hadrons. At 1 GeV, the resolution varies between 16% (15%) for electrons and 38% (49%) for neutrons for the Back (Side) HCal. The resolution is significantly worse for the Side HCal as the particles are emitted at a large angle towards the lateral face of the device, effectively crossing a greater amount of absorber material.

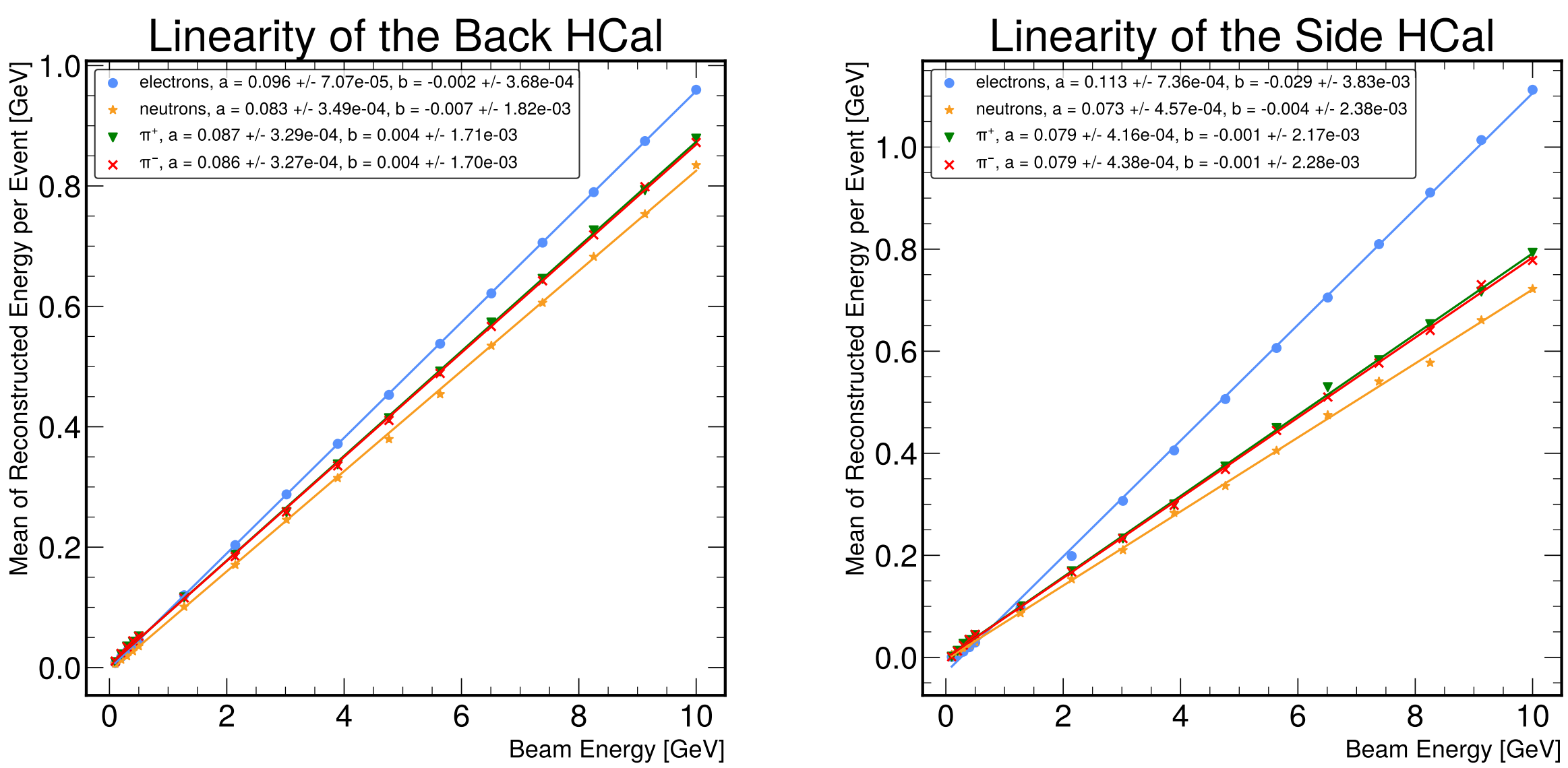

Figure 3.125: Energy response as a function of the beam energy derived from MC simulations for the (left) Back HCal and (right) Side HCal.

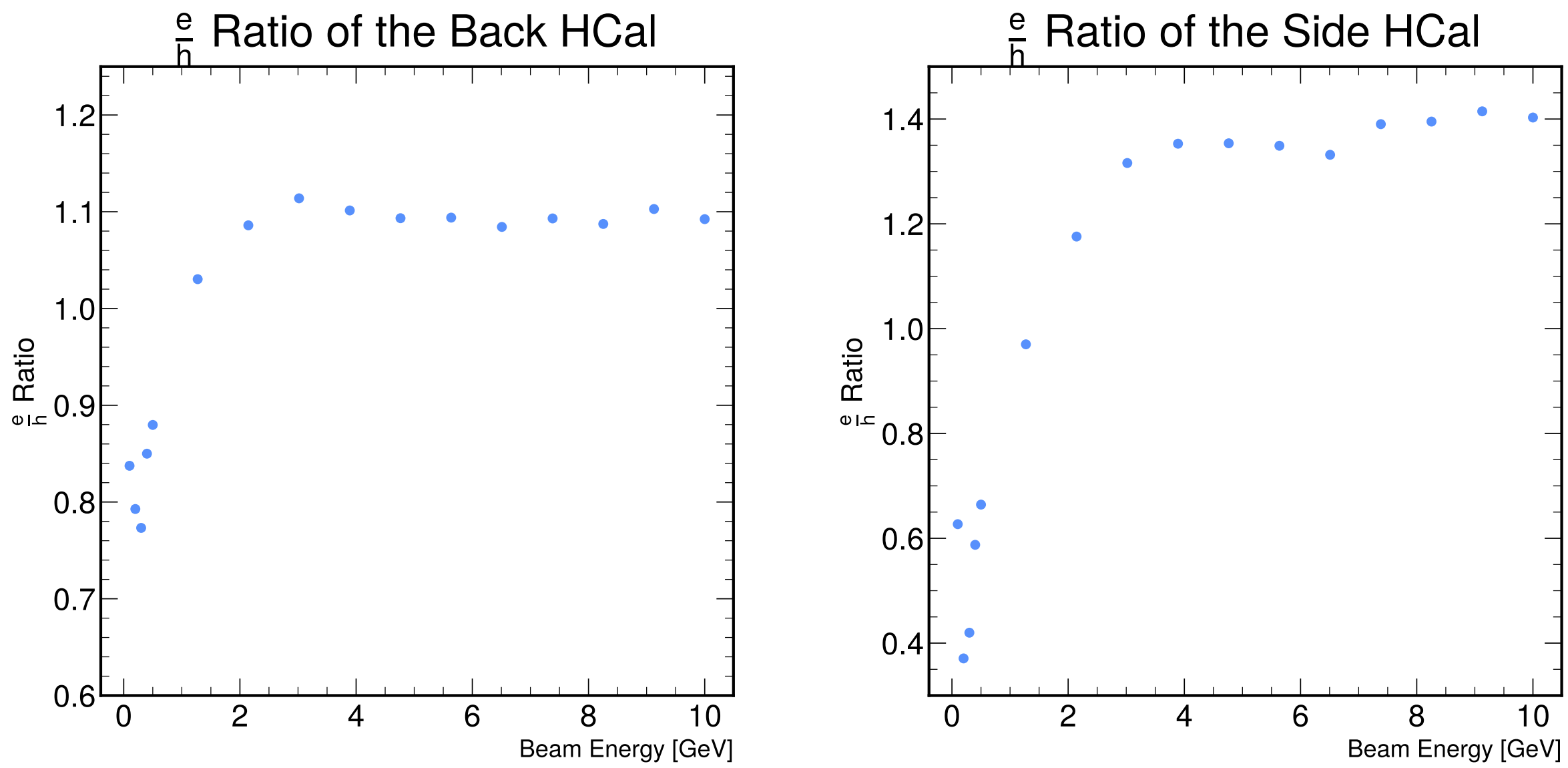

Figure 3.126: The ratio of the response for electrons to hadrons (e/h ratio) as a function of the beam energy for the (left) Back HCal and (right) Side HCal.

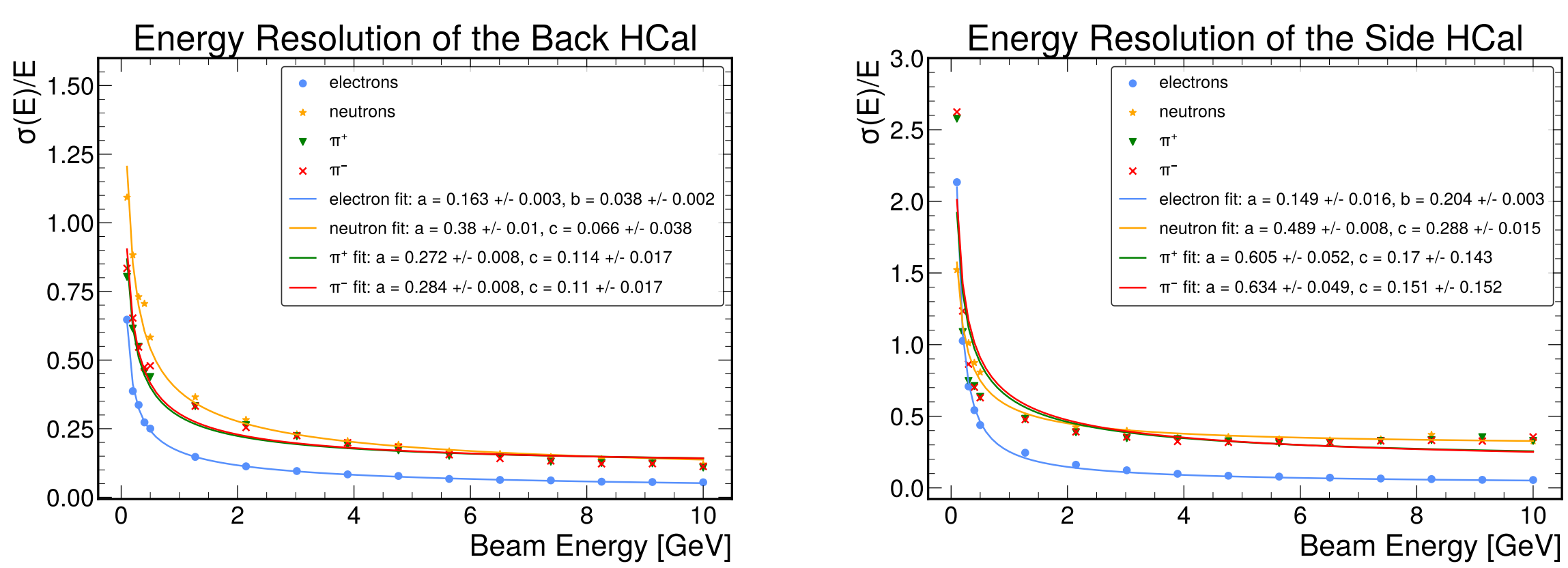

Figure 3.127: Energy resolution as a function of the beam energy derived from MC simulations for the (left) Back HCal and (right) Side HCal.

#### 3.8.7.6 Timing resolution

While the timing resolution has yet to be studied in detail with the HCal prototypes, the performance should be similar to the timing resolution of Mu2e bars studied with a test beam at Fermilab [110]. Using minimum ionizing protons directed at the center of prototype counters, a single-channel time resolution of 1.7 ns was measured by comparing the times recorded on both ends of the counter. As such, we can expect a timing resolution from a single hit in a single bar of the HCal to be slightly better than 2 ns, corresponding to a longitudinal position resolution of about ±20 cm for counters read out on both ends.

#### 3.8.7.7 Photon/neutron separation

The photon/neutron separation power is evaluated using large samples of photons and neutrons generated with energies $500 < E_{\gamma,n} < 2000\,\text{MeV}$. The two species are classified using a Boosted Decision Tree (BDT) based on the following shower shape variables: the number of reconstructed hits; the average longitudinal position of reconstructed hits; the RMS of longitudinal positions of reconstructed hits; the energy deposited per unit length in Z; the average energy deposited in reconstructed hits.

The BDT is trained on sets of photons and neutrons of energies 200 MeV, 500 MeV, 1 GeV and 2 GeV. The signal and background efficiencies as a function of $S_{BDT}$ are displayed in Fig. 3.128. For example, a selection criterion of $S_{BDT} > 0.9$ results in a signal efficiency of 54% and a background efficiency of 1.0%. Further optimization of the BDT or deep learning methods could provide better efficiencies and will be investigated in the future. In addition, the impact of the HCal geometry on the separation power was investigated by repeating the study with different absorber thicknesses. It was found that the performance was only marginally affected by using thicker or thinner absorber plates.

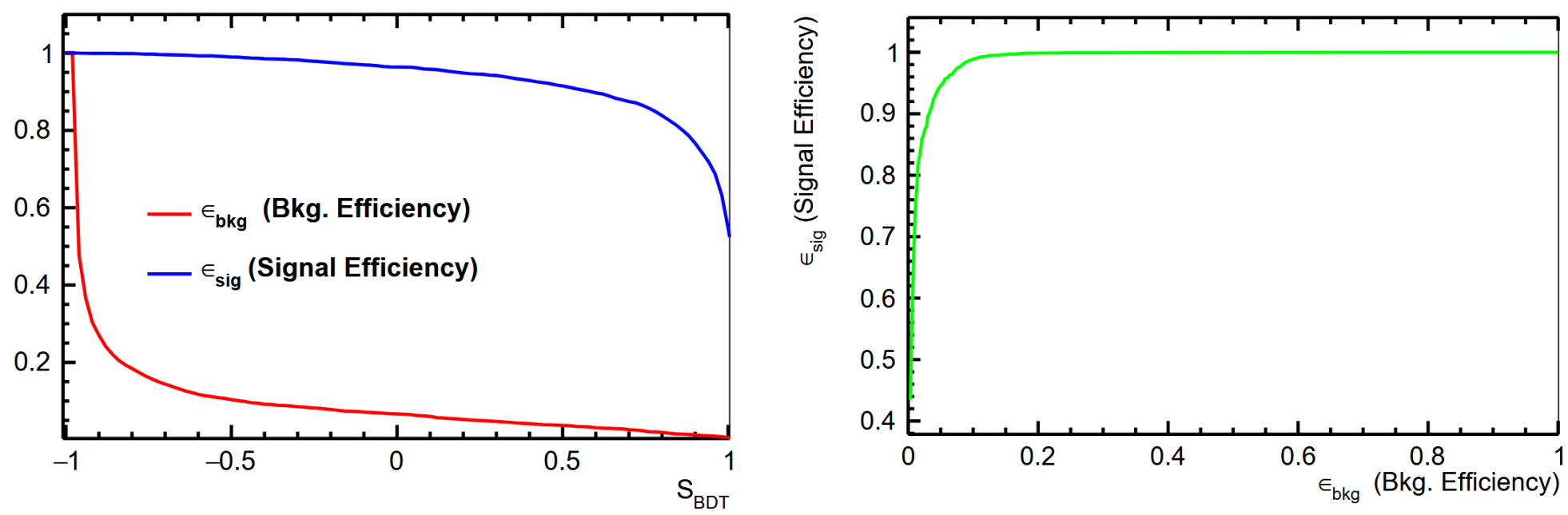

Figure 3.128: Signal and background efficiencies as a function of $S_{BDT}$ for 1 GeV photon ('signal') and 1 GeV neutron ('background').

#### 3.8.7.8 Radiation hardness

The ECal and HCal radiation levels were estimated using FLUKA [111, 112] and are described in Appendix A. As can be seen in Fig. A.11, the maximal doses for $10^{15}$ electrons on target with an 8 GeV beam energy, are ∼1.6 Gy for the Side HCal and ∼0.5 Gy for the Back HCal. These levels were estimated for the scintillator layers at the most exposed positions. The dose calculation in the Side HCal does not include the contribution from particles produced by electro-nuclear interactions or wide-angle bremsstrahlung. A conservative estimate indicates that this component would be at the level of $\mathcal{O}(0.01)$ Gy. These modest radiation doses do not pose a threat to the performance of the polystyrene scintillators or their wavelength-shifting fibers as levels of radiation damage effects typically do not appear until $\sim 10^3$ Gy [113]. This remains true up to $10^{16}$ electrons on target. The radiation levels at the positions of the on-detector electronics and the SiPMs are several orders of magnitude below these values.

### 3.8.8 Fabrication and integration

The HCal construction is distributed to five different sites: Caltech, FNAL, Lund University, the University of Minnesota, and the University of Virginia. The design and procedure for the component production and module assembly have, to a significant degree, already been performed for the construction of the large-scale prototype 3.8.6. The different steps are described below.

#### 3.8.8.1 Scintillator and quad-counter assembly

The scintillator bars are extruded at the FNAL-NICADD Extrusion Line Facility, which has been in activity for more than two decades. The bars are sent to the University of Virginia (UVA) in several shipments of manageable-sized crates (about 200 counters per shipment).

The quad-counter production takes place in the same facility and by the same people who already performed a corresponding production for the Cosmic Ray Veto (CRV) for the Mu2e experiment, where more than 2500 di-counters were successfully constructed. A significant fraction of the tools, equipment, and procedures devised for the CRV construction can be readily repurposed for the LDMX case, reducing R&D efforts and costs.

The production starts by unpacking the scintillator bar from the crates sent from FNAL, followed by cutting extrusions to length using a chop saw fixture. A vertical fixture then holds four bars together while the epoxy cures to bond them together. Springs apply pressure to the stack to ensure adequate contact between the bonding surfaces and to minimize the overall curvature of the assembly. While the epoxy cures, WLS fibers are inserted into the hole of every extrusion, and a fiber guide bar is installed at the end of each quad-counter. A fast-curing epoxy is used to fix the fibers within the holes of the fiber guide bar. The excess fiber protruding from the fiber guide bars is cut off using a hot knife, and the surface is finally polished using an automated fly-cutting machine. After final QC checks, the quad-counters are packed and sent to Caltech in several shipments.

#### 3.8.8.2 Module assembly

The Back HCal modules are assembled at the Caltech Synchrotron building, leveraging the large crane-covered space used in the past for the construction of many large detector systems. The high bay assembly area is configured with three tables, each capable of holding a full eight-plane module. Crane coverage extends over these tables as well as over an adjacent area to store completed modules. The assembly procedure starts by measuring the response of each quad-counter with a set of standard Counter Mother Boards (CMBs). Two tables are used to attach good quad-counters onto the steel plate in parallel. Completed planes are then assembled into modules on the third table, fastened by steel pipes inserted through the ears and secured with a threaded flange. Feet to secure the module to floor-mounted concrete blocks are also installed. The CMBs are mounted, and the module is finally tested using cosmic rays before being moved to the storage area and readied for transport by truck to SLAC.

The Side HCal module fabrication starts by assembling the external steel frame (bottom, rear, and side plates) before welding 17 mm thick absorber plates to the rear module face, leaving an opening for the insertion of the quad-counter assemblies. In parallel, the quad-counters are epoxied to 3 mm thick steel plates, followed by the installation of a 0.5 mm aluminum cover to ensure light tightness. The quad-counter plates are finally glued onto the steel absorber plates and secured by corner bars that allow access to the CMBs. After the last QC tests, the modules are transported to SLAC by truck.

Fixtures needed to handle and install the Back HCal and Side HCal modules will be constructed at Caltech and shipped to SLAC as well.

#### 3.8.8.3 Readout electronics

The readout electronics include four components: the backplane board, the ECON Mezzanine board, the lpGBT Mezzanine board, and the HGCROC boards. The HGCROC and backplane boards are fabricated at Lund University, while the ECON and lpGBT Mezzanine boards are produced by the University of Virginia and the University of Minnesota, respectively.

The electronics assembly proceeds as follows. The ECON (lpGBT) Mezzanine boards are produced and tested at the University of Virginia (Minnesota) to check the functionality of the slow control and basic communication with the ASIC; functional boards are shipped to Lund. The Mezzanine boards are mounted on the backplane board, and the communication with the Mezzanines is verified. The four HGCROC boards are then installed on the backplane board. An automatic full verification, including internal charge injection, is performed for each channel. The connectivity with the CMB is verified with a device having 16 HDMI connectors (one HGCROC-board). This device will check all signals normally established with the CMB. The whole unit is finally mounted in its labeled casing, followed by a final verification of the complete box, before being shipped to SLAC for installation.

#### 3.8.8.4 Components Delivered to ESA

The HCal will be delivered to ESA in multiple shipments. Given the constraints on the Caltech high bay area and the module weight, each Back HCal module will be shipped individually to SLAC right after completion. The Side HCal modules weigh much less; all four will be shipped together at the end of the construction phase. Fixtures to transport the module will be shipped to ESA as well. The electronics will be sent by separate shipment. The final installation and integration are described in Sec. 3.11.7.1.

### 3.8.9 Quality Control

As with Quality Assurance, the outcome of Quality Control (QC) will be monitored and continuously reported to ensure early detection of failures to specifications, so that corrective measures and appropriate improvements are implemented in a timely manner. The institutions involved in the HCal construction will hold regular production meetings where QC performance will be reviewed as a standing item on the agenda.

#### 3.8.9.1 Scintillator bar extrusion

FNAL-NICAAD will be responsible for the procurement and quality of the material to extrude the scintillator bars. They will perform quality checks on the extruded scintillator and monitor the QC results to ensure that appropriate measures are taken in the operation and maintenance of the Extrusion Line Facility. More specifically:

- The lot numbers for the material used in the extruded scintillator bars are recorded. The transparency of the polystyrene pellets is checked, and the purity of the dopants is analyzed before their use.
- Each scintillator bar is barcode labeled in a way that allows its position in the production sequence to be determined.
- Samples are taken every 10-15 bars for dimensional and light yield tests. The samples will be 15-20 cm long and uniquely identified with the date and position in the production sequence. The width, height, and diameter of the hole are measured and recorded for each sample. Samples shall be taken more frequently if the dimensions indicate any production issues so that the process parameters can be corrected before the tolerances are exceeded.
- The light yield of scintillator bar samples is compared to the light yield of a reference sample. The acceptable light yield for the production scintillator is set to at least 90% of that of the reference sample.
- The QC results from dimensional and light yield tests are recorded in a database and provided upon request.

#### 3.8.9.2 Quad-counters

Since quad-counters are sealed into the modules at Caltech, they cannot easily be repaired or replaced after production. Quality control of the scintillator, fibers, and quad-counters before module fabrication is therefore critical.

Several tests are planned during production. Upon reception from FNAL, a visual inspection is conducted when the scintillator bars are unpacked from the crate. These bars are first cut in stacks of four at a time, and it is certified from the Mu2e production that this cutting is done with a precision of better than 0.5 mm length tolerance. The widths are measured in two places with a caliper, their lengths with a dial indicator, and these values are recorded in a spreadsheet so they can be monitored over time. The bars forming a quad-counter are then epoxied together in a vertical jig with supports spaced every 0.25 meters. It has been certified by the Mu2e production that this support spacing guarantees the required flatness.

Upon delivery, the diameter, light yield, and attenuation at different wavelengths of the WLS fibers are measured and recorded using the test jig shown in Fig. 3.129. It consists of the fiber spool, a large diameter transfer pulley, and a take-up spool carrying the optical readout hardware. The transfer pulley is driven by a stepping motor and controlled for gentle acceleration and deceleration. A blue LED light source illuminates the fiber and is read out by a USB-enabled spectrophotometer and a photodiode. The apparatus can increment at any predetermined distance, stopping for measurements of the light spectra and intensity. The measurements are performed on the first 25 m of every spool, each of which contains 300 m of the fiber. The tested fiber is rewound onto the spool.

The fibers are inserted into the bars and epoxied into the FGB. The alignment of the FGB is inspected to ensure it does not extend beyond the extrusion perimeter. The width of the quad-counter is measured in two locations using a caliper and recorded in the database.

Four QC measurements are performed on the completed quad-counters. Firstly, a photograph is taken of each fiber end, using a high-resolution digital camera. Secondly, the roughness of the FGB surface is measured using a Mitutoyo SJ-210 roughness tester. Thirdly, a fiber transmission test is conducted to ensure that the fiber didn't sustain damage through handling. One end of the fiber will be flashed with a calibrated light source and the yield measured on the other end. Finally, the quad-counters will be placed in a dark box and illuminated with a CS(137) source. The currents produced in the SiPMs are recorded and monitored. Only one quad-counter will be tested at a time, but each test takes less than one minute. A threshold requirement will be set and counters below this level are rejected.

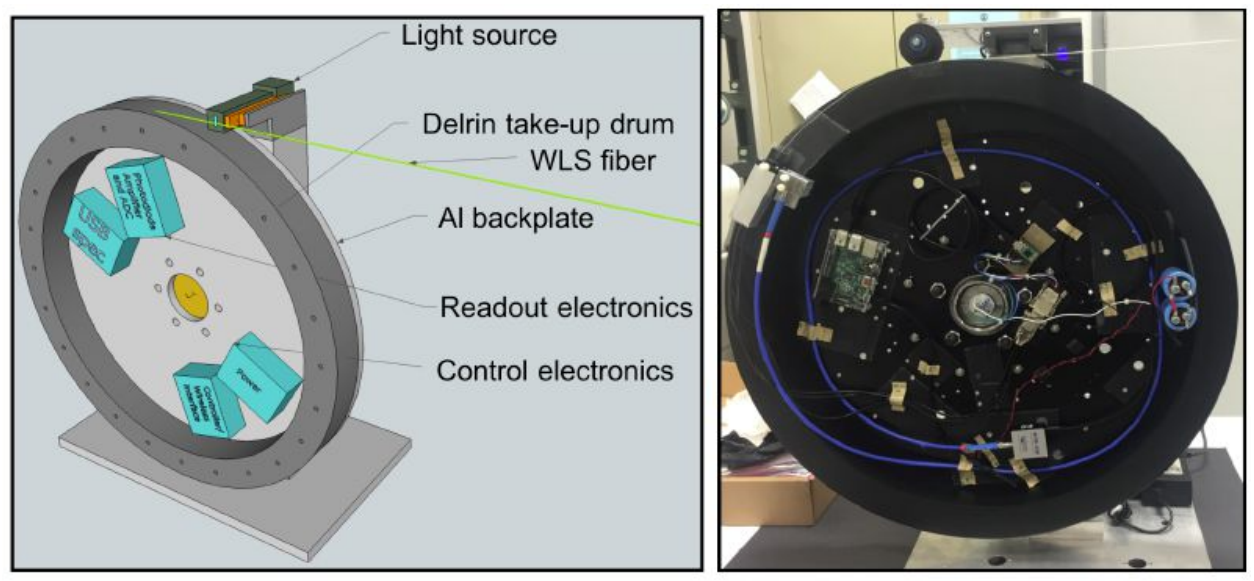

Figure 3.129: The QC test jig for the fibers (JINST. 2018;13(12):P12028).

#### 3.8.9.3 Module assembly

The quad-counters received from UVA are visually inspected for damage during transport. A custom test stand is used to measure the light output of each scintillator bar with a Sr(90) source placed at the center of the bar. The temperature will be monitored to derive corresponding corrections. The MIP response is measured with cosmic rays, using a set of small plastic scintillators to tag cosmic rays crossing the middle of the bar. Finally, the dimensions and curvature of each quad-counter are individually determined before assembly to minimize dead spaces.

The CMBs are checked with a standalone test stand before being mounted on the quad-counters. The test stand collects data from 16 CMBs (64 SiPM channels) simultaneously with a set of reference SiPMs and flash LEDs to measure the single PE response. In addition, spot tests are conducted to characterize the SiPMs gain, breakdown voltage, and dark count rate as a function of the temperature. Previous measurements from Mu2e have shown that the SiPM performance is very stable across different production batches, with a rejection rate less than 1%; spot testing is therefore sufficient for QC purposes.

The absorber plates are received at Caltech. Their dimensions, flatness, and weights are measured before the instrumented quad-counters are glued to them.

After assembly, the planes are tested for potential light leaks by measuring the dark current of each SiPM with a standalone data acquisition system. The internal labels on the CMBs are read and registered together with the outer label on the quad-bar. This will enable the experiment data acquisition to identify a CMB with its physical location in the final experiment. After sealing potential leaks, the light output of each quad-counter is measured once again with a Sr(90) source placed at the center of the bars. The corresponding data will be saved for calibration purposes.

#### 3.8.9.4 Readout electronics

Each step towards the assembly of the complete electronics includes QC tasks:

- All boards of each type are tested to verify basic functionality (like correct voltages and currents) before being assembled together.
- Each ECON mezzanine is tested to check the functionality of the slow control and basic communication with the ASIC with signals sourced from an FPGA before assembly. The testing of the lpGBT Mezzanine boards is described in 3.7.8.2.
- An automatic full verification, including internal charge injection, is performed for each channel of the HGCROC board. The unique number of each HGCROC board is registered together with the external label on the Backplane board.

- The connectivity with the CMB is verified with a device containing 16 HDMI connectors (one HGCROC-board). This device checks all signals normally established with the CMB. At this stage, the individual channel/board differences (e.g., bias voltages) are measured.
- A final verification of the whole unit (one backplane, one mezzanine, and four HGCROC-boards) mounted in its labeled casing is performed. The label of the casing is added to the register with the backplane and HGCROC board identifiers.

### 3.8.10 Calibration

To ensure optimal performance, the HCal must be carefully calibrated and monitored during data-taking periods. In particular, the determination of the noise level and the scintillator response are instrumental in ensuring a high veto efficiency. While a precise characterization of the energy response is less critical, an accurate calibration of the energy and angular response of the detector could be valuable in the context of electro-nuclear measurements (see Sec. 4.9) or searches for visible signatures of new physics (see Sec. 4.6). The calibration strategy of the HCal relies on measurements performed during construction and data-taking periods. The different steps are detailed below.

#### 3.8.10.1 Pre-calibration

An initial calibration of the scintillator response is performed during the quad-counter fabrication and module assembly. The response of the scintillator is measured with a Sr(90) source placed at the center of the bar. The average energy deposited by the electrons emitted by the source is $\sim 0.9$ MeV. A custom test stand measures the response of each scintillator bar in each quad-counter independently. The single photo-electron peaks are extracted from the data, providing an initial calibration of the scintillator light yield. Complementary measurements using cosmic ray muons are also performed for a subset of quad-counters to validate the measurements performed with the Sr(90) source.

#### 3.8.10.2 Internal calibration

The detector will collect regular calibration data using functionalities built into the readout electronics or using random triggers during periods without beam delivered to the detector.

**Scintillator response**
The scintillator response can be monitored internally by measuring the single photo-electron spectrum when the LED system integrated into the CMB is flashing. The distance between neighboring peaks provides an absolute determination of the number of ADC counts per photo-electron. Combined with a measurement of the number of ADC counts per MeV of energy deposited using cosmic muons (see below), the number of photo-electrons per MeV can be calculated. This information can be used to track the scintillator performance over time. The temperature sensor embedded in the CMB is also used to correct the SiPM response.

**Linearity**
The HGCROC pre-amplifier includes charge injection functionalities to study the response linearity in the ADC and TOT regimes, as well as to establish the matching between the two components so that the conversion to MIP equivalent energy can be done continuously over the entire dynamic range. The charge injection itself is expected to be linear to a very good degree and no significant uncertainty is introduced in the calibration from this step.

A charge injection scan below the pre-amplifier saturation level is used to establish the conversion between the injected charge and the sum of ADC counts by means of a linear fit. The results serve as input to the calibration of the TOT response, determining the conversion between the sum of ADC counts and TOT response via fits in two different regimes. Well above the ADC-to-TOT transition point, the TOT response exhibits a linear behavior, while close to the transition point, some non-linearity is expected that can be fitted with a polynomial of higher degree. In combination, these results provide a linear translation between the amount of injected charge into a sum of ADC counts (see Sec. 3.8.6.3).

**Pedestals and Noise**
During periods without beam delivered to the experiment, dedicated runs will be taken using random triggers to establish the noise level of each channel. The measurements performed with the HCal prototype demonstrated that the noise level can be obtained by performing Gaussian fits to the corresponding distributions (see Fig. 3.115). The resulting values will be stored in a condition database. These runs will be performed regularly to monitor the noise levels over time and update the database if necessary. In addition, data taken during physics runs can be used to cross-check the pedestal and noise levels.

#### 3.8.10.3 In situ calibration

The individual channel calibration relies mainly on cosmic muon data, but events in which a bremsstrahlung photon created in the target converts into a pair of muons could also be employed. Possible triggers to select such events are described in Sec. 3.9.5.4. Simulation studies indicate that several minutes of recording cosmic data will be sufficient to gather on the order of 1000 calibration hits per bar. Such a sample could be collected at any time, including during physics runs, since little activity is expected in the Back HCal even at that time.

Cosmic muons will hit every bar, generally from the top, but at varying angles. The track trajectory will have to be determined by a fit to correct the energy depositions as a function of the path length through the scintillator. Conversion muons will be concentrated in the center of the Back HCal and arrive at shallow angles from the front. This sample will provide a cross-check of the MIP response for the bars hit by muons from both sources. Once a clean sample of muon data has been selected, the conversion from ADC counts to MIP equivalents can be performed in the same way as for the test beam data (Sec. 3.8.6.5).

#### 3.8.10.4 Energy scale calibration

Since the HCal is situated in the shadow of the ECal, a full determination of the energy calibration in situ might prove difficult, and the procedure partially relies on simulations. We foresee, however, the possibility of having data-driven methods of cross-checking the calibrations, for example, using photo-nuclear reactions in which only a single energetic neutron is produced that carries most of the bremsstrahlung photon energy (which can be determined from measuring the recoiling electron energy). A second option would be to use a tag-and-probe method based on the production of $K^0_S/K^0_L$ pairs in the target. Within some acceptance, the $K^0_S$ decay products can be fully reconstructed [114] and serve as the tags, while the $K^0_L$ are the probes.

This data-driven calibration can then be used to verify or adjust the simulations, allowing us to cross-calibrate the Side HCal, taking into account the thinner absorber layers. Neutral pions produced in electro-nuclear reactions might also provide a probe to determine the Side HCal energy response.

### 3.8.11 ES&H

Several considerations from an environment, safety, and health perspective during both installation and operation need to be considered for the HCal:

- Mechanical – handling of the detector during installation will require appropriately trained riggers and careful verification of the mounting of the detector. For work at Caltech and UVA, approval of necessary procedures is overseen by the corresponding safety team, while at SLAC, the calorimeter installation will be handled by the trained SLAC team.
- Low-voltage – The silicon photomultipliers require a voltage slightly above 50 V to operate, and the power supply will be programmed to limit the current to avoid damage to the front-end electronics. The temperature inside the readout electronics casings will be monitored to detect abnormal conditions (e.g., the fans stop working). If the temperature is above a certain threshold, the power to the readout boards will be interrupted.
- Hazardous materials – Polystyrene, the scintillator base and fiber core material, is classified according to DIN4102 as a "B3" product, meaning highly flammable or easily ignited. It burns and produces a dense black smoke. At temperatures above 300°C it releases combustible gases. This will be taken into account during the HCal production, assembly, storage, and operation phases. Polystyrene scintillator and fibers are commonly used in experiments at SLAC, and the required safety precautions are well documented and understood. Small quantities of adhesive will also be used in fabricating counters and modules. Ventilation appropriate for these quantities will be installed in the quad-counter and module assembly factory, and personnel working with adhesives will wear the appropriate personal protective equipment. The FNAL-NICADD Extrusion Facility has a documented set of ES&H procedures that will be followed.

## 3.9 Trigger and Data Acquisition

### 3.9.1 Overview

The purpose of the LDMX trigger and data acquisition (TDAQ) system is to enable efficient and high-quality data selection and storage in real-time. Its main components are a timing and synchronization system for fast control, an online trigger subsystem for real-time event filtering and selection, a data acquisition and slow control subsystem for aggregating and organizing data selected for offline analysis, and infrastructure to configure and monitor detector subsystems.

The TDAQ system relies on existing electronics hardware platforms to reduce risks and required resources. For the data acquisition path, we choose an off-the-shelf PCIe-based FPGA system being developed by the SLAC electronics group based on the Bittware XUP-VV8 card [115]. It is a modern and flexible DAQ system with processing power that meets LDMX DAQ requirements. The hardware-level trigger of LDMX will read out trigger primitives from the ECal, HCal, and Trigger Scintillator (TS) systems at full rate and reduce the event rate to approximately 10 kHz using information from those subsystems. The primary task of the trigger system is to save events for which one of the incoming electrons has lost a significant fraction of its energy. However, a wide range of secondary triggers is expected in order to collect background side-band data, calibration data, and physics data for other dark matter and nuclear measurement final states. Because the trigger system must perform a wide range of tasks under challenging latency constraints, based on readout buffer size in the tracker, we require electronics for triggering with a powerful processing FPGA and low-latency, high bandwidth communication from FPGA-to-FPGA to aggregate all detector signals in a single processing node for subsystem-wide data processing. For this task, we chose the ATCA-based APx (Advanced Processing) board developed for the CMS L1 Trigger upgrade. The same platform will be used for fast control and timing signal distribution across the trigger and DAQ subsystems and to the detector front-ends. The system backend DAQ system consists of Central DAQ CPU servers. They integrate slow control and run control functionalities and have event building capabilities with interfaces to subsystems for online monitoring.

A high-level software trigger (HLT), using off-the-shelf, conventional computing systems, is additionally being considered. The HLT is an option to mitigate any risks from the hardware trigger by more efficiently selecting events online and reducing overall data volume offline. This is discussed more in Sec. 3.9.7.2.

The organization of this chapter is as follows. First, we discuss the physics requirements and technological constraints of the system in Sec. 3.9.2. This leads to a discussion of the system architecture in Sec. 3.9.3, including hardware processing technology and interfaces within and external to the TDAQ system. Then, we will detail each of the components of the TDAQ system: the timing and fast control component, the trigger component, and the data acquisition and slow control component. Finally, we discuss the project plan, including the risk mitigation and opportunity for a high-level trigger, and the installation and commissioning of the TDAQ system.

### 3.9.2 Requirements

The LDMX experiment TDAQ design is driven by the search for light dark matter via a missing-momentum signature. It must identify events in which a beam electron produces little or no visible energy in the detector – while simultaneously limiting data rate – over a broad range of ECal energies. At the same time, the system can also be tuned to capture displaced electromagnetic showers that develop deep in the calorimeter, as well as rare electron–nuclear scattering events in which the incoming electron transfers a large fraction of its momentum to the target. Beyond single-particle signatures, the TDAQ system may also remain sensitive to multi-particle final states arising from dark-sector decays, ensuring that high-multiplicity visible topologies are efficiently recorded.

Equally important is the characterization of backgrounds and the continual calibration of the detector. The TDAQ must sample sideband regions in the hadronic calorimeter to monitor energy deposits across a range of thresholds, tag wide-angle bremsstrahlung photons that fall outside the primary electromagnetic acceptance, and track penetrating muons originating from both cosmic rays and on-target interactions. A flexible, prescaled trigger stream for zero-bias and calibration signals enables continual monitoring of detector performance and stability, providing the necessary control samples to validate background models and maintain reliable physics reach throughout the experiment.

The main technological drivers of the TDAQ system are accepting data from the detector at the accelerator frequency of 37 MHz and processing that data in real-time. All but one detector subsystem, the tracking system, will provide inputs to the trigger system for informing the decisions made about whether each event is saved. The tracking system's buffering capabilities constrain both the maximum latency of each trigger decision and the readout rate of the trigger and DAQ systems. In addition, DAQ rates are coordinated with planned disk and computing capabilities. For the total latency for trigger decisions, the requirement is driven by the maximum pipeline depth of the tracker readout ASIC, which is 3.8 $\mu s$ after including a safety margin of 0.5 $\mu s$. We assume here that front-end systems have a latency of approximately 1 $\mu s$, including transmission time. This leaves 2.8 $\mu s$ for trigger decisions to be computed and transmitted back to detector front-ends. For readout rates, the TDAQ requirement is driven by the tracker readout ASIC absolute maximum bandwidth, 100 kHz. The DAQ must be capable of reading out maximum rates with a safety factor of 2 lower than this, 50 kHz. Alternatively, we balance the overall DAQ rate with the capacity of the computing and storage system and thus require an average readout rate of 10 kHz, providing a safety margin with respect to the maximum rates of each system. For DAQ bandwidth, we rely on modern, though common, technological capabilities and require DAQ links to be 10 Gb/s. Given these considerations, the key technical requirements for the TDAQ system are listed below.

- TDAQ1: Provide an average (maximum) of 10 kHz (50 kHz) DAQ rate.
- TDAQ2: 10 Gb/s DAQ data rate
- TDAQ3: Deliver each trigger decision to the detector front-end controllers with a 3.8 $\mu s$ latency.

### 3.9.3 Design and System Architecture

The overall system architecture is shown in Fig.3.130 and provides a high-level schematic of the system, including how the various internal components are connected and the interfaces with the other subsystems. We lay out the various conceptual elements:

- The white boxes in the schematic are external interfaces to the detector subsystems (Tracker, ECal, HCal, and Target) as well as the accelerator timing and conditions information. The detector subsystem upstream inputs to the TDAQ system are presented in Sections 3.4.3.9,3.5.3.3,3.7.3.7, and 3.8.4. We summarize those interfaces from the point of view of the TDAQ system in Sec. 3.9.5.1.
- The green box is the "Timing Hub" which synchronizes our TDAQ system with the accelerator timeline and distributes clock signal and fast control signals to the rest of the TDAQ and also the detector subsystems through the green "Fast Control" lines. This will be discussed further in Sec. 3.9.4.
- The red boxes define the trigger system consisting of FPGA boards, described further below, and the red lines represent the communication speeds and number of links into the trigger system. Three detector subsystems provide data to the trigger (ECal, HCal, and Target), and those provide inputs into the Global Trigger, which defines event selection and filtering to the Timing Hub. The trigger inputs, trigger subsystem functionality, and trigger menu are described in Sec. 3.9.5.
- The blue boxes define the DAQ and slow control system. This includes the front-end FPGA cards, described below, for the Tracker, ECal, and DAQ, which zero-suppress and align subdetector data before transmitting them through a 10 GbE switch to the Central DAQ backend machines (blue lines). This consists of two physical machines that perform run and detector control (Run Control) over the slow control interfaces (pink lines) and an event building machine, which ultimately stores built events on the offline S3DF cluster and provides inputs to the subdetector data quality monitoring systems. The purple cylinders are data volumes that are operated by the LDMX Software and Computing teams. This is described further in Sec. 3.9.6.

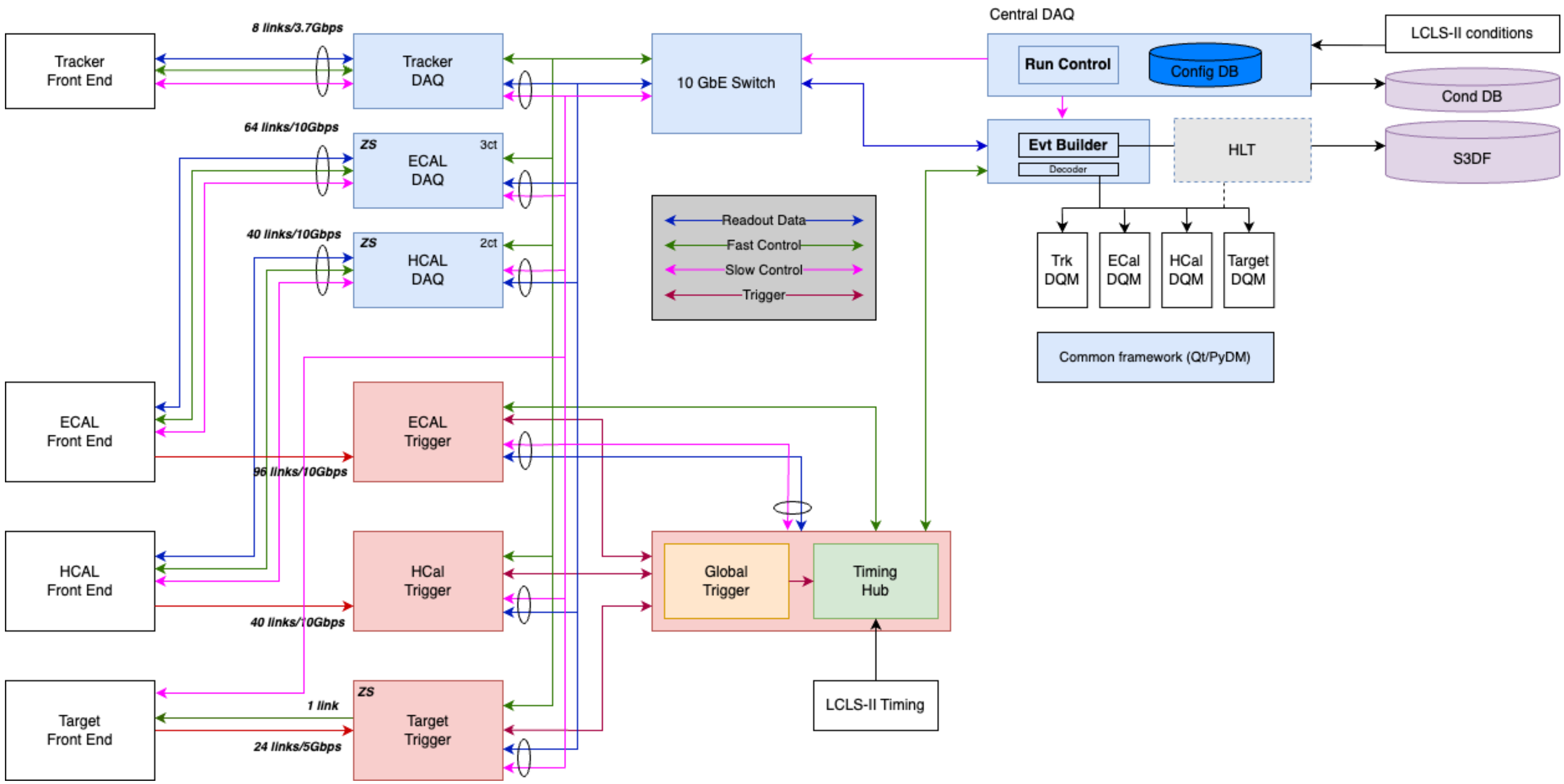

Figure 3.130: A high-level schematic of the system including how the various internal components are connected and the interfaces with the other subsystems: white boxes are external interfaces; the green box is the timing and fast control hub; the red boxes make up the trigger system; and the blue boxes make up the DAQ system. The blue, green, red, and pink arrows indicate the communication interfaces within and external to the TDAQ system.

#### 3.9.3.1 TDAQ System Hardware and Setup

The LDMX TDAQ system is illustrated in Fig. 3.130, and the front-end consists of two types of FPGA-based processing hardware that communicate with the control and DAQ backend through a GbE networking switch. One exception is that the event-building backend machine will also have to receive timing and fast control signals from the timing hub.

The technology choices for the FPGA-based hardware are as follows:

- **Advanced Processor board (APx) for Triggering:** We use the Advanced Processor board developed for the HL-LHC upgrade of the CMS experiment for the trigger data processing path for LDMX. These cards adhere to the ATCA standard and their main processing unit is a single high-performance (Virtex Ultrascale+, VU13P) FPGA for data processing through which up to 120 optical links that can run up to 25 Gbps can be used for data receiving and transmission. Because trigger algorithms proposed for the ECal and HCal, in particular, would benefit from having the entire sub-detector's trigger data in one processing FPGA, we propose the use of the APx technology. For example, whole subsystem tracking (e.g. cosmic ray detection), clustering, and sums are necessary for the primary dark matter trigger as well as several secondary physics triggers. The block diagram of the APx card, with an ATCA form factor, is shown on the left of Fig. 3.131.
- **Off-the-shelf PCIe-based Bittware cards for DAQ:** We use the Bittware XUP-VV8 off-the-shelf processor board the front-end DAQ data path and processing. These have a single high-performance (Virtex Ultrascale+, VU13P) FPGA for data processing through which up to 32 optical links (4× QSFP-DD) that can run up to 25 Gbps can be used for data receiving and transmission. Because the DAQ path does not require all subsystem data in the same processing card, the data DAQ path can proceed on parallel processing cards. Using off-the-shelf cards, when possible, provides more flexibility for the project. The block diagram of the Bittware XUP-VV8 card is shown on the right of Fig. 3.131.

The FPGA-processing hardware will be housed in electronics racks near the LDMX detector as shown in Fig. 3.150. Subdetector readout fibers will be approximately 3-7 m in length, depending on the subdetector. The Ethernet communication, as well as the timing fiber, will be run to central DAQ machines in ESA which will be controlled in the DAQ operations control room at SLAC Building 84, approximately 1000 ft away.

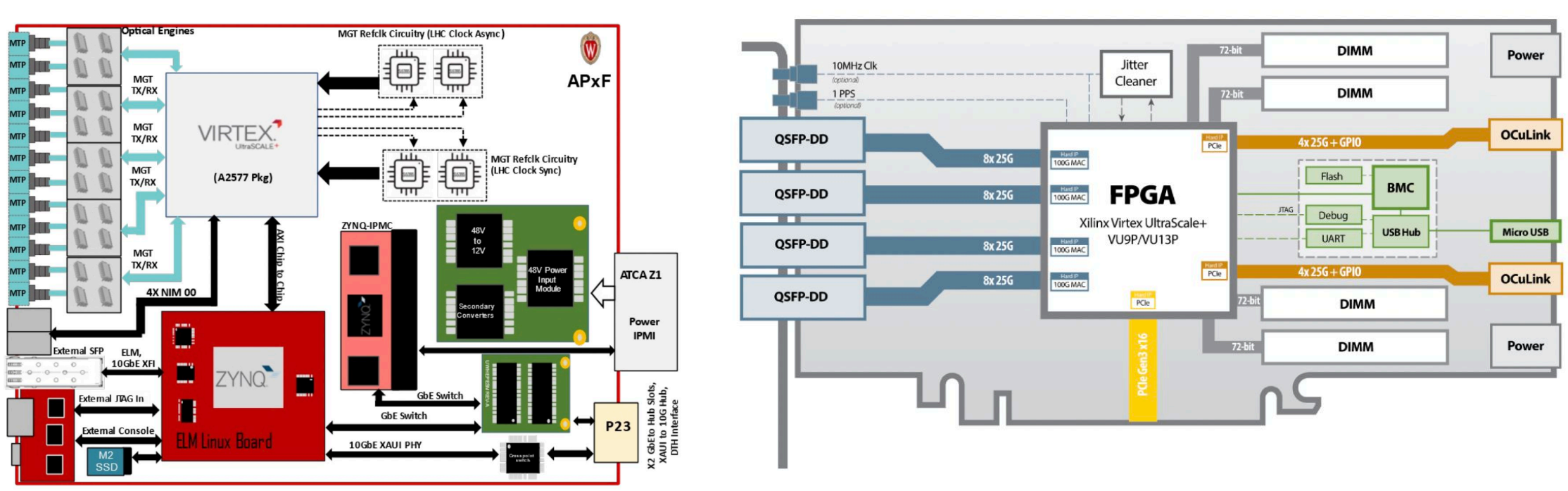

Figure 3.131: Left: Trigger (APx) board block diagram and Right: Data Acquisition card (Bittware) block diagram

### 3.9.4 Timing and Fast Control

The LDMX TDAQ system requirements depend on the timing and structure of the electron beam to the experiment. Based on decisions from the trigger system, discussed below in Sec. 3.9.5, a fast control message called a "ReadOut Request" (ROR) is distributed to the subdetector data readout and trigger paths. In this section, we detail the LDMX beam timing structure and the design and implementation of the fast control signal.

#### 3.9.4.1 Beam timing overview

LDMX will operate using trains of smaller electron bunches that are injected between the larger 928.564 kHz (period of 1.0769 µs) LCLS-II bunches. This is often rounded to 1 MHz/1 µs for informal discussion. The overall reference clock of the LCLS-II system is 185.712855 MHz. These Pulse Trains will have a finer structure of beam bunches of 37.14 MHz, which is 185.713 MHz divided by five. The timing scheme described is illustrated in Fig. 3.132. We also show an example of a fast control signal in this timing scheme diagram in purple text. For example, an event of interest occurs (e.g. PulseID = 5, BunchCount = 10) and a trigger decision should come within the overall trigger latency of 2.8 $\mu s$. At that time, a ROR is issued within the fast control message, which takes approximately 100 ns to be propagated to the subdetector readout paths.

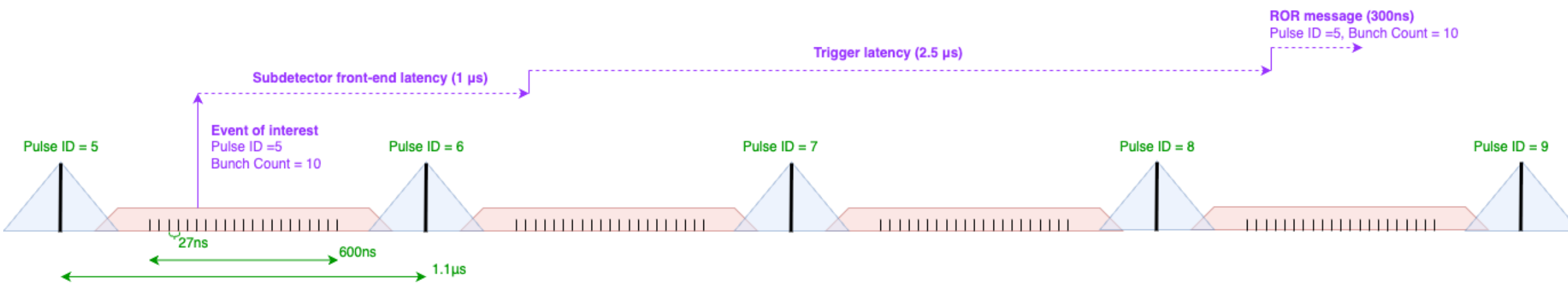

Figure 3.132: Example of fast control signals within the LCLS-II timing structure

#### 3.9.4.2 LCLS-II Timing Distribution Overview

The LCLS-II accelerator distributes an 8B10B encoded Timing Stream. This Timing Stream distributes the 185.712855 MHz reference clock, which is recovered from the encoded stream by the Timing Receiver. The LCLS-II Timing Stream sends a Timing Frame at each pulse interval (∼1 µs), which contains (among other values) a 64-bit PulseID. Each PulseID is unique over the lifetime of the accelerator, typically incrementing by 1 with every Timing Frame (sometimes also referred to as a Pulse Train). Larger increments may occur following accelerator downtime.

#### 3.9.4.3 PGP2FC fast control specification and protocols

The fast control system utilizes the PGP2-FastControl (PGP2FC) protocol to distribute clock, data, and fixed-latency fast control messages. PGP2FC, an extension of SLAC's PGP2 protocol, is a bidirectional, point-to-point, 8B10B-encoded protocol with a 16-bit word size and configurable line rate. It supports data streaming across four virtual channels and can interrupt streaming data transactions to send fixed-latency fast control messages.

PGP2FC ensures fixed-latency clock transmission and recovery at the configured word rate. FPGA high-speed transceivers, operating in a fixed-latency mode, guarantee a consistent phase offset between transmitting and receiving word clocks. This offset, influenced by transceiver type and fiber length, remains stable when the link locks.

PGP2FC's four virtual channels manage streaming data with independent flow control states, indicated by `FULL` and `ALMOST_FULL` flags. Although streaming data functionality is not currently utilized by LDMX, it remains available if required.

Fast control messages of configurable length (up to eight 16-bit words) can interrupt ongoing data transmissions, ensuring immediate transmission with minimal latency. Messages include minimal protocol overhead (one extra word), allowing fast control messages every (N+1) clock cycles, where N is the message word length.

#### 3.9.4.4 PGP2FC configuration for LDMX

LDMX will distribute the 185.712855 MHz accelerator timing clock via PGP2FC, resulting in a 3.714 Gbps line rate. LDMX will employ a four-word (64-bit) fast control message, enabling fast control messages every five cycles. This aligns precisely with the beam bunch rate, facilitating back-to-back Readout Requests.

Two primary fast control messages exist: Timing Messages and ROR. Fig. 3.133 illustrates the structure of these messages.

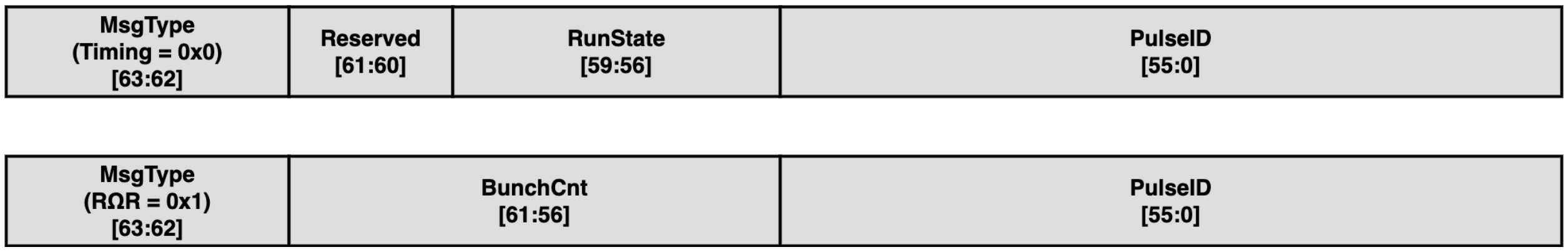

Figure 3.133: Fast Control message structure.

#### 3.9.4.5 Timing Messages

Timing Messages (i.e. `MsgType = 0x0`) are transmitted whenever run-state changes occur, as well as at regular intervals of 1 s. The 56-bit LDMX `PulseID` originates from the LCLS Timing Receiver, and consists of the 56 least significant bits of the latest LCLS-II system PulseID. Timing Messages are always transmitted co-incident with Bunch Count of 0. In this manner, all systems receiving Fast Control are able to establish a common time, consisting of 56-bit LCLS PulseID and 8-bit Bunch Count. The Bunch Count increments every 5 cycles, from 0 to 39. It is generated by common fast control receiver firmware at each subsystem endpoint. This type of message also informs the nodes of any run-state changes. Run states will primarily be used during run start, to synchronize pre-start procedures between all of the various subsystems and their components. The run-state distribution in this manner ensures that firmware processes can execute in perfect synchronicity across every FPGA with a fast control interface.

#### 3.9.4.6 Readout Requests

Readout Requests (ROR) represent a trigger accept decision (also referred to as Level-1 accept decision), issued from within the system's trigger decision processing elements. The ROR is transmitted with the `PulseID` and the `BunchCount`, which it is associated with (i.e., the PulseID of a ROR is a past PulseID). Fig. 3.132 illustrates this, as a ROR is processed in 2.5 μs and distributed in 100 ns.

### 3.9.5 Trigger

The overall trigger design philosophy emphasizes modularity, with the data flow developed at the register-transfer level (RTL) while the core algorithms are implemented using the AMD/Xilinx high-level synthesis (HLS) C-based compiler flow. This design approach ensures that both hardware and software components are optimized concurrently by firmware and physics experts.

A dedicated firmware structure project has been established and will be shared across each of the four trigger boards. This project builds on well-validated firmware tools, such as the SURF framework developed at SLAC, and establishes standard AXI interfaces between the various intellectual property (IP) blocks. These standardized interfaces promote interoperability and streamline integration across different trigger system components. Custom firmware is also being developed to process the inputs from each subsystem. However, this development effort is focused only on the timing system (TS), electromagnetic calorimeter (ECal), and hadronic calorimeter (HCal) systems, with the ECal and HCal sharing an identical data format. This focused approach allows for efficient and consistent data handling across multiple detector subsystems.

The physics-based reconstruction and selection algorithms will be developed using HLS, which not only facilitates co-design between firmware developers and physics experts but also enables bit-wise validation within the LDMX software framework, given the C-based nature of HLS. This is important for validation, both at the individual algorithm level and across the full menu.

In a few of the cases below, preliminary HLS algorithms have already been developed to understand if critical path trigger algorithms meet the resource and latency constraints of the overall trigger system. The algorithms are relatively simplistic based on sums, hit energies, and consecutive layers hit for MIP triggers. One exception is the electron $p_T$ trigger, which requires clustering for position and energy information. We discuss potential mitigation approaches for firmware complexity in the Risks discussion in Sec. 3.9.7.2. A preliminary version of the APx-based LDMX firmware shell has been successfully implemented and tested with the trigger scintillator detector front-end during the test beam campaign. This initial success validates the design approach and provides a solid foundation for further development and integration efforts.

As a reminder, the trigger system takes new inputs at the 37 MHz beam crossing rate (27 ns pipeline interval) and has to complete the trigger computations in $2\,\mu s$. Below, we describe in further detail the different subdetector inputs and the tasks of each of the different trigger boards – TS, ECal, HCal, and global trigger (GT).

#### 3.9.5.1 Subdetector interfaces

The trigger system receives inputs from 3 subsystems: the TS, the ECal, and the HCal. The upstream electronics, which deliver those trigger primitives, are described in Sec. 3.5.3.3,3.7.3.7, and 3.8.4. We summarize that information from the point of view of the trigger as inputs to the trigger algorithms and decision.

**The TS readout electronics system** sends data from QIE charge integrator chips with 144 scintillator bar channels for the entire system. The QIEs output 8-bit ADC and 6-bit TDC at a deadtime-less rate of 37.2 MHz. No zero suppression is done in the readout electronics. After data aggregation on-detector, the FPGAs are used to transmit 24 links of 4.8 Gbps of data each. Within the trigger FPGA on the APx, electron counting is performed through the formation of clusters from bar hits, particle tracking, and duplicate removal. The algorithms for these tasks are described in Sec. 3.5.3.3, and the firmware demonstration is discussed further below in Sec. 3.9.5.2.

**The ECal trigger readout electronics system** sends data to the Trigger system via the ECON-T ASIC. For each of the 32 layers in the ECal, multiple ECON-T ASICs transmit data to the trigger over 3 lpGBT links – 1 lpGBT aggregates data for the center module of the layer, while the other 2 lpGBTs each transmit data for 3 peripheral modules (of 6 total). This totals 96 links, which are inputs to the ECal Trigger board.

Each ECON-T will transmit a sum of energy across the module, as well as the location and energy for the highest energy trigger cells. For the trigger cells data, 4 algorithms are available in the ECON-T, and as a baseline, we use the fixed-latency "Best Choice" algorithm. The energy sum across the module is encoded into an eight-bit floating-point format. The energies for the trigger cells are encoded into a seven-bit floating-point format. For the peripheral modules, the top 4 energy trigger cells are transmitted while the center module sends the top 18 energy cells. The trigger data format is given in Table 3.8. The data is sent to the lpGBTs where it is encoded and transmitted at 10 Gbps.

Within the trigger FPGA which takes as input the ECal trigger data, energy sums are calculated, electromagnetic energy clusters are formed, and track reconstruction from MIPs, primarily charged pions and muons, should be performed. This will be discussed further below in Sec. 3.9.5.3.

**The HCal trigger readout electronics system** will also use ECON-T ASICs to transmit trigger data from the HCal scintillator bars to the trigger system. The difference with respect to the ECal is in the number of channels aggregated and the ECON-T algorithm chosen. The HGROC in the HCal will aggregate 4 bars into a single trigger primitive for the ECON-T, and the ECON-T will use a simple summing algorithm of 4 trigger primitives to create a "Super Trigger Cell" (STC). Thus, the STC, the basic unit of triggering in the HCal, consists of 16 scintillator bars. In the back HCal, the configuration is such that one STC unit will read out one end of 8 bars in 2 consecutive layers as shown in Fig. 3.134. Because of this, each end of the bars in the back HCal will need to be summed in the HCal trigger system. There is no timing (TDC) information available in the trigger data payload, so the hit position along the bar is not available as it is in the full data path. The side HCal bars are read out only on one side. The readout geometry is illustrated in Fig. 3.134.

Similar to the ECal, within the trigger FPGA, which takes as input the HCal trigger data, energy sums are calculated, energy clusters are formed, and track reconstruction from MIPs, primarily charged pions and muons, should be performed. This will be discussed further below in Sec. 3.9.5.4.

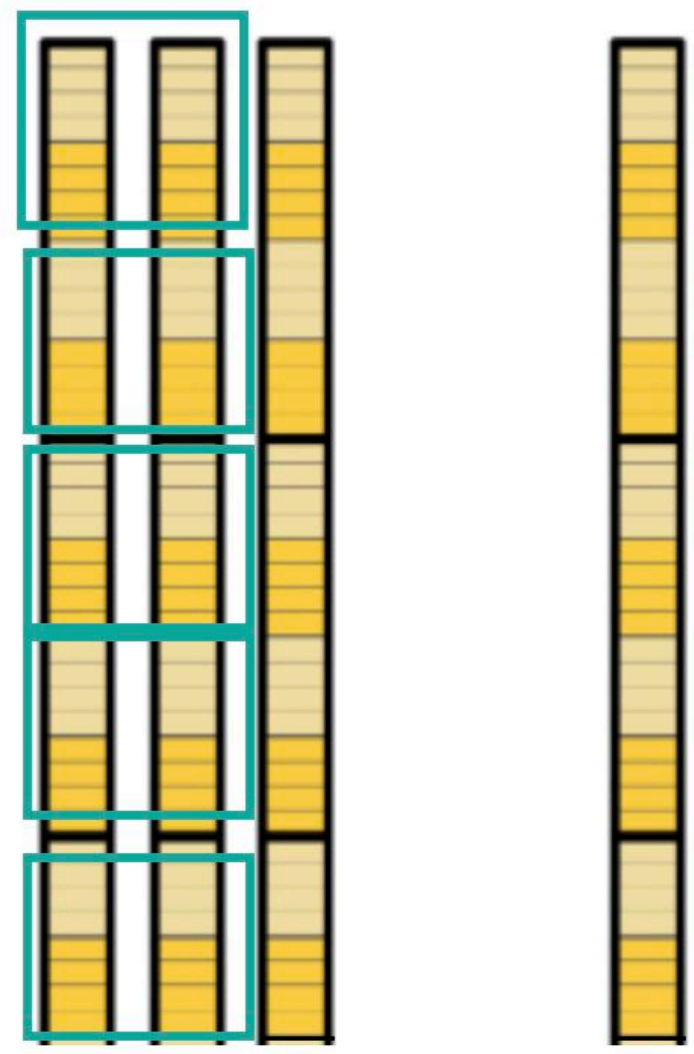

Figure 3.134: Back HCal Trigger primitives – 8 bars in 2 adjacent, same-orientation layers are combined into a “Super Trigger Cell”

Each of the detector subsystems sending inputs to the trigger will have its own data formats and payloads. In the TS scenario, the non-zero-suppressed QIE data is sent over 4.8 Gbps links while for the ECal and HCal data is sent over 10 Gbps links from the lpGBTs. The ECal will send trigger cell energy and locations, and module sums, while the HCal will send summed “Super Trigger Cells” and the full 256 bar sum as well. Each of the subsystems in the LDMX trigger system will output objects to be used in the Global Trigger. The specific inputs and bandwidths will be described in Sec. 3.9.5.5. The following sections describe plans and prototypes for how the trigger objects are reconstructed, including their trigger bandwidths and efficiencies. This will be compiled into inputs that determine the physics menu. These trigger inputs from the TS, ECal, and HCal are all simulated in the `ldmx-sw` software framework so that physics performance and expected trigger rates can be estimated. This makes the total bandwidth of several triggers, including their overlaps, straightforward to estimate.

#### 3.9.5.2 Trigger Scintillator

The primary function of the target trigger system is to estimate the number of electrons impinging upon the target in each time sample and provide their positions. The presence of at least one well-reconstructed electron is a requirement for each of the main physics analysis and support triggers. The critical path algorithms for the trigger system are the electron counting algorithm in the TS and the ECal energy sum described below. Because the ECal total energy firmware is a simple sum of module energies and fits easily into FPGA resources, it is important to illustrate that the TS electron counting algorithm is also able to meet the requirements of the system.

Unbiased data samples (for, e.g. trigger performance studies and detector calibration) can also be recorded by randomly selecting events based solely on the presence or absence of a reconstructed electron. The position of reconstructed tracks at the target plane is also computed, enabling the Global Trigger to correlate these measurements with the position of calorimeter deposits to infer particle momenta transverse to the beam direction. A detailed treatment of the TS reconstruction algorithm and physics performance can be found in Sec. 3.5.4. The TS FPGA board will send the reconstructed candidate tracks and hit clusters to the Global Trigger for further combined subdetector reconstruction.

**Firmware demonstration** The TS Firmware is written to emulate `ldmx-sw` reconstruction software (described in Sec. 3.5.4); test objects are created in the ldmx-sw framework, which are then fed into an HLS test-bench. The firmware output is directly compared with `ldmx-sw` output for validation. Firmware models are validated by correctly reproducing 4000 events of output test vector information.

The LDMX firmware repository and testbench uses Vitis HLS version 2022.2. The TS firmware is separated into 4 firmware modules: the hit maker, cluster engine, track engine, and duplicate removal module. A summary of the physics performance and structure of these modules is given in Sec. 3.5.4.

We present here the firmware demonstration results and the FPGA resources of the current version. While we expect the algorithms to evolve and continue to be optimized, we demonstrate that a straightforward TS tracking algorithm can reasonably fit with the resources of the system and meet the latency constraints. For 48 hit producers (i.e., one scintillator module), the total number of resources used is 8457 FF and 68695 LUT, which represent $\sim 0$ and 5%, respectively, of the total amount of these resources on the board. The total usage of resources should scale linearly with the number of hit producers, which implies that this firmware module would require 15% of the LUT on the board. The LUTs are required to retain 5 clock cycles worth of charge for hit integration, and for this reason, the latency is also 5 clock cycles or 27 ns to integrate a hit.

A single firmware clustering module utilizes 7713 FF and 52012 LUT, which corresponds to $\sim 4\%$ of the available resources on the chip. This implies that 3 TS layers should require $\sim 12\%$ of the onboard LUT. It is fully pipelined, so while it has a latency of 4 clock cycles, it can be read in continuously with each clock cycle.

Finally, the track production and removal module utilizes 6% of the DSP, 19% of the FF, and 38% of the LUT on the FPGA. It is fully pipelined with a latency of 127 cycles ($0.6\,\mu s$). This should work within the constraints of the overall trigger.

#### 3.9.5.3 ECal Trigger

The ECal trigger system receives the charge measured in each module of the sub-detector, in addition to a limited number of individual trigger cell energies that vary according to the module position. From these inputs, energy sums can be constructed to identify invisible and long-lived particle signatures. Lower-level trigger cell information can be used to form clusters of electromagnetic or hadronically-interacting particles that shower in the calorimeter. Energy sums, clusters, and track candidates will be computed in the ECal Trigger FPGA board and sent to the Global Trigger for further combined subdetector reconstruction and final event selection.

**ECal Energy Sums** The module sums received from ECal as discussed in Sec. 3.7.4.2 are used to construct ECal energy sums for use in the global trigger. The ECal trigger first applies calibration constants to the module sums to put them all onto a common linear scale with a least-significant-bit of 1 MeV. The calibration and linearization from the floating point scale is carried out in a single step using a per-module lookup table to minimize the impact of rounding errors.

From these calibrated module sums, several sums are prepared for use in global trigger lines. The sums differ in terms of which cells are included. For some triggers, the sum over all modules is most appropriate. For others, it is most relevant to sum the energy only in the peripheral modules of the flower or only in the core modules. For displaced decays and other cases, it can be more appropriate to consider only the deep layers of the calorimeter.

Triggering on the full ECal energy sum is critical to delivering the main missing momentum dark matter analysis for LDMX. We need to identify events in the trigger with high efficiency, where a significant amount of the initial electron energy is lost while keeping the overall trigger bandwidth manageable. To demonstrate this, we show the missing energy trigger in the ECal in Fig. 3.135. The true $E_{miss}$ is defined by reconstructing the ECal information available offline from the data path, while the online ECal trigger data is used to form the trigger ECal energy sum. Using a sample of simulated inclusive electron events and assuming an incoming electron energy of 8 GeV, we first show the overall rate of events, Fig. 3.135 (left), that pass a cut on $E_{miss}^{trigger}$. For a $E_{miss}^{trigger}$ of 3500 MeV, the trigger rate is a manageable 1 kHz. The efficiency, defined as $E_{miss}^{trigger}/E_{miss}^{true}$, is shown in Fig. 3.135 (right) as a function the true $E_{miss}$. For true $E_{miss}$ greater than 5000 MeV, the efficiency is over 90%.

Beyond the total energy sum in the ECal, we will also compute energy sums over subsets of the ECal layers further downstream of the beam. For example, we consider ECal energy sums beyond the 20th and 27th layers to search for displaced electromagnetic energy deposits, which can be used in displaced dark sector visible decays.

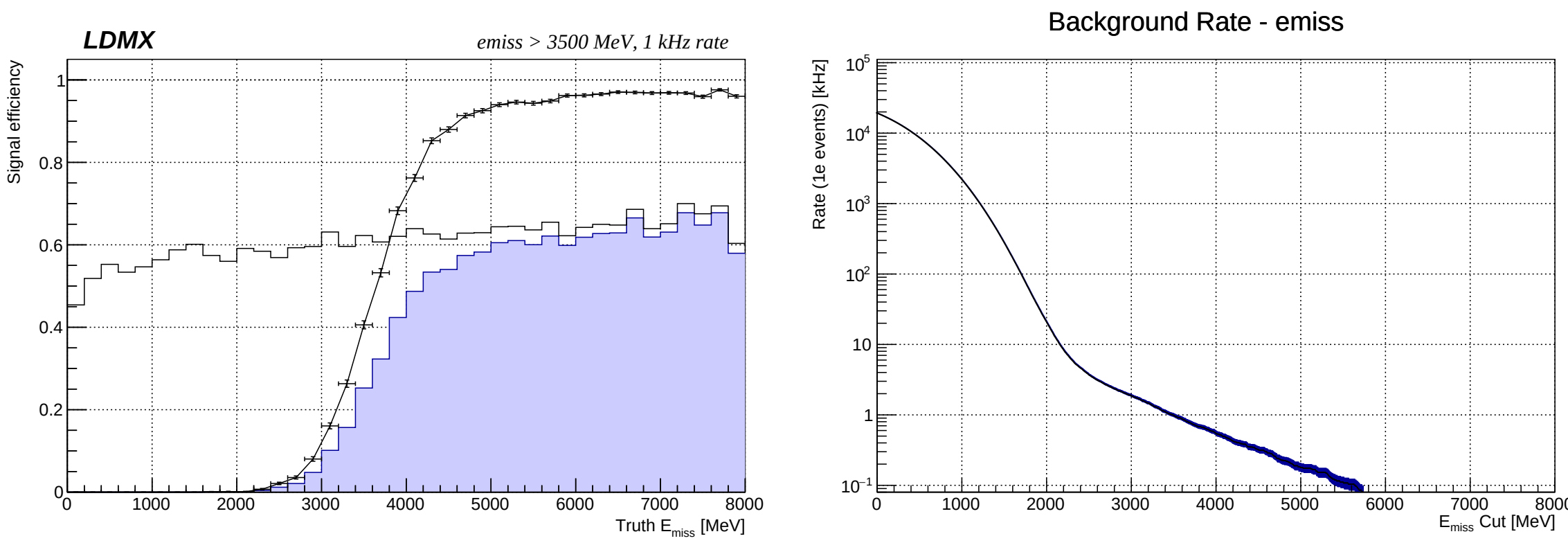

Figure 3.135: Trigger performance for incoming 8 GeV electrons – Left: The efficiency for a 1 kHz rate trigger with a 3500 MeV cut as a function of truth $E_{miss}$. The number of events that pass the trigger is shaded in, while the total number of events is unshaded. Right: The background rate of events as a function of the $E_{miss}$ trigger cut. These events were made with a simulated electron beam.

**Electromagnetic clusters** Along with the total energy sum recorded in a given module, the ECal trigger concentrator ASIC will send a limited amount of 'trigger cells', composed of a $3 \times 3$ group of adjacent calorimeter sensors. The ECal trigger will receive the 18 highest energy trigger cells in the central (peripheral) modules of the calorimeter, whose energies are calibrated using the same lookup-based approach done for each module's sums.

Electromagnetic clusters are built from these trigger cells in a sequential algorithm that first identifies groups of adjacent cells within the same layer, working in parallel across all 32 layers. These $2d$ proto-clusters correspond to seed cells with a local maximum of energy and their immediate neighbors. Full clusters are formed by linking $2d$ clusters with adjacent seed cells across layers.

These clusters will be used to identify high transverse momentum electron events by correlating information between the ECal and TS triggers. This will be discussed further below.

**ECal track/MIP identification** Another interesting signature to identify for the trigger is tracks from MIPs. For example, muons from virtual photons produced in the target and traveling along the beamline or from cosmic rays can help us to understand critical backgrounds for the experiment as well as calibrate all channels of the detector. The $\gamma^* \to \mu^+\mu^-$ process itself is an important background to understand in certain parts of kinematic phase space so understanding those events better is valuable to the experiment.

For triggering on these events, we will focus on a simple initial implementation for firmware based on consecutive layer counting. For example, if consecutive layers have localized energy deposits consistent with MIP energies (within some energy window). From a firmware perspective, this will be straightforward to implement and require few resources. While work on this trigger is on-going, we can tune this algorithm's parameters, such as number of consecutive layers hit, in order to fit in with the overall trigger menu bandwidth.

#### 3.9.5.4 HCal Trigger

The HCal trigger system receives "super trigger cells" which are the summed energy on *one end* of 16 HCal scintillator bars. The HCal trigger processor must combine the energies from both ends of the bars, including application of any necessary intercalibration, which was not possible in the ECON-T ASICs. This step may also need to correct for missing readouts on only one end of the bar or different length bars as is true in the case of the Side HCal as compared to the Back HCal detector. Calibrations will be based on the definition of a "perpendicular MIP" which is the amount of energy deposited in a scintillator bar when the incoming particles are perpendicular to the HCal face.

The HCal trigger processor must also carry out bunch assignment for the signals coming from the HCal ECON-T ASICs. The ECON-T is primarily designed for use with fast sensors operated in the high-pileup environment of the HL-LHC. It does not carry out any shaping or bunch assignment for the signal beyond that carried out in the HGCROC. The HGCROC does suppress out-of-time extensions of the pulse shape in the case when the time-over-threshold circuit operates, but not for signals that are read in only the ADC range. As a result, the HCal trigger processor applies a digital filter to the signal, which sums the sample over several crossings and uses the pulse shape to assign energy to a specific bunch. Carrying out the filtering after summing both ends of the bar minimizes the impact of positional dependence on the pulse shape. After calibration, summing of ends, and bunch assignment, the rest of the HCal trigger is then able to operate using calibrated trigger blocks, which represent the sum over sixteen scintillator bars in either the horizontal or vertical direction. Typically, these blocks combine two subsequent vertical or horizontal layers in the $z$ direction, meaning that eight bars in the $x$ or $y$ direction are summed. This is illustrated in Fig. 3.134.

Similar to the ECal Trigger, quantities for energy sums, clusters, and track candidates are computed in the HCal Trigger FPGA board and are sent to the Global Trigger for further combined subdetector reconstruction and final event selection.

**HCal Energy Sums** Total energy in HCal STCs is summed to get total energy across 4 layers in the HCal. For the HCal trigger, it is critical to understand the full spectrum of energies in the HCal from high energy deposits from hadronic events to very low-energy signals, which could be close to the signal region of the dark matter search. It is therefore important to keep events at many total HCal energy thresholds with varying trigger prescaling factors so as not to inflate the overall data bandwidth. Using a sample of simulated inclusive electron events and assuming an incoming electron energy of 8 GeV, we compute the total sum of energy in the HCal using HCal Trigger inputs. This is shown in Fig. 3.136 as a function of the total energy (in ADCs) for the HCal. Using this information, we can set various values of threshold cuts for a range of energies in the HCal. A proposed set of HCal energy sum trigger thresholds and their prescales will be discussed more below when discussing the overall trigger menu. It is also important to note that the HCal energy sum trigger is correlated with the missing energy trigger in the ECal. So while the rates are considered for each trigger path individually, we will discuss the overlaps between various trigger paths in the physics menu (Sec. 3.9.5.5).

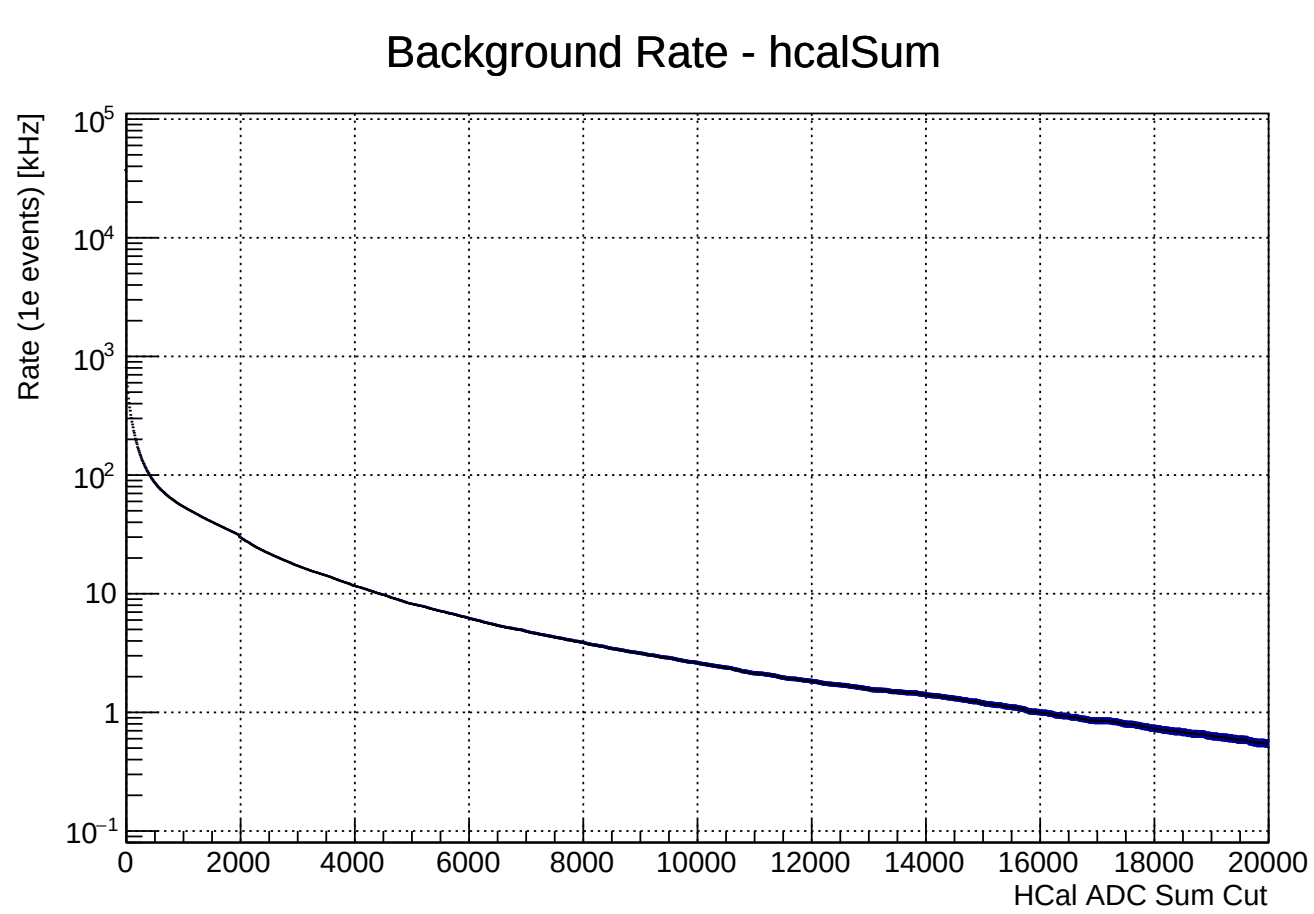

Figure 3.136: The background rate of events as a function of the calorimeter ADC sum trigger cut. These events were made with a simulated sample of inclusive electron events.

Beyond the total energy summed in the back HCal, we consider other energy sums in the HCal for physics and calibration purposes. For example, we will compute side HCal energy sums to understand energy leakage from the ECal to the HCal and wide-angle bremsstrahlung events. We can also study energy sums in regions of the back HCal, particularly for the very back, where there is less activity. Alternatively this could also be used for triggering on displaced particles signatures.

**HCal clusters and tracks** Energy locality, either as clusters or tracks, can be used to study specific signatures in the HCal. The granularity of the HCal STCs, as shown in Fig. 3.134, does not provide very high precision spatial information at the trigger level, but is sufficient for capturing important events for study. For example, a trigger based only on high-energy STCs, potentially isolated from its neighboring STCs, can indicate electromagnetic energy clusters in the HCal from displaced dark sector signatures or high-energy photons hitting the side HCal (wide-angle Bremsstrahlung).

Tracks from MIPs in the HCal are also a valuable signature for understanding physics processes and calibrating our detector. The predominant physics processes that provide MIP tracks in the HCal are cosmic ray muons and $\gamma^* \to \mu\mu$ conversions in the target and ECal. The $\gamma^* \to \mu\mu$ process can be an important background for the primary missing momentum dark matter search, so understanding the rates and kinematics of that process is critical for LDMX. Cosmic ray muons have a significant rate for the HCal, approximately 1 kHz, so it will be important to veto a large rate of cosmic muons while the beam is on. When the beam is off, it will be valuable to collect a significant sample of cosmic muons to monitor and calibrate the HCal, particularly those channels that do not have much activity when the beam is on. While work is ongoing, the current HCal MIP trigger algorithm is envisioned to be formed by placing thresholds on the blocks which are compatible with an energy deposit from one or two through-going MIP particles, allowing for expected energy down-fluctuations as well as up-fluctuations and longer path lengths. The count and sequence of blocks that meet this requirement *and/or* the count and sequence of bilayers that contain blocks which pass this requirement, is provided to the global trigger as an input. Additionally, as an unbiased MIP trigger path to study cross-calibrations across the HCal, we will include a trigger based on the number of HCal STCs above a fraction-of-MIP threshold. This $N_{STC\,hits}$ will be a valuable alternative trigger for studying our MIP track trigger.

#### 3.9.5.5 Global Trigger and Menu

The global trigger defines trigger bits using a set of inputs from the subdetector-level trigger processors and from technical sources such as the beamline. Based on the subdetector trigger objects described above, we collate the trigger data that will be transmitted to the global trigger FPGA board. This is presented in Table 3.12. From the TS trigger board, we send the number of tracks, hits, and clusters, as well as track information for up to 6 tracks. From the ECal trigger board, we send energy sums, clusters (up to 6), and MIP tracks (up to 4). From the HCal trigger board, we send energy sums for both the back and side HCal. In lieu of clusters, we send the top energy STCs (4) in the back and side HCal as well as up to 4 MIP tracks in the back and side HCal. We tally the number of bits sent per event to compute the overall data rate from the subdetector trigger systems to the Global Trigger. Given the capabilities of APx trigger boards, these data rates can fit into one fiber if necessary – the APx is capable of transmitting data at 25 Gbps, though we anticipate operating at 16 Gbps. While this input table is an example, it is clear that the number of fibers between the TS, ECal, and HCal to the GT is rather limited, and there is plenty of overhead for transceivers available.

The GT board is responsible for information correlation between the different detector subsystems and implementing the trigger menu. While there are many potential ways to combine information from the subsystems, we give two critical examples here as an illustration of the task of the global trigger: the missing energy trigger and the electron $p_T$ trigger. Other examples include MIP track matching and clustering between the ECal and HCal. However, we do not detail those here as they are subjects for future study and not critical path to delivering the physics program of LDMX.

**Missing energy trigger** The missing energy trigger is the main trigger path for the missing momentum dark matter search and is a combination of the number of electrons (tracks) determined by the TS system and the total energy in the ECal. It is relatively simple to implement this combination from a logic point of view and to compute signal trigger efficiencies and background rates. The selection simplicity feature of our baseline trigger allows us to understand trigger events more easily. The missing energy thresholds in the ECal are based on the average number of electrons ($\mu$) and the counted number of electrons (TS tracks) in a given event. For example, in a $\mu = 1$ number of electrons scenario, and allowing events where the number of tracks is less than or equal to 3, we consider the following missing energy triggers in Table 4.9 optimized for a signal mass of 0.1 GeV:

- `MISSING_ENERGY_1E` == `TS_NTRACK==1 .AND. ECAL_TOTAL<3010 MeV`
- `MISSING_ENERGY_2E` == `TS_NTRACK==2 .AND. ECAL_TOTAL<10790 MeV`
- `MISSING_ENERGY_3E` == `TS_NTRACK==3 .AND. ECAL_TOTAL<18540 MeV`

Table 3.12: Set of inputs available to the global trigger for defining trigger bits

| Input Name | Size | Description of the global trigger input quantity |
|---|---|---|
| **Technical Inputs** | | |
| `LCLS_ACTIVE` | 1b | Indicates that a bunch could be received in this crossing, based on the expected kicker condition and timing |
| `BUNCH_IN_TRAIN` | 6b | Number of the current bunch in the train, can be used for special monitoring triggers which require no beam before or after the event |
| **Trigger Scintillator Information** | | |
| `TS_NTRACK` | 4b | Number of tracks reconstructed by the trigger scintillator system in this event |
| `TS_NHIT` | 6b | Number of hits above threshold by the trigger scintillator system in this event |
| `TS_NCLUSTER` | 4b | Number of clusters reconstructed by the trigger scintillator system in this event |
| `TS_TRACK` | 6b × 6 | Reconstructed track information (position in $y$) for up to 6 tracks, 1 mm LSB |
| **ECal information** | | |
| `ECAL_TOT` | 16b | Total energy in ECal, summed over all modules in all layers, 1 MeV LSB |
| `ECAL_DEEP12` | 16b | Energy sum over last ten layers in ECal, summed over all modules in these layers, 1 MeV LSB |
| `ECAL_DEEP4` | 16b | Energy sum over last four layers in ECal, summed over all modules in these layers, 1 MeV LSB |
| `ECAL_CLUSTER` | 32b × 6 | Energy and position (x,y,z) of ECal clusters for up to 6 clusters, 1 MeV LSB |
| `ECAL_TRACK` | 32b × 4 | Trajectory and position of MIP track in ECal, 1 cm LSB |
| **HCal information** | | |
| `HCAL_TOT_BACK` | 16b | Total energy in back HCal, summed over all modules in all layers, 4 ADC LSB |
| `HCAL_TOT_SIDE` | 16b | Total energy in side HCal, summed over all modules in all layers, 4 ADC LSB |
| `HCAL_STC_BACK` | 16b × 4 | Top 4 STC (by energy) in back HCal, 4 ADC LSB |
| `HCAL_STC_SIDE` | 16b × 4 | Top 4 STC (by energy) in side HCal, 4 ADC LSB |
| `HCAL_NTRACK` | 4b | Top 4 STC (by energy) in back HCal, 4 ADC LSB |
| `HCAL_TRACK_BACK` | 16b × 4 | Trajectory and position of 4 MIP track in back HCal, 10 cm LSB |
| `HCAL_TRACK_SIDE` | 16b × 4 | Trajectory and position of 4 MIP track in side HCal, 10 cm LSB |
| Total TS bits | 50b | Rate at 37.2 MHz = 1.9 Gbps |
| Total ECal bits | 368b | Rate at 37.2 MHz = 13.7 Gbps |
| Total HCal bits | 292b | Rate at 37.2 MHz = 10.9 Gbps |

A full description of trigger optimization as a function of incoming electron number is presented in Sec. 4.3.3. From the Global Trigger firmware point of view, these thresholds will be made configurable and accessible from the software run control system. This will be discussed in more detail below when discussing the trigger menu.

**High-$p_\mathrm{T}$ electron trigger** Another example of a trigger path that requires correlation of information between the TS and the ECal is the electron $p_T$ trigger, which is used to identify events that are useful for characterizing electronuclear processes. These are used to understand the modeling of nuclear processes, including Deep Inelastic Scattering, which is critical for the neutrino oscillation program. More information about the physics motivation is presented in Sec. 4.9.

Because the TS system only gives a measurement in the $y$ direction, we compute only $p_y$ by comparing the displacement in the electron between the TS and the ECal. That information, combined with the electron energy in the ECal provides a measurement of $p_y$. For simulated single-electron events, we explore the bandwidth for a $p_y$ trigger and the efficiency for such a trigger. This is shown in Fig. 3.137. The left plot shows overall data bandwidth for a trigger threshold on $p_y$ as computed using trigger inputs. Given the position granularity of the TS and ECal trigger primitives, the data rate is quite significant. As an example, using a cut of 680 MeV on $p_y$, we get a data rate of 1 kHz. If we consider this 680 MeV $p_y$ cut, we can plot the efficiency as a function of true $p_T$ in Fig. 3.137 (right). For example, we find that for a true $p_T$ of 700 MeV, the efficiency is roughly 20% and grows to nearly 50% at 1000 MeV. While the efficiency is relatively low, given the amount of electronuclear events expected, this will give us a significant sample of interesting high $p_T$ events to study these signal processes.

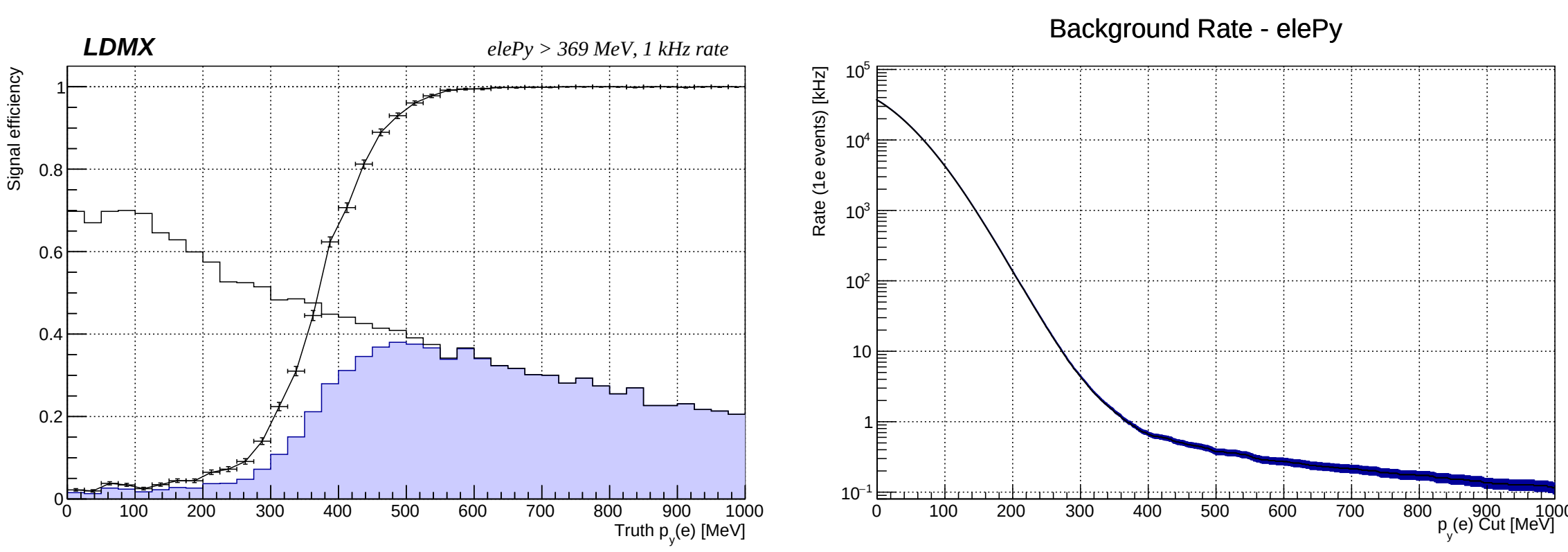

Figure 3.137: Left: The efficiency for a 1 kHz rate trigger with a 369 MeV cut on $p_\mathrm{Y}$ as a function of truth $p_\mathrm{Y}$. The number of events that pass the trigger is shaded in, while the total number of events is unshaded. Right: The background rate of events as a function of the $p_\mathrm{Y}$ trigger cut. These events were made with a simulated electron beam.

**Trigger Menu** Finally, we collate all the discussed trigger paths into a full menu to understand the total data rate expected at LDMX with a proposed set of cuts and how those trigger paths may overlap. Our implementation strategy for the Global Trigger firmware will be to provide several potential paths as defined in the firmware with configurable thresholds. Those will be accessible from the slow control infrastructure, discussed in Sec. 3.9.6. Versions of the trigger menu will be tagged and emulated in the LDMX software framework to understand trigger efficiencies in offline physics analyses.

An example trigger menu is presented below based on the trigger level objects we have discussed in the previous sections. The trigger menu is defined by simulating a sample of 1-electron events. While we do not consider a distribution of numbers of electrons per event, e.g. $\mu = 1$, this provides an example illustrating the driving factors in the overall data rate. The trigger menu is presented in Table 3.13.

The main trigger for the missing momentum dark matter search is the missing energy trigger, which has a data rate of approximately 1 kHz. Examples of missing energy trigger paths for a distribution of the number of incoming electrons, which totals 1 kHz in rate is shown in Sec. 4.3.3. Beyond that, we have a number of trigger paths with less stringent thresholds that are prescaled by some factor to be able to capture some sideband events for studying regions of phase space adjacent to the primary trigger. For the electronuclear program we utilize the electron $p_T$ trigger, and similarly, have lower threshold triggers that are prescaled. We also have a number of support triggers to study signatures in the HCal based on energy sums in the back and side HCal. Finally, we have other objects that are used to explore exclusive processes. For example, MIP track triggers are used to study muons in the ECal and HCal coming from cosmic muons or $\gamma* \to \mu\mu$. The remaining triggers are focused on energy localization in the calorimeters. This is in the form of energy sums for some part of the calorimeter or clusters themselves. Finally, additional calibration triggers are available, including a zero-bias trigger. Not shown in this menu is the collection of data when the beam is off. This is very valuable for collecting beam-off cosmics to calibrate detector elements that are not very active.

By laying out this example trigger menu, we illustrate how the physics signatures of interest can fit within the resources of the trigger and beyond. The total trigger menu rate is limited to be less than 10 kHz, as shown in Table 3.13.

| Trigger path | Cut value | Rate (hz) | Prescale |
|---|---|---|---|
| Missing energy | $> 3.0$ GeV | 3000 | 1 |
| Missing energy, support | $> 1.9$ GeV | 400 | 200 |
| Electron $\lvert p_y \rvert$ | $> 370$ MeV | 1000 | 1 |
| Electron $\lvert p_y \rvert$, support | $> 700$ MeV 200 | 1000 | |
| HCal sum | $> 12900$ ADC | 200 | 1 |
| HCal sum, support 1 | $> 4500$ ADC | 100 | 50 |
| HCal sum, support 2 | $> 180$ ADC | 100 | 1000 |
| Side HCal sum | $> X$ ADC | 500 | 1 |
| MIP track in ECal | $\geq 1$ | 500 | 1 |
| MIP track in HCal | $\geq 1$ | 500 | 1 |
| Displaced ECal 1 | $\sum_{\text{layer}>20} E > 4.7$ GeV | 350 | 1 |
| Displaced ECal 2 | $\sum_{\text{layer}>28} E > 3.0$ GeV | 200 | 1 |
| Calibration and monitoring | | 500 | |
| Total | | 7000 | |

Table 3.13: LDMX example trigger menu for the case of 1 incoming electron

We also illustrate the trigger menu rates, including overlaps, in Fig. 3.138. This menu includes prescales for high-rate triggers. In this menu, the dominant triggers in terms of rate are the missing energy, electron $p_{\mathrm{T}}$, and HCal sums. While this is not the final trigger menu planned for the experiment, it does capture the relevant, critical physics paths needed for the primary LDMX physics drivers.

While we show the unique rates in Fig. 3.138, it is important to understand the overall data bandwidth of the full trigger, which requires studying how many events overlap between various triggers. Figure 3.139 shows the overlap between different trigger selections. On the left, we show the overlap between triggers without trigger prescales applied, while on the right, the prescales are applied. As can be seen from these overlap plots, certain trigger paths are highly correlated. For example, triggers with different HCal energy sums have larger overlaps. Figure 3.139 (right) also shows that the HCal sum and missing energy triggers are highly overlapping. This is because there is a high correlation between large missing ECal energy and energy in the HCal.

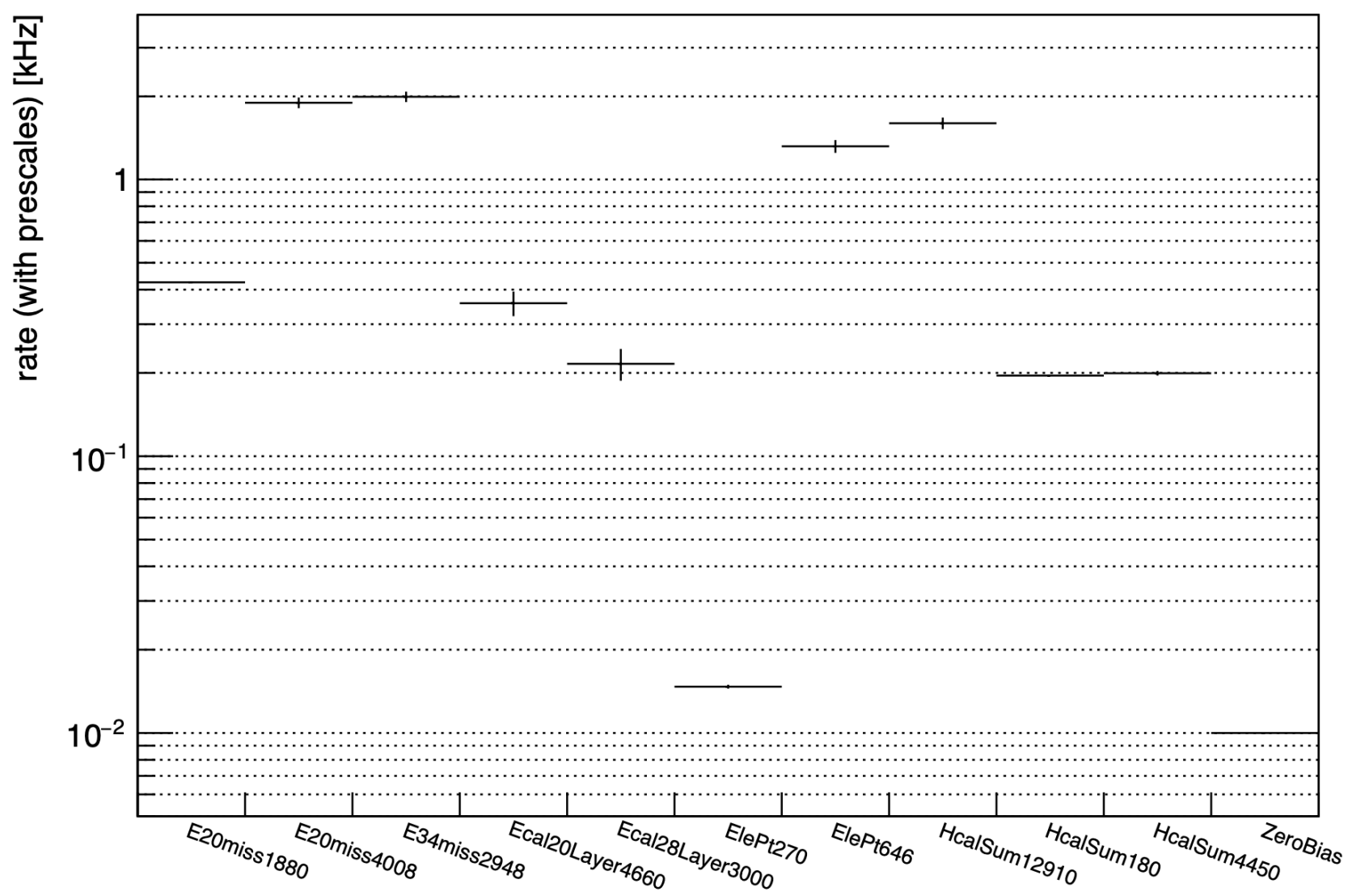

Figure 3.138: Full trigger menu rates

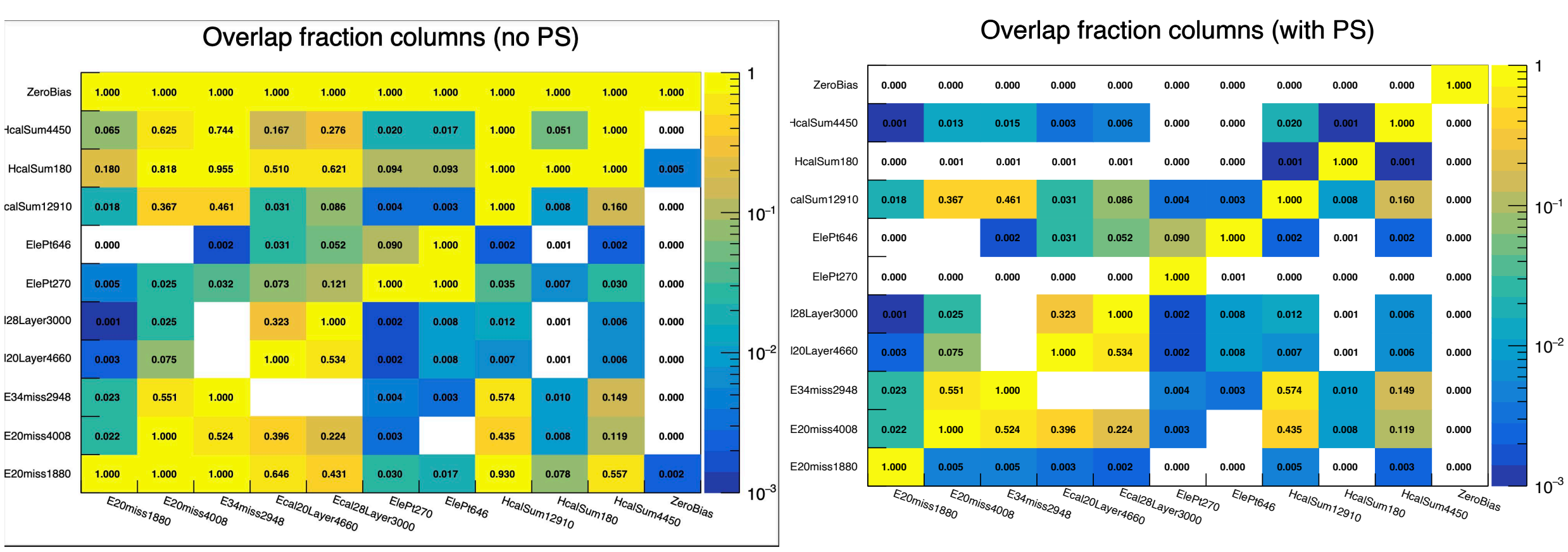

Figure 3.139: Overlaps between various trigger paths – left: no prescale of triggers, right – prescale of triggers.

### 3.9.6 Data acquisition and Run Control

In this section, we describe the data acquisition path and the software and firmware infrastructure that will be developed to configure, commission, and operate the experiment—from individual subdetector systems to the full detector. We begin by outlining the common firmware and software components required to read out data from the detector subsystems, assemble events, and enable near real-time data monitoring for offline analysis. Next, we describe the infrastructure and tools for slow control, configuration, and monitoring of the detector.

Fig. 3.130 illustrates this architecture. DAQ and trigger components acquire data and transmit it over the network to a central DAQ and event-building machines. These systems are expected to be physically separate from the run control infrastructure. The event builder will process and decode events for storage in the offline S3DF system and forward them to data quality monitoring (DQM) applications for each subsystem. To maintain synchronization with the fast control timing system, the event builder will interface with the timing hub located on the Global Trigger board. Finally, the central DAQ will include an interface to the LCLS-II accelerator to access and incorporate accelerator conditions into run control.

#### 3.9.6.1 Central DAQ Path

**Common Firmware Logic** In Fig. 3.130, each of the blue and red boxes – representing the FPGA boards in the system – will output data to be stored offline for software and computing. An example of the interface between the DAQ Bittware FPGA cards and the subdetector front-ends is illustrated in Fig. 3.140. Data from the tracker, ECal, and HCal front ends is transmitted to the Bittware cards over any number of fiber optic links. The number of links and protocol they carry will be specific to each subsystem. The data is then processed and transmitted over PCIe to the DAQ host server. That host server transmits data to the central event building machine over Ethernet. The payloads for the trigger subsystem boards and the TS system will be formatted in the APx boards and transmitted over Ethernet to the central event building machine as well.

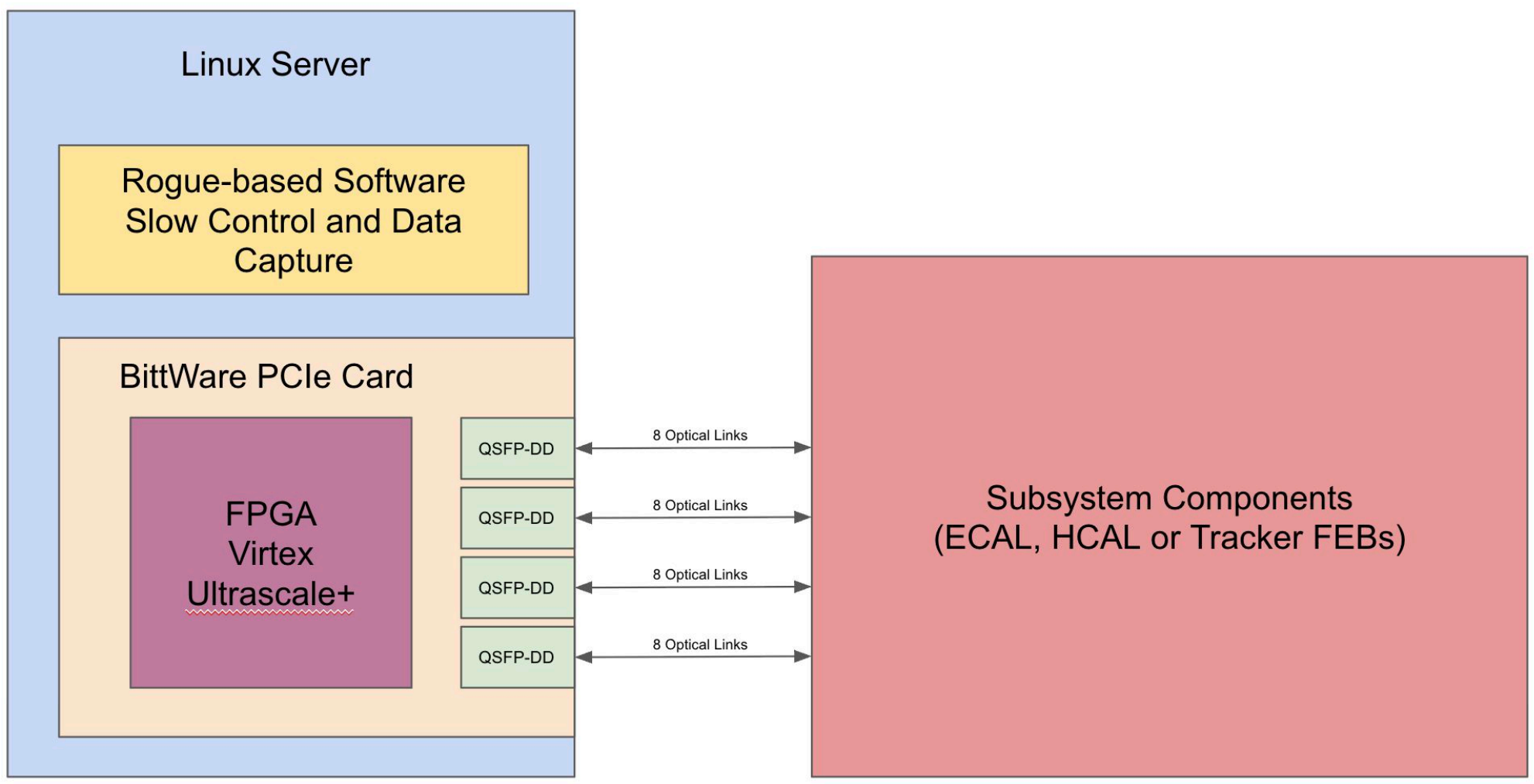

Figure 3.140: Diagram of DAQ common interface

For each of the data sources, the interface between the systems (TS, Tracker, ECal, HCal, and Trigger) is in the firmware. Each subsystem has its own dedicated firmware for decoding its data. For example, zero suppression is done for the ECal and HCal subsystem in the firmware while the zero suppression for the tracker is done in the detector front-end. The data for each of the subsystems is sent to a common PCIe DMA block using AXI-Streams. This is illustrated in Fig. 3.141.

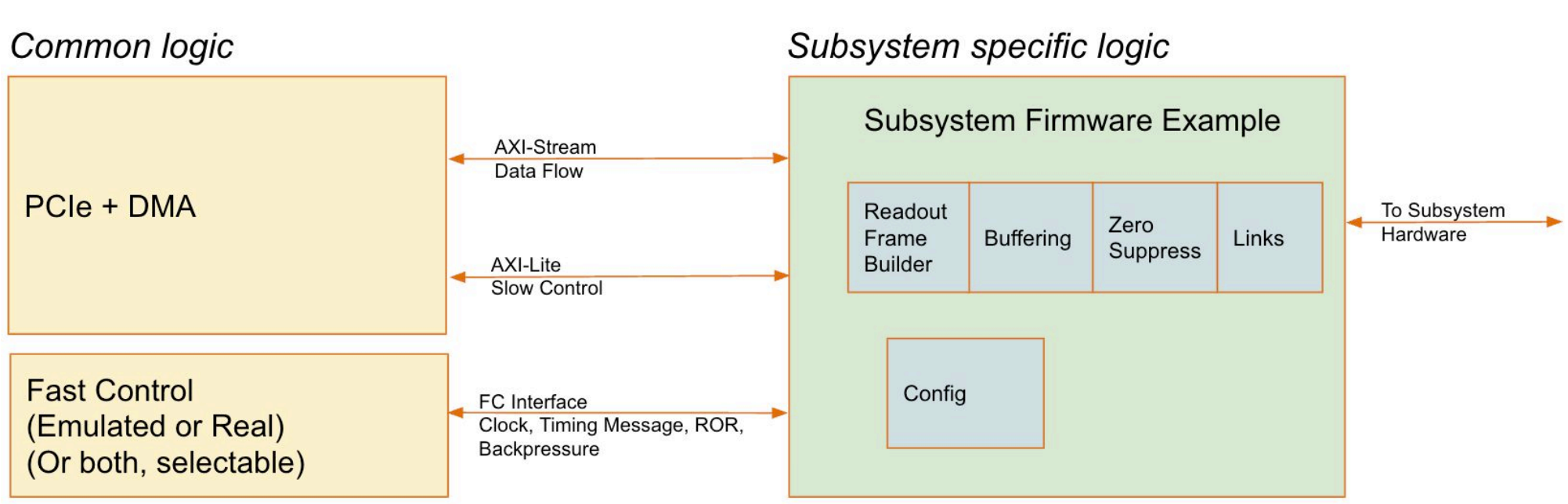

Figure 3.141: Schematic of DAQ common firmware block and interface with subsystem

**Event building** The purpose of the Event Builder is to bring together information received in different contributors (sub-systems and their sub-components) and assemble this information into a collection of coincident “events" which can then be used for analysis. The current LDMX Event Builder is based on that developed for the Heavy Photon Search. The code base is written in C++. The Event Builder receives the event fragments from each contributor, combines these into events, and then transmits those complete events to the onward trigger DAQ architecture. Once decoded, some events will be sent to a Data-Quality Monitor (DQM), where various quality metrics will be checked to understand systematic issues in the data. The High-Level Trigger (HLT) will utilize the event information to make a software-based trigger decision.

Contributors to an event come from different sub-systems, and each sub-system can also have multiple contributors. Each sub-system/contributor combination has a unique ID. Contributions are assembled into “events" which all contain the same fast control timestamp. The Event Builder checks that all possible contributors provide a contribution for each event.

#### 3.9.6.2 Slow Control, Run Control, and Monitoring

**Rogue platform and Slow Control** We will use the Rogue[6] package for slow control, run control, and monitoring. Rogue is a system designed to facilitate hardware development and the creation of data acquisition (DAQ) systems for interfacing with various hardware components. Its primary goals include supporting a wide range of hardware and software interface technologies and providing easy-to-understand mechanisms for connecting independent management and data processing modules through well-defined interfaces. Rogue aims to enable data paths in independent high-performance threads while also allowing Python-based data access for visualization. The architecture of Rogue is characterized by a mixed C++/Python codebase that utilizes the `boost::python library`. This allows for rapid development in Python while retaining the ability to drop into C++ for performance-critical sections. Most development within Rogue is done in Python. Rogue features a flexible structure for creating hierarchical system components that are independent of network and hardware hierarchies. It allows for things like FPGA firmware register parameters to be cleanly described in Python and quickly integrated into application software. Rogue also facilitates loading of common configurations via YAML files that describe values for registers in the hardware tree. Example Rogue GUIs for register peek-and-poke, configuration loading and run control are shown in Fig. 3.142. These can be further customized with elements specific to LDMX.

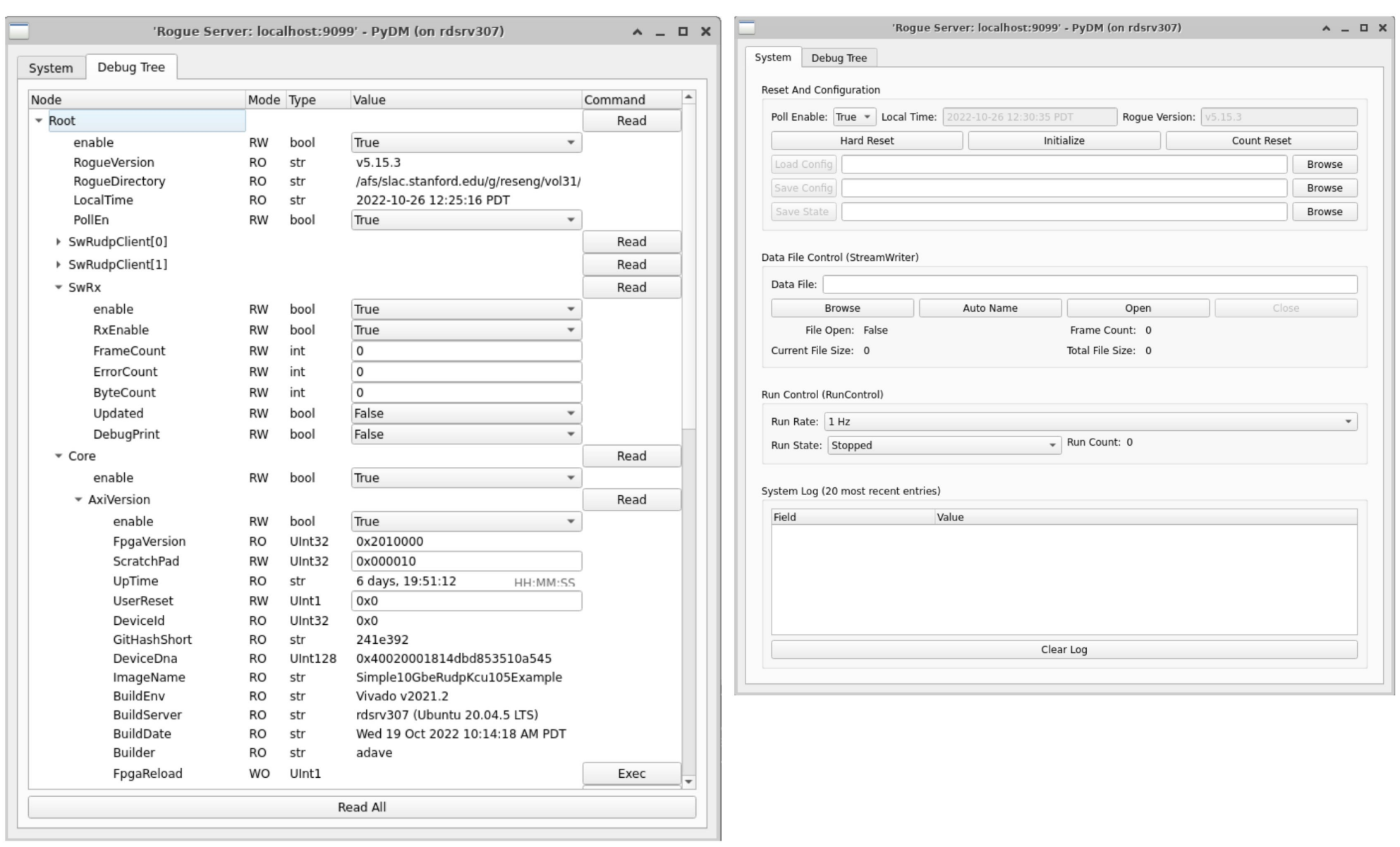

Figure 3.142: Example Rogue GUIs for register peek-and-poke, configuration loading, and run control

[6] `https://slaclab.github.io/rogue/`

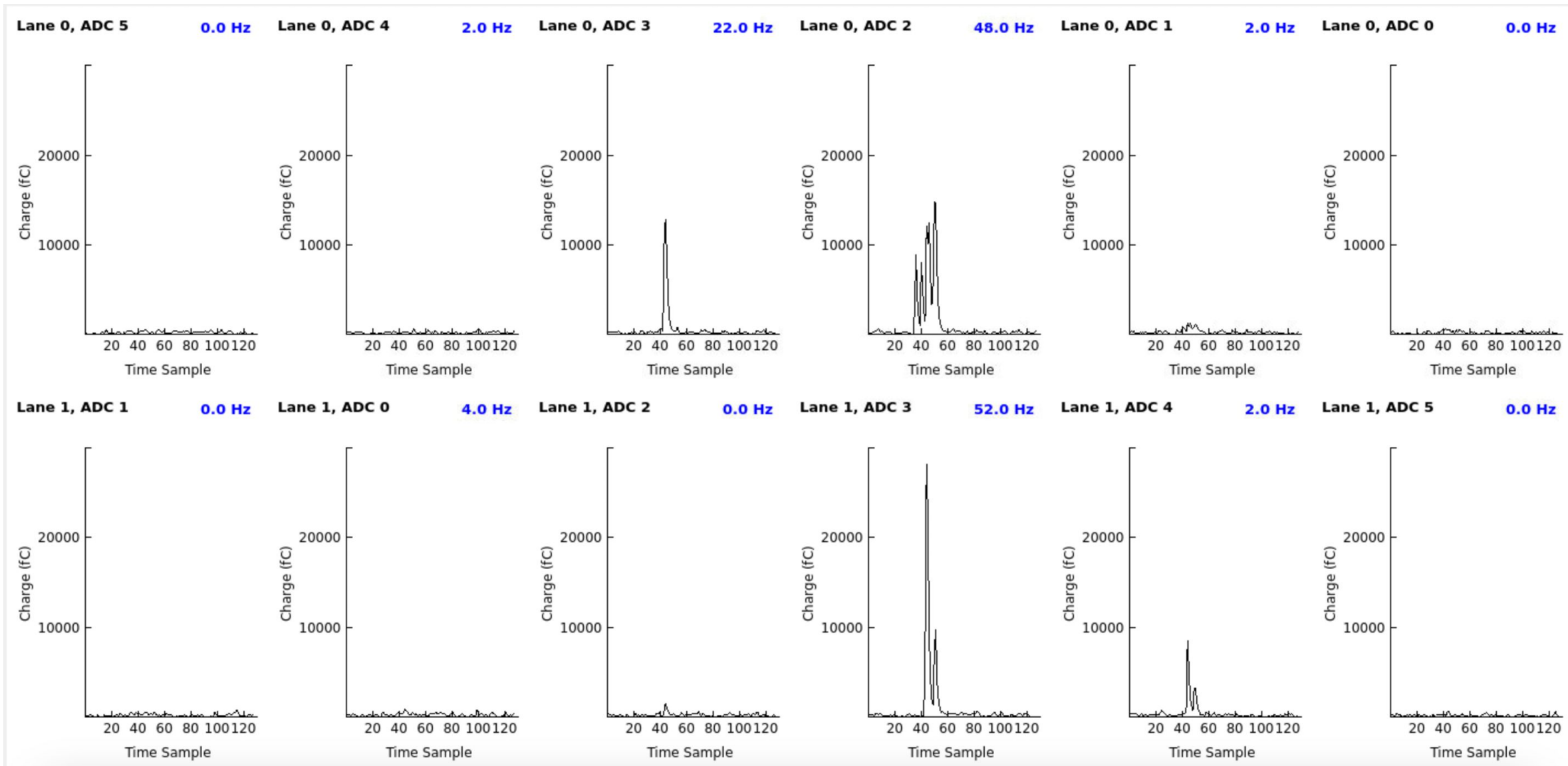

Figure 3.143: An example live display GUI created for the S30XL test beam of a TS prototype using PyDM.

**Slow Control** Rogue will facilitate the creation of slow control software. This software will allow configurations to be loaded to the various subsystems, and for any errors to be reported back. Configuration data will be contained in YAML files, which Rogue is natively able to process and load to the global slow control device tree. Status variables in each subsystem firmware can also be automatically monitored by configuring polling at various intervals (typically 1 second). Historical monitoring data will be written to the DAQ S3DF back end. A run control interface will also be constructed with Rogue, providing a GUI to facilitate starting and stopping data taking, and reporting system status.

**Monitoring** Rogue APIs allow event data from the various subsystem contributors to be branched off and sent to monitoring processes, which can analyze and plot the data to display relevant metrics in real-time. Each monitoring process will receive only the event data relevant to it. Real-time plot and display GUIs from event streams will be built using the PyDM framework. PyDM is a PyQt-based framework for building user interfaces for control systems. The goal is to provide a no-code, drag-and-drop system to make simple screens, as well as a straightforward Python framework that allows the user to also build complex applications, if needed. PyDM[7] is developed by SLAC and is layered on top of QT. It allows display elements such as plots and widgets to be easily connected to variables published by Rogue processes. Fig. 3.143 shows an example of part of the monitoring interface that has been developed using PyDM and Rogue for the S30XL test beam, which is discussed in Sec. 3.9.7.1. Each plot shows the linearized charge output against time sample for each channel of the prototype Trigger Scintillator module. The blue text above each plot is the live hit rate for that channel. The twelve plots are laid out to follow their channel real positions based on cable mapping - from Fig. 3.143, the beam spot appears to be relatively centered on the TS prototype.

#### 3.9.6.3 Luminosity

Luminosity monitoring is a crucial component of the LDMX experiment, as it directly impacts the precise determination of both signal and background rates and physical cross-sections. Precise knowledge of the luminosity is important to monitor over an LCLS-II pulse train for accelerator stability and performance and over the full run for inclusion in the final physics results.

For online, real-time monitoring, we plan to collect information about every bunch crossing to understand any fine-grained fluctuations in the beam performance. For this, we will use the TS system and the primary monitoring information with supplemental information from the ECal system. The data will be histogrammed to manage the amount of monitoring data. The following quantities will be monitored with an entry in the histogram *per pulse train* with histograms over a luminosity section (more below). The monitored quantities include, but are not limited to: the number of live bunches from the accelerator; the number of hits in the TS system per bunch crossing as well as the number of tracks in the TS system; the number of hits and total energy in the ECal system; and the average position in $x$ and $y$ of the hits in the ECal and TS (in this case only $x$). Furthermore, the pulse ID for the initial and final pulse in the luminosity section will also be recorded.

[7] `https://github.com/slaclab/pydm/`

Operationally, the luminosity data are organized into defined sections spanning approximately one minute, during which stable luminosity conditions are anticipated – we define this as a *luminosity section*. Within these intervals, the event data sizes remain manageable, approximately 3 GB per luminosity section. This structured approach ensures efficient handling and processing of data, thereby optimizing the analysis workflow and improving the overall precision and robustness of luminosity monitoring within the experiment.

Offline luminosity determination is facilitated through the use of a zero-bias prescaled trigger, enabling unbiased sampling of beam conditions. This approach allows for the correlation of luminosity data across multiple detector subsystems, including the tracker information, which is available offline, significantly improving the reliability of luminosity measurements. Furthermore, incorporating detailed positional information and full hit distributions in the offline analysis can provide valuable insights into luminosity conditions, enhancing both accuracy and diagnostic capability. We will correlate the online luminosity determination with the prescaled zero-bias events used for offline luminosity to ensure fine-grained luminosity measurements are consistent with offline calculations.

### 3.9.7 Project Plan and Status

The coordination of the development of the various elements of the TDAQ project includes dividing and sharing tasks across institutions participating in the project (FNAL, SLAC, Stanford University, and University of Michigan), so as not to have a critical bottleneck.

The work breakdown is organized into three main sections. The first two are focused on the development of the firmware and software: the *Trigger System* and the *Data Acquisition (DAQ) System*, with each further divided into development, integration, and commissioning phases. The final section is hardware acquisition, which is focused on the procurement of FPGA boards, networking, and computing.

For the Trigger system, the tasks cover multiple subsystems and infrastructure areas: ECal, HCal, TS, Global Trigger (GT), Fast Control/Timing Hub, and overall trigger firmware and hardware infrastructure. In the development phase, each subsystem builds initial `V0` firmware prototypes to implement essential algorithms and basic infrastructure. This is followed by an integration phase where `V1` firmware is implemented to incorporate key interfaces and communication pathways between subsystems, such as trigger primitive interfaces, transmission paths to the Global Trigger (GT), and board-to-board tests. Finally, the commissioning phase focuses on finalizing the full operational paths (`V2` of the firmware) for each subsystem, ensuring that the complete trigger chain from initial detection to GT configuration and control is robustly tested and ready for use.

The second section focuses on the DAQ system and related control mechanisms. Here, the process is similarly divided into development, integration, and commissioning phases. Initially, a common firmware DAQ block is developed to test subsystem interactions across Tracker, ECal, HCal, Trigger, and TS. This block is then integrated into each subsystem with a dedicated interface that transmits data to an event-building machine. In parallel, a slow control system is implemented through a prototype (using the Rogue platform), which is later integrated to manage front-end controls for individual subsystems and, eventually, commissioned to provide a centralized slow control interface. To pull everything together, the run control system and configuration database is incrementally integrated and commissioned to facilitate both standalone subsystem testing and multi-system global runs, ensuring synchronized and efficient operation across the entire DAQ framework.

Finally, the third area of the work is on the procurement of computing and electronics hardware and is coordinated with the other two work breakdown sections to make sure testing and production hardware are available for development and the experiment's operations.

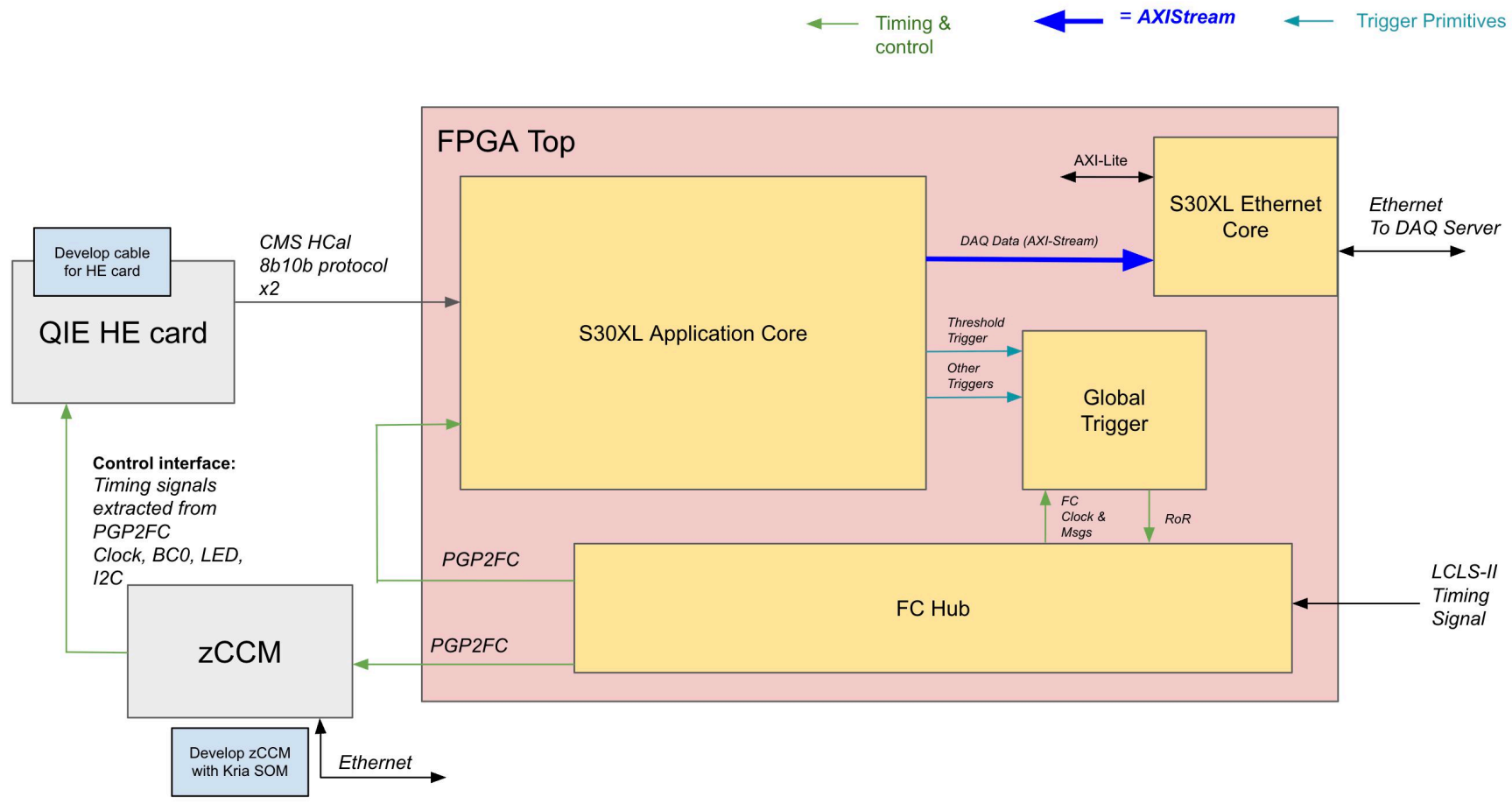

Figure 3.144: S30XL test beam TS system readout, trigger algorithm, fast control hub, and data readout path

#### 3.9.7.1 TDAQ R&D at CERN and S30XL

Two test beam campaigns were used as milestones to develop the TDAQ system concept and test that the system design meets requirements that were possible to test during the test beam. The first test beam was at CERN in 2022 and was largely used to test existing technology. This was used as input to the second test beam which was at SLAC in 2025.

The S30XL test beam setup utilized a layer of the TS subsystem to measure the dark current at the S30XL (Sector 30 Transfer Line). This includes the LCLS-II beam structure (185.71 MHz RF structure) and was a valuable test of the backend of a small part of the TS system. While only a fraction of the channels were read out of the TS subdetector, it provided an opportunity to evaluate a complete test readout system. This is shown as a block diagram in Fig 3.144.

The different blocks required in the FPGA firmware readout were tested for the TS detector in S30XL. This includes the interface to the TS detector itself using an 8b10b protocol. That data was processed by the S30XL application core which linearized hits for transmission to the global trigger. Based on the global trigger decision, the Fast Control (FC) hub would send a readout request for the data to be read out over Ethernet. While the interfaces between the blocks may be different in the final experiment, this allows us to test many aspects of the TDAQ system. This includes the firmware shell and environment needed to build the single FPGA readout system, a FC hub which can process the LCLS-II Timing Signal and produce PGP2FC messages. On the software side, the monitoring tools were developed and the Rogue tools were evaluated. While there are many other milestones for the full project, being able to test these elements with the S30XL beam provides key experience for the project, which helps us to better understand the risk and timelines of the TDAQ system.

#### 3.9.7.2 Risks and Opportunities

We identify sources of potential risk in the TDAQ system of the experiment which could cause design changes or failure to meet requirements. Based on the system requirements in Sec. 3.9.2, we identify two main potential risks: (i) the trigger calculation latency is longer than expected due to algorithm complexity, and (ii) the trigger bandwidth is higher than expected due to higher backgrounds than expected, e.g. from detector subsystem performance challenges. To address the former, we discuss simplified trigger algorithms which can meet the demands of the physics program. For the latter, we consider additional computational power to augment the FPGA-based trigger.

**Simplified trigger algorithms** The trigger menu, as defined in Sec. 3.9.5.5, presents an example of the primary trigger paths we will implement in FPGA firmware that must meet the latency requirements of the system. Given the relatively tight latency constraints, we consider that the system should be able to achieve its primary goals with fairly simplistic algorithms.

For the primary physics trigger for missing energy, we have performed feasibility studies that demonstrate that the algorithm will fit into system specifications. The electron counting algorithm for the TS system is found to fulfill the requirements for FPGA resources and latency. It is critical to demonstrate the TS electron counting trigger capability meets the desired performance goals, as the other element of the missing energy trigger (layer sums) is simpler from a firmware perspective.

Beyond the primary trigger, the rest of the trigger selection is also relatively straightforward to implement in firmware – based on prior experiences in implementing trigger algorithms in CMS and ATLAS. The energy sums and clustering in the ECal are based on max values and sums, and the MIP tracking algorithms are based only on consecutive layer counts.

The exception is the electron $p_T$ trigger. The current implementation requires 3D clustering of trigger primitives to extract high precision cluster energy and position information in the ECal. While this is the basis for our high efficiency electron $p_T$ trigger for studying electronuclear physics and background sideband kinematics, if this algorithm cannot be implemented in firmware, we will be able to simplify the trigger at the sacrifice of overall performance. We will also consider simple 2D cluster projections if necessary.

While we have designed the system to meet the requirements of the experiment, there are also opportunities for improved performance beyond what is possible currently. This could potentially improve the overall signal efficiency or reduce backgrounds even further to reduce computing resources and provide more bandwidth for other physics goals.

**High Level Trigger extension** In the scenario that the hardware trigger cannot meet the 10 kHz rate requirement, a potential mitigation strategy to restore our rate could be to implement a software-based High Level Trigger (HLT). An HLT would allow for software-based object reconstruction in a simplified "fast reconstruction" implementation. This would enable the experiment to utilize the full granularity of the detectors including tracker information in a downstream trigger filter. The HLT would both reduce the size of the data we need to store and also increase the rates we can accept from the initial L1 hardware trigger.

In a situation where the trigger rate exceeds the planned 10 kHz rate – e.g., 10-50 kHz – we could consider an off-the-shelf commodity CPU-based system to reduce the DAQ rate. As an example, we consider that the primary physics missing energy trigger rate needs further reduction to get to the target 10 kHz rate.

As an initial study of the feasibility of an HLT, we use the current implementation of offline reconstruction in `ldmx-sw` to deduce the required processing power. HLT reconstruction software would include optimized versions of the trigger scintillator (TS) hit reconstruction and clustering, ECal hit reconstruction, and tracking algorithms, to comply with the strict time requirements of the online selection. Timing studies on 10000 electron events using a single CPU core show that the combined current offline TS + ECal + tracking in both the Tagging and Recoil trackers takes around 4.07 ms or $\sim 250$ events per second for single-electron events. For multi-electron events, with a mean of $\mu = 2$, this is reduced to $\sim 140$ events per second.

Fig. 3.145 shows the relative time taken for each part of the reconstruction. The tracking algorithms dominate; further optimization of these will reduce the time per event significantly, and we could envisage running only one (the recoil tracking) to reduce the time by a factor of 2.

The single-core throughput can be used to obtain the hardware needs of such a system. Given the 250 ev/second rate found above, a 100 CPU core system would have a 25 kHz throughput. Such a system is feasible even before any streamlining from optimized tracking algorithms or other HLT parameters.

The HLT potentially presents an opportunity as well for other physics priorities of the experiment. For example, the experiment could accept a higher electron $p_{\mathrm{T}}$ trigger rate at the hardware trigger level and use the tracker information to capture a higher efficiency of high $p_{\mathrm{T}}$ events.

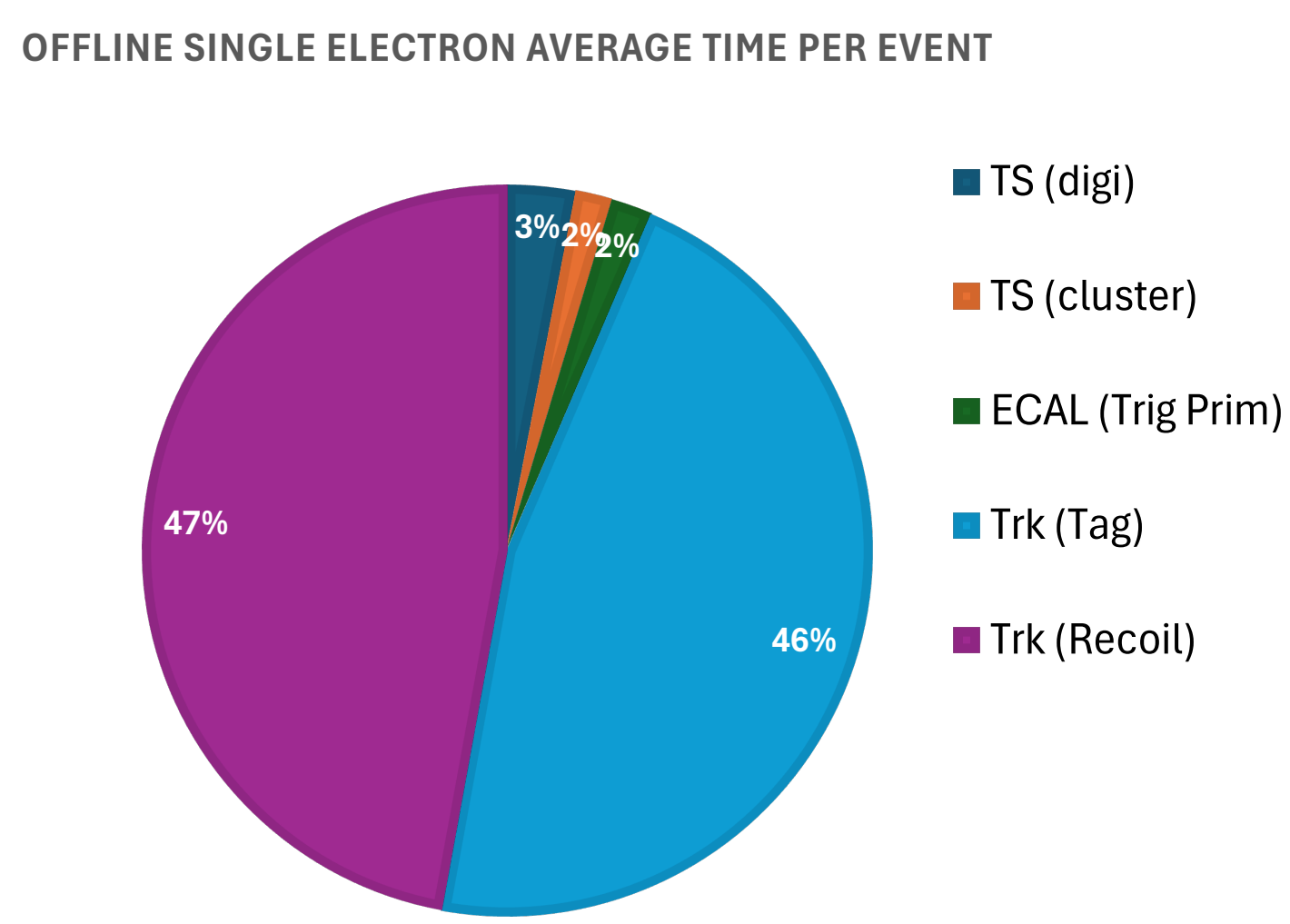

Figure 3.145: Fraction of average times per event for single 8 GeV electron sample on single core.

## 3.10 Software and Computing

### 3.10.1 Introduction and overview

The LDMX computing and software infrastructure serves to process data and perform event reconstruction, produce Monte Carlo (MC) simulations, facilitate data management, storage, and access, handle conditions, calibration, and configuration data for offline applications, and support data analysis. The design of the LDMX computing model is driven by the high-statistics nature of the experiment, which results in the production of relatively large volumes of experimental data and simulation. The computing resource needs are based on the requirements for processing and storing the experimental data and for producing, storing, and processing simulated data.

### 3.10.2 Requirements

The high-level requirements for the LDMX computing and software system are defined in Sec. 3.2. The technical requirements that the LDMX computing infrastructure and software need to provide to support the LDMX experimental program are the following:

- **Technical computing infrastructure requirements:**

COMP1 sufficient tape storage to maintain a copy of raw data collected in the first run of LDMX

COMP2 sufficient tape storage to maintain a copy of the associated reconstructed data (supporting up to 5 reconstruction iterations for $\sim$10% of the data) and the equivalent amount of legacy Monte Carlo simulations

COMP3 sufficient disk storage to maintain a copy of the reconstructed data and MC needed for analysis

COMP4 sufficient computing resources (CPUs) to run reconstruction of the raw data within 3-4 months and to produce the full suite of supporting Monte Carlo simulations within 2-3 months

- **Technical software requirements:**

SW1 robust and flexible software framework that provides the ability to perform event reconstruction, simulation, and support analysis within the above timescale

SW2 auxiliary services to support data collection like data catalogs, and databases or other systems to manage detector conditions, calibration and configuration data.

### 3.10.3 Software

The LDMX software performs event simulation, reconstruction, and analysis, and is deployed using a containerized approach. LDMX uses a C++-based software framework (further described in Sec. 3.10.3.2), known as `ldmx-sw`, for event processing and simulation. A software bus model is implemented for communication between data processing modules in sequence. The data processing pipeline is configured using an embedded python interpreter at run-time, which allows for the dynamic loading of simulation, digitization, reconstruction, and analysis modules. The framework was designed to be lightweight, with flexibility and ease of use in mind.

The propagation of particles through the detector, illustrated in Fig. 3.146, as well as their interaction with material, are described using a custom version of the `Geant4` toolkit [116, 117], 10.02.p03. This version includes modifications to the Bertini Cascade model, improved matrix elements for muon conversions, and the incorporation of dark-bremsstrahlung signal generation. A description of the detector geometry and materials is provided in a modular way using the XML-based Geometry Description Markup Language (GDML) [118]. The implementation of the LDMX detector description in GDML allows for the study of different full detector configurations, as well as subsets or prototypes. Sub-detector specific algorithms are used to digitize and reconstruct data, add noise, perform track-finding and clustering, and implement vetoes for the rejection of backgrounds.

The computing resources needed for the processing and storage of experimental and MC data will primarily be hosted at the SLAC Shared Scientific Data Facility (S3DF). In addition, LDMX can take advantage of computing resources available at other collaborating institutes for the production of MC samples via a lightweight distributed computing system, LDCS (see Sec. 3.10.4.3). Computing clusters at Caltech, UCSB, SLAC, and Lund University are integrated in LDCS to run simulation jobs. The simulated data produced at these sites can be transferred to SLAC for storage and further analysis access.

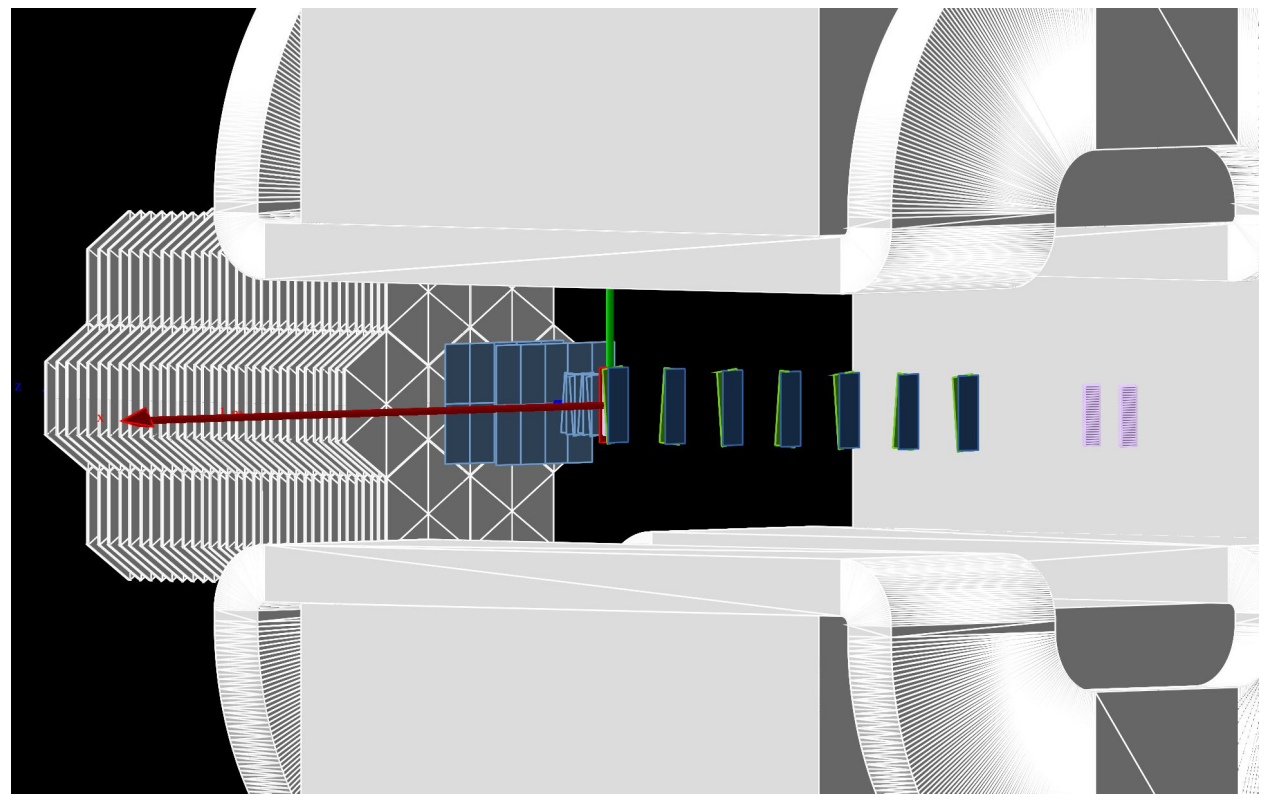

Figure 3.146: GEANT4 visualization of detector elements in the LDMX simulation. For better visibility, the HCal is not included.

#### 3.10.3.1 Containerized approach for software distribution

The use of "containers" is a technology that puts together various Linux kernel features, essentially enabling a process to be run within a specific environment (down to but not including the kernel itself) defined within an "image". The containerization of intricate software stacks has rapidly grown in popularity in recent years due to the stability the technique provides. Containers ensure that the developer's computer and the production server on which it is run share the same (containerized) environment. From the software's point of view, there is no difference between the computer where it is being developed, built, or run. Many experiments in HEP have adopted containers as a means for distributed computing to avoid the tedious task of aligning software versions across many servers and clusters.

The LDMX collaboration has chosen to standardize the software stack using container images and has chosen to go a step forward in their use. Not only are container images used for distributing `ldmx-sw` to clusters and servers for running large campaigns of data processing, but images are also used for distributing the dependencies of the software to developers. This ensures that the software is always developed for the target environment and simplifies both setup and development. Instead of installing all of the software dependencies, many of which are required to be built from scratch, developers can simply install a single "container runner" like Docker, Podman, or Apptainer in order to gain access to the software and begin development. The popularity of containers outside of the HEP ecosystem also provides another benefit – LDMX software can be built and run on non-Linux computers due to their support by Docker and Podman. Besides facilitating contributions from a wider group of LDMX collaborators, this feature reduces the load on core developers as, for instance, they do not have to maintain support for a range of operating systems.

Finally, containerization simplifies reproducibility, since all of our software, down to which operating system and core development libraries are used, is strictly versioned.

#### 3.10.3.2 Framework

The event processing framework designed, developed, and maintained by LDMX is focused on allowing for the flexibility necessary to perform the wide range of tasks necessary for the experiment. The C++ framework uses CERN's ROOT [119] for data serialization, Boost [120] for logging, and custom configuration specifications implemented in Python [121] for dynamic run-time configuration. The design involves a sequential model: each event[8] is passed to individual processors in a certain sequence. The individual processors can add data such as reconstruction objects, veto decisions, or data quality flags to the event for

[8] defined as the full processing of the aftermath of an incoming (beam) particle in simulation, and in detector data as the pre-defined time window where data are recorded after a positive trigger decision

later processors to use, which are eventually serialized into the output ROOT file. The processors can be built separately from the framework and then dynamically loaded and created at run-time by an abstract factory. This design choice allows for all of the computationally-intensive software tasks necessary for LDMX (simulation, reconstruction, and some analysis tasks) to use this framework and be organized into separate modules that are only loaded into memory when that module is being used.

The serialization portion of the framework is similarly dynamic, focusing on enabling users' code to add data structures ranging from simple (e.g., individual booleans) to considerably more complex (e.g., containers of custom classes). This wide array of data types is supported by ROOT's dictionary system during the serialization stage and abstract wrapper classes with partially-specialized template derivatives during run-time. This complexity within the framework is necessary to allow a simple interface – one where the user interacts with simple and complex types in the same way.

Combining this highly dynamic serialization library with the sequential-processing model configured at run-time gives a strong foundation for all of the software needs of LDMX. Written in C++, this software framework enables high performance for all of the major data processing tasks necessary for the experiment. A wide variety of different data processing tasks, including simulation, detector (readout) emulation, event reconstruction, analysis calculations, and reading and writing files, are unified under one framework. This approach enables the software to be well organized while also centralized in one location.

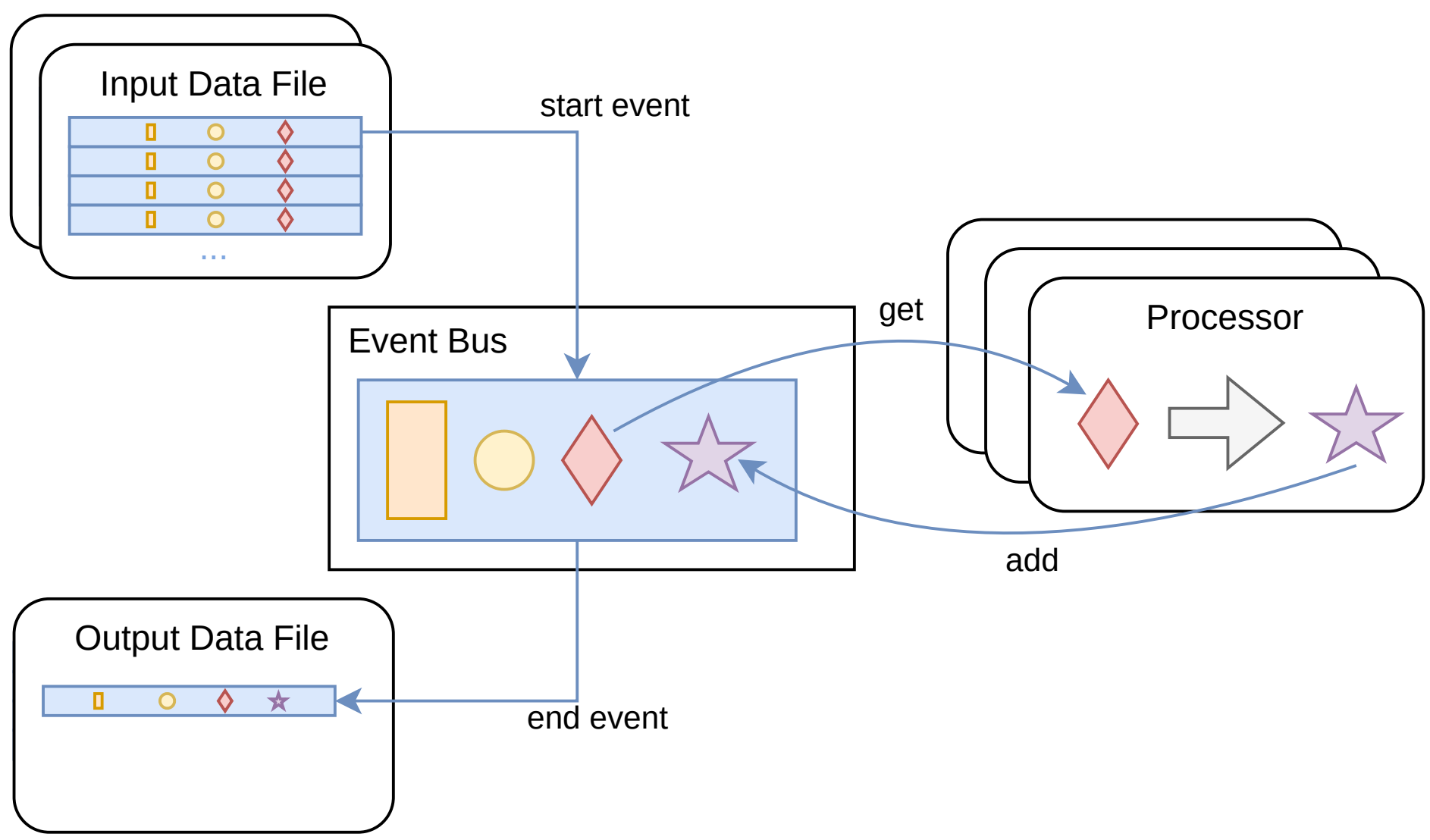

Figure 3.147: Flow chart of data processing in the LDMX framework. Each processor has the ability both to "add" data to and "get" data from the event. The processors are run in a user-defined sequence. Event data can also be loaded into memory from one or more input data files before the first processor is run. After all processors are done with an event, it is saved to the output data file.

Many tasks and data structures can be shared between different stages of data processing and different physical subsystems during the same stage, which ensures coherent definitions. For example, the various data processing modules that need detector information can all use the same code linking to the same library, allowing for consistency between these different modules.

The structure of the LDMX data processing software allows experts to focus on developments in individual areas and integrate updates made in other areas seamlessly. Table 3.14 lists the currently active modules in `ldmx-sw` and a short description of their purposes. Notice that a module for reading in real data taken by a DAQ system already exists, and has been used to process data from testbeams and teststands. This ensures that `ldmx-sw` has all the capabilities necessary to fully support LDMX's experimental run. While further refinements of the code will be needed, the core structure of the entire data (real and MC) pipeline is mature and well-established. Subsequent sections describe some of these modules in more detail.

| Module | Description |
|---|---|
| SimCore | Interface with Geant4 to simulate energy depositions in the detector |
| Recon | Common reconstruction tasks shared across different subsystems |
| Packing | Reading from and writing to a custom binary format for real data DAQ |
| HCal | Simulation and reconstruction of HCal subsystem |
| ECal | Simulation and reconstruction of ECal subsystem |
| Tracking | Simulation and reconstruction of Tagger and Recoil Tracking subsystems |
| TrigScint | Simulation and reconstruction of Trigger pad subsystems |
| Trigger | Emulation of firmware trigger algorithms |
| DQM | Shared Data Quality Monitoring analyses |

Table 3.14: Examples of the different processing modules *currently in use* by LDMX, showing the flexibility that is accomplished by the processing framework.

The following paragraphs highlight specific processing capabilities that are commonly used within the normal procedure of the experiment.

**Event Selection** The framework enables output data reduction through the concept of "storage hints", allowing multiple processors from disparate modules to vote on whether a specific event should be saved within the output data file. In its simplest and most common form, the framework is configured to only "listen" to one processor, typically a trigger emulator or vetoing algorithm, which is in charge of making the decision on whether an event should be saved into the output file or not. This ability to "skim out" interesting events from a larger sample is extremely helpful to efficiently study specific events in more detail. Additionally, the skimming can be done in the same program execution as the simulation to avoid writing events to disk that would inevitably be removed in further analysis.

**Object Selection** Similar to event selection, users can inform the framework which event objects should be stored in or "dropped" from the output file. Dropped objects still exist within memory so that they can be used throughout the processing sequence; however, dropping them from the output file can lead to substantial space savings.

**Multiple Passes** Each of the objects that are stored within the output file includes a "pass name" associated with the processing sequence that produced it. The pass name is configurable by the user and shared across all objects created during that run of the program. The pass name can be used to accumulate multiple runs of similar algorithms within the same output file, enabling easier comparison between changed parameters.

**Real Data Formatting and Persistence** The goal of the DAQ system is to acquire real data in a known way and save this data in a known format. We intend for the event builder in the DAQ system to produce files directly readable by `ldmx-sw`, since this framework's event file schema is a thin and understandable layer on top of the ROOT file format. Until then, a DAQ system will write out custom binary files that are less structured than ROOT files. As mentioned above, given the experience with data from testbeam and teststands, there are proof-of-concept processors within this framework focused on unpacking[9] custom binary file formats into structured event objects, and programs that align these data blobs offline according to timestamps embedded within them. This code is a foundation on which the last stages of the DAQ system could be built.

**Random number seeding** Monte Carlo event generation relies on random sampling of probability distributions. While `Geant4` has its own random number generator, `ldmx-sw` uses ROOTs `TRandom3`

[9]both encoding `ldmx-sw` data to detector specific binary data format, and conversely, decoding

generator alongside standard library generators, with all random number generator seeding based on either

- the run number, which is unique for each file (and common to all events therein),
- the generation time stamp,
- or an explicit seed passed by the user through the run configuration.

Run number seeding is the default mode to set up the main seed. Each process uses a seed that is deterministically based on the main seed and a hash of the processor name. This mode guarantees both simulation reproducibility and seed uniqueness, as long as run numbers are chosen to be unique.

In addition, at the start of each attempted GEANT4 event, the full random number generator state is recorded into the event header such that the same event can be trivially reproduced (resimulated) regardless of the run configuration. This allows us to store less information on disk for most events since those of particular interest can be re-generated, when needed, with more detailed information. One pertinent example is full particle histories that would require prohibitive amounts of disk space if kept for every event.

#### 3.10.3.3 Event generation and simulation

As mentioned above, `ldmx-sw` interfaces to the `Geant4` toolkit to simulate the interaction of particles with the material of the LDMX detector. This interface is constructed within the SimCore module of `ldmx-sw` and is designed to enable a lot of the flexibility from `Geant4` while keeping control of the data within the centralized framework described above. This is accomplished by using `Geant4`'s "user actions" to observe the simulation unfold and pass the created data structures to the LDMX framework while also helping `Geant4` decide whether a certain simulation event should be kept within the sample ("filter" the simulation for specific types of events).

Besides using these user actions for data handling, we also use `Geant4`'s GDML parser to read in the detector description, biasing framework to create different types of samples, and primary generators to begin the simulation in different ways. All of these interfaces are built upon an abstract factory, so defining new classes of user actions, biasing operators, and primary generators is relatively simple and allows all of them to be configured by our dynamic Python configuration system, similar to how data processors themselves are configured at the beginning of a run.

Below, more specifics are given on how different simulation samples are generated; all of these are done from within this unified interface, making it simple for them all to share tools and be well-validated.

##### 3.10.3.3.1 Background sample generation & modeling

Being able to reject background events to a level of $O(10^{14})$ implies studying very rare Standard Model processes. The leading backgrounds for LDMX arise from either a rare interaction of a hard bremsstrahlung photon or a rare hard interaction of the primary electron. Simulation samples used by LDMX to study background processes are generated directly in `Geant4`. Version 10.02.p03 is currently used, with modifications to the photonuclear (PN), electronuclear (EN), and photon-conversion processes to achieve better accuracy in modeling.

The `Geant4` Bertini Cascade model is based on a cascade of individual particle-nucleon and particle-dinucleon interactions, with total cross-sections and 2-body final state kinematics taken from data. The Bertini model has been broadly validated, and is the default model in `Geant4` for PN reactions initiated by $< 3.5$ GeV incident photons. However, it is believed to be accurate up to $\sim 10$ GeV incident energies. For LDMX simulations, we use a modified version of the `FTFP_BERT` physics list, with photo-nuclear reactions of photons with energies of up to 10 GeV always modeled using the Bertini cascade. EN reactions, which at these energies proceed through a virtual photon, are closely related to the PN ones, and are simulated in `Geant4` using the equivalent-photon approximation, with Bertini cascade as the default generator. A detailed study of how the `Geant4` modeling with Bertini compares to other generators is discussed in Sec.4.1.2.1.

In order to ensure that we accurately sample the phase space of PN reactions that are the most challenging to veto for LDMX, the rates and phase-space distributions of event types that are design drivers for LDMX have been validated. These reactions include single forward neutrons, moderate-angle neutron pairs, and single forward charged kaons and pions. Compared to data (as described in Ref. [78]), the single- and di-neutron reactions, as well as backscattered hadrons in the so-called "cumulative" region – the kinematic phase space only accessible through scattering off a multi-nucleon initial state – were found to be significantly overpopulated by the default Bertini model. In contrast, single-pion and single-kaon final states were found to be underpopulated. After identifying the origin of these discrepancies, corrections were implemented in our customized version of `Geant4`. Most of these have since been adopted by `Geant4`.

A combination of *filtering* – physics process selection, such as interaction type and detector volume in which it occurs – and the `Geant4` *occurrence biasing* toolkit is used to efficiently simulate Monte Carlo samples corresponding to these rare event types. In addition, we have developed a method to resample events until a desired final state is reached.

**Filtering** Filtering is a requirement on the physics process being simulated that avoids spending simulation time on events that will not be retained in downstream event selection steps. A filter can ensure e.g. that a certain physics process occurs in a specific detector volume, that a certain initial or final state particle is involved, or that some kinematic criteria are fulfilled.

As an example, an important class of background processes for LDMX begins with a hard bremsstrahlung reaction in the target, with the resulting high-energy photon subsequently undergoing a rare interaction in the target area, recoil tracker, or ECal. To efficiently simulate hard bremsstrahlung events that would survive the baseline tracking and ECal selection (see Sec. 4.2.4), a filter restricts the simulation to events where the electron undergoes bremsstrahlung in the target and leaves the target with an energy less than approximately the missing energy trigger threshold. For events surviving this filter, we place the electron track in a waiting stack until all other interactions have been simulated, and apply additional selections on the bremsstrahlung photon and its interactions before incurring the computational cost of simulating the electron shower.

**Occurrence biasing** In order to efficiently simulate rare photon interactions such as photo-nuclear interactions or muon conversions that are relevant for LDMX, and have cross sections $10^{-3}$ to $10^{-5}$ times the photon to electron pair conversion cross section, we use the `Geant4` occurrence biasing toolkit that was introduced in version 10.0. This facilitates the simulation of particle interactions with a biased interaction law, e.g., enhancing the cross section for rare processes. All occurrence biasing follows the basic scheme

$$\sigma^{\text{biased}} = B\sigma^{\text{physical}}, \tag{3.8}$$

where $\sigma$ denotes cross-section and $B$ is the biasing factor. Event weights are assigned to account for the effect of the cross-section biasing on both the probability of occurrence of the biased interactions and the survival probability of the interacting particle.

**Event resampling** For studies of rare PN interactions `ldmx-sw` can be configured to resample the specific PN interaction outcome until a desired one is found. One example is resampling the Bertini Cascade model until a specific final state topology is produced, while keeping the incoming photon kinematics fixed. To account for the resampling, the event weight is naïvely multiplied by the inverse of the number of resampling attempts. While the resulting electron-on-target estimate for such samples is slightly inaccurate, the method is satisfactory given the large resource savings, which enable studying these rare events to sufficient detail.

#### 3.10.3.3.2 Pileup modeling

In addition to the various specific physics processes, the energy deposited in the detector by additional electrons in the same or neighboring bunches in the accelerator bunch train is modeled using an event overlay technique. A sample of "inclusive" beam electron interaction events, i.e., with no biasing, filtering or other selection, is used for the additional beam electrons. The pileup events can be shifted in time with respect to the interaction of interest, both as a time smearing within the same bunch (modeled by sampling a Gaussian with the expected width) and by being placed in some interval of earlier or later bunches. Both a fixed and Poisson-fluctuated electron multiplicity can be used. The minimum multiplicity is always 1 (no pileup), i.e., at least the interaction in the target in the event of interest, defined to occur at $t = 0$.

The combination of events happens at "`SimHit`" level, which is the deposited energy combined at the level of granularity of every readout channel in a detector system. Digitization, emulation, and further reconstruction happen on the event after pileup energy deposition is added. This mimics the situation in real data.

This technique allows for separately biasing the process of interest and reusing a smaller sample of pileup events. The pileup events are taken from a separate input file in the process, and can be added on the fly as the simulation proceeds to reconstruction, or applied to an existing simulation input file. To minimize systematic effects, the pileup event number to start from is randomized (using the seeding scheme described in Sec. 3.10.3.2) and thus different for different input files, even if the same pileup file would be used more than once.

#### 3.10.3.3.3 Dark bremsstrahlung generation

Accurate kinematic simulation of the outgoing electron is required for optimal experimental sensitivity measurements and appropriate design of search strategies. To accurately describe the kinematics of the dark bremsstrahlung process, the simulation must account for the possibility of energy loss through bremsstrahlung or multiple scattering before the dark matter interaction. This is particularly important in thick targets. The process must thus be included within the experimental simulation rather than being imported from initial state event generators.

For this reason, we utilize G4DarkBreM [122] which embeds the dark bremsstrahlung process into `Geant4`. G4DarkBreM calculates the cross section using numerical integrals of the Weizsäcker-Williams approximation. The kinematics simulation employs the technique of scaling MadGraph/MadEvent event libraries. The accuracy of the total cross section and kinematics is validated using MadGraph/MadEvent samples at a range of incident lepton energies.

For the study of dark bremsstrahlung events in LDMX, the code package used to develop and validate G4DarkBreM was also used to generate the input reference libraries for sample generation. This code is a custom version of MadGraph/MadEvent4 with the following updates.

- Introduction of basic dark sector particles (massive boson and spin-1/2 fermion) which act as the representatives of the dark sector, interacting with the standard model particles.
- Definition of a nuclear particle (electrically neutral, spin-1/2 fermion) with new couplings to the dark photon, including the nuclear form factors.
- Updating the definition of the electron to include its small (but non-zero) mass to prevent divergence of the cross section at lower energies.

These updates, along with some wrapping code, enables the generation of dark bremsstrahlung events for a range of target nuclei, incident energies, and either incident electrons or muons.

Libraries are generated for incident electron energies starting at the beam energy, and decreasing in 10% steps, as suggested by the validation of G4DarkBreM [122].

Since the dark photon is required to be simulated with at least 50% of the beam energy, the library generation halts at 50% of the beam energy. We generate libraries for the primary nuclei that could generate dark bremsstrahlung within the target or ECal: tungsten, silicon, oxygen, and copper. The remaining elements within the simulated ECal (sodium, calcium, carbon for example) have atomic numbers close to those nuclei already sampled and can be faithfully simulated with these libraries. The random number seeding scheme described in Sec. 3.10.3.2 yields a unique reference library for each simulation run, even when keeping the dark photon mass and beam energy the same.

#### 3.10.3.3.4 Use of other generators for `Geant4`'s photonuclear interaction modeling

Starting from `Geant4` version 11.3, it is possible to run the PEANUT [123] hadronic interaction model from FLUKA@CERN [124] as an alternative to native `Geant4` hadronic interaction models. This integration feature has been backported into the version of `Geant4` used by `ldmx-sw`. In particular, `ldmx-sw` is designed to allow the simple exchange or instrumentation of the default photonuclear interaction model in a desired energy range.

#### 3.10.3.3.5 Pythia8 integration

In the context of `ldmx-sw`, a library is being developed as an independent module that can be compiled and linked to in any `Geant4` application that allows for the usage of the Pythia8 event generator as part of the application. In particular for experiments like LDMX, this allows for incorporating BSM processes, such as dark bremsstrahlung, directly into the simulation.

Since Pythia8 already supports incorporation of MadGraph/MadEvent event libraries into its event generation, this integration enables *native* MadGraph/MadEvent in `ldmx-sw`.

### 3.10.3.4 Reconstruction

Reconstruction aims to create higher-level objects, useful for physics analysis, by combining and interpreting the readout from the individual readout channels.

**Digitization** In `ldmx-sw`, both simulated and recorded detector data ultimately undergo the same reconstruction sequence. To that end, an emulation of detector digitization is applied only to simulated data. Detector digitization samples a charge or current pulse over a given time interval, resulting in a series of integers across a detector-specific number of time samples, along with, e.g., measures of arrival time, pulse saturation, and other pulse characteristics obtainable from the specific readout electronics. To mimic this procedure in simulated data, all energy deposited by particles traversing a detector element that would be read out by a given single channel is combined into a total energy deposited, to which pathway-dependent modifications and detector noise emulation can be applied. The resulting energy is converted to a system-specific signal pulse (with some amplitude and shape over time), which is sampled and digitized according to the known behavior of the system's readout electronics.

The digitized time samples from either data taking or simulation are then processed in the same way to reconstruct the energy deposited in each detector element. This subsection broadly describes both digitization emulation and reconstruction as implemented for each detector system in software. For detailed descriptions, refer to each system section.

**Trigger Scintillator** The data from the trigger scintillator are mainly used for the trigger decision. A relatively high level of reconstruction sophistication is thus required already from the firmware implementation (described in Sec. 3.5.4). This development was preceded by software algorithms for hit, cluster and track reconstruction, which are currently used for electron counting in the trigger decision emulation.

Digitization emulation in software produces a sequence of time samples with ADC and TDC values for each channel, by digitizing a piece-wise exponential response pulse with an amplitude given by deposited energy in the corresponding bar (as shown in Fig.3.35). Simulated noise is added as additional pulses and included in digitization.

Hit reconstruction integrates over the charge (linearized ADC values) in a time interval large enough to contain the full pulse, producing one hit per channel. The hit's main characteristics are the number of photo-electrons (PEs) corresponding to the readout signal and a time stamp given by the sample of interest and the TDC value (0.5 ns precision subsample) of the rising edge. If the rising edge is in a different time sample, the TDC is not a subsample timestamp but instead a value signaling early or late arrival, which makes hit reconstruction discard it.

Hits in adjacent trigger scintillator bars can be combined into clusters. Contributions from noise are suppressed with threshold cuts on the number of PEs (efficiently filtering out low electronics noise) and hit timing (suppressing delayed energy depositions from secondaries).

Finally, tracking combines clusters across the three modules that are compatible with a horizontal beam electron trajectory within some tolerance. Correlating the spatial information across all three pads is a powerful way to suppress fake electron counts from spurious activity in the TS.

**Tracker** The reconstruction of simulated events in the tracker systems is organized in multiple stages. First, the energy deposits in the tracker sensors, simulated by the GEANT4 engine, are smeared according to the expected position resolution of the tracker sensor to emulate the detector response. Only information in the direction orthogonal to the orientation of the strip is used in the definition of a hit in subsequent tracking reconstruction steps.

The reconstruction of charged particle trajectories is based on the A Common Tracking Software (ACTS) [125] library and is performed separately for the Tagger and Recoil trackers. We use the framework implemented in ACTS to describe the geometry and material of the trackers and to provide track finding and fitting algorithms.

Track finding starts with grouping combinations of five hits, collected on different layers, which are used as seeds for the track finding algorithm. An initial combined linear and parabolic fit is performed on each group to extract an initial estimate of the track parameters. The group of hits and initial parameters are then fed into the Combinatorial Kalman Filter (CKF) algorithm in ACTS that performs track fitting and attempts to extend the track in layers not included in the seed. This procedure can make different tracks with very similar hit content so after running the CKF we trim the track list by choosing between tracks that share more than 1 hit based on their number of hits on track and track $\chi^2$. The performance of the trackers and reconstruction software is discussed in Sec. 3.4.5.

**ECal** The ECal reconstruction has been broken into two main stages: emulation of the digitization process (which only occurs in the simulation chain) and reconstruction of digitized data into physical signals.

The digitization emulation is done by constructing a simulated analog pulse from the simulation information and then passing that pulse through an emulation of the readout chip. The readout chip emulation mimics both the low-energy readout mode, where simple pulse height measurements are done, and the high-energy readout mode, where the chip times how long it takes for the accumulated charge to drain off. The parameters controlling the pulse shape and the readout chip emulation are all configurable and integrated into the centralized conditions system so measurements derived from data and test benches can be applied.

The reconstruction of the digitized data can also be roughly broken down into two steps. First, an amplitude in charge equivalents is estimated from the digitized samples. Both the low- and high-energy readout modes are scaled to this amplitude using configurable pedestals and gains that are different for the two modes. After this amplitude is determined, further parameters can be applied to convert this into energy units, apply sampling-fraction constants, and other second-order corrections. Higher-level ECal variables (like shower-shape features) are then determined from these reconstructed hits.

**HCal** The HCal reconstruction structure is similar to the ECal reconstruction. Since both systems share the same readout chip, the steps included in the digi-emulation stage are similar to the ECal. From the simulated calorimeter hits in a scintillator bar, digitized samples are formed for each instrumented end of the bar. This process includes summing the energy for all SimHits within each detector bar before configuring the emulation of the readout with the HGCROC chip. The emulation parameters are configured separately from the ECal and are integrated into the central conditions system. To convert to charge, we estimate the total number of photo-electrons associated with each energy deposit using standalone measurements with the scintillator and muons. Once the number of photo-electrons associated with each energy deposit is computed and noise is accounted for, a pulse is sent to the shaper to create a digitized sample.

Currently, the time of the simulated hits used in the digitization step does not account for photon transport through the scintillator and wavelength-shifting fibers. The arrival times of these photons at the SiPMs will be simulated using the method described in Ref. [126]. This method employs lookup tables to determine both the probability of a photon being detected by one of the four SiPMs in anHCal CMB and its corresponding arrival time.

The reconstruction step uses the digitized data to form hits. The digitized samples provide an amplitude at each end of the bar. The amplitude is summed for both ends and then translated to an estimate of deposited energy using energy and photo-electron conversion factors. The time of arrival information is also extracted from the digitized samples and used to get a rough estimate of the position of a hit along the scintillator bar. The other two position coordinates are extracted from the location of the bar inside the detector geometry, using the half-length of these shorter sides (effectively, positioning all hits on a scintillating fiber).

**Trigger emulation** The LDMX trigger system makes use of logic implemented in FPGA firmware to select events for permanent storage and offline study. In parallel with this system, a series of software *emulators* are maintained, which mirror the internal trigger logic as well as external interfaces. These software implementations are useful for a variety of purposes, including rate estimates, rapid prototyping of new algorithms, and serving as 'golden model' targets for firmware development. Emulators also allow efficient and bit-accurate evaluation of the trigger logic within the standard `ldmx-sw` simulation framework.

Traditionally, emulators are written by hand in conjunction with the trigger firmware and are tested to achieve perfect agreement on the full space of possible inputs to be encountered by the trigger. Parts of this process may be simplified through the use of *high-level synthesis* (HLS) tools, which compile circuit descriptions written in C++ directly to register transfer level (RTL) code. In addition to these enhanced automated design checking capabilities, directly calling the HLS implementation of algorithms within `ldmx-sw` allows for greater coverage of the trigger logic by rapid, bit-accurate emulators. These capabilities are currently utilized to emulate logic across the TS, ECal, HCal, and global trigger processors.

#### 3.10.3.5 Offline validation

At any given time, `ldmx-sw` is being actively maintained and developed by many collaborators. This calls for a thorough validation procedure of any suggested code changes.

As the code base is hosted on GitHub, its existing tools are used for code review and approval, and continuous integration testing. The tests verify that the code base builds after the proposed changes and trigger a set of short simulation runs that target different types of physics and use cases. The resulting data are used to produce a suite of histograms. These histograms are compared to a set of "golden" histograms generated by the same workflow using the most recent, validated code release, which are kept in the repository. Any significant differences (as judged by a Kolmogorov-Smirnov test) trigger a "failure" and have to be manually checked. Similarly, log printouts from the jobs are compared to reference logs, which primarily catch changes to workflows. While code changes are often intended to change certain distributions, this validation procedure prevents unintended changes to the code to the extent possible. Once a change is understood to be desirable, the updated code is merged into `ldmx-sw`. Updates of reference logs and validation histograms are triggered at every new version release.

Some of the validation histograms display low-level quantities and could be implemented for quality monitoring of real data, similarly using a set of histograms from well-understood, real data as a reference.

#### 3.10.3.6 Analysis workflow

The analysis of both experimental and simulated data involves accessing and loading the data, processing it in order to extract the relevant information and calculate higher-level variables, and writing analysis outputs in the desired format, e.g. ROOT files containing variables or histograms of analysis-level quantities. Two possible workflows are currently envisaged, one of which involves running C++ processors with `ldmx-sw`, while the other involves running stand-alone code implemented in Python, using packages like `uproot` for loading data and `awkward` for efficient data manipulation.

Both approaches have their respective use cases and advantages. The implementation of an analyzer directly in `ldmx-sw` allows for the long-term preservation and maintenance of the analysis code, and also facilitates analyses that might be inefficient to implement via Python (e.g., chains of object associations). On the other hand, the use of Python in scientific programming is increasingly widespread, making this method more accessible to a wider group of collaborators. In addition, the availability of packages that implement "vectorized" operations allows for performant data manipulation, while interactive tools like `Jupyter Notebook` enable users to easily test code and visualize outputs.

### 3.10.4 Computing

The computing resources required by LDMX are determined by the needs of processing and storing the experimental data, as well as the production, processing, and storage of MC simulation samples. The projected needs for the $4 \times 10^{14}$ EoT first data run are summarized in Table 3.15.

| Sample | Disk Storage (TB) | Tape Storage (TB) | Processing (CPU-hrs$\times 10^6$) |
|---|---|---|---|
| Raw data | — | 1900 | — |
| Reconstructed data | 670 | 4800 | 4 |
| $5 \times 10\%$ data reprocessing | — | 2400 | 2 |
| Monte Carlo (one iteration) | 300 | 500 | 3.5 |
| **Total** | 970 | 9600 | 10.5 |

Table 3.15: Data storage and processing requirements for the $4{\times}10^{14}$ EoT data collected in LDMX's first run. Data processing requirements will be met with a planned allocation of 1500 CPUs. Staging raw data and Monte Carlo to and from tape will utilize 300 TB of storage. These estimates include the full safety factor for DAQ operation at up to 25 kHz.

With $4{\times}10^{14}$ 8 GeV electrons on target, LDMX will collect a total of $2.7{\times}10^{11}$ events at a conservatively estimated maximum trigger rate of 25 kHz. Table 3.16 describes the event size estimated from the full simulation along with the design of the readout for each subsystem. The storage of raw data from the first run is accordingly estimated to require $\sim$ 1900 TB of space.

| Subsystem | Size (kB) | Notes |
|---|---|---|
| Trigger Scintillator | 0.1 | 12x6b=4.5B for header + 11B per hit, 8 hits/event |
| ECal | 2.8 | 10B/hit for TDC hits, 8B/hit for low-amplitude hits, assuming zero-suppression. |
| HCal | 3.7 | 20B/hit for channel id and data, assuming zero-suppression. |
| Tracker | 0.9 | 20B/hit for channel id and data, 45 hits/event. |
| Trigger&DAQ | 0.3 | Event accounting, ECal trigger sums, trigger counter, HCal trigger info. |
| **Total** | 7.8 | |

Table 3.16: Estimated event size for each subsystem for running with an average of one electron per pulse ($\mu = 1$). All HCal hits are assumed to have TDC information.

The reconstruction of physics objects for analysis adds $\sim$ 20 kB/event to the total event size, estimated with the current digitization and reconstruction pipeline with a factor 2 safety margin. Including the raw data as usable analysis objects, the reconstructed data for the first run are therefore expected to require a total of $\sim$ 7 PB of space. All raw and reconstructed data will be transferred to tape, with 10% ($\sim$ 700 TB) being kept on disk for analysis. Backups of the raw data will also be maintained on tape outside SLAC. Furthermore, we anticipate that 10% of the data will be processed multiple times as improvements to the reconstruction and calibrations are made. Based on previous experience, we estimate that these data could be reprocessed as many as 5 times. This will add $\sim$ 2.4 PB (50%) of tape storage to our needs. Taking all of the above into account, processing of the raw data from the first LDMX run will require 1 PB of disk and $\sim$ 10 PB of tape storage.

In addition to the experimental data, large Monte Carlo samples are required to compare against data and for performance studies. Table 3.17 summarizes these samples. The sample size is minimized by careful use of event biasing and filtering for the most challenging backgrounds, since the generation and simulation of $> 10^{14}$ events is impractical. For samples not used to evaluate the performance of triggering itself, size is further reduced by at least a factor O(10) by only keeping events that pass the emulated triggers. We estimate the size of the MC samples needed for the first run to be $\sim$ 400 TB, based on our ongoing physics studies.

Reconstruction of the raw data dominates the processing required. The full reconstruction chain has been profiled and benchmarked at 15 ms/event using digitized Monte Carlo from physics studies on the currently available CPU. Adding a factor two safety margin, 4 million CPU hours are required to fully reconstruct the data from $4 \times 10^{14}$ EoT. With a planned allocation of 1500 dedicated CPU cores, the full dataset can be reconstructed in about three months, and a 10% reconstruction pass can be completed in about two weeks. Meanwhile, generation, simulation, digitization, and reconstruction of the Monte Carlo required for the first run will require approximately 3.5 million CPU hours, and a full suite can be produced in roughly three months. We however anticipate that Monte Carlo samples will be regenerated continuously as our understanding of the detector matures. Our containerized software model with frozen releases ensures reproducibility of earlier Monte Carlo samples, ensuring both a way to recover lost simulated data and allowing us to replace earlier versions of simulations on disk over time.

| Monte Carlo Sample | Total Size (TB) |
|---|---|
| $1 \times 10^{14}$ ECal PN sample – unskimmed | 170 |
| $4 \times 10^{14}$ ECal PN sample – skimmed | 55 |
| $4 \times 10^{14}$ ECal PN sample multi-electron sample - skimmed | 140 |
| $4 \times 10^{14}$ ECal as target sample | 20 |
| $4 \times 10^{14}$ Target photo/electro-nuclear sample | 15 |
| $4 \times 10^{14}$ Muon pair conversion sample | 20 |
| Inclusive + signal | 20 |
| **Total** | 440 |

Table 3.17: Monte Carlo samples and expected size needed for the first run.

The data storage and processing needs described above will be hosted by the SLAC Shared Scientific Data Facility (S3DF), a shared high-performance computing facility that supports the computing needs of SLAC experimental facilities and programs. In addition, the LDMX Lightweight Distributed Computing System (LDCS) [127], has been developed by the collaboration to leverage computing resources available at other sites. This system is a significant contributor to our resources for off-site MC production and analysis. These resources are described below.

#### 3.10.4.1 SLAC Shared Scientific Data Facility

The computing and storage requirements discussed in Sec. 3.10.4 will be hosted at the SLAC Shared Scientific Data Facility (S3DF). S3DF is SLAC's next-generation compute, large throughput storage, and network architecture aimed at serving the massive scale analytics needs of all experiments, including LCLS-II and the Vera Rubin Observatory, as well as smaller experiments such as LDMX. Currently, S3DF consists of 31,000 AMD Milan and Rome cores, 22 PB of spinning disk and 2 PB of nonvolatile memory express (NVMe) storage all interconnected via a high-speed networking fabric. As shown in Fig. 3.148, the total computing and storage capacity is expected to steadily increase beyond the date that LDMX is expected to begin taking data.

The current S3DF compute model grants all SLAC users access to baseline computing resources, including 25 GB of storage and access to the shared CPU node partitions. To ensure dedicated access to larger partitions of storage and CPUs, groups may contribute hardware to S3DF. A group can then allocate contributed resources to "repos" which, in turn, are specified by users when submitting jobs to the SLURM-based batch system. The requirements discussed in Sec. 3.10.4 assume the full compute and storage needs have been contributed to S3DF.

#### 3.10.4.2 Interface with DAQ and services

The offline computing system will receive raw data from the central DAQ, which is then transferred to S3DF for storage and processing. The interface between the DAQ system and Computing is illustrated in Fig. 3.130.

In addition, the computing system is also responsible for the storage and access to various conditions, calibration, and configuration data for online monitoring and offline processing applications such as recon-struction and calibration workflows. Finally, the computing system will also be responsible for cataloging all data files. Currently, it is envisaged that these services can be hosted on a virtual cluster at SLAC.

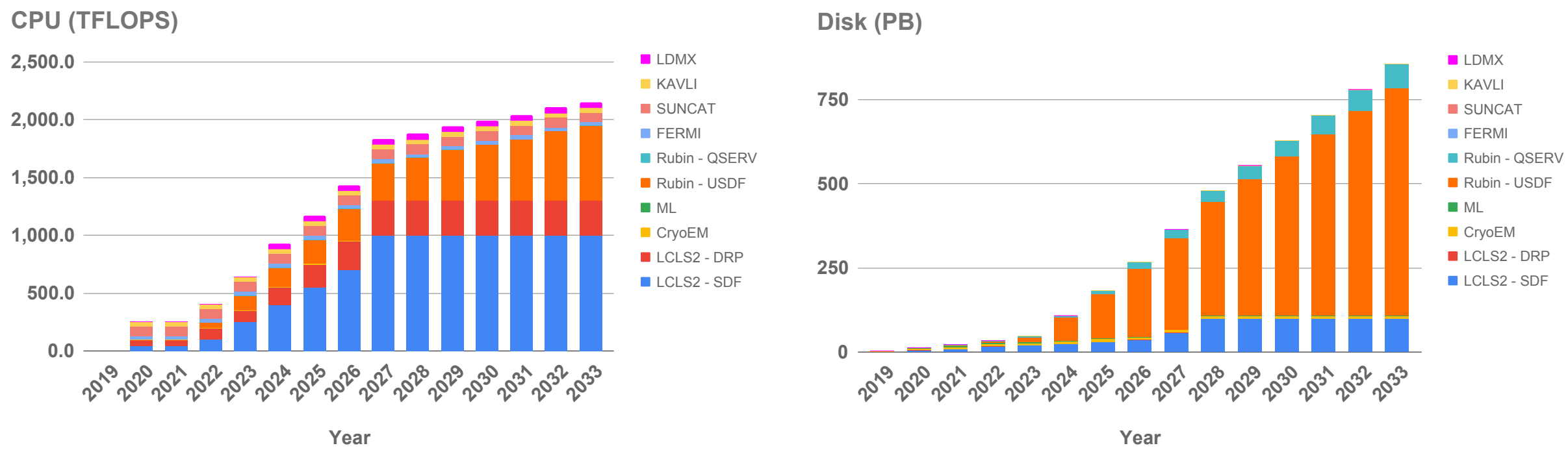

Figure 3.148: Current and projected computing (left) and storage (right) capacity of S3DF.

During the installation period of LDMX, a small control room will be set up at SLAC in Building 84, with several workstations for experts and shifters to monitor and control operations of the experiment.

#### 3.10.4.3 The Lightweight Distributed Computing System, LDCS

Since 2020, four LDMX institutes (Caltech, Lund University, SLAC and UCSB) have been contributing computing power and data storage to a distributed system described in detail in Ref. [127]. Utilizing the Advanced Resource Connector (ARC) [128] grid computing components developed and maintained initially for the World LHC Computing Grid [129, 130], LDCS is lightweight in resources needed for deployment, development, and operations. It uses ARC for authentication and job submission control, Rucio [131] for cataloging, and a small number of LDMX-specific scripts to manage job configuration and metadata extraction. The needed LDMX software is deployed as Singularity/Apptainer [132] images built from `ldmx-sw` and its dependencies. Lund and SLAC provide storage that is read/write accessible to the entire system through GridFTP. The system is illustrated schematically in Fig. 3.149.

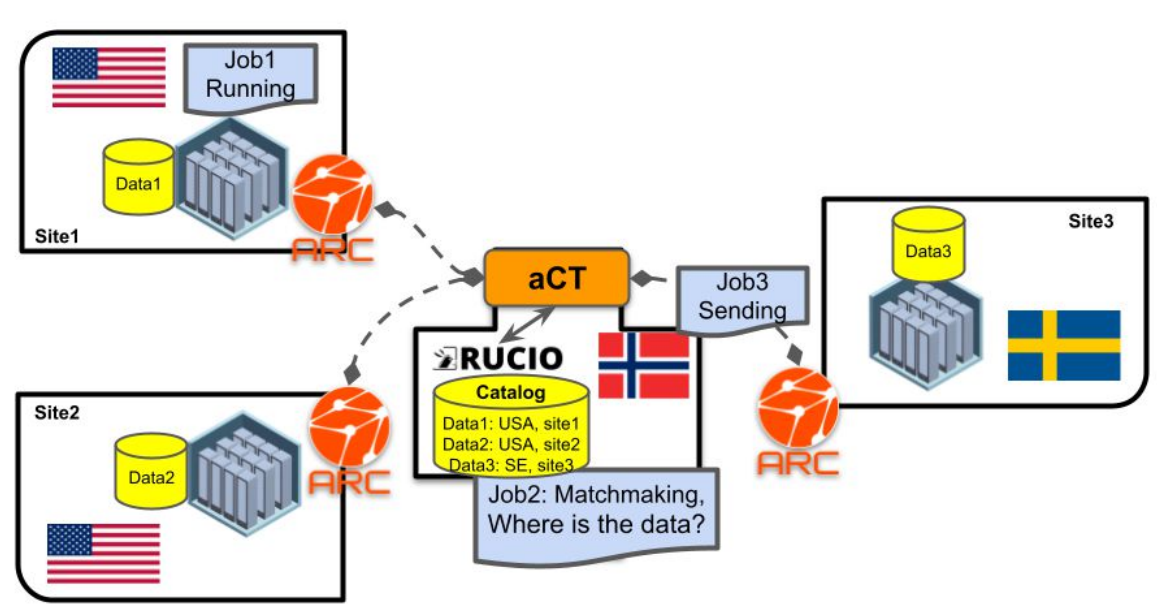

Figure 3.149: Schematic showing the components of LDCS.

A distributed system has several advantages over a local batch system, beyond increased access to resources. Specifically, implementing LDCS brought a data catalog (using Rucio) for bookkeeping file location and simulation job metadata, higher reproducibility through metadata preservation and containerized software, automatic failed job resubmission, data organization into datasets, and data-driven job brokering.

Jobs are submitted centrally from the ARC Control Tower (aCT) [133], physically located in Oslo, Norway, taking a job batch configuration file and an `ldmx-sw` job template steering file as input. The batch configuration file specifies system-level details such as the number of jobs to submit in the batch, a unique batch name, and the final storage location, as well as simulation-specific details such as which `ldmx-sw` version and steering file to use. The aCT manages individual job numbering and submission, ensuring each job has a unique choice for random number generator seeding. The jobs are distributed according to a queue depth at each site, with priority given to sites with a local replica of any input data needed, minimizing data transfer. Each job is followed by aCT through waiting, queuing at an individual site, running, and data transfer, including cataloging. Only successfully finished jobs will be cataloged in Rucio. Unsuccessful jobs automatically get resubmitted (depending on failure mode) up to 3 times.

The job submission progress can be monitored via a Prometheus+Grafana dashboard interface published (password protected) to the web, as well as a published table of jobs in different states per site and batch that aCT generates. This dashboard also monitors the used disk space on the different sites through queries to Rucio.

The Rucio metadata contains information about the individual job, such as production site and walltime used, dataset information used for cataloging, as well as detailed simulation setting information extracted from all used `ldmx-sw` processor parameters, and which version of `ldmx-sw` was used. Any input files used are listed and their metadata is copied over to become part of the output file metadata, ensuring that knowledge about provenance is retained. Metadata can be queried for information about a given file or used for listing files satisfying certain criteria. All output files are given unique names by a combination of file naming conventions and a time stamp. Both data and metadata are immutable by design, meaning adding or deleting files are the only allowed operations. On the two storage disks accessible via GridFTP, Rucio can be used to delete data from disk as well as from the catalog (when no more replicas exist), ensuring consistency between files available on disk and listed in the catalog.

The Rucio concepts of *scope* and *dataset* are used both for structuring the data, and for specifying input data for subsequent processing or analysis. Scopes are used to distinguish test jobs (`validation`), simulation production campaigns (`mc21`, `mc23`) and user analysis job output (`user.[user name]`). A script at the end of each job collects metadata and organizes data on disk according to scope and batch ID, and in case of simulation data, also by paths specifying detector version, beam energy, and sample ID.

LDCS is used for all large simulation campaigns, submitted by the production manager. As an example, producing an ECal photonuclear reaction sample of $10^{14}$ EoT equivalent takes about two weeks, with $\sim 1000$ jobs running in parallel. LDCS can also be used by other LDMX collaborators for analysis jobs, using a custom Docker container of `ldmx-sw` and its dependencies that has been uploaded to a repository. Since Docker requires running as `root`, it is typically not allowed on computing clusters. An initial job converts the Docker container to an immutable Apptainer image, collects and catalogs its metadata, and transfers it from the production site to storage. Subsequent analysis jobs request this specific image and the input dataset by a query to the Rucio catalog, and the image is distributed to sites along with any other input files. The aCT manages splitting the input dataset across jobs according to the requested number of files per job. Since an image is typically used for a large number of jobs at a site, caching is used for the image, reducing the number of data transfers needed. As each aCT user is also a Rucio user, job output is associated with their scope and organized accordingly on disk. Each user can only write to and delete files within their own user scope.

### 3.10.5 Data management plan

The data distribution and access policy is set by the Collaboration Board on behalf of the LDMX collaboration. The policy can be revised by the Collaboration Board at any time after a review process with input from the collaboration.

#### Data description and processing of products

LDMX will produce data from the following sources:

- Raw data from testing and calibration of detector prototypes at collaborating institutions
- Monte Carlo data generated using S3DF and LDCS as described in Sec. 3.10.4
- Raw and reconstructed data from the experimental run, as well as from test beams.

As discussed in Sec.3.10.4, `ldmx-sw` will be used to build the reconstruction and analysis pipelines needed to process and persist the data. All data is persisted to a ROOT-based data model. All data will be centrally stored at SLAC and made available to all members of the LDMX collaboration.

### Plan for serving data to the Collaboration and community

Before being released to the LDMX collaboration, data is tagged using the framework version used to produce it. These tagged releases will serve as the standard data sets that will be used for analysis and publication. Dissemination of the data beyond collaborators will be cost-prohibitive.

### Plan for making data used in publications available

In all cases of publications, data in the plots, charts, and figures, and Digital Object Identifiers will be made available in accordance with the policy at the time of publication by using mechanisms provided by the publisher, hosting by a collaborating institution, or services provided by INSPIRE. This includes publications resulting from research data from experiments, simulations, and research and development projects, such as detector prototype data.

### Responsiveness to Office of Science Statement on Digital Data Management

The data management plan adheres to the data management plan guidelines of the DOE Office of Science: `https://www.energy.gov/datamanagement/doe-requirements-and-guidance-digital-research-data-management`.

## 3.11 Installation and Integration

### 3.11.1 Overview

Upon completion of the individual detector systems in laboratory spaces, installation of the LDMX detector in End Station A can begin, followed by the integration of those systems into the completed LDMX apparatus. The materials and effort required for this process are captured in WBS 1.8 – Installation and Integration – which begins with the transport of equipment to End Station A and ends with a fully integrated and tested detector apparatus ready for operation with the beam. This is the final step in the construction project, where subsequent work to commission the apparatus with the beam is the first step in the Operations phase of the experiment, which is discussed in Sec. 3.12.

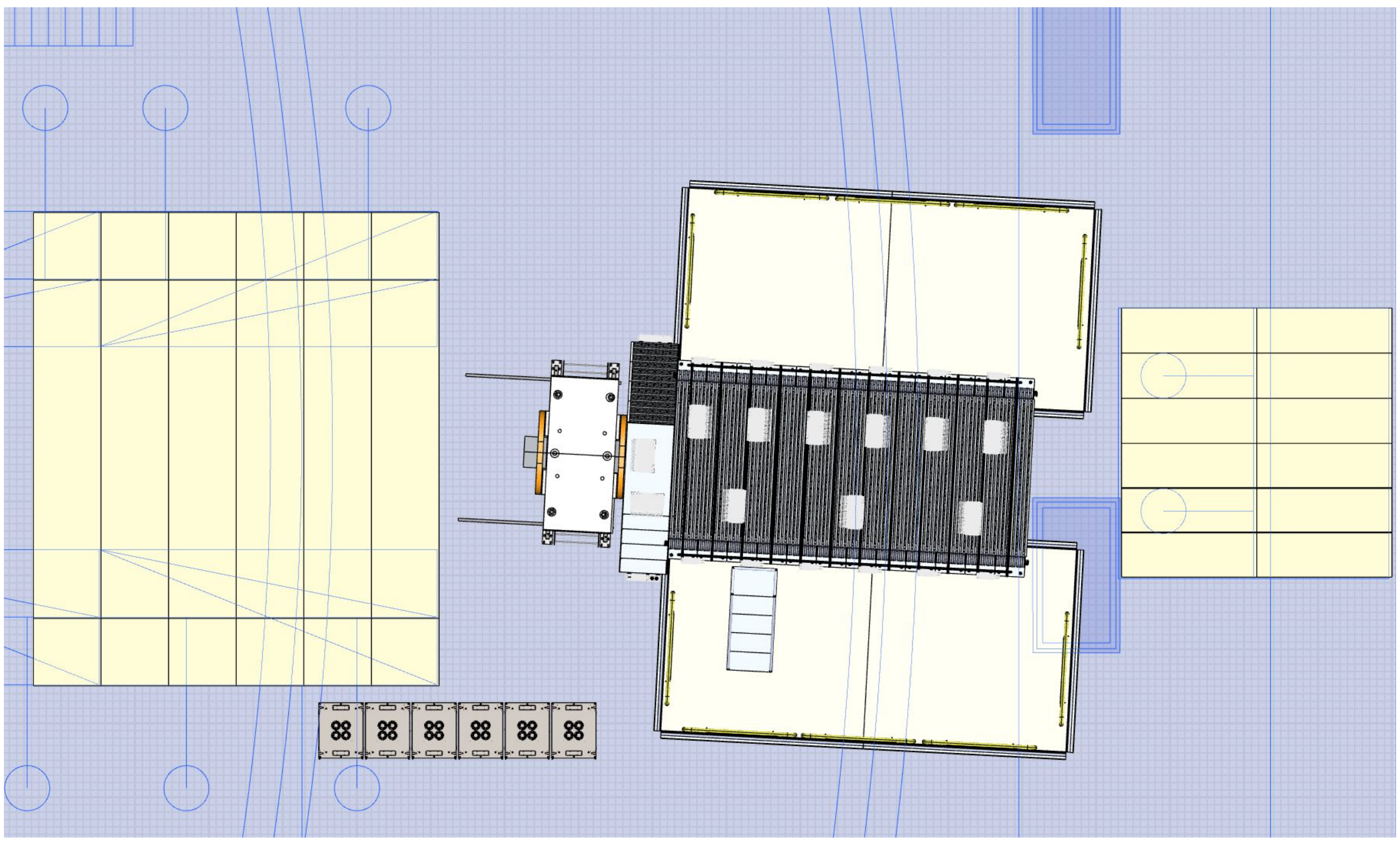

Figure 3.150: Plan view of LDMX in End Station A. The tunnel and beam dump shielding blocks are shown. The magnet is shown in the running position; it can be moved forward on rails for access to the ECal

The beamline components and the magnet will be completed, tested, and installed in End Station A well in advance of the completion or arrival of the other detector systems at SLAC, together with common infrastructure required to support power, cooling, and communications to the detector. Next, the Tracker and Trigger Scintillator will be integrated in the lab at SLAC and moved to End Station A for installation into the magnet. Finally, the ECal will be installed onto the magnet. With the magnet in the retracted position, the installation of the HCal can proceed largely independently of work involving the magnet. Similarly, the trigger and data acquisition infrastructure can be installed simultaneously.

Once all of these individual systems have been installed, their connections to power and cooling, as well as the interconnections among them, will be made to complete the integrated detector apparatus and allow for full system testing prior to beam delivery.

### 3.11.2 Common Infrastructure

The TDAQ electronics and the power supplies for the detectors are hosted in a set of six racks, which will be mounted on the blocks that support the HCal. The preliminary plan for the utilization of these racks is shown in Fig. 3.151. The racks are assumed to support up to 42U of components, though no rack is currently fully occupied in the plan. Given the large volume of End Station A, we expect that the heat dissipation in the racks can be released directly into the air volume of the detector hall. As discussed elsewhere, the heat dissipation from the dipole magnet and its power supply, as well as from the chiller for the ECal, will be transferred to process cooling water through fluid heat exchange.

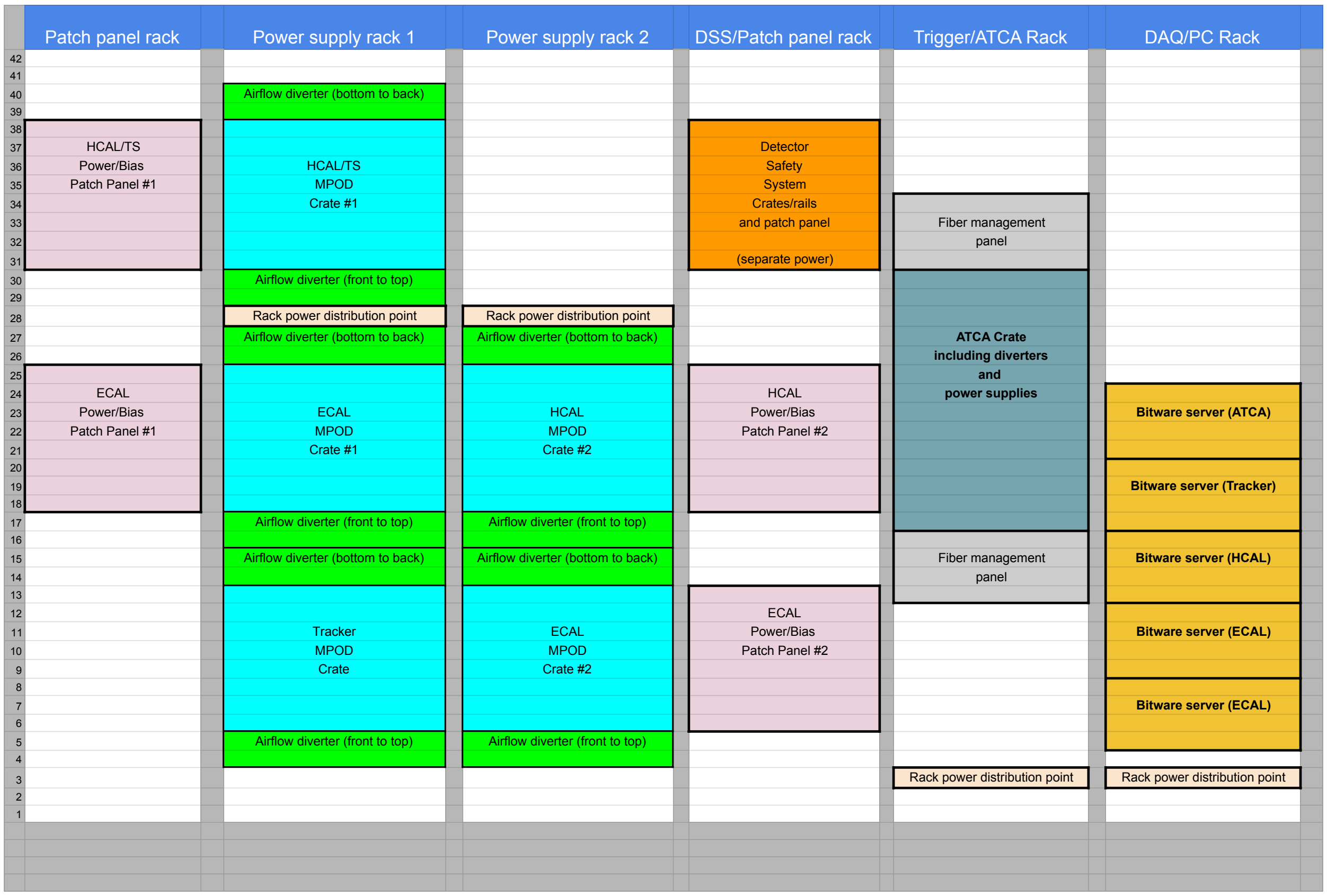

Figure 3.151: Rack Layout for the off-detector electronics, showing the allocation of space for power supplies, readout hardware including ATCA crate and Bittware services. Two racks are allocated for patch panels, and four racks contain electronics and power supplies.

Cables and fibers from the racks will be carried in cable trays above the level of the racks. The cables and fibers for all subdetectors which move with the magnet (all subdetectors except HCal), will be held in a cable tray which is eight feet above the floor of the ESA. The cables will descend in a vertical cable chain to the magnet cart, with appropriate levels of slack to allow the detectors to remain connected to the racks regardless of the position of the magnet along the rails. The cables and fibers for HCal will cross the detector access path on an elevated cable tray and be distributed around the detector using HCal-specific cable trays.

### 3.11.3 Magnet

Installation of Beamline and Magnet components in End Station A will be performed in advance of the rest of the detector. For the Beamline components, this begins with the installation of beampipe supports, followed by the beampipe itself, and finally the vacuum pumps and connections to monitoring.

The first step of installing the magnet is the installation of the rails and earthquake bracing in the floor of ESA, followed by installation of the magnet support stand on the rails, and finally the magnet itself on the magnet stand. The fully tested magnet power system and cooling hoses will then be connected to the magnet. Finally, the beam halo monitoring and point beam loss monitors (PBLM) of the beam containment system (BCS) will be installed on the magnet, connected, and tested. The BCS and fast shutoff systems of LESA will be tested with beam, including with high-charge LCLS-II pulses, to test these safety systems without risk to any of the detector systems. Once the magnet installation is complete, installation of the Tracker, Trigger Scintillator, and Target can begin.

Table 3.18: Estimated rack power requirements (supply from AC mains) and heat dissipation within the racks. Low-voltage power system efficiency is estimated conservatively as 75%.

| **Item** | **Input Power** | **Rack Dissipation** |
|---|---|---|
| ECal Low Voltage | 4.3 kW | 1.1 kW |
| ECal Bias Voltage | 0.1 kW | 0.1 kW |
| HCal Low Voltage | 1.3 kW | 0.3 kW |
| HCal Bias Voltage | 0.1 kW | 0.1 kW |
| Tracker Low/Bias Voltage | 0.3 kW | 0.1 kW |
| Bittware servers | 2.5 kW | 2.5 kW |
| ATCA crate | 0.5 kW | 0.5 kW |
| *Total* | 9.1 kW | 4.7 kW |

### 3.11.4 Tracking Systems

There are three areas of activity involved in installing and integrating the tracking systems with the rest of the LDMX apparatus in End Station A. These are the installation of the Tracker Support Box into the magnet, the installation of the infrastructure and cabling for readout and power, and the installation of the cooling, slow control, and interlock hardware.

While the tracker support box is designed to be installed from the upstream end of the magnet, it will be important to have access to both ends of the magnet when installing for the first time, so tracker installation will take place prior to mounting the ECal on the downstream face of the magnet. The Tracker Support Box will be lifted with the crane to a platform upstream of the magnet and will engage rails that facilitate sliding it into the magnet bore and onto the kinematic mounting points that position it within the magnet. The tracker support box will then be positioned and surveyed within the magnet using a laser tracker. The supports for the FEBs and Optoboards will be installed on the magnet, and all cable connections will be made. The power supplies will be installed in racks, and the power and back-end DAQ will be connected to the tracker readout for baseline testing of each individual tracker module. The chiller will be installed, and all cooling lines will be connected, along with the connection of flow and temperature sensors for monitoring and interlocks to allow for full system testing. Following all connections, the system will be brought up element by element, for full system testing and commissioning without beam, before integration with other subsystems for commissioning and operation with beam.

### 3.11.5 Trigger Scintillator and Target

The TS mechanics, frontend electronics and modules with scintillators, SiPMs, and the target will be transported to SLAC all at once, since there are no heavy or large components involved. All of these components will have been tested and qualified prior to shipment. See Sec. 3.5.6 for more details on the quality control plan for the TS hardware.

The TS will be located within the magnet bore along with the tagger and recoil trackers. As such, it will first be installed into the magnet box prior to any installation work in ESA. The primary component relevant to the TS that is located within this box will be an aluminum support plate designed to host both tracker and TS modules and the TS frontend electronics. Installation work within the magnet box will occur in a clean room at SLAC.

Electronics for reading out the TS will be set up in this clean room, along with all necessary power supplies. Once received, the electronics and SiPMs will be tested to ensure they were not damaged during transportation. These tests will be simply to confirm slow control communication and to confirm data validity by performing runs to calibrate the gain of the SiPMs, which can be done with dark current from the SiPMs.

Once the cold plate has been installed in the box, the TS modules will be installed along with their frontend electronics onto the plate. Each module has a group of 4 readout boards that are arranged into a card pack. These card packs will be preassembled before shipping. All cables will be attached to electronics and routed to a patch panel at the edge of the magnet box.

Thermal testing of the box will be done to ensure that proper contact of all relevant surfaces is maintained. Electronics testing will be performed to ensure no issues with cabling. Source testing will also be done to ensure that the coupling between scintillators and SiPMs is appropriate. Once testing is complete all active components of the TS will be surveyed along with the tracker components. Once surveyed, the box will be made light tight and tests of the dark current will be performed.

After completion of the installation of TS components within the support box, the box will then be installed into the magnet bore. See Sec. 3.11.4 for more details on this installation work. Power supplies, control hardware, and readout electronics for the TS will be installed into rack within ESA. Cables between racks and the magnet box patch panel will be connected and final tests of the TS will be performed, including measurement of the SiPM gains and dark currents.

### 3.11.6 ECal

#### 3.11.6.1 ECal Preparation and Mounting

In advance of ECal installation, low-voltage, bias-voltage, and safety system trunk cables will be prewired from the detector racks through the cable trays described in Sec. 3.11.2 and down to the magnet cart. Half of the service connections will be made on either side of the ECal, but the cables and other services will arrive on one side or the other, with services passing under the magnet on services trays to the other side when necessary for connection to the ECal. As a result, cable lengths will be somewhat longer (by 2-3 meters) for cables which connect on the away side of ECal.

In advance of ECal arrival, the safety PLC and all its connections to the ECAL patch panels will be verified to confirm that all sensor channels operate properly and that the actions are properly programmed for any over-temperature or other actions.

The low-voltage and bias-voltage supplies will be installed in the detector electronics racks in advance of ECal delivery. The installation will include the full integration of the supplies with the EPICS-based detector control system. The supplies will be fully checked out for operation, including a minimum one week burn-in with resistive loads. The detector chiller will also be installed in advance of ECal delivery.

As discussed in Sec. 3.7.10.1, the ECal will be delivered to ESA as a fully-integrated and sealed detector through to bulkhead connectors at the periphery of the detector for all signals and services. After delivery from UCSB, the detector will be checked out on a temporary stand nearby the magnet before mounting onto the magnet. This testing will be carried out near room temperature (above the ambient dew point), allowing the use of temporary flexible pipes to connect the detector to the chiller. A set of temporary extension cables will be used to connect the safety system sensors to the detector and the detector will be checked out one bilayer at a time using a single set of power supply and bias channels as well as a single set of fibers. After the detector is fully checked out and any interventions required after transport have been completed, the detector will be craned into position on the magnet and the full set of trunk cables and fibers will be connected, along with the final cooling pipes.

#### 3.11.6.2 ECal Testing and Pre-beam Commissioning

After connection to the final detector services, a set of baseline noise/pedestal and charge-injection runs will be carried out to confirm that the in-situ noise levels are as-expected and that the detector readout chips are operating properly. A set of channel-mapping runs will be carried out to confirm the expected mapping for the detector safety system, low-voltage cabling and power supplies, bias-voltage cabling and supplies, readout and slow control fibers and electronics, and trigger fibers.

Self-triggered charge-injection runs will be used to establish the relative timing between the ECal readout path and the trigger path through to the global trigger. With these parameters established, it will be possible to carry out self-triggered cosmic runs for additional commissioning activities until beam is available. Depending on the amount of time spent in this configuration, it may be possible to carry out useful pre-calibration of the detector using these cosmic events. Cosmic muon events can also be used to establish the ECal-HCal inter-detector timing.

### 3.11.7 HCal

#### 3.11.7.1 Installation

The HCal modules are assembled and fully characterized at Caltech, starting with the Back HCal (see section 3.8.9 detailing the QA/QC plan). Given the constraints on the Caltech high bay area and the module weight, each Back HCal module will be shipped individually to SLAC right after completion. The Side HCal modules weigh much less; all four will be shipped together at the end of the construction phase.

Upon reception in ESA, each module will be unpacked, visually inspected, and installed into its final position using a dedicated fixture. The installation sequence will begin with the module at the rear of the Back HCal (i.e. farthest away from the target area), and proceed forward to the front of the detector. The module feet will be fastened with Hilti anchors to 9' x 9' x 3' concrete blocks located on each side of the detector, leaving sufficient space to access the bottom of the device. The Side HCal modules are then fastened onto the first absorber plate of the Back HCal module (the first plate includes specific extensions for this purpose), starting with the bottom module, the side modules, and finally the top one. The ECal will be inserted inside the rectangular cavity defined by the four Side HCal modules. Each module will be surveyed at the end of the installation procedure - a precision at the level of a millimeter is sufficient.

The readout electronics consist of 38 crates mounted on the detector. Each crate houses four HGCROC boards and three mezzanine boards, all mounted on a large backplane board. One crate can read out 64 CMBs. The boards are produced and tested at Lund and the University of Minnesota (UMN), and they are assembled into crates in Lund. The crates will be tested in Lund before being shipped to SLAC, but a longer burn-in test will be conducted at SLAC.

The crates are mounted to a support structure surrounding the HCal. This structure also supports the HDMI cables that connect the HGCROC boards to the CMBs. The positions of the crates are selected to minimize cable length to the CMBs. Power supplies located in racks next to the HCal provide power to the crates, and optical fibers are routed from the crates to the DAQ. To a large extent, the installation and cabling of the HCal modules can proceed in parallel with other subsystems.

#### 3.11.7.2 HCal Testing and Pre-beam Commissioning

The installation of the HCal modules will begin from the far end of the complete experiment. Cabling the modules, connecting them, and starting to test the system can proceed on the installed modules in parallel with the installation of additional modules, provided there is a buffer zone between the two work areas.

This parallel activity can begin once the first five Back HCal modules have been put into place. At that point, the mechanical structure can be mounted on the four modules at the far end, their HDMI cables connected to the CMBs, and the crates can be mounted on the mechanical structure. Afterward, the crates are sequentially connected and tested one by one (see below). This procedure is repeated after the next four modules are in place. The procedure for the final four Back HCal modules must wait until the Side HCal is installed. Only then can the remaining HCal be connected and tested.

**Sequential testing of each crate:** Before the HDMI cables are connected, the crate to be tested is powered on and connected to its TDAQ optical fibers. The DAQ will read the internal ID number of each HGCROC board and its corresponding HGCROC to confirm that the two IDs match the expected values. It is checked that the internal readings of the power to be transmitted to the CMBs are reasonable, and that communication between the DAQ/DCS and the crate is functioning properly. Pedestals are read for all channels, and internal pulsing is performed to scan all channels.

Following this step, the HDMI cables are connected with output power to the CMBs still switched off—and pedestals are again read for all channels. Power is then enabled for all connected CMBs. The internal ID of each CMB is read and verified, and the temperature of each CMB is measured. Pedestals are established for all channels. All channels are read during LED flashing, which tests the entire chain from the WLS fibers in the scintillator to the DAQ. Finally, the SiPMs are read using random triggers to check for any light leaks.

**Testing of each module:** The left-right and up-down pairing of readout channels will be cross-checked by LED flashing one end of each scintillator bar while reading the other end, and vice versa.

**When the entire HCal is connected and all of the above tests are complete:** The procedure of establishing pedestals for all channels, internally pulsing each channel, LED flashing all channels, and reading the CMB temperatures is repeated several times per day. Data is also recorded over extended periods using the HCal trigger on cosmic radiation to accumulate events with muons with different impact points and angles of incidence, and random triggers to monitor the variation of the pedestal distributions over time.

### 3.11.8 Trigger and data acquisition

Prior to experiment installation, Trigger and DAQ teststands will be set up at SLAC and FNAL to test integration of subsystem firmware with the fast control. Milestones will be defined by the TDAQ project to ensure that functionality is commissioned systematically and so that subsystems can interoperate.

Commissioning of the fast timing and control system has started with the TS in S30XL test beam effort, in which we have tested integration with the LCLS-II timing system and the timing hub firmware. The DAQ CPU endpoint will be installed in B84, the trigger will be in an ATCA crate in ESA as specified in Fig. 3.150.

Once detector installation begins in ESA, TDAQ racks will be populated, and milestones for detector subsystems will be timed in as available. Installation milestones will be defined by TDAQ in collaboration with the subdetector groups.

### 3.11.9 Software and Computing

The installation activities for Computing will involve setting up the control room at SLAC in Building 84 for control and monitoring of the experiment, with workstations for experts and shifters. This includes setting up desktops and installing the necessary software stack for monitoring. Additionally, we will commission the interface between Central DAQ and offline storage and processing on S3DF, as well as services such as the management of conditions, calibration, and configuration data, and data cataloging, as mentioned in Sec. 3.10.4.2. The associated computing infrastructure will be installed to support TDAQ activities as soon as they are needed.

In terms of the software needs, for data quality monitoring we plan to use a similar workflow as the one already used in the continuous integration of `ldmx-sw`, using reference histograms to flag deviations, as outlined in Sec. 3.10.3.5.

We will develop several databases to manage calibration and configuration data, hardware registry, and data cataloging. Using LDCS as a production system has already given us experience with using Rucio for the latter.

## 3.12 Operations

When the LDMX apparatus is fully tested and ready to operate with beam, the LDMX Construction Project is complete and the Operations and Analysis phase begins. Prior to production data-taking, the apparatus must be commissioned with beam; to bring it safely into operation, to take important calibration data, and to establish and exercise the process of recording and monitoring the data. Once this commissioning step is complete, production operations begins, and continues until the experiment has collected the full dataset. In addition to operating the experiment, the principal task during this period is the analysis of the data to produce results. The following sections describe the work and resource requirements during this Operations and Analysis phase.

### 3.12.1 Commissioning with Beam

The step of commissioning the experiment with beam achieves several goals. First, it allows for a carefully controlled exposure of the apparatus to first beam, where ramping up from very low currents allows any operational issues to be identified before they could present any hazard to the detector. Second, it allows for necessary testing and calibration of detector response to beam particles with known attributes. In some cases, data will be collected during special runs with specific conditions to provide unique calibration handles. Finally, it exercises the fully integrated trigger and data acquisition systems with the LESA timing and control systems, to align trigger and readout windows, coordinate and exercise the data paths, and establish monitoring of data quality. Plans for commissioning the different subsystems with beam are described below.

#### 3.12.1.1 Beamline and Magnet

There will be extensive experience operating LESA as a test beam facility prior to the installation of LDMX. As a result, the process of setting up the beam for LDMX will be well established, with only fine tuning of the beam position, profile, and intensity required.

Only a few simple tests of beamline systems are needed after installation. The magnet will be powered to establish the trajectory to the target for commissioning of the detector subsystems, and the beam halo monitoring will be tested and calibrated with intentionally mis-steered and mis-focused beam, alongside tuning the beam position and profile using the Trigger Scintillator, Trackers, and ECal.

#### 3.12.1.2 Tracking Systems

The principal task of commissioning the STT and SRT with beam is "timing in" – establishing the correct latency for reading out data from the pipelines of the tracker readout that correspond to the triggered event. Timing in will be performed with 0.93 kHz beam from LESA – with only one filled bunch per LCLS-II cycle. Once the trackers have been timed in, their data can be used to help monitor the beam position and profile.

In order to aid the task of offline alignment of the STT and SRT, data will be taken during commissioning with the magnetic field off. The beamspot will be expanded to the largest size that fits within the beampipe for these runs to illuminate the largest possible area of the detector with beam energy electrons.

#### 3.12.1.3 Trigger Scintillator

The main goals of commissioning for the TS will be to time in with the rest of the system, demonstrate electron tracking, and align the modules with the tagger tracker. It is foreseen that a significant portion of the TS commissioning will have occurred with an ESA and/or cosmic test stand prior to full detector commissioning with beam.

#### 3.12.1.4 ECal

Once beam is available, the fine timing of the ECal modules with respect to the beam will be determined, to ensure that the amplifier peaking times are well-matched to the arrival of in-time particles. Given the size of the detector, the peaking time phase will vary by up to two nanoseconds through the detector.

Once the detector timing with respect to beam has been established, the full intercalibration of ECal can be performed using the upstream target described in Sec. 3.7.11.3. This intercalibration will establish the MIP scale for each channel, while the calorimetric energy scale will be determined with full-energy electron events. Once the energy scale has been established, the ECal will be ready to operate as a triggering detector for physics measurements.

#### 3.12.1.5 HCal

The fine timing of the HCal modules with respect to the beam will be determined to ensure that the amplifier peaking times are well matched to the arrival of in-time particles. Given the size of the detector, the peaking time phase will vary by up to 20 nanoseconds across the detector. Initially, this will be measured for the channels closest to ECal and estimated using calculated offsets for the more distant channels.

The timing will be further fine-tuned using the upstream target described in Sec. 3.7.11.3, which will deliver muons distributed over a larger cross-section of the HCal. This intercalibration with beam will also establish the MIP scale for most channels, as described in Sec. 3.8.10.3.

#### 3.12.1.6 TDAQ

There will be extensive experience operating LDMX with the LESA beam timing structure and distributing the fast control and timing signal to all the individual subdetectors based on test stand validation and test beam efforts with S30XL and LESA. Commissioning of individual subsystems will start with cosmic data and a key aspect of commissioning will be the operation of admixtures of the subdetectors. This will also be a critical opportunity to test the DAQ paths and slow controls system including subsystem calibration and monitoring.

With beam commissioning, we will validate much of the same functionality that is tested during cosmic ray commissioning. Critical validation tests with the beam will be to test various trigger selections and configurations, and align metadata with software and computing. Commissioning with beam will also be a crucial test of the backpressure functionality and fast control distribution.

#### 3.12.1.7 Software and Computing

During the commissioning phase we will have the opportunity to exercise the data processing and reconstruction pipelines, as well as the databases established to manage conditions, calibration and configuration data. In addition, software development will be ongoing throughout the lifetime of the experiment.

### 3.12.2 Production Operations

The operations plan for LDMX is closely tied to LCLS-II, since LESA beam is provided via normal operations for LCLS-II. The operations plan for LCLS-II calls for 5000 hours of beam per year, and LESA plans to allocate 80%, 4000 hours, to operating LDMX, with the rest allocated to test beam activities. With 37.14 MHz bunches, and accepting 600 ns long bunch trains in each 1.077 $\mu$s LCLS-II cycle, LDMX will receive $4.25 \times 10^{14}$ bunches per year.

The operations plan for LDMX to collect the complete dataset of the experiment is summarized in Table 3.19.

| **LDMX Operations Plan** | | | | |
|---|---|---|---|---|
| Run Period | Beam Energy | $\langle n_{e^-} \rangle$/bunch | target | duration (years) |
| EaT Run | 8 GeV | 1 | 0.1 $X_0$ W | .1 |
| LDMX Pilot | 8 GeV | 2 | 0.1 $X_0$ W | 1 |
| eN | 8 GeV | 1 | 0.1 $X_0$ Ti | .1 |
| LDMX Extended | 8 GeV | 2 | 0.4 $X_0$ Ti | 4.7 |

Table 3.19: Details of the LDMX operations plan.

During the first period, the nominal target – 10% $X_0$ tungsten – will be used to collect data for the ECal-as-a-target analysis described in Sec. 4.5, and provide a low-luminosity sample for an initial missing momentum analysis. Following this, a short run will be taken with a titanium target to provide data for analysis of electron-nuclear events as discussed in Sec. 4.9. Finally, in order to collect the full dataset to achieve the ultimate sensitivity of the experiment, a thicker target – 40% $X_0$ titanium – will be used for the duration of operations.

The experimental reach demonstrated in Sec. 2.2 utilizes $1.6 \times 10^{15}$ electrons on a 40% $X_0$ target. We anticipate being able to use events with at least two electrons without significantly impacting background rejection, suggesting operation with an average number of electrons per bunch of two ($\mu = 2$), requiring roughly 5 years of additional operation in this configuration. Extensive operational experience with its subsequent data analysis, may well allow us to increase the average number of electrons per bunch beyond two ($\mu > 2$).

# Chapter 4

# Physics Capability

The primary physics goal of LDMX is to test the possibility of a light thermal relic across a wide range of potential DM masses and couplings to the SM. The high duty factor 8 GeV electron beam delivered by LCLS-II allows for a range of search strategies to be conducted by the experiment based on missing-momentum and missing-energy signatures, which are optimal for a range of potential DM production mechanisms. The success of such analyses crucially depends on the ability to efficiently reject a range of background sources from Standard Model processes. These capabilities have been extensively studied using simulations [134, 135] in addition to test beam data. This chapter goes considerably beyond previous publications in presenting new studies that also incorporate an improved detector simulation and more sophisticated analysis strategies. In particular, studies of events with multiple electrons are presented for the first time.

Opportunities beyond the search for invisible particles will also be discussed. For example, new mediators of SM-DM interactions may be sought directly via their decays back into SM particles, which are typically displaced from the point of their production due to the small dark sector portal coupling. The experiment's capabilities are also demonstrated to extend to novel SM measurements, including probes of electron-nucleon scattering that may provide unique insights into features of sister interactions in neutrino-nucleus interactions.

## 4.1 Simulated event samples

The design of LDMX has been driven by the ability to conduct the missing-momentum search for sub-GeV dark matter. The optimization of the detector apparatus itself and the development of analysis techniques needed to uncover potential signals require extensive calculations performed with a Monte Carlo simulation. This section reviews the details of simulated event samples corresponding to various processes induced by the incoming electron beam. This includes potential DM signals as well as a range of SM backgrounds.

### 4.1.1 Signal benchmark processes

Simulation of dark photon signal events is performed using the G4DarkBreM [122] package, which embeds a MadGraph/MadEvent calculation of the dark bremsstrahlung process into the complete Geant4 simulation of the experimental setup. The technique, described in Sec. 3.10.3.3.3, allows taking into account beam energy losses due to material interactions upstream of the target, while preserving the full matrix-element accuracy of the calculation. Simulated samples of events are produced for $A'$ masses ranging from 1 MeV to 1 GeV, and by default make use of filters to select events where the dark photon energy is at least half the beam energy. This setup is used to generate events with dark photon production in the target as well as the ECal, for the missing momentum and missing energy analyses, respectively.

### 4.1.2 SM background processes

The main backgrounds are discussed in Sec. 3.2 and shown (along with the detector signature used to detect them) in Fig. 3.3. The primary simulation for each of them is provided by `Geant4`, applying the enhanced simulation efficiency techniques described in 3.10.3.3. In particular, all simulations except the unbiased electron interactions employ the filtering and occurrence biasing techniques described in Sec. 3.10.3.3.1. All biasing factors have been chosen as the maximum which doesn't change the interaction specifics, such as where in the detector material the interaction occurs.

**Unbiased electron interactions** This "inclusive" sample of events includes any process occurring after an electron is delivered from the initial beam position upstream of the complete detector apparatus (outside of the magnetic field). It is dominated by electrons that do not interact in the (10% $X_0$ thick) target. These samples are used for trigger studies and the development of reconstruction algorithms in the TS and trackers, as well as the validation of the biased and filtered samples described below. Events in which the electron retains most of its initial energy will not be selected by the missing-energy trigger requirement enforced for invisible searches.

**Photo-nuclear (PN) interactions** The main background processes giving rise to unmeasured energy start from a hard bremsstrahlung event, where the radiated high-energy photon subsequently interacts with a nucleus. While this can happen anywhere in the detector, the events from this type of background are hardest to differentiate from signal when occurring in the target or ECal. As described in Sec. 3.10.3.3.1, simulated events are biased and filtered to efficiently generate exclusive samples of events with photo-nuclear interactions occurring in either the ECal or the target area, respectively. The `Geant4` biasing factor $B_{\rm PN}$ enhances the physical (unbiased) cross section as

$$\sigma_{\rm PN}^{\rm biased} = B_{\rm PN}\sigma_{\rm PN}^{\rm physical}. \tag{4.1}$$

Furthermore, to ensure that the total photon interaction cross section is unaffected by the biasing, the following scheme to reduce the photon pair production cross section is used:

$$\sigma_{\rm conv}^{\rm biased} = \sigma_{\rm conv}^{\rm physical} - (B_{PN} - 1)\sigma_{\rm PN}^{\rm physical}, \tag{4.2}$$

where $B_{PN}$ is again the biasing factor.

For the 8 GeV beam, the bremsstrahlung photon is required to have at least $E_\gamma = 5$ GeV and the PN cross sections are modified by a `Geant4` biasing factor $B_{\rm PN}$ of 550.

**Electro-nuclear (EN) interactions** Another class of background events originates from electro-nuclear (EN) reactions occurring in the target area. For this class, the event yields are lower than for photo-nuclear reactions, and an accurate modeling of the interacting particle's survival probability is less crucial. A biasing scheme is also used in this case:

$$\sigma_{\rm EN}^{\rm biased} = B_{\rm EN}\sigma_{\rm EN}^{\rm physical}. \tag{4.3}$$

A combined sample of electro- and photo-nuclear events is also produced for use in the ECal-as-target analysis, where the two backgrounds are experimentally indistinguishable. For this setup, a common biasing factor is used for both process classes. Finally, alternative samples of electro-nuclear events are generated using GENIE for the electro-nuclear measurement program at LDMX. These samples, which allow for direct comparisons with measurements of neutrino-nucleus interactions, are described in more detail in Sec. 4.9.1.

**Muon conversions** Another potential source of background originates from photon conversions to $\mu^+\mu^-$ pairs via coherent scattering off a nucleus in the target or in the ECal. The default implementation of this process in `Geant4` was found to have an unphysical form factor, and approximations to the phase-space distribution of outgoing muons were inaccurate in the tails. Therefore, di-muon samples are generated with a modified version of the `G4GammaConversionToMuons` class using the full phase space distribution of Eq. 2.3 of Ref. [136], assuming elastic recoil of the nucleus and, considering the heavy target nucleus, keeping only the W2 component of the structure function result. The muon kinematics from the resulting procedure have been validated against a MadGraph/MadEvent event sample with the same approximations in modeling of the nuclear couplings, and found to be in good agreement.

The muon conversion biasing factor is substantially larger ($\sim 10^4$), and is chosen to be as high as possible without sculpting the $z$ distribution of where the process takes place.

#### 4.1.2.1 Comparisons of different generators for photo-nuclear reactions

As discussed in Sec. 3.10.3.3.1, the default modeling of photo-nuclear reactions from hard bremsstrahlung-photons uses the Geant4 implementation of the Bertini model with modifications. In order to ensure that conclusions drawn about potentially limiting final states are not model- or implementation-dependent, the congruence between outcomes from the Bertini model in Geant4 and other general purpose Monte Carlo generators has been studied. The rates of challenging final states from a 3 GeV photon interacting with a tungsten nucleus have been compared between hadronic event generators that use three different models (similar results are obtained for a 6 GeV photon): events were produced using the JAM [137] hadronic cascade model from PHITS [138, 139], the Cascade-Exciton Model (CEM) [140] model in MCNP [141], and the Pre-Equilibrium Approach to Nuclear Thermalisation (PEANUT) model in FLUKA [111, 112].

The relative rates of specific final states that have the potential to mimic the missing-momentum signal are shown in Fig. 4.1 for a sample of 25000 photo-nuclear events. The threshold for a particle to be counted as part of the final state is a kinetic energy above 200 MeV. Events in which no particle in the final state fulfills this requirement are classified into the "Nothing hard" category. The events are taken directly from the event generator in question (Geant4/FLUKA) or extracted from the output of a simple simulation with a photon beam incident on a thin tungsten target (PHITS/MCNP). For the latter case, only events where the photo-nuclear reaction was the first interaction of the incident photon are considered.

Despite the relatively significant differences in the approaches used in the models, the fraction of events that result in key final states relevant for the design of LDMX – such as single energetic neutral hadrons or charged kaons – differ at most by a factor of a few when comparing the different generators. For the single particle case, the largest deviation to Geant4 occurs for PHITS/JAM where the rate of single hard neutrons occurs about three times as often, and events with hard kaons occurring about twice as often. The most striking difference lies in the "Nothing hard" category, where MCNP and PHITS produce an order of magnitude more events than Geant4.

The origin of the events in this category has been investigated in detail in the Bertini model. They originate from interactions where the target nucleus receives an unphysical amount of excitation energy causing the de-excitation stage to evaporate nucleons one by one until the nucleus appears to have exploded. This is not a feature of the particular de-excitation code that is used in the Bertini model by default, as it also happens when switching the Bertini-internal de-excitation code to the native Geant4 PreCompound model. All of the photonuclear models investigated feature a statistical evaporation code like this, which suggests a common origin of this event category between the different generators.

While these events could be rejected outright at the simulation stage because they represent an edge case of cascade models coupled to statistical evaporation rather than a physical phenomenon, the "Nothing hard" category of events are generally rejected either by a clear local energy deposition in the electromagnetic calorimeter or by the hadronic veto. In particular, the side HCal plays an important role in catching the isotropically emitted low-energy neutrons.

In general, the comparison between the different generators yields results similar enough to justify basing further discussion on the default simulation Geant 4.

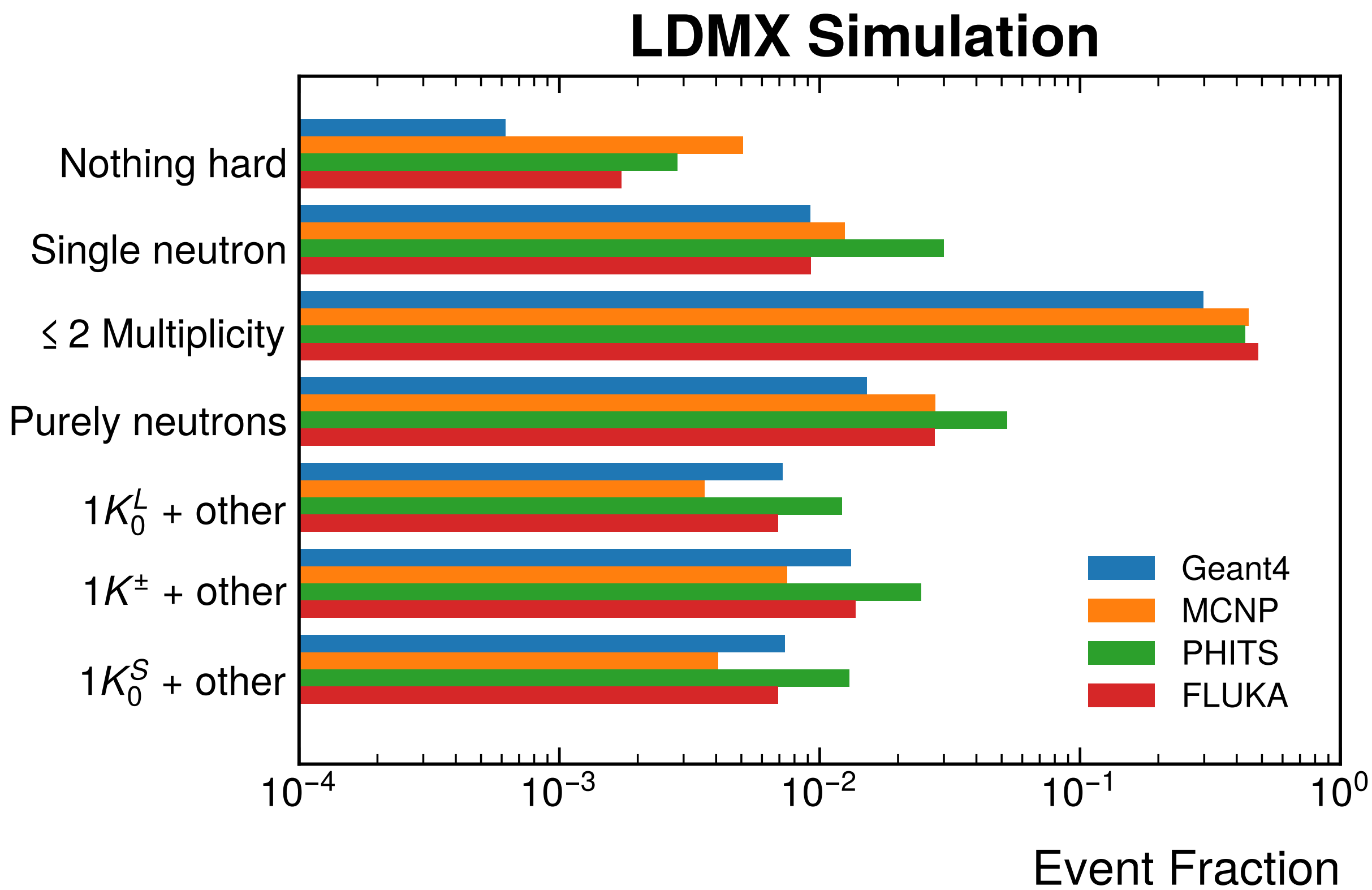

Figure 4.1: Predicted relative rates of photo-nuclear final states from 25000 events generated by the Bertini model in Geant4 (blue), the CEM model in MCNP (orange), the JAM model in PHITS (green), and the PEANUT model in FLUKA (red). Events are classified based on the multiplicity of different particle species with kinetic energy above 200 MeV.

## 4.2 The baseline missing-momentum search strategy

In the initial phase of running, LDMX will conduct a missing-momentum search for dark photon production in the target based on a data sample of $4 \times 10^{14}$ EoT from an 8 GeV electron beam. To that end, studies based on simulated event samples have been undertaken to establish the experiment's ability to reject SM background processes to a negligible level, while retaining high signal efficiency.

The following sections proceed by noting the key features of the DM signal in LDMX. We describe analyses targeting DM production in the thin target using the missing momentum techniques. The background veto is accomplished through a series of requirements, beginning with a significant missing-energy signal seen at the trigger level. Events passing the trigger are required to be consistent with a beam-energy track in the tagger tracker and a single, low-energy track in the recoil tracker from the recoil electron. Information about the energy depositions and patterns in the electromagnetic and hadronic calorimeters is used to further distinguish signal from background processes. Our baseline approach achieves this with a Boosted Decision Tree (BDT) analysis, though alternative results based on rectangular cuts and neural networks are also shown to demonstrate the full robustness and power of the calorimeter system.

Furthermore, we describe the prospects of inferring parameters of the underlying DM model from the recoil electron spectrum in the event of an observation.

Finally, we comment on how the veto procedure for single-electron events can be adapted in order to effectively analyze events with multiple beam electrons.

### 4.2.1 Characteristics of the DM signal

The sensitivity of LDMX to scenarios with light thermal relics is primarily explored through the dark photon model. In this setup the SM is augmented by a new dark hypercharge symmetry group $U(1)_D$ whose gauge boson $A'$ couples to electromagnetically charged SM particles through a feeble kinetic mixing term, parameterized by $\epsilon e$. An additional new particle $\chi$ composes the DM itself, and interacts with the $A'$ via a coupling $\alpha_D = g_D^2/4\pi$, typically assumed to be of $O(1)$. While invisible decays of the $A'$ are dominant when $2m_\chi < M_{A'}$ for all potential values of the DM spin and coupling structure, these can impact the coupling values preferred by the observed relic abundance. In the remainder of this chapter, we will assume the representative values of $\alpha_D = 0.5$ and $3m_\chi = M_{A'}$, expressing thermal targets in terms of the normalized coupling $y = \epsilon^2 \alpha_D (m_\chi/m_{A'})^4$. These are important assumptions for the mass-coupling estimation described in Sec. 4.4.1.

Dark photons may be produced through several potential channels at LDMX:

- In **dark bremsstrahlung**, a beam electron scatters off a target nucleus and produces a pair of DM particles either through the production and decay of an A' or an effective four-particle interaction. In a similar process, electrons that interact in the electromagnetic calorimeter can produce DM particles either directly or through their secondaries, mid-shower.
- **Photo-production of vector mesons**, originating from hard bremsstrahlung photons produced in the target, can provide complementary sensitivity through invisible decays mediated by their $A'$ mixing [3].
- Finally **positron capture on atomic electrons**, sourced by $\gamma^* \to e^+e^-$ conversions within EM showers of the beam electrons, provides a narrow window of sensitivity around $\sqrt{2m_e E_{\text{beam}}}$, 90 MeV for the 8 GeV LDMX beam [142].

Each of these production mechanisms leads to a missing momentum signature.

Fig. 4.2 shows the recoil electron's $p_{\text{T}}$ and angle given by $\cos\theta = p_Z/|\mathbf{p}|$ as measured at the target for four masses (1 MeV, 10 MeV, 100 MeV, 1000 MeV) of the dark photon. The incoming electron is simulated with an 8 GeV energy, and it is required that the dark photon has at least half of the beam energy ($E_{A'} > 4$ GeV). An angle $\theta = 0$ indicates momentum vectors of the recoil electron that are parallel to the $z$ (beam) axis. One may observe that the greater the mass of the dark photon, the larger the kick the electron receives in the transverse plane. However, most of the momentum is still in the beam direction ($+z$).

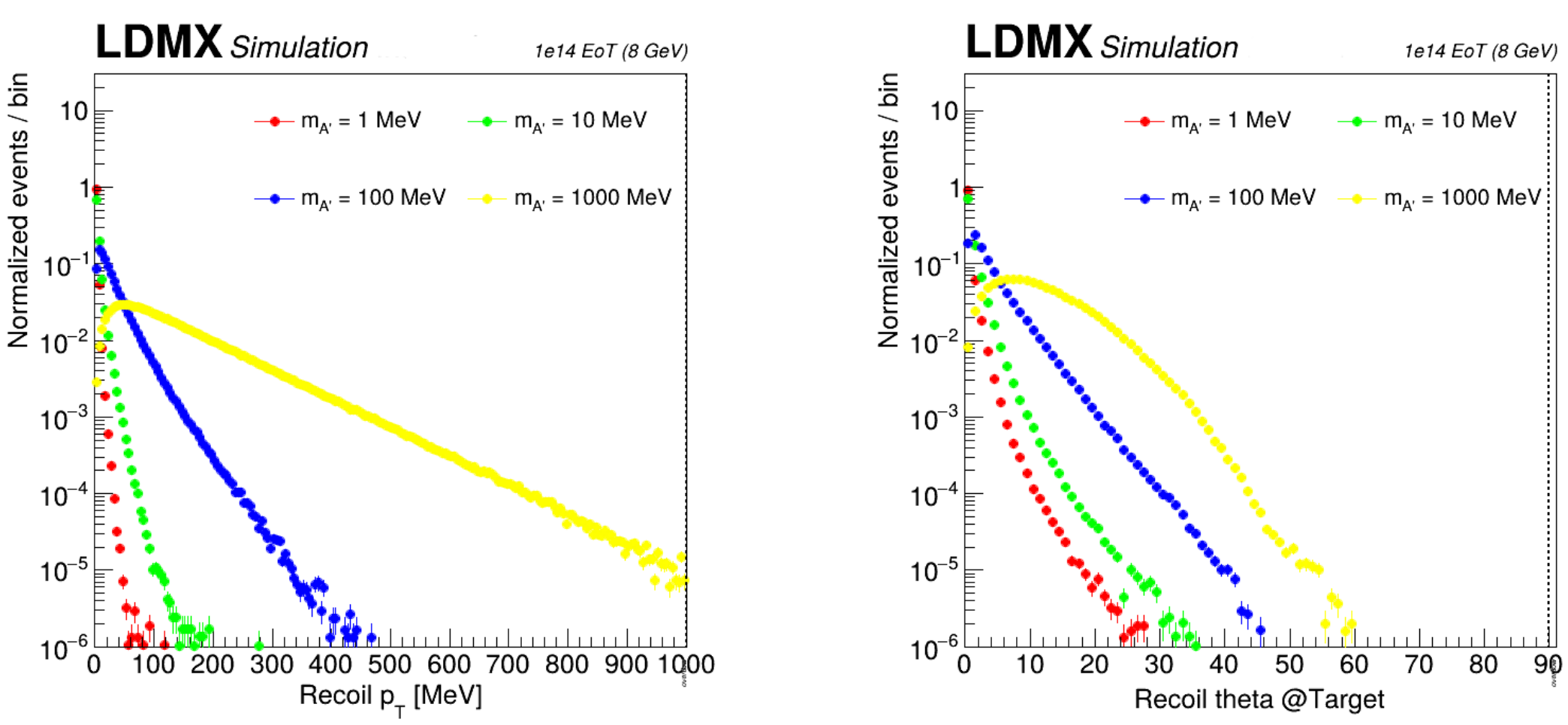

Figure 4.2: Recoil electron's $p_{\text{T}}$ (left) and angle (right) for different masses of the dark photon shown with different colors.

A fiducial region for analysis is defined by requiring that the electron traverses at least five layers of the recoil tracker before reaching the ECal face. This threshold ensures the minimum track length necessary for a quality helical fit. Fig. 4.3 shows the fraction of signal events where the recoil electron leaves hits in either at least five ("loose") or all ("tight") layers of the recoil tracker as a function of a cut on the recoil energy and angle. For this study, we employed dedicated signal samples that do not require that the A' has at least half the beam energy. The acceptances vary with dark photon mass due to energy-angle correlations in the underlying distributions.

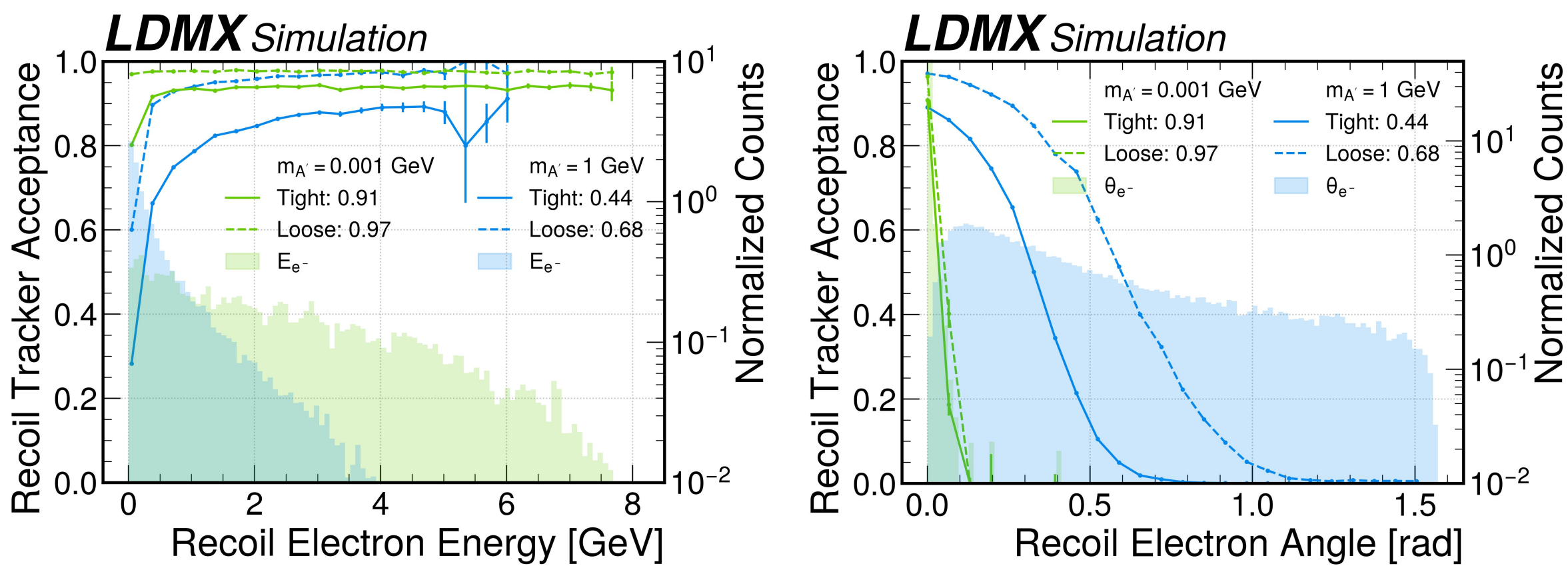

Figure 4.3: The acceptance for recoil electron tracks is shown for two A' signal masses, as a function of energy (left) and angle (right). The distribution of all recoil tracks is shown (filled histogram) in addition to the fraction of tracks leaving hits in at least five ("loose") or all ("tight") recoil tracker layers. The values in the legend indicate the fraction of the signal after the given requirement.

Fig. 4.4 shows the constituents of the fiducial (acceptance) definition: 2nd bin shows the percentage of events passing the minimum electron energy requirement, the 3rd bin shows the "loose" track hit requirement (as defined above), the 4th bin the ECal hit requirement, the 5th bin shows the possibility of an HCal hit requirement, and the last bin is the logical AND of the 2-4 bins, leading to the fiducial definition described above. The overall trend is that the heavier the dark photon, the bigger kick the recoil electron receives, leading to fewer fiducial events. In this figure, the original signal simulations, with the minimum momentum transfer discussed earlier, are used to be consistent with the rest of the section.

### 4.2.2 Trigger and missing energy requirements

The trigger decision is based on calculating the energy deposited in the first 20 layers of the ECal. The event is kept if the energy sum is less than 3 GeV in these layers for a single recoil electron as counted by the Trigger Scintillator. Fig. 4.5 shows the trigger efficiency as a function of missing ECal energy as calculated in all ECal layers. We reach the plateau very sharply at 5 GeV for all masses of the signal models. This is because the recoil electron deposits most of its energy in the first 20 layers, and that is also why only 20 layers are required in the trigger decision. The turn-on curve is less sharp for the background processes, and less efficient in the case of backgrounds that deposit energy deep in the ECal, such as the ECal PN sample, or the ECal conversion sample. Table 4.1 summarizes the global trigger efficiencies.

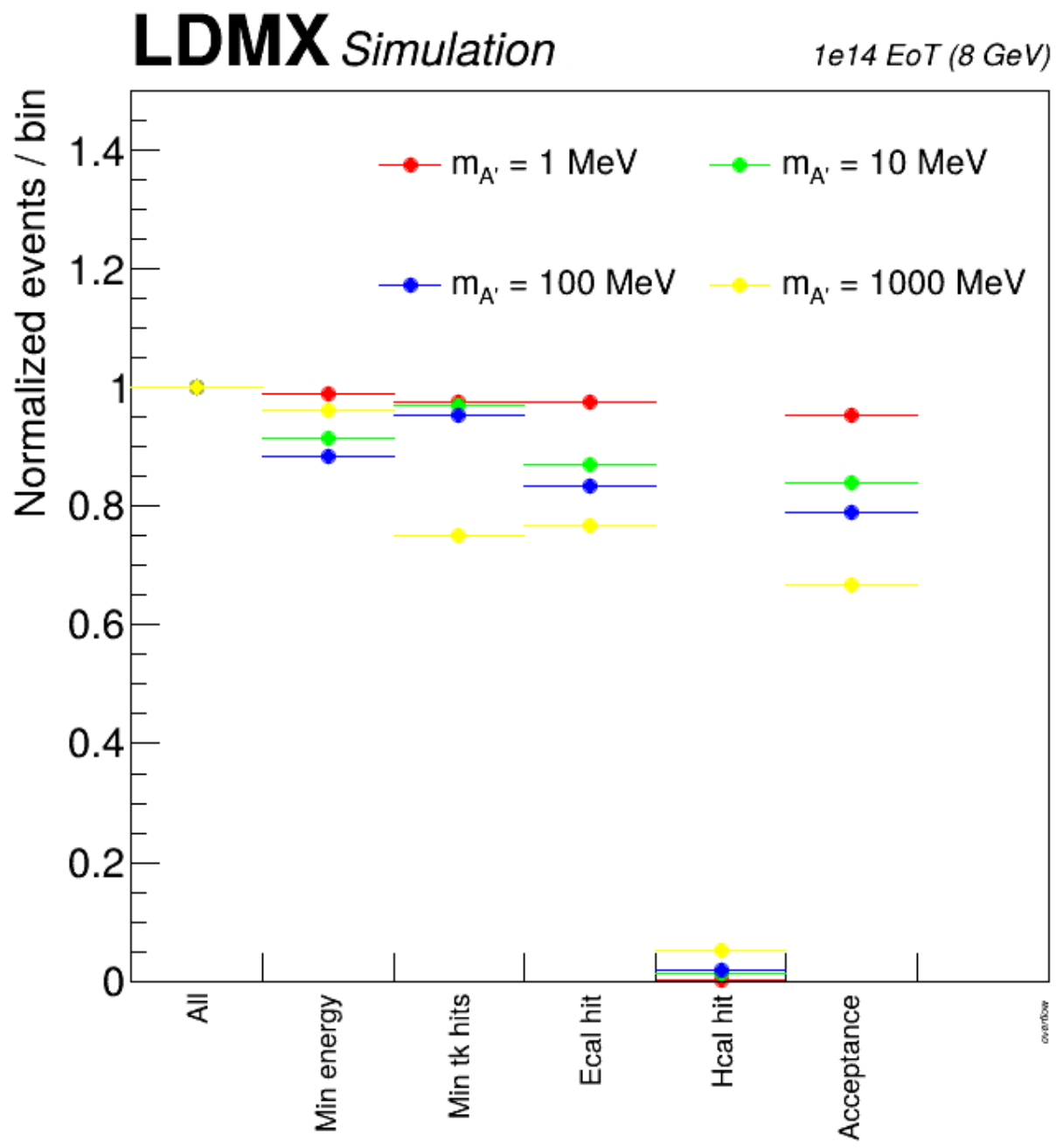

Figure 4.4: The constituents of the fiducial (acceptance) definition: 2nd bin shows the percentage of events passing the minimum momentum requirement, the 3rd shows the minimum number of hits in the tracker requirement, the 4th bin shows the ECal hit requirement, the 5th bin shows the possibility of an HCal hit requirement, and the last bin ("acceptance") is the logical AND of the 2-4 bins (i.e. min energy, number of tracker hits, and ECal hit).

| Sample | Trigger efficiency |
|---|---|
| Inclusive background | 0.0028% |
| ECal PN | 7.6% |
| ECal conversion | 80.2% |
| Target PN | 37.0% |
| Target conversion | 67.8% |
| Target EN | 0.4% |
| Signal with $m_{A'} = 0.001\,\text{GeV}$ | 78.2% |
| Signal with $m_{A'} = 0.01\,\text{GeV}$ | 89.6% |
| Signal with $m_{A'} = 0.1\,\text{GeV}$ | 91.3% |
| Signal with $m_{A'} = 1\,\text{GeV}$ | 94.3% |

Table 4.1: Trigger efficiency for different background and signal samples. These numbers are after the generation level cuts (detailed in Sec. 4.1) applied.

### 4.2.3 Observables

The missing energy requirements in the trigger decision eliminate five orders of magnitude for the inclusive background, and one order of magnitude for the photo-nuclear processes. For further background reduction, we use the tracker detectors' track multiplicity and momentum measurement. Furthermore, information from the ECal and HCal are critical to remove the SM background events that cannot be eliminated using missing energy and track momentum requirements alone. In this section, we discuss the observables as measured in these sub-detectors. All the results are based on the latest detector geometry (`v14`). The following figures show normalized event counts after the trigger decision and after ensuring that the event passes the fiducial selection. The last bin in the plots, denoted with a vertical dashed line, is the overflow bin. Features of the signal process are shown alongside the background processes discussed earlier.

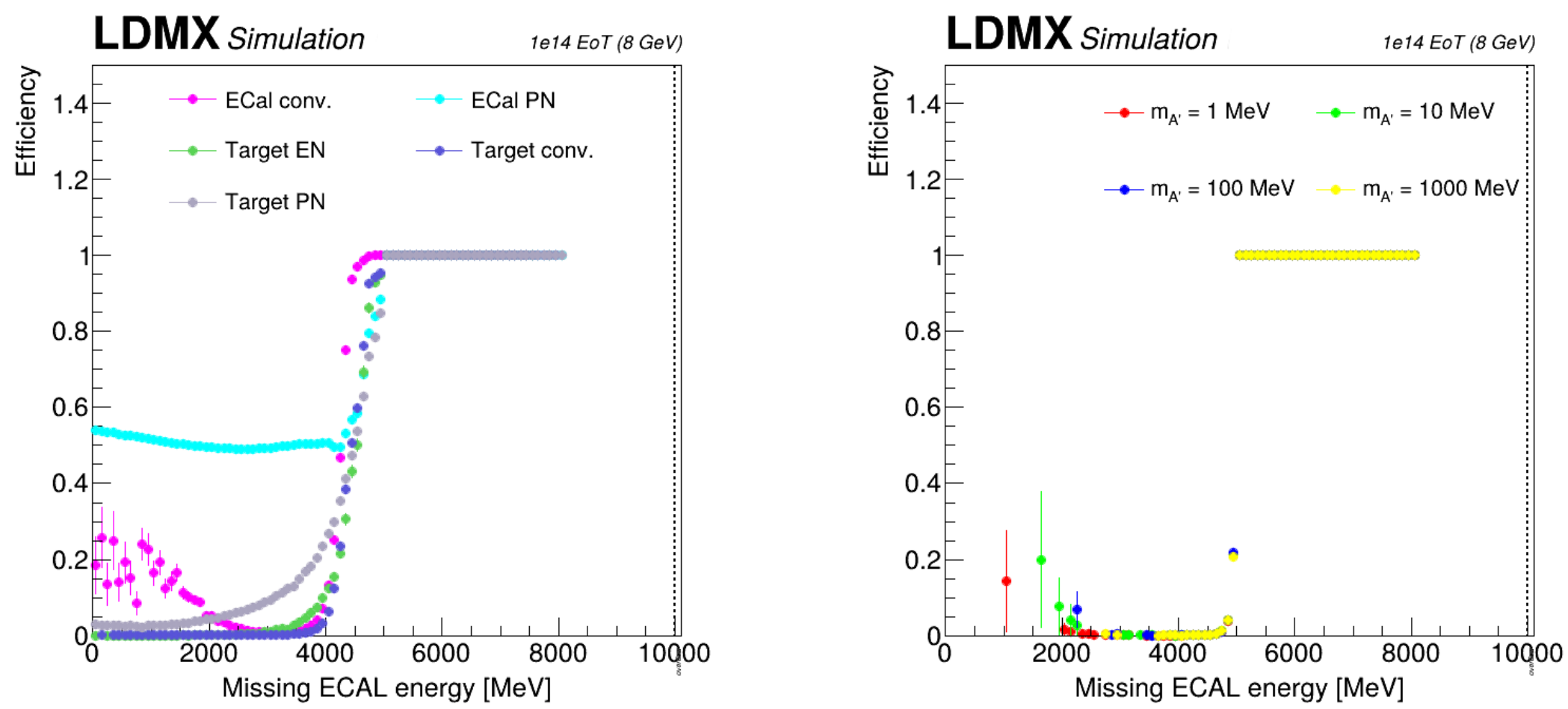

Figure 4.5: Trigger efficiency as a function of missing energy for the background samples (left) and signal samples (right).

#### 4.2.3.1 Tracker observables

Tracks are reconstructed with the Combinatorial Kalman Filter algorithm, followed by a Greedy Ambiguity Resolution algorithm. The impact parameter of the tracks at the target is required to match the beamspot size, i.e. $d_0 < 10$ mm and $z_0 < 40$ mm. Fig. 4.6 shows the total momentum in the tagger tracker (left) and recoil tracker (right). In the tagger tracker, this is centered around the beam energy (8000 MeV), and this measurement can help us reject off-beam energy electrons that interacted with the material early on. The target PN sample was produced with a generator filter, where the requirement on the tagger track was at least 3800 MeV, while for the signal this threshold is at 7000 MeV, and for the rest of the sample it is 7600 MeV. For each of these we see events below these thresholds, which is due to the reconstruction resolution effects. In the recoil tracker, some of the beam's energy is converted into other particles created in the target interaction, leaving a smaller total momentum for the recoil electron. The cutoffs in the recoil electron momentum are artifacts of the simulation settings, requiring a certain energy to go into the photon.

Fig. 4.7 shows the transverse momentum in the recoil tracker (left), and the number of tracks in the recoil tracker (right) with a normalized logarithmic scale. This shows that in the case of the signal samples, we dominantly have one track. Since the track multiplicity is independent of the dark photon mass, the histograms for signal overlap. On the other hand, in the case of the target PN, target EN, and target conversion samples, we have two or more tracks. The transverse kick to the recoil electron is characteristic to the dark matter mass, and is generally small for the background processes, but also for the low signal masses.

#### 4.2.3.2 ECal observables

This section revisits key calorimeter observables used to construct the veto logic. The recoil electron from the recoil tracker is extrapolated to the ECal and used in the veto through electron and photon shower matching.

Fig. 4.8 shows the total energy deposited in the ECal (left) and the number of read-out hits in the ECal (right). In the left plot we can observe the effect of the trigger leading to a sharp drop at 3 GeV for the signal and a cut-off at 4 GeV. Background processes happen deeper in the ECal, beyond the first 20 layers used for the trigger, which then leads to no obvious cut-off like it did for the signal. The conversion samples deposit more energy than a single recoil electron in the signal events, while the EN and PN events have even higher tails. In the right plot we see that the background events, containing both an electron and additional interacting particles, create more hits in the detector.

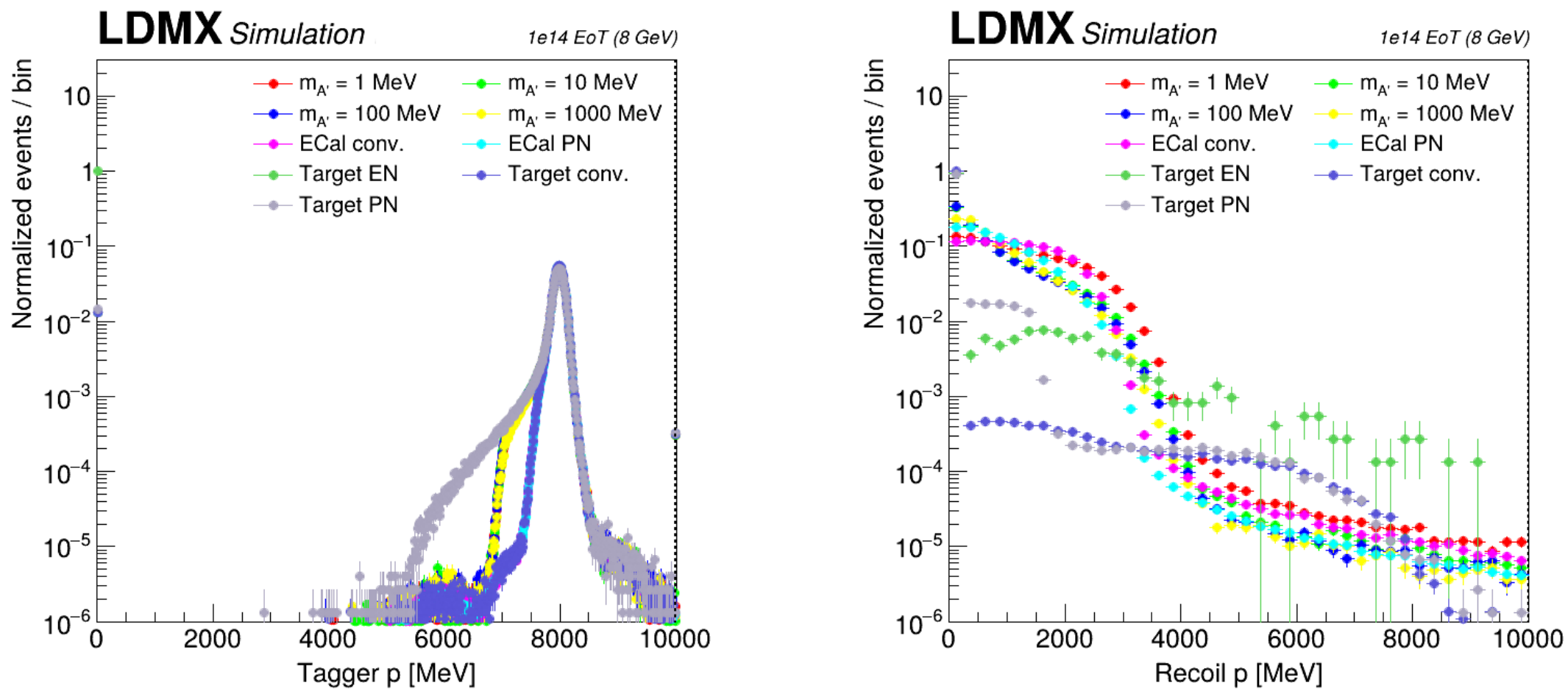

Figure 4.6: Plots showing the total momentum in the tagger tracker (left) and recoil tracker (right) with normalized logarithmic scale.

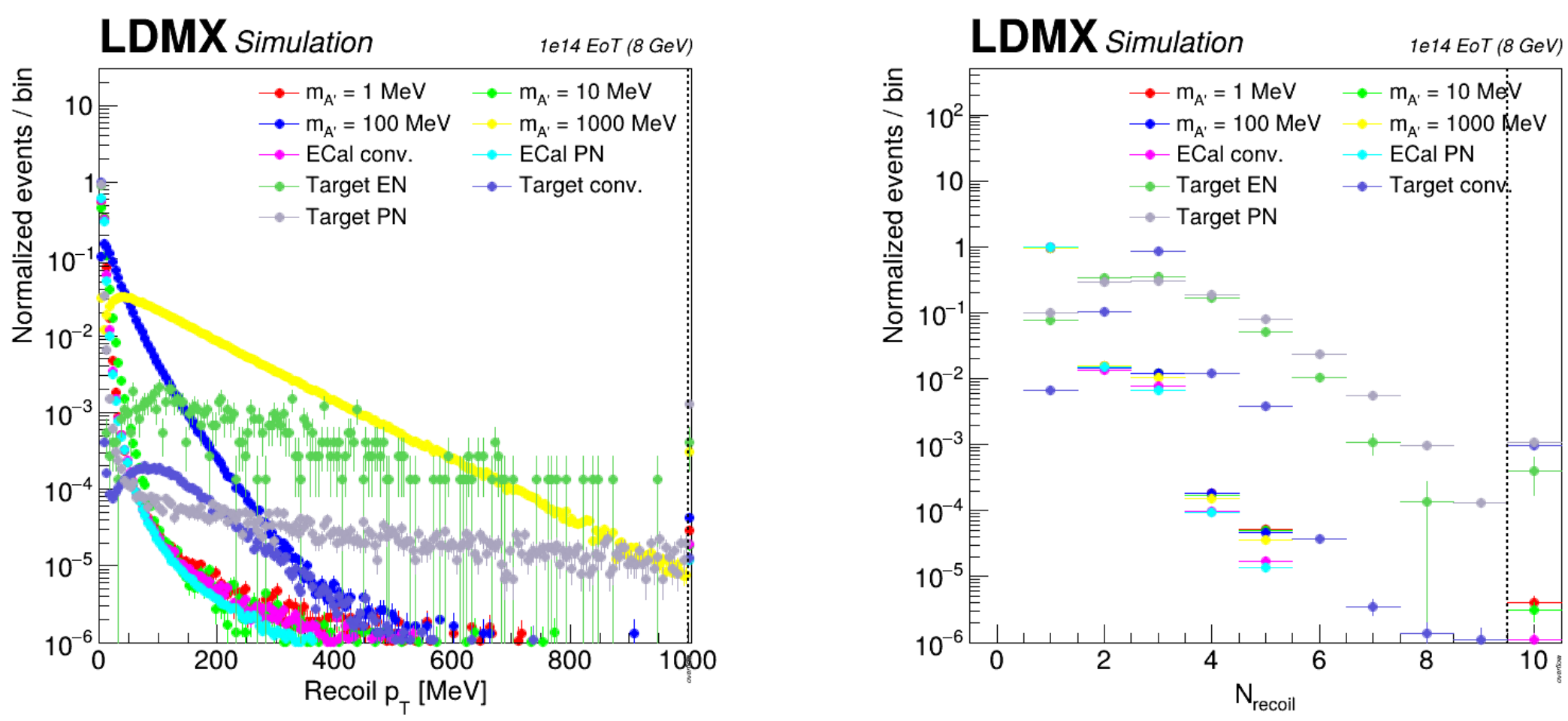

Figure 4.7: Plots showing the transverse momentum in the recoil tracker (left), and the number of tracks in the recoil tracker (right) with normalized logarithmic scale.

Fig. 4.9 shows the standard deviation of the shower size (left) and the deepest layer where a hit is measured (right). The left plot shows us that the shower for EN and PN background processes is wider than for the signal processes, which is expected since there is more activity in the event. These shower RMS get wider for the larger dark photon mass since the recoil electron is deflected more, which leads to arriving at the ECal at an angle. The target conversion sample leads to a larger shower RMS as the angular difference between the muons is higher. The right plot suggests that the background showers penetrate deeper in the ECal. We see that most of the background is in the last bin, suggesting that these particles leave the ECal and end up in the back-HCal.

The granularity of the ECal allows us to develop simplified tracking algorithms within the ECal. Given that there is no magnetic field in the ECal volume, the Minimum Ionizing Particles (MIPs) in the shower will travel on linear paths. Some particles, especially kaons and pions, can live long enough to pass through several layers of the ECal. We are looking for MIP tracks along the projected photon trajectory, so first we calculate the trajectory of the photon based on the recoil electron's trajectory and conservation of momentum. This procedure does not require a photon to be produced; the projection is based solely on the electron information. Then we look for hits that are in the same X and Y coordinate, i.e. perpendicular to the ECal face, in consecutive layers and call these "straight tracks".

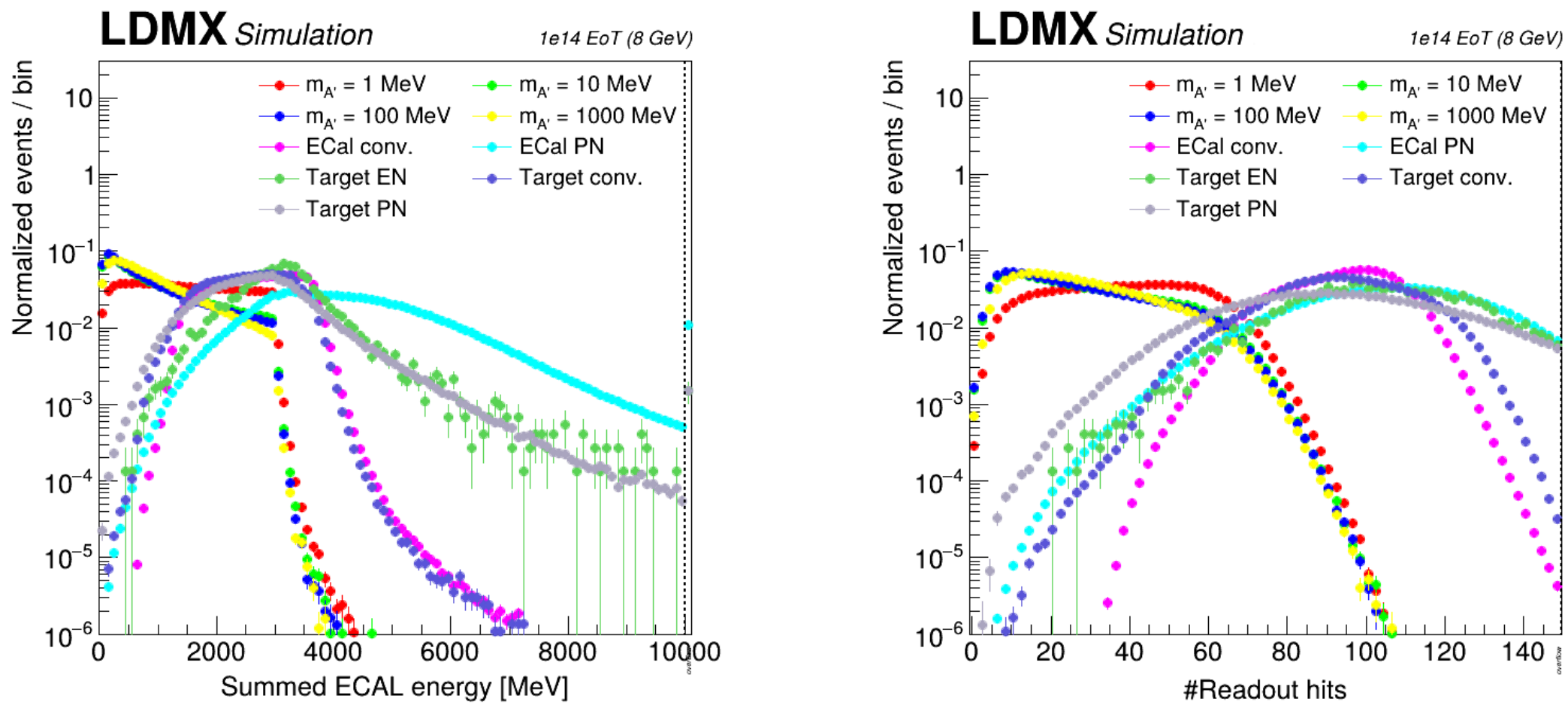

Figure 4.8: The total energy deposited in the ECal (left) and the number of read-out hits in the ECal (right).

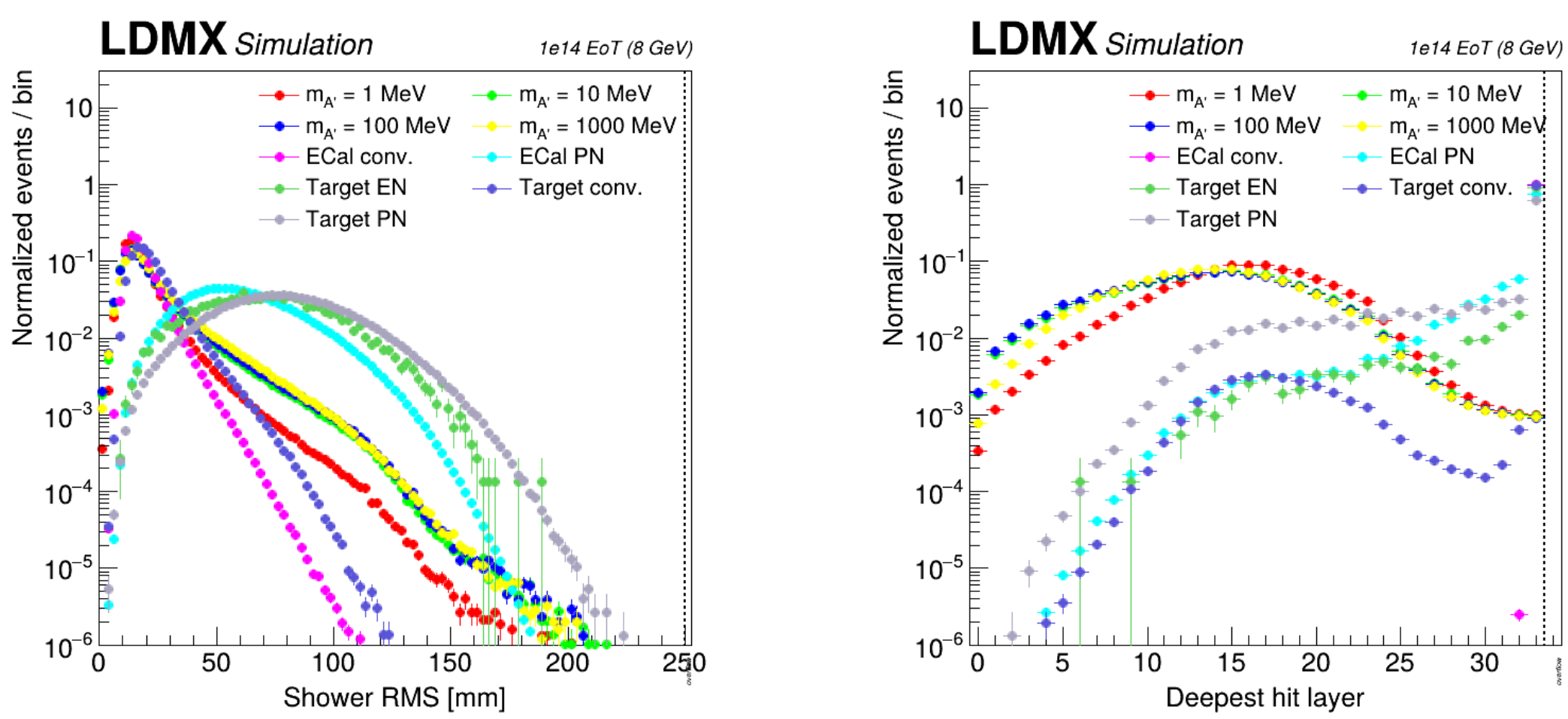

Figure 4.9: The standard deviation of the shower size (left) and the deepest layer where a hit is measured (right).

Figure 4.10 shows the number of straight tracks (left) and the angle between the electron and the photon direction (right) in the ECal. The angle is determined using the scalar product of the projected photon's and the measured electron's trajectory vector as obtained in the ECal face. The overflow bin in the electron-photon angle is for the case when the electron was not found in the ECal, so the angle cannot be determined. These variables are found to be useful discriminators for long-lived particles, like kaons, since for the DM signal we do not expect genuine MIP tracks in the ECal around the photon trajectory.

#### 4.2.3.3 HCal observables

For background events that are not removed based on ECal information, the HCal provides additional rejection handles. We do not expect any energetic secondary particles from signal events, so any sign of activity from a secondary particle is indicative of a background process.

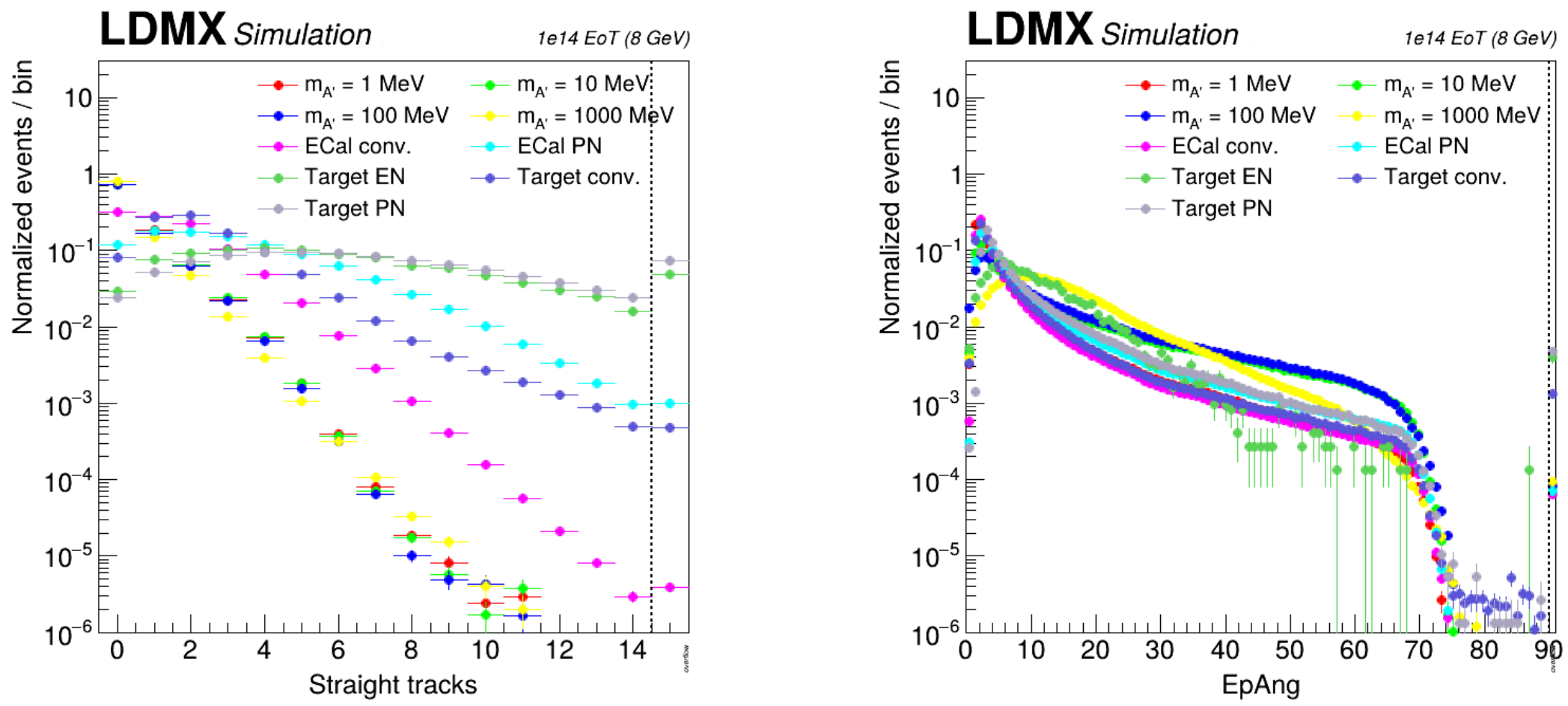

Figure 4.10: The number of straight tracks (left) and the electron-photon angle in the ECal (right).

Fig. 4.11 shows the largest hit amplitude recorded in a single bar in the full HCal (left) and the location of that bar (right). The largest hit amplitude recorded in the HCal is expressed in photo-electrons (PEs), assuming a response of 16 PE/ MeV of energy deposited in the scintillator bar, which is equivalent to about 60 PE/MIP. However, due to the double-sided read-out, this doubles and we expect around 120 PE/MIP. The electronics noise is at the level of a few PEs. As earlier, we see that background produces more activity, therefore more hits in the HCal as well, and they are more likely to end up in the Side-HCal.

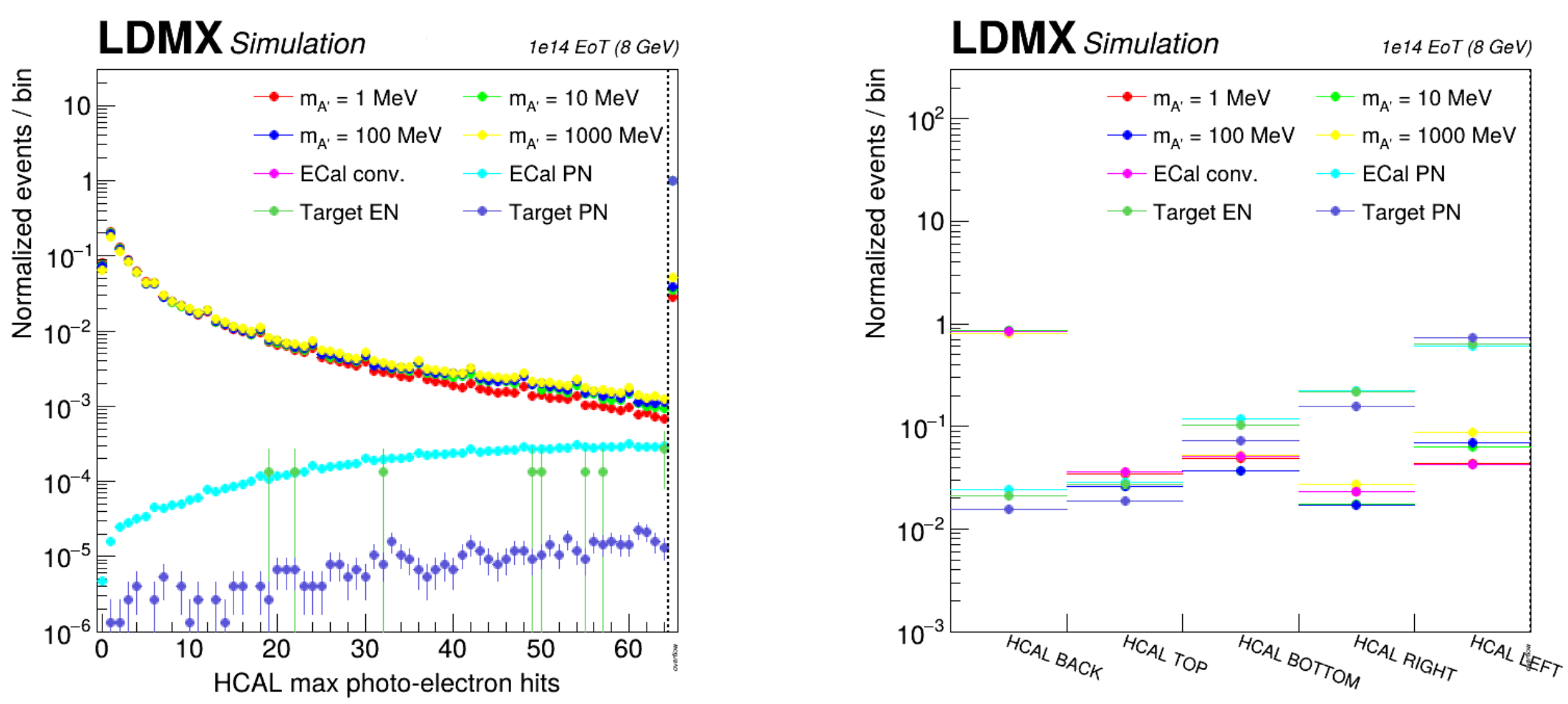

Figure 4.11: The amplitude of the most energetic hit in the full HCal ($PE_{max}$) expressed in photo-electrons (left) and their location (right).

Fig. 4.12 extends the range of max PE to 600 PE (left), and shows the total number of PE in the detector (right). The excesses at 250 PE and 500 PE are features of the reconstruction algorithm: when the readout mode changes from the ADC mode to the time-over-threshold mode, the ADC values enter overflow in ADC, that translates to 500 PE in the back-HCal and 250 PE in the side-HCal. From all the samples the ECal PN is most likely to produce particles that reach the HCal and produce showers there. The conversion samples contain muons that act as a MIP and will go through the ECal and many layers of the HCal as well. These will deposit around 60 PE in a single bar, which then is read out as around 120 PE, leading to the ECal conversion sample starting at 120 PE. If the muons end up in the side-HCal this number is halved, and the geometrical effects can lead to the dip as visible on the target conversion curve. These muons will also hit several HCal bars, leading to a total PE higher than 200 PE and showing up in the overflow bin.

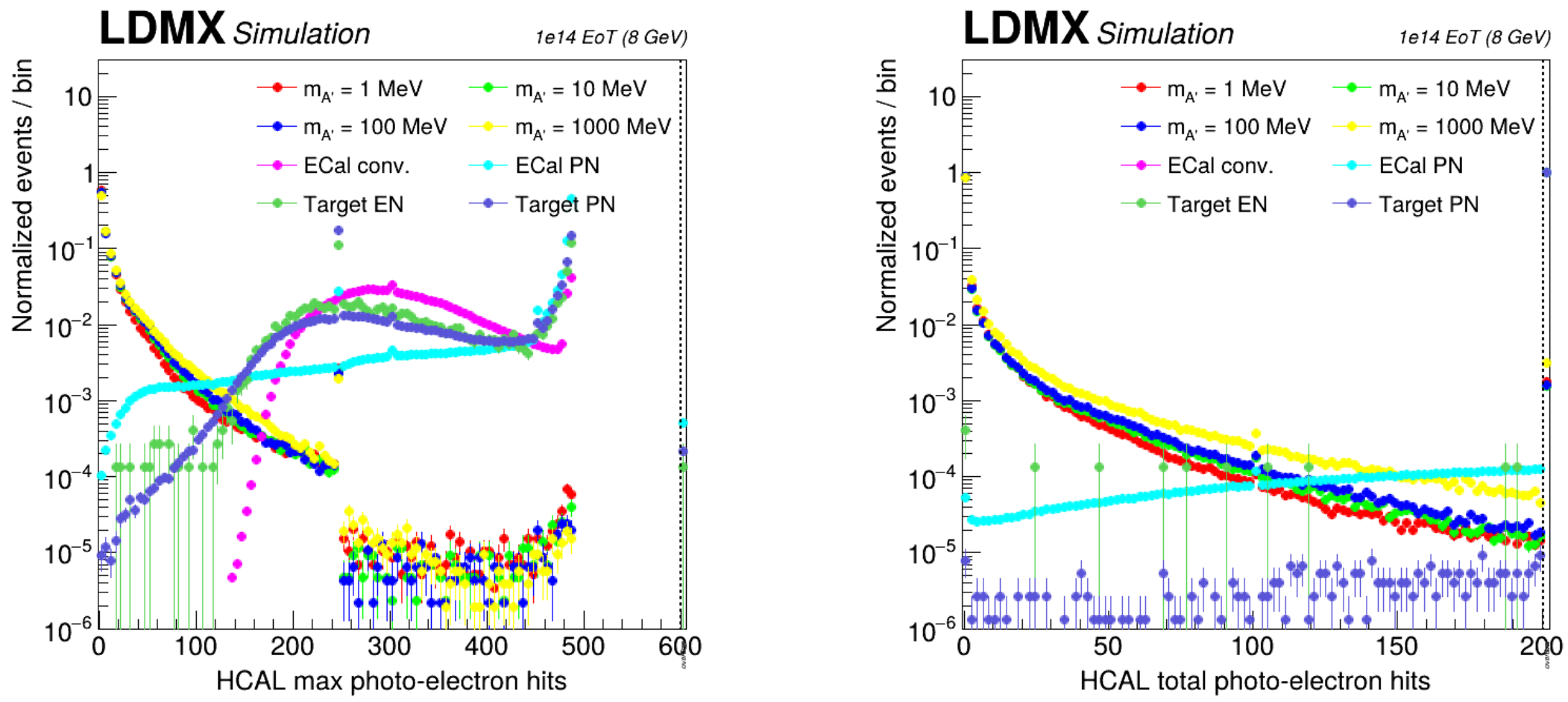

Figure 4.12: The amplitude of the most energetic hit in the full HCal expressed in photo-electrons up to 600 PE (left) and the total number of them (right).

### 4.2.4 Baseline background veto strategy

The baseline analysis strategy utilizes the observables introduced above to construct a background veto. This combination of requirements leverages each of the sub-detector systems to reduce all known SM backgrounds to a negligible level.

#### 4.2.4.1 Tracker veto

The tracker veto includes both the tagger and recoil tracker. To reject off-beam energy electrons in the tagger tracker, we require the track momentum to be at least 70% of the beam energy, i.e. $p_{\text{tagger}} > 5.6\,\text{GeV}$. To reject non-interacting electrons and events where several tracks originate from the target, we require a single, high-quality track in the recoil tracker, i.e. $N_{\text{recoil}} = 1$.

#### 4.2.4.2 ECal veto

Several potential implementations of an ECal calorimeter veto have been studied, which optimally distill the information summarized in the previous ECal observables section. Here we present the baseline strategy, building a linear multivariate discriminant from a more extensive list of observables ($\sim 50$) using a boosted decision tree (BDT). Two alternative approaches are presented in later sections: a simple 'rectangular cut' analysis (Sec. 4.3.1.1), and a more aggressive multivariate strategy, using a graph neural network (GNN) to directly analyze calorimeter hits as 'point cloud' images (Sec. 4.3.1.2).

The BDT approach, implemented with XGBoost [143], exploits the high granularity of the ECal by using the same variables discussed in the Sec. 4.2.3.2 together with extended features exploiting longitudinal shower structures and more detailed MIP tracking. Concretely, the ECal is divided into 3 longitudinal segments (Layers 0 - 5, Layers 6 - 16 and Layers 17 - 33) and the following 8 features are calculated for each segment:

- Total energy deposited
- Number of recorded hits
- Energy-weighted average position of recorded hits (X, Y, Layer)
- Energy-weighted standard deviation in position of recorded hits (X, Y, Layer)

To parameterize the electromagnetic shower, we calculate the containment radii $r_c$. This variable is defined as the radius containing 68% of the shower energy on average. The values are different for each layer and depend on the detector geometry. These parameters are determined in signal simulations, ensuring that only a single (recoil) electron enters the ECal. Based on this information ten energy sums are calculated for use in the BDT, by summing deposits within $r_c, 2r_c, \ldots, 5r_c$ of both the recoil electron and projected photon trajectories.

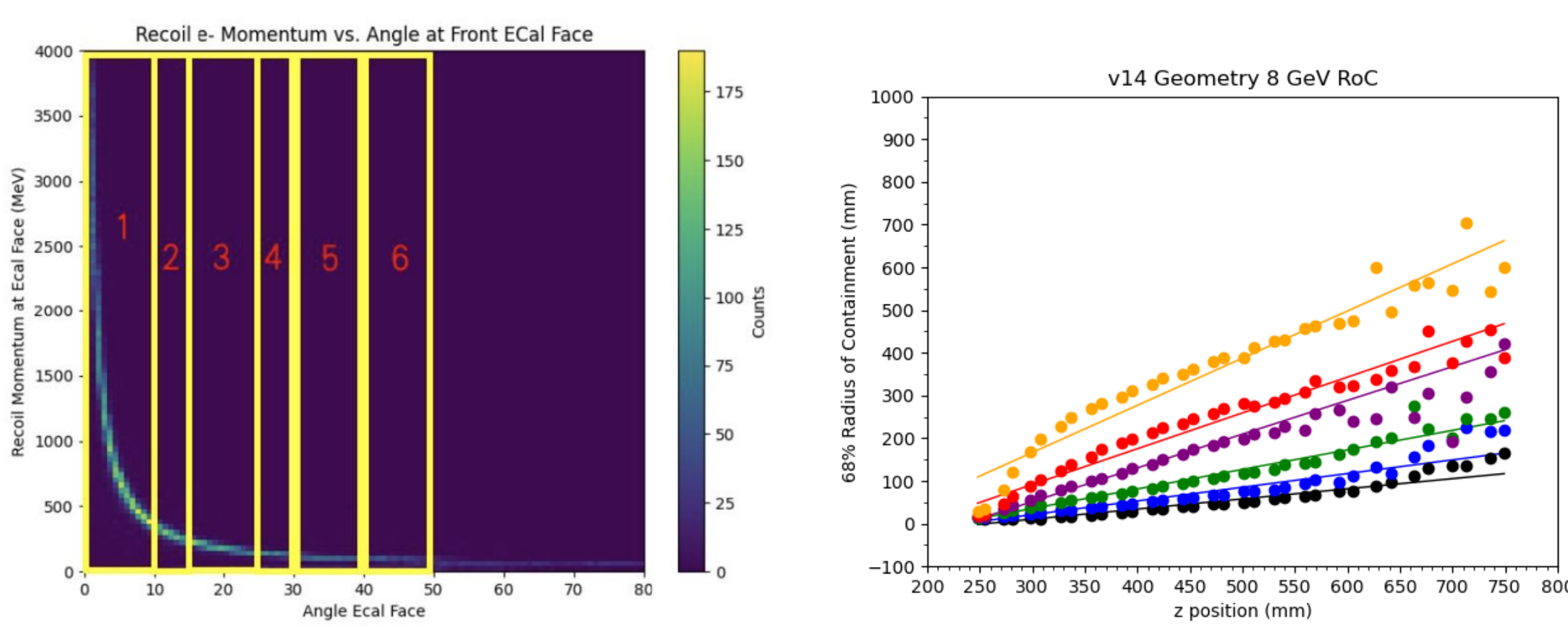

Figure 4.13: Binning choice for the radii of containment (left) and the values of it as a function of ECal depth (right). Colors on the right plot correspond to the different angle bins.

Fig. 4.13 on the left shows the recoil electron's momentum together with its angle as measured at the ECal face ($\theta$). The showers at 8 GeV are narrower, leading to a simpler binning than was presented for 4 GeV in Ref. [134]. Here we find that a single momentum category with $p < 3.160$ GeV is sufficient, and derive separate radii for angles in the ranges: (1) $\theta < 10$, (2) $10 \leq \theta < 15$, (3) $15 \leq \theta < 25$, (4) $25 \leq \theta < 30$, (5) $30 \leq \theta < 40$, (6) $40 \leq \theta < 50$. Fig. 4.13 (right) shows the radii of containment as a function of the ECal depth, where a linear fit is employed for simplicity.

All together, these calculations produce 395 possible BDT input features. This was reduced to 47 by removing 'redundant' variables whose omission had no impact on the final significance. The BDT was trained to distinguish the DM signal process from photo-nuclear interactions, using a small fraction of each of these MC samples (including several $A'$ masses), corresponding to events that pass the trigger. The distribution of "BDT scores", resulting from this optimized linear map, is shown in Fig. 4.14.

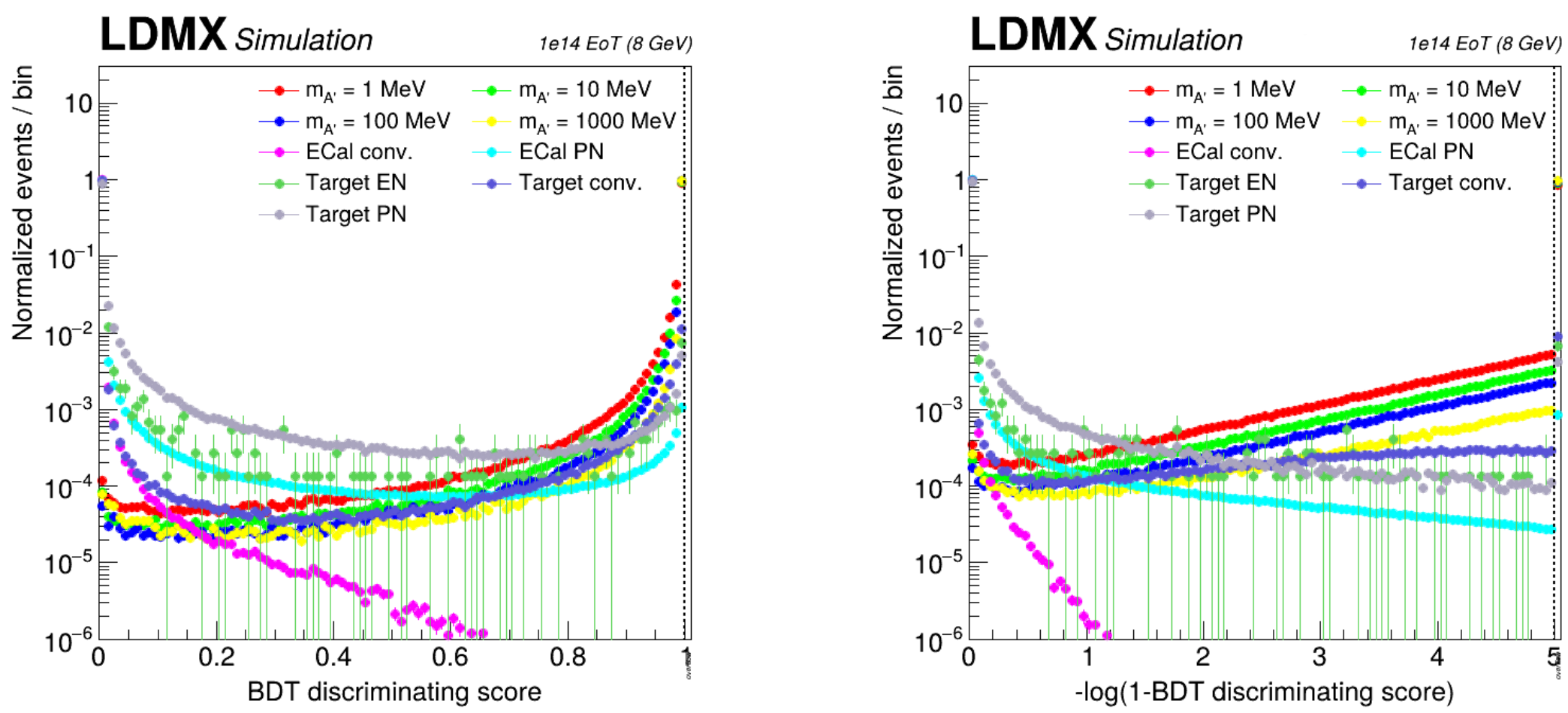

Figure 4.14: The BDT scores for signal and background MC events passing the trigger and fiducial selections (left) together with its logarithmic transform (right).

To understand the dominant factors in the BDT performance, the SHAP algorithm [144] is used to determine the feature importance. Fig. 4.15 shows the ordered SHAP absolute values, with the highest number being the most important feature. The most important feature is found to be the energy sum in isolated cells (those without a neighboring hit). Following this, are the number of read-out hits, the deepest layer hit, the total energy sum, and the shower RMS, seen in Figs. 4.8 and 4.9.

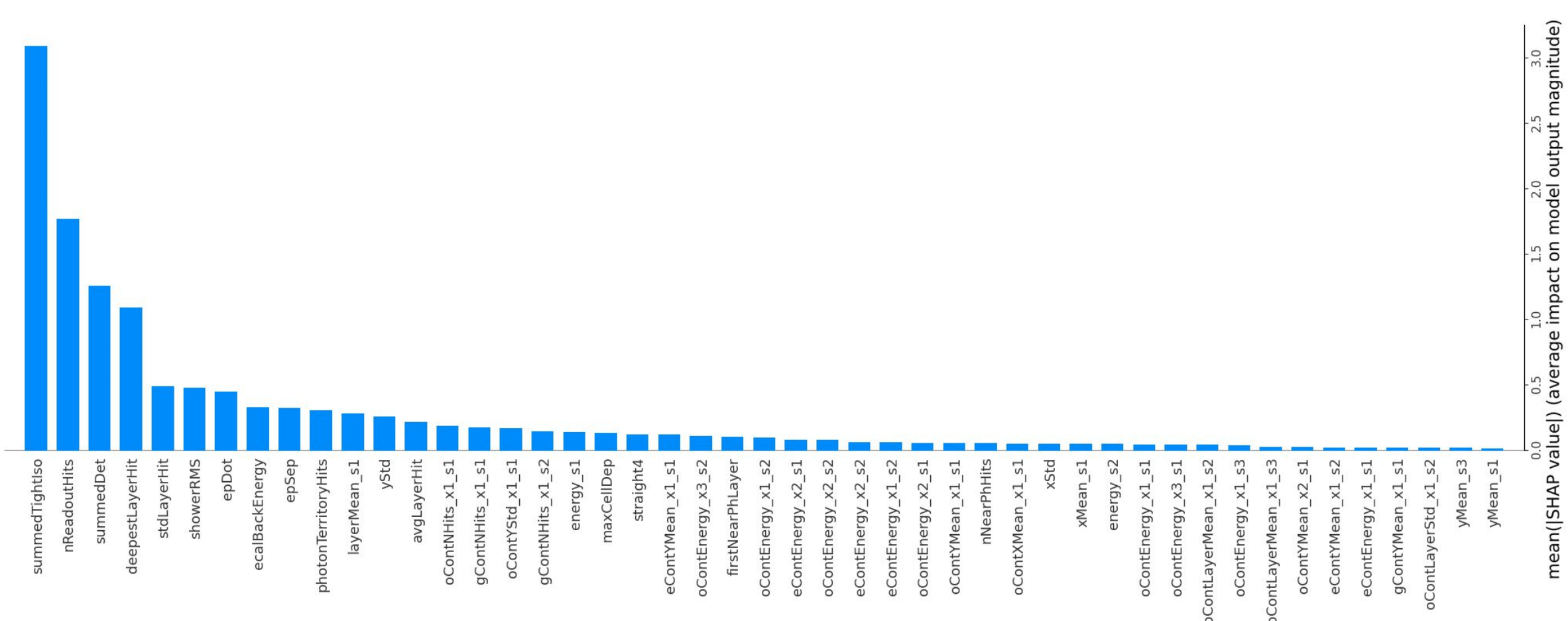

Figure 4.15: Ordered SHAP values for feature importance in the BDT.

To assess the veto's power, we adopt a BDT score requirement that removes all but two MC events from the PN sample (0.99741) when combined with the HCal veto. We present results with this cut so that the last events can be studied, and facilitate projections to higher EoT. The value of 0.99741 corresponds to a transformed score of 2.586 in the right panel of Fig. 4.14.

#### 4.2.4.3 HCal veto

After the ECal BDT selection, the PN background sample is largely composed of events involving the production of a single forward neutron, which accounts for a significant portion of the 92.6% dominant background. These neutrons typically have kinetic energies peaking around 6 GeV, with some contribution from two lower-energy neutral particles (6.9%). In cases where only neutral particles are produced, suppression relies on the HCal veto.

The 1D distributions of the HCal max PE and total PE were already shown in Sec. 4.2.3 in Fig. 4.11 and Fig. 4.12. Fig. 4.16 shows the BDT score vs the maximum number of HCal photo-electrons for both signal and the PN background, after applying the ECal preselection cuts. The two veto selections are essentially uncorrelated and each add significant discriminating power.

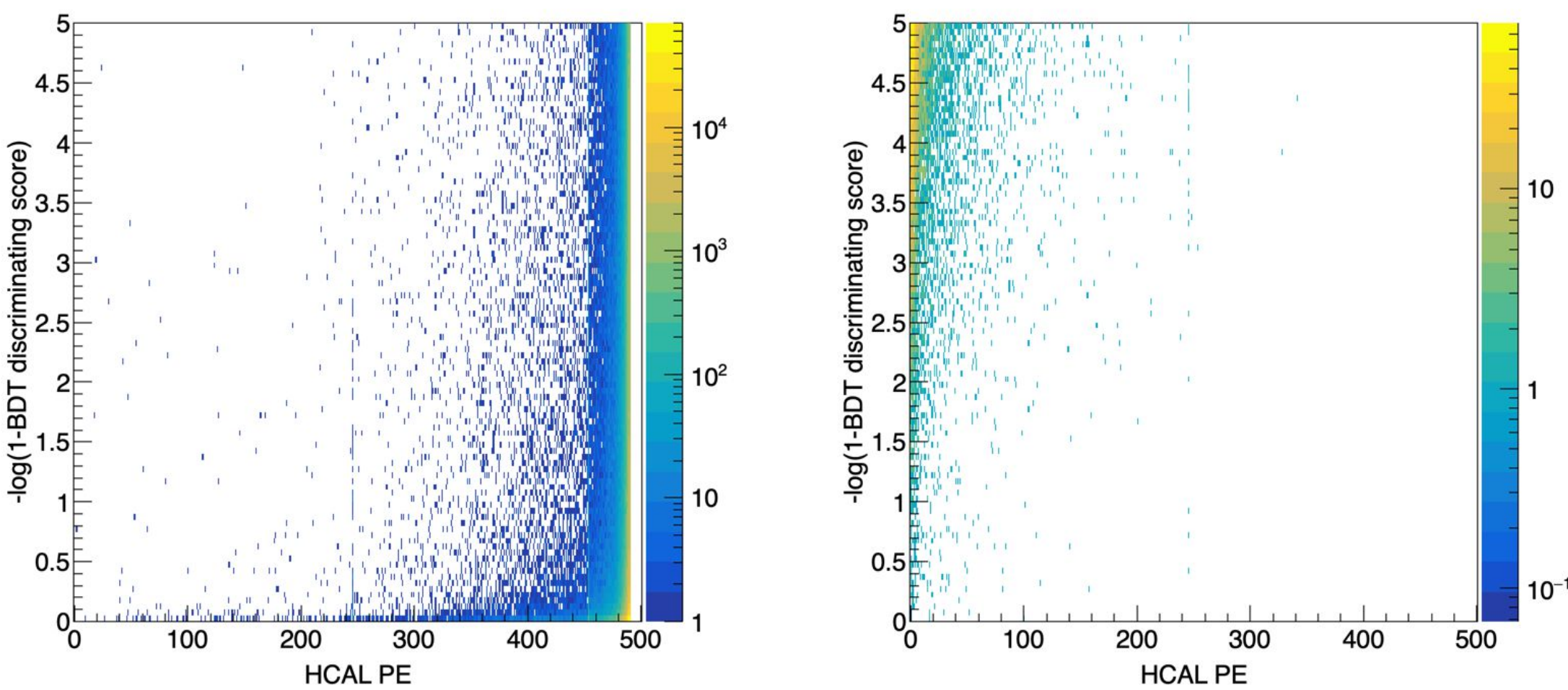

Figure 4.16: The transformed BDT score as a function of the maximum number of HCal photo-electrons for the ECal PN sample (left) and the 1 GeV signal sample (right).

All but a few events have a hit above 200 PEs, well above the noise level of a few PEs. The few entries below that threshold arise from events in which all particles but low-energy neutrons (with kinetic energies typically between $200 - 600$ MeV) are contained in the ECal or, in the most extreme case, only neutrinos reach the boundaries of the HCal. These are, for example, the signature of reactions in which a single charged kaon (possibly with additional soft neutrons) is produced in PN reactions, and neutrinos carry almost all the energy in the subsequent decay chain. These topologies are mainly identified using dedicated MIP tracking algorithms in the ECal. A sub-dominant contribution also originates from reactions in which the tungsten nucleus emits many low-energy neutrons without protons. Restricting the maximal number of photoelectrons (HCal maxPE), as shown in Fig. 4.11, to be less than eight removes all neutrons; in the few remaining MC events, only neutrinos enter the HCal.

#### 4.2.4.4 MIP tracking veto

To complete this analysis, a dedicated MIP tracking algorithm leverages the fine ECal granularity to identify backgrounds that may escape the tracking, HCal, and ECal BDT vetos. A significant part of the dominant ECal PN background component after the other vetos is forward-going $K_L$ mesons. These $K_L$ mesons have kinetic energies peaking around 6 GeV. Some of them go on to interact within the ECal, and the resulting final states often include charged particles. These are mitigated using MIP tracking selections, which limit the number of straight MIP tracks ($N_{\text{straight}}$). A further selection to reduce these is a tighter cut on the electron-photon angle.

### 4.2.5 Results and Discussion

The results of the veto requirements described above are shown in Table 4.2. The veto performance is shown for the most challenging background processes, each of which drive the experiment's physics requirements. After all selections, the $10^{14}$ EoT sample of backgrounds is reduced to a negligible level, while maintaining signal efficiencies of 50-68%.

Two Monte Carlo events from the PN sample remains by construction, as discussed in Sec. 4.2.4.2. This event contains an energetic neutrino from kaon decay: a challenging process that is nevertheless reducible with ECal MIP-tracking capabilities.

While we take these results to indicate that LDMX is well positioned to conduct a background-free search with $4 \times 10^{14}$ EoT, the impact of a modest remaining background is shown in Fig. 4.17. Though some effect on the projected limit is visible, only the lowest DM masses are impacted, where LDMX anyhow tests the complete range of thermal relics due to the large production rate of light $A'$s.

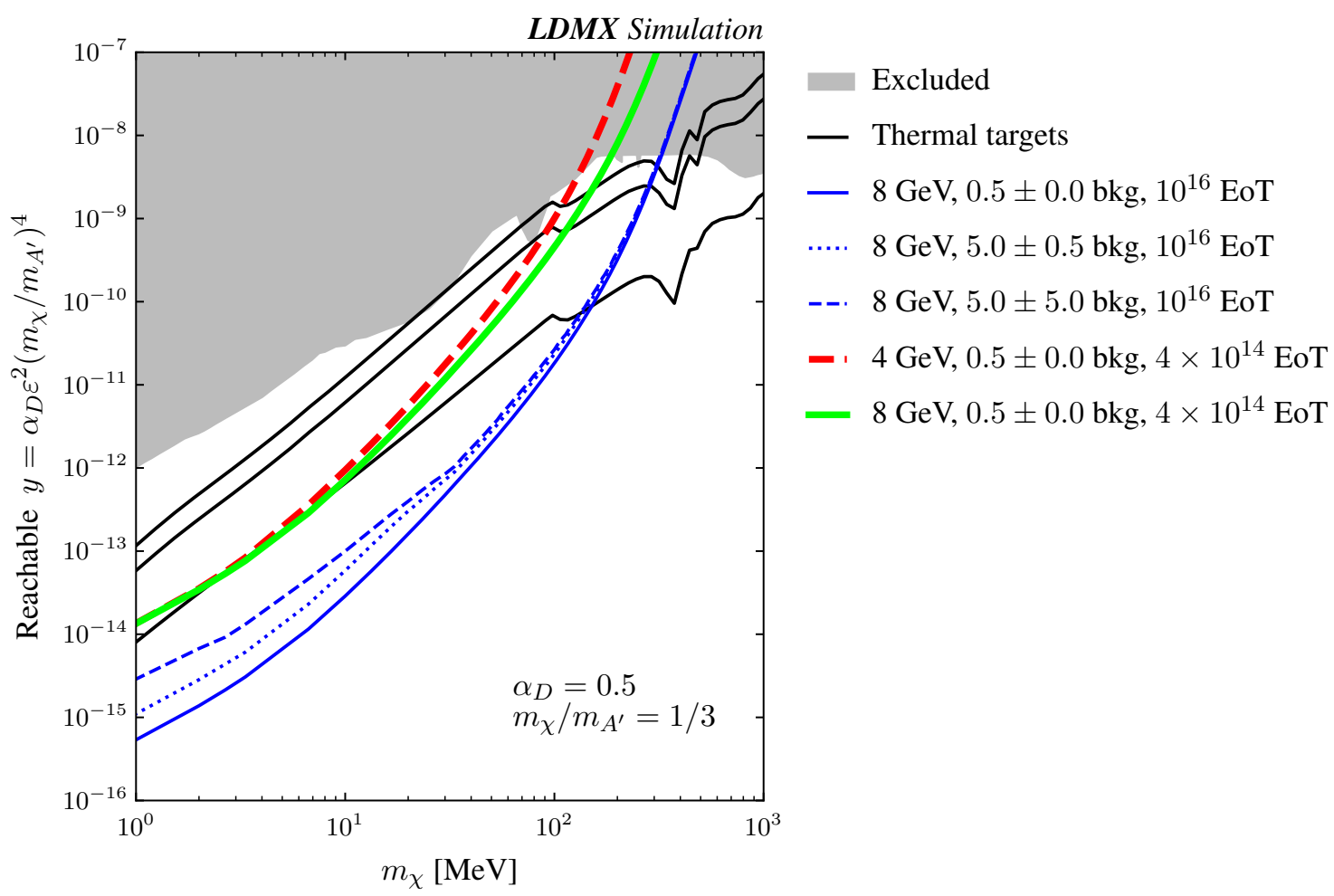

Figure 4.17: Projected reach of the missing-momentum search for several possible background levels.

From this study, we conclude that LDMX will be capable of carrying out a robust missing momentum search with the Phase 1 data set, delivering order-of-magnitude improvements over the current sensitivity to light DM. In the section to follow we build on this work with a number of supporting studies, which collectively aim to demonstrate how this search technique may be extended to the ultimate Phase-2 $10^{16}$ EoT equivalent data sample.

| Cut | Number of Background Remaining | | | | Signal Efficiency for different $m_{A'}$ | | | |
|---|---|---|---|---|---|---|---|---|
| | target PN | target conv. | ECal PN | ECal conv. | 1 MeV | 10 MeV | 100 MeV | 1000 MeV |
| EoT equivalent | $10^{14}$ | $10^{15}$ | $10^{14}$ | $10^{14}$ | | | | |
| All / Acceptance | 1,941,296 | 7,091,907 | 299,299,867 | 18,720,663 | 100% | 100% | 100% | 100% |
| Triggered | 756,661 | 4,280,536 | 170,993,364 | 14,042,193 | 78.2% | 89.6% | 91.3% | 94.3% |
| $p_{\text{tagger}} > 5.6\,\text{GeV}$ | 745,494 | 4,224,191 | 168,725,637 | 13,856,953 | 77.1% | 88.4% | 90.0% | 93.0% |
| $N_{\text{recoil}} = 1$ | 72,927 | 21,625 | 154,191,349 | 12,685,692 | 72.2% | 82.4% | 83.0% | 72.4% |
| Ecal BDT > 0.9974 | 1,529 | <1 | 64,353 | <1 | 52.9% | 68.6% | 73.1% | 67.8% |
| $N_{straight} < 3$ | 1,412 | <1 | 61,771 | <1 | 51.8% | 67.2% | 71.7% | 64.3% |
| HCal maxPE < 8 | <1 | <1 | 2 | <1 | 49.9% | 64.3% | 68.3% | 59.3% |

Table 4.2: Cutflow with trigger, tracker veto, ECal BDT, HCal veto and MIP tracking cuts, in the target / ECal conversion and photo-nuclear events, together with a target electro-nuclear sample as backgrounds, and signal samples with masses from 1 MeV to 1 GeV.

## 4.3 Further studies of the missing-momentum search methodology

The previous section described the baseline missing momentum analysis strategy, based on large simulated samples representing the key background processes motivated in Sec. 3.2. The following sections extend this work in several key ways, with a particular view towards the ultimate $10^{16}$ EoT-equivalent search.

First, two alternate ECal veto methodologies are discussed. The first is a 'cut-based' analysis (without multivariate discriminants), allowing one to construct well-understood 'sideband' regions that complement simulation-based predictions with data-driven background estimates. The second leverages the full power of the ECal granularity with graph neural network to construct a more aggressive veto. Next, two important background processes are studied in greater detail using specialized simulations: wide-angle bremsstrahlung and kaon production through PN interactions. Finally, the nominal analysis strategy is extended to events with multiple electrons, a critical aspect for rapid collection of a $10^{16}$ EoT-equivalent dataset. In particular, we show that excellent trigger performance and PN rejection can be achieved in this busier environment with additional electrons.

### 4.3.1 Alternate ECal veto techniques

#### 4.3.1.1 Cut and count approach

The strategy taken to this point has been to demonstrate the feasibility of the missing momentum search through the analysis of large samples of simulated events. However, the collection of real collision events can provide an improved understanding of background processes that can in turn enable the development of more powerful search strategies. As one example, 'sideband' event selections can be defined, which mimic the search region selections except for a key inverted cut to enrich the process of interest.

To develop this capability, an alternate ECal veto strategy has been developed, based on 'rectangular cuts'. Thus, instead of combining many features into a multivariate discriminant (a BDT, e.g.), individual requirements are set on each observeable. Here, these selections are optimized using the Punzi figure of merit (FOM) [145], which provides a balanced optimization for both discovery and upper limit settings at the same time. The FOM, $f$ is calculated as shown in Eq. 4.4, where $\epsilon_S$ is the signal efficiency, $B$ is the background yield and $a = 3$ is in the unit of standard deviations for evidence.

$$f = \frac{\epsilon_S}{\frac{a}{2} + \sqrt{B}} \tag{4.4}$$

Fig. 4.18 shows some examples of the FOM as a function of a given cut on the variable: the total energy deposited in the ECal (left), and the number of read-out hits (right). The cut value is chosen around where the FOM reaches its plateau, or sometimes having a looser cut to enhance the sensitivity to the higher mass point. In this optimization, the ECal PN sample is used for the background yields.

The variables used were selected based on the most important features in the BDT, as shown in 4.15. The $E_{\mathrm{sum}}$ is the energy sum of the ECal hits, the $E_{\mathrm{sumTight}}$ is the energy sum requiring that there are no hits in the neighboring cells on a given layer, while the $E_{\mathrm{back}}$ is the energy sum in the back ECal, i.e., in the last 16 layers of the ECal. The $N_{hits}$ counts the number of read-out hits in the calorimeter, the $\mathrm{RMS}_{\mathrm{shower}}$ measures the size of the showers, and the $E_{\mathrm{cell,max}}$ is the maximal energy deposited in a single cell. $\mathrm{RMS}_{\mathrm{Layer,hit}}$ is the RMS of the number of hit layers, which is useful to study the longitudinal development of the showers. The $N_{\mathrm{straight}}$ is the number of straight MIP tracks, and the HCal veto uses the same value as earlier. These variables convey the fact that signal tends to have shorter, more isolated showers, leading to fewer hits and activity in the ECal than the ECal backgrounds. For some of the distributions of these variables, see Sec. 4.2.3.

The results of this optimization are shown in the cutflow, Table 4.3, which shows the yields after each cut for the ECal photo-nuclear and conversion background samples, and the selection efficiencies for the signals considered. Here we focus on the ECal samples that are most significantly affected by the ECal veto.

With these selections, signal efficiency of $\sim 60\%$ or larger is achieved for the benchmark $A'$ signals, with seven PN and zero conversion events remaining. First, this demonstrates how the powerful design of LDMX allows us to carry out an effective DM search with simple methods. Second, the table indicates several promising requirements that may be inverted to define large event samples enriched in key backgrounds. For example, selections requiring ECal tracks or energetic cell hits can be used to select muons, leading to a sample enriched in charged kaon decays. Similarly, inverted HCal cuts and late ECal activity can be used to measure the rate of neutron production in PN reactions with many thousands of events.

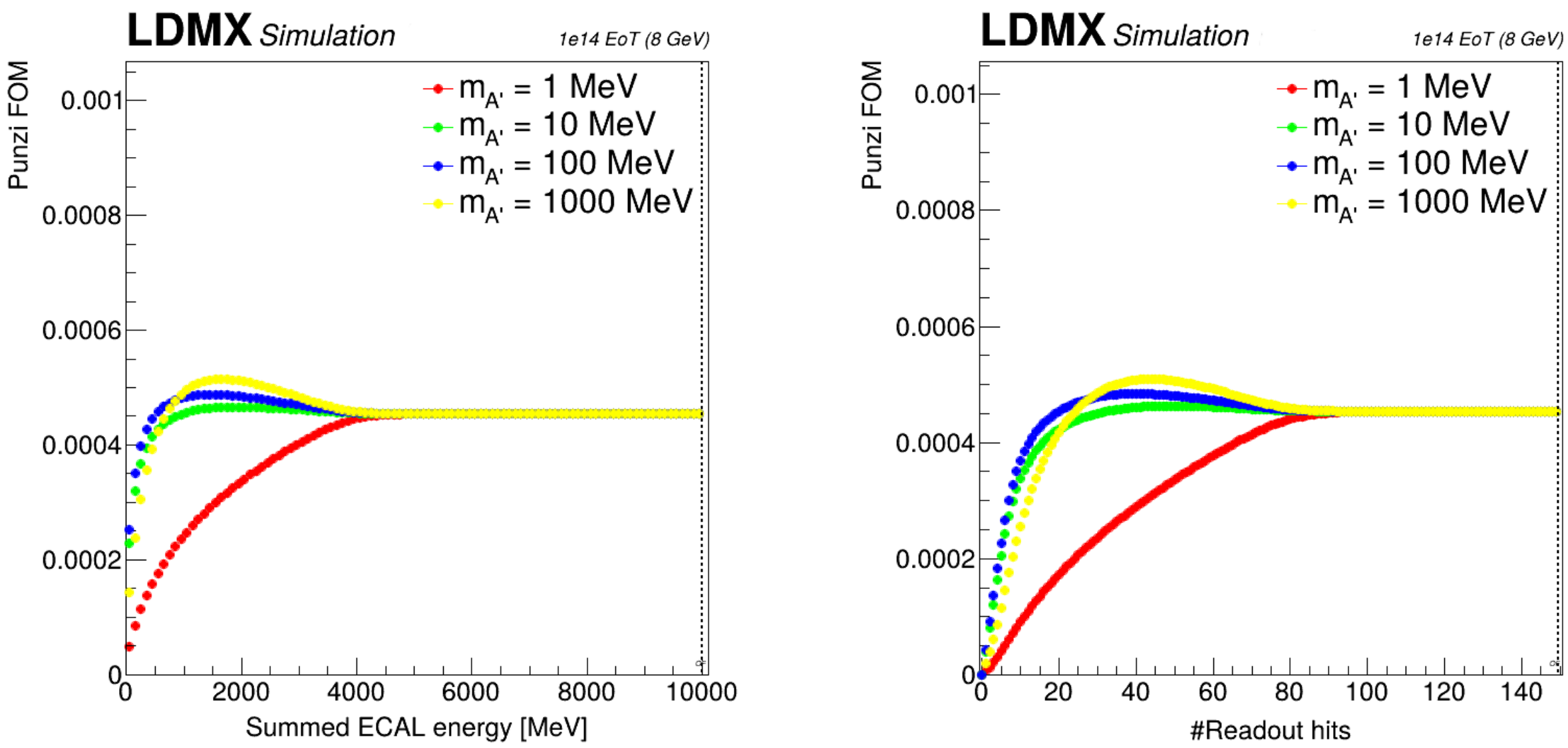

Figure 4.18: Example Punzi FOMs as a function of the total energy deposited in the ECal (left) and the number of ECal read-out hits (right).

| Cut | Number of Background Remaining | | Signal Efficiency for different $m_{A'}$ | | | |
|---|---|---|---|---|---|---|
| | ECal PN | ECal conversion | 1 MeV | 10 MeV | 100 MeV | 1000 MeV |
| EoT equivalent | $10^{14}$ | $10^{14}$ | | | | |
| All / Acceptance | 299,299,867 | 18,720,663 | 100% | 100% | 100% | 100% |
| Triggered | 170,993,364 | 14,042,193 | 78.2% | 89.6% | 91.3% | 94.3% |
| $E_{\mathrm{sum}} < 3500$ MeV | 56,037,910 | 12,258,095 | 78.2% | 89.6% | 91.3% | 94.3% |
| $E_{\mathrm{sumTight}} < 800$ MeV | 2,488,172 | 1,123,523 | 73.8% | 86.6% | 88.3% | 91.5% |
| $E_{\mathrm{back}} < 250$ MeV | 1,851,644 | 296 | 73.7% | 86.5% | 88.2% | 91.5% |
| $N_{\mathrm{hits}} < 70$ | 1,472,425 | 116 | 71.6% | 85.3% | 87.2% | 90.8% |
| $RMS_{\mathrm{shower}} < 110$ mm | 1,466,579 | 116 | 71.5% | 85.1% | 86.9% | 90.3% |
| $E_{\mathrm{cell,max}} < 300$ MeV | 1,194,631 | 114 | 65.7% | 81.0% | 83.0% | 85.5% |
| $\mathrm{RMS}_{\mathrm{Layer,hit}} < 5$ | 1,001,030 | <1 | 64.4% | 79.0% | 81.0% | 83.6% |
| $N_{\mathrm{straight}} < 3$ | 933,886 | <1 | 62.1% | 76.2% | 79.2% | 73.2% |
| HCal maxPE < 8 | 7 | <1 | 59.4% | 72.6% | 75.0% | 65.0% |

Table 4.3: Cutflow with trigger, a cut and count analysis in the ECal and the HCal veto.

#### 4.3.1.2 Graph neutral network approach

While we have shown above that the BDT-based analysis is sufficient to perform a powerful DM search in single electron events, graph neural networks offer an even more powerful alternative in principle. The granularity of the ECal permits a natural representation of detector hits as unordered ensembles of points, or 'point clouds'. Hence, we may leverage the spatial profiles of calorimeter showers to classify them as being characteristically 'signal-like' or 'background-like'. To this end, we adapt the jet-tagging algorithm, ParticleNet [146], to the task of shower classification. ParticleNet is a Dynamic Graph Convolutional Neural Network (DGCNN), capable of casting point clouds in position space to a phase-space representation in which semantically-related events tend to coalesce into distinct bunches. This capability is especially useful for resolving the highly boosted, and often overlapping, showers expected in LDMX. We find ParticleNet to outperform both the cut-based and BDT analyses when applied to our 8 GeV simulated signal and background samples with v14 detector geometry. ParticleNet reduces the $1\times10^{14}$ EoT equivalent photo-nuclear (pn) background sample to a single remaining background event, while maintaining >75% signal efficiency for all four $A'$ mass points ranging from 1 MeV to 1 GeV.

We preprocess raw event data to ensure they are suitable as ParticleNet inputs. We also determine a number of computed quantities at this stage, which would be too computationally expensive to include in the training loop. Preprocessing is also responsible for applying the preselection cut: ECal hits $<90$ and total isolated energy in ECal $<$ 1100 MeV. These lead to 8M ECal PN events, while losing less than 0.5% signal efficiency.

ParticleNet performance for each signal mass-point is shown in Fig. 4.19. The four receiver operator characteristic (ROC) curves represent the true positive rate (signal efficiency) as a function of false positive rate (background mistag rate). A perfect classifier would be a point sitting in the top left corner at (0, 1), with an area under the curve (AUC) of 1, while a random classifier would be a line of slope 1 (in linear scale) with an AUC of 0.5. Performance increases monotonically with increasing dark-photon mass. This does not necessarily imply any true $p_{\mathrm{T}}$ bias, as the monotonicity is predicted by detector acceptance considerations alone. That is to say, higher signal masses, since they are associated with larger transverse kicks, will yield overall better separation between electrons and photons. Larger electron-photon separation angles reduce the rate at which shower components land in the same ECal cell, making showers more readily resolvable.

For the results shown, the ECal is treated as a single region. A multi-region extension of ParticleNet adaptively defines containment regions around electron and photon trajectories. Performance gains from this approach are modest in the 1-electron case, but we expect this technique to be of greater importance in cases with more than one electron on target.

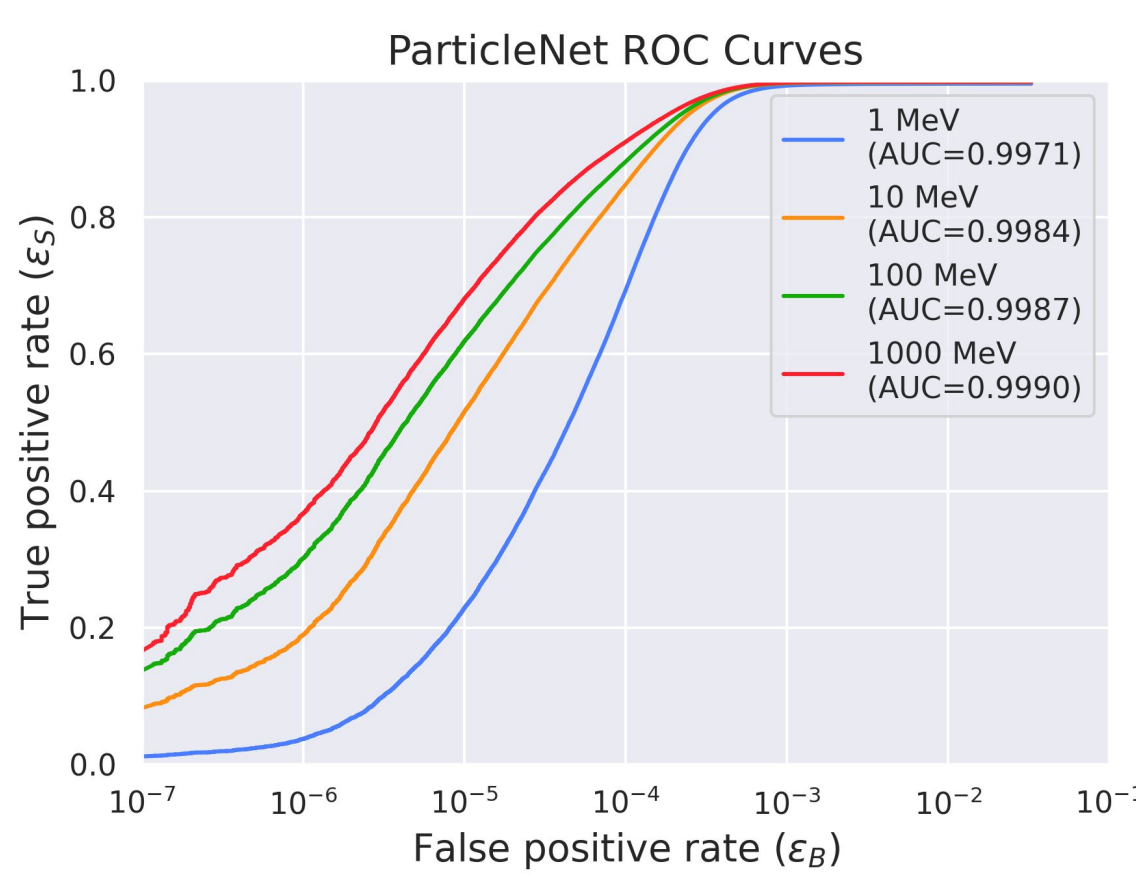

Figure 4.19: ROC curves showing ParticleNet performance.

The final event cutflow includes the preselection cut, the ParticleNet discrimination threshold cut, and the HCal veto. Results are shown in Table 4.4, with net efficiencies in the last row. As ParticleNet is not evaluated on events seen during training, the number of PN events is extrapolated to the full $1\times10^{14}$ EoT equivalent sample. Only the ECal PN is shown since this is the most challenging background for the ECal veto.

| Cut | Number of Background Remaining | | Signal Efficiency for different $m_{A'}$ | | | |
|---|---|---|---|---|---|---|
| | eval sample (PN) | at $10^{14}$ EoT | 1 MeV | 10 MeV | 100 MeV | 1000 MeV |
| Preselection | 8,929,094 | | 99.5% | 99.8% | 99.8% | 99.8% |
| ParticleNet > 0.74 | 74,805 | | 92.2% | 95.9% | 96.4% | 97.1% |
| HCal maxPE < 8 | 1 | 1.1 ± 0.3 | 87.6% | 89.7% | 89.4% | 75.7% |

Table 4.4: Cutflow including the preselection cut, the ParticleNet discrimination threshold cut, and the HCal veto.

### 4.3.2 Additional studies of rare background processes

To demonstrate how rare background processes can be mitigated at the level of $10^{-16}$, this section buttresses the Geant4 study presented in Sec. 4.2.4 with a deeper look into dedicated calculations of potentially-challenging backgrounds. First, an ECal PN sample enriched in kaon events is studied, verifying performance of the ECal veto in the most challenging phase space. Second, dedicated calculations of wide-angle bremsstrahlung and electron scattering are considered to ensure the robust veto capabilities of the side HCal sub-detector.

#### 4.3.2.1 Studies of processes with kaons

To understand which physical processes could be most challenging to distinguish from signal, we look at those that remain after the cut-based selection shown in Sec. 4.3.1.1. Here, generator-level information shows that six of the seven remaining events contain kaons in the final state. The momenta of these kaons and the hard bremsstrahlung-photon indicate that most (85-90%) of the photon energy is transferred to the kaons. (Similar conclusions may be drawn from those events with the most signal-like BDT and ParticleNet scores.)

To further investigate these kaon processes, we generated a custom sample of ECal PN events that contain kaons, but reducing the kaon lifetime by a factor of 50 and forcing their leptonic decay. This strategy intentionally over-samples the most challenging phase space with large neutrino momenta and short Kaon tracks. While it is very useful to study the kaon events and to develop algorithms to remove them, the final yields should not be taken directly as a realistic estimate of the kaon yield at the end of the run, but need to be converted.

For completeness, we compare results for this sample with the cut-based, BDT, and ParticleNet analysis strategies. Table 4.5 shows the cutflow after the cut and count approach (left) and the BDT approach (right). The cut and count approach removes all but 293 MC events from this sample, corresponding to $< 5$ expected background events in a $10^{14}$ EoT, and leaving a large population of interesting events whose features we can study. The BDT analysis leaves 85 MC events, or $< 2$ in $10^{14}$ EoT. Here we note that the BDT was not trained with any of this data; it is simply reevaluated on the new sample of events. The estimated yields shown in the bottom row of the table here approximate the realistic yields in this phase space by including the biasing factor used in the generation of the sample. To further characterize these remaining background events, we note that applying a photon-electron angle cut of more than 3 degrees reduces this count to 3, leading to an estimated yield less than 0.06 at $10^{14}$ EoT.

| **Cut** | **Events** | **Cut** | **Events** |
|---|---:|---|---:|
| All / Acceptance | 12,931,424 | All / Acceptance | 12,931,424 |
| Triggerred | 7,547,806 | Triggerred | 7,547,806 |
| $p_{\text{tagger}} > 5600$ | 7,447,962 | $p_{\text{tagger}} > 5600$ | 7,447,962 |
| $N_{\text{recoil}} = 1$ | 7,215,834 | $\lvert z_0 \rvert < 40$ | 7,215,834 |
| $E_{\text{sum}} < 3500$ | 3,069,020 | Ecal BDT | 19,348 |
| $E_{\text{SumTight}} < 800$ | 197,603 | $N_{\text{straight}} < 3$ | 18,816 |
| $E_{\text{back}} < 250$ | 137,124 | HCal max PE $< 8$ | 85 |
| $N_{\text{hits}} < 70$ | 122,113 | | |
| $\text{RMS}_{\text{shower}} < 110$ | 121,639 | | |
| $E_{\text{cell,max}} < 300$ | 100,700 | | |
| $\text{RMS}_{\text{Layer,hit}} < 5$ | 80,475 | | |
| $N_{\text{straight}} < 3$ | 77,561 | | |
| HCal maxPE $< 8$ | 293 | | |
| Estimated yield at $10^{14}$ EoT | $< 5 \pm 3$ | Estimated yield at $10^{14}$ EoT | $< 2 \pm 1$ |

Table 4.5: Cut and count cutflow (left) and the BDT cutflow (right).

Analyzing this sample using ParticleNet, we quote the results in Table 4.6. Again we note that, similarly to the BDT case, the network did not train on this data. The second and third rows show the kaon yields and the signal efficiencies at the same discriminator value as used for the ECal PN sample in Table 4.4. This already leads to better results than the cut and count and the BDT approach, while it maintains very good signal efficiency. The last three rows show the results for a discriminator value that retains only one kaon event in this sample, leading to an estimated yield of $< 0.02$ for $10^{14}$ EoT (and thus $< 2$ in $10^{16}$ EoT equivalent). In this case, the signal efficiency is still higher than 72% efficiency. This indicates that, while the kaons are a challenging background, our techniques should be suitable to reduce them to a negligible level, even over the full lifetime of the experiments.

| Cut | Background Remaining | Signal Efficiency | | | |
|---|---|---|---|---|---|
| | Kaon enhanced sample | 1 MeV | 10 MeV | 100 MeV | 1000 MeV |
| Preselection | 687882 | 99.5% | 99.8% | 99.8% | 99.8% |
| ParticleNet $> 0.74$ | 24723 | 92.2% | 95.9% | 96.4% | 97.1% |
| HCal Veto (maxPE $< 8$) | 27 | 87.6% | 89.7% | 89.4% | 75.7% |
| Estimated yield at $10^{14}$ EoT | 0.54 | | | | |
| ParticleNet $> 0.85$ | 6279 | 79.6% | 89.7% | 91.6% | 93.6% |
| HCal maxPE $< 8$ | 1 | 75.7% | 83.9% | 84.8% | 72.6% |
| Estimated yield at $10^{14}$ EoT | 0.02 | | | | |

Table 4.6: Cutflow for the kaons using ParticleNet with different discriminator values: the nominal cut (top) and a tighter cut that removes all but one simulated event, corresponding to 0.02 events at $10^{14}$ EoT.

#### 4.3.2.2 Studies of wide-angle bremsstrahlung

In the following study, we consider a class of purely electromagnetic background processes: wide-angle bremsstrahlung (WAB), wide-angle electron scatters (WAS), or their combination (WASB). Bremsstrahlung photons emitted at very large angles may not enter the ECal, spoiling the ability to accurately measure their energy. The Side HCal detector that surrounds the ECal extends LDMX's angular coverage beyond 60 degrees from the beam, is thus a critical instrument in identifying such showers. Wide angle electron scatters can transfer significant energy into the target, leaving a relatively low momentum recoil electron. At larger momentum transfers, such events overlap with the electron-nucleus scattering process studied in Sec. 4.9. WASB events may be doubly-challenging, due to the combined energy loss mechanisms for the recoil electron and a photon that may require special techniques to reconstruct. Because of these qualitative differences with the main background processes considered in Sec. 4.2, we perform a dedicated study based on specialized calculations of this process.

To study these backgrounds, custom models of these interactions were built in MadGraph MG5_aMC-v3.5.4. Three specific categories of wide-angle backgrounds have been studied (assuming an incoming 8 GeV electron beam):

- **WAB** Wide-angle bremsstrahlung, in which the outgoing electron has energy $< 2$ GeV, significantly less than the beam electrons, and the photon is emitted at an angle $> 30°$. This has the potential to fake a signal event if the outgoing photon is not correctly identified.
- **WASB** Wide-angle scatter of the electron, followed by a bremsstrahlung event in which the electron has $< 2$ GeV and the photon with energy $> 0.01$ GeV emitted at an angle $> 30°$. The signature is similar to the WAB type event, however, the bremsstrahlung photon has the potential of being much lower energy than in the cleaner WAB case. Here, energy is transferred to the target nuclei, potentially leading to additional activity in the recoil tracker.
- **WAS** Wide-angle scatter of the electron, there is no outgoing primary photon, but we must be able to correctly identify and measure the outgoing electron, assuming it scatters at an angle $> 30°$. Again, such processes will include a large energy transfer to the target nuclei.

Table 4.7 summarizes the definition of each process within the MadGraph simulation.

Table 4.7: Selection criteria applied to background samples in the MadGraph generator

| Selection | WAB | WASB | WAS |
|---|---|---|---|
| Electron Lab Frame Energy | $< 2$ GeV | $< 2$ GeV | - |
| Photon Lab Frame Energy | $> 2$ GeV | $> 0.01$ GeV | - |
| Electron Angle | - | - | $> 30°$ |
| Photon Angle | $> 30°$ | $> 30°$ | - |

Fig. 4.20 shows, from simulation, where the simulated electron and photon from the wide-angle processes end. The large majority of both particles end in the ECal or the Side HCal, showering in z. The more problematic case is where both the electron and photon (if present) deposit most of their energy in the side HCal.

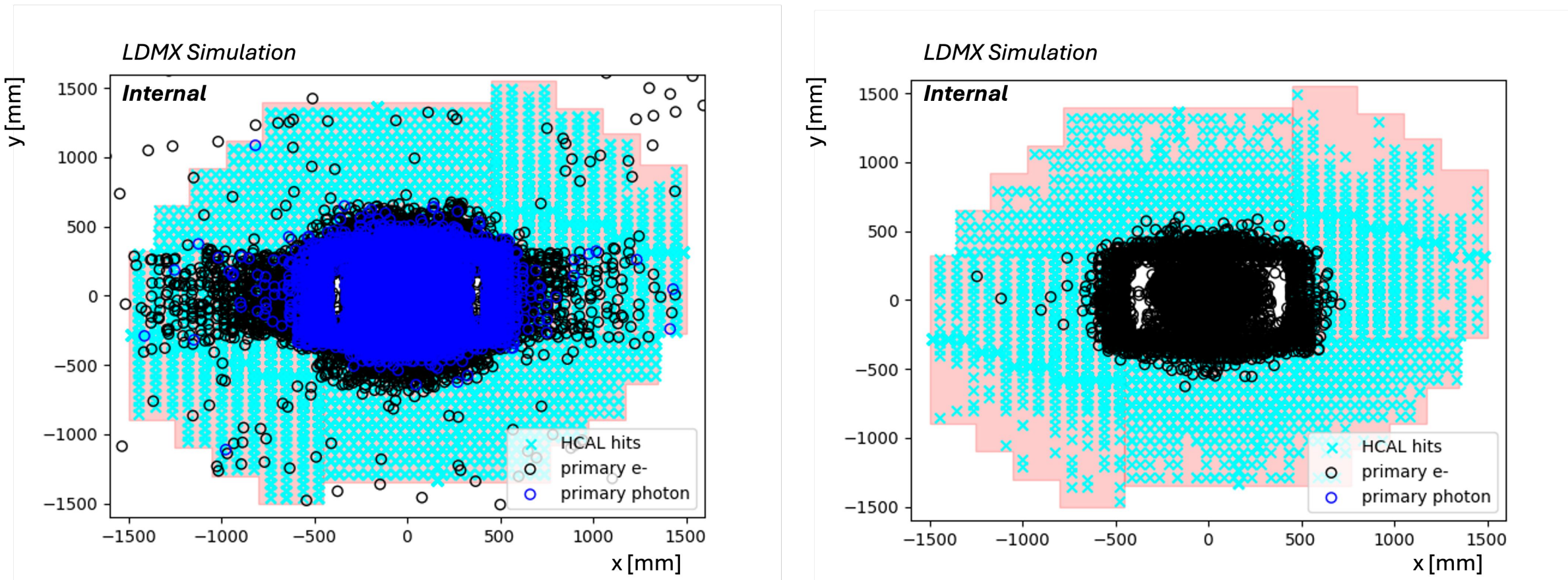

Figure 4.20: End position location for simulated particles from WAB/WASB (left) and WAS (right) events. Both electrons and photons from the two-body processes spread over a wide angle, and many deposit some or all of their energy in the side HCal. Reconstructed HCal hits are indicated, which are mostly a result of secondary processes (regular bremsstrahlung, ionization and photon conversions).

Table 4.8: Yields for wide-angle events when no veto is applied. Cross-sections ($\sigma$) and errors are taken from MadGraph.

| Process | $\sigma$ [pb] | Events per $1 \times 10^{16}$ EoT |
|---|---|---|
| WAB | $364.5 \pm 1.6$ | $5.93 \times 10^5 \pm 35.2$ |
| WASB | $363.12 \pm 1.7$ | $5.90 \times 10^5 \pm 37.4$ |
| WAS | $459456 \pm 510$ | $7.38 \times 10^8 \pm 1.1 \times 10^4$ |

A “WAB-veto" has been developed, including information from the total energy deposited in the HCal and ECal, cluster information in the HCal, as well as information from the recoil tracker. The veto allows for full exclusion of all simulated WAB and WAB-like events.

Figure 4.21 shows the total energy deposited in the ECal and HCal by inclusive wide-angle background events and a combined set of $A'$ events (with $m_{A'} = 1, 10, 100, 1000$ MeV). There are clear differences in the energy deposits in the two calorimeters for the different types of events. In the signal scenarios, the recoil electron leaves most of its energy in the ECal, and little in the Side HCal (or indeed the Back HCal). The backgrounds deposit some energy in the ECal but also a large amount of energy in the Side HCal, which surrounds the ECal at wide angles.

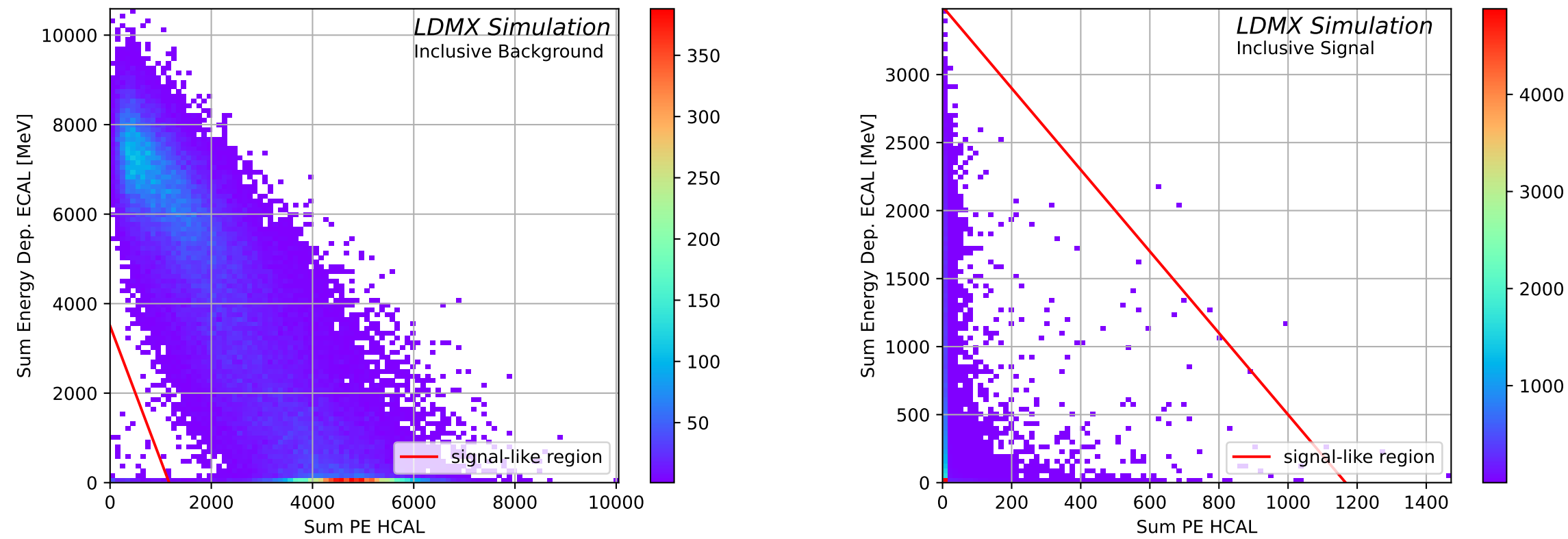

Figure 4.21: The wide-angle backgrounds deposit large amounts of energy in the HCal. The electrons recoiling for an $A'$ signal produce little activity in the HCal but activity in the ECal.

In the Sec. 3.8.7.2, it is shown that close to all wide-angle events produce activity in the side HCal. In wide-angle scatter processes 38 % of events produce some activity in the side HCal and ECal; and $\sim 7$ % of the 2-body events produce no activity in the ECal. This means that the primary particles deposit all their energy in the Side HCal and shower through its layers. Only a small fraction of events completely miss the HCal (Back or Side).

Clustering in the HCal can be employed to distinguish the two-body wide-angle backgrounds from the signal. It may also provide help with distinguishing the higher energy electron shower from wide-angle scatter events. Fig. 4.22 shows examples of the DBScan clustering method applied to WAB events. In cases in which both primary particles enter the side HCal, two clear clusters are apparent. On average, the two-body process provides 2 distinct clusters, however, there are some events that produce more than that. Most signal events produce no clusters. Further enhancements in the clustering algorithms are expected, but for the sake of vetoing wide-angle backgrounds, this is sufficient to assist in discriminating the wide-angle backgrounds from our signal.

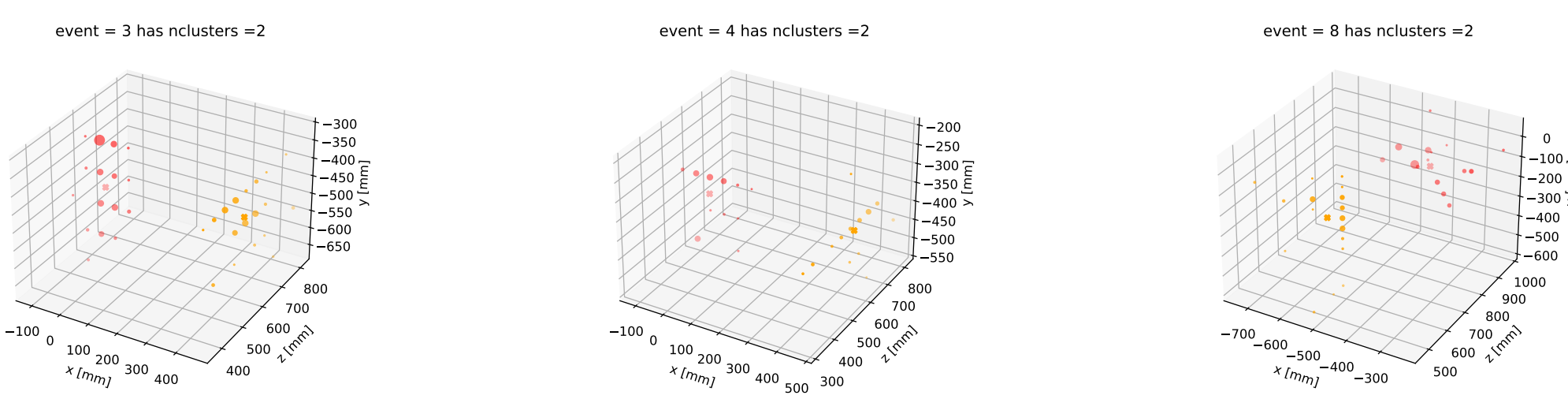

Figure 4.22: Examples of clusters for 2-body processes. The yellow and red hits represent those in the two clusters. The cluster centroid is indicated by the “x" of the same color. The size of the hits is proportional to the energy deposited.

To provide efficient rejection of these backgrounds, a BDT is built and trained on an inclusive set of signals, which contains equal contributions from $m_{A'} = 1, 10, 100, 1000$ MeV and an equal mixture of WAB, WASB, and WAS wide-angle backgrounds. This relatively simple BDT considers the total energy deposit in the ECal and HCal, the number of HCal clusters, and the angular parameters $\phi$ and $\theta$ of the recoil track. These features are compared in Figures 4.23 and 4.24 for the inclusive signal and background distributions.

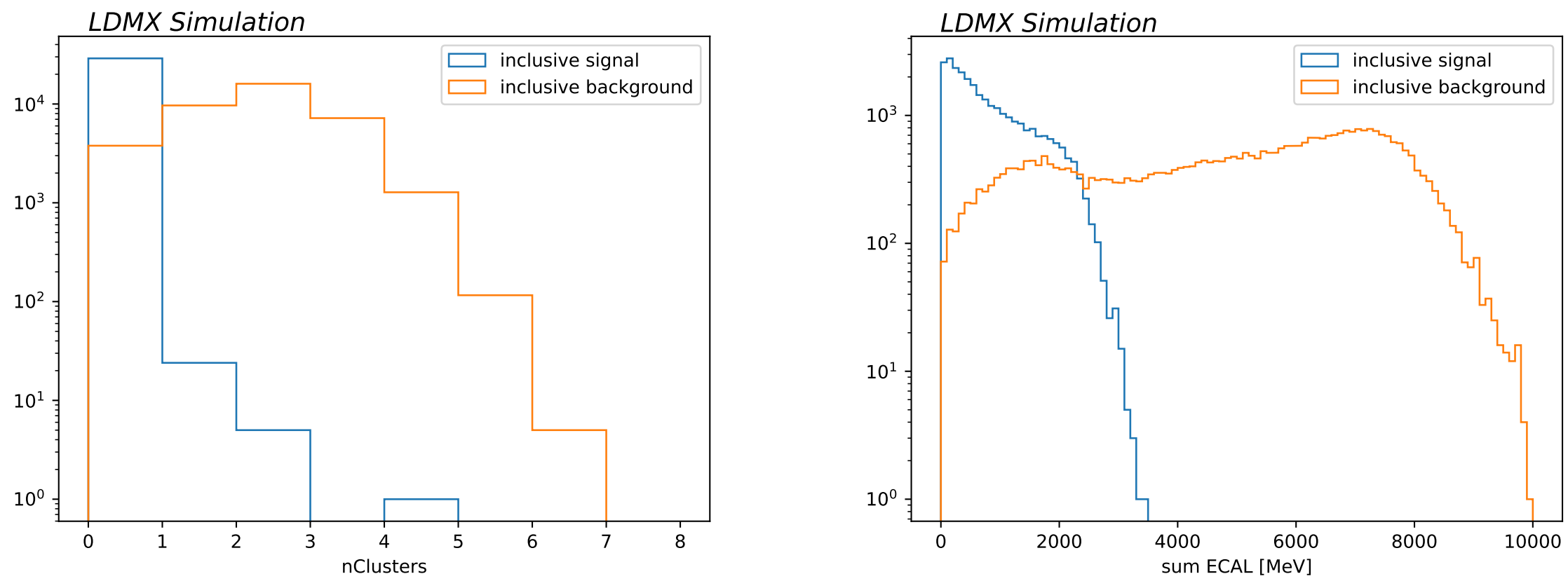

Figure 4.23: The HCal cluster multiplicities and ECal energy sums are compared for the merged wide angle backgrounds and signal samples.

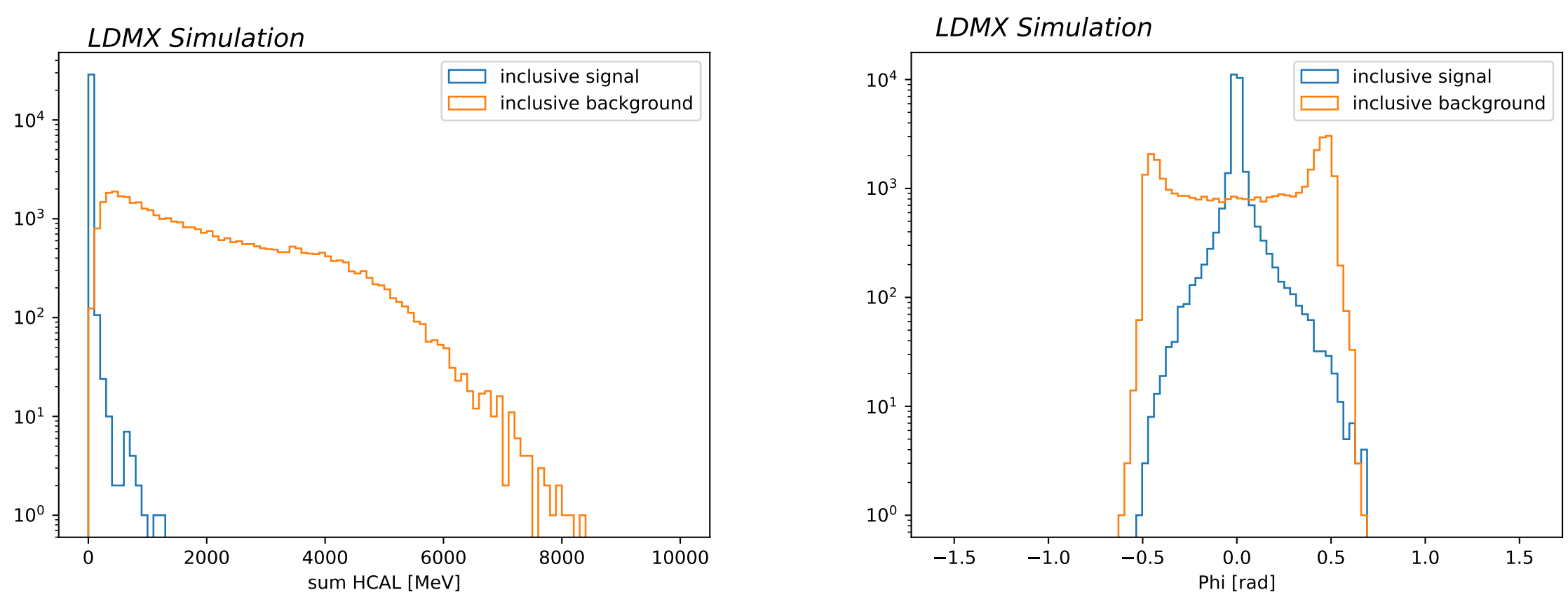

Figure 4.24: The HCal energy sums and recoil track declination angles are compared for the merged wide angle backgrounds and signal samples.

This distilled information from the ECal, HCal, and tracker sub-detectors proves to be powerful in removing wide-angle backgrounds. A BDT trained on only the first three features retains on average 96% of the inclusive signal while reducing the background by more than three orders of magnitude. Introducing recoil tracking allows the correlation of HCal clusters with recoil tracks, exploiting the clear difference in electrons producing wide-angle events and those which have recoiled off of new physics particles of a given mass. The full BDT allows for the rejection of all simulated wide-angle background events, while retaining 100 % of the signal for each mass hypotheses included in this study. From this, we conclude that WAB and WASB can be reduced to a negligible level in the $10^{16}$ EoT-equivalent analysis. Because of the larger cross section, it is possible that rare WAS (i.e. eN scattering) events may require a more sophisticated veto strategy. Here, ample handles may be provided by more detailed cluster information and the forward products of the electrons' hard interactions with target nuclei, many of which are expanded upon in Sec. 4.9.

### 4.3.3 Considerations for multi-electron events

While the STT can be used to efficiently select only those events containing exactly one beam electron, the ability to analyze events with multiple electrons in the same RF bucket has the potential to substantially accelerate the number of EoT collected. LDMX eventually aims to acquire a sample of O($10^{16}$) EoT. To reach this target, higher beam intensity will be required, with multiple electrons per time sample. LDMX will therefore need to have the capability to efficiently analyze events with multiple electrons.

To capitalize on these events while maintaining the ability to separate the DM signal from the background, the single-electron analysis described previously needs to be adapted. Most notably, it requires extending the trigger logic and designing a veto of photonuclear interactions in the calorimeter that is robust to the presence of additional 'pileup' electrons. The various challenges arising from the presence of pileup electrons are addressed below.

**Multi-e Trigger** At trigger level, a triple-space-point tracking algorithm is used to count electrons with a very low fake rate, exploiting positional and timing information from the trigger scintillator system. The fact that the three trigger scintillator pads are laid out to follow a beam electron's expected trajectory allows for fitting a constant to the hit positions (in a suitable coordinate system), which is a necessary constraint for making a track from only three points. Moreover, the time of flight from the first to the last of the three scintillator pads is greater than the time resolution of the system, allowing suppression of tracks formed with delayed hits. Electron counting efficiencies and fake rates for various electron multiplicities are summarized in Fig. 3.39.

The missing-energy trigger will rely on a count of the number of incoming electrons ($n_e$) provided by the trigger scintillator system to determine the expected total energy of the incoming electrons. The missing-energy threshold varies with $n_e$ in order to maximize signal efficiency while keeping the overall trigger rate within the allocated bandwidth. Critically, an overestimate of $n_e$ would lead to an overestimate of incoming energy and thus produce a fake missing energy signal.

Energy thresholds for the multi-track case are calculated by maximizing signal efficiency, under the constraint that the total trigger rate (over all tracks) remains $\leq 1$ kHz. The total trigger rate $R_{tot}$, considering events with up to four tracks, is given by

$$R_{tot} = \sum_{n=1}^{4} \epsilon_n(E_n) R(n;\mu), \tag{4.5}$$

where $n$ indexes the track count, $\epsilon_n(E_n)$ is the fraction of accepted events of track count $n$ as a function of energy threshold $E_n$, and $R(n;\mu)$ is the rate of events with track count $n$ for a given Poisson parameter $\mu$. We then maximize our total signal efficiency $\epsilon_{sig}$, given by,

$$\epsilon_{sig} = \sum_{n=1}^{4} \epsilon'_n(E_n) f(n;\mu), \tag{4.6}$$

where $\epsilon'_n(E_n)$ denotes the fraction of accepted signal events of track count $n$ as a function of energy threshold $E_n$, and $f(n;\mu)$ is the fraction of signal events with track count $n$. The factors $\epsilon_n(E_n)$ and $\epsilon'_n(E_n)$ are functions of the shape of the underlying inclusive and signal distributions. The optimized thresholds correspond to the energies $E_n$ that maximize $\epsilon_{sig}$, under the conservative constraint $R_{tot} = 1$ kHz.

The corresponding energy thresholds and signal efficiencies for a signal mass of 0.1 GeV, with energy summed over 20 ECal layers, are listed in Table 4.9. In these tables, the calculated fraction of bunches corresponds to the quantity $R(n;\mu)$, relative to all bunches, while the fraction of signal is $f(n;\mu)$. Efficiencies for a given $N_{track}$ correspond to the parameter $\epsilon'_n$, while total efficiencies, given in the last row, correspond to $\epsilon_{sig}$. We also calculate energy thresholds and efficiencies for cases where we limit triggering to some number of tracks $N_{track} \leq 4$. In these cases, we suffer a penalty in $\epsilon_{sig}$ to increase $\epsilon'_n$ for a given $n$. The first two configurations in Table 4.9 show optimized energy thresholds for a beam profile of $\mu = 1$, assuming that we trigger on events with a single track and up to four tracks, respectively. We see that triggering on multi-track events increases $\epsilon_{sig}$ from 0.58 to 0.88, while single-track efficiency remains high ($> 90\%$). The last configuration in Table 4.9 shows energy thresholds for a beam profile for $\mu = 2$, where we again see that introducing acceptance to multi-track events does not significantly decrease efficiency for single-track events. Fig. 4.25 summarizes these optimized energy thresholds for each track multiplicity, with vertical solid lines showing missing energy thresholds for $\mu = 1$, and vertical dashed lines for $\mu = 2$. The black curves show trigger rates of background processes. The red curves show signal efficiency for $m_{A'} = 0.1$ GeV, corresponding to the last columns in Table 4.9.

Table 4.9: Optimized energy thresholds for $m_{A'} = 0.1$ GeV, with energy summed in the first 20 ECal layers. Sub-tables show various trigger configurations.

| $\mathbf{N_{track}}$ | **Fraction of Bunches (Signal)** | **Threshold (GeV)** | **Trigger Rate (Hz)** | **Efficiency 0.1 GeV** |
|---|---|---|---|---|
| 1 | 0.393 (0.616) | 4.41 | 1004 | 0.95 |
| 2 | 0.176 (0.284) | 0.00 | 0 | 0.00 |
| 3 | 0.052 (0.084) | 0.00 | 0 | 0.00 |
| 4 | 0.010 (0.016) | 0.00 | 0 | 0.00 |
| Total | 0.632 (1.00) | | 1004 | 0.58 |

(a) $\mu = 1,\ \mathrm{N_{track}} \leq 1$

| $\mathbf{N_{track}}$ | **Fraction of Bunches (Signal)** | **Threshold (GeV)** | **Trigger Rate (Hz)** | **Efficiency 0.1 GeV** |
|---|---|---|---|---|
| 1 | 0.393 (0.616) | 3.01 | 188 | 0.90 |
| 2 | 0.176 (0.284) | 10.79 | 383 | 0.86 |
| 3 | 0.052 (0.084) | 18.54 | 300 | 0.81 |
| 4 | 0.010 (0.016) | 26.25 | 131 | 0.84 |
| Total | 0.632 (1.00) | | 1004 | 0.88 |

(b) $\mu = 1,\ \mathrm{N_{track}} \leq 4$

| $\mathbf{N_{track}}$ | **Fraction of Bunches (Signal)** | **Threshold (GeV)** | **Trigger Rate (Hz)** | **Efficiency 0.1 GeV** |
|---|---|---|---|---|
| 1 | 0.327 (0.370) | 2.19 | 54 | 0.80 |
| 2 | 0.299 (0.349) | 9.74 | 233 | 0.70 |
| 3 | 0.177 (0.208) | 17.15 | 365 | 0.57 |
| 4 | 0.062 (0.073) | 25.20 | 352 | 0.66 |
| Total | 0.865 (1.00) | | 1004 | 0.74 |

(c) $\mu = 2,\ \mathrm{N_{track}} \leq 4$

The trigger efficiency with the threshold cuts indicated in Fig. 4.25 is shown as a function of dark photon momentum for the lightest and heaviest dark photon signal benchmark samples in Fig. 4.26, for simulated electron multiplicities between 1 and 4. The dark photon momentum, as obtained from simulation, is a measure of the “true" missing momentum in the event and removes any impact of misreconstructed energy. The efficiency increases with true missing momentum, and plateaus later with higher electron multiplicity. There is only a weak dependence on the dark photon mass.

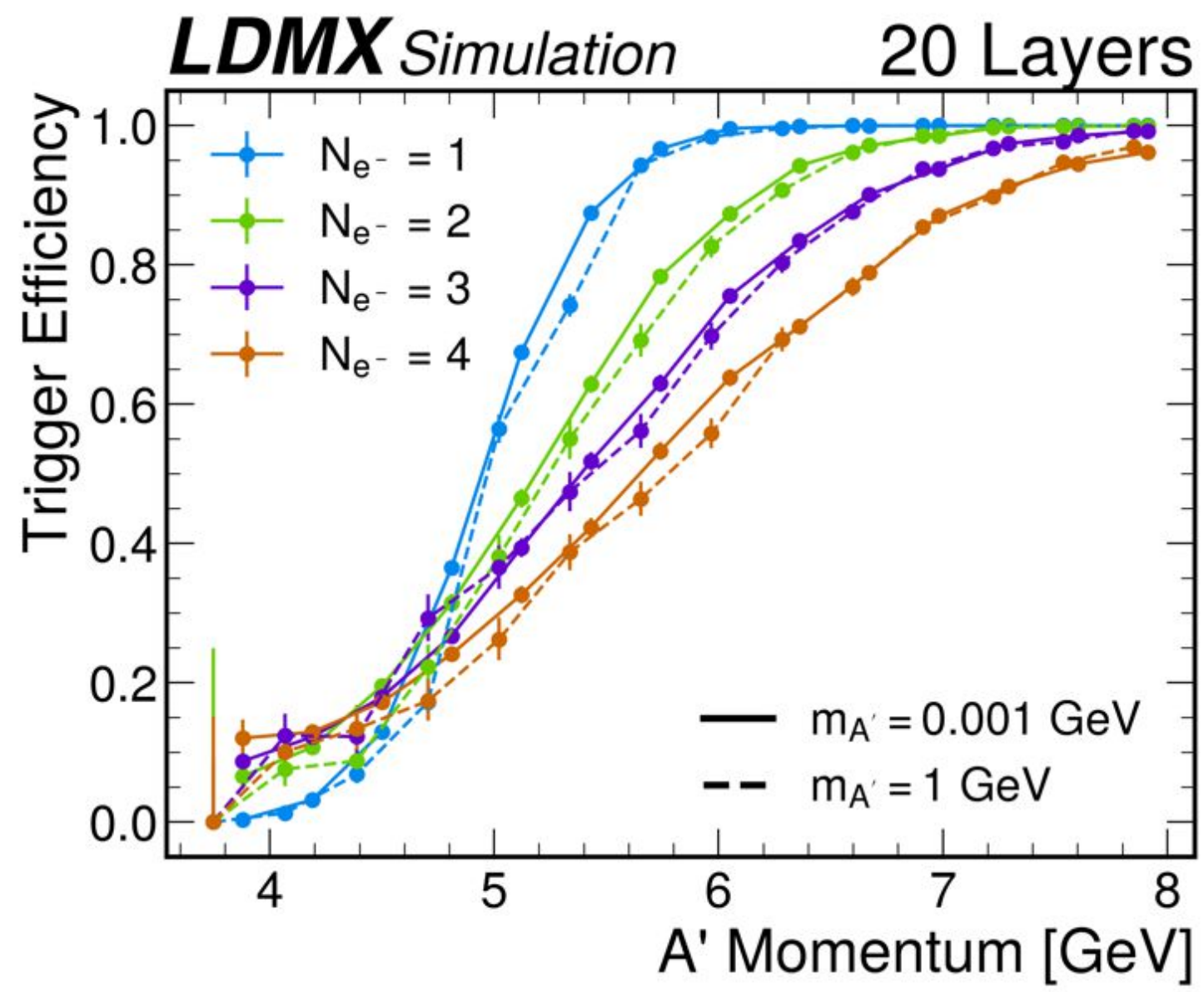

Figure 4.26: Signal trigger efficiency at the optimized thresholds as a function of the momentum carried by the dark photon, for dark photon masses $m_{A'} = 1\,\text{MeV}$ (solid) and 1000 MeV (dashed), shown for 1 to 4 electrons (colors).

**Multi-electron ECal veto** Beyond the task of designing a trigger strategy that maximizes the advantage of a high-current beam, dedicated reconstruction schemes will be required to analyze events with multiple interacting electrons. As the rate of multiple photonuclear interactions occurring in the same time sample is negligible, the task reduces to performing the analysis described in Sec. 4.2.4 in the presence of one or more additional common electron interactions. The major concern in this regard is adapting the ECal veto to account for additional showers. The excellent segmentation of the ECal, together with the large size of the beamspot, enables an optimized multi-electron analysis to perform this task. To demonstrate the ability of the ECal to maintain good signal vs background discrimination in the case of multi-electron events, we have trained a version of the ECal BDT and ParticleNet veto algorithms, described in Sections 4.2.4.2 and 4.3.1.2, respectively, with 2-electron simulations containing an additional pileup electron. Similar to the single-electron ECal veto algorithms, the 2-electron BDT and ParticleNet vetoes are trained using a mixture of four different dark photon mass hypotheses for the signal (0.001 GeV, 0.01 GeV, 0.1 GeV, and 1 GeV) against the ECal PN background. For each simulation sample, hits from the interactions arising from a pileup electron arriving in the same bucket as the one giving rise to the dark bremsstrahlung (for signal) or hard SM bremsstrahlung (for background) are overlaid on the original event using the technique described in Sec. 3.10.3.3.2. Of the two methods, the ParticleNet approach shows better performance, as will be detailed below. This is expected, as ParticleNet is better able to exploit the high-granularity information from the ECal in the presence of additional showers. The ParticleNet method is therefore chosen to be the default ECal veto method at present for the multi-electron case; however, we also present results for the 2-electron BDT for comparison.

The 2-electron ParticleNet algorithm is trained with the 2-electron samples using a very similar approach as for the single-electron ParticleNet (described in Sec. 4.3.1.2). The preselection criteria applied to improve the computational efficiency of the training are adjusted for the 2-electron case as follows: total ECal hits $< 300$ and total deposited energy in ECal $<$ 14000 MeV. These requirements remove less than 0.15% of the signal and reduce the size of the background training sample to a manageable level, while ensuring that the training concentrates on the most challenging background topologies. In addition, all events are required to pass the trigger selection described above. The combined set of preselection requirements rejects 90% of the ECal PN background while retaining 76-85% of the signal. The performance of the 2-electron ParticleNet is shown in the ROC curves displayed in Fig. 4.27. While the performance does not perfectly match the level attained in the single-electron case (shown in Fig. 4.19), the achieved discrimination is still promising and supports the case that with further optimization, multi-electron events can be included in the DM search.

For a more intuitive understanding of how higher-level ECal observables evolve from the single-electron case, Fig. 4.28 shows a comparison of distributions of some of the ECal features that provide the best discrimination between signal and background, for 2-electron events. Even with the presence of an additional inclusive electron, these variables still display good separation between the DM signal and background. To validate the performance of the ParticleNet-based veto, and also to facilitate comparisons with the single-electron analysis, a version of the BDT-based ECal veto is trained for the 2-electron case. As the statistical power of the 2-electron simulation samples available at this time for training is limited, the 2-electron BDT is trained with a reduced subset of the input features deployed for the single-electron BDT. Specifically, the following features are used:

- the total energy deposited in the ECal
- the total energy deposited with a tight isolation requirement
- the total energy deposited in the back of the ECal
- the maximal energy deposited in a single cell
- the number of read-out hits
- the average number of layers hit and its standard deviation
- the deepest layer hit
- the shower size and its standard deviation, in $x$ and $y$ dimensions

In addition, variables defined using the radii of containment, computed as defined in Sec. 4.2.4.2, are used. They are computed using hits in the concentric containment regions defined by integer multiples (1 through 5) of the radii of containment computed for each ECal layer, around the projected electron and photon trajectories. Here, the projected electron trajectory is calculated for the electron undergoing the signal or background process of interest (i.e., the dark bremsstrahlung, or a hard SM bremsstrahlung giving rise to a photonuclear interaction), using truth information. Ultimately, this would be determined based on recoil tracking information. The total energy calculated for each of the 5 electron containment regions and the 5 photon containment regions, as well as the total energy, total number of hits, and energy-weighted standard deviations of the $x$- and $y$- positions of hits in each of the regions of the ECal defined as being outside both the electron and photon containment regions for a given multiple of the containment radii, are used as inputs to the BDT.

The distribution of the BDT output for signal and background events is shown in Fig. 4.29 (left), while the ROC curves showing the performance of the BDT are displayed in Fig. 4.29 (right). For a background rejection of $10^{-4}$ after a trigger preselection, the BDT selection provides a signal efficiency of $38 - 63\%$.

While there is still a lot of room to optimize the analysis for multi-electron events, the observation of the performance of the ECal veto methods adapted from the single-electron analysis to the 2-electron case supports the feasibility of LDMX to operate efficiently with multiple electrons per bucket.

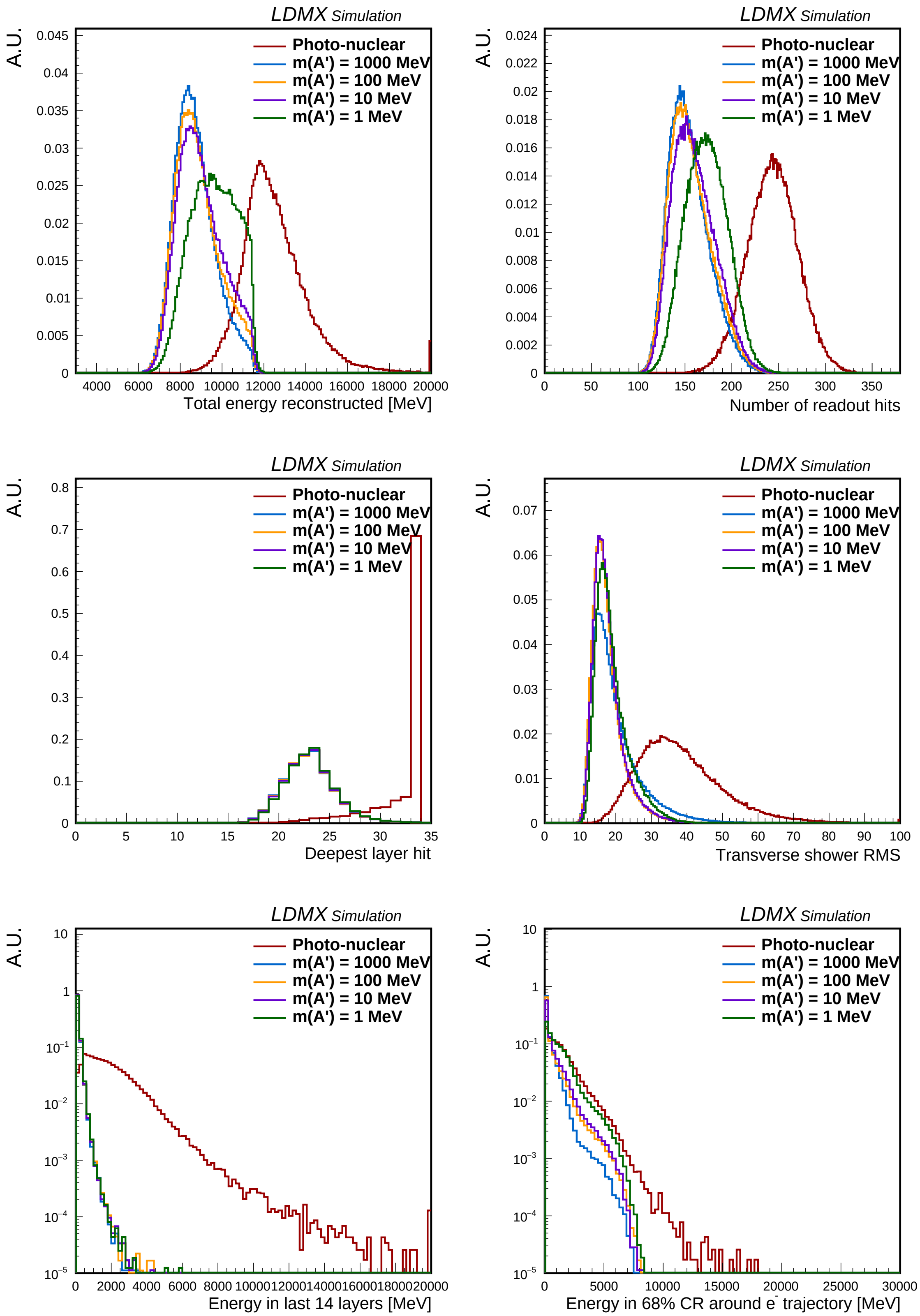

Figure 4.28: Distribution of selected ECal veto BDT input variables in 2-electron ECal PN and signal events. From top left to bottom right: total energy deposited in the ECal, number of ECal read-out hits, deepest ECal layer hit, ECal shower RMS, total energy in the last 14 ECal layers, and energy in the first ECal containment region around the projected electron trajectory.

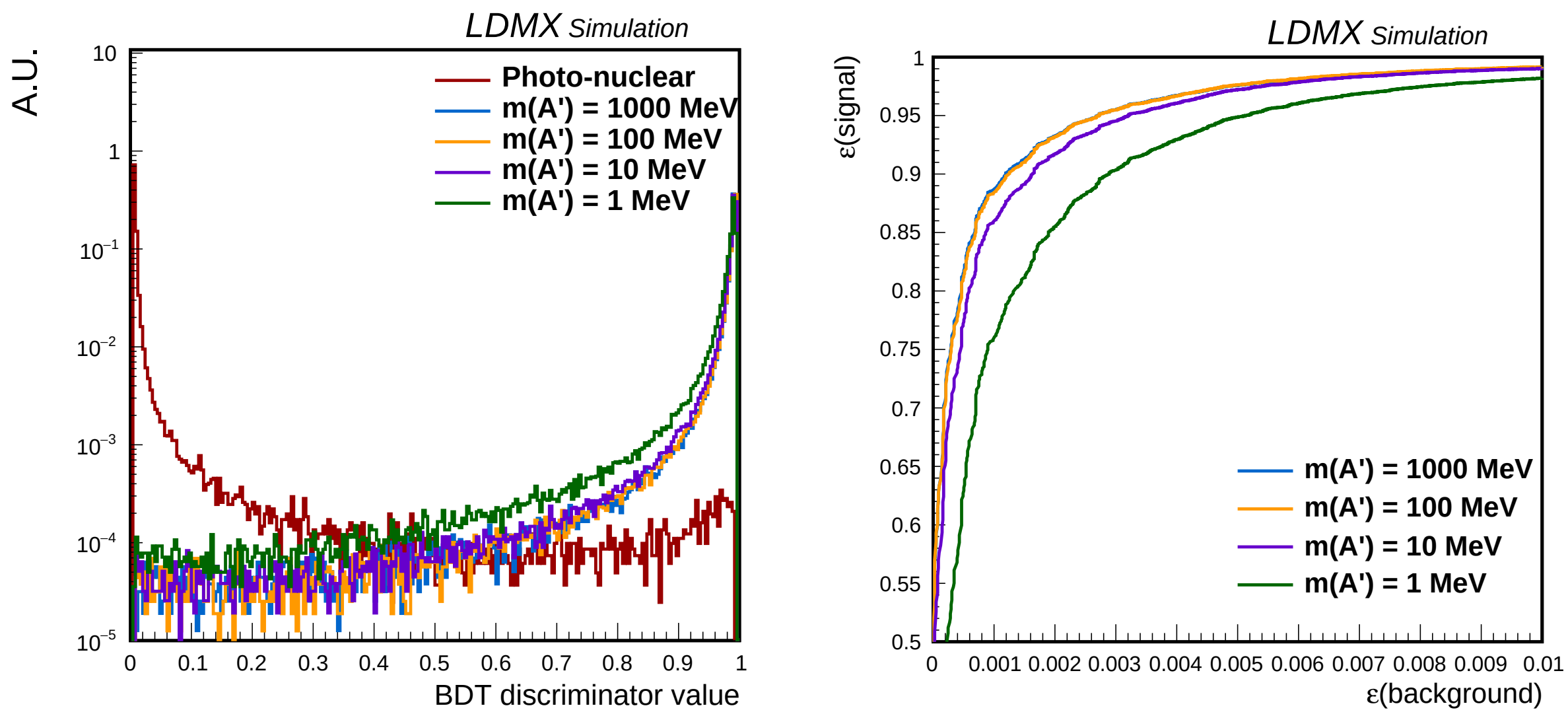

Figure 4.29: Distribution of ECal veto BDT output in 2-electron ECal PN and signal events (left), and ROC curves showing the BDT performance (right).

## 4.4 Ultimate outlook for the missing-momentum search at LDMX

Our motivating goal in designing LDMX is to realize a 100-1000-fold increase in sensitivity to DM in the MeV-to-GeV mass range. This ultimate aim sets physics requirements on the detector design; in short, we must develop the capability to perform a near-background-free search while retaining high efficiency for potential DM signals. Sec. 3.2 laid out these requirements in detail; Sections 3.3-3.10 described the experimental sub-systems developed to meet these challenges; Sections 4.2 and 4.3 described simulation-based studies that establish the ability of this detector to meet our ultimate physics goal.

Here we briefly summarize the key results of these studies. The vast majority of background events leave a purely electromagnetic signature in the detector apparatus. Electrons that pass through the target without interacting, bremsstrahlung, and trident events are filtered away by the missing energy trigger and similar offline cuts. The ECal energy is bolstered by efficient track-finding in the recoil tracker (for tridents) and a Side HCal with broad angular coverage (wide-angle scatters and bremsstrahlung).

Sec. 4.2 focused on demonstrating how photonuclear interactions and muon conversions could be removed with minimal loss in signal. Analyzing samples equivalent to $10^{14}$ or more EoT, the BDT-based analysis achieves 50% signal efficiency while rejecting all but two PN events. (This extends our previously-published methods [134, 135] with complete track reconstruction and the latest detector geometry). By using a more powerful Graph Neural Network discriminant, the same background is reduced to a single event with signal efficiencies from 76-89%.

These studies of inclusive PN production processes with $10^{14}$ EoT samples are further reinforced by exclusive simulations of the most challenging background processes at the level of $10^{16}$ and beyond. Sec. 4.3.2.1 showed that PN reactions enhanced in short-lived kaons decaying to energetic neutrinos could be removed to the level of part-per-$10^{16}$ while retaining $> 70\%$ DM efficiencies. PN reactions that dominantly yield energetic neutrons were shown to be reduced to a similar level in Sec. 3.8.7.1. While all of these studies use simulations with GEANT4, it was demonstrated in Sec. 4.1.2.1 that other simulation programs predict comparable yields of challenging final states. Sec. 4.3.3 demonstrated the feasibility of these analyses in multi-electron events, enabling the rapid collection of a $10^{16}$ EoT equivalent sample together with a thicker aluminum target.

Taken together, these studies demonstrate the capability of LDMX to perform an efficient DM search with negligible backgrounds. We further note that nowhere in the above analysis has the recoil electron $p_{\mathrm{T}}$ spectrum been exploited to remove background. As a consequence, we hold this final $p_{\mathrm{T}}$ discriminant as an 'ace up our sleeve' that can always be played in the event that an unexpected background arises in the real data. We close the discussion of the missing momentum search by considering the possibility of a DM discovery at LDMX, and how this recoil $p_{\mathrm{T}}$ distribution may be employed to characterize the signal.

### 4.4.1 Signal characterization in case of an observed excess

To this point, we have shown that we can reduce our backgrounds to a handful of events, while maintaining excellent signal efficiencies. In the case of an observed excess, the natural next steps will be to determine whether or not the events that survived the background vetoes are consistent with a potential DM signal and, if they are, to attempt to extract information about the dark mediator. Given the good momentum and energy resolution of the apparatus and the intentionally minimal $p_T$ biases of the veto algorithms, the missing transverse momentum at the target, $p_T$, and total ECal energy, $E_{\mathrm{ecal}}$, contain considerable information about the dark mediator mass $m_{A'}$. Signal events have a characteristic $p_T$ "kick" that depends on $m_{A'}$, making the $p_T$ spectrum an effective tool to check whether an excess could be produced by dark bremsstrahlung and to then use it together with the $E_{\mathrm{ecal}}$ distribution to reconstruct the dark mediator mass itself.

Fig. 4.30 shows the $p_T$ distributions before and after cuts, and their ratio for 1 MeV signal (left), and 1000 MeV signal (right) with a normalized logarithmic scale. In the ratio inset, we can see that the cuts do not introduce any $p_T$ bias. The BDT cut value was chosen so that we have 10 background events left, however, other choices do not introduce a bias either.

We perform a non-parametric curve fitting procedure for the signal $p_T$ spectrum. However, we need a mechanism that would generate a $p_T$ distribution for any arbitrary choice of $m_{A'}$. We start with simulating roughly, 40 different mass points ranging from 0.1 MeV to 1000 MeV. We then utilize a Gaussian Process

Regression (GPR), which will estimate the heights of the probability density function given a specific $p_T$ value and, most importantly, the mass $m_{A'}$. This fitting includes systematics effects such as smearing the $p_T$ distribution with a Gaussian of 5 MeV width. Furthermore, we performed checks by using mass points that were not included in the GPR training, and the results prove the robustness of the GPR fitting method.

The number of observed DM events in an experiment is directly proportional to the DM production cross section, which is proportional to the coupling strength $y$ as shown in Eq. 4.7 [147]. This proportionality can be used to reconstruct the coupling strength.

$$\sigma v\left(\chi\chi \to A'^{*} \to ff\right) \propto \epsilon^2 \alpha_D \frac{m_\chi^2}{m_{A'}^4} = \frac{y}{m_\chi^2}, \quad y \equiv \epsilon^2 \alpha_D \left(\frac{m_\chi}{m_{A'}}\right)^4. \tag{4.7}$$

The number of expected signal events depends on the luminosity $L$, the dark matter mass, and our veto efficiency $\varepsilon$. This can be summarized in Eq. 4.8.

$$N(y) = \varepsilon L \sigma(y_{\text{ref}}) \left(\frac{y}{y_{\text{ref}}}\right) \tag{4.8}$$

Assuming that the kinematics of the recoil electron do not depend heavily on the spin of the DM particle, the only significant difference between models comes from the coupling strength. The expected number of signal events for $4 \times 10^{14}$ and $1 \times 10^{16}$ EoT for each model are shown in Fig. 4.31.

Now knowing the event yields, we can apply a statistical treatment and construct the likelihood as a function of mass for a given observation as defined in Eq. 4.9, where $m$ is the mass, the $n_{\text{sig}}$ is the number of signal events, $n_{\text{bkg}}$ is the number of background events, $\mathbf{x}$ is the observed data (i.e. the mixture of signal and background).

$$L(m, n_{\text{sig}}|\mathbf{x}) = \text{Pois}\left(n_{\text{sig}}|\langle n_{\text{sig}}\rangle(m)\right) \cdot \text{Pois}\left(n - n_{\text{sig}}|\langle n_{\text{bkg}}\rangle\right) \cdot \prod_{i=1}^{n} \left(n_{\text{sig}} p_{\text{sig}}(x_i|m) + (n - n_{\text{sig}}) p_{\text{bkg}}(x_i)\right). \tag{4.9}$$

We then run a maximum likelihood estimation to find the parameter ( $\hat{m}$ and $\hat{n}_{\text{sig}}$) that maximizes the likelihood by minimizing the negative log likelihood.

After running the ensemble tests using the methods outlined above, we can compute the average reconstructed mass and coupling strength of the ensemble. Further, by calculating the covariance matrix of mass and coupling and its eigenvalue/vector, we can define the 68% and 95% $(1\sigma, 2\sigma)$ confidence ellipses on the mass-coupling parameter space.

Utilizing the GPR and the defined likelihood in Eq. 4.9, we find the tabulated reconstructed mean and standard deviation in Table 4.10 for $4 \times 10^{14}$ EoT and Table 4.11 for $1 \times 10^{16}$. We simulate masses at the standard $m_{A'} = 1\,\text{MeV}, 10\,\text{MeV}, 100\,\text{MeV}$ but also at mass points not used in the GPR training. Further, the exclusion of $m_{A'} = 1000\,\text{MeV}$ is purely from the lack of expected number of signal events we expect to observe, at $1 \times 10^{16}$ EoT, we expect $\langle n_{\text{sig}}\rangle(1000\,\text{MeV}) \approx 1$. These results show that even in the presence of background events, ranging from a handful to dozens, we have excellent mass reconstructions abilities.

| | $\langle n_{\text{bkg}}\rangle = 0$ Events | | $\langle n_{\text{bkg}}\rangle = 3$ Events | | $\langle n_{\text{bkg}}\rangle = 10$ Events | | $\langle n_{\text{bkg}}\rangle = 20$ Events | |
|---|---|---|---|---|---|---|---|---|
| Simulated Mass | $\hat{m}$ [MeV] | $\sigma_{\hat{m}}$ [MeV] | $\hat{m}$ [MeV] | $\sigma_{\hat{m}}$ [MeV] | $\hat{m}$ [MeV] | $\sigma_{\hat{m}}$ [MeV] | $\hat{m}$ [MeV] | $\sigma_{\hat{m}}$ [MeV] |
| 2.818 MeV | 2.955 | 0.667 | 2.794 | 0.554 | 2.872 | 0.628 | 3.091 | 0.583 |
| 10.000 MeV | 10.110 | 3.670 | 10.285 | 4.124 | 10.363 | 4.611 | 9.086 | 3.373 |
| 28.184 MeV | 30.331 | 6.224 | 30.805 | 5.763 | 26.671 | 4.418 | 30.500 | 7.392 |
| 100.000 MeV | 98.359 | 17.680 | 105.051 | 23.3822 | 107.728 | 21.843 | 95.945 | 31.452 |

Table 4.10: Simulated and reconstructed mass value, and their uncertainties for $4 \times 10^{14}$ EoT.

Data collected at $1 \times 10^{16}$ EoT could be used to distinguish between models, as shown in Fig. 4.32. This figure shows the comparison of simulated and reconstructed masses and couplings, including the mean and the $1\sigma$ (solid ellipses) and $2\sigma$ (dashed ellipses) confidence ellipses. We observe that the confidence ellipses do not have overlaps between models and we have very good sensitivity in distinguishing the three different models. As expected, the Pseudo-Dirac DM is the most challenging model, but the reconstructed values are still very well within the simulated values.

| | For $\langle n_{\text{bkg}} \rangle = 0$ Events | | $\langle n_{\text{bkg}} \rangle = 75$ Events | | $\langle n_{\text{bkg}} \rangle = 100$ Events | | $\langle n_{\text{bkg}} \rangle = 500$ Events | |
|---|---|---|---|---|---|---|---|---|
| Simulated Mass | $\hat{m}$ [MeV] | $\sigma_{\hat{m}}$ [MeV] | $\hat{m}$ [MeV] | $\sigma_{\hat{m}}$ [MeV] | $\hat{m}$ [MeV] | $\sigma_{\hat{m}}$ [MeV] | $\hat{m}$ [MeV] | $\sigma_{\hat{m}}$ [MeV] |
| 1.000 MeV | 0.995 | 0.041 | 0.997 | 0.054 | 0.985 | 0.040 | 1.001 | 0.053 |
| 2.818 MeV | 2.814 | 0.102 | 2.825 | 0.87 | 2.844 | 0.098 | 2.832 | 0.131 |
| 10.000 MeV | 9.918 | 0.695 | 9.853 | 0.803 | 9.776 | 0.737 | 9.885 | 0.615 |
| 28.184 MeV | 28.700 | 1.250 | 28.787 | 1.184 | 29.050 | 1.113 | 28.594 | 1.646 |
| 100.000 MeV | 99.054 | 3.773 | 101.704 | 4.895 | 98.289 | 4.138 | 100.929 | 5.001 |
| 223.872 MeV | 222.772 | 6.017 | 224.855 | 6.737 | 226.507 | 7.033 | 227.307 | 8.629 |
| 446.684 MeV | 449.423 | 13.629 | 447.584 | 13.505 | 445.767 | 15.825 | 454.856 | 17.372 |

Table 4.11: Simulated and reconstructed mass value, and their uncertainties for $1 \times 10^{16}$ EoT.

## 4.5 Early Dark Matter Searches using the ECal as a target

One of the primary strengths of the LDMX detector design is its ability to use the tagging and recoil tracker system to reject a large number of background events by separating a nominal beam with non-standard energy loss from a nominal beam with standard energy loss or even a low-energy beam. Moreover, the tagging and recoil tracker system gives LDMX the potential to further suppress backgrounds or potentially study DM properties by studying the transverse momenta of electrons recoiling from dark-bremsstrahlung candidate events.

While this design is optimal for a large number of EoT, the strategy has some limitations for a low EoT data run. To limit the effects of multiple scattering, the baseline detector configuration requires a thin ($\approx 0.1\,X_\circ$) target. As such, in the earliest stages of LDMX before a large number of total EoT will be collected, a different analysis strategy and detector configuration may be optimal to probe new territory in the $y$–$m_\chi$ plane. The alternative strategy would ignore the dedicated target inside the tracker volume and instead use the ECal as an active target. Using the ECal as Target (EaT) for this early-running phase increases the potential number of dark-bremsstrahlung events and, as long as backgrounds can be suppressed relative to signal, allows a stronger search capability for a fixed EoT.

The prospective sensitivity of an EaT analysis channel was studied in Ref. [148] and is briefly summarized in the remainder of this section. simulation technique (see Sec. 3.10.3.3.3 for further details). The target $10^{13}$ EoT was chosen since it can be collected in an $O$(week) timescale, can be analyzed without needing to consider the rare charged-current backgrounds, and allows us to reach a scale where specific thermal relic models predict enough events to allow for their potential discovery ($\gtrsim 5$ – Fig. 4.33). For the purpose of this study, both 4 GeV and 8 GeV beam energies were considered, and the updated dark bremsstrahlung simulation technique described in Sec. 3.10.3.3.3 was adopted.

Comparing various signal hypotheses against the combined photo-nuclear and electro-nuclear background (here called "Enriched Nuclear") and photon conversion into muons ("Di-Muon") shows that a simple missing energy cut in the ECal, a requirement on the size of the ECal shower, and a maximum activity cut in the HCal eliminates enough background events to then reach into previously-unexplored territory. Additionally, the EaT analysis channel utilizes the same trigger as the nominal missing-momentum channel, but an orthogonal requirement within the recoil tracker. This allows the EaT analysis channel to easily strengthen any DM search analysis that LDMX does in the future. Table 4.12 shows the cutflow of this preliminary analysis with the selected events on the basis of reconstructed ECal energies and shower shapes. Fig. 4.34 shows these distributions along with the edges of the final analysis bins that are used to perform the upper limit estimate.

Since EaT is expected to be the first physics analysis done by LDMX, we also incorporate more realistic techniques for the analysis of real data. Namely, we fit the background distribution with a functional form constrained by a control region (in this case, a simple exponential function constrained by the region between the upper analysis bin and the trigger threshold) and include estimates of the systematic uncertainties of the analysis. The final background prediction, systematic uncertainties, and signal efficiencies for each of the three analysis bins are then used to perform a 3-bin expected exclusion estimate using the CLs criterion with the modified profiled likelihood ratio as a test statistic and using a lognormal model for nuisance parameters affecting the yields within the COMBINE[149] statistical framework. This limit on the signal yield is then scaled to a limit on $y$ using an estimate of the expected production rate of dark photons within the EaT analysis channel, which is the systematic uncertainty with the largest effect on the resulting limit. Fig. 4.35 shows these limits as a function of the DM candidate mass $m_\chi$ and demonstrates that this analysis channel can explore new phase space in either beam case.

| Analysis Stage for 4 GeV Beam | Background Event Yield | Signal Efficiency (%) | | | |
|---|---|---|---|---|---|
| | | 1 MeV | 10 MeV | 100 MeV | 1 GeV |
| ECal Trigger ($E_{20} < 1.5$ GeV) | $5.11 \times 10^7$ | 58 | 67 | 71 | 83 |
| Tracker Requirement ($E_e > 3.5$ GeV) | $4.60 \times 10^7$ | 52 | 60 | 64 | 75 |
| ECal Energy ($E_{\text{ECal}} < 1.1$ GeV) | $1.95 \times 10^6$ | 32 | 43 | 48 | 65 |
| $\max(\text{PE}_{\text{HCal}}) < 10$ | $1.15 \times 10^3$ | 31 | 42 | 47 | 62 |
| RMS Event Size < 20 mm | 126 | 25 | 33 | 37 | 30 |

| Analysis Stage for 8 GeV Beam | Background Event Yield | Signal Efficiency (%) | | | |
|---|---|---|---|---|---|
| | | 1 MeV | 10 MeV | 100 MeV | 1 GeV |
| ECal Trigger ($E_{20} < 3.16$ GeV) | $6.78 \times 10^7$ | 66 | 74 | 79 | 89 |
| Tracker Requirement ($E_e > 7$ GeV) | $6.10 \times 10^7$ | 59 | 67 | 71 | 80 |
| ECal Energy ($E_{\text{ECal}} < 2.76$ GeV) | $6.88 \times 10^6$ | 47 | 57 | 62 | 76 |
| $\max(\text{PE}_{\text{HCal}}) < 10$ | 31.8 | 45 | 55 | 60 | 73 |
| RMS Event Size < 20 mm | 7 | 39 | 47 | 50 | 47 |

Table 4.12: Yields are summarized at each stage of the analysis, comparing the total background and various signal hypotheses for the simple cuts used in this analysis. The event yield for the background sample is calculated using the event weights and represents the number of events out of $10^{13}$ EoT equivalent. The signal efficiency is relative to the total number of simulated events. The efficiency and event yield values on a given row are reported *after* the analysis stage of that row. The first table corresponds to analysis of the 4 GeV beam, and the second table to the 8 GeV beam.

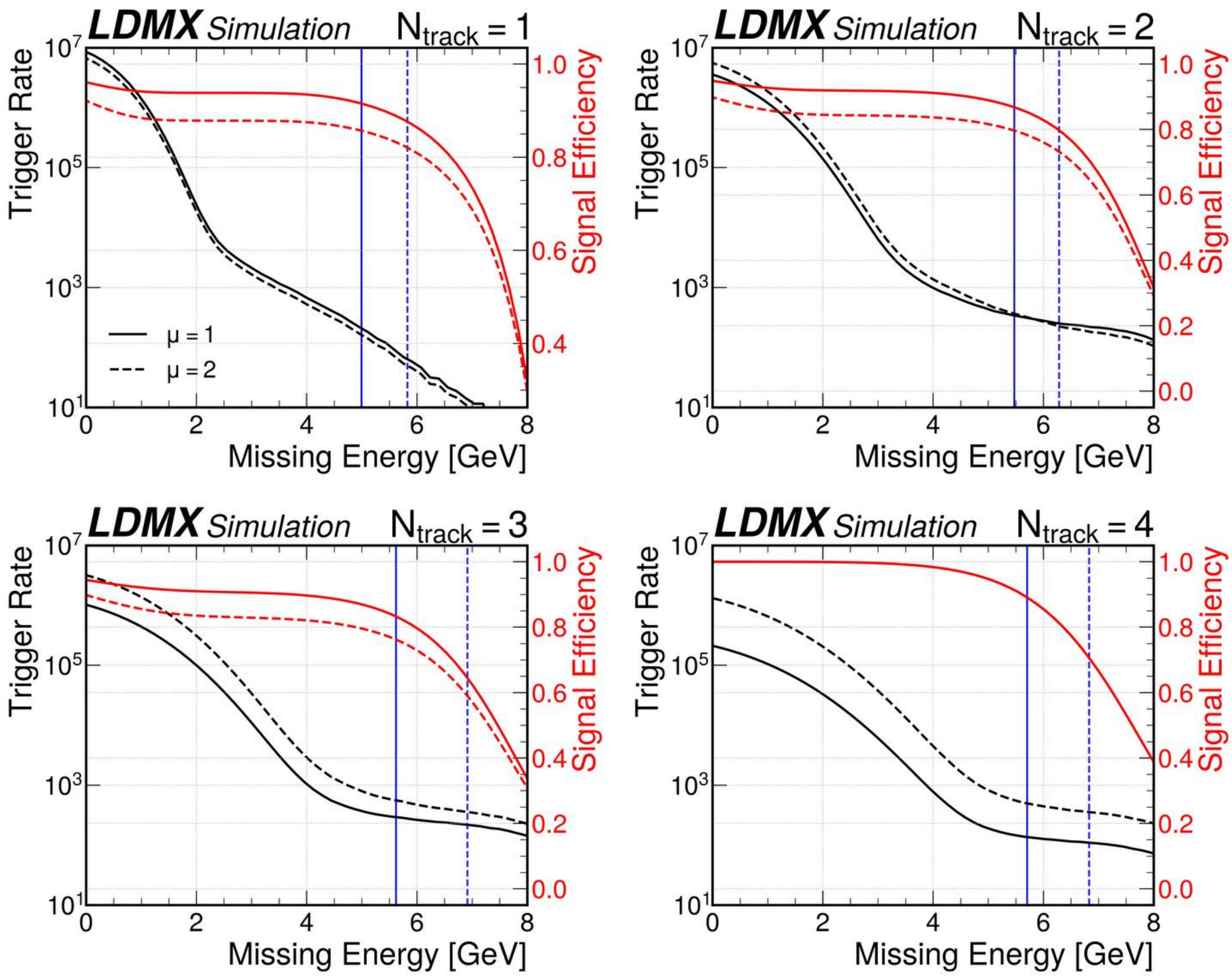

Figure 4.25: The trigger rate of accepted background events (black) is shown as a function of the total ECal energy, for events with one (top left), two (top right), three (bottom left), and four (bottom right) reconstructed tracks in the TS. Each sample of simulated events contains a Poisson-distributed number of simulated electrons in each time-sample with an average of either $\mu = 1$ (solid) or 2 (dashed). The total trigger rate is obtained by adding each $N_{\text{track}}$ category. The efficiency for a dark photon signal (red) with $m_{A'} = 0.1\,\text{GeV}$ is shown on the right axis for the same total ECal energy requirement. The vertical blue lines indicate optimized trigger thresholds for $\mu = 1$ (solid) and 2 (dashed).

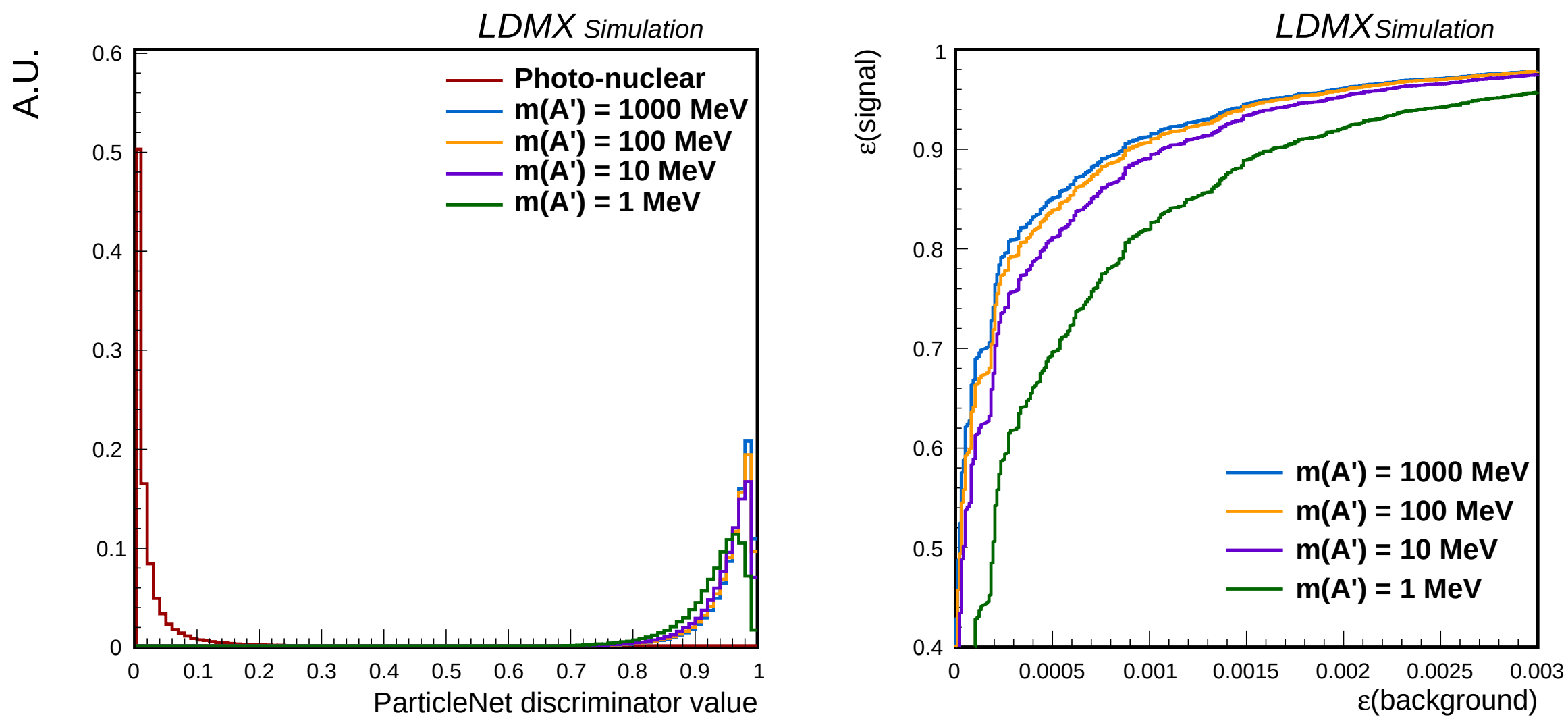

Figure 4.27: Distribution of ECal ParticleNet output in 2-electron ECal PN and signal events (left), and ROC curves showing the performance of the ParticleNet ECal veto (right) in 2-electron events passing the training and trigger preselections.

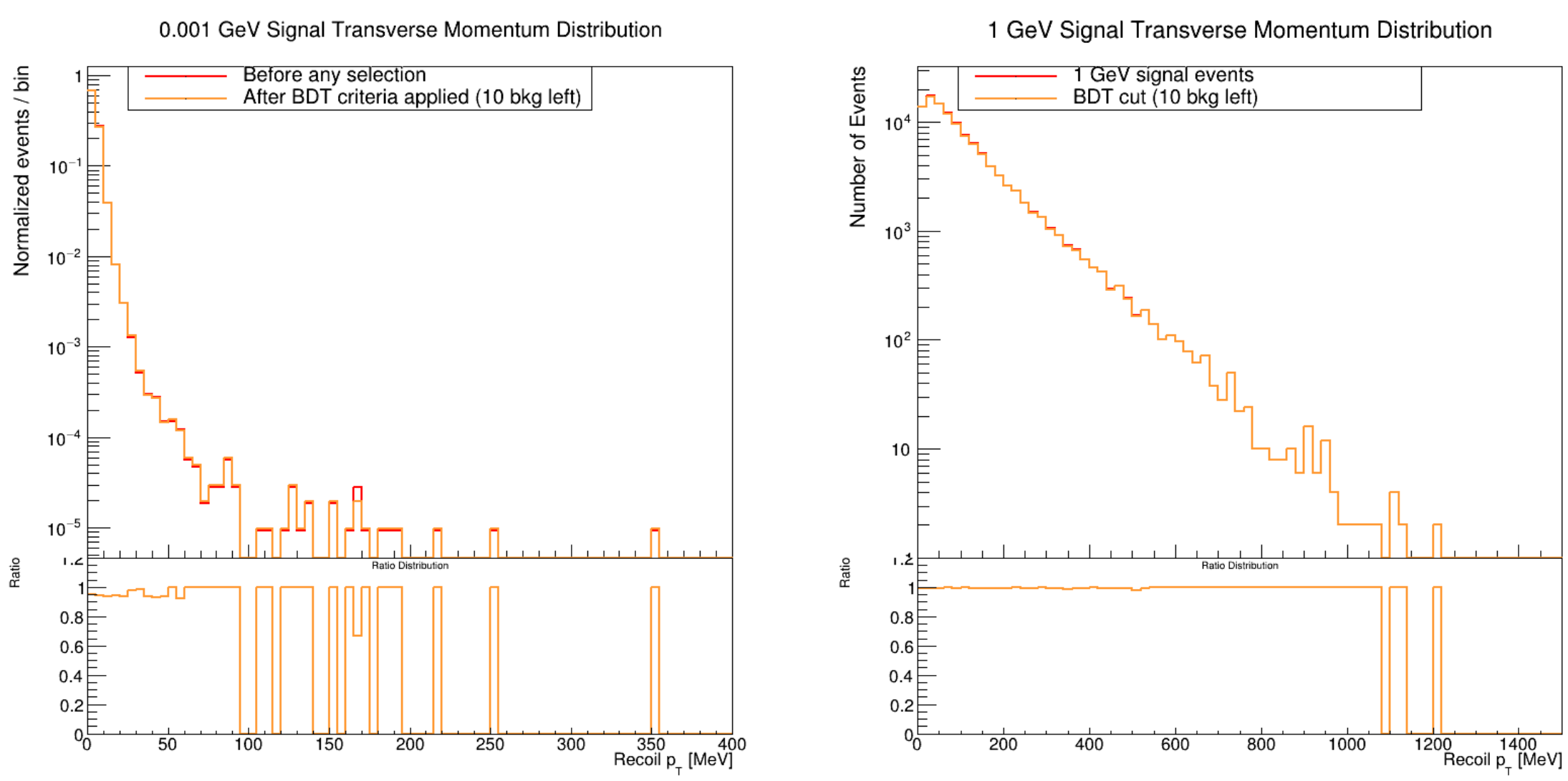

Figure 4.30: The $p_T$ distributions before and after cuts, and their ratio for 1 MeV signal (left), and 1000 MeV signal (right) with normalized logarithmic scale.

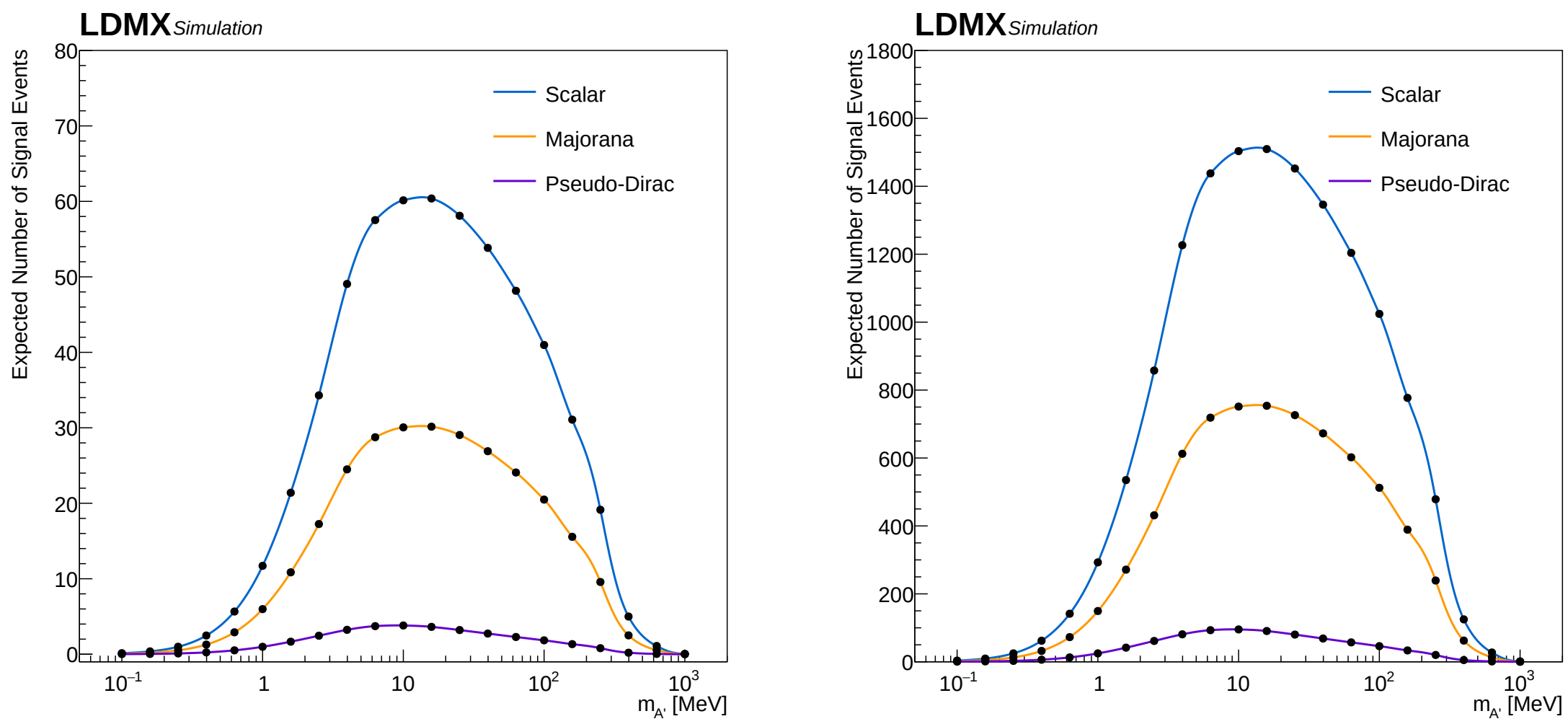

Figure 4.31: The number of expected signal events after the cuts for $4 \times 10^{14}$ (left) and $1 \times 10^{16}$ EoT (right) for each model

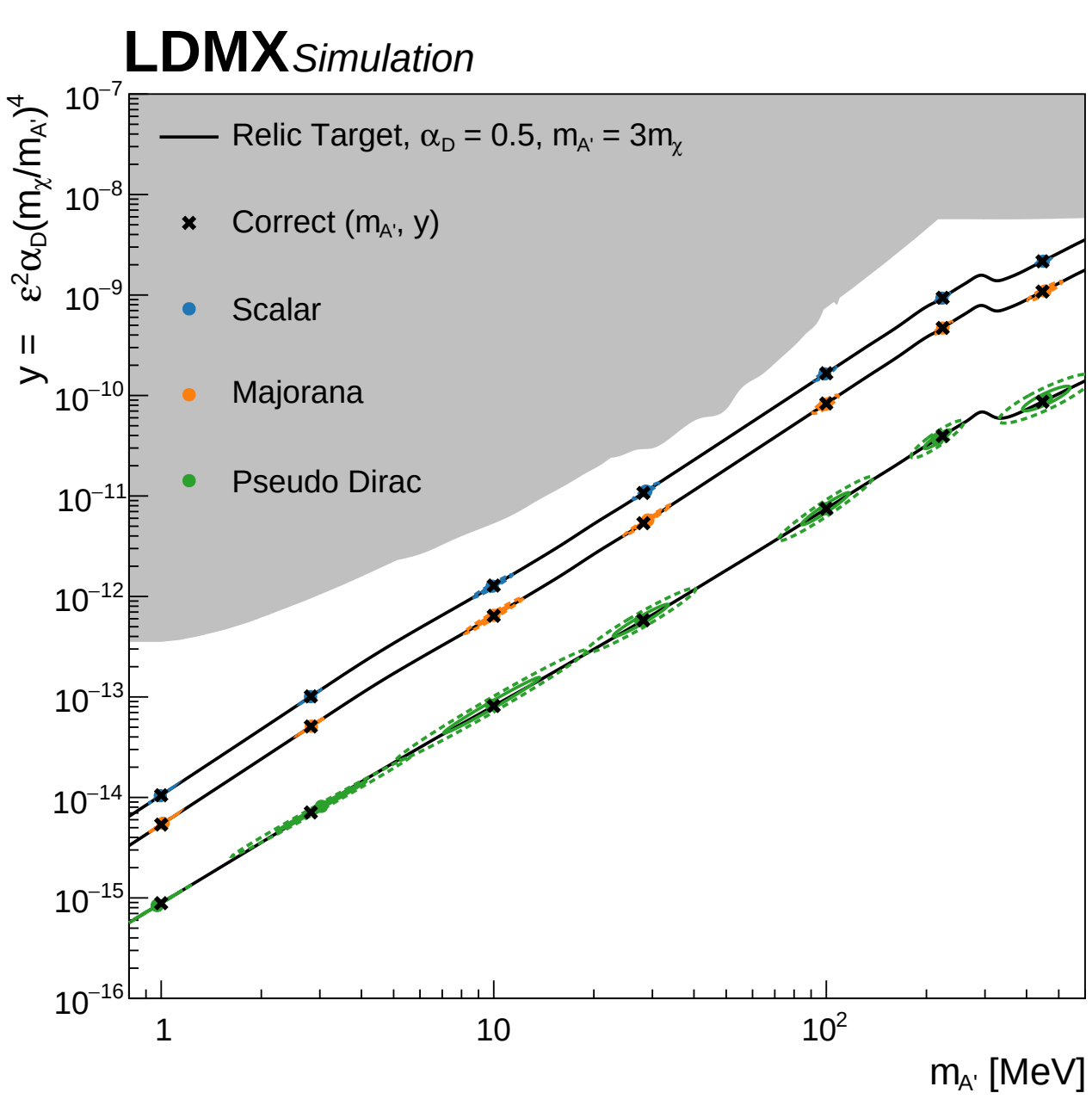

Figure 4.32: Comparison of simulated and reconstructed masses and couplings, including the mean and the $1\sigma$ (solid ellipses) and $2\sigma$ (dashed ellipses) confidence ellipses for different model assumptions at $1 \times 10^{16}$.

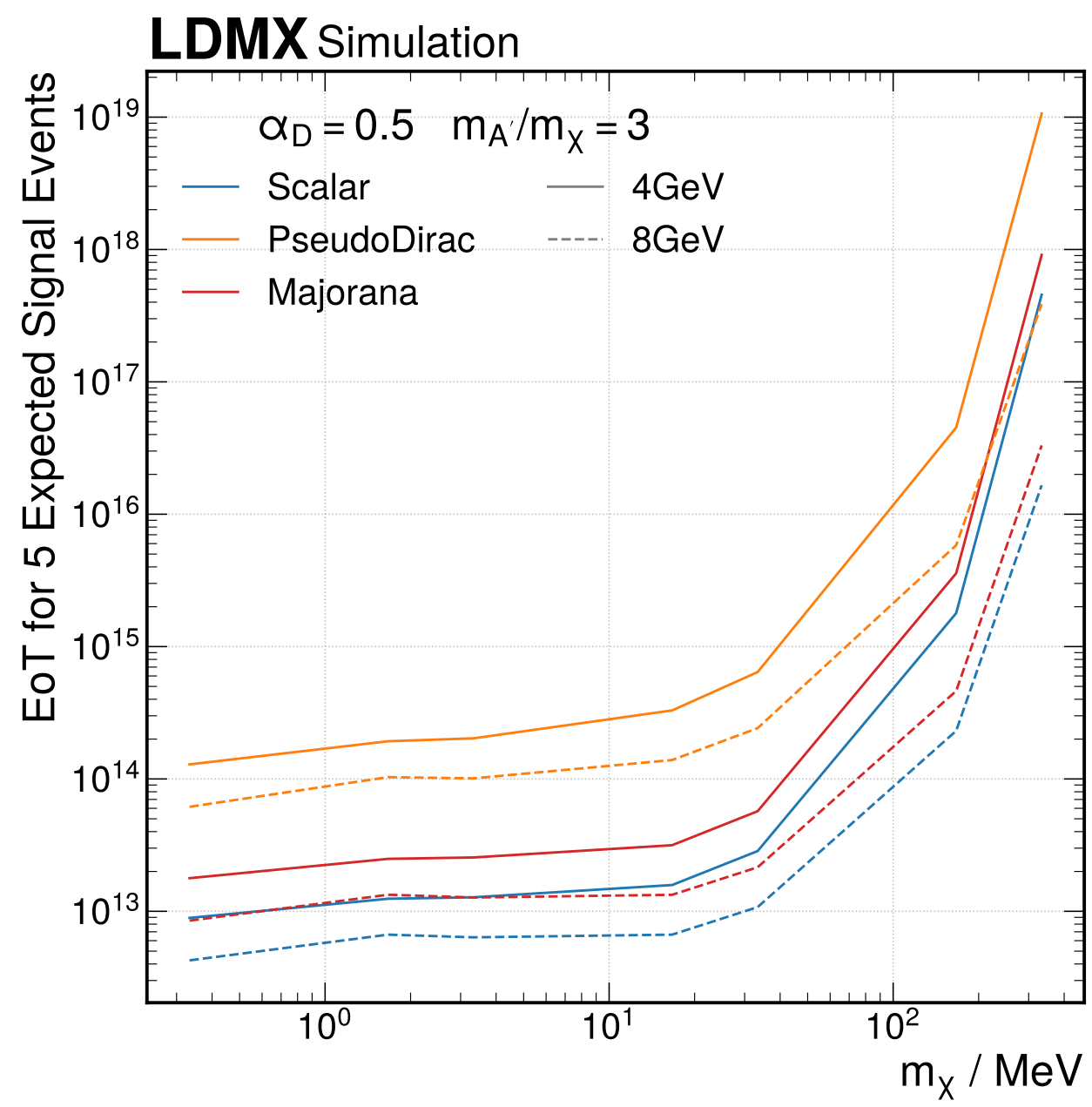

Figure 4.33: The necessary EoT for specific models (varying colors) expected to produce five events in the EaT signal region for both beam cases (4 GeV solid, 8 GeV dashed) as a function of the DM candidate mass $m_\chi$.

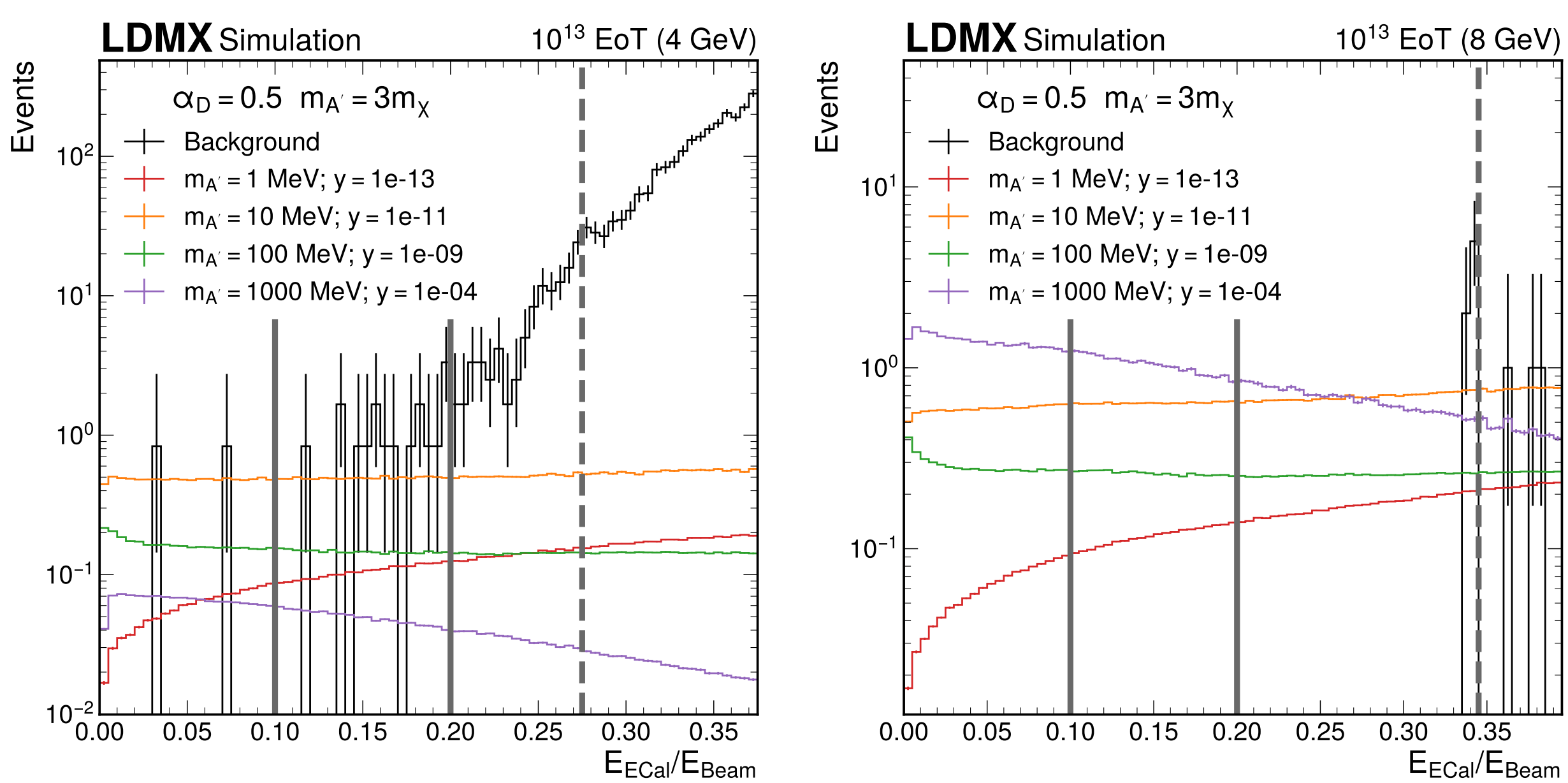

Figure 4.34: The total reconstructed energy in all layers of the ECal ($E_{\mathrm{ECal}}$) as a fraction of the beam energy ($E_{\mathrm{Beam}}$) for all samples that pass the selection criteria except the selection on ECal energy. The 4 GeV (8 GeV) beam is shown on the left (right). The gray lines mark the edges of the analysis bins used to estimate the expected exclusion limit, and the black line is the upper limit on the ECal energy, which also serves as the upper limit of an analysis bin.

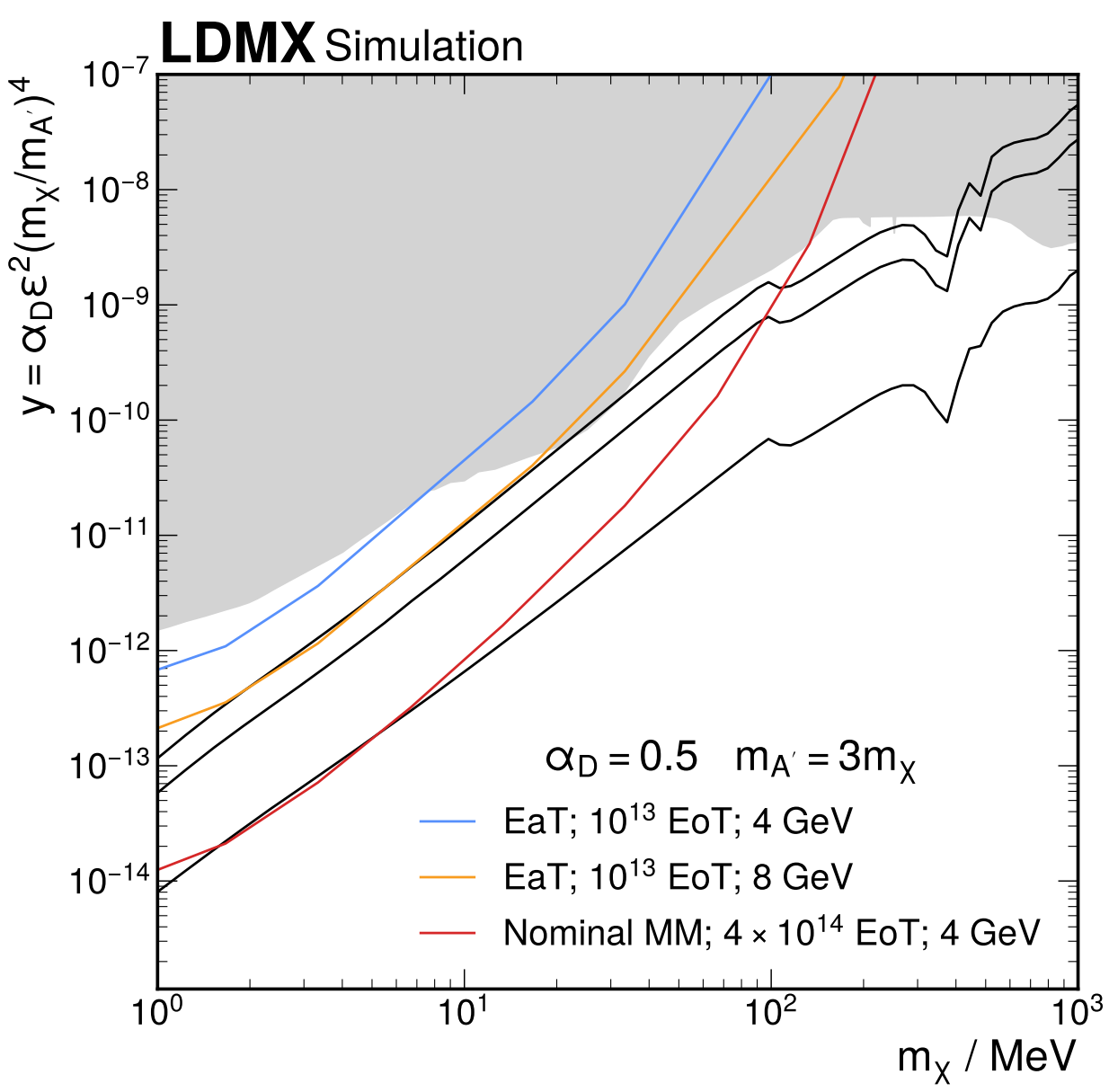

Figure 4.35: The reach of the EaT analysis channel compared to other experiments led by BaBar [18], NA64 [8, 37], and COHERENT [20] (gray); Scalar, Majorana, and Pseudo-Dirac thermal-relic targets (black, top-to-bottom); and our own projections (colors). The blue (orange) line corresponds to the 4 GeV (8 GeV) beam.

## 4.6 Searches for long-lived particles

In scenarios where the dark photon mass $m_{A'}$ is less than $2m_\chi$, the invisible decay channel exploited by the missing momentum search closes, and visible decays of the $A'$ back to SM particles must be considered. When $m_{A'} < m_\chi$, the relic abundance of this "secluded dark matter" is solely determined by dark sector interactions, dispensing with the convenient predictivity of the inverted mass hierarchy. The off-shell and resonant regimes $m_\chi \lesssim m_{A'} \lesssim 2m_\chi$ require careful treatment: visible and invisible channels compete and DM annihilation can be enhanced in the early universe, allowing the viability of smaller values of $\epsilon$. In these scenarios, weakly-coupled mediators present the possibility of searches for long-lived particles that travel macroscopic distances before decaying into SM particles. Though optimized for a missing-momentum search, LDMX is effectively a fully-instrumented, short baseline beam-dump experiment and can probe a variety of dark sector models in motivated and untouched parameter space through searches for displaced decays. The minimal dark photon model has been investigated most thoroughly, as it is a ubiquitous benchmark in the community. This search channel has been deemed the visibles analysis, and will be referenced as such in the following sections. The analysis was completed at a beam energy of 8 GeV and $4 \times 10^{14}$ EoT (with these results extrapolated to $10^{16}$ EoT), following the planned run schedule of LDMX.

### 4.6.1 Analysis strategy for the minimal dark photon

It is possible to search for long-lived particles that decay into SM particles (including light quarks, charged leptons, and photons) by searching for displaced energy depositions in either the ECal or HCal. As will be discussed in Sec. 4.6.2, due to the late photon conversion $\gamma \to e^+e^-$ background in the ECal, this search presented here has been restricted to looking for $A'$ signals in the HCal only. It is feasible to do a zero-background search for visible, displaced decays at LDMX at $10^{16}$ EoT with a carefully selected analysis strategy, even with a large number of events producing an EM shower near the beam energy and displaced SM interactions. This calls for a specific analysis strategy that focuses on suppressing as many of these backgrounds as possible, since the analysis will be carried out targeting $4 \times 10^{14}$ EoT and extrapolated to $10^{16}$ EoT. Many important backgrounds stem from the same physical processes as those in the missing momentum search, though in this case the concern is late energy deposition in the HCal instead of undetected particles escaping the HCal.

The expected signal events for the visibles search can be qualitatively described as containing a single electromagnetic shower in the front of the ECal from the recoil electron and a second, more energetic electromagnetic shower in the HCal, which can be seen in Fig. 4.36. A visibly decaying $A'$ that decays after passing through the ECal fiducial volume will leave the same signature in the ECal as in the missing momentum search, so the same trigger selection and analysis cuts can be used. The selection cuts diverge from the missing-momentum search in the HCal, where a large amount of energy deposition in the Back HCal is required instead of none. Sec. 4.6.2 will describe the HCal-specific event selection cuts, including a *Visibles* boosted decision tree (BDT) to distinguish the electromagnetic shower of $A' \to e^+e^-$ from hadronic showers originating from photo-nuclear-induced backgrounds. Also discussed in this section is how the selection cuts veto specific background processes expected in the analysis: photo-nuclear (PN) reactions and di-muon production. Finally, a sensitivity plot was generated based on the expected background and signal efficiency after the full analysis cutflow.

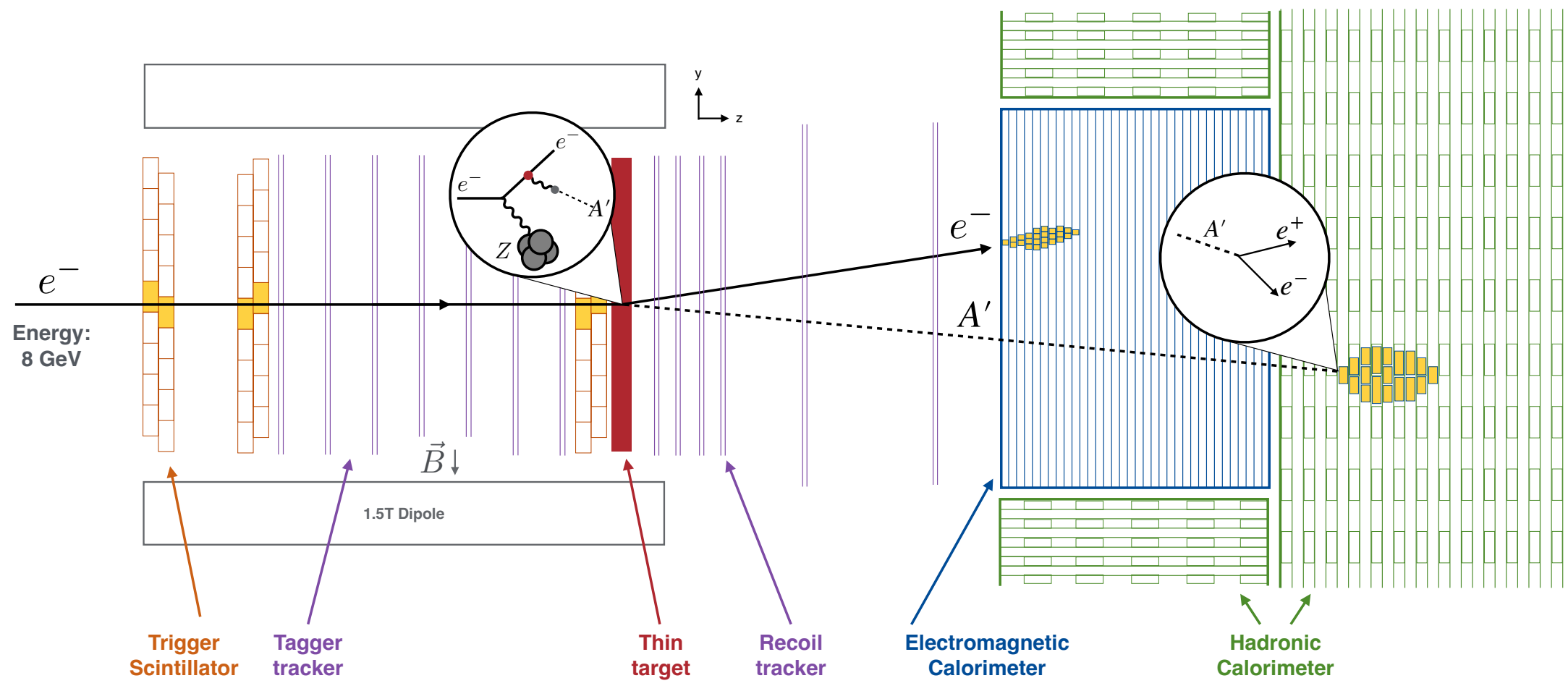

Figure 4.36: Cross-section cartoon view of the LDMX apparatus with the visible search sketch overlaid.

### 4.6.2 Background rejection techniques

The kinematics of $A'$ production with an $A'$ decaying to an $e^+e-$ pair (visible decay) are exactly the same as for an invisibly decaying $A'$, a process LDMX is optimized to detect. The leading-order background process, if bremsstrahlung occurs in the target, is a high-energy photon leaving an electromagnetic shower in the calorimeter. If the photon travels sufficiently far into the calorimeter before interacting, the electromagnetic shower will be displaced from the recoil electron shower. This is an indistinguishable background from the visible $A'$ signal, so it is critical for the analysis to ensure that this process does not occur in the signal region. Due to this background, there is not any expected competitive sensitivity to visible $A'$ decays in the ECal, which leads to the decision to pursue an HCal only analysis.

Therefore, since the signature in the ECal for this analysis is identical to an invisibly decaying $A'$, the same trigger selection as the main search can be used. After the trigger, three additional cuts are applied to select for signal events. The first is a more stringent missing energy requirement in the whole ECal, which requires less than 3160 MeV deposited in all ECal layers –the same energy requirement as the first 20 ECal layers from the trigger. The next cut is a single, low-momentum track in the recoil tracker; exactly one track reconstructed in the recoil tracker with a momentum less than 2400 MeV/c is necessary for this condition to be met. This cut is to ensure that momentum was lost in the target via dark bremsstrahlung.

With the above event selection cuts, remaining background events will consist of events in which the electron loses most of its energy to a bremsstrahlung photon that doesn't deposit all of its energy visibly in the ECal. This can occur in PN reactions or muon pair production. The granularity of the ECal can be used to distinguish whether photon-induced energy depositions accompany the recoil electron shower, which should be the only feature in a signal event. The BDT developed for the missing-momentum analysis to distinguish between signal and background events based on shower shape features in the ECal easily fills this role and consists of the third cut after the trigger.

At this point, all of the same selection cuts as the missing-momentum search have been used, up to requirements on the HCal activity. An invisibly decaying $A'$ will require no activity in the HCal– but a visibly decaying $A'$ will leave an energy deposition in the HCal, with the amount lost by the incident electron. Thus, a requirement to have greater than 4840 MeV deposited in the Back HCal to complement the trigger requirement of less than 3160 MeV deposited in the ECal is necessary. Additionally, the shower must be fully contained in the Back HCal. To mitigate backgrounds originating from particles decaying in the gap between the ECal and HCal, less than 5 PEs deposited in the first layer of the HCal is required - indicated in Table 4.13 as “Containment cut”.

Background events that leave very little activity in the ECal and a large amount of energy in the Back HCal will originate from PN reactions producing neutral hadrons (such as $K_S, K_L, n$) and muon pair production. These backgrounds will primarily leave MIP tracks or hadronic shower signatures in the Back HCal, while the visibles signal will leave an electromagnetic shower. To mitigate this remaining background, a BDT was trained to distinguish between electromagnetic and hadronic showers in the Back HCal. This Visibles BDT was trained on shower shape features and other kinematic features to discern signal from background, described in detail below.

The Visibles BDT used twelve features:

- Total Back HCal layers hit
- Energy-weighted hit standard deviation (x, y, and z)
- Energy-weighted hit mean position (x, y, and r, where $r = \sqrt{x^2 + y^2}$)
- Isolated hits
- Isolated energy
- Total Back HCal hits
- Total energy deposited in the Back HCal
- Energy-weighted mean hit distance from the projected photon line

For the HCal, a hit is defined as “isolated” if there are no additional hits in a neighboring bar in the same layer. The distance from the projected photon line feature is introduced for the kinematics of signal events. For dark bremsstrahlung in the target, there are two outgoing particles: a recoil electron and a dark photon. The recoil tracker can be used to measure the momentum of the recoil electron and infer the momentum of the dark photon. The momentum vector of the dark photon is projected to the HCal, and the distance is measured in the $(x, y)$ plane of the reconstructed hits from the momentum projection. Signal events from an $A'$ decay will have an electromagnetic shower centered on this projected axis, so the mean distance of all hits from this axis is expected to be small.

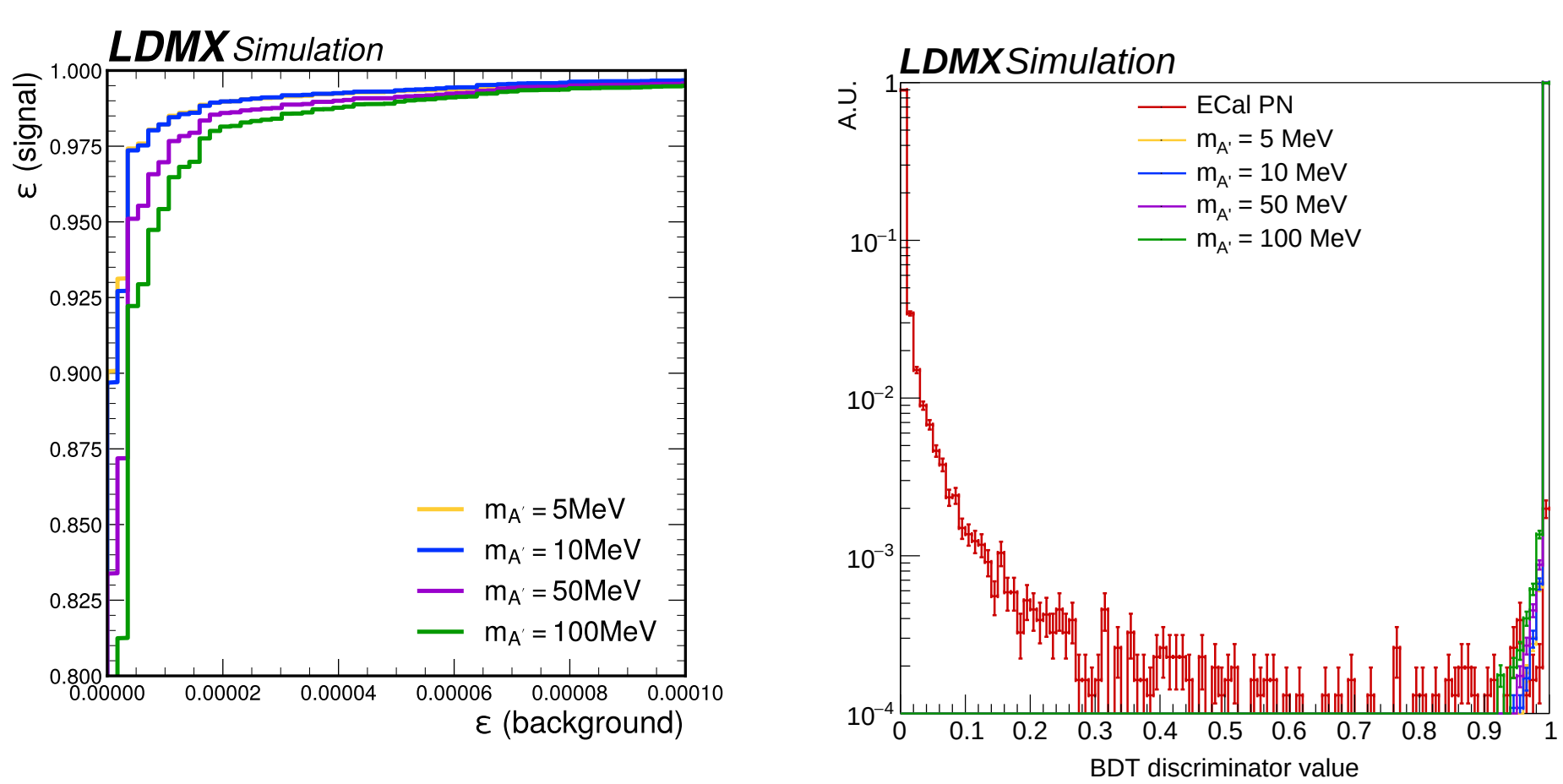

(a) (b)

Figure 4.37: (a) Visibles BDT ROC curve and (b) discriminator value output plot.

The performance of the Visibles BDT can be seen in Fig. 4.37. Although it was only trained/tested on ECal PN events as the background, the Visibles BDT performed very well on the three other backgrounds of concern: target PN, and muon conversions occurring in the target and the ECal. PN events originating in the target will produce the same type of challenging events for this analysis as ECal PN processes, such as charged pions. A muon conversion event in the target or ECal can typically be easily vetoed by the MIP tracks left in the ECal or Back HCal.

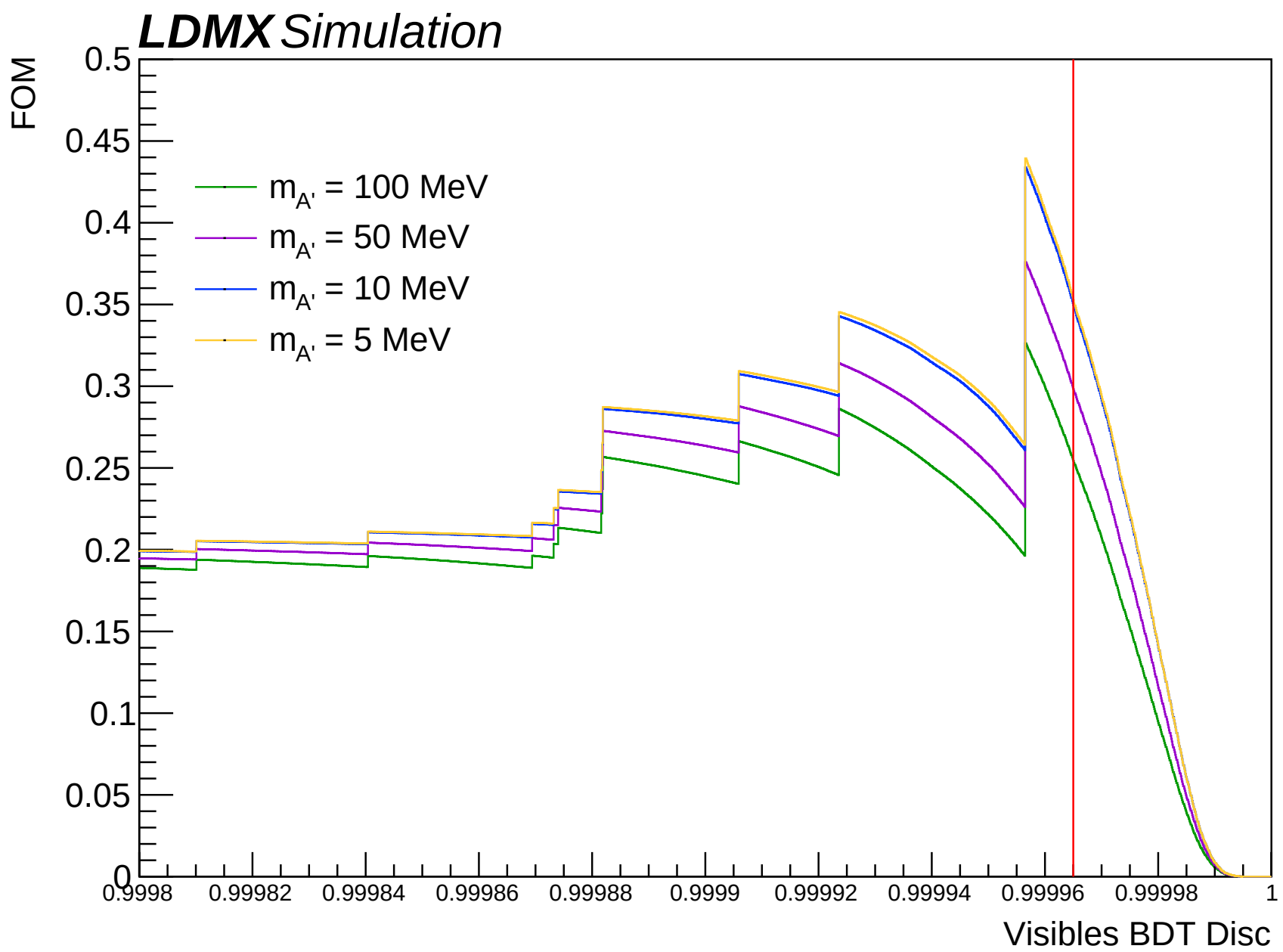

Figure 4.38: The FOM as a function of the discriminator value on the Visibles BDT for the four test masses. The red line indicates the thresholds chosen for the BDT, with everything above this threshold passing the veto.

The discriminator threshold on the Visibles BDT was optimized for this analysis based on a Figure of Merit (FOM). Since the expected number of signal events is not known a priori, and depends on both the mass of the $A'$ and kinetic mixing $\varepsilon$, a traditional FOM like $S/\sqrt{B}$ or $S/\sqrt{S+B}$ cannot be used. Instead, a FOM proposed by Punzi [145] is more applicable - the main missing-momentum analysis also uses this, shown in Equation 4.4.

The FOM for each of the four test masses in this analysis is shown in Fig. 4.38, along with a red line that shows the chosen discriminator cut. One can see the discontinuities in the FOM where the number of background events decreases. The point at which the FOM becomes a smooth function again (above ∼0.99996) is the region of zero background events. Although slightly lower discriminator thresholds on the BDT would optimize the FOM, this is on the border of having one or zero background events from the sample events tested. Discriminator thresholds were chosen slightly away from this point to ensure background rejection for other random samplings of background events.

| | **Photo-nuclear** | | **Muon conversion** | | **Signal efficiency (%)** | |
|---|---|---|---|---|---|---|
| | **Target** | **ECal** | **Target** | **ECal** | **10 MeV $A'$** | **100 MeV $A'$** |
| EoT equivalent | $1.03 \times 10^{14}$ | $\sim 1.00 \times 10^{14}$ | $1.00 \times 10^{15}$ | $1.00 \times 10^{14}$ | | |
| Trigger (front ECal energy $< 3160$ MeV) | $1.77 \times 10^{7}$ | $1.61 \times 10^{8}$ | $4.56 \times 10^{6}$ | $1.47 \times 10^{7}$ | 75.4 | 69.1 |
| Total ECal energy $< 3160$ MeV | $1.22 \times 10^{7}$ | $3.87 \times 10^{7}$ | $3.46 \times 10^{6}$ | $1.02 \times 10^{7}$ | 75.2 | 69.0 |
| Single track with p $< 2400$ MeV/c | 69 282 | $3.50 \times 10^{7}$ | $< 1$ | $9.86 \times 10^{6}$ | 69.0 | 65.0 |
| Missing-momentum ECal BDT | 3919 | 61 145 | $< 1$ | $< 1$ | 64.1 | 62.5 |
| Back HCal energy $> 4840$ MeV | 2190 | 55 283 | $< 1$ | $< 1$ | 61.7 | 58.9 |
| Containment cut | 722 | 30 665 | $< 1$ | $< 1$ | 61.7 | 58.9 |
| Visibles BDT | $< 1$ | $< 1$ | $< 1$ | $< 1$ | 33.4 | 27.0 |

Table 4.13: Number of simulated background events remaining at different selection stages. The background classes are separated by where in the detector the reaction took place.

Background events due to photons undergoing muon conversion pass the trigger and event selection at a rate higher than that of photons undergoing PN conversions, but they are significantly less challenging to veto. These events can originate from either the target or ECal. Backgrounds like this can most easily be vetoed by the MIP tracks they leave in the ECal and HCal. As seen in Table 4.13, after the missing-momentum ECal BDT is applied, the muon conversion backgrounds are fully suppressed. One of the other major background processes that will pass the trigger requirement and often leave a significant amount of energy deposited in the Back HCal is PN reactions, both in the target and in the ECal. These events will create hadrons that decay or interact in the HCal. No ECal or target PN events are left after all selection cuts.

One type of background from PN reactions that is of particular concern for this analysis is those that produce neutral kaons, $K_S$ or $K_L$, that decay to $\pi^0$s. These events will leave electromagnetic showers in the Back HCal that are hard to distinguish from the signal. There is work ongoing to understand the rates of these backgrounds at $10^{16}$ EoT and if there is a risk of any of these events passing all selection cuts.

After using the FOM to optimize the BDT discriminator value, and combining with the other cuts shown in Table 4.13, it is clear that a zero-background analysis can be completed at 8 GeV beam energy and $10^{14}$ EoT. With all of this information in mind, a sensitivity plot can be made for the visibles analysis, as shown in Fig. 4.39. No new parameter space is covered, but the opportunity to verify FASER's results via a different production method is still useful to the community.

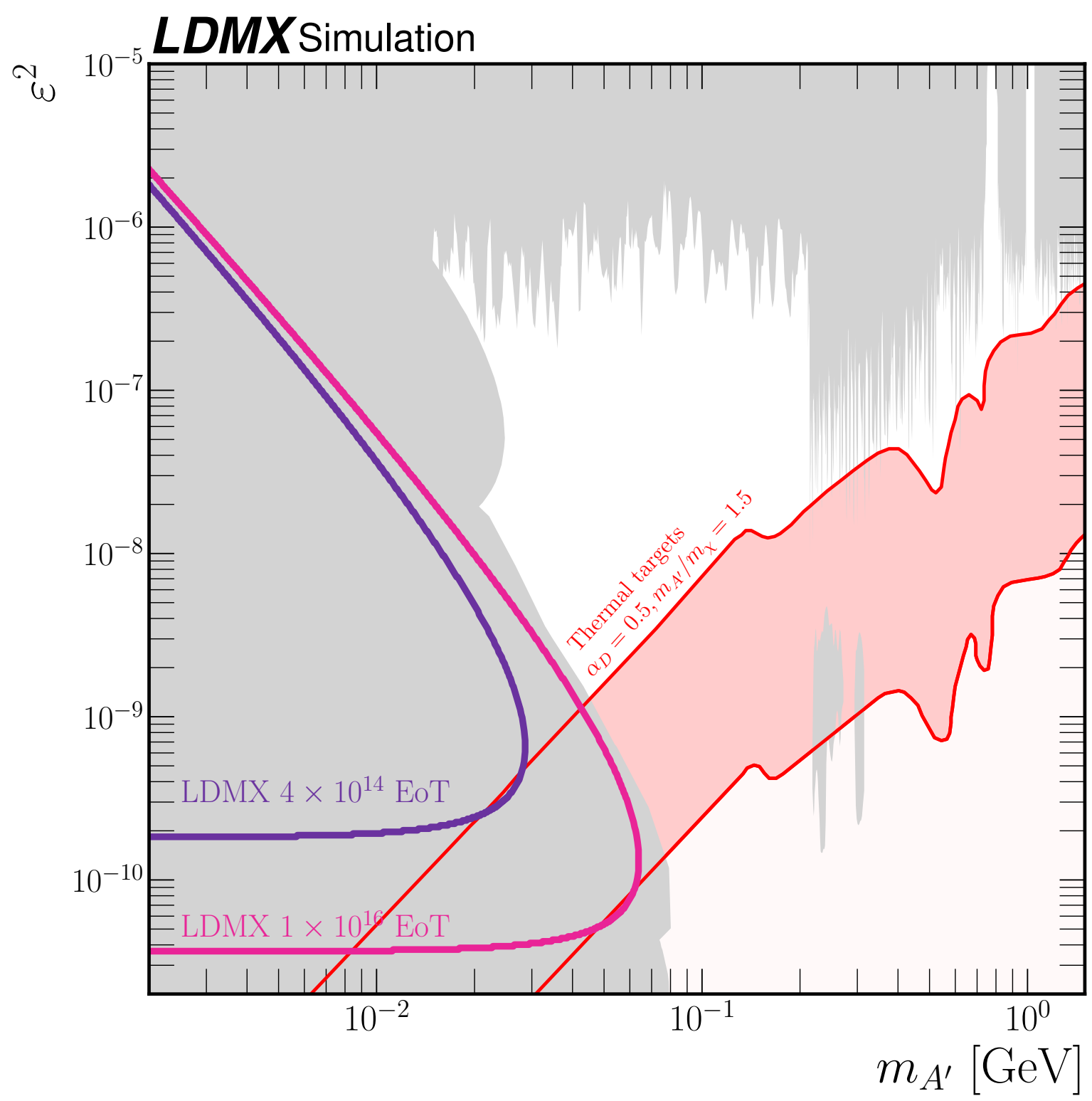

Figure 4.39: 90% $CL_s$ sensitivity for an LDMX Visibles search projected for $4 \times 10^{14}$ EoT and $10^{16}$ EoT. Existing leading limits from colliders [150, 151, 152, 153, 154, 155]; beam dump [156, 157, 158, 159, 160, 161, 162]; measurements of the electron $g-2$ [163, 164, 165, 166]; and fixed target [167] experiments are shown for reference.

### 4.6.3 Other long-lived particle signatures

While not fully realized (which may not be attainable without a larger EoT sample), it should be possible to achieve a zero-background search for visibly decaying $A'$s in the Back HCal at $10^{16}$ EoT. Additional changes could be made to improve the analysis and make it more "life-like", such as using a more advanced machine learning algorithm instead of a BDT or including multi-electron bunches, instead of a single electron. For example, while this analysis uses an up-to-date version of the ECal BDT, this may be improved in the future through the use of a neural network such as ParticleNet.

One sub-category of the PN background that this analysis struggled with was the deep bremsstrahlung conversion, as discussed briefly in Sec. 4.6.2, as the primary reason for switching the analysis to using only the Back HCal. As discussed, an easy way to mitigate this background is to essentially ignore the ECal as a sensitive volume and just use the Back HCal to search for visible $A'$ decays. While the visibles analysis described previously still relied on electrons hitting the target, there is another option. This type of search focuses on electrons that do not interact with the target, and instead start their interaction at the ECal- in the standard search, this analysis is known as Ecal-as-Target (EaT), described in Sec. 4.5. A similar search could be done for the visibles channel at 8 GeV, and a brief proof-of-concept was completed at 4 GeV to bolster this claim.

## 4.7 Searches for Visible Signatures of Axion-like Particles

Axion-like particles (ALPs) are very light, weakly interacting particles that can couple to photons. ALPs are predicted by many beyond the Standard Model theories and are cold dark matter candidates. In many well-motivated models, the coupling to photons is dominant:

$$\mathcal{L} \supset \frac{g_{a\gamma}}{4} a F_{\mu\nu} \tilde{F}^{\mu\nu} \tag{4.10}$$

describing the photon interaction (with $g_{a\gamma} = \frac{1}{\Lambda_\gamma}$ being the ALP photon coupling). The decay width ($\Gamma_a$) is related to the coupling ($g_{a\gamma}$) and the ALP mass ($m_a$) by the following relation:

$$\Gamma_a = \frac{g_{a\gamma}^2 m_a^3}{64\pi}. \tag{4.11}$$

The lifetime is then: $\tau = \frac{1}{\Gamma_a}$.

To understand the LDMX sensitivity to ALPs we consider the photon and electron couplings to be independent and investigate the limiting cases where the photon coupling dominates. As with the visible $A'$ search, we can use the Hadronic Calorimeter (HCal) as the primary mode of identifying displaced ALP decays to photons.

For the HCal to be a means of identifying ALP decay photons we need to ensure good photon/neutron separation in the HCal to discriminate against dominant photo-nuclear (PN) backgrounds. As with the $A'$ visible decay, the main background will arise from PN events, in particular those originating from the ECal. There are also potential backgrounds emanating from PN events at the target, electro-nuclear, muon conversions, and bremsstrahlung processes, but we expect these to be subdominant.

This study is carried out in two parts. Firstly, we evaluate the theoretical sensitivity of LDMX to displaced ALP to photon decays, factoring in the geometric acceptance of the back HCal to understand "optimal couplings and masses" that we will be sensitive to for our pilot run (4e14 EoT) and our main physics run (1e16 EoT). Once we understand our theoretical sensitivity we devise an analysis method, very similar to that used for the visible $A'$ search. The aim is to mitigate Standard Model backgrounds while preserving high ALP signal efficiency.

### 4.7.1 ALP Production Modes

Two dominant ALP production processes can take place within LDMX:

1. The dominant **Primakoff production (direct-photo production).** ALPs are produced via secondary photons (from the primary electron) interacting with the nuclei in the target:

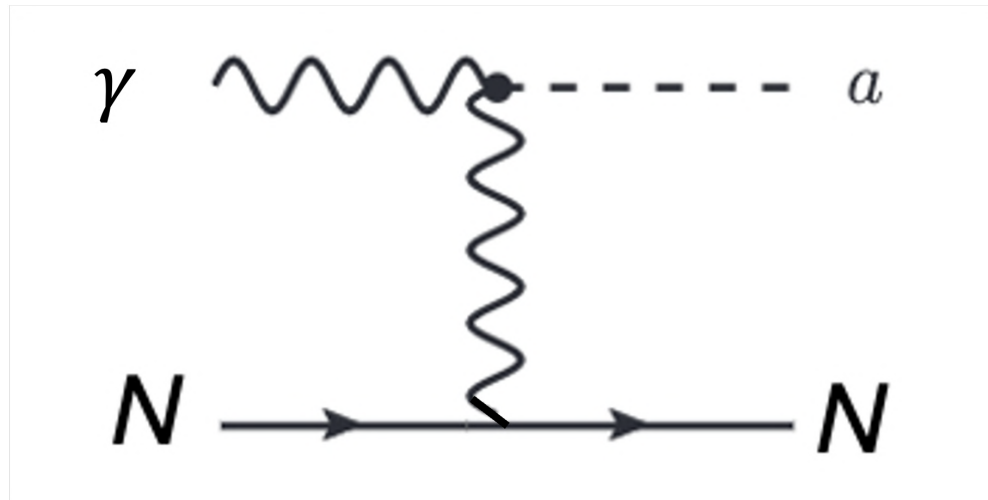

   We focus on the case were the ALPs are produced in the target. In this scenario, a beam electron radiates a real secondary photon, this photon travels some small distance through the material, and then interacts with a photon from an atomic nucleus to produce an ALP ($a$).

2. The sub-dominant, quasi-elastic, **photon fusion production.** ALPs are produced directly from the fusion of two virtual photons.

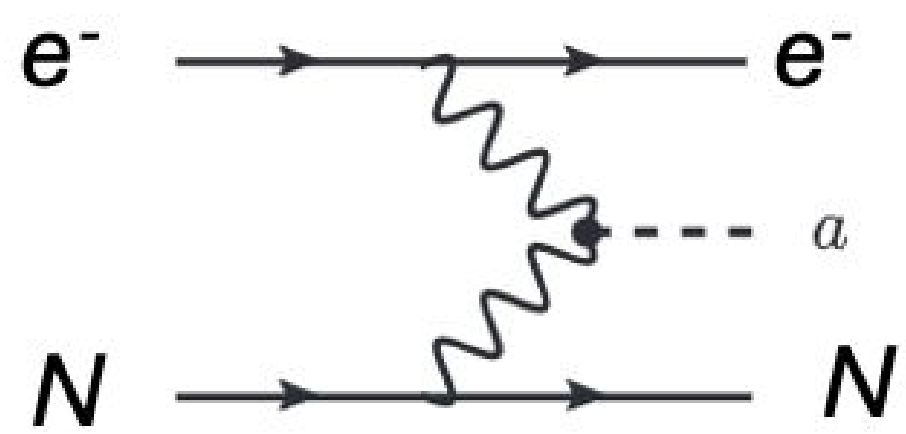

   In this scenario, the electron produces bremsstrahlung as it travels in a nuclear field. The photons are virtual in this case.

As a result of the large flux of secondary photons in LDMX, the Primakoff production mode dominates, but photon fusion can produce ALPs with a slightly higher boost, enabling sensitivity to shorter lifetimes. For both processes, the production is coherent over the nucleus for ALP masses less than a GeV. Compared to the Primakoff process, heavier ALPs produced in photon fusion tend to be more boosted. This results in decay photons that are more forward and energetic, and this effect leads to more collimated photons for larger ALP masses.

Due to the differing kinematics of the photons in these two cases, we expect the outgoing ALP and its decay products to potentially have different characteristics.

### 4.7.2 Signal Simulations

- **Photon Fusion** MadGraph was used to simulate ALP production from photon fusion, and the ALP decay to two photons. A coupling of $g_{a\gamma} = 10^{-3} GeV^{-1}$ is used as the nominal choice. From earlier studies, this was found to be well within the potential sensitivity region.

  ALP masses of $m_{ALP}$ = 10 - 500 MeV are simulated. At higher masses, the ALP decay rate is fast, and we lose the displaced signature for masses above a few hundred MeV. Therefore, we stop at 500 MeV. In MadGraph, we took our nuclear model for the tungsten nucleus from [168].
- **Primakoff**

  A custom generator, adapted from that presented in [169], is used to simulate ALPs from Primakoff production. To produce ALPs for the Primakoff process a sample of secondary photons, from electrons at the LDMX target, was extracted using LDMX-sw v14. An 8 GeV electron gun was fired into the tungsten target. Photons were collected at the back of the target, and the spectrum of these photons was sampled and input into the ALP generator. To produce an ALP, the photon must have $E_\gamma \geq m_{ALP}$. This means we are more likely to produce lower mass ALPs, and the majority of photons will be too soft to produce ALPs with heavier masses.

### 4.7.3 Acceptance

Based on simple kinematics we can see how many ALPs from each production mode produce something within our back HCal acceptance, assumed to be $750 < z < 5500$ mm. We scan through couplings for a given mass to find couplings producing at least 1 ALP within the back HCal acceptance for our 1e16 EoT main run.

Fig. 4.40 shows the projected number of events in the back HCal acceptance for 1e16 EoT for two selected ALP masses. We initially simulated 10-500 MeV in 100 MeV intervals, however, we established we had no sensitivity to displaced vertices at masses $\gtrsim$ 100 MeV, so we expanded in the region 10 - 130 MeV at 10 MeV intervals. We find that ALP masses > 125 MeV do not produce viable signal rates, so we focus on the range 10 - 125 MeV in the remainder of the study.

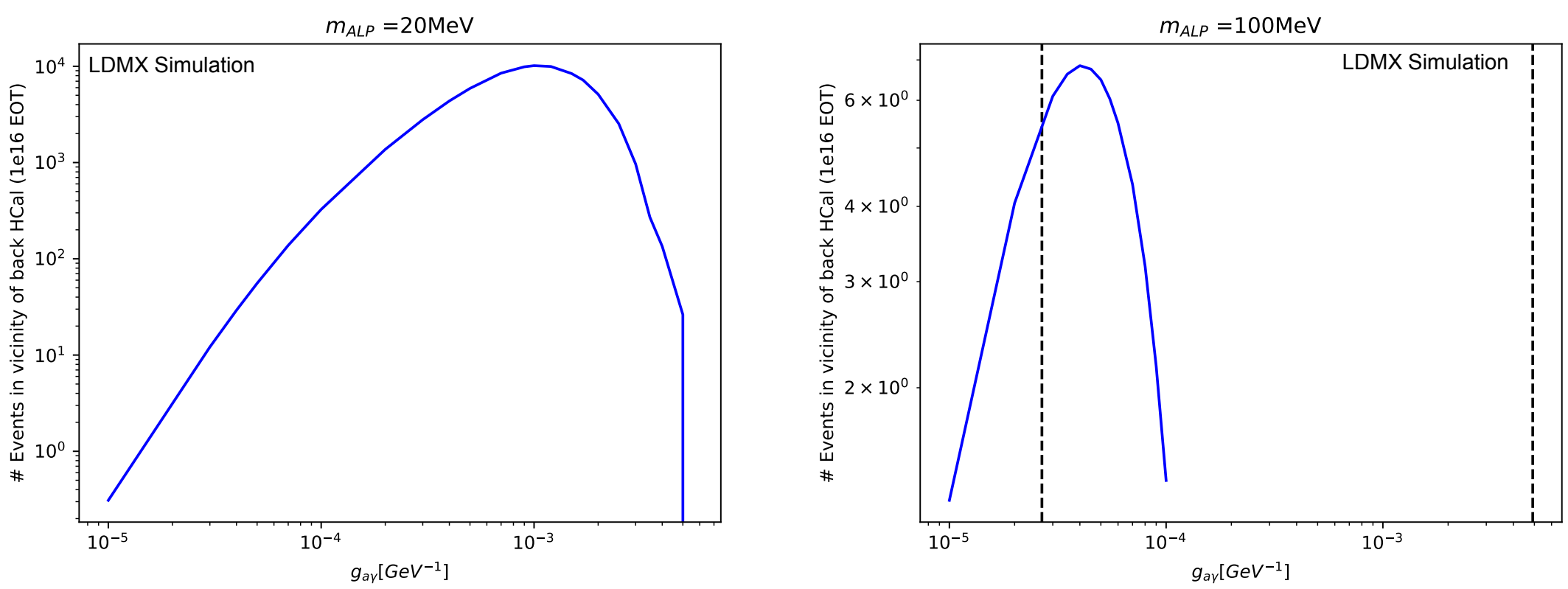

Figure 4.40: Acceptance for a selection of masses, dashed lines indicate current limits (space between is unexcluded).

### 4.7.4 Reconstruction

Samples of ALPs at masses 10 - 125 MeV, in a range of couplings found to provide non-zero events in the back HCal, are passed into our LDMX-sw simulation and reconstruction software. We find that if the ALP decays in the acceptance of the HCal, we can reconstruct it.

To understand if we can viably identify a signal we need to be able to veto backgrounds. To do that, we follow a similar strategy as for the visible A′ study. The main sources of background are:

- **PNs in the ECal**
  The primary source of background to isolating the photons from the ALP decay in the HCal is a hard bremsstrahlung photon from the target that undergoes a PN interaction in the ECal and produces hadrons. These hadrons will interact with one of the calorimeters and deposit energy near the beam energy. We use a sample that represents an equivalent of 1e14 EoT.
- **PNs in the Target**
  Beyond interactions in the ECal PN interactions can also occur in the target.

  For the target PN background, a custom simulation was made as part of the dark photon visible decay analysis. We utilize these samples which represent 1e14 EoT.
- **Muon Conversions**
  Background events due to photons undergoing muon conversion $\gamma \to \mu^+\mu^-$ pass our trigger and event selection at a rate higher than that of photons undergoing PN conversions, but they are significantly less challenging to veto. These events can originate from either the target or ECal and can most easily be vetoed by the MIP tracks they leave in the ECal and HCal.

For this study we again utilize samples made for the dark photon visible decay search. These represent 1e15 EoT (target conversions) and 1e14 (ECal conversions).

We train a Boosted Decision Tree (BDT) on the same set of features used for the A′ analysis. However, we re-train the BDT; our signal sample is inclusive, with an equal mix of each mass from 10 - 125 MeV (for couplings found to provide non-zero yields in the HCal acceptance). Fig. 4.41 shows the resulting BDT score, clear signal and background discrimination is achieved. The Punzi figure of merit is used to determine the optimal cut on the BDT score at 0.99; this provides a signal efficiency of 89 % and background rejection of close to 100 %. The remaining background will be removed from further selection cuts.

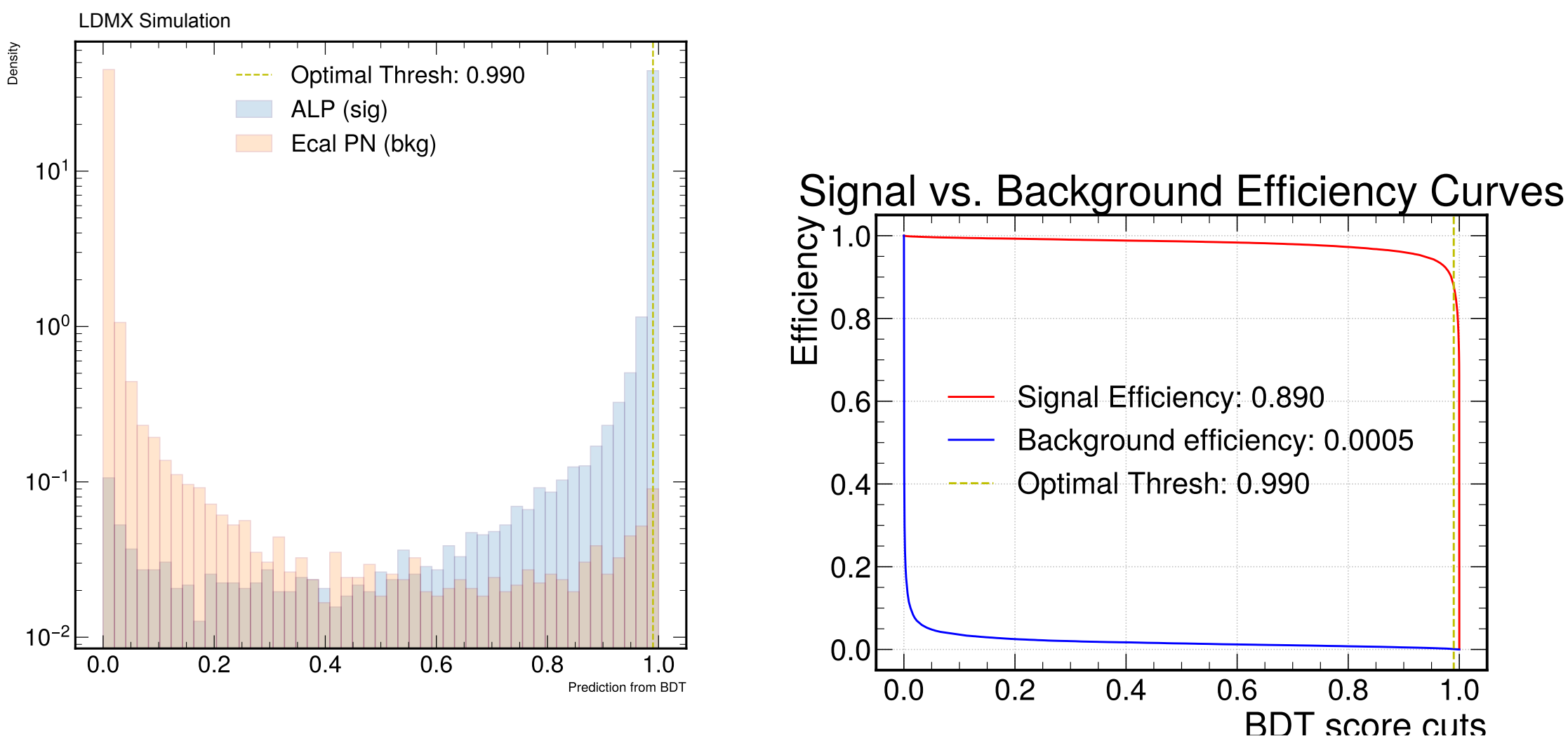

Figure 4.41: (left) Score from BDT trained on 120K ECal PN background events and 120K mixed signal events. The BDT provides good signal to background separation. (right) Using a Punzi figure or merit an optimal selection of BDT Score > 0.99 provides a signal efficiency of 89 %, while rejecting most background.

### 4.7.5 Possible Sensitivity

Fig. 4.42 shows the projected 90 % CL contours for the 4e14 EoT and 1e16 EoT scenarios. It is assumed that we can be background-free using the strategy outlined above. LDMX is sensitive to ALPs of mass 10 - 125 MeV for the 1e16 EoT scenario when factoring in an assumed trigger, selection and reconstruction efficiency similar to that of the visible $A'$. This analysis is still under development, and further refinements could change this result.

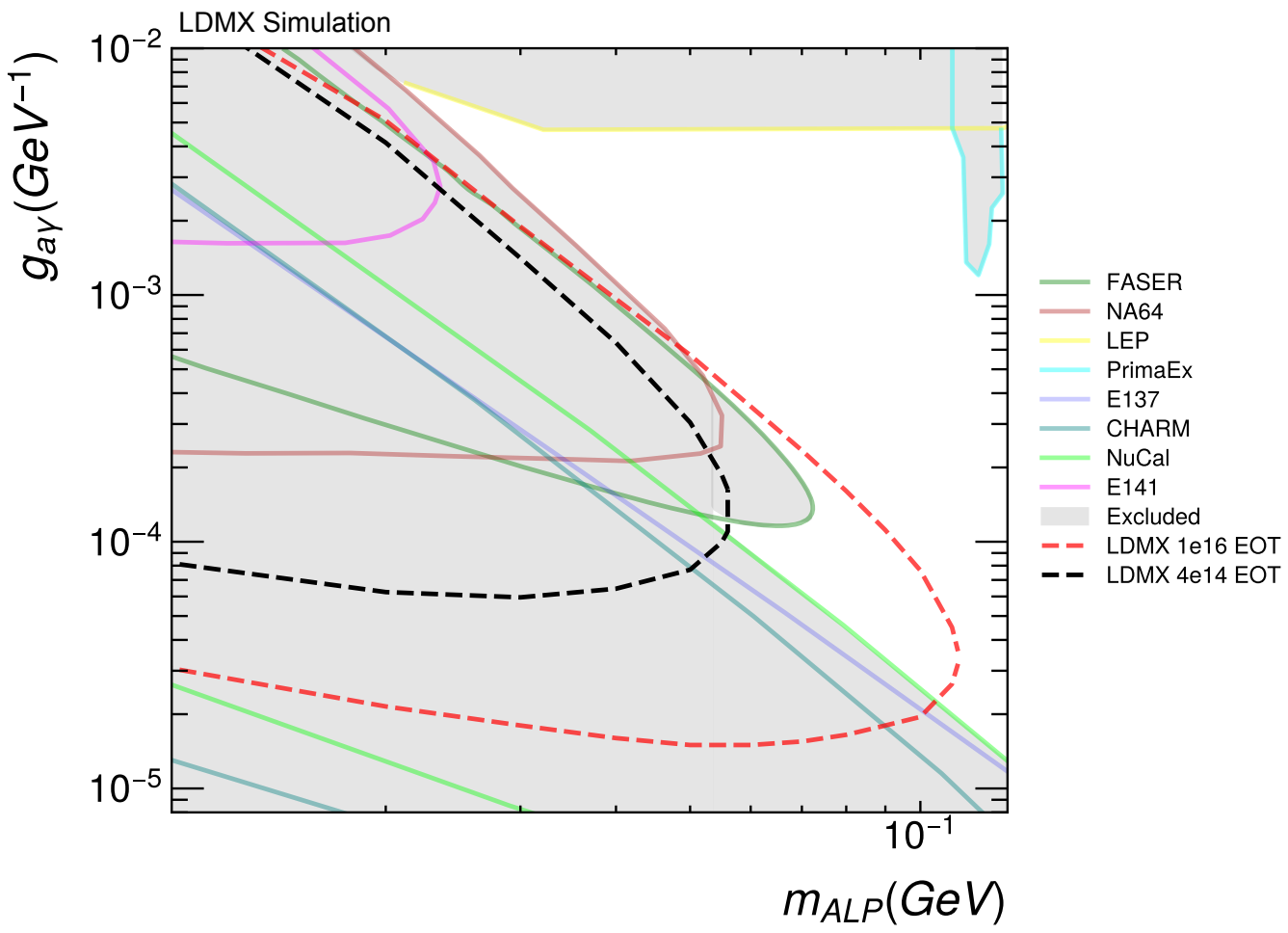

Figure 4.42: LDMX sensitivity (at 90 % CL) to ALP photon decays, with displaced vertices. The results factor in the reconstruction, trigger and selection efficiencies detailed in the earlier discussion. Two EoT assumptions are shown (black) shows the projected sensitivity at 4e14 EoT and (red) 1e16 EoT. LDMX has unique sensitivity to a range of couplings at masses > 53 - 125 MeV.

### 4.7.6 Conclusions

This study shows that LDMX has sensitivity to visibly decaying ALPs with a dominant photon coupling for ALP masses 10 - 112 MeV. The analysis strategy uses the back HCal as the primary detector. LDMX has unique sensitivity to a range of photon couplings for ALP masses 53 - 112 MeV.

## 4.8 Photophobic ALPs and dedicated shower techniques for long-lived particles

In this section, we extend the ALP studies both in decay mode and lifetime. We assume the ALPs to decay to electrons in the ECal. The backgrounds in the ECal are much higher than in the HCal so new analysis techniques had to be developed.

### 4.8.1 Production Modes

Photophobic ALPs are light pseudoscalar particles such which, to leading order, couple not to the usual photons, but to an electron-positron pair.

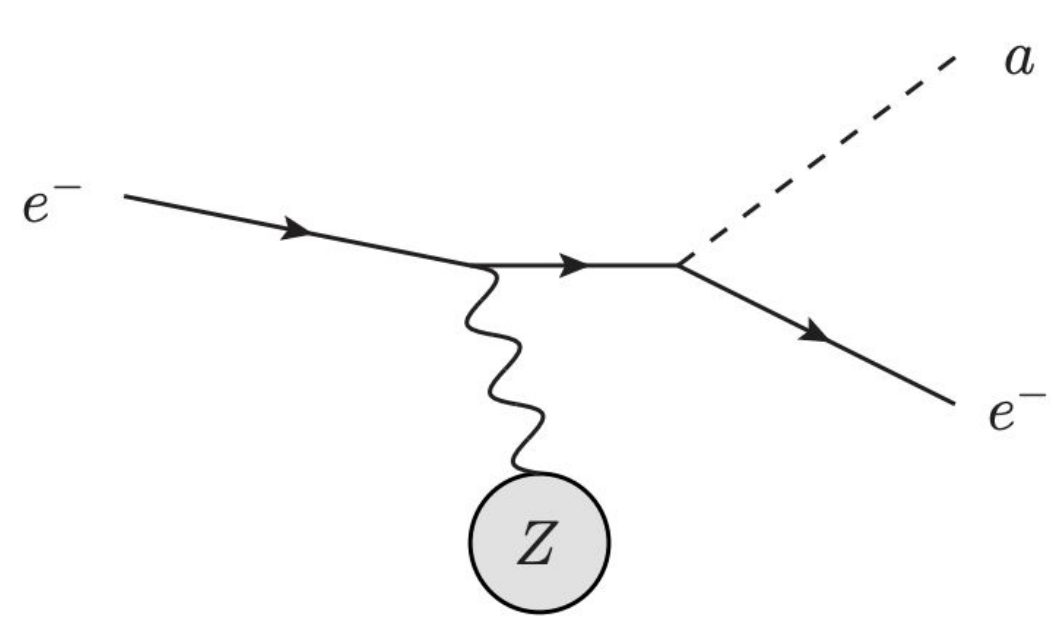

Figure 4.43: Production of an ALP through bremsstrahlung radiation at an electron fixed-target experiment.

The main production mechanism of the ALP is via a bremsstrahlung radiation in an electron-nucleus collision as shown in Fig. 4.43, which then decays into an electron-positron pair with the lab-frame decay length given as:

$$\gamma c\tau_a = 15\,\text{cm} \times \left(\frac{E_a}{8\,\text{GeV}}\right) \times \left(\frac{\Lambda_e}{10^2\,\text{GeV}}\right)^2 \times \left(\frac{100\,\text{MeV}}{m_a}\right)^2. \tag{4.12}$$

Both the production and the decay of the ALP were done using `MadGraph5` utilizing a Photophobic ALP + Nucleus Universal Feynman Output at 4 different ALP mass points of 0.005, 0.025, 0.05, and 0.5 GeV. The resulting LHE file was then fed into the existing `Geant4` full detector simulation for LDMX. For this initial study, we manually decayed the ALP at $z_{\text{decay}} = 350\,\text{mm}$, which is at around the 8th layer of the ECal, ignoring all trigger effects to understand the underlying kinematics and event structure of such ALPs.

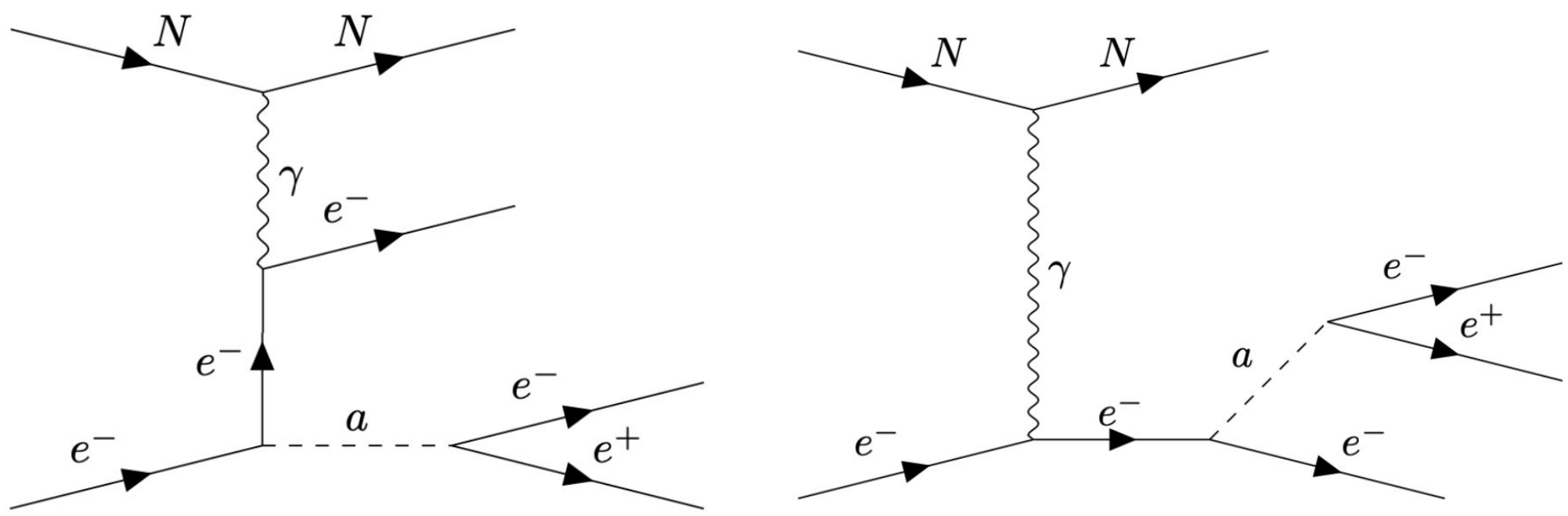

Figure 4.44: Feynman diagrams generated from `MadGraph5` where $N$ is the tungsten nucleus and $a$ is the ALP. Background events are equivalent to the above diagrams by exchanging $a \leftrightarrow \gamma$.

Primary background events considered were hard photons ($E_\gamma > 5\,\text{GeV}$) produced at the target that undergo the same electron-positron conversion inside the detector. Such events were generated utilizing a filter in `ldmx-sw` that forces the decays at a given $z$ position. However, the efficiency of generating events drops significantly as $z_{\text{decay}}$ increases. Thus, for this initial study, we have chosen $z_{\text{decay}} = 350\,\text{mm}$, which had a reasonable event generation efficiency.

### 4.8.2 Discriminating Observables

In this section, we describe the techniques developed for the ALPs, but they could be used for many long-lived signals.

Due to the production of a massive ALP, the recoil electron receives a significant transverse kick, resulting in two spatially separated ECal showers. On the other hand, for the background events, the recoil electron is nearly collinear with the produced photon. While still producing two ECal showers from the recoil electron and photon, we find the two showers overlapping.

The primary kinematic differences can be observed in some ECal variables as shown in Fig. 4.45 where we observe the distribution between signal and background to differ.

On top of these three variables, we have aimed to construct further observables that increase the discriminating power of signal and background by implementing jet clustering and substructure techniques from LHC [170, 171] to create shower-level observables.

Setting the coordinate system origin in front of the ECal and calculating the angles with the axis as described in the rest of this document (i.e. z axis is the beam axis), results in a large spread of the hits in the azimuthal angle $\phi$, causing any kind of jet clustering to fail.

To counteract such a problem, we have slightly modified our definition of pseudorapidity ($\eta$) and $\phi$ where we define the polar angle with respect to the global $x$-axis and utilize $E_T \to E_L$ as our energy term for the jet clustering. After this redefinition of the polar angle from $\eta - \phi \to \eta' - \phi'$, we find that the hits are more localized, enabling the use of a jet clustering algorithm. Utilizing this newly defined representation of ECal hits, we introduce two common jet substructure observables.

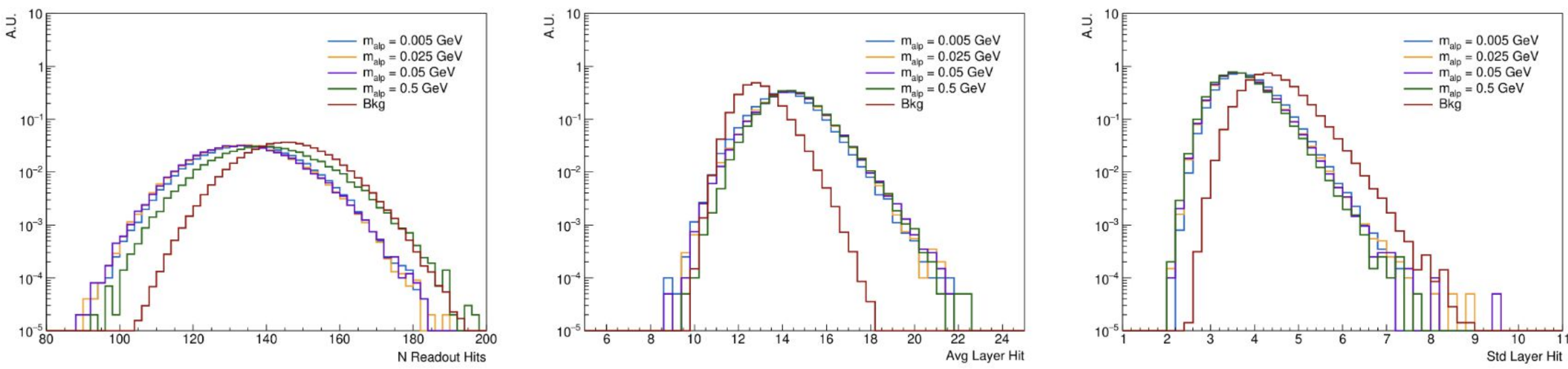

Figure 4.45: The ECal variables; Number of readout hits (left), average layer of hits (center), and standard deviation of layer of hits (right).

We define $N$-jettiness $\tau_N^{(\beta)}$ to quantify the global event shape between signal and background, defined as shown in Eq. 4.13.

$$\tau_N^{(\beta)} = \sum_i E_{L,i} \min\left(\Delta R_{1,i}^{\beta}, \ldots, \Delta R_{N,i}^{\beta}\right), \tag{4.13}$$

where for $\beta = 2$ the $\Delta R_{ij}^2 = (\eta_i' - \eta_j')^2 + (\phi_i' - \phi_j')^2$. The shower axis is defined using an exclusive $k_T$ jet clustering algorithm with $R = 0.4$ using `FastJet` [172]. We then define the ratio $\tau_{2,1} = \tau_2/\tau_1$ as our discriminating variable such that a large $\tau_{2,1}$ indicates that the event is well described by two showers and a small $\tau_{2,1}$ indicates that the event is well described by one shower. Fig. 4.46 shows the N-jettiness ratio $\tau_{2,1}$ on the left, for different signals and the background. We choose $\beta = 2.5$ by finding the value that maximizes the background rejection rate with 50% signal efficiency.

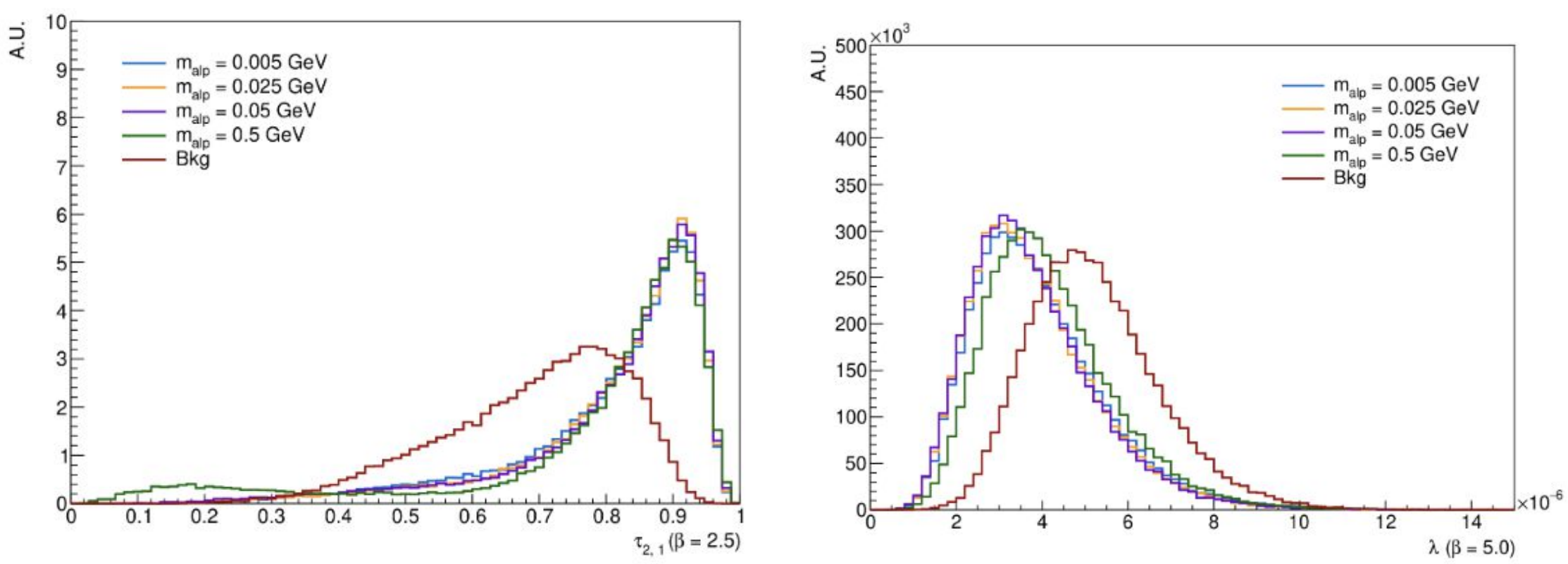

Figure 4.46: Distribution of $N$-jettiness ratio $\tau_{2,1}$ with $\beta = 2.5$ (left). Distribution of angularity with $\beta = 5.0$ (right).

The second observable we have constructed is the angularity $\lambda^{(\beta)}$ to quantify the shower shape and width of the ALP and photon shower, defined as:

$$\lambda^{(\beta)} = \frac{\sum_{i \in S} E_{L,i} \Delta R_{i,S}^{(\beta)}}{\sum_{i \in S} E_{L,i}}.$$

In the sum, we only consider hits contained in the highest $E_L$ inclusive anti-$k_T$ jet area with $R = 0.15$. The change from exclusive $k_T$ to the inclusive anti-$k_T$ is motivated by the soft "jets" in the event. We have opted for a smaller jet radius of $R = 0.15$ to reduce recoil electron hit contamination in background events. Since we are using $E_L$ in the definition of angularity, a small $\lambda^{(\beta)}$ implies that the energy is concentrated about the shower axis, while a large $\lambda^{(\beta)}$ implies that the energy is spread away from the shower axis. Fig. 4.46 shows the angularity $\lambda$ on the right, for different signals and the background. Similar to $N$-jettiness, we optimize the discriminating power of the angularity by finding the $\beta$ that maximizes the background rejection rate with 50% signal efficiency.

### 4.8.3 Cut-based analysis

In total, we have five discriminating observables, three ECal variables, and two jet observables. In the next subsection, we have evaluated the total discriminating power between signal and background using these 5 variables in two different methods: a binary hypothesis test and a cut-based analysis. For simplicity, we show the cut-based results.

We optimize each cut through the Punzi Figure of Merit, as defined in Eq. 4.4. We were successfully able to remove all 350k background events, which is approximately equivalent to $10^{13}$ EoT with the cutflows shown in Table. 4.14.

| | 0.005 GeV | 0.025 GeV | 0.05 GeV | 0.5 GeV | Background |
|---|---|---|---|---|---|
| $N_{\text{readout hits}} < 141$ | 68.4% | 72.3% | 72.3% | 53.1% | 33.1% |
| Avg Layer Hit > 14.26 | 34.8% | 41.9% | 42.6% | 33.3% | 1.8% |
| Std Layer Hit < 3.74 | 20.6% | 27.0% | 27.7% | 23.9% | 0.034% |
| $\ln\left(\tau^{(\beta=2.5)}\right) > -0.11$ | 10.8% | 12.2% | 12.5% | 7.9% | $8.5 \times 10^{-4}$% |
| $\ln\left(\lambda^{(\beta=5)}\right) < -12.39$ | 9.4% | 10.5% | 10.8% | 6.3% | $2.8 \times 10^{-4}$% |

Table 4.14: Cutflow efficiencies for ALP samples with different masses and the background sample.

The signal efficiency could be improved by implementing further sophisticated methods like a BDT and ParticleNet while maintaining the same background rejection.

Fig. 4.47 shows the number of expected ALPs as a function of the mass and coupling, after the cuts and luminosity at $10^{16}$ EoT. The solid colored lines show the specific mass-coupling values in which we have 1 (10) observed ALP event(s). With this, we conclude that we can show sensitivity using this model and decay modes.

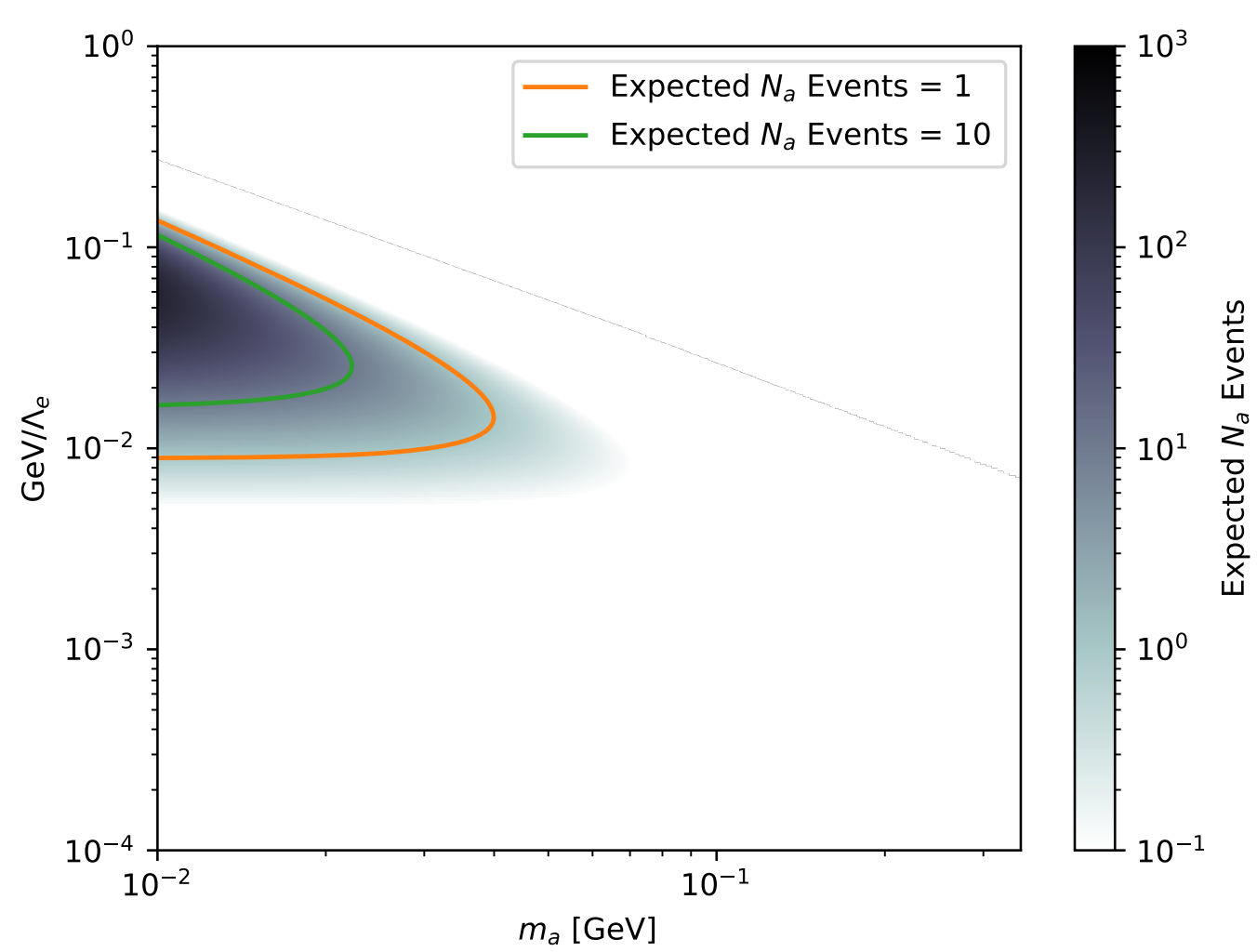

Figure 4.47: Expected number of photophobic ALP events as a function of the ALP mass and coupling. The orange and the green show the contour where 1 and 10 events are left.

## 4.9 Studies of electron-nucleus scattering

Long-baseline neutrino oscillation experiments rely on measuring neutrinos through their interactions with the nuclei in detectors. Due to the complexities of nuclear physics, current generation experiments - NO$\nu$A and T2K - already suffer from large uncertainties on these interactions, which lead to large uncertainties in oscillation parameters. Future experiments such as DUNE and Hyper-K need to achieve a significant reduction in these uncertainties.

There are a few reasons these uncertainties have a large impact on these experiments, but the primary reason is neutrino energy estimation. Neutrino beams are naturally broadband, and each neutrino's energy must be reconstructed from the products of an interaction. A combination of undetectable energy (from particles below threshold, neutral particles, and energy lost to binding energy) and detector response effects (for example, having different energy resolution for pions and protons) will inevitably lead to biases in the neutrino energy estimation, which must be corrected for based on modelling.

Historically, electron scattering data have played a crucial role in understanding the impact of nuclear effects on neutrino scattering. In an electron scattering experiment, the incoming electron energy and direction (as well as the electron beam intensity) can be precisely known. To first order, nuclear effects such as Fermi motion, binding energy, and rescattering/absorption of hadrons, are independent of the probe (whether it be a photon or a W/Z boson), but determining their impact generally requires some knowledge of the transferred 4-momentum, making an electron beam much more useful than a neutrino beam.

Traditional electron scattering experiments place a spectrometer at a fixed scattering angle from a target and measure the outgoing electron energy, thus measuring cross sections as a function of energy transfer. In some cases, additional detectors can be placed to identify scattered protons and neutrons. LDMX falls into a small set of experiments with broad angular acceptance for all final state particles - the only other experiment currently in this category is CLAS/CLAS12. This broad acceptance will allow LDMX to make measurements of not only the scattered electron, but a large portion of the final state hadrons, too. In this way, reconstructed energy biases can be directly mapped for a variety of final states. Additionally, targeted measurements of specific variables can pin down more specific aspects of the interaction model for further scrutiny.

The primary intent, given the electron beam energy, is to provide measurements useful to the future DUNE program. For its primary oscillation measurements, DUNE will detect neutrinos in its near and far detectors from the LBNF neutrino beam. LBNF will produce a broadband muon and anti-muon neutrino beam at Fermilab, with DUNE's on-axis position optimized to maximize discovery potential for the neutrino mass hierarchy regardless of oscillation parameters, and for sensitivity to CP-violation measurements in the neutrino sector. Neutrino energies produced in the beam extend to the 10 GeV range, though the bulk of the beam is within 0-5 GeV.

Interactions of these neutrinos can be generally characterized by their momentum and energy transfer. The range of events in that 2D-phase-space are shown in Fig. 4.48 using the GENIE neutrino generator. While DUNE will see a large number of neutrino interactions at lower energy and momentum transfers, dominated by quasi-elastic interactions and resonant production, deep-inelastic-scattering (DIS) events will be the dominant interaction modes at higher initial neutrino energies, and will be critical to understanding neutrino oscillations.

DUNE will use argon as the active detector material, while LDMX plans to run initially with a tungsten target. Tungsten, being larger than argon, will have more significant nuclear effects. Additional data collection with alternative targets would be more directly beneficial to this measurement program. Titanium has the same number of protons as argon has neutrons, allowing one an approximate map of the neutrons in argon—for the purposes of studies shown here, we assume the use of a thin titanium target. Other target materials may be of interest; for example, a liquid argon cell would also be an interesting prospect, being the exact nucleus planned for use in DUNE, though this is more challenging practically.

### 4.9.1 Simulation and expected event rates for electronuclear measurements

To study eN interactions in LDMX, we have integrated the GENIE neutrino generator [173] into `ldxm-sw`. GENIE, like many other neutrino generators, allows for running in an electron scattering mode, leveraging the similar physics. GENIE is used by many neutrino experiments, including DUNE, due to the many factorized model sets that are available to be used, and the ability to tune parameters of many of these models and apply event re-weighting to probe systematic uncertainties. To interface with `ldmx-sw`, we make use of the HepMC3 event format and NuHepMC standard [174, 175], which will make integration with other generators also possible.

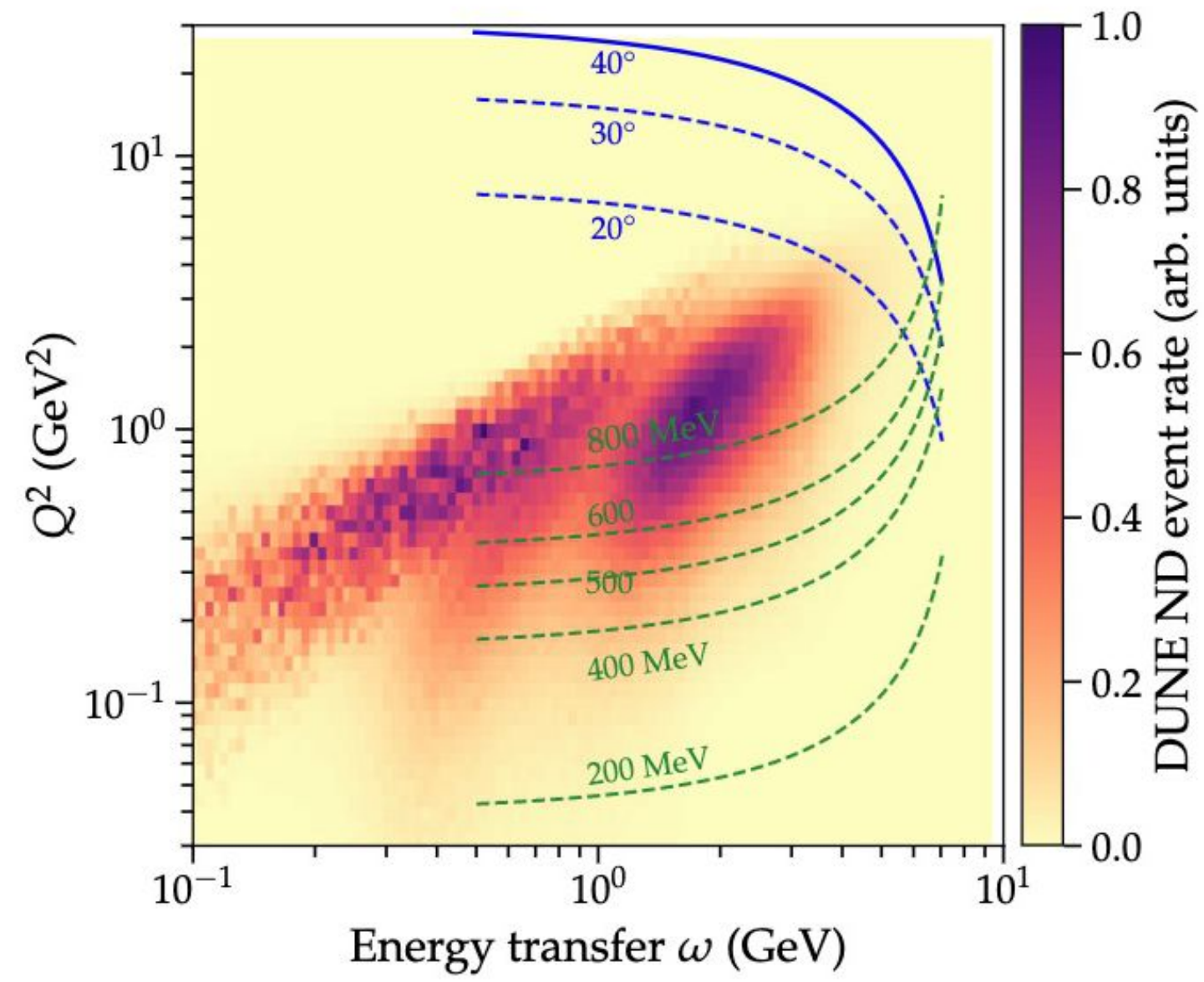

Figure 4.48: Momentum transfer vs. energy transfer for DUNE neutrinos, overlaid with lines of constant final state lepton $p_T$ (green dashed) or scattering angle (blue solid) for a lepton with 8 GeV initial energy. There is significant overlap in phase space between electron scattering measurements in LDMX (which are likely to require the final lepton $p_T > 500$ MeV/$c$) and DUNE.

In the studies presented in this section, we generally use the `G18_02a_02b_11b` GENIE model set and tune [176] as a base model, which uses a relativistic Fermi Gas model for the initial nuclear ground state, the Llewellyn-Smith quasi-elastic interaction model, an empirical model for correlated nucleon interactions ("MEC"), the Rein and Sehgal model for resonant production and coherent pion production, and the Bodek-Yang model for inelastic scattering. Final state interactions in this model set use the effective intranuclear transport model in INTRANUKE. Where different models are used they are described below. As stated above, while tungsten will be the primary target for the LDMX dark matter searches, a titanium target would present a nucleus more similar to argon, of particular interest for future DUNE neutrino cross sections, and so we assume a titanium target for these studies. Measurements of eN interactions would still be possible to do with a tungsten target, and could be incorporated into global fits to improve neutrino generators, but the difference in nuclear size may lead to larger uncertainties in those results.

We can estimate the expected statistics for an inclusive measurement, and a variety of semi-inclusive and exclusive measurements, assuming an 8 GeV electron beam incident on a titanium target. We assume triggering capabilities laid out in 4.9.2, with a cut on the final electron $p_T > 500$ MeV/$c$ for an inclusive trigger. With such a cut, we expect roughly half a million events from eN scattering processes per $10^{12}$ electrons on target, corresponding to only a few days of data-taking. With expected sample sizes possible of tens to hundreds of times that, this will be a significant dataset that would allow for many multi-dimensional cross-section measurements.

For semi-inclusive measurements, we define a 'reconstructable' categorization where we require final-state particles to be within 80° of the beam axis, with a momentum threshold of 100 MeV/$c$ for charged pions, 800 MeV/$c$ for protons and kaons, and an energy threshold of 1 GeV for neutrons. For $\pi^0$, see Sec. 4.9.3.2 for details of the event selection, as it requires reconstruction of both photons, and with well-reconstructed invariant mass, the efficiency for reconstructable photons is noticeably lower than for other particles. The summary of expected number of events in various semi-inclusive final state 'reconstructable' topologies is shown in Tab. 4.15. A confusion matrix for final states with protons, charged pions, and neutral pions is shown in Tab. 4.16. Note that due to the limited acceptance in LDMX, it is often the case that the total true final state is not properly reconstructed. However, that does not imply that the partial reconstruction cannot make meaningful measurements, as by modeling the acceptance, well corrections can be made on extrapolations to the overall cross sections.

| | Trigger Events | Reconstructable Events | Reconstructable Efficiency |
|---|---|---|---|
| $N\pi^0 + X$ | 230000 | 57883 | 25.2% |
| $N\pi^\pm + X$ | 296116 | 246464 | 83.2% |
| $N\pi^0 + Np + X$ | 183403 | 16391 | 8.9% |
| $N\pi^\pm + Np + X$ | 240418 | 198032 | 82.4% |
| $N\pi^0 + Nn + X$ | 191761 | 36594 | 19.1% |
| $N\pi^\pm + Nn + X$ | 235296 | 193969 | 82.4% |
| $NK^\pm + X$ | 12512 | 5901 | 47.2% |
| $NK^0_s + X$ | 6739 | 6739 | 100.0% |
| $NK^0_l + X$ | 6733 | 6733 | 100.0% |

Table 4.15: Expected number of observed events for $1 \times 10^{12}$ electrons on target that are likely to be triggered, categorized by the expected final-state categorization based on reconstructable particle requirements. $N$ is taken to be 1 or more, while $X$ may be anything additional (or nothing additional).

| Truth \ Reco | $0\pi^0 + X$ | $1\pi^0 + X$ | $2\pi^0 + X$ | $\geq 3\pi^0 + X$ |
|---|---|---|---|---|
| $\geq 3\pi^0 + X$ | 3388 | 2552 | 794 | 111 |
| $2\pi^0 + X$ | 11053 | 5716 | 760 | 0 |
| $1\pi^0 + X$ | 49235 | 11481 | 0 | 0 |
| $0\pi^0 + X$ | 112388 | 0 | 0 | 0 |

| Truth \ Reco | $0p + X$ | $1p + X$ | $2p + X$ | $\geq 3p + X$ |
|---|---|---|---|---|
| $\geq 3p + X$ | 43284 | 11761 | 1133 | 80 |
| $2p + X$ | 18712 | 9259 | 613 | 0 |
| $1p + X$ | 49051 | 27772 | 0 | 0 |
| $0p + X$ | 35813 | 0 | 0 | 0 |

| Truth \ Reco | $0\pi^\pm + X$ | $1\pi^\pm + X$ | $2\pi^\pm + X$ | $\geq 3\pi^\pm + X$ |
|---|---|---|---|---|
| $\geq 3\pi^\pm + X$ | 101 | 636 | 1442 | 1525 |
| $2\pi^\pm + X$ | 1932 | 8807 | 14132 | 0 |
| $1\pi^\pm + X$ | 33837 | 93254 | 0 | 0 |
| $0\pi^\pm + X$ | 239290 | 0 | 0 | 0 |

Table 4.16: "Confusion" matrix, showing how often events may be misclassified based on expected reconstructable final states, relative to the "true" final state. Diagonal elements show properly classified events, while off-diagonal elements show where the reconstructed and truth differ.

### 4.9.2 Trigger strategies

The main trigger for eN studies in LDMX will be an "inclusive" electron trigger based on a calculation of the outgoing electron $p_y$ magnitude (the momentum transverse to the beam axis, in the non-bending direction). See Sec. 3.9.5.5 for more details of the implementation of this inclusive eN trigger[1]. Using this trigger, we show the efficiency for eN interactions as a function of the $p_y$ magnitude in Fig. 4.49. We expect that above a $|p_y|$ of 500 GeV/c that the inclusive trigger will be highly efficient across interaction types and final states, with little expected bias. We do see some decreased efficiency at higher $|p_y|$ (and increased efficiency at lower $p_T$ ) in final states with $\pi^0 \to \gamma\gamma$ into the ECal where the presence of additional photons can interfere with the electron $p_T$ calculation. When comparing an alternative model set (see Fig. 4.50), we see very similar performance in the trigger turn-on curves, and still see very little bias when breaking down by interaction type, building confidence that this trigger won't induce a significant model-dependent efficiency correction.

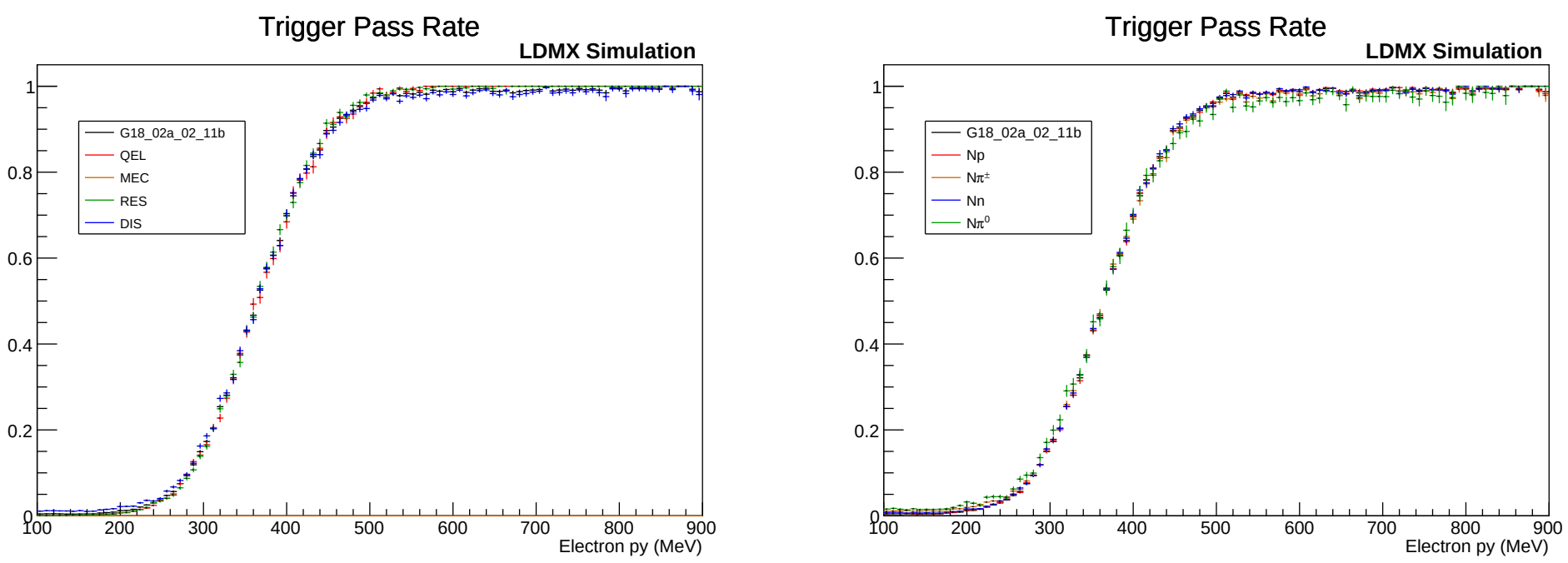

Figure 4.49: Efficiency for the inclusive eN trigger as a function of $p_y$, broken down by interaction type (left) and reconstructible final states (right). There are no large biases in interaction type or final state induced by the trigger, with some minor deviations seen in final states with $\pi^0$'s.

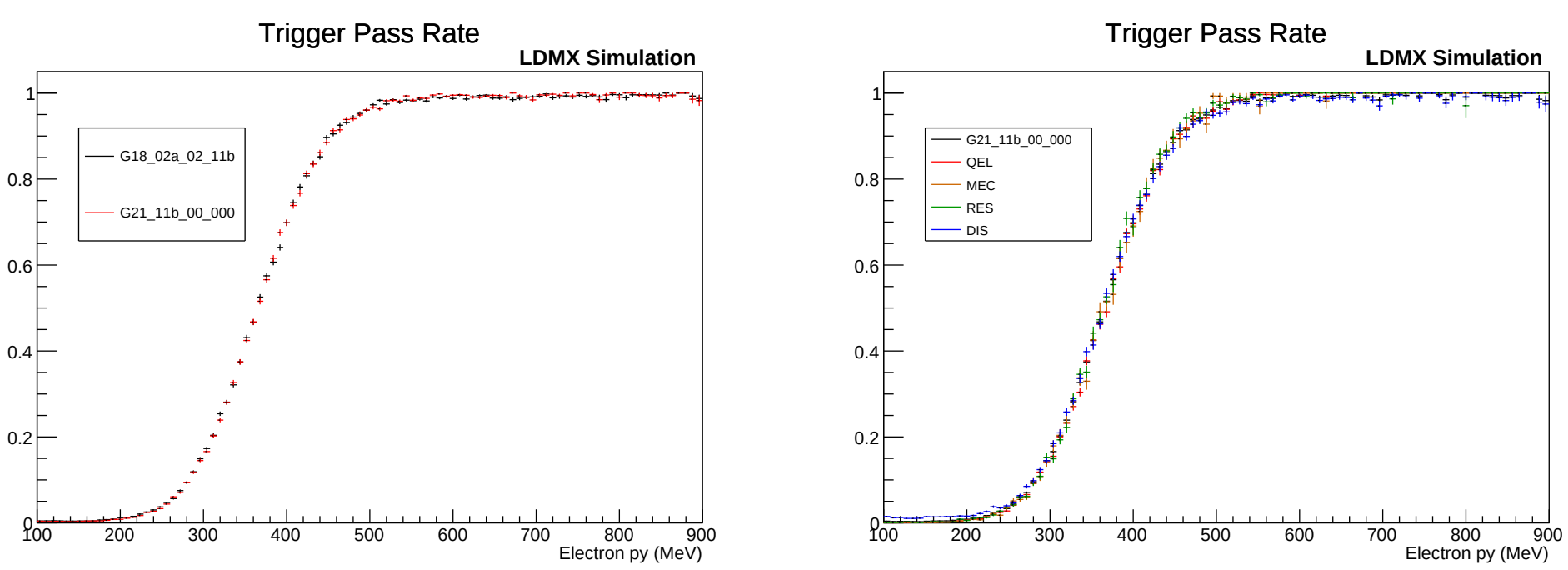

Figure 4.50: Efficiency for the inclusive eN trigger as a function of $p_y$, comparing two GENIE model sets (left) and with a breakdown by interaction type in the `G21_11b_00_000` model set (right). There is no significant difference between the model sets.

The primary backgrounds requiring a $p_y$ threshold of 500 GeV/c are from hard bremsstrahlung interactions, which should have distinct signatures relative to eN interactions when considering the tracking detectors and the HCal. Thus, there is potentially additional reach for eN interactions with final-state electron $p_y$ below 500 GeV/c with the inclusion of additional information in the trigger, either through use of the HCal in the hardware trigger, or with the additional of tracking and something like full event reconstruction in a high-level-trigger. We will continue to explore these possibilities, which would allow for interesting samples of eN interactions in semi-inclusive final states at lower energy transfer than the inclusive eN trigger would allow on its own.

[1]Measurement of the electron $p_x$ at trigger level is not possible with the baseline design, but presents a future opportunity that may be achieved with modifications to the TS.

### 4.9.3 Particle reconstruction

#### 4.9.3.1 Charged particles

Detailed descriptions of the tracking detectors and tracking reconstruction are found in Sections 3.4.3 and 3.4.5. The tracking detectors will be essential for identifying charged particles within an acceptance range of $\pm 40^\circ$ relative to the incident electron.

Current tracking reconstruction is still being developed, and among other quality cuts, requires a minimum of 7 hits across the layers of the recoil tracker. The tracking reconstruction efficiency as a function of momentum for single-particle simulation samples is shown in Fig. 4.51, comparing reconstruction between truth track-seeding and energy depositions with purely reconstructed quantities. Current reconstruction is quite performant for electrons and charged pions, even down to low momenta. For protons, the current reconstruction performance struggles below 800 MeV/c due to the requirement on the number of hits; however, truth-tracking shows that a good efficiency to 200 MeV/c in momentum is achievable, and we can anticipate improvements in performance with reconstruction.

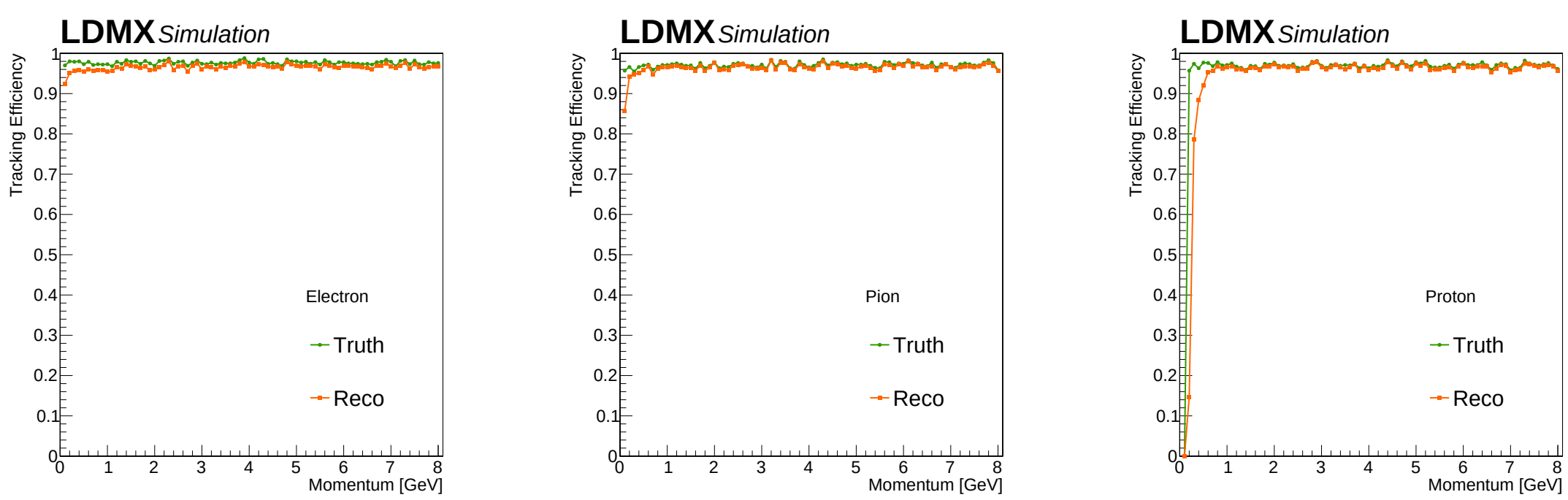

Figure 4.51: Tracking efficiency as a function of particle momentum for electrons (left), charged pions (center), and protons (right) from single-particle simulation samples. Performance of the current reconstruction is shown in orange, with performance based on truth quantities (representing what may be achievable with future work) shown in green.

In addition to the use of the recoil tracker as a tracking detector, we can leverage dE/dx measurements in the recoil tracker to aid in particle ID. We construct a dE/dx estimate [177] based on the hits in the recoil tracker, defined as

$$I_h = \left( \frac{1}{N} \sum_j^N \left( \frac{dE}{dx} \right)_j^{-2} \right)^{-\frac{1}{2}}$$

We show in Fig. 4.52 a comparison of this quantity on truth tracks versus particle momentum for simulated charged pions and protons. For high momentum tracks, when both particles are mostly minimum ionizing, there is little discriminating power, but below 1 GeV/c in momentum, we can see higher dE/dx for protons. Since most charged hadrons in eN interactions will be at GeV momenta or lower, we expect that the recoil tracker will be able to aid in particle identification in eN events. Further studies are ongoing to develop a single-value estimator for the particle mass based on the dE/dx vs. momentum, as well as extending and validating these studies with reconstructed tracks.

Inverting the Bethe-Bloch formula we calculate a mass estimate by

$$m = p \times \sqrt{\frac{I_h - C}{K}}$$

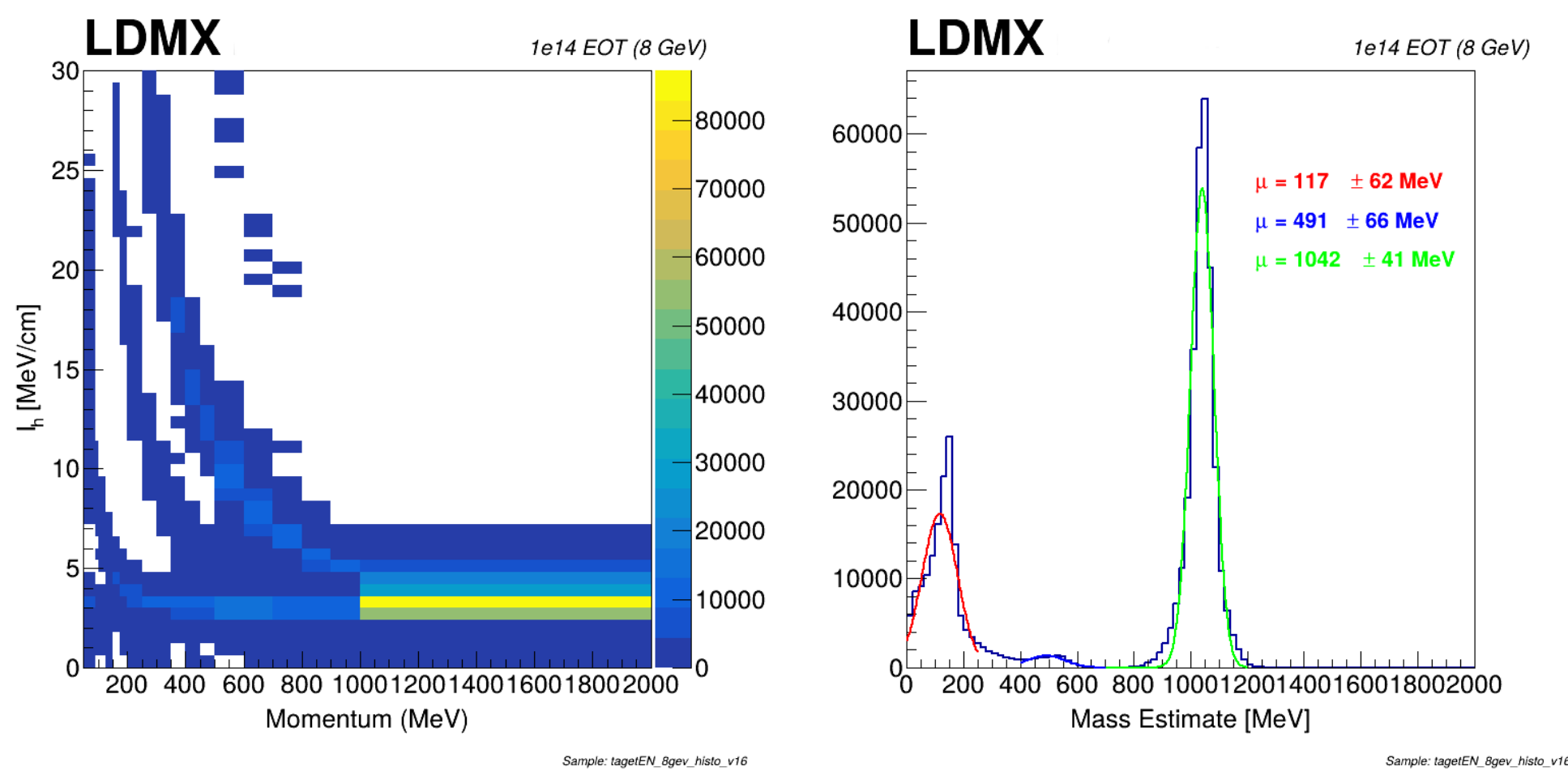

Figure 4.52: An estimate of dE/dx from the recoil tracker versus particle momentum for all tracks in the electro-nuclear sample (left). We see clear discrimination power below 1 GeV/c momentum, indicating the ability to use the recoil tracker as an aid in particle identification. The extracted mass estimate for all tracks below 1 GeV momentum (right).

where $K = 1.862$ and $C = 3.094$ are empirical constants that we determined with single-pion particle guns. We can see a very good separation for the pions and protons, and we can see the kaon peak as well. The estimated masses are within the uncertainty of the real values for the kaons and pions, and it is very close to the proton too.

#### 4.9.3.2 Neutral pions

Neutral pions can be produced in neutrino interactions through resonant production, coherent production, inelastic scattering processes, and through final-state interactions of hadrons propagating through nuclear material. This makes neutral pions a particularly interesting final state to reconstruct in eN interaction as well, but it is challenging to do so in the LDMX detector due to the limited acceptance, where the photons produced in $\pi^0$ decays may not be captured in the detectors.

To investigate our ability to reconstruct neutral pions, we have studied eN interactions of 4GeV electrons scattering off of a titanium target, grouping the decay photons into three topologies defined by which calorimeter the photons deposit at least 90% of their reconstructed energy: ECal-ECal, ECal-HCal, and HCal-HCal. Here, HCal most often refers to the side HCal, which significantly extends LDMX's ability to detect photons from neutral pions. Events where at least one photon does not deposit energy into either the ECal or HCal are not considered in this study.

We show the reconstructed neutral pion's invariant mass in Fig. 4.53, after calibrations based on the differing response of the ECal and HCal are applied. We see a clear invariant mass peak, and can use this reconstructed di-photon mass to form an event selection for $\pi^0$s, represented by dashed lines in Fig. 4.53. The better energy resolution of the ECal allows for a tighter energy window on ECal-ECal events. $\pi^0$ decays in the ECal-HCal and HCal-HCal categories show a poorer resolution, but may still be reconstructed and utilized in event selections.

Using these reconstructed $\pi^0$ mass cuts as part of an event selection, we show in Fig. 4.54 the efficiency as a function of the pion's kinematics from our simulation. From this, it is clear that neutral pions below about 30° relative to the beam axis need to have sufficient energy to boost the decay photons forward enough to be measured by the ECal. Another important feature to note is that, since the side HCal has such a large angular coverage, lower energy pions can be measured with relatively high efficiency, overcoming the hurdle of the larger opening angle between the decay photons. There is a notable drop in efficiency in the region where the $\pi^0$ angle relative to the beam is between 30° and 45°, but this due to the photon selection and categorization, which requires that *either* the ECal or HCal contains 90% of the deposited energy—in this angular range, there is overlap between the ECal and HCal, and so there is more energy sharing between the two and will require cross-calorimeter clustering and energy reconstruction. This effort is underway, and we expect to recover efficiency in this region.

In conclusion, LDMX will be able to precisely measure neutral pions with high efficiency for pions above about 2 GeV. The angular extent of the side HCal significantly improves the phase space of neutral pions that LDMX can detect, highlighting its importance in maximizing the reach of eN measurements in LDMX.

#### 4.9.3.3 Neutrons

Of particular interest for the LDMX eN program is the use of the large back and side HCal detectors to identify neutral hadrons, particularly neutrons for which there is limited available data from electron scattering experiments. As the primary purpose of the HCal system in LDMX is to act as a veto for rare PN events with minimal deposited energy in the ECal, the HCal will be highly efficient at tagging neutral hadrons. Beyond simple tagging of neutral hadron activity, the HCal is capable of performing energy and position reconstruction, and may have a timing resolution that will aid in the PID of neutrons, as described in Sec. 3.8.7. Further development of the reconstruction of neutrons and neutral hadrons is in progress, including studies of depositions in both the ECal and HCal (important for lower energy neutrons), and the use of shower development profiles to improve energy resolution.

### 4.9.4 Observable distributions in LDMX

As described in Sec. 4.9.1, even with a modest exposure of electrons on target, LDMX will see a large number of electro-nuclear interactions. With such a high level of statistics available, LDMX will be capable of a set of inclusive, semi-exclusive, and exclusive electron scattering measurements that will inform neutrino interaction models in the regions most relevant to DUNE, particularly in resonant and DIS interactions.

For inclusive electron measurements, measurements that explore the energy and momentum transfer as a function of outgoing electron angle (or vice versa) will be highly valuable to neutrino modeling efforts.

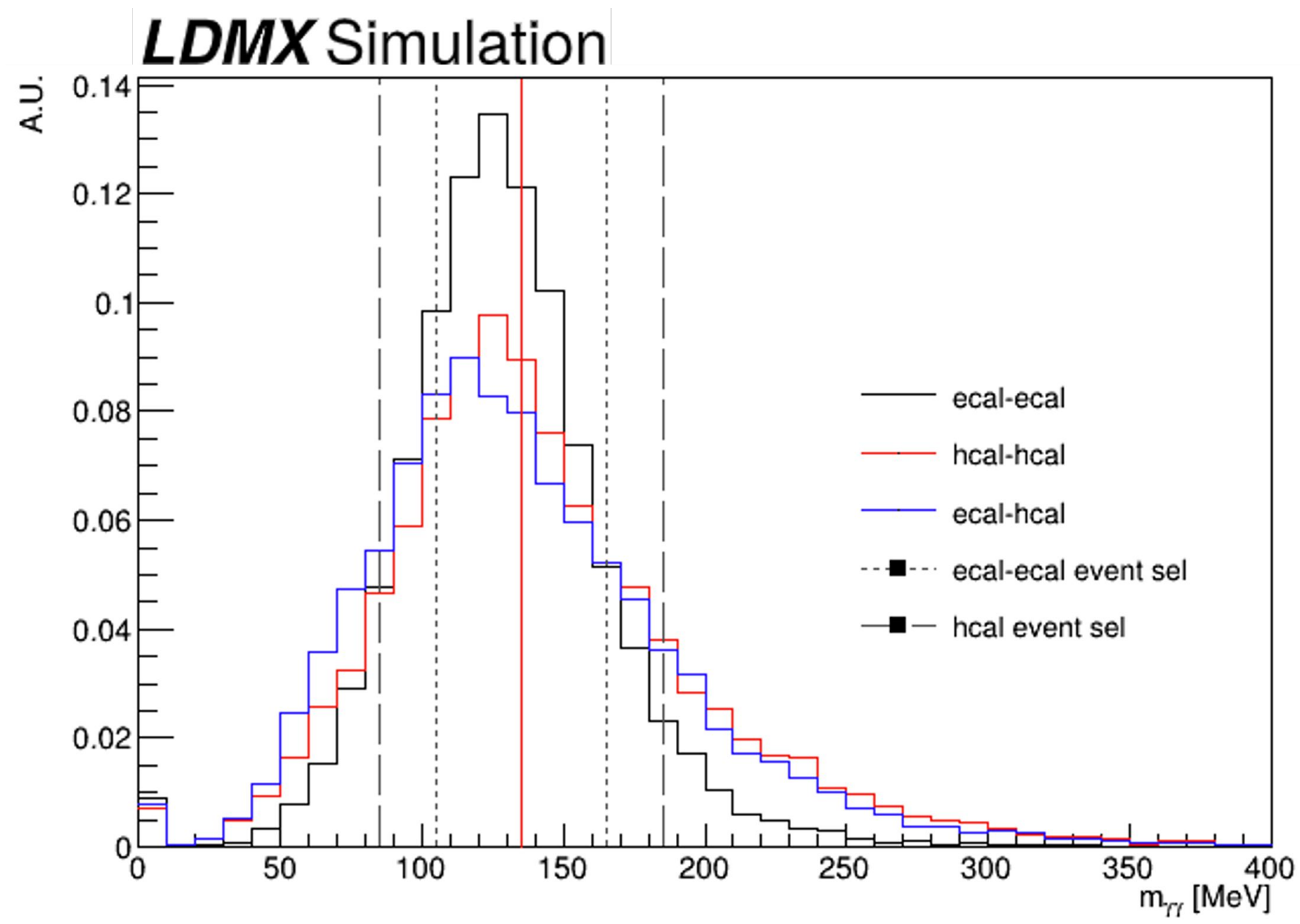

Figure 4.53: The invariant mass of two decay photons from neutral pions for the three topologies: ECal-ECal, ECal-HCal, and HCal-HCal shown with the true neutral pion mass in red and the two event selections. The resolution for ECal photons is better, leading to a narrower event selection window. Since the resolution for photons in the HCal is lower, the other two topologies have a wider event selection.

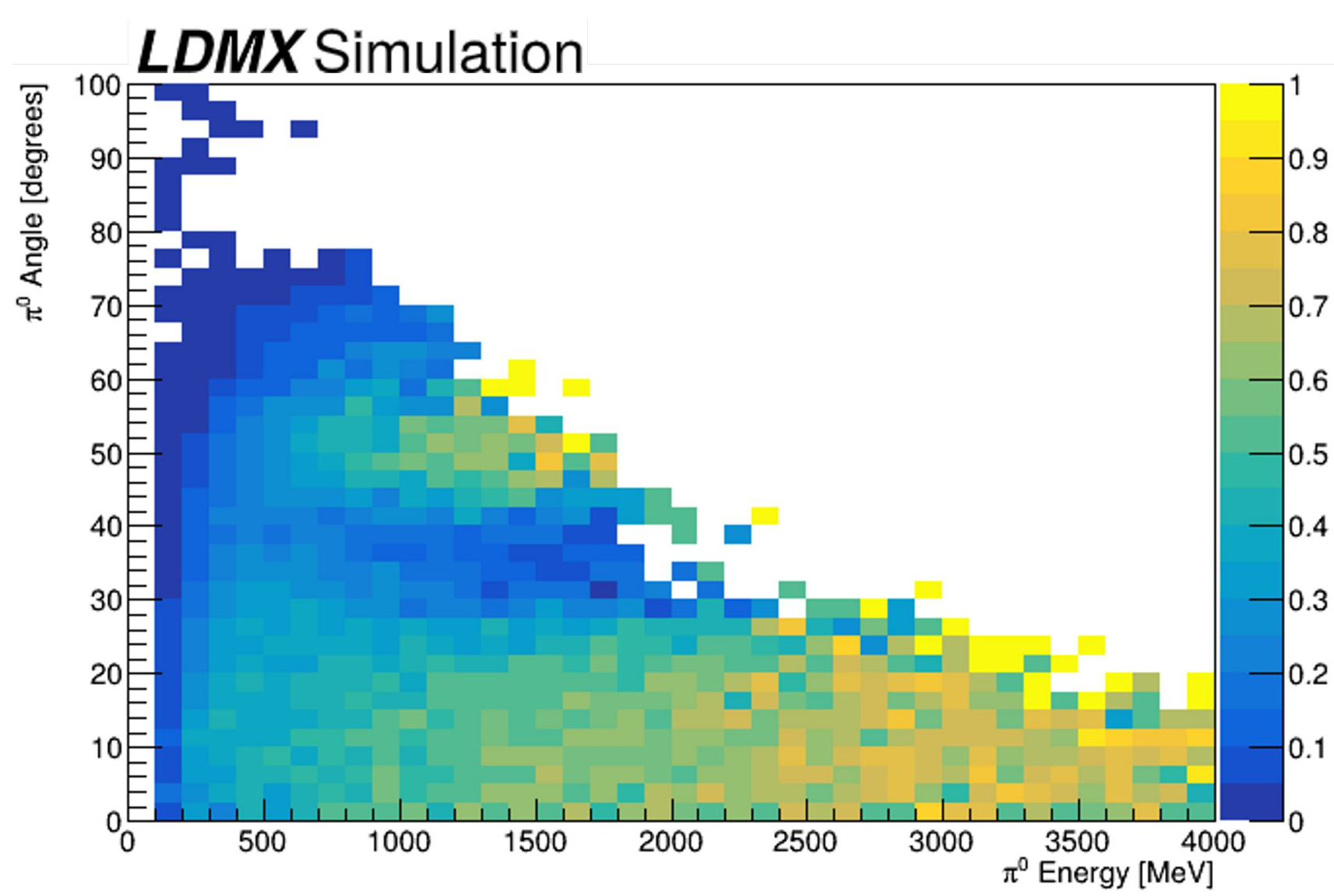

Figure 4.54: The combined efficiency of neutral pions, as a function of true energy and angle, for all three topologies. Efficiency here is defined as the ratio of neutral pions that meet the previously shown event selection to all neutral pions, regardless of whether or not the decay photons can be detected. Below 30°, the efficiency is dominated by the ECal-ECal topology with only a marginal contribution from the ECal-HCal events. The highest efficiency actually occurs at angles greater than 30° in the HCal-HCal topology. This study assumed a sufficiently high resolution on HCal photons, which the real detector may not meet.

Ankowski *et al.* [69] show large differences in standard neutrino model generators in such distributions, such that LDMX will be able to readily disambiguate between available interaction models. Similar distributions, comparing two model tunes from the most recent version of GENIE (v3.4), are shown in Fig. 4.55, where differences in both shape and normalization are readily apparent.

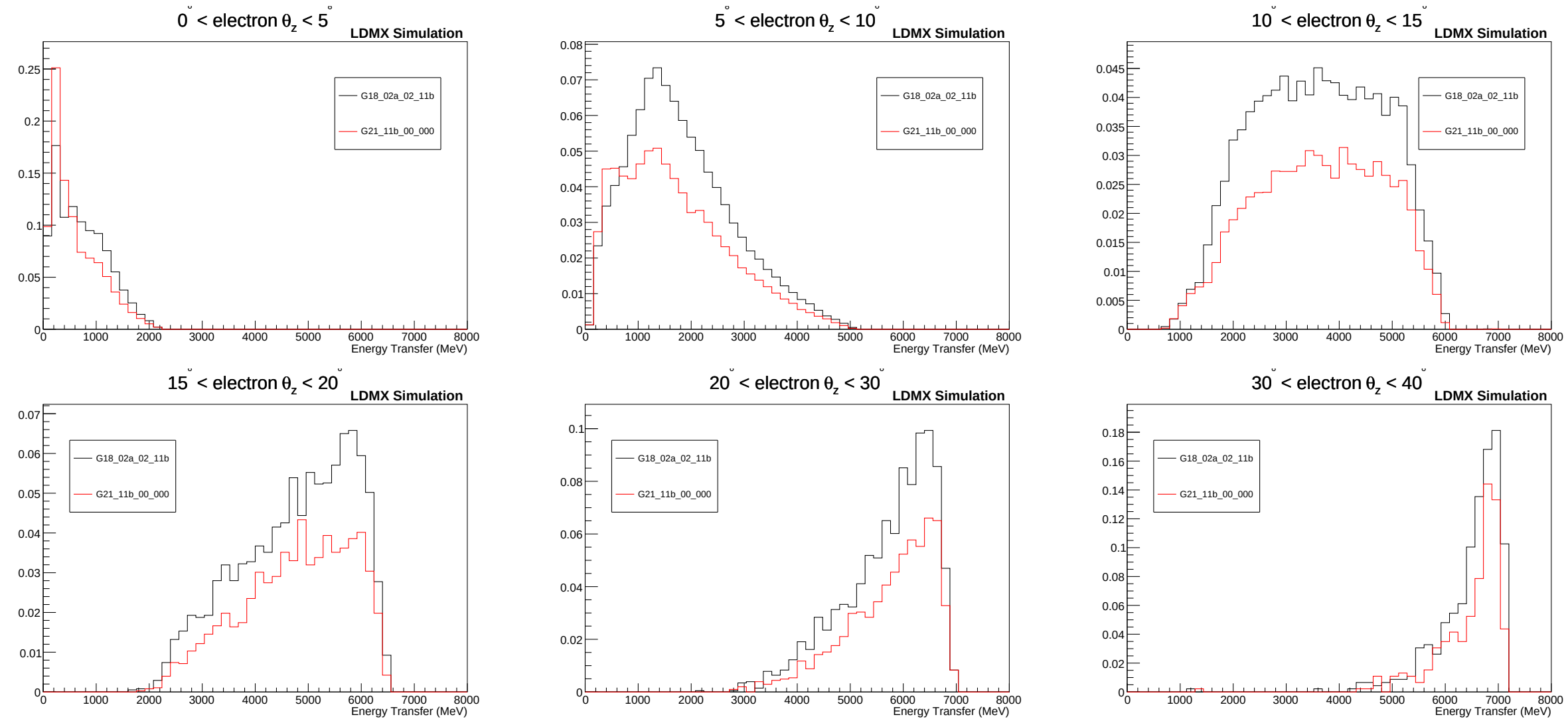

Figure 4.55: Distribution of energy transfer in eN interactions in bins of outgoing lepton angle for two different GENIE model tunes. A minimum electron $p_T$ of 500 MeV/$c$ is required to match expected trigger conditions.

Beyond measurements of the outgoing lepton, observations of outgoing hadrons and measurements of their kinematics also provide important handles on both interaction types and parameters, as well as final state interactions of hadrons as they traverse and exit the nuclear environment. Fig. 4.56 shows a simple track multiplicity comparison between GENIE model sets and tunes, counting just the number of protons and the sum of protons and charged pions that we may expect to observe within LDMX. Multiplicity of hadrons is sensitive to both interaction models but also final state interaction models, and so can already provide some discrimination between models. Fig. 4.57 shows a more detailed comparison of the sum of kinetic energy for neutrons, protons, and pions within an angle of 40° of the initial electron beam—where reconstruction and particle ID is expected to be of highest quality—between two different GENIE model sets that differ in their handling of the initial nuclear state, QE and MEC models, and their treatment of final state interactions. (Fig. 4.57 uses an initial electron beam energy of 4 GeV, though the results will be expected to be similar for 8 GeV.) While the differences between protons and neutrons appear more subtle, the large statistical samples available from LDMX will be able to measure the differences in the spectrum at low and high kinetic energies. The differences in pion energies, driven by the handling of FSI effects, is more clearly visible and would be easily probed by LDMX.

More challenging, but potentially highly sensitive to initial and final nuclear state modeling, are measurements using momentum imbalance variables. An example of shape comparisons of $\delta p_T$, defined as the magnitude of the difference in vector $p_T$ between the outgoing electron and hadronic system, across two GENIE model sets is shown in Fig. 4.58. (Fig. 4.58 also uses an initial electron beam energy of 4 GeV, though again, the results are expected to be similar for 8 GeV.) In this variable, the position of the peak of the $\delta p_T$ distribution is sensitive to the initial nuclear momentum, while the shape of the distribution, particularly in the tails, is sensitive to final state interactions. Shown are both assumptions of a "perfect" detector where all final-state hadrons can be observed, versus the case where observed hadrons are restricted to the angular range of $\theta_z < 40°$. The large differences in the models are clear in the case of a perfect detector, but the $\delta p_T$ distributions are significantly sculpted by the acceptance for well-reconstructed final-state hadrons. Still, there are notable differences in these distributions between the two model sets, and with both careful modeling and potential improvements in the detector and reconstruction performance at high angles, the ability to utilize such measurements to parameterize nuclear modeling is very promising.

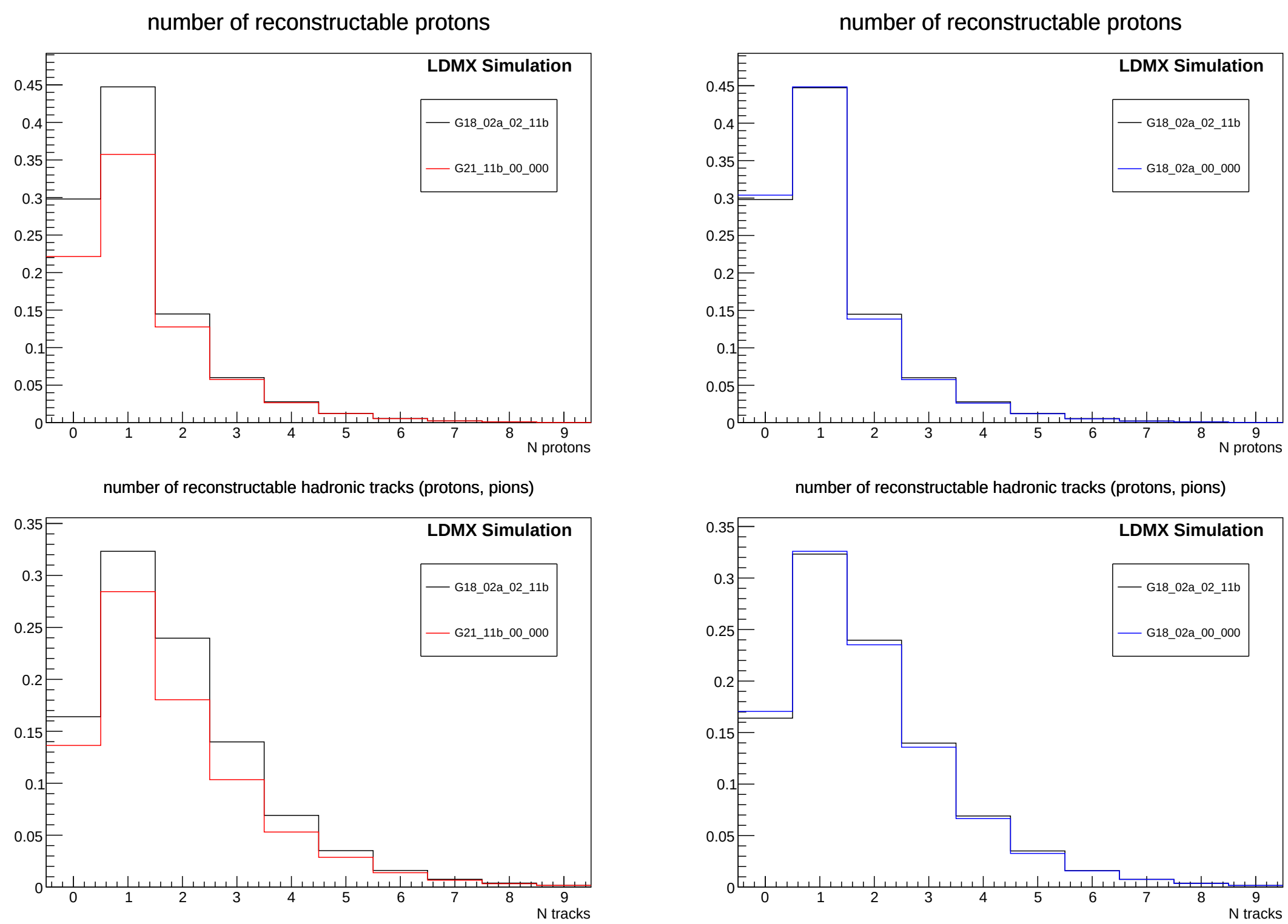

Figure 4.56: Comparison of proton (top) and "charged hadron" (protons and charged pions, bottom) multiplicity between GENIE models and tunes. Distributions are after an electron $p_T > 500$ MeV/$c$ cut modeling an inclusive trigger, and include hadrons expected to be "reconstructable". Distributions are normalized relative to unit area of GENIE G18_02a_02_11b. Even in comparison between two tunes of the same model (right-hand side), we see differences that, with the statistics of LDMX, would be observable.

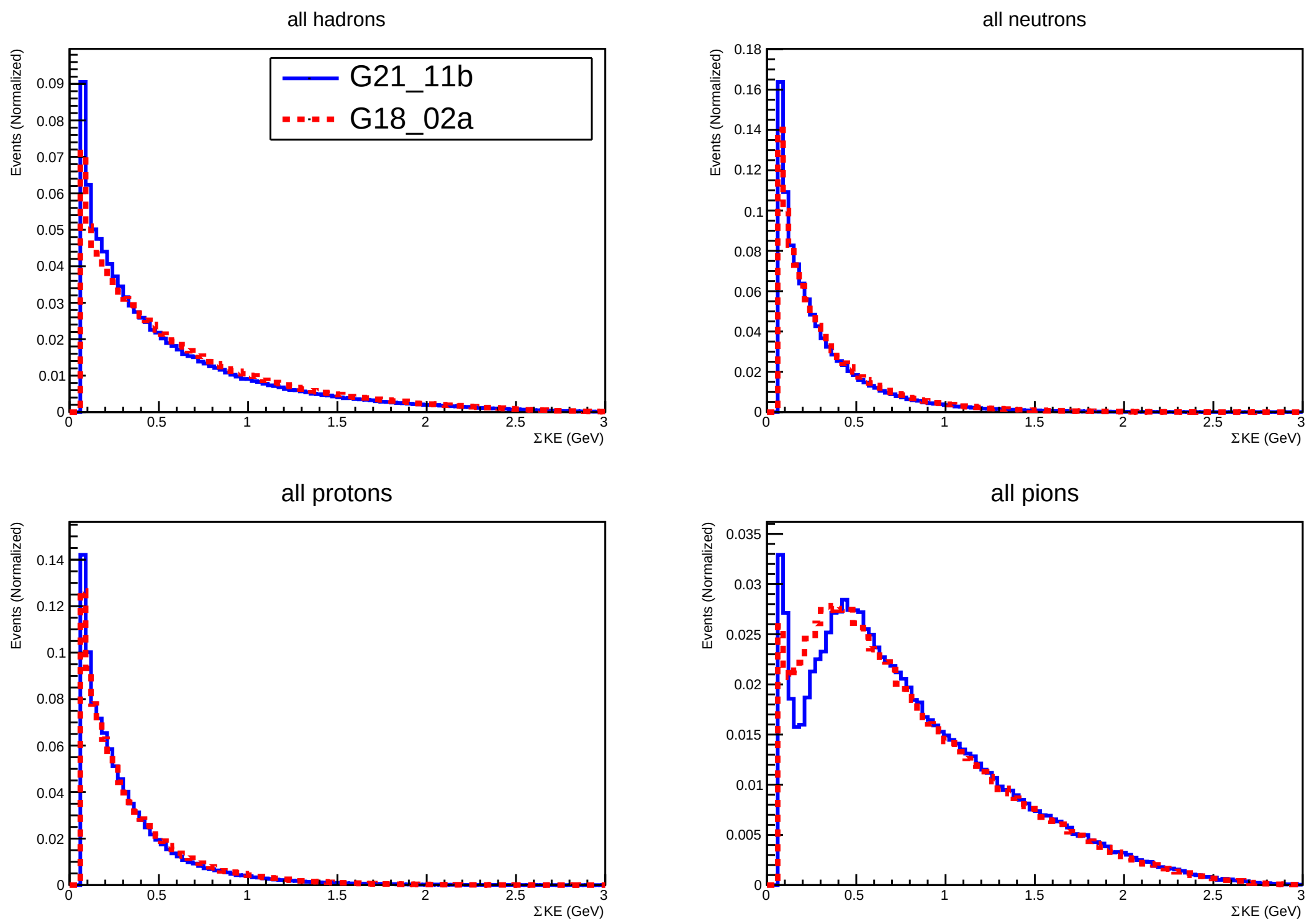

Figure 4.57: Comparison of the sum of kinetic energies for outgoing hadrons produced in eN interactions (with initial electron beam energy of 4 GeV) in two different GENIE model sets for all hadrons (top left), neutrons (top right), protons (bottom left), and pions (bottom right). Events are required to have a lepton $p_T > 400$ MeV$/c$, $\theta_e < 40°$, and require all hadrons to also have $\theta_z < 40°$ (except for neutral pions, where the decayed photons are required to be within that range). Distributions are normalized to area.

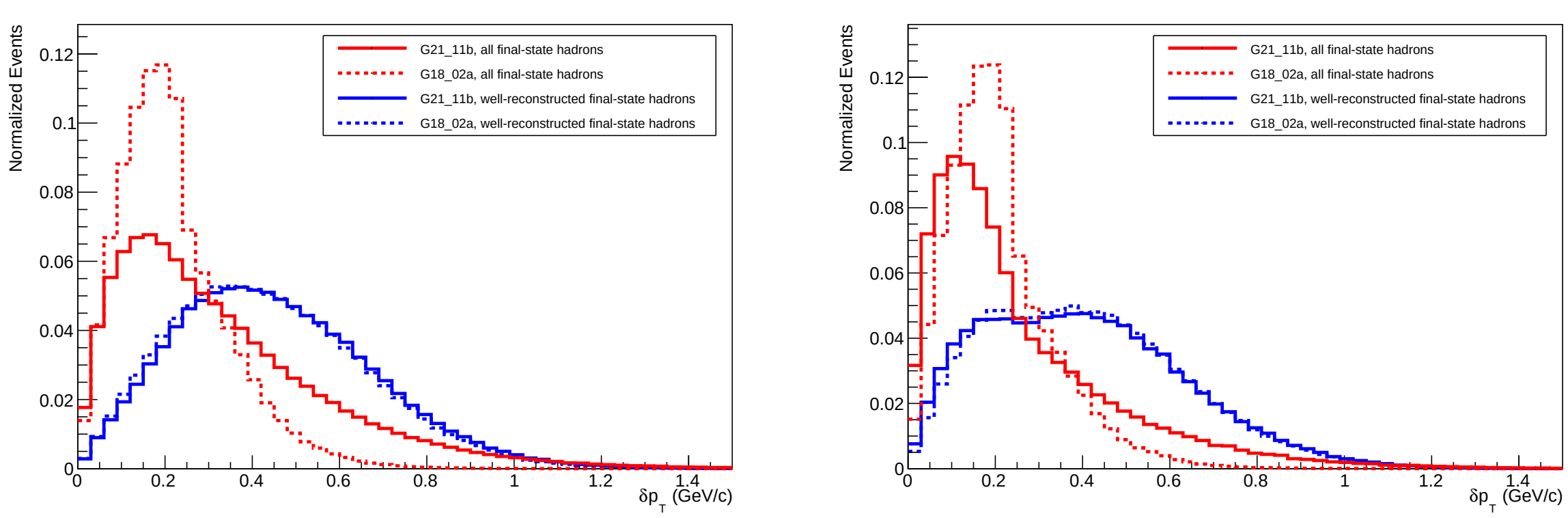

Figure 4.58: The transverse momentum imbalance (the $p_T$ difference between the outgoing electron and reconstructed hadronic system) for eN interactions (with initial electron beam energy of 4 GeV) with no pions (left) and with pions (right) in the final state for two different GENIE model sets (solid and dashed). Red curves show the imbalance assuming a 'perfect' detector where all final state hadrons are observed and reconstructed, while blue curves limit require hadrons be within $\theta_z < 40°$. Distributions are normalized to area.

# Chapter 5

# Conclusions

The search for MeV-GeV scale thermal relic dark matter at accelerator-based experiments is highly motivated as part of a comprehensive global program aiming to unveil the particle nature of dark matter. In this report, we have demonstrated that LDMX, with its unique missing-momentum approach, will play an important part in this, being able to achieve improvements of factors 10-1000 in sensitivity, compared to existing constraints. This is equivalent to exploring couplings well beyond the thermal targets over a wide range of dark matter masses.

The detector design is driven by the goals to exploit the LCLS-II accelerator to collect up to $10^{16}$ electrons on target within a few years, maintaining a high efficiency for the dark matter signal while reducing backgrounds to a negligible level — without making use of the recoil $p_\mathrm{T}$. This is possible in principle since any irreducible backgrounds to the dark-matter signal occur only at rates well below the targeted sensitivity for the planned beam energy of 8 GeV. These goals translate into stringent requirements on the experiment's capabilities to be able to individually measure several incoming electrons at the same time and efficiently detect reactions in which a significant fraction of the incoming electron energy is transferred to a small number of final-state particles, e.g. single-neutron final states of photo-nuclear reactions or two MIPs from photon conversions into muons. These requirements are introduced in Sec. 3.2 where they are illustrated based on Fig. 3.3. Chapter 3 develops these requirements further and describes the solutions adopted by LDMX, many of which are based on proven designs from other experiments, reinforcing the viability of the detector design. Enabled by the advanced simulation and analysis software suite described in Sec. 3.10.3.3, Chapter 4 demonstrates that the developed design comfortably meets the specified requirements for $10^{14}$ EoT and beyond, and outlines further developments and strategies to achieve up to $10^{16}$ EoT. It also shows how in the case of an observed excess, information about the mass scale of the new particles can be extracted. The same chapter also gives a glimpse of the broader physics potential of LDMX beyond the thermal relic dark matter search, including searches for long-lived, visibly decaying particles and measurements of electro-nuclear processes.

In summary, this report presents a mature, viable detector, TDAQ, and software design to successfully perform a missing-momentum search that significantly extends the reach into the parameter space of light dark matter models, making the timely realization of LDMX an important milestone for the field.

# Acknowledgments

Support for UCSB is made possible by the Joe and Pat Yzurdiaga endowed chair in experimental science. Use was made of the UCSB computational facilities administered by the Center for Scientific Computing at the California NanoSystems Institute and Materials Research Laboratory (an NSF MRSEC; DMR-1720256) and purchased through NSF CNS-1725797, and from resources provided by the Swedish National Infrastructure for Computing at the Centre for Scientific and Technical Computing at Lund University (LUNARC), as well as LUNARC's own infrastructure. Contributions from Caltech, CMU, Stanford, TTU, UMN, UVA, and UCSB are supported by the US Department of Energy under grants DE-SC0011925, DE-SC0010118, DE-SC0022083, DE-SC0015592, DE-SC00012069, DE-SC0007838, and DE-SC0011702, respectively. Support for Lund University is made possible by the Knut and Alice Wallenberg foundation (project grant Light Dark Matter, Dnr. KAW 2019.0080), and by the Crafoord foundation (Dnr 20190875) and the Royal Physiographic Society of Lund. RP acknowledges support through the L'Oréal-UNESCO For Women in Science in Sweden Prize with support of the Young Academy of Sweden, and from the Swedish Research Council (Dnr 2019-03436). LB acknowledges support from the Knut and Alice Wallenberg Foundation Postdoctoral Scholarship Program at Stanford (Dnr. KAW 2018.0429). JE, GK, CH, WK, CMS, and NT are supported by the Fermi Research Alliance, LLC under Contract No. DE-AC02-07CH11359 with the U.S. Department of Energy, Office of Science, Office of High Energy Physics and the Fermilab LDRD program. CB, PB, OM, TN, PS, and NT are supported by Stanford University under Contract No. DE-AC02-76SF00515 with the U.S. Department of Energy, Office of Science, Office of High Energy Physics.

# Appendix A

# The LDMX Radiation Environment

The LDMX detector will experience roughly $10^{15}$ 8-GeV electrons on target (EoT) over the course of its operational lifetime. As seen below, this means that some areas of LDMX will integrate significant radiation doses, albeit much lower than the expectations for the CMS HL-LHC upgrade calorimeter (HGC). This can have implications for the design of LDMX. For example, as for the CMS HGC, the LDMX ECal will require the radiation tolerance provided by silicon. On the other hand, while the CMS HGC peak exposure regions will require the use of very thin silicon operated at -35°C to ensure full depletion with reasonable charge collection and low noise, LDMX will use 400$\mu$m thick sensors for a higher signal-to-noise ratio, and operate at a less extreme low temperature. Exactly what operating temperature would be optimal requires a detailed simulation of the LDMX radiation environment, as presented here.

Similarly, for other parts of the apparatus, such as the HCal where the sensing layers are plastic scintillators with lower radiation tolerance than silicon, knowledge of the detector's radiation environment has potential implications for detector design.

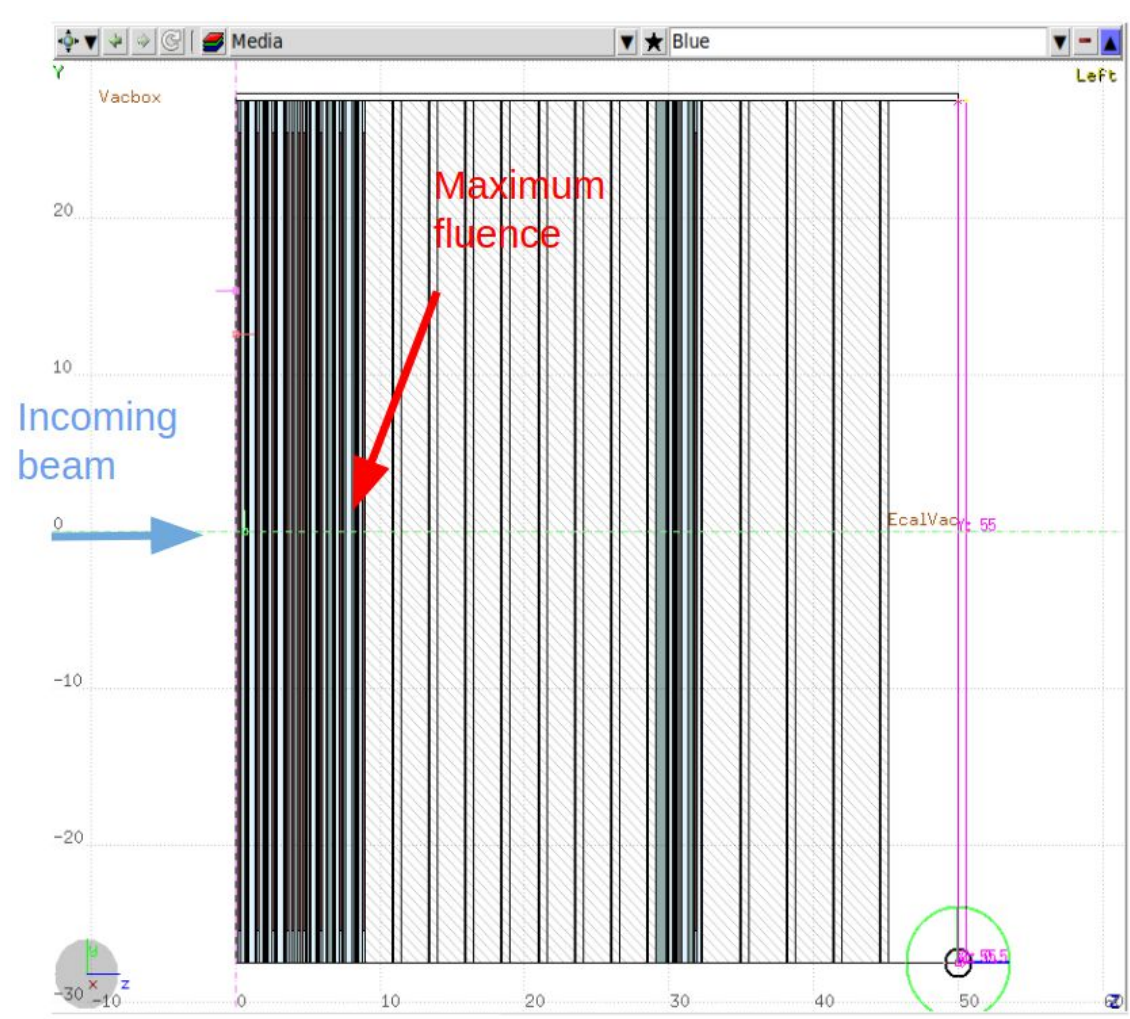

Figure A.1: The FLUKA ECal model as depicted by FLAIR, a user interface for FLUKA. White regions contain copies of C-type and D-type layers, except for the back-most region at right, which is left empty. A red arrow highlights the shower-max region, which occurs in the vicinity of layer 8, which is at the transition from B-type to C-type layers.

For our radiation studies, both the construction of the detector model and the simulations made use of the multi-particle-transport code FLUKA in combination with the FLUKA Advanced Interface (FLAIR) [178, 179]. The ECal model created in FLUKA is similar to the full design with a few simplifications. Most notably, the flower-shaped sensor plane of 7 hexagonal modules was replaced by a square region that just contains the flower. The tungsten (W) absorber layers and the carbon layers of the two sides of the cooling plane were extended beyond the sensor region to the edges of the ECal box. Furthermore, small items like readout chips were not included, since they add complexity to the model without absorbing a significant amount of radiation. All empty spaces were filled with air.

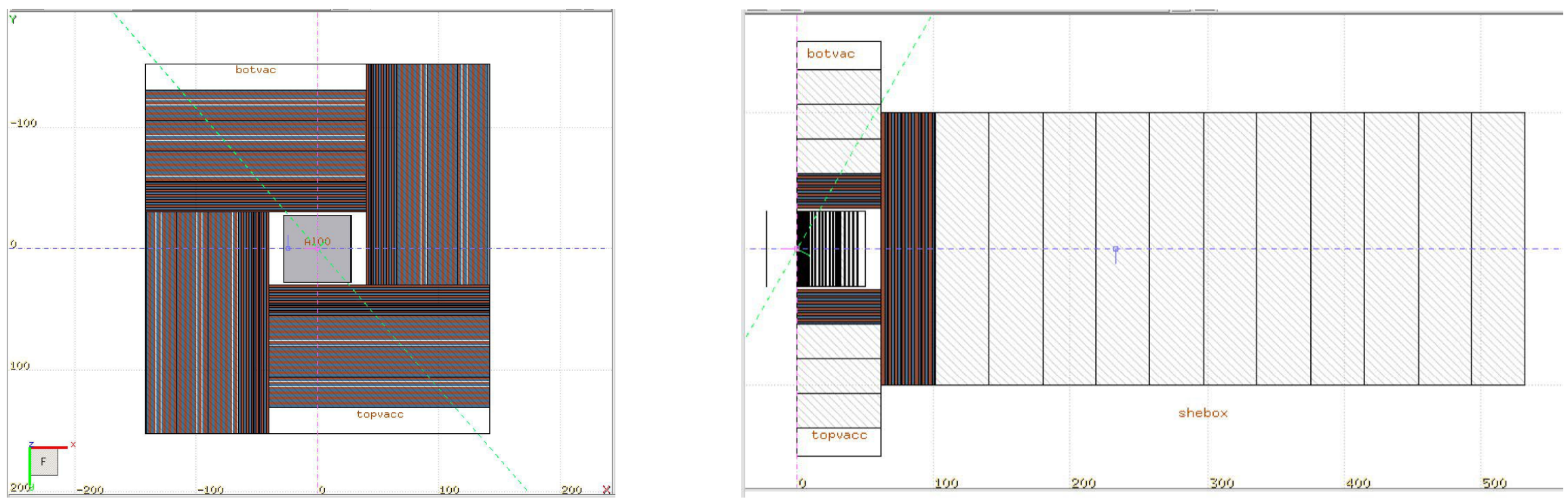

Figure A.2: The combined FLUKA ECal and HCal model created with FLAIR. Left: The ECal inside the side HCal modules as seen from the beam-axis. Right: The HCal surrounding the ECal as seen from above. The HCal layers included in the study are the dark layers nearest to the ECal.

For all plots shown in this Appendix, a baseline of $10^{15}$ electrons on target was selected as an intermediate value between the $4 \times 10^{14}$ EoT expected during a Phase I operation and the $\geq 10^{15}$ EoT expected during a Phase II. A 2 cm $\times$ 8 cm rectangular beamspot was used. Due to computational constraints, it was not possible to simulate a full $10^{15}$ events with FLUKA. Instead, smaller samples of order $10^6$ to $10^7$ EoT were used, with all measured quantities scaled linearly to $10^{15}$ events. In each region of interest we measured fluence [1-MeV-neq/cm$^2$] and dose [Rad]. The ECal simulation assumed a silicon sensor thickness of 200 $\mu$m. For the baseline of 400 $\mu$m the fluence is unchanged but the dose must be doubled. A side view of the ECal model is presented in Fig. A.1. Results were obtained and plotted for both layers in every bilayer of the ECal in the preshower (PS) region (layers 1 and 2), the bilayers of type A and B (silicon layers 3-4 and 5-6), as well as for the first bilayers of type C (layers 7- 8) and D (layers 25 - 26). All other type C and D bilayers have only their second layer included in the plots. Table 3.7 contains absorber thicknesses and multiplicities for ECal bilayers of the various types mentioned here.

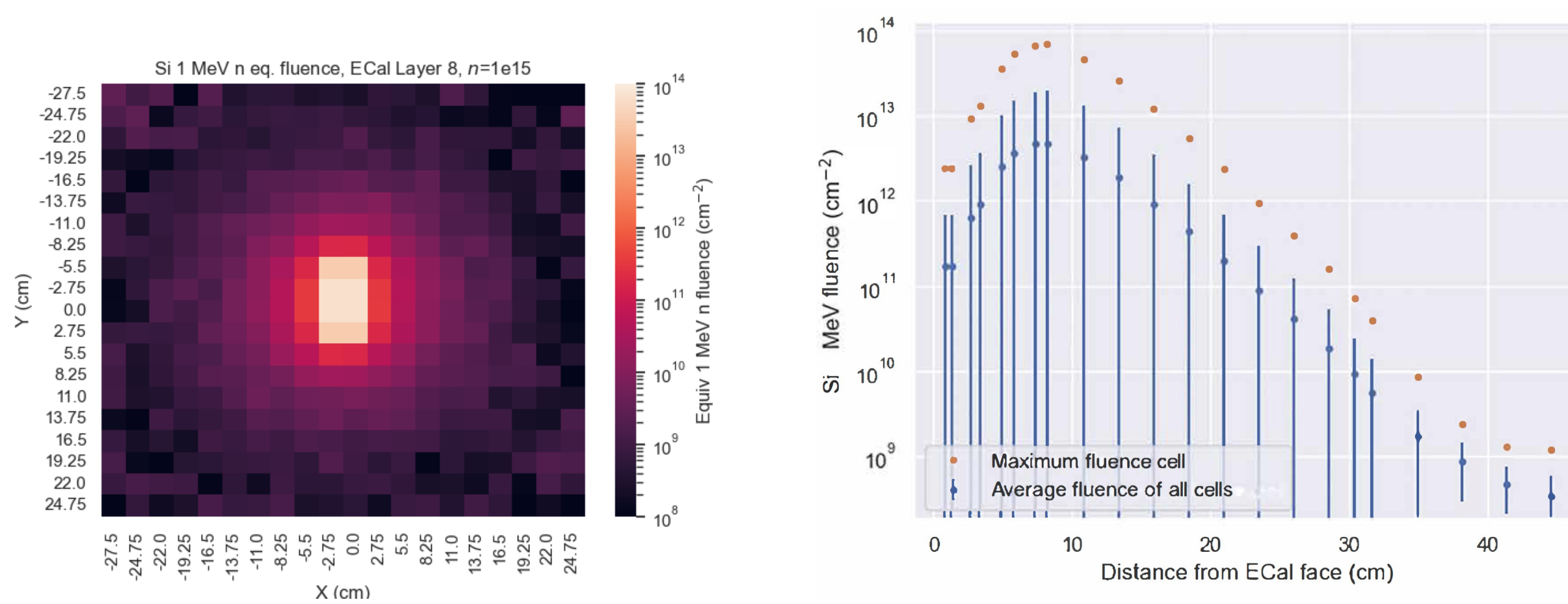

Figure A.3: Left: Silicon 1-MeV-neq fluence across the ECal full sensor plane at shower-max (Layer 8). Right: Maximum and average 1-MeV-neq fluence in individual readout cells for sensor planes versus depth.

The simulation was also extended to include portions of each major section of the HCal nearest to the ECal. The absorber and sensing layers of the back-HCal each consist of a 25 mm thick $Fe$-absorber and a 20 mm thick plastic scintillator, respectively. The side HCal consists of four supermodules similar in structure to the back HCal, with each layer made up of 20 mm thick $Fe$-absorber and 20 mm thick scintillator layers. The side-HCal supermodules are arranged in a pinwheel configuration around the ECal. Fig A.2 displays both front and top views of the combined ECal and HCal model with the dark regions of the HCal being those included in the simulation.

## A.1 ECal Results

Fig A.3 shows a map of the 1-MeV-neq/cm$^2$ fluence across ECal layer 8, the sensing plane at shower-max, as well as a plot of the average and maximum fluences in individual silicon cells as a function of layer depth in ECal. The cell fluence peaks for cells in the central part of the 2 cm $\times$ 8 cm beamspot in each layer, with a maximum value approaching $\sim 10^{14}$ 1-MeV-neq/cm$^2$ in layer 8. The fluence falls dramatically away from the beamspot in the plane and further away from shower-max in either direction along the beam axis. Note that the one standard deviation bars on the average fluence data points extend much farther below than above the average, reflecting the dramatic drop in activity outside of the beamspot region. Fig A.4 maps the fluence in a 10 cm $\times$ 10 cm region containing the beamspot for four sensors centered on the beam axis that receive high fluences in layers 6, 7, 8, and 10.

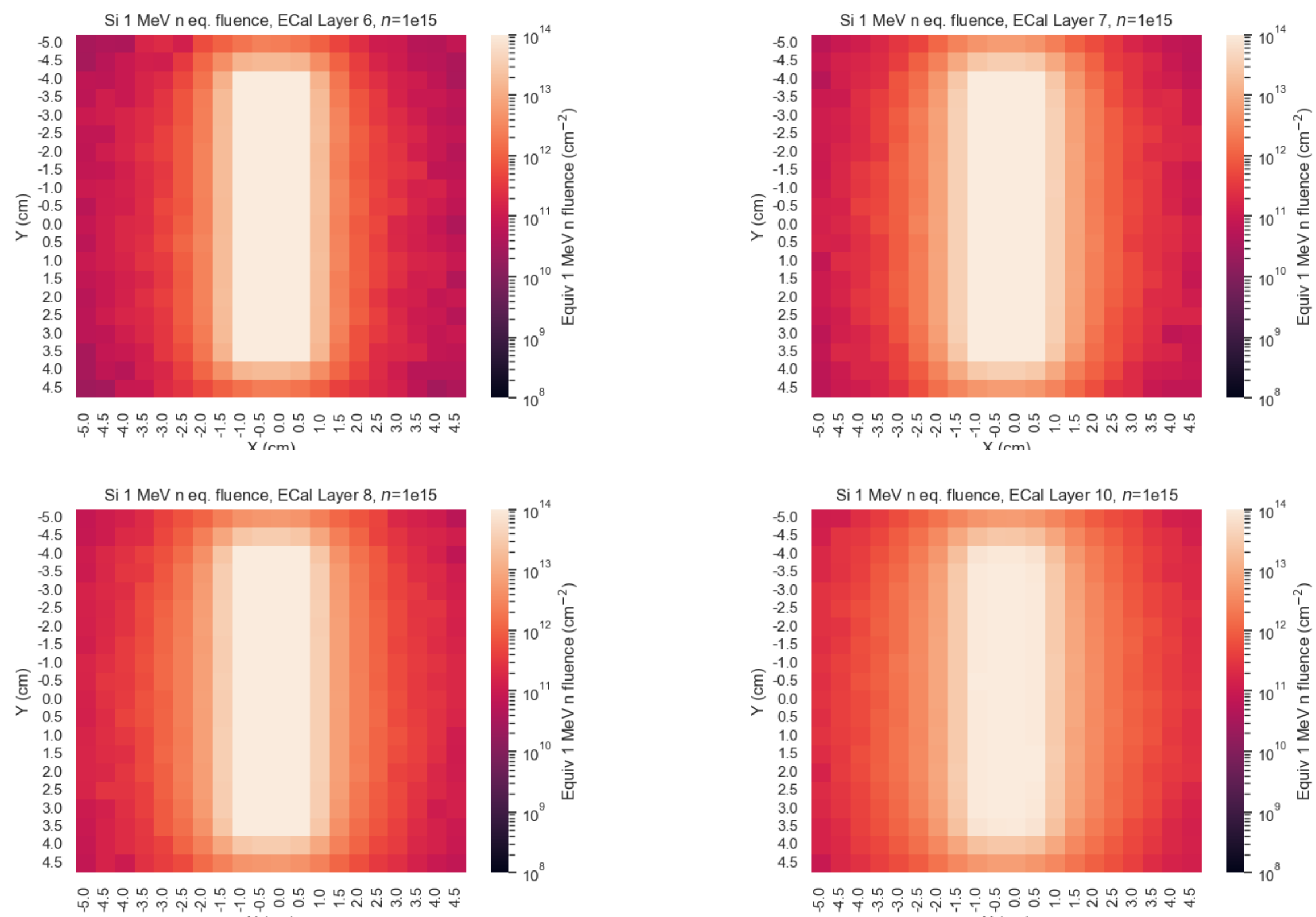

Figure A.4: Silicon fluence maps for 10 cm $\times$ 10 cm regions of sensors centered on the beam axis near shower-max in layers 6, 7, 8 and 10.

Next, the dose in 200 $\mu$m sensors as a function of ECal depth for $10^{15}$ EoT is mapped in Fig A.5. The dose is roughly proportional to the fluence everywhere in the ECal. It is completely dominated by the electromagnetic component with all other components contributing less than 0.1% to the total.

**Implications for the design and operation of ECal**

For high fluences, silicon sensors can incur several types of damage that have implications for performance and operation. Damage to the bulk of the silicon leads to higher leakage current and charge accumulation in surface layers that can alter the field configuration in ways that inhibit charge collection, leading to the need for higher bias voltages. As discussed in Sec. 3.7.3.8 the LDMX ECal sensors will be nearly identical to the CMS HGC 432-cell High Density (HD) sensors. The main difference is that while CMS uses 120$\mu$m and 200$\mu$m active regions, the lower radiation environment of the LDMX ECal allows for 400$\mu$m sensors to be used. This has two positive impacts. First, the path length of particles passing through the sensors, and hence the charge generated for signal, increases linearly with thickness. Second, the capacitance of individual readout cells drops linearly with thickness, reducing the capacitive load on the front-end readout ASIC.

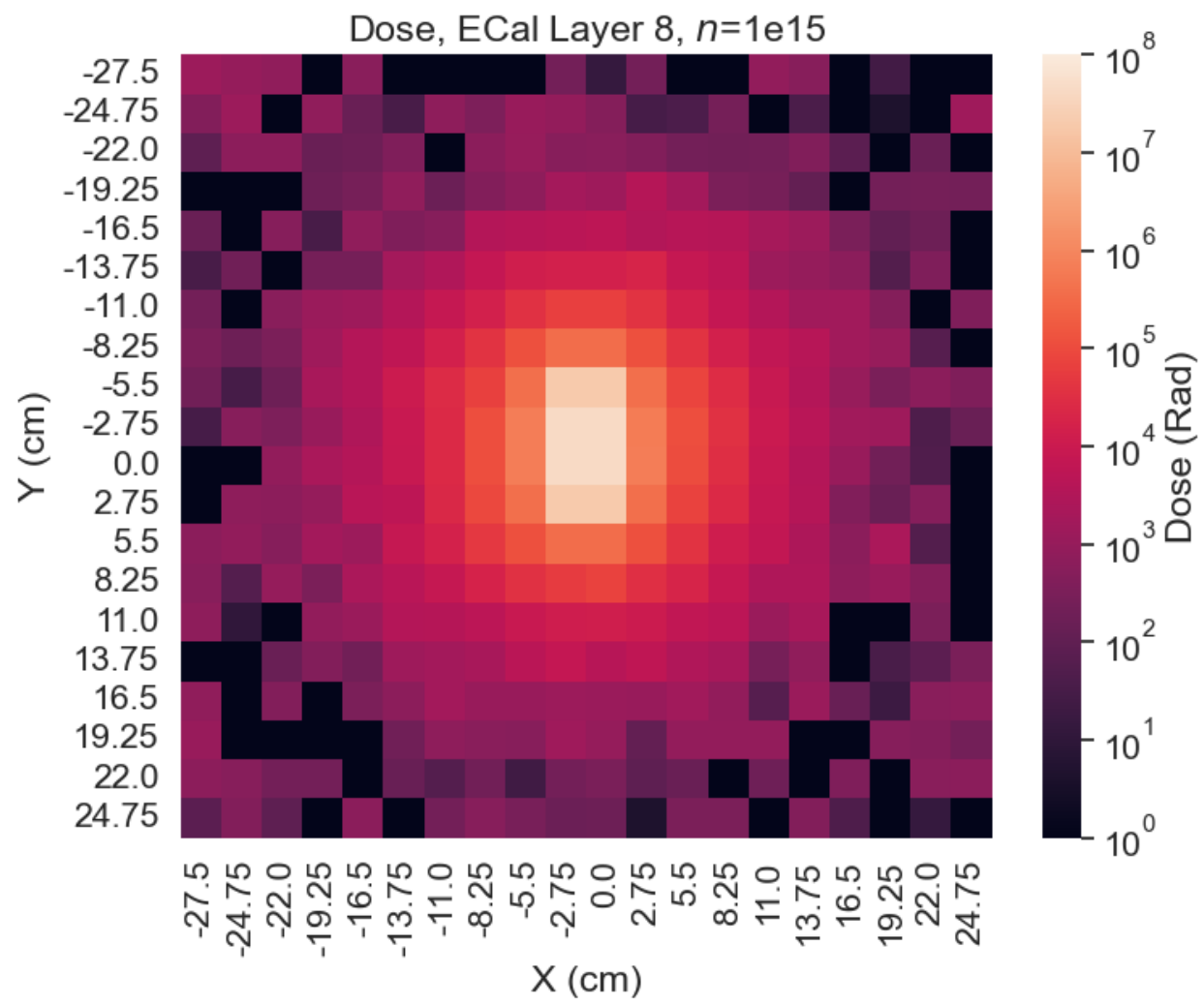

Figure A.5: Silicon dose map for 200 $\mu$m sensors in ECal Layer 8 at $10^{15}$ EoT.

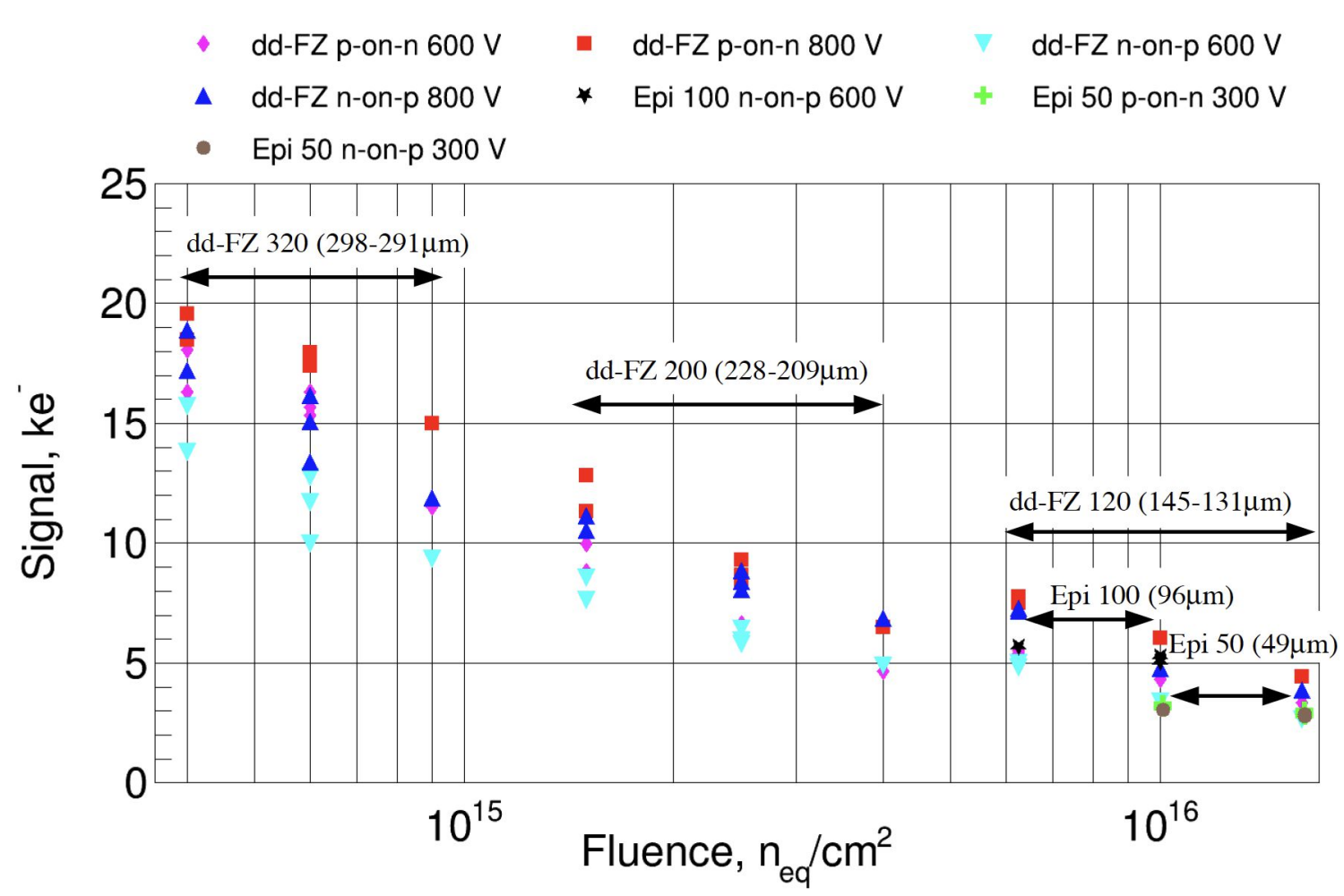

Figure A.6: Fig. 2.1 of the CMS HGC TDR: Signal [ke$^-$] vs. fluence [1-MeV-neq/cm$^2$] for various sensor types, thicknesses, and bias voltages. The lowest fluence plotted at $4\times10^{14}$ is about 4 times the maximum expected in the LDMX ECal.

For LDMX, the CMS HD mask-set will be used with p-type float zone sensors that have both active and physical thickness of 400 $\mu$m. Fig. A.6, taken from the CMS HGC TDR (Fig. 2.1) [180], shows signal [ke-] versus fluence for a variety of sensor types, thicknesses, and bias voltages. The lowest fluence of $4\times10^{14}$ 1-MeV-neq/cm$^2$ is at least 4 times the maximum expected in any cell at shower-max in the LDMX ECal. The black double-arrow lines in the plot indicate the signal charge obtained for unirradiated sensors. For the lowest fluence plotted, it is seen that the charge collected for the dd-FZ p-on-n 320 $\mu$m thick sensors operated at 800V and 600 V for MIPs are around 95% and 85% of their unirradiated values, respectively.

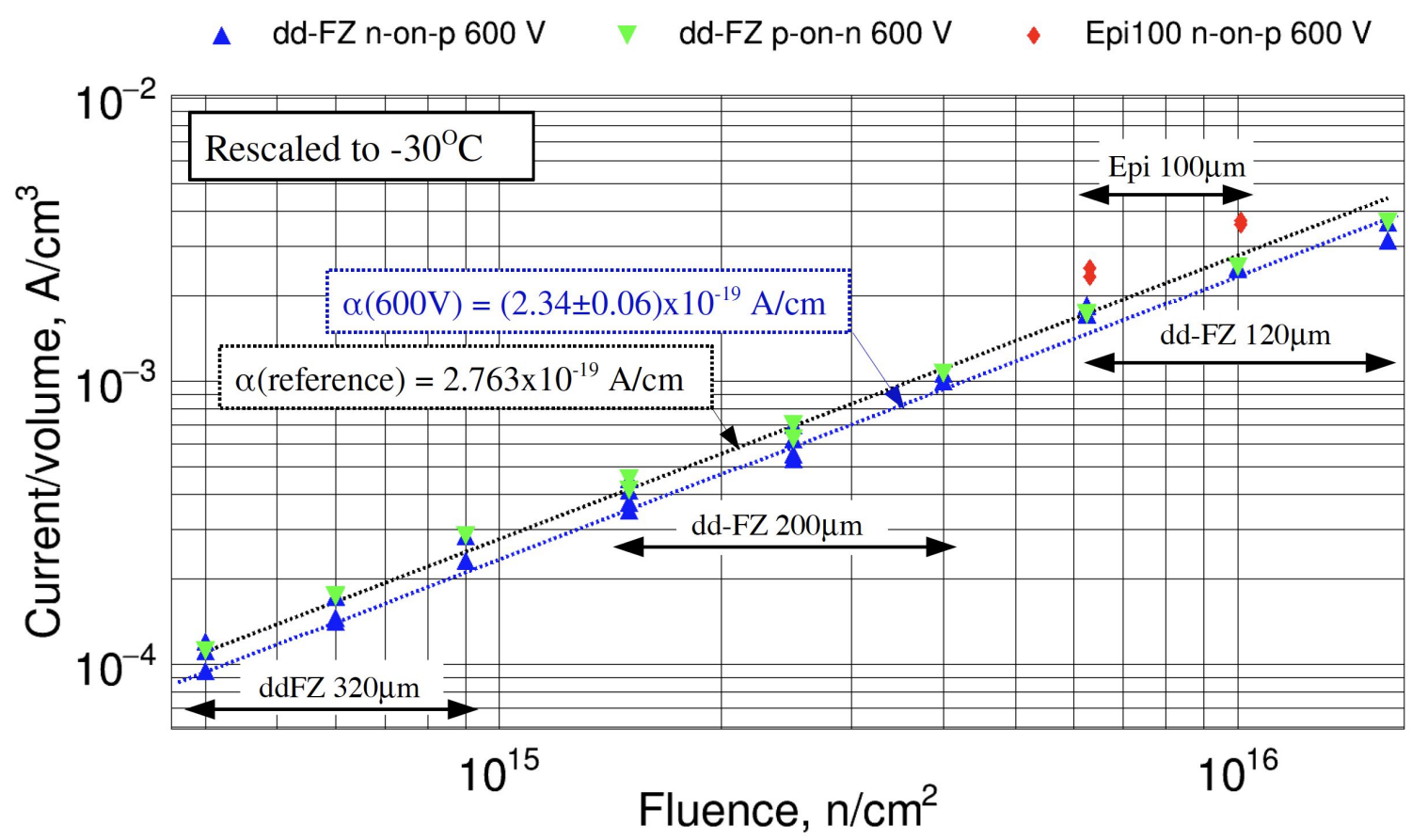

Figure A.7: Fig. 2.2 from the CMS HGC Technical Design Report: Volume current [A/cm$^3$] versus fluence for a variety of sensor types, thicknesses, and bias voltages. The lowest fluence plotted at $4\times10^{14}$ 1-MeV-neq/cm$^3$ is at least 4 times the maximum expected for LDMX.

Extrapolating down to $10^{14}$ 1-MeV-neq/cm$^2$, we could reasonably expect close to 100% charge collection in LDMX for 400$\mu$m sensors.

Fig. A.7, taken from the CMS HGC TDR (Fig. 2.2) [180], shows the leakage current per unit volume of silicon as a function of fluence with the lowest fluence being again at $4 \times 10^{14}$ 1-MeV-neq/cm$^2$, which is at least four times the maximum expected at shower-max in the ECal. Extrapolating down to $10^{14}$ 1-MeV-neq/cm$^2$ we would expect a current density of $\sim 3 \times 10^{-5} A/cm^3$ if operated at -30°C. However, the dark current at -20°C where we plan to operate the ECal, is a factor of 4 higher than at -30°C, canceling the effect of lower fluence. Nevertheless, this is a very low level of dark current.

In summary, the LDMX ECal radiation environment requires the radiation tolerance of silicon but is not expected to lead to significant enough radiation damage that very thin sensors are required. Rather, 400 $\mu$m active depth is a functional choice with higher signal and lower cell capacitance. Moreover, as seen above in Fig. A.3, outside of the ∼10 sensors near shower-max that are located on the beam axis, there is no cell in ECal sensors that this study predicts will reach a fluence above $10^{13}$ 1-MeV-neq/cm$^2$ for $10^{15}$ EoT. This means that if we encounter any performance issues, it will be with only a very small fraction of the silicon modules in the system. In view of this, the ECal is being designed to facilitate swapping out poorly performing layers.

**Radiation Safety**

Finally, in the interest of radiation safety, the radiation doses produced by $10^{15}$ EoT in materials just beyond the ECal were considered. To this end, we added 0.5 cm thick silicon layers on the back and sides of the ECal. The dose observed in these layers is reported in Fig. A.9. As the maximum total dose at each of these locations is under 1 kRad, even after $10^{15}$ EoT, it appears that LDMX will be largely self-shielding.

We then considered the radiation that exits via the ECal face. Another dedicated study scaled to $10^{15}$ EoT was conducted to examine the potential dose from this source. A 2 cm thick volume of water matching the lateral dimensions of the ECal was placed 2 cm in front of the ECal face. The electron beam origin was placed midway between the face and the water layer. As figure A.10 illustrates, the peak dose in the water for radiation coming off the ECal face is 10 kRad - an order of magnitude higher than what was seen on the sides of the ECal but still very low, suggesting that the full area around ECal will be a relatively safe working environment.

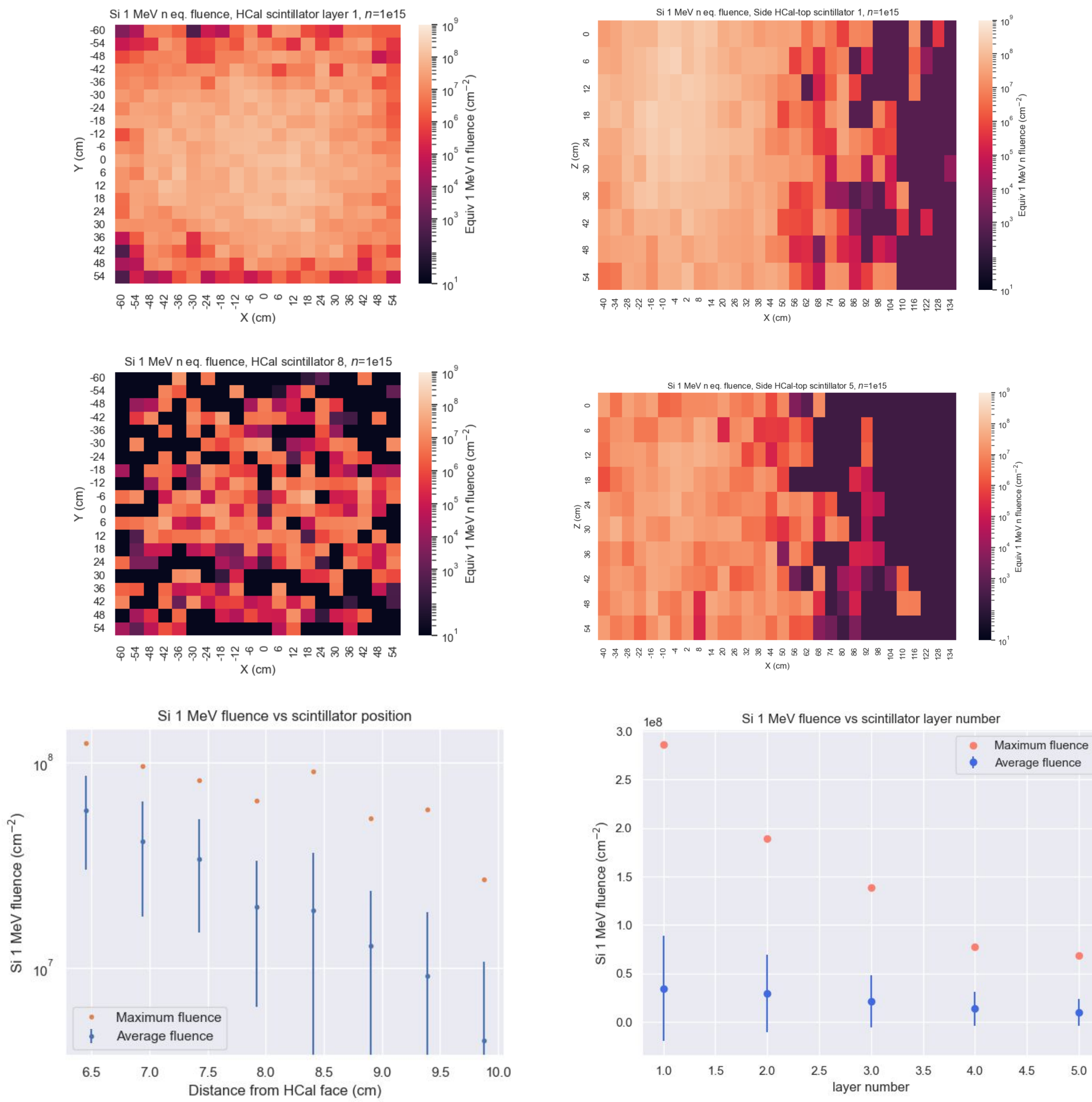

Figure A.8: Scintillator 1-MeV-neq/cm$^2$ fluence maps for the back and top HCal layer 1 (first row), layer 8 for the back and layer 5 for the top HCal (second row) and plots of the peak and average fluences in the back layers 1-8 and top HCal layers 1-5 (third row).

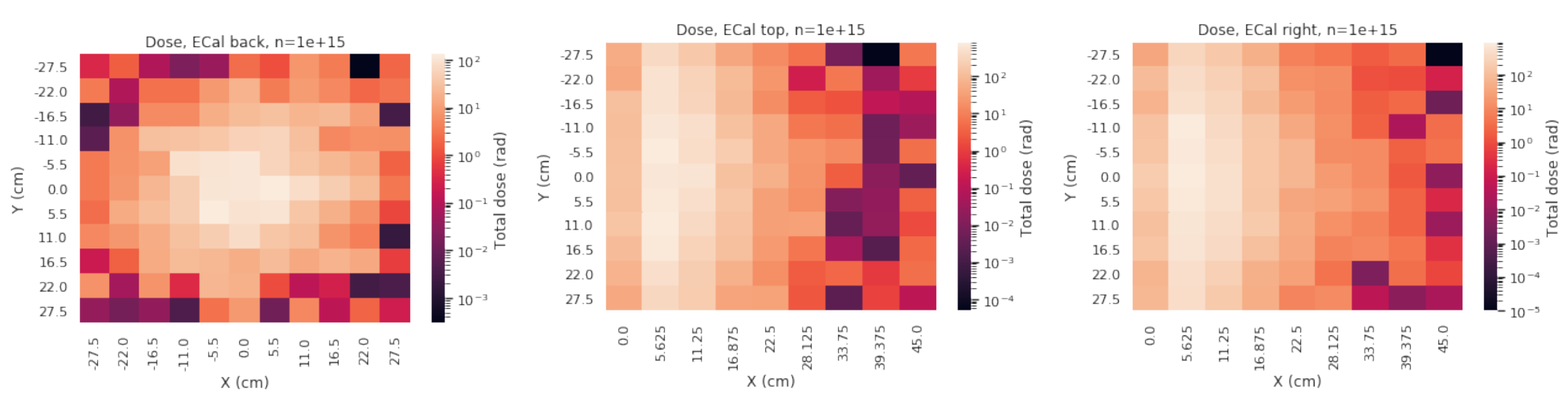

Figure A.9: Dose maps in 0.5 cm silicon thick layers covering the back, top, and side of the ECal used to help assess radiation safety around ECal after $10^{15}$ electrons on target.

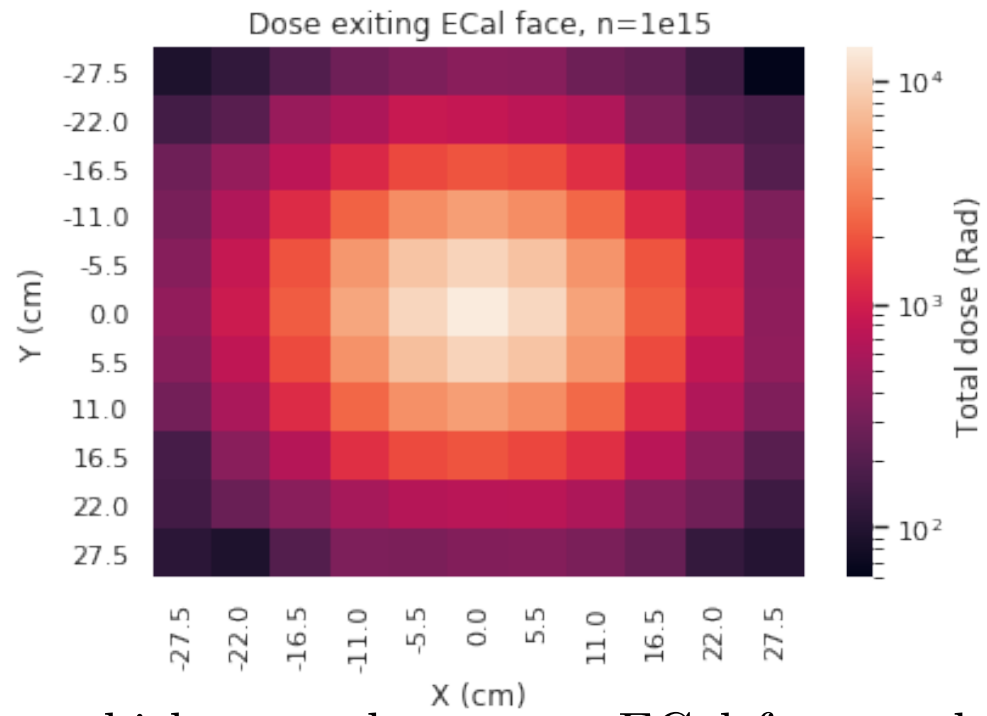

Figure A.10: Dose map in a 2 cm thick water layer near ECal face used to assess radiation safety around ECal after $10^{15}$ electrons on target.

## A.2 HCal Results

As noted above, the simulation of the HCal only includes the regions of the back and side that are closest to the ECal. This is adequate for our purposes because the radiation levels are relatively low and drop rapidly with distance from the ECal as seen in Fig. A.8. The peak fluences for the back and top HCal are around $10^8$ to $10^9$ 1-MeV-neq/cm$^2$ in the first scintillator layers. In the top HCal, the average fluence remains fairly constant while for the back it falls steeply. This is because the top HCal can be close to the shower-max of recoil electrons that have upward trajectories. This is seen even more clearly in Fig. A.11, which shows the dose maps for the first layers of scintillator in the back and top HCal along with the peak and average dose per layer. The dose is extremely low in the back except in the region of the beam spot. The top catches showers for upward going electrons that exit the ECal and enter the top HCal, in some cases near to shower-max. Note that the bright spot in the top HCal is shifted to negative X values as a result of the deflection of recoil electrons by the vertical magnetic field. Peak doses in the HCal are seen to be in the neighborhood of 100 Rad for $10^{15}$ EoT.

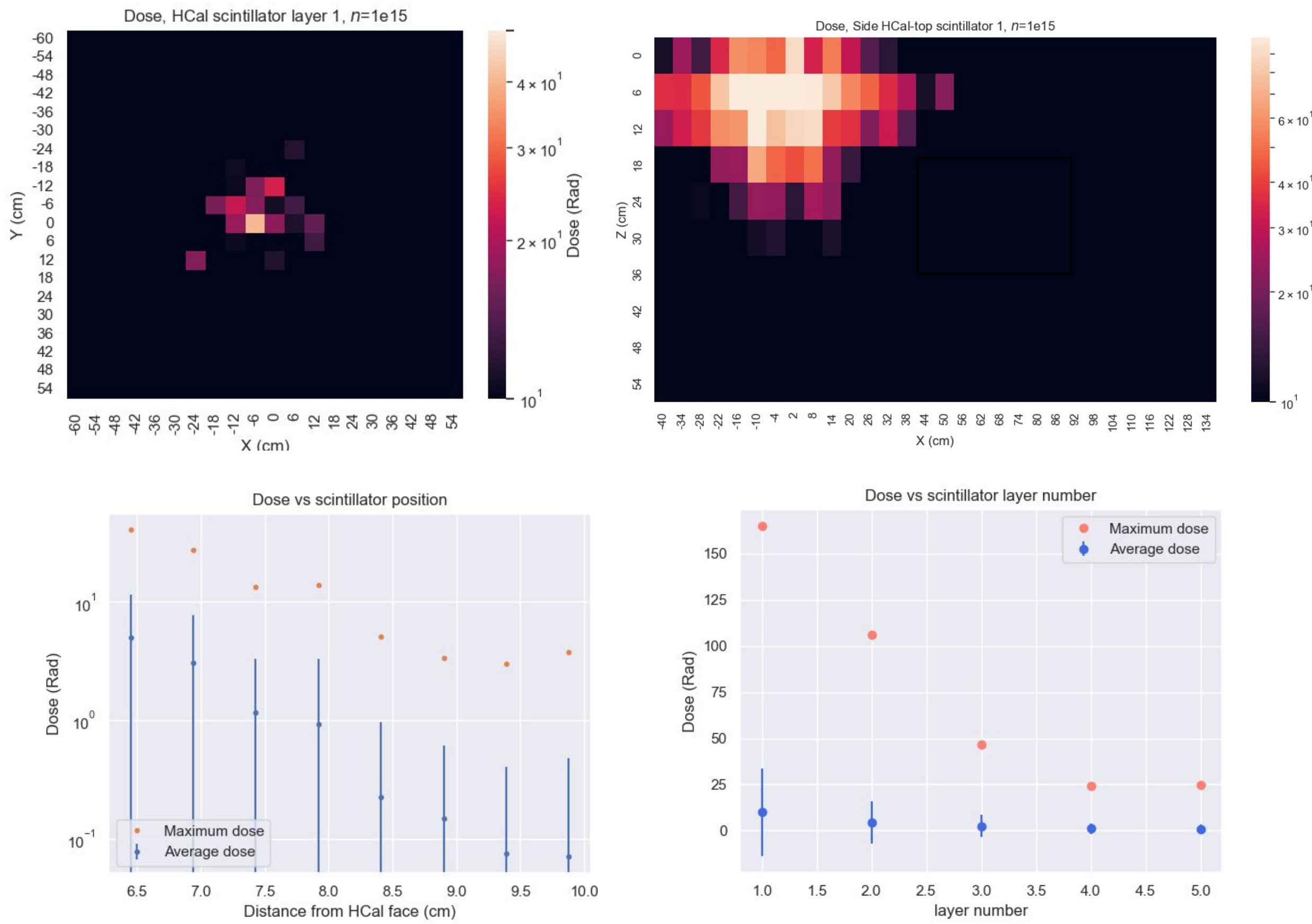

Figure A.11: Scintillator dose maps for layer 1 of the back and top HCal (top row) and plots of the peak and average doses in layers 1-8 of the back and layers 1-5 of the top HCal (bottom row).

# Appendix B

# Details of the Active Target

This section provides details on the active target design and performance.

## B.1 Triplet readout

A novel triplet readout configuration (one SiPM reads out three LYSO bars) has been designed to collect photons produced in the active LYSO target. Twenty-four Hamamatsu S13360-2050VE SiPMs are used to read out central twenty-four LYSO bars in the first layer and twenty-five LYSO bars in the second layer. As shown in Fig. B.1, each SiPM reads three LYSO bars. This design ensures 100% coverage and provides 1.6 mm positioning capability for incident electrons while requiring half the number of photosensors compared to the traditional one-to-one readout configuration.

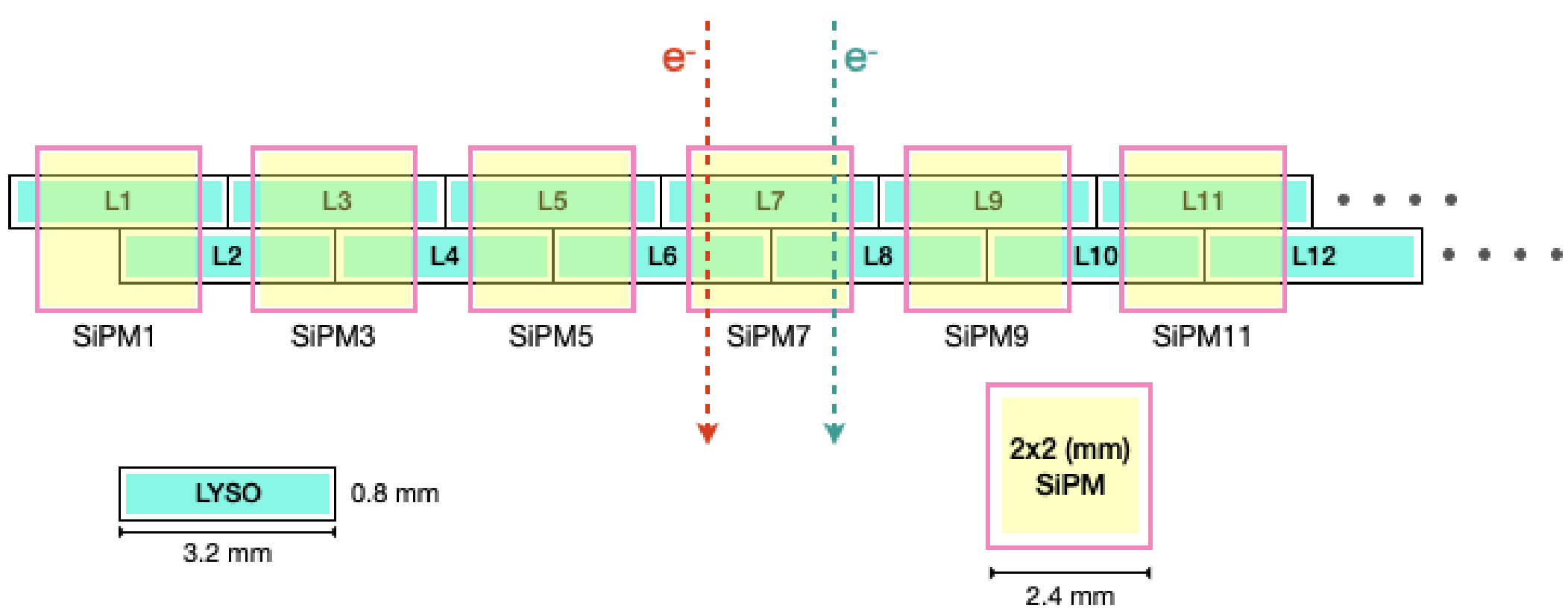

Figure B.1: The first 12 LYSO bars and 6 SiPMs in the proposed active LYSO target.

Using the left-right asymmetry of the triplet readout design, we can determine the position where an electron passes through with 1.6 mm precision (half of the LYSO bar width). When an electron passes through the left half of L7 and the right half of L6 (the red line), the photons generated in L7 will be detected by SiPM7, while the photons generated in L6 will be detected by SiPM5 and SiPM7. SiPM9 will see only background noise or light that leaks from adjacent LYSO bars.

However, when an electron goes through the right half of L7 and the left half of L8 (the green line), the photons generated in L7 will be detected by SiPM7, the photons generated in L8 will be detected by SiPM7 and SiPM9, while SiPM5 will see only background noise or light leaking from adjacent LYSO bars.

The performance of the active LYSO target with this triplet readout design was verified through cosmic ray tests, beam tests at CERN in the fall of 2021 and spring of 2022, and GEANT4 Monte Carlo simulations [181]. The test results show that the active LYSO target with the triplet readout will produce more than 300 photoelectrons (PE) when a minimum ionizing particle passes through the LYSO target, and it can determine the position of the particle with a precision of 1.6 mm (half-width of a LYSO bar), or even 0.2 mm when the particle passes through the gap of 0.2 mm between two adjacent LYSO bars. Furthermore, the 300 PE signal will effectively distinguish a target photonuclear event from a regular MIP event, thereby providing additional rejection of false dark matter candidates.

## B.2 Prototype Performance

The LDMX Collaboration conducted a beam test at CERN in April 2022. The performance of LDMX's prototype trigger scintillator, active LYSO target, and Hadronic Calorimeter (HCal) were studied using electron, muon, pion, and proton beams at various energies ranging from 100 MeV to 4 GeV. Fig. 3.42(a) shows LDMX's prototype in CERN's beam-test area T9. The trigger scintillator prototype and the active LYSO target prototype were combined to form a hybrid TS-LYSO module and positioned directly in front of the first HCal iron plate, as shown in Fig. 3.42(b).

The hybrid TS-LYSO module consists of one layer of 2.0 x 3.0 x 30 (mm) plastic trigger scintillators followed by a double-layer LYSO array. Each array of LYSO bars consisted of six 0.6 x 3.0 x 30.0 (mm) LYSO bars, covered by 0.1 mm ESR on the top and bottom with 0.2 mm gaps between the bars. As no ESR reflector was inserted between the LYSO bars of the hybrid TS-LYSO module, crosstalk was expected due to the light leakage between adjacent LYSO bars.

As illustrated in Fig. B.2, only the top six scintillator bars of each layer were coupled to the SiPMs. Since the width of each trigger scintillator unit is 3.3 mm while the width of the LYSO bar unit is 3.2 mm, some misalignment is introduced during the beam test. We will discuss the effect of this misalignment in the last part of this section.

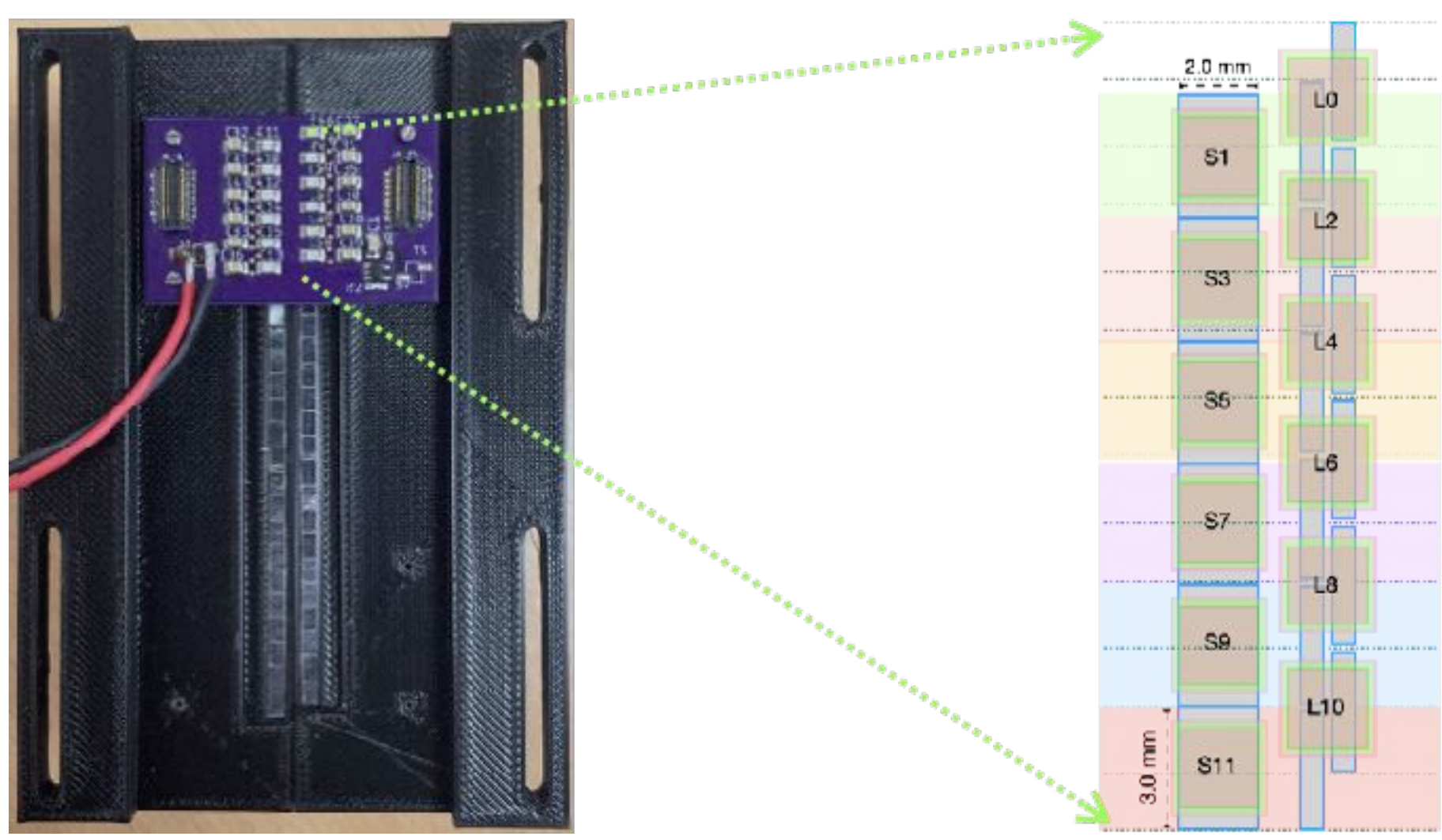

Figure B.2: (a) The hybrid TS-LYSO module used in the Spring 2022 beam test. (b) Illustration of a two-layer LYSO array with a triplet readout design (L0, L2, ..., L10) positioned behind a trigger scintillator array (S1, S2, ..., S11).

The results of the hybrid TS-LYSO prototype obtained from the 4 GeV electron beam test are discussed below. In our study, trigger scintillator signals were used to identify the passage of the electrons and determine which LYSO bars were hit. In total, 20,000 events were collected and analyzed.

Fig. B.3(a) shows the light-yield spectrum of LYSO bar L6 when 4 GeV electrons pass through. It is plotted by applying the following conditions:

1. The plastic trigger scintillator S7 has an MIP signal greater than 18,000 fC (see Fig. B.4).
2. L6 has a signal greater than 5,000 fC, which is one-tenth of the low end of the fitted Landau distribution.
3. L6 signal is greater than the L8 signal, ensuring that only events where electrons hit the upper half of S7 are included.

The spectrum is then fitted with a Landau distribution, and the Most Probable Value (MPV) of the fit is used to estimate the number of photoelectrons (PE). The ADC-to-PE conversion factor of $300 \pm 5$ fC per PE is calibrated using a light-yield spectrum produced by random trigger events, shown in Fig. B.3(b), where the beam electrons pass through other trigger scintillators (S1, S3, S9, S11) that do not overlap L6. As a result, we observed on average 336 PE when 4 GeV electrons hit the LYSO bar L6. This beam test demonstrates that the light yield of the proposed active LYSO target reaches its design goal that the peak of the MIP distribution must be at least 3 $\sigma$ away from the pedestal.

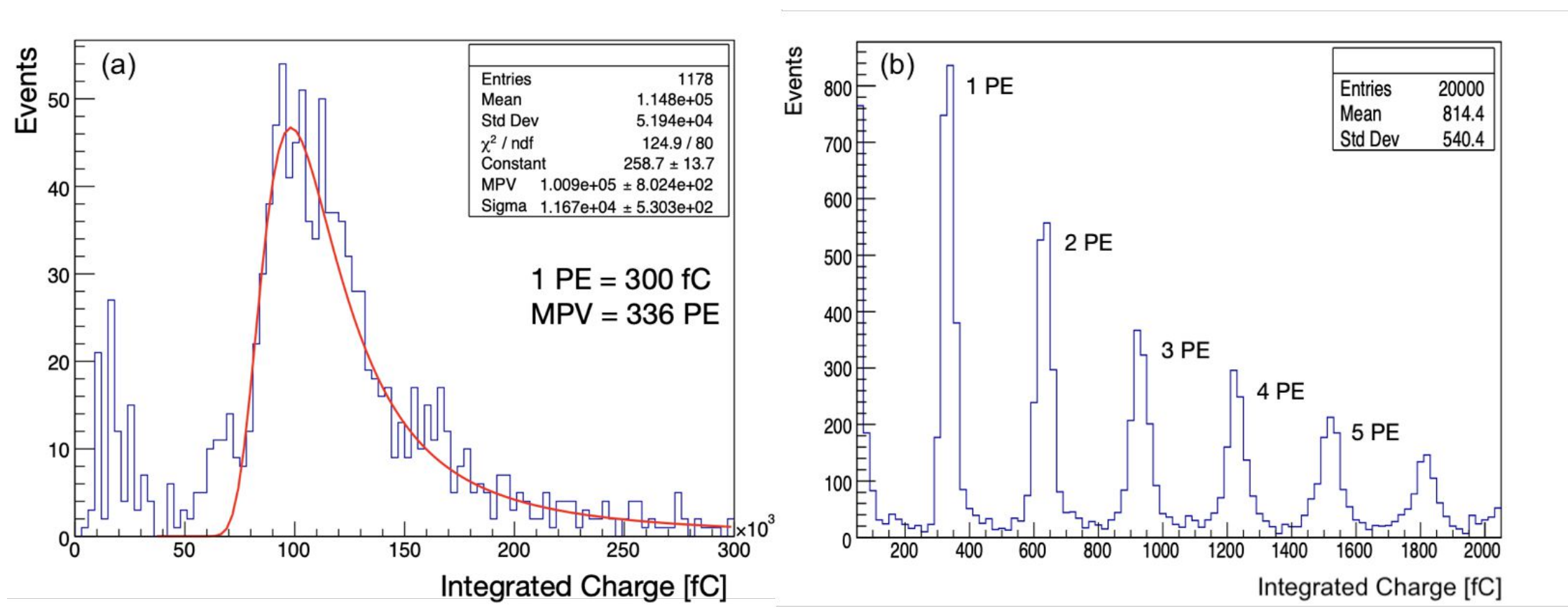

Figure B.3: (a) Light yield spectrum generated by 4 GeV electrons in the LYSO bar L6. The trigger scintillator was used to provide triggers and identify MIP events. (b) Light yield spectrum of random trigger events, which was used for SiPM calibration.

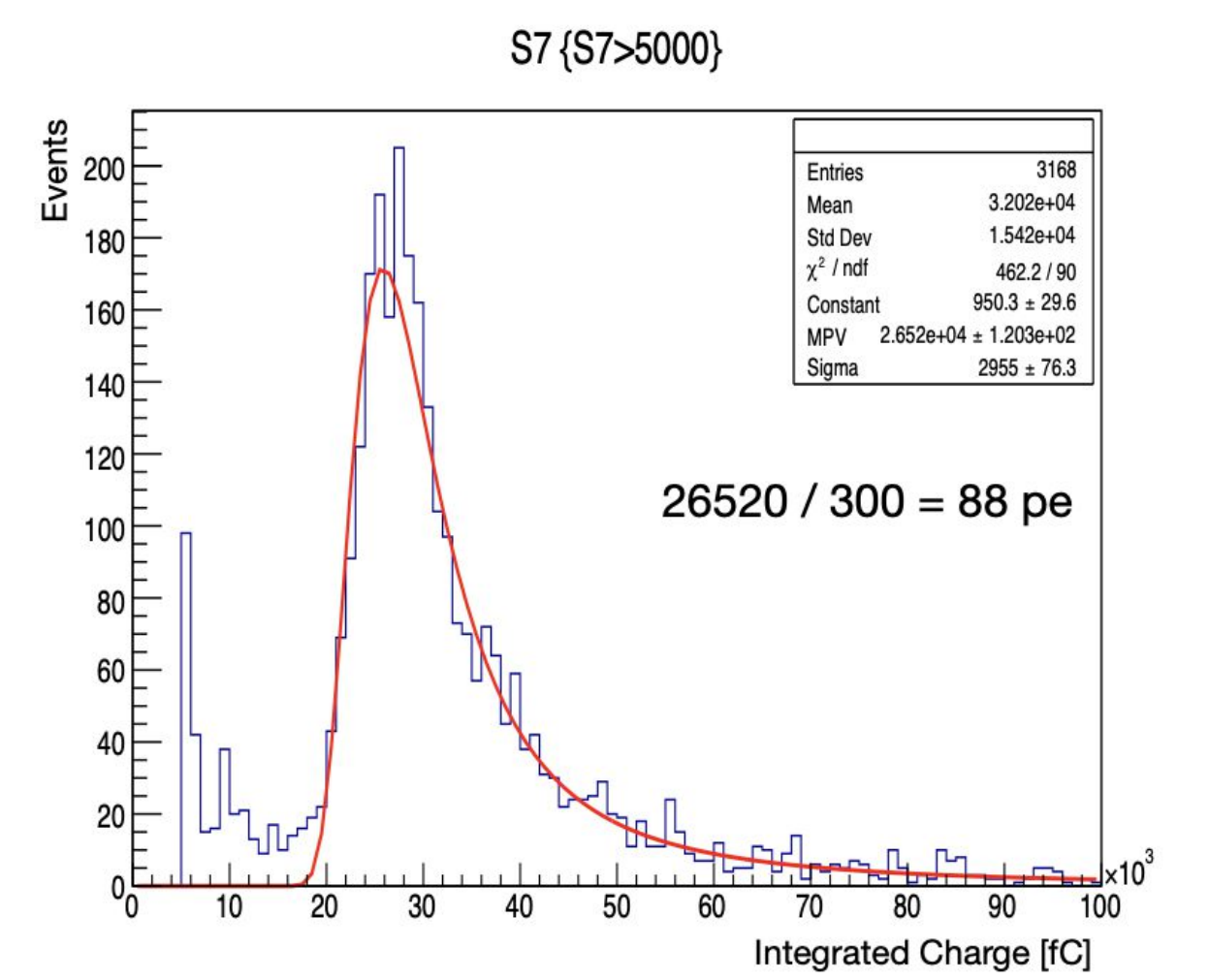

Figure B.4: Light yield spectrum generated by 4 GeV electrons passing through the plastic scintillator S7.

In addition to measuring light yield, we further examine how well the design of a two-layer LYSO target with triplet readout can determine where beam particles pass through. In Fig. B.5(a), the light spectrum of LYSO bar L6 has two bumps when condition (3) of the L6 signal larger than L8 is not imposed. One bump maximizes around 150 PE; the other peaks at about 336 PE. The sources of these two peaks can easily be identified by carefully examining the triplet readout setup. As discussed in the first section, when an electron passes through the lower half of the trigger scintillator S7, photons created in the first-layer LYSO bar will be seen (or shared) by both SiPM L6 and SiPM L8 while the photons generated in the second-layer LYSO bar will only be collected by SiPM L8.

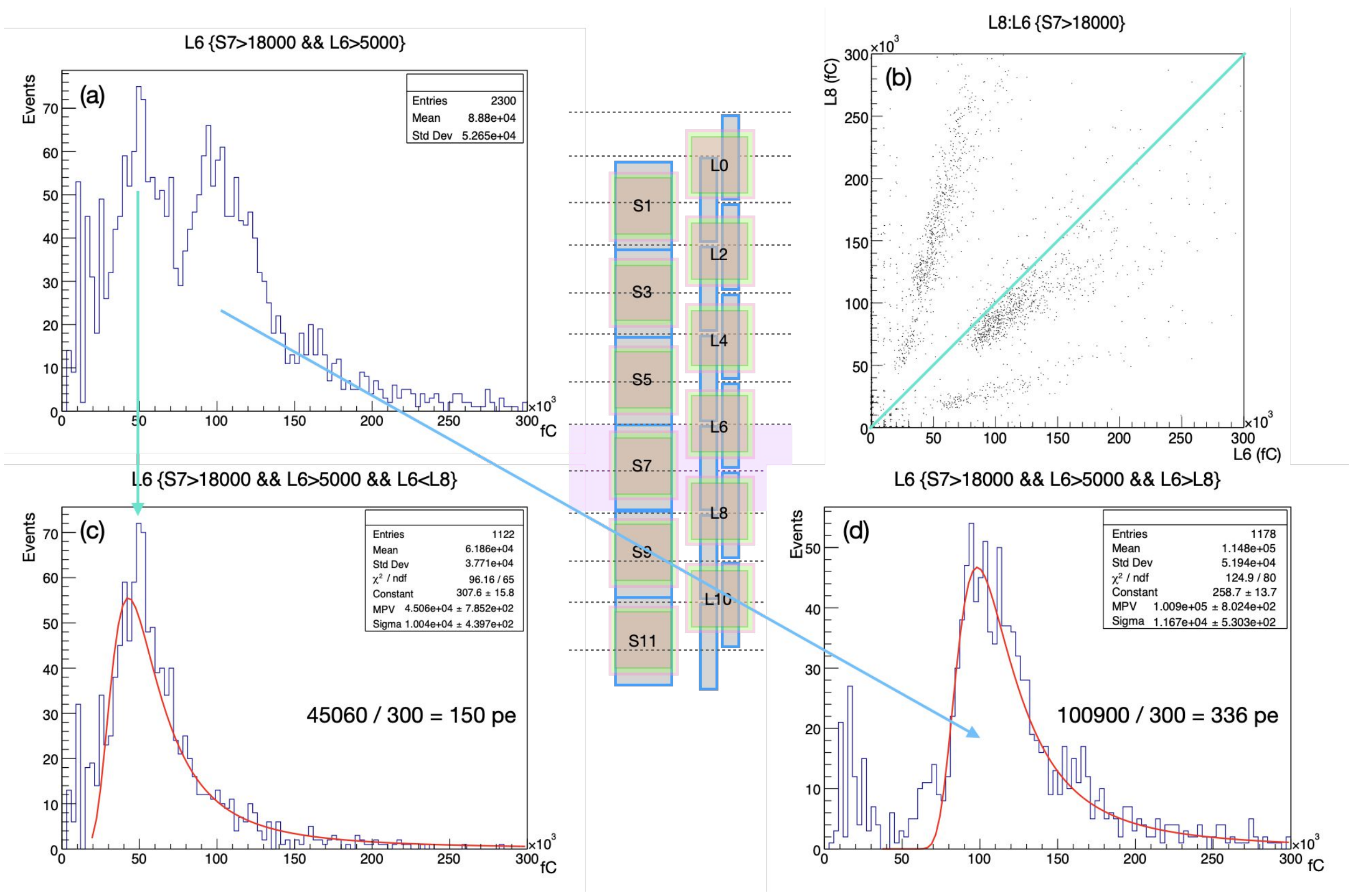

Figure B.5: (a) L6's light yield spectrum without specifying L6's signal > L8's signal. (b) Light yield correlation between adjacent readout L6 and L8. (c) L6's light yield spectrum if L6 < L8. (d) L6's light yield spectrum if L6 > L8.

Fig. B.5(b) plots the light yield correlation between L6 and L8 (L8:L6). Four clusters can be easily identified in the L8:L6 correlation plot.

To better understand positioning capability, light yield performance, and sources of the correlation clusters, the TS-LYSO hybrid module for the beam test is modeled and simulated in GEANT4, as shown in Fig. B.6, to compare with the beam test results. The GEANT4 simulation of the hybrid TS-LYSO module is constructed with the following settings:

1. a modified Birks' law [91] response is used for photon yields in each step,
2. triplex readout's efficiency is simply based on geometric coverage,
3. crosstalk between adjacent LYSO bars is set to 20%,
4. the center of the two-layer 1x6 LYSO array is shifted down by 0.4 mm relative to the center of the 1x6 plastic scintillator array
5. overall photon detection efficiencies are set to be 2.4% for the LYSO and 2.5% for the plastic trigger scintillator

In total, 50,000 simulated `Geant4` events were generated and compared with beam test data. As shown in Fig. B.7(a) and (b), the cluster highlighted in green corresponds to electrons going through the lower gap (the 0.2-mm gap in the second layer directly above L8), while the cluster highlighted in pink corresponds to electrons passing through the upper gap (the-0.2 mm gap in the first layer covered by L6). The position of the two-layer LYSO target relative to the readout SiPM array affects the correlation pattern of two neighboring LYSO bars. This simulation study indicates that the correlation pattern is sensitive to a 0.2 mm gap between the LYSO bars, as shown in Fig. B.7 (c) and (d).

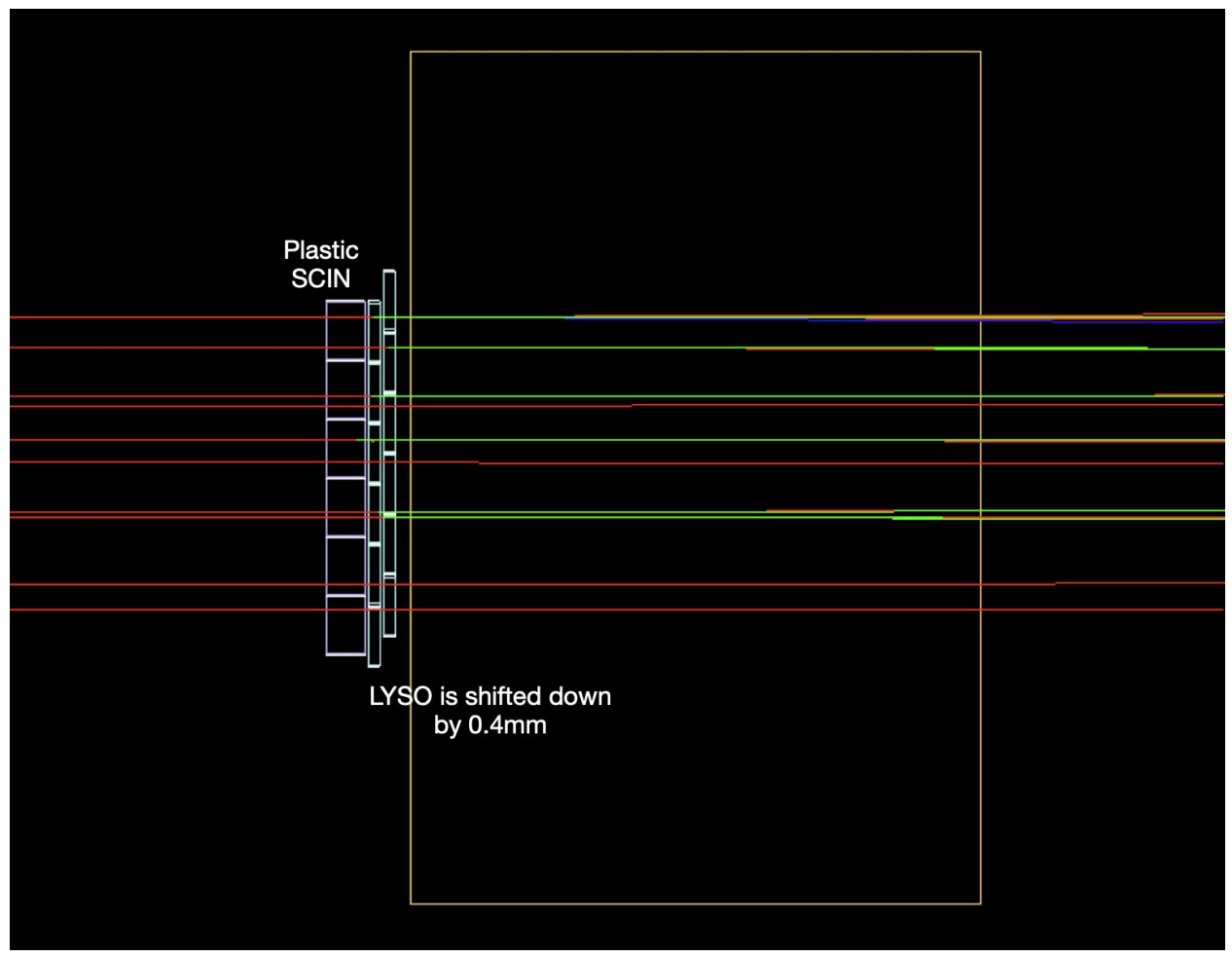

Figure B.6: GEANT4 simulation of the hybrid TS-LYSO prototype for the beam test.

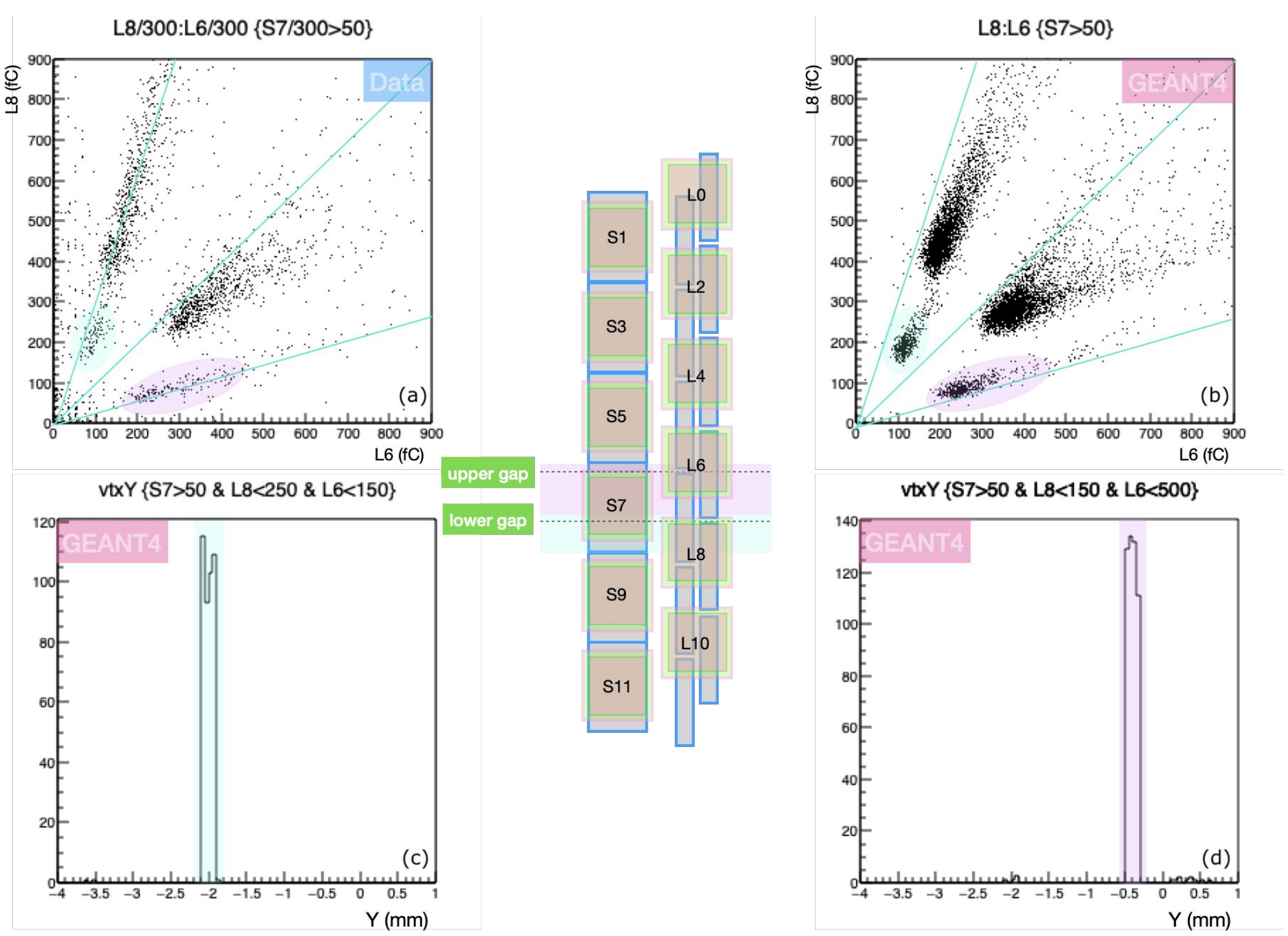

Figure B.7: Comparison of the light yield spectra and L8:L6 correlations of (a) beam test data and (b) GEANT4 simulations. Subfigures (c) and (d) display the corresponding beam positions of the green cluster and the pink cluster, respectively, in the Y direction from the simulation data.

In summary, the active LYSO target achieves its design goals with a radiation length of 10.3%, approximately the same as the passive tungsten target. The prototype studies show that it can (1) collect a sufficient number of photoelectrons ($\geq$ 300 PE) so that the peak of the MIP distribution is more than $3\sigma$ away from the pedestal (2) effectively distinguish target photonuclear events from MIP events thereby providing additional background rejection for the light-dark-matter search, (3) provide a position determination of the incident electron with 1.6 mm precision, and (4) cover the effective acceptance area equivalent to that of the tungsten target.